\documentclass[12pt,preprint]{aastex}

\newcommand{\LX}{\ensuremath{L_{\mathrm{X}}}}

\newcommand{\NH}{\ensuremath{N_{\mathrm{H}}}}

\newcommand{\arcs}{\ensuremath{^{\prime\prime}}}

\newcommand{\erg}{\ensuremath{\mbox{erg}}}

\newcommand{\cm}{\ensuremath{\mbox{cm}}}

\newcommand{\cmsq}{\ensuremath{\cm^2}}

\newcommand{\nm}{\ensuremath{\mbox{\nm}}}

\newcommand{\ps}{\ensuremath{\s^{-1}}}

\newcommand{\s}{\ensuremath{\mbox{s}}}

\newcommand{\ergps}{\ensuremath{\erg~\ps}}

\newcommand{\etal}{{\it et al.\thinspace}}

\newcommand{\CHANDRA}{\emph{Chandra}}

\newcommand{\chisq}{\ensuremath{\chi^2}}
\newcommand{\rchisq}{\ensuremath{\chi^2_\nu}}



\begin{document}

\title{Deep \CHANDRA\ Monitoring Observations of NGC 3379: Catalog of 
 Source Properties  \\}

\author{N. J. Brassington, G. Fabbiano, D.-W. Kim, A. Zezas}
\affil{Harvard-Smithsonian Center for Astrophysics, 60 Garden
Street, Cambridge, MA 02138}
\email{nbrassington@head.cfa.harvard.edu}
\author{S. Zepf, A. Kundu}
\affil{Department of Physics and Astronomy, Michigan State University, East Lansing, MI 48824-2320}
\author{L. Angelini}
\affil{Laboratory for X-Ray Astrophysics, NASA Goddard Space Flight Center, Greenbelt, MD 20771}
\author{R. L. Davies}
\affil{Sub-Department of Astrophysics, University of Oxford, Oxford
OX1 3RH, UK}
\author{J. Gallagher}
\affil{Department of Astronomy, University of Wisconsin, Madison, WI 53706-1582}
\author{V. Kalogera, T. Fragos}
\affil{Department of Physics and Astronomy, Northwestern University, Evanston, IL 60208}
\author{A. R. King}
\affil{Theoretical Astrophysics Group, University of Leicester, Leicester 
LE1 7RH, UK}
\author{S. Pellegrini}
\affil{Dipartimento di Astronomia, Universita di Bologna, Via Ranzani 1, 40127 Bologna, Italy}
\author{G. Trinchieri}
\affil{INAF-Osservatorio Astronomico di Brera, Via Brera 28, 20121 Milan, Italy}
\author{}
\affil{}
\author{}
\affil{}

\shorttitle{Catalog of the X-ray sources in NGC 3379}
\shortauthors{Brassington \etal}
\bigskip

\begin{abstract}
 
We present the properties of the discrete X-ray sources detected in
our monitoring program of the `typical' elliptical galaxy, NGC
3379, observed with {\em Chandra} ACIS-S in five separate pointings,
resulting in a co-added exposure of 324-ks. From this deep
observation, 132 sources have been detected
within the region overlapped by all observations, 98 of which lie within the
$D_{25}$ ellipse of the galaxy. These 132 sources range in \LX\
from 6$\times 10^{35}$ \ergps\ (with 3$\sigma$ upper limit $\le$4$\times 10^{36}$ \ergps) to $\sim2~\times 10^{39}$\ergps,
including one source with \LX$>1~\times 10^{39}$\ergps, which has been
classified as a ULX. 
From optical data, 10 X-ray sources have been determined to be
coincident with a globular cluster, these sources tend to have high X-ray luminosity, with
three of these sources exhibiting \LX$> 1\times10^{38}$\ergps. From X-ray source
photometry, it has been determined that the majority of the 132 sources that
have well constrained colors, have values that are consistent with
typical LMXB spectra. Additionally to this, a sub-population of 10
sources has been found to exhibit very hard spectra and it is expected
that most of these sources are absorbed background AGN.
There are 64 sources in this population that exhibit long-term
variability, indicating that they are accreting compact objects. 5 of
these sources have been identified as transient candidates, with a further 3
possible transients. Spectral variations have also been identified in
the majority of the source population, where a diverse range of
variability has been identified, indicating that there are many
different source classes located within this galaxy.

\end{abstract}

\section{Introduction}

Low-mass X-ray binaries (LMXBs) are the only direct fossil evidence of
the formation and evolution of binary stars in the old  stellar
populations of early-type galaxies. First discovered in the Milky Way
(see Giacconi 1974), these binaries are composed of a compact
accretor, neutron star or black hole, and a late-type stellar
donor. The origin and evolution of Galactic LMXBs has been the subject
of much discussion, centered on two main evolution paths (see Grindlay
1984; review by Verbunt \& Van den Heuvel 1995): the evolution of
primordial binary systems in the stellar field, or formation and
evolution in Globular Cluster (GC). 

With the advent of {\em Chandra} (Weisskopf \etal\ 2000), many LMXB
populations have been discovered in early-type galaxies (see review
Fabbiano 2006), and the same evolutionary themes (field or GC
formation and evolution) have again surfaced, supported and stimulated
by a considerably larger and growing body of data. These {\em Chandra}
observations  have provided important results on the spatial
distributions and X-ray luminosity functions of  LMXB populations
(e.g., Kim \& Fabbiano 2004; Gilfanov 2004), on their average spectra,
and on their association with GCs (e.g. Angelini \etal\ 2001; Kundu
\etal\ 2002; White \etal\ 2002; Sivakoff \etal\ 2006). However,  most of the {\em Chandra}
observations of LMXB systems so far consist of fairly shallow
individual snapshots for each observed galaxy, with limiting
luminosity ($\sim0.3-$8.0~keV) of a few $10^{37}$~erg~s$^{-1}$. These
data give us information on the high luminosity LMXB sources, but do
not cover  the typical luminosity range of the  well studied LMXB
populations of the Galaxy and M31, which extends down a decade towards
dimmer luminosities. 
Moreover, apart from rare exceptions, these
observations do not have the time sampling that would permit
variability studies and the identification of X-ray
transients. Although, from this limited sample of multi-epoch
observations with higher limiting luminosities, already a variety of different
variability behaviours of LMXBs have been observed (e.g. Irwin 2006; Sivakoff \etal\ 2007). 
Both multi-epoch observations and low luminosity thresholds are
important aspects of the observational characteristics of 
Galactic LMXBs and are needed for constraining the evolution of these
populations (e.g., Piro \& Bildsten 2002, Bildsten \& Deloye
2004). For these reasons we proposed (and were awarded) a very large
program of monitoring observations of nearby elliptical galaxies with
{\em Chandra} ACIS-S3.

NGC3379, in the nearby poor group Leo (D=10.6~Mpc (Tonry \etal\ 2001)) was chosen for this study because  is a relatively isolated unperturbed `typical' elliptical galaxy, with an old stellar population (age of 9.3~Gyr, Terlevich \& Forbes 2002) and a poor globular cluster system ($S_{GC}=1.3 \pm 0.7$, Harris 1991; where $S_{GC}$ = No.GC$\times 10^{(0.4(M_V+15))}$). These characteristics make NGC3379 ideal for exploring the evolution of LMXB from primordial field binaries.

Observationally, NGC3379 is an ideal target for LMXB population
studies, because of its proximity, resulting in a resolution of
$\sim30$~pc with {\em Chandra}, and the lack of a prominent hot
gaseous halo, demonstrated by a previous short {\em Chandra}
observation (David \etal\ 2005). These characteristics optimize the
detection of fainter LMXBs, and minimize source confusion; because of
its angular diameter ($D_{25} = 4.6$ arcmin, RC3), NGC3379 is entirely
contained in the ACIS-S3 CCD chip, and is not affected by the
degradation of the {\it Chandra} PSF at large radii. 

Here we publish the catalog of LMXBs with their properties
resulting from the entire observational campaign of NGC3379 (four
observations between January 2006 and January 2007, for a total of $\sim300$~ks), which has been
recently completed, and including the first 30~ks observation taken in
2001, from the {\it Chandra} archive. In the companion paper (Fabbiano
\etal\ 2007) we summarize our results
relative to GC-LMXB associations and discuss their implications
for our understanding of LMXB formation. 

In addition to these two
papers, further highlights from the X-ray binary population of NGC
3379 will be presented in Brassington \etal\ (2008, in prep), where
the properties of the transient population of NGC 3379 will be presented.
Forthcoming papers will also present: the properties of the ULX, the
X-ray luminosity function and the diffuse emission of the galaxy, as well
as the properties of the nuclear source and the intensity and spectral variability
of the luminous X-ray binary population. Preliminary results from the first {\it Chandra} observations
of our program were reported in Kim \etal\ (2006) and Fabbiano \etal\ (2006).

This paper is organized as follows: \S 2. details the
observational program and describes the data analysis methods and
results, including pipeline processing of the data, source detection,
astrometry and matching of sources from the different observations,
X-ray photometry and overall population results, variability analysis
and optical counterpart matching (GC and background objects); \S
3. is the source catalog, including the results from the individual
observations and the co-added data; \S 4 presents the discussion
of the properties of the sources catalog; \S 5 summarizes the conclusions of this work.

\section{Observations and Data Analysis}
\label{sec:process}

The five separate \CHANDRA\ observations of NGC 3379 have been carried
out over a six year baseline, with the first of these,
a 30 ks pointing, being performed in February 2001. This observation has been followed by four
deeper pointings, all carried out between January
2006 and January 2007, resulting in a total exposure time of 337-ks.
 
The initial data processing to correct for the motion of the
spacecraft and apply instrument calibration was carried out with the
Standard Data Processing (SDP) at the {\em Chandra} X-ray Center
(CXC). The data products were then analysed using the CXC CIAO
software suite (v3.4)\footnote{http://asc.harvard.edu/ciao} and
HEASOFT (v5.3.1). The data were reprocessed, screened for bad pixels,
and time filtered to remove periods of high background.
Following the methods of Kim \etal\ (2004a), time filtering was done
by making a background light curve and then excluding those time
intervals beyond a 3$\sigma$  fluctuation above the mean background
count rate, where the mean rate was determined iteratively after
excluding the high background intervals. This resulted in a total
corrected exposure time of 324-ks, the log of these exposures is
presented in Table \ref{tab:log}.

From the five individual data sets, a combined observation has been
produced. This has been created by using the {\em merge\_all }
script\footnote{http://asc.harvard.edu/ciao/ahelp/merge\_all.html}, where
the reprocessed level 2 event files from each observation were
reprojected to a given RA and Dec, and then combined, and a combined exposure
map was also created. The methods that were applied to correct for the
astrometry of
the individual observations, used to create the co-added observation,
are discussed in \S \ref{sec:astrometry}.

\begin{figure}
\begin{centering}
  \begin{minipage}{0.7\linewidth}
  \vspace{0.7cm}
\includegraphics[angle=-90,width=\linewidth]{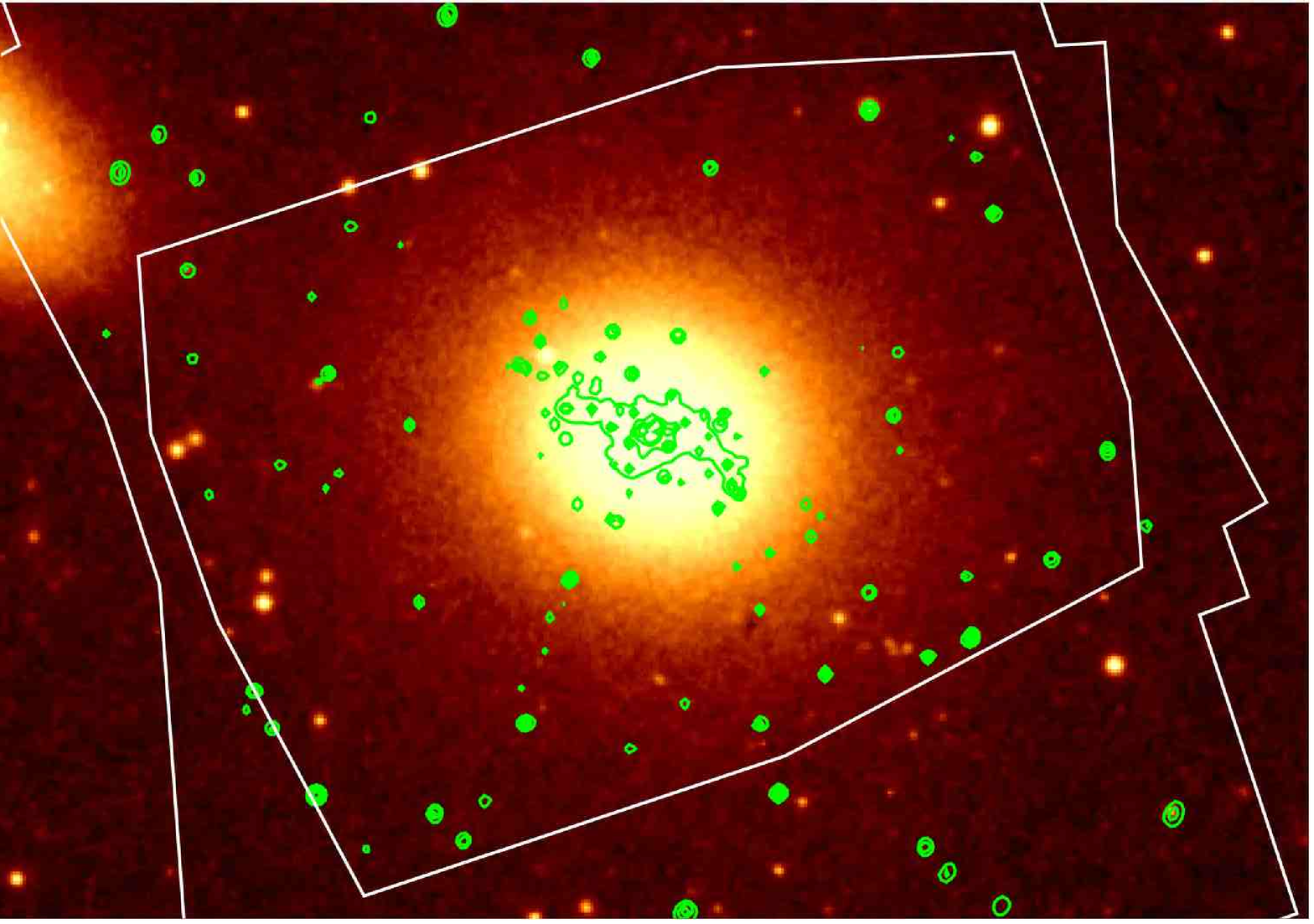}
\end{minipage}
\vspace{1cm}

  \begin{minipage}{0.7\linewidth}

\includegraphics[angle=-90,width=\linewidth]{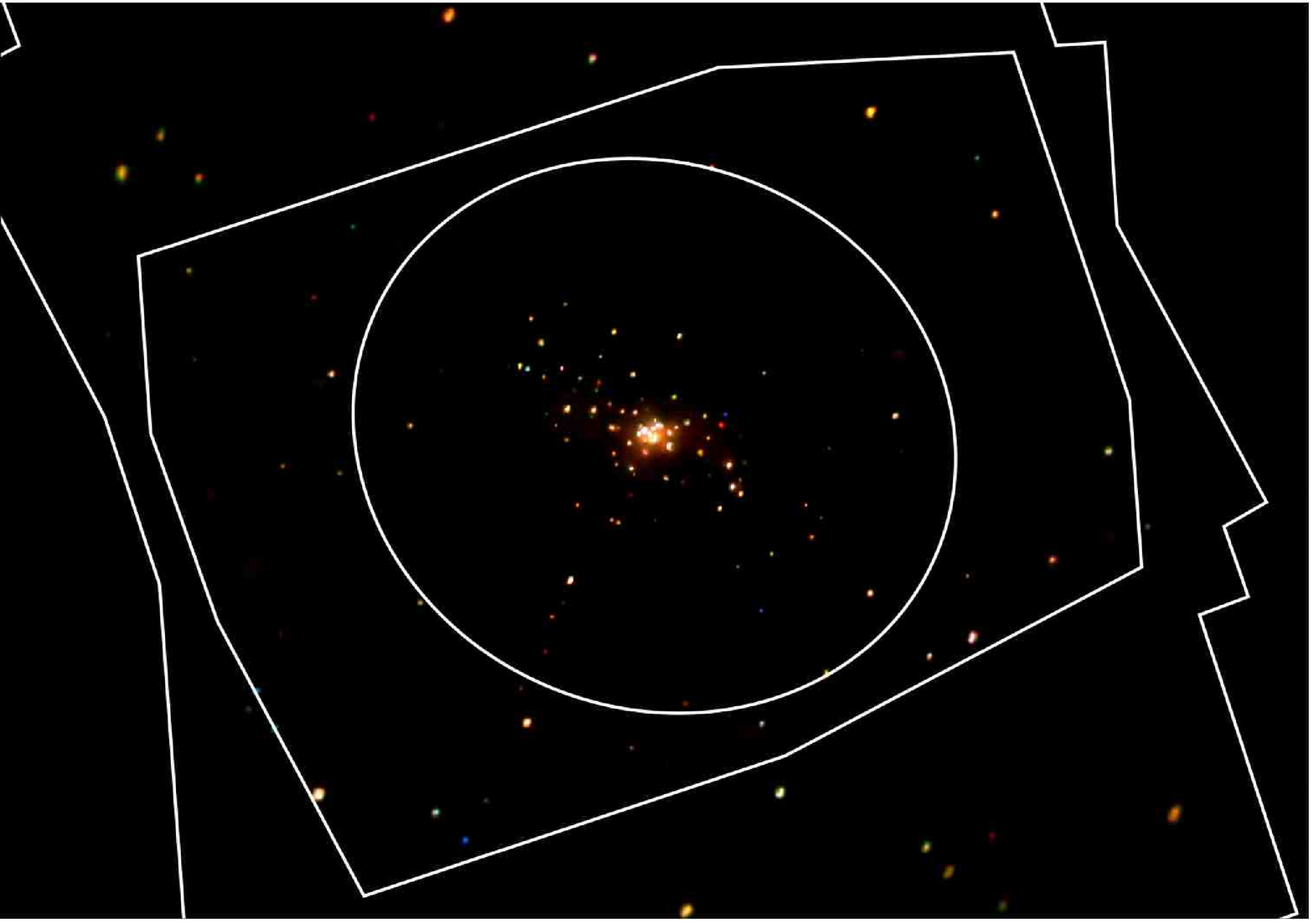}
\end{minipage}
\caption{Top: An optical image of NGC 3379, with full band, adaptively
smoothed, X-ray contours overlaid. Also shown is the outline of the
total area covered by the ACIS-S3 chips, from the five separate
observations, and the smaller region overlapped by all five
of the pointings. Bottom: A `true color' image of the galaxy, where red
corresponds to 0.3$-$0.8 keV, green to 0.8$-$2.5 keV and blue to
2.5$-$8.0 keV. The $D_{25}$ ellipse of this galaxy, the coverage from all
observations and the overlapping region are also shown.}\label{fig:image}
\end{centering}
\end{figure}

From this combined dataset, a
0.3$-$8.0\,keV (from here on referred to as `full band') {\em Chandra}
image was created and adaptively smoothed
using the CIAO task {\em csmooth}. This uses a smoothing kernel to
preserve an approximately constant signal to noise ratio across the
image, which was constrained to be between 2.6$\sigma$ and 4$\sigma$. In Figure
\ref{fig:image}, both the optical image, with the
full band X-ray contours overlaid (top), and the `true color' image
of the galaxy system (bottom) are shown. The `true color' image was
created by combining three separate smoothed, and exposure corrected,
images in three energy bands; 0.3$-$0.9 keV, 0.9$-$2.5 keV and
2.5$-$8.0 keV, using the same smoothing scale for each image. These
energy bands correspond to red, green and blue respectively. 

\subsection{Source Detection and Count Extraction Regions}
\label{sec:src_props}

Discrete X-ray sources were searched for over each observation (the
five single observations and the combined observation) using the
CIAO tool {\em wavdetect}, where the full
band, with a significance threshold parameter of 1$\times10^{-6}$,
corresponding to roughly one spurious source over one CCD, was searched
over. This CIAO tool searches for localized
enhancements of the X-ray emission, and does not set any apriori
thresholds on the SNR of each source (in contrast to sliding cell
algorithms: Freeman \etal\ (2002)). In Kim \etal\ (2004a) simulations were carried out to investigate the number of false detections compared to the expected $\sim$1 false source per image provided by the threshold significance of 1$\times10^{-6}$. These simulations and results are detailed in \S4.4.1 of the paper, where they find that the performance of {\em wavdetect} is as expected, resulting in $\sim$1 spurious source per images. These simulations cover the values of background ($\sim$0.2 counts/pixel) of our co-added observation. Further to this, these simulations were compared to \CHANDRA\ observations with relatively long exposures ($\sim$100 ks), where 0.3 spurious sources per exposure were detected, fully consistent with the simulation results. A similar approach was also used by Kenter \etal\ (2005).

Following the results of these simulations it is clear that, when setting a detection threshold of 1$\times10^{-6}$ in {\em wavdetect}, only one of the formally identified sources is expected to be a false detection per image and so this prescription has been followed here. 
We reiterate that using this method does not set any apriori thresholds on the SNR of each source, it is therefore possible to include sources that have a high detection significance but at the same time a low flux significance, or SNR, therefore resulting in sources with poorly constrained flux.

When running {\em wavdetect} a range of 1, 2, 4, 8,
16 and 32 pixel wavelet scales were selected (where pixel width
is 0.49\arcs), with all other parameters set at the default
values. Exposure maps were created for the S3 chip from each
observation, at 1.5 keV. The {\em wavdetect} tool was used in
preference to other source detection software, as this detection
package can be used within the low counts regime, as it does not
require a minimum number of background counts per pixel for the
accurate computation of source detection thresholds. Further to this,
{\em wavdetect} also performs better in confused regions, which is the
case in the nuclear region of elliptical galaxies (Freeman \etal\ 2002).

Once the X-ray sources had been detected, and their position had been
determined by {\em wavdetect}, counts were extracted from a circular
region, centered on the {\em wavdetect} position, with background counts determined
locally, in an annulus surrounding the source, following the
prescription of Kim \etal\ (2004a). The extraction radius 
for each source was chosen to be the 95\% encircled energy radius at
1.5 keV (which varies as a function of the off-axis angle\footnote{see
http://cxc.harvard.edu/cal/Hrma/psf/index.html}), with a
minimum of 3\arcs\ near the aim point. Similarly, background counts for
each source were estimated from a concentric annulus, with inner and
outer radii of two and five times the source radius respectively. 

When nearby sources were found within the background region, they were
excluded before measuring the background counts. Net count rates were
then calculated with the effective exposure (including vignetting) for
both the source and background regions. Errors on counts were derived
following Gehrels (1986). For cases where sources have fewer than 4 counts, the Gehrels
approximation begins to differ to Poissonian errors. However,
these error values are still accurate to 1\%, and, if anything,
provide a more conservative estimate as Gehrels approximation
does not account for the smaller error value at the lower limit.
When the source extraction regions of nearby
regions were found to overlap, to avoid an overestimate of their
source count rates, counts were calculated from a pie-sector, excluding
the nearby source region, and then rescaled, based on the area ratio
of the chosen pie to the full circular region. Once the correction
factor was determined, it was applied to correct the counts in all
energy bands. For a small number of sources that overlapped with
nearby sources in a more complex way (e.g. overlapped with more than 2
sources), instead of correcting the aperture photometry, the source
cell determined by {\em wavdetect} was used to extract the source
counts.

From these source counts, fluxes and luminosities were calculated in
the 0.3$-$8.0 keV band, with an energy conversion factor (ECF)
corresponding to an assumed power law spectral shape, with $\Gamma$ =
1.7 and Galactic \NH\footnote{\NH=2.78$\times10^{20}$\cmsq\ (from
COLDEN: http://cxc.harvard.edu/toolkit/colden.jsp).} (see Figure
\ref{fig:cc_pop} for a justification of this assumption). The ECF was
calculated with the {\em arf} (auxiliary 
response file) and the {\em rmf} (redistribution matrix file) generated for
each source in each observation. For each source, the temporal quantum
efficiency variation
\footnote{See http://cxc.harvard.edu/cal/Acis/Cal\_prods/qeDeg/
 for the low energy QE degradation.} was accounted for by calculating
the ECF in each observation and then taking an exposure-weighted mean
ECF. The ECF over the 0.3$-$8.0~keV band varied by $\sim$14\% between
2001 and 2006, and by only $\sim$0.3\% between the four observations
taken in 2006\footnote{http://asc.harvard.edu/ciao/why/acisqedeg.html}. This procedure was applied to each single observation
and to the total co-added exposure. 

In the instances where {\em wavdetect} did not formally identify a
source in a single observation, source counts have been extracted from a
circle with a 95\% encircled energy radius, centered on the position
from the co-added observation (or in cases where the source was not
formally detected in the co-added observation, the source position from
the single observation was used). The definition of background regions and the treatment
of overlapping sources are outlined above. From these extracted source counts, a Bayesian
approach, developed by Park \etal\ (2006), has been used to provide
68\% source intensity upper confidence bounds on the full band counts. These values have
then been used to calculate upper limits on the flux and luminosity of
these sources.

\subsection{Astrometry and Source Correlation}
\label{sec:astrometry}

Prior to merging the five exposures into a single co-added observation,
the astrometry of the individual pointings were checked, to correct
for any systematic shifts in the coordinate systems of the event
lists. This was done by selecting the longest single observation, obs-7073, as a
point of reference, and then comparing the positions of the twenty brightest
point sources, detected in all five separate observations, to the
source positions of these 20 sources in each of the individual observations.  From
these comparisons it was found that the
fifth observation, obs-7076, showed a significant declination offset of
0.44\arcs, compared to values less than 0.2\arcsec\ in all other
observations. It is believed that this systematic offset is a result of a
change in thermal environment, following the procedure to cool the
ACA CCD from -15C to -19C, which took place between late Nov-2006 and
early Jan-2007\footnote{http://cxc.harvard.edu/cal/ASPECT/celmon/}.

At the time of data processing there were no calibration files to
correct for this offset and therefore offsets had to be defined and
corrected for individually\footnote{Data between these dates have now
been reprocessed to correct for this offset, see:
http://cxc.harvard.edu/bulletin/bulletin\_64.html}.
This was done by producing a co-added file of the first four
individual observations, all of which show offset values of
$\le$0.2\arcs\ in both RA and Dec from the single reference
observation. With this co-added observation, the CIAO tool {\em
reproject\_aspect} was used calculate offset values for the fifth
observation, and from these, produce a new, corrected, aspect
solution file, which was used to create a new
level 2 event file. This new corrected event file,
alongside the four other individual observations, was then combined to
produce a co-added observation, using the {\em merge\_all} script, as
described in \S \ref{sec:process}. From this co-added file the
astrometry was once again checked, a summary of
these offset values is given in Table \ref{tab:pointings},
where it can be seen that all five of the individual observations have
offsets of $\le$0.12\arcs\ in both RA and Dec when compared to the corrected,
co-added observation.

\begin{figure}
\begin{centering}

   \begin{minipage}{0.48\linewidth}

  \includegraphics[width=\linewidth]{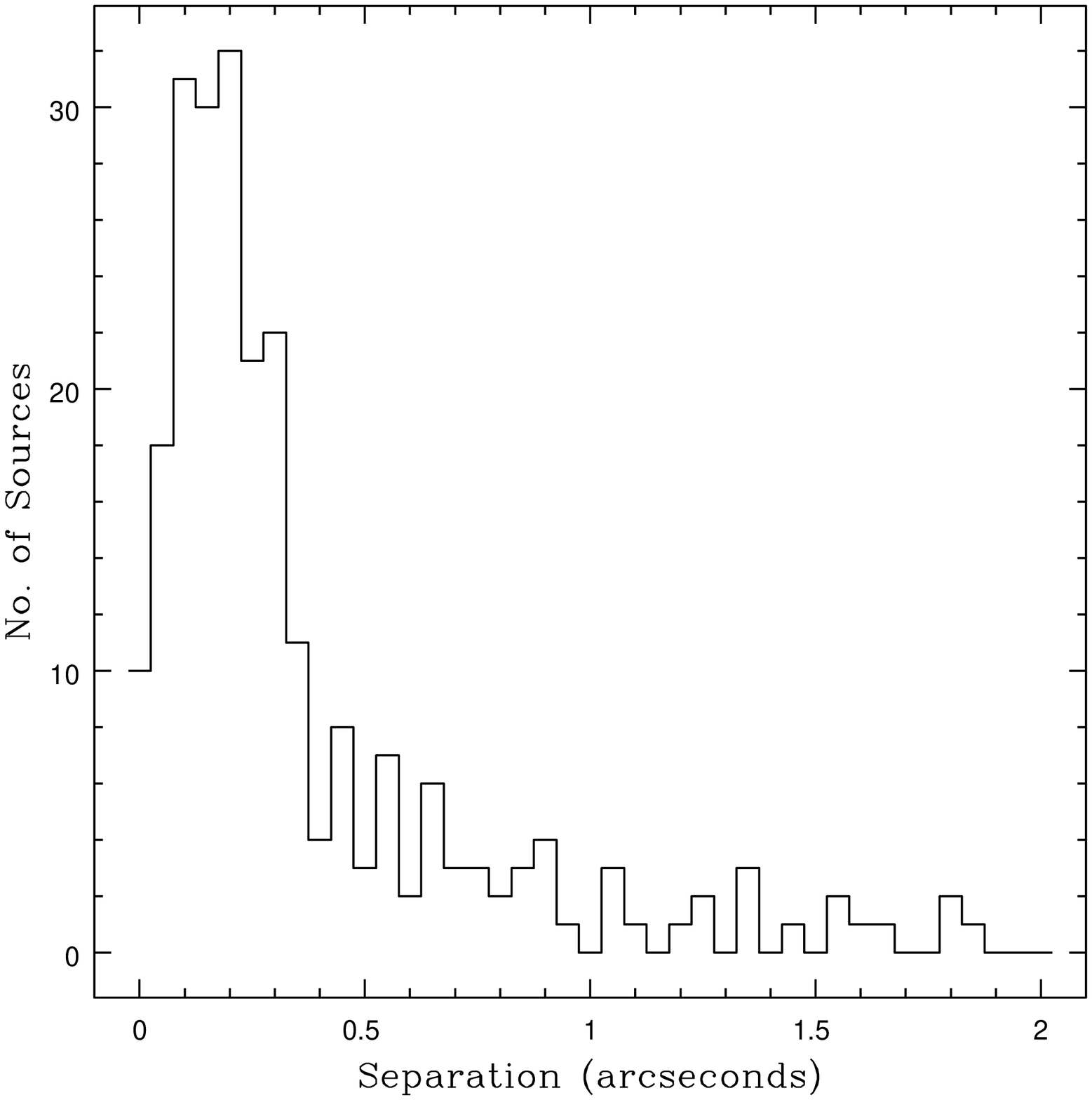}
 
  \end{minipage}
  \begin{minipage}{0.48\linewidth}

  \includegraphics[width=\linewidth]{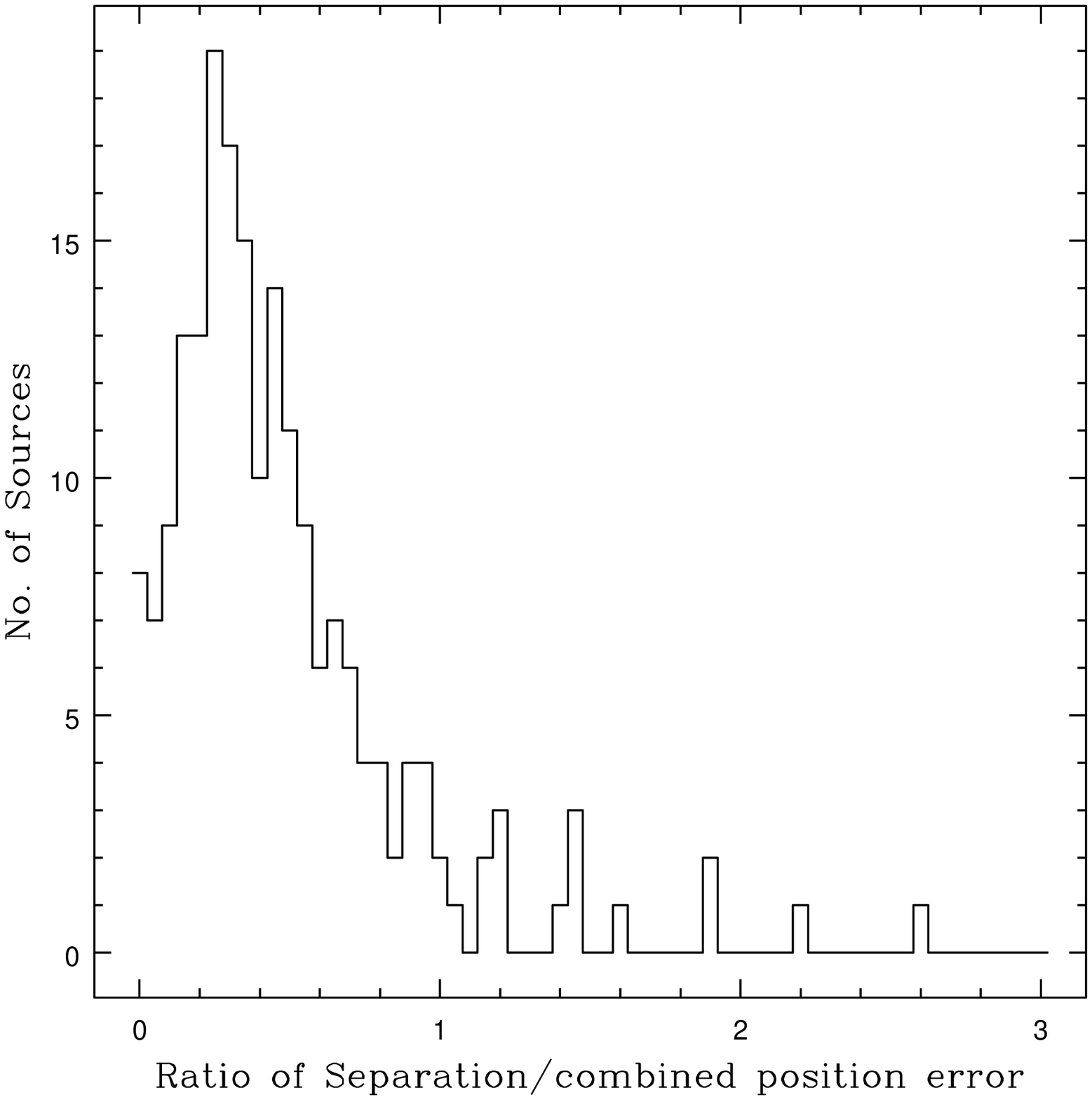}
	\end{minipage}
  \caption{Left: Histogram of the separation between sources detected in the
co-added observation and sources detected in single
observations. Right: Histogram of the ratio of separation between sources
detected in the co-added observation and sources detected in single
observations, divided by the combined position uncertainty of these sources.}
  \label{fig:sep_histo}

\end{centering}
\end{figure}

From the co-added observation only, 164 sources were detected by {\em
wavdetect}. From this list, sources external to the  overlapping area covered by
the S3 chip in
all five individual observations were excluded, reducing this total
number to 125 point sources. Using this source list from the co-added
observation, sources detected
in the individual observations were matched with this combined
observation source list, where source correlations were
searched for up to a separation of 3\arcs. In the cases where multiple
matches were detected for a source, the closer correlation was
selected. From these matches, a histogram of source separations, shown
in the left panel of Figure \ref{fig:sep_histo}, was produced. In this figure it
is clear that the peak separation between sources lies $\sim$0.2\arcs,
with the number of correlated sources dropping at $\sim$1.6\arcs, and this
is therefore the value we set for maximum separation when cross-correlating sources.

Once a cut of 1.6\arcs\ had been applied to the cross-matched source
list, the remaining unmatched sources, detected in the separate
pointings only, were investigated individually, resulting in further
matches being established. These matches correspond to sources with
fewer counts, and hence a greater positional uncertainty, leading to
larger values of separation. From the list of sources detected in
individual observations, seven of these point sources were determined to be well
separated from the sources detected in the co-added observation, and
were therefore included in the final list of detected sources,
increasing the total number of detected sources to 132.

These source correlations were then further investigated by
calculating the ratio of
the source separation and the combined position uncertainty. Where the
position uncertainty at the 95\% confidence level has been defined by
Kim \etal\ (2007a), as:
\begin{equation}
\label{equ:pu}
\mathrm{log PU} = \left\{ \begin{array}
{r@{\quad \quad}l}
0.1145\times \mathrm{OAA}- 0.4958 \times \mathrm{log C} +0.1932, & 0.0000 < \mathrm{log C} \le 2.1393\\
0.0968\times \mathrm{OAA} -0.2064 \times \mathrm{log C} -0.4260, & 2.1393 < \mathrm{log C} \le 3.3000 
\end{array}\right.  
\end{equation}
where the position uncertainty, $PU$, is in arcseconds, and the off
axis angle, $OAA$, is in arcminutes. Source counts, $C$, are as
extracted by {\em wavdetect}. Using this ratio of source
separation and position uncertainty allows low \LX\ source
correlations to be identified. Often these sources, particularly at
greater off axis angles, cannot be matched by source separation cuts
alone, due to the increasing PSF spread out and asymmetry at larger
$OAA$\footnote{See \S 5 in Kim \etal\ (2004a) for a full
discussion}. Therefore, by using this source separation - $PU$ ratio, the greater
position uncertainties in these weak sources can be accounted for, resulting in
smaller ratios, and thereby identifying correlations that would
otherwise be missed with source separation cuts alone.

In the right panel of Figure \ref{fig:sep_histo}, a histogram of the ratio of separation and the
combined position uncertainty is shown, where sources with a ratio of
greater than 1 were investigated individually. In all but two
instances it was found that these higher ratio sources lie in the
central region of the galaxy, where both source confusion is likely
and diffuse gas is present. This
emission results in higher background fluctuations, which can lead to
the $PU$ of these sources to be underestimated, therefore resulting in
a falsely high ratio value. For the two sources
that were detected outside the central region, both are too faint ({\em net
counts} $< \sim$ 100 counts) to allow their radial profiles to be
compared with corresponding model PSF profiles, generated  for
the position of each source using the CIAO tool {\em mkpsf}, and have
been flagged as possible double sources.

\begin{figure}
\begin{centering}

  \includegraphics[angle=-90,width=\linewidth]{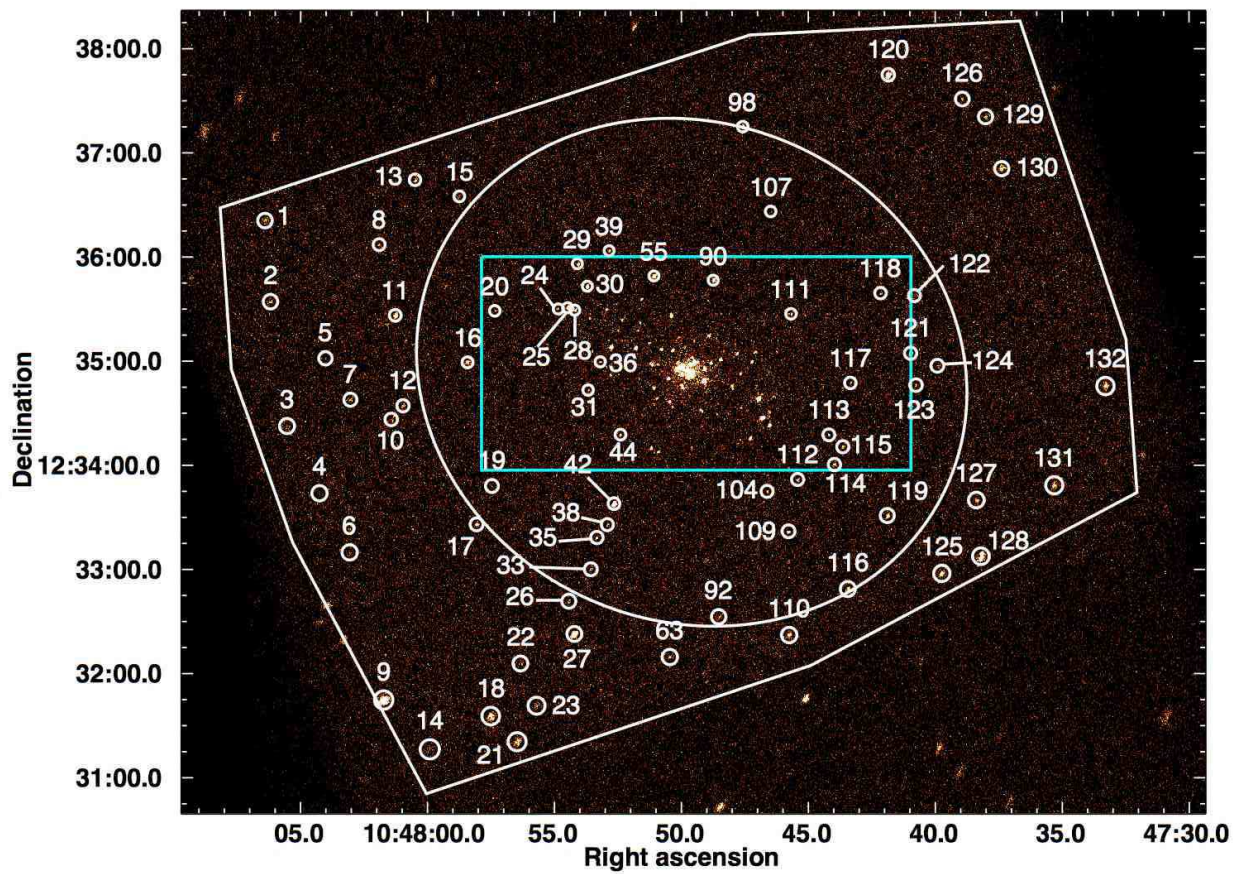}
  \caption{The main image, presents a full band, raw (unsmoothed, with no exposure
correction) image from the co-added
observation of NGC 3379, with the $D_{25}$ ellipse and the region
overlapped by all five observations overlaid. Source region
numbering corresponds to the 
naming convention in Table \ref{tab:Mainprops} and regions represent
the 95\% encircled energy radius at 1.5 keV. The cyan box in the central region
indicates the area shown in the next image, 
the central region of the galaxy, with sources labeled with the same convention
as in the main image. The cyan box shown here encloses the
nuclear region of the galaxy, where there is a dense population of
sources. This is presented in the final image, where these
individual sources can be more clearly seen.}
  \label{fig:rgb}
\end{centering}
\end{figure}

From this complete list, light curves were produced for
sources with {\em net counts}$<$ 20, which were detected in single
observations, to screen for the possibility of false sources by cosmic
ray afterglows. None of the 24 sources that
were examined exhibited light curves consistent with cosmic ray
afterglows, as is expected from the S3 chip (a back-illuminated CCD), due
to this problem mostly occurring in the front-illuminated chips. From this
screening, the total number of detected point sources remains at
132. These sources are presented in Figure \ref{fig:rgb}, where the
unsmoothed full-band image from the co-added dataset, with regions
overlaid in white, is shown.

\subsection{Hardness Ratios and X-ray colors}
\label{sec:HR}

Within NGC 3379, the range of net counts for the pointlike sources in
the co-added observation is
$\sim~2-7200$ (with signal-to-noise ratio (SNR) values ranging from
0.5 to 83.7), corresponding to 0.3$-$8.0 keV luminosities of 6
$\times 10^{35}$ \ergps (3$\sigma$ upper limit $\le$4$\times 10^{36}$ \ergps) $-~2~\times 10^{39}$\ergps, when using the
energy conversion factor described in \S
\ref{sec:src_props}. Most of these sources are too faint for detailed
spectral analysis, therefore their hardness ratio and X-ray colors
were calculated in order to characterize their spectral
properties. The X-ray hardness ratio is defined as
HR~$=$~(Hc$-$Sc)$/$(Hc+Sc), where Sc and Hc are the net counts in the
0.5$-$2.0 keV and 2.0$-$8.0 keV band respectively. Following the
prescription of Kim \etal\ (2004b), the X-ray colors are defined as
C21~=~log(S$_1$/S$_2$) and C32~=~log(S$_2$/H), where S$_1$, S$_2$ and
H are the net counts respectively in the energy bands of 0.3$-$0.9
keV, 0.9$-$2.5 keV and 2.5$-$8.0 keV (energy bands and definitions are
summarized in Table \ref{tab:bands}). These counts were corrected for
the temporal QE variation, referring them all to the first, recent
observing epoch (Jan. 2006, Table \ref{tab:log}), and for the effect
of the Galactic absorption, using \NH=2.78$\times10^{20}$\cmsq\ (from
COLDEN: http://cxc.harvard.edu/toolkit/colden.jsp). 

By definition, as the X-ray spectra become
harder, the HR increases and the X-ray colors decrease. For faint
sources with a small number of counts, the formal calculation of the
HR and colors often results in unreliable errors, because of negative
net counts in one band and an asymmetric Poisson
distribution. Therefore a Bayesian approach has been applied to derive the
uncertainties associated with the HR and colors. This model was developed by
Park \etal\ (2006) and calculates values using a method based on the
Bayesian estimation of the `real' source intensity, which takes into
account the Poisson nature of the probability distribution of the
source and background counts, as well as the effective area at the
position of the source (van Dyk \etal\ 2001), resulting in HR and color values that are
more accurate than the classical method, especially in the
small-number-of-counts regime (less than 10 counts), where the Poisson
distributions become distinctly asymmetric. More details of this technique
and the source and background counts used in the derivation of the
HR and color values and the 1$\sigma$ confidence bounds are provided in Appendix \ref{apen}.


\subsection{Source Variability}
\label{sec:var}

Due to the monitoring approach that has been used when observing NGC
3379, both long-term and short-term variations have been
able to be searched for in the galaxy's LMXB population.
Long-term variability was defined by the chi-squared test, where 
a straight line model was fitted to the luminosities derived for each
individual observation, with errors based on the Gehrels approximation
(Gehrels 1986). For
the cases where sources only had upper limit values of \LX, the
associated error was defined to be the standard deviation of 
the upper limit from the mode value attained from the Bayesian
estimates method, resulting in a conservatively large error, due to
the nature of the Poissonian statistics. From these best fit models,
sources were determined to be variable if \rchisq$>$1.2, and those with
fits with \rchisq$<$1.2 were defined as non-variable sources. For
sources that were only detected in the co-added observation,
long-term variability was not searched for. This long-term behaviour
will be further investigated in a forthcoming paper, where full Poissonian
error treatment will be applied to sources with very low observed counts.

In addition to the chi-squared test variability criterion, transient candidates (TC), sources that either appear or disappear, or are only detected for a limited amount
of `contiguous' time during the observations, were searched
for. Typically, sources are defined to be TCs if the ratio between the
`on-state', the peak \LX\ luminosity, and the `off-state', the lower
\LX\ luminosity or non-detection upper limit, is greater than a certain value (usually between 5$-$10; e.g. Williams \etal\ (2008)). However, such a criterion can overestimate the number of transient candidates, when the `on-state' X-ray luminosity is poorly constrained. To address this, the Bayesian model developed by Park \etal\ (2006) was used to derive the uncertainties associated with the ratio between `on-state' and `off-state'. In this model, source and background counts from both the peak \LX\ luminosity and the non-detection observations were used to estimate the ratio, where the differences in both the exposure and ECF values were also accounted for. From this Bayesian approach a value of peak \LX/non-detection upper limit was calculated, along with a lower bound value of this ratio. This lower bound value was then used to determine the transient nature of the source, where a ratio of greater than 10 indicated a TC and sources with a ratio between 5 and 10 were labeled as possible transient candidates (PTC). This transient behavior was only searched for in sources that were only detected for a limited amount of `contiguous' time during the observations and were determined to be variable using the chi-squared test.

Further to these four long-term variability classifications, the
variation of the source luminosity between each observation was also
investigated, by comparing the significance (in $\sigma$) of the change in
luminosity between exposures, where the significance has been estimated by:
\begin{equation}
\label{equ:sig}
sign = \frac{|\ensuremath{L_{\mathrm{X1}}}-\ensuremath{L_{\mathrm{X2}}}|}{\sqrt{(\sigma_1^2 + \sigma_2^2)}},
\end{equation}
where $\sigma_n$ is the error value of the luminosity from that
individual observation, based on the Gehrels approximation, or, where upper
limits have been used, the standard deviation of the estimated luminosity. 

Short-term variations in each source were investigated when $net$ 
$counts > 20$ in a single observation. In these instances, the
variability was identified by using the Kolmogorov-Smirnov test
(K-S test), where sources with variability values $>$90\% confidence were
labeled as possible variable sources and sources with values $>$99\% confidence were
defined as variable sources. This short-term variability was also
quantified by using the Bayesian blocks method (BB) (Scargle 1998;
Scargle \etal\ 2008, in prep). This method searches for abrupt changes
in the source intensity during an observation, and therefore is very
efficient for detecting bursts or state changes. Because it is based
on the Poisson likelihood it can be used on the unbinned lightcurves
of sources with very few counts. The implementation of the method used
in this analysis is the same as in the ChaMP pipeline (see
\S3.3.2 in Kim \etal\ 2004a). This assumes a
prior of $\gamma=4.0$ which {\em roughly} translates to a significance
level of $\sim99$\% for each detected block (however see Scargle \etal\ (2008, in prep)\footnote{see also
http://space.mit.edu/CXC/analysis/SITAR/functions.html}, for a caveat
on this interpretation of the  value of the prior). 

\subsection{Radial Profile}
\label{sec:prof}

\begin{figure}

\begin{centering}
\includegraphics[width=\linewidth]{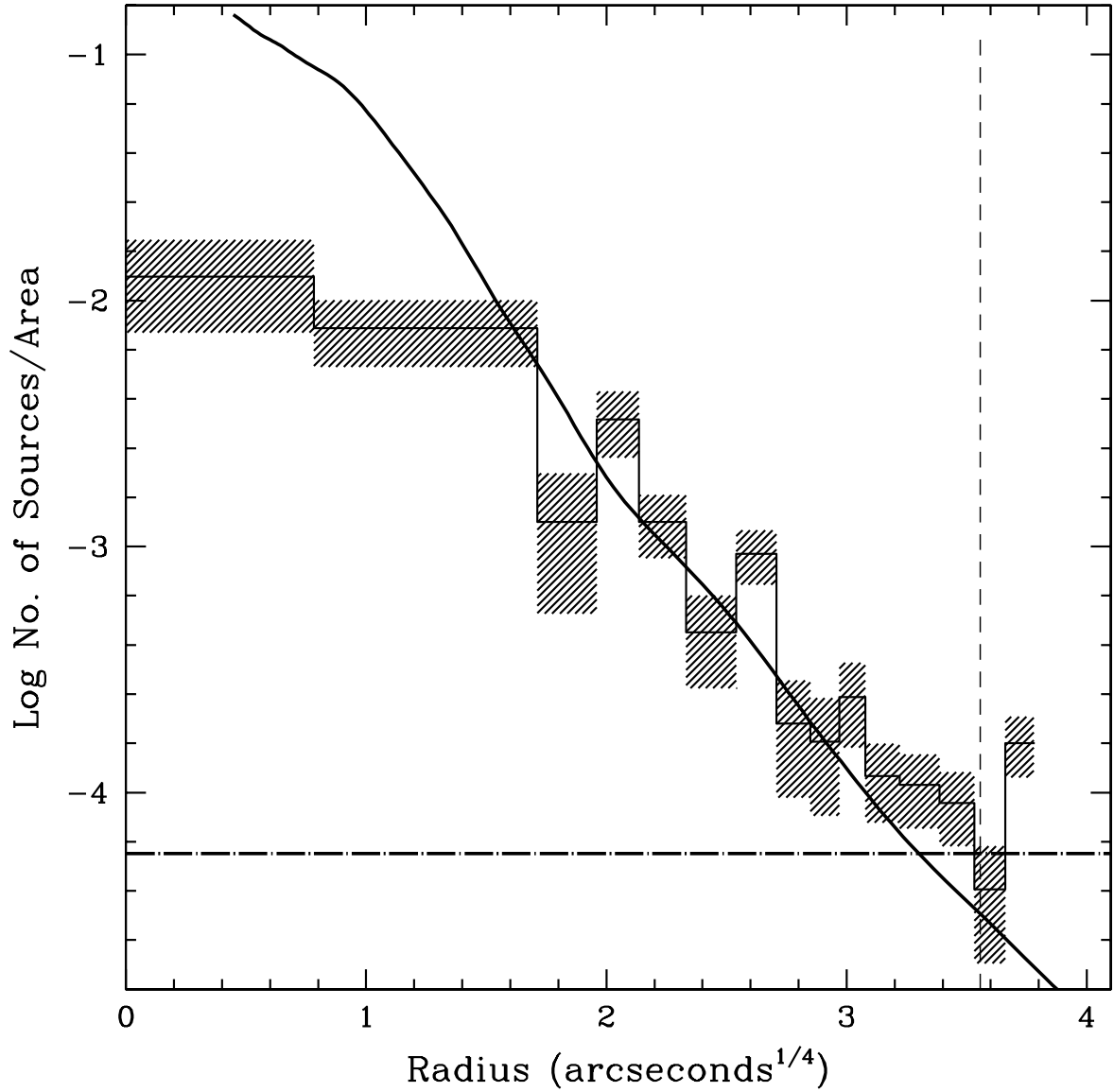}
\caption{X-ray source density profile compared to the optical
profile. The histogram indicates the X-ray data and the thick black
line is the I-band surface brightness best fit of Cappellari \etal\
(2006). The vertical dashed line is the $D_{25}$ ellipse and the
horizontal dot-dashed line indicates the expected number of background
sources. }\label{fig:profile}
\end{centering}
\end{figure}

From the complete source list from the co-added observation a radial
distribution of LMXBs has been created, using annuli centered on the
nucleus of the galaxy (source 81). This profile has been compared to a
multi-Gaussian expansion model of the I-band optical data (Cappellari
\etal\ 2006), which is assumed to follow the stellar mass of the
galaxy (Gilfanov 2004). This X-ray source density profile is presented
in Figure \ref{fig:profile}, where the optical profile has been normalized
to the X-ray data by way of a \chisq\ fit. Also indicated in this
figure is the $D_{25}$ ellipse and the number of background sources,
which has been estimated from the hard-band $ChaMP+CDF$ $logN - logS$
relation (Kim \etal\ 2007b), where $\sim$36 sources are expected to be
objects not associated with  NGC 3379. 
From this figure it can be seen that the X-ray profile follows the
optical surface density profile at larger radii, with the flattening
in the central region (r$\le$10\arcs) a consequence of source confusion. This
indicates that the number and spatial distribution of LMXBs follows
that of their parent population, the old stellar population.

\subsection{Optical Counterparts}
\label{sec:opt}

The globular cluster system of NGC 3379, observed with WFPC2, on-board
{\em HST}, is reported in Kundu \& Whitmore (2001), where images in
both the {\em V} and {\em I} bands have been analyzed. 
In addition to this GC system identified in the {\em HST} data,
background objects have also been classified (Kundu, A. 2007, private
communication). These have been identified as objects that were well
resolved in the {\em HST} images and were clearly more extended than any
known globular cluster. Further to this, these background objects often had other features,
such as visible disks, indicative of a galaxy rather than a globular
cluster. 
In addition to the {\em HST} data, radial velocities and
{\em B$-$R} values of spectroscopically confirmed GCs within this
galaxy have been reported in Bergond \etal\ (2006), Puzia et
al. (2004), and Pierce \etal\ (2006). With further {\em BVR}
photometry information, provided by images obtained with the Mosaic
detector on the Kitt Peak 4-m telescope (Rhode \& Zepf, 2004). 

Right ascension and declination corrections have been applied to the
astrometry of these
data-sets, relative to the co-added {\em Chandra} observation. These
offsets were calculated and corrected for using the methods described
in \S \ref{sec:astrometry}, where correlations between
the X-ray and optical sources were made, and systematic offsets in
RA and Dec were removed. In the case of the Rhode \& Zepf (2004) data,
only two correlations were found between the X-ray and optical sources,
therefore the corrected {\em HST} data were used to make these offset
adjustments. 

\begin{figure}
\begin{centering}
   \begin{minipage}{0.48\linewidth}
  \includegraphics[width=\linewidth]{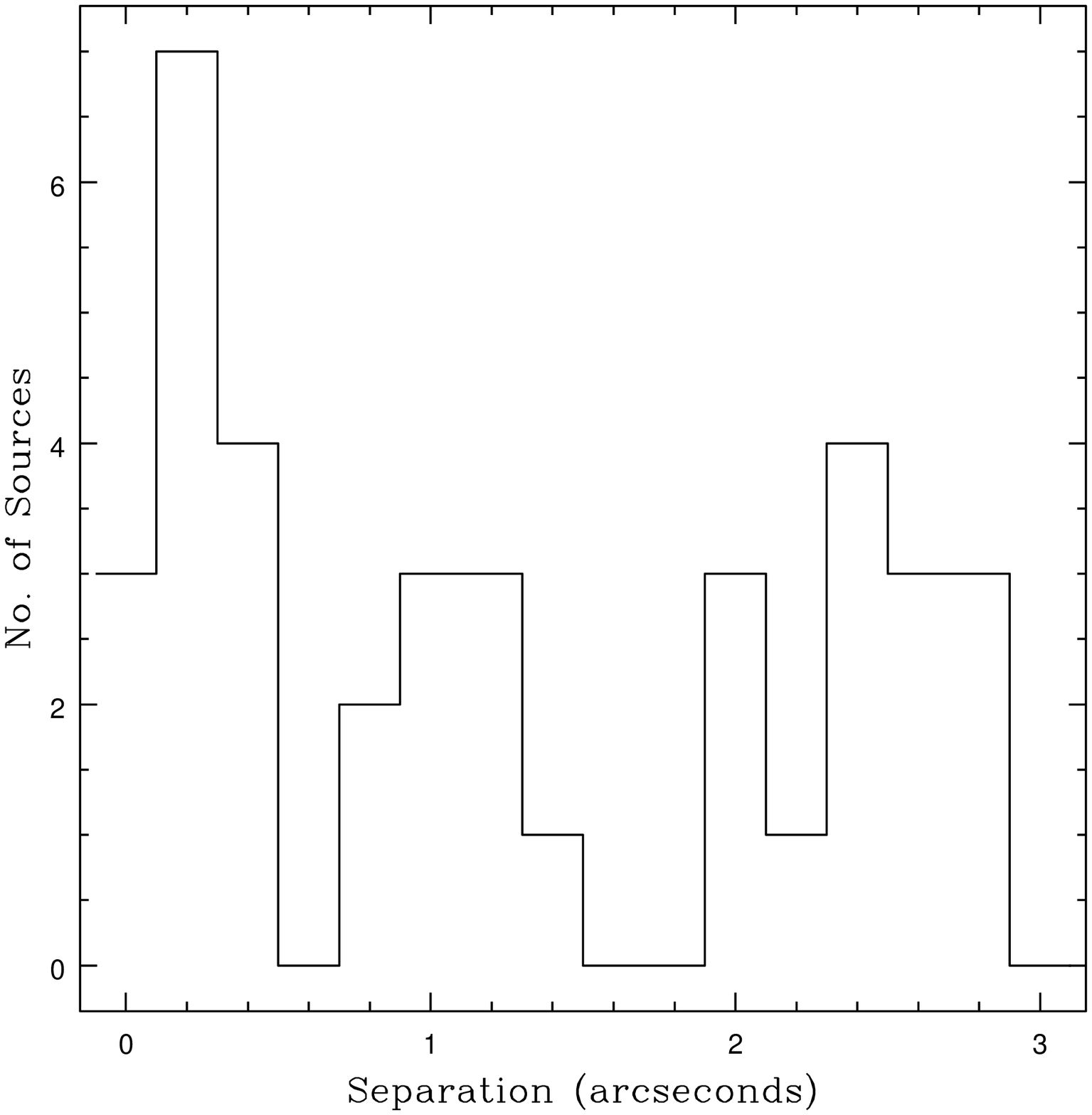}

  \end{minipage}
  \begin{minipage}{0.48\linewidth}

  \includegraphics[width=\linewidth]{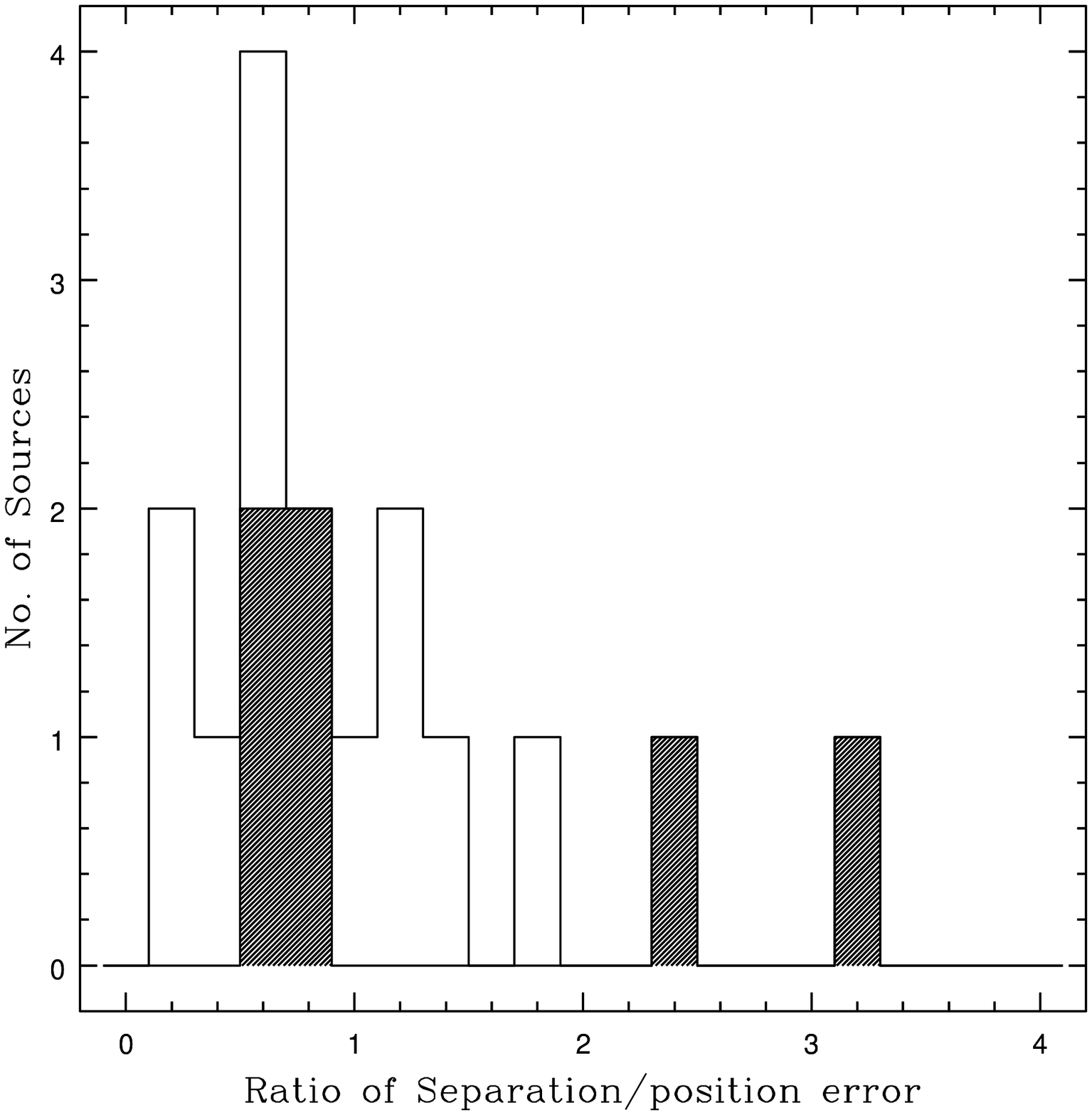}
	\end{minipage}

  \caption{Left:A histogram of the separation between the co-added X-ray source
position and the optical counterpart. Right: Histogram of the ratio of separation divided by the
position uncertainty from the X-ray point source for all optical-X-ray
correlations with separations smaller than 1\arcs. Shaded regions
indicate correlations with optical objects that have been classified
as background sources (details of this classification are given in the
text).}
  \label{fig:pser_GChisto}

 \end{centering}
\end{figure}

After correcting the astrometry of the optical data, correlations up
to an offset of 3\arcs\ with the X-ray sources, were searched
for. When multiple matches were found, the closer matching object was
selected. In the left panel of Figure \ref{fig:pser_GChisto}, a histogram of these matches
is shown, where it can be seen that, due to the poor statistics, it
is not clear where the separation cut off should be made. Initially,
this was set to 1\arcs, and sources between 1\arcs\ and 3\arcs\ were
defined as `excluded matches'. This cut off value was then tested by
comparing these correlations with the ratio of the separation divided
by the combined position
uncertainty from the co-added X-ray point sources (the definition of
this is given in equation \ref{equ:pu}) and the
uncertainty in the astrometry in the optical data, which has been
conservatively set at 0.2\arcs.

These ratios are shown in the right panel in Figure \ref{fig:pser_GChisto}, where a
histogram of all optical-X-ray correlations is presented, with a
shaded histogram of the background correlations only, overlaid. From
this figure, it is shown that the correlated sources with a
separation-position uncertainty ratio of greater than 2 are background
objects. Both of these objects have separations $>$0.9\arcs, and, from
both of the histograms presented in Figure
\ref{fig:pser_GChisto}, the separation value cut off was redefined to
be 0.6\arcs. This results in 14 X-ray-optical correlations, 4 of which
have been classified as background objects, leaving 10 GC-X-ray source
correlations, one of which lies external to the {\em HST} FOV,
although this source has
been detected in two separate studies (Bergond \etal, 2006; Rhode \&
Zepf, 2004). The optical properties of these GC-LMXB sources, and the
`excluded matches' are shown in Tables \ref{tab:GCprops_corr} and
\ref{tab:GCprops_near} respectively (Full descriptions of these
tables are given in \S \ref{sec:atlas}). 

In order to estimate the chance coincidence probability of the sources
within the {\em HST} FOV, the same method as in Zezas \etal\ (2002) was
followed, where the positions of the globular clusters were randomized
by adding a random shift between 0.6\arcs and 30\arcs, and for
each new fake dataset the cross-correlation was performed using the
same search radius as for the observed list of globular clusters. The
limits of the shifts were chosen so that the new positions did not fall
within the search radius and that they follow the general spatial
distribution of the globular clusters. 500 such simulations were
performed, resulting in 0.35$\pm$0.59 associations expected by
chance. If the cross-correlation radius is increased to 1\arcs, the
chance associations rises to 0.5$\pm$1.1. Increasing this radius to
3\arcs\ results in 8.4$\pm$14.2 associations expected by chance, which compares well
with the nine `excluded matches' that have been found within this radius.

\begin{figure}

\begin{centering}
\includegraphics[angle=-90,width=\linewidth]{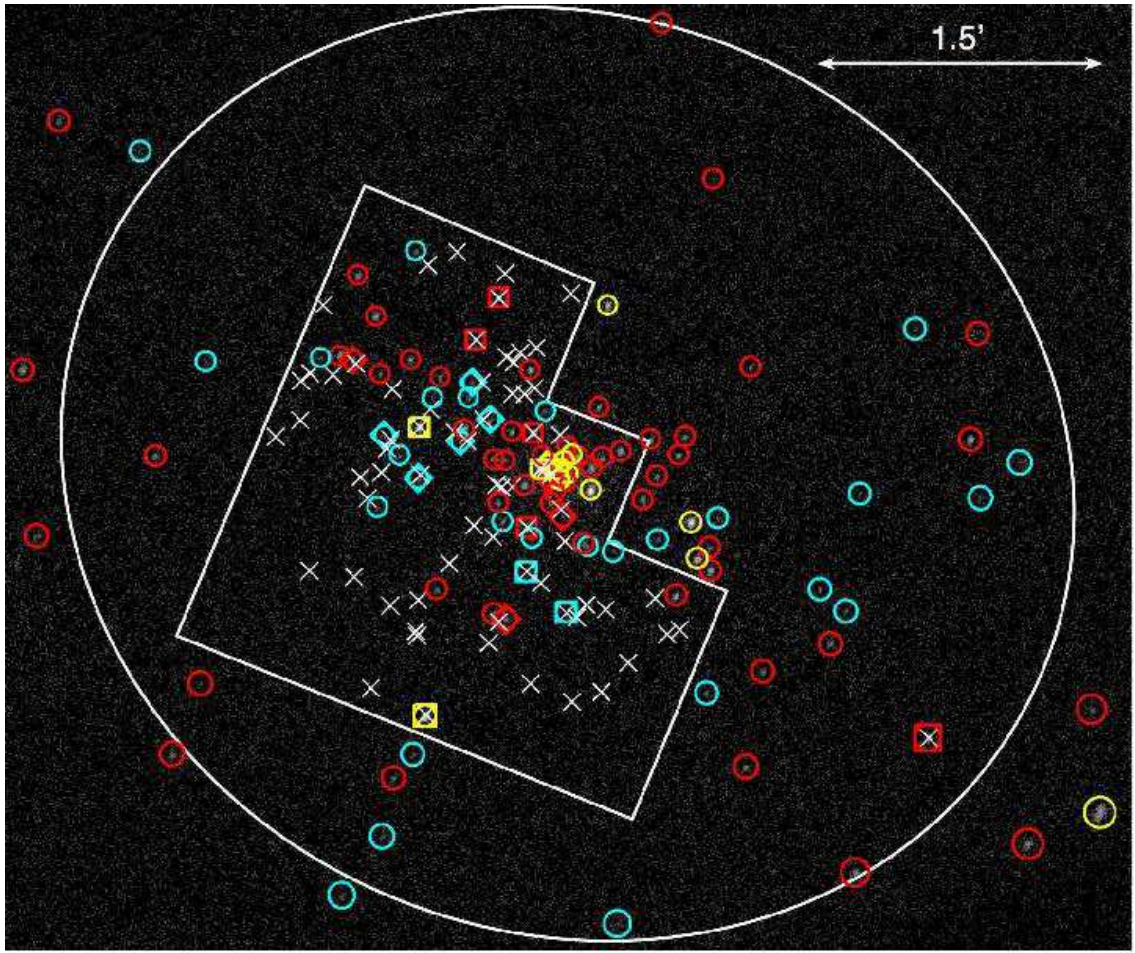}
\caption{A full band X-ray image from the co-added observation of NGC 3379,
with confirmed GCs from the {\em HST} observation indicated by white
`X' marks (with the GC position of the additional LMXB-GC match,
external to the {\em HST}
Field of View, also shown). X-ray sources that
have correlated GCs are indicated by box regions, sources that
are `excluded matches' are indicated with a diamond region and X-ray
sources without a GC counterpart are shown by circular regions. X-ray
regions are colored to indicate the 0.3$-$8.0 keV luminosity of
the source from the co-added observation;
yellow regions indicate \LX$\ge$1$\times
10^{38}$\ergps, red regions have
1$\times 10^{38}\ge$\LX$\ge$1$\times 10^{37}$\ergps, and cyan regions
show sources with \LX$\le$1$\times 10^{37}$\ergps. Also shown are the
$D_{25}$ ellipse and the {\em HST} FOV. The GC X-ray correlation
shown external
to the {\em HST} FOV was detected in two other independent GC studies
(see text for more details). }\label{fig:gccorr}
\end{centering}
\end{figure}

The 10 GC-X-ray correlations that have been found in NGC 3379 are shown in Figure
\ref{fig:gccorr}, where an X-ray image with confirmed GCs is shown. In
this figure the GCs are indicated by white `X' marks and the
corresponding X-ray sources are indicated by box regions. The
`excluded matches' are indicated by diamond regions and X-ray sources
with no matches are shown as circular regions. X-ray luminosities are
also indicated in this image, where sources with \LX$\ge$1$\times
10^{38}$\ergps\ are shown in yellow, sources with 1$\times
10^{38}\ge$\LX$\ge$1$\times 10^{37}$\ergps\ are shown in red and
sources with \LX$\le$1$\times 10^{37}$\ergps\ are indicated in cyan. 

\section{Source Catalog and Variability Atlas}
\label{sec:atlas}

Table \ref{tab:Mainprops} presents the properties of the master list
of the 132 X-ray sources detected within NGC 3379, from the co-added
observation of 324-ks. This table has been divided into two sections, where the first part presents all sources with SNR$>$3 in at least one observation, and the second part lists all sources with SNR$<$3. In this table column (1) gives the source number used
through out this series of papers, column (2) gives the IAU name
(following the convention ``CXOU
Jhhmmss.s$+/-$ddmmss''), columns (3) and (4) give the R.A. and
Dec. of the source aperture, columns (5) and (6) give the
radius and the position uncertainty ({\em PU}) of the source
(both in arcseconds), column (7) gives the SNR,
column (8) gives the log value of the co-added luminosity in the 0.3$-$8.0 keV
energy band (for sources with SNR$<$3, 3$\sigma$ upper limit values are also presented in brackets). For sources detected in a single observation only, 1$\sigma$ upper limit from the co-added observation are shown, with 3$\sigma$ upper limit values from the detected observation presented in brackets.
Column (9) provides information about the long-term variability of the
source,
indicating if the source is non-variable (N), variable (V), a
transient candidate (TC) or a possible transient (PTC). In all other
cases the source was only detected in the combined observation,
providing insufficient information to investigate long-term
variability. In columns (10)
and (11) the short-term variability of the source is indicated from both
Bayesian block analysis (BB) and the Kolmogorov-Smirnov test (K-S),
where `V' indicates that the source is variable in at least one
observation and `N' indicates that is has been found to be non-variable
in all five observations. In the K-S column, sources have  also been
labeled as possible variable sources (P) (see \S \ref{sec:var}
for further information). In all other cases there were insufficient counts
to investigate the short-term variability. In column
(12) the optical associations with the X-ray source are indicated, where `GC'
indicates that the associated optical sources has been confirmed as a
globular cluster, and `BG' indicates that the sources has been
classified as a background object. `corr' denotes matches that have
been defined as correlations, and `exmt' denotes the `excluded matches', between
0.6\arcs\ and 3\arcs\ in separation. Sources with a `none' label were
inside the field of view of the {\em HST} observation, but have no
optical counterpart, and sources denoted with an `X' were within the {\em
HST} FOV, but were also within 5\arcs\ of the nucleus, and were therefore not
considered for optical associations. All other sources were external to the {\em HST}
FOV. Column (13) gives the distance
from the galactic center (in arcseconds), where values in bold type face
indicate sources that lie within the $D_{25}$ ellipse. Column (14) provides
source flag information, indicating sources that have been detected in
a single observation only (X), overlapping sources (O1 for single
overlaps and O2 for more complicated cases), possible background
objects (BKG?) (see \S \ref{sec:xcol} for details of this
classification) and possible double
sources (double?).

In this table, the 132 sources presented are the complete list
detected by {\em wavdetect}, for which we estimate that $\sim$1 source
is a spurious detection (see \S \ref{sec:src_props}). Since this
catalog of X-ray sources is intended to 
be as complete a study as possible, all detected sources are
included in the complete list, although for sources with SNR$<$3 source parameters such as flux, hardness ratio and color values are not as well constrained as sources with higher flux significance. We have therefore separated the table into two sections, where the first part presents sources with SNR $>$3 in at least one observation and well constrained properties, and the second part lists the sources with low SNR values.

Table \ref{tab:counts_main} presents the detailed source parameters
from the co-added observation; column (1) gives
the source number, columns (2)$-$(8) give the net counts, in each of
the 7 energy bands (see Table \ref{tab:bands} for definitions of these
bands), column (9) indicates the hardness ratio, columns (10) and (11)
show the color-color values and column (12) gives the log value of the
luminosity in the 0.3$-$8.0 keV energy band. Where sources were not
detected in the co-added observation, upper limits for net broad band
counts and \LX\ are given. 

Tables \ref{tab:counts_ob1}$-$\ref{tab:counts_ob5} present the source parameters,
measured for each observation, where columns (1)$-$(11)
provide the same information presented in Table \ref{tab:counts_main},
but further provided in this table is source variability information,
where columns (12)$-$(14) present results of Bayesian block analysis (BB), the
Kolmogorov-Smirnov test (K-S) and the significance of the change in
\LX\ between the previous observation and the current observation
respectively. Column (15) indicates the log value of the
luminosity in the 0.3$-$8.0 keV energy band.

Table \ref{tab:GCprops_corr} presents the optical properties of the
counterparts found from the
optical data of NGC 3379, where 10 GCs and 4 background objects have
been found to be coincident with X-ray sources. Table
\ref{tab:GCprops_near} summarizes the results for the `excluded matches'
sources. In both tables
column (1) gives the X-ray source number, column (2) the {\em V} band
magnitude, column (3) the {\em I} band magnitude, column (4) {\em
V$-$I} colors, column (5) {\em B$-$V} colors, column (6) {\em V$-$R}
colors, column (7) {\em B$-$R} colors, column (8) gives the radial
velocity, column (9) the separation between the X-ray source and the
GC, column (10) the ratio between separation and the combined position
error and column (11) gives a references to where the GC information
comes from; 1. Bergond \etal\ 2006, 2. Puzia \etal\ 2004, 3. Pierce et
al. 2006, 4. Kundu \& Whitmore (2001), confirmed GCs, 5. Kundu \&
Whitmore (2001), background objects, 6. Rhode \& Zepf, 2004.
The horizontal line in both tables separates the confirmed GCs (top
section of table) from the background objects (bottom section of table).

\begin{figure}
  \begin{minipage}{0.485\linewidth}
  \centering
  
    \includegraphics[width=\linewidth]{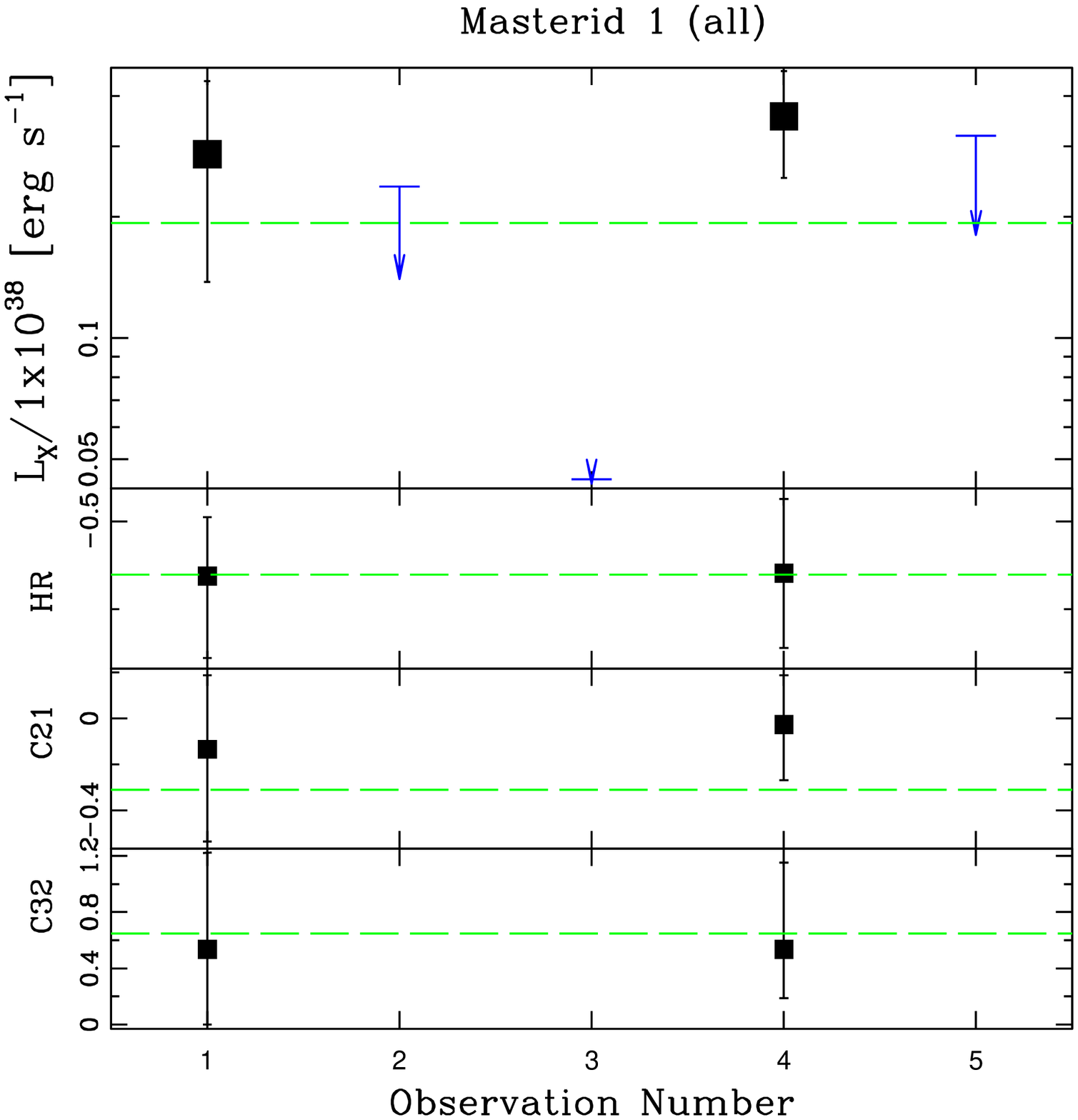}
  
  \end{minipage}\hspace{0.02\linewidth}
  \begin{minipage}{0.485\linewidth}
  \centering

    \includegraphics[width=\linewidth]{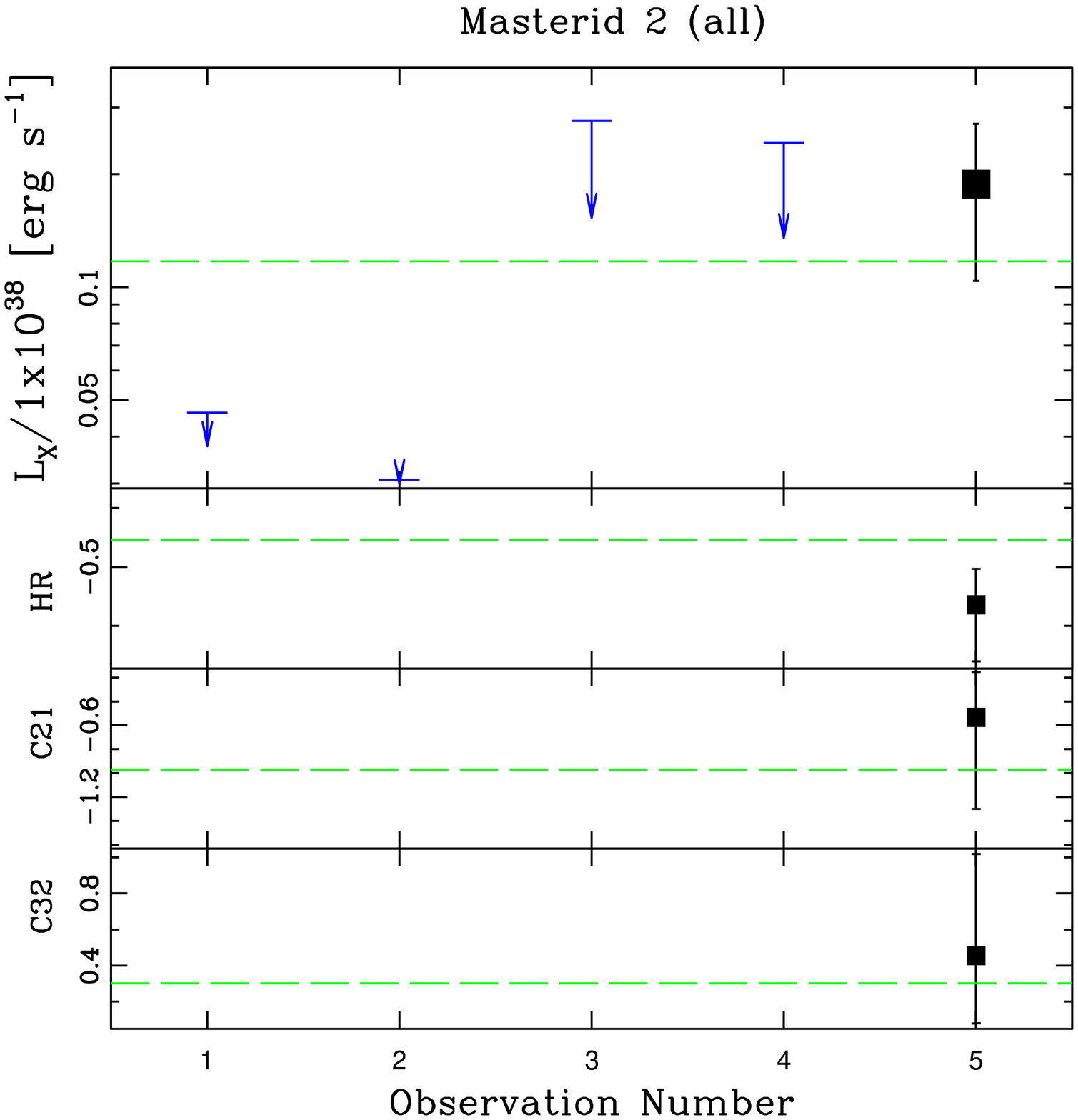}

\end{minipage}

\begin{minipage}{0.485\linewidth}
  \centering

    \includegraphics[width=\linewidth]{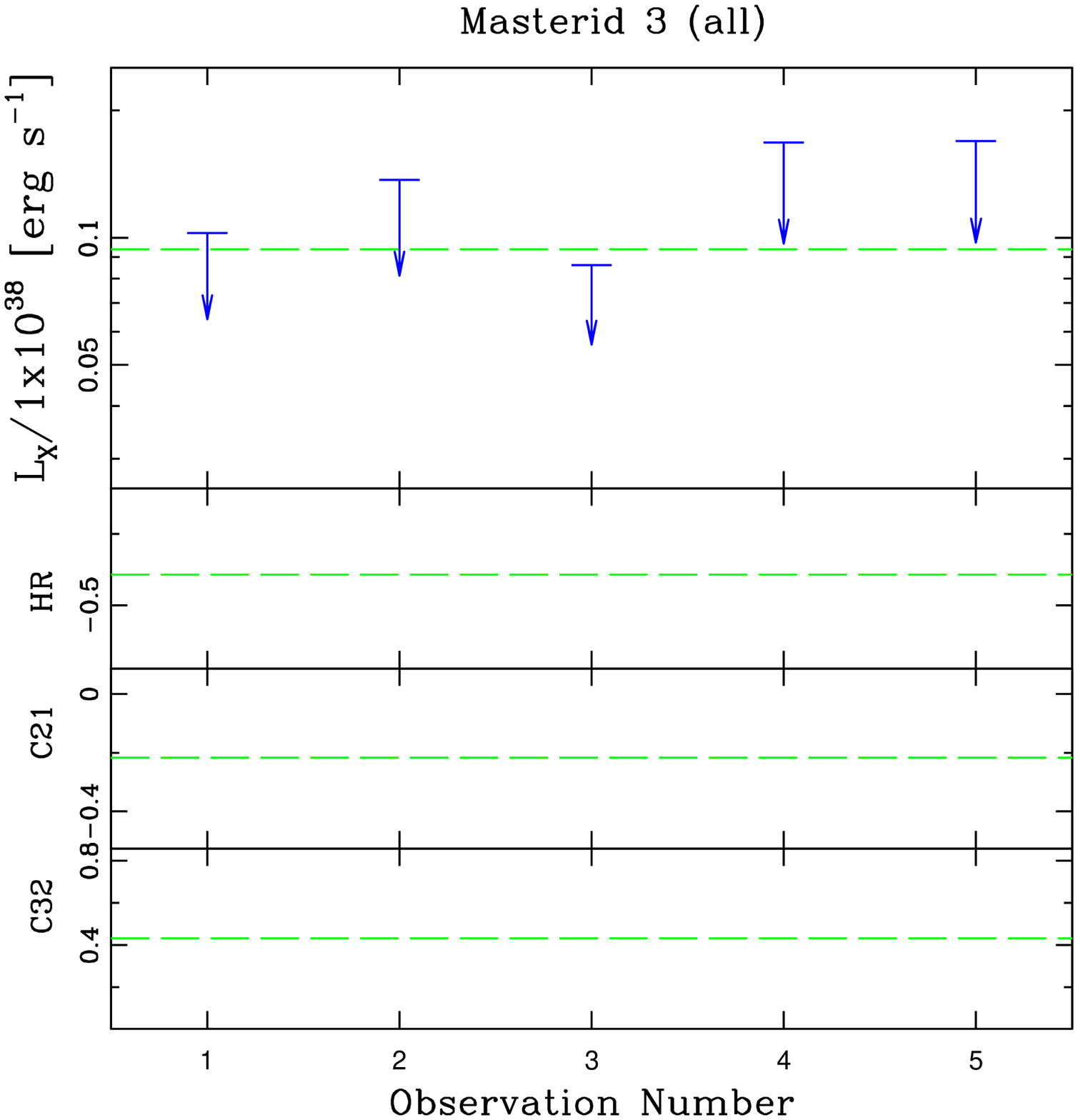}

 \end{minipage}\hspace{0.02\linewidth}
\begin{minipage}{0.485\linewidth}
  \centering
  
    \includegraphics[width=\linewidth]{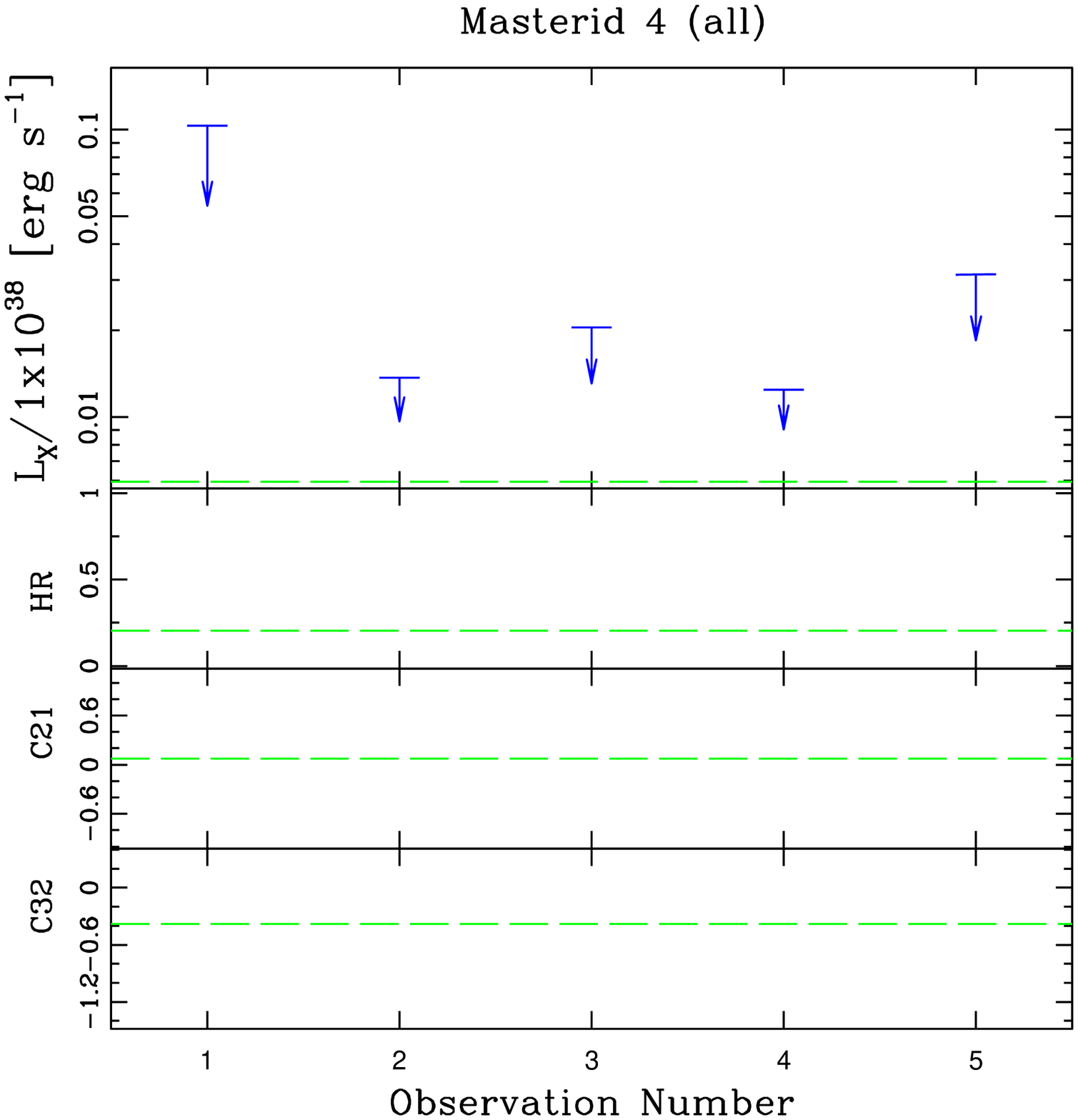}
  
  \end{minipage}\hspace{0.02\linewidth}

\caption{Plots of the the 132 detected sources, summarizing the
variations in properties of each source between each observation. In
the main panel the long-term light curves are shown. In the second
panel down, the hardness ratios are indicated. These are defined to be
HR = H$-$S / H+S, where H is the number of counts in the
hard band (2.0$-$8.0 keV) and S is the number of counts in the soft
band (0.5$-$2.0 keV). In the third and
fourth panels the color ratios; C21 and C32, are
plotted, where C21={\em log}S2+{\em log}S1 and C32=$-${\em log}H+{\em log}S2. For
the color ratios the bandwidths are defined to be S1=0.3$-$0.9
keV, S2=0.9$-$2.5 keV and H=2.5$-$8.0 keV. 
In cases where a source was not detected in a single observation, an
upper limit of the X-ray luminosity has been calculated, details of
which are discussed in section \ref{sec:src_props}. In all of the panels, the green horizontal
line indicates the value derived from the co-added observation. In
cases where the source was not detected in the co-added observation, a blue
horizontal line indicates the upper luminosity calculated for that source.}
\label{fig:all_LC}

\end{figure}

Figure \ref{fig:all_LC} presents the intensity and spectral
variability of each of the 132 X-ray sources, over all five 
pointings, where the
temporal properties of each point source are shown in four separate
panels. In the top panel the long-term light curve of each source is
presented, with errorbars indicating the 1$\sigma$ uncertainty in the
intensity of the source, with upper limit values provided for sources
that were not detected in a single observation. The second panel shows
the hardness ratio variation of each source, and panels three and four, show
 the temporal properties of C21 and C32 respectively. In all four
panels, the co-added values are also indicated, by a horizontal
dashed green line. In instances where the source was not detected in
the co-added observation, a blue line indicates the upper limit of the
source luminosity.

\begin{figure}
  \begin{minipage}{0.32\linewidth}
  \centering
  
    \includegraphics[width=\linewidth]{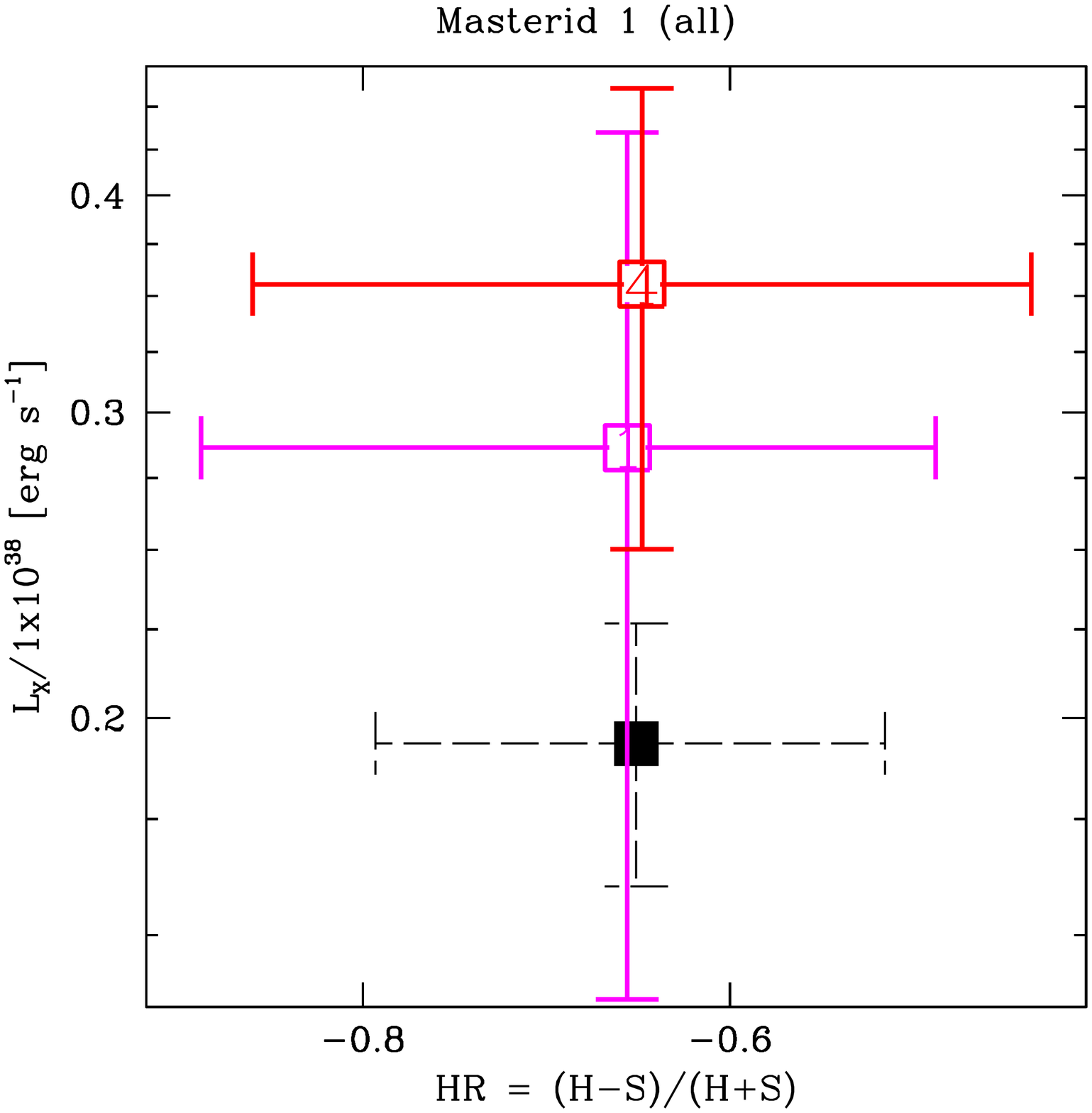}
  
  \end{minipage}
  \begin{minipage}{0.32\linewidth}
  \centering

    \includegraphics[width=\linewidth]{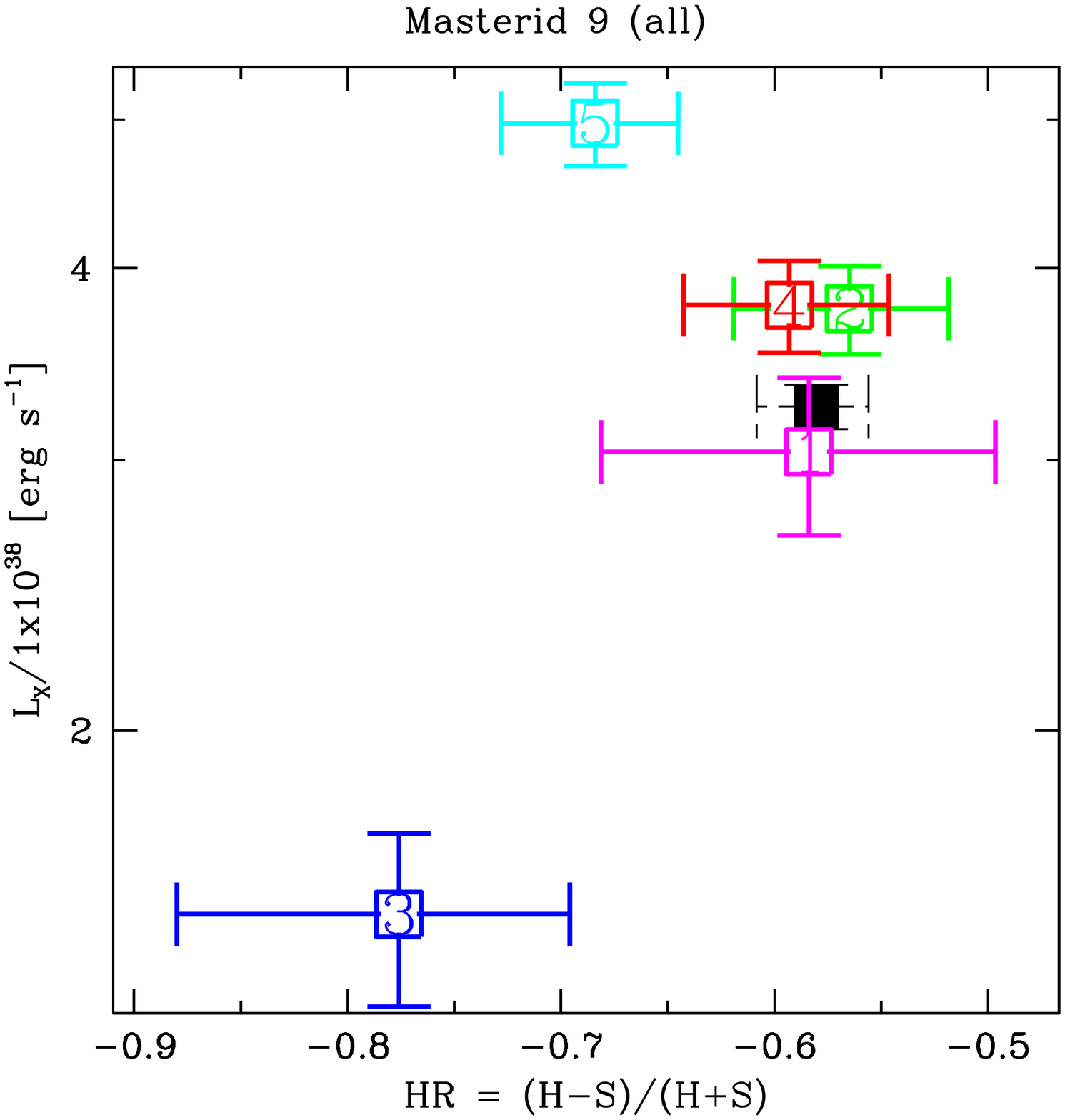}

\end{minipage}
\begin{minipage}{0.32\linewidth}
  \centering

    \includegraphics[width=\linewidth]{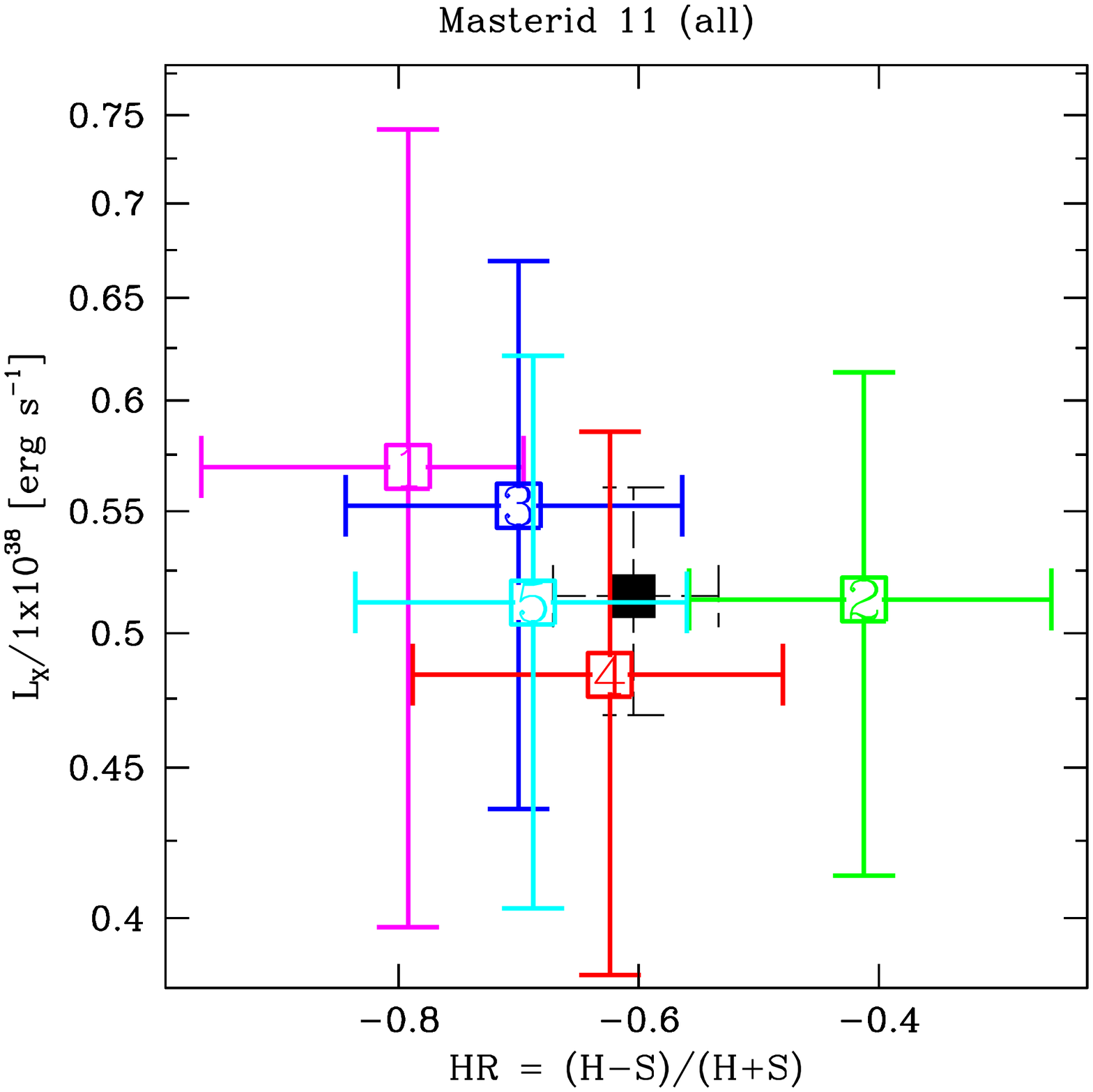}

 \end{minipage}

\begin{minipage}{0.32\linewidth}
  \centering
  
    \includegraphics[width=\linewidth]{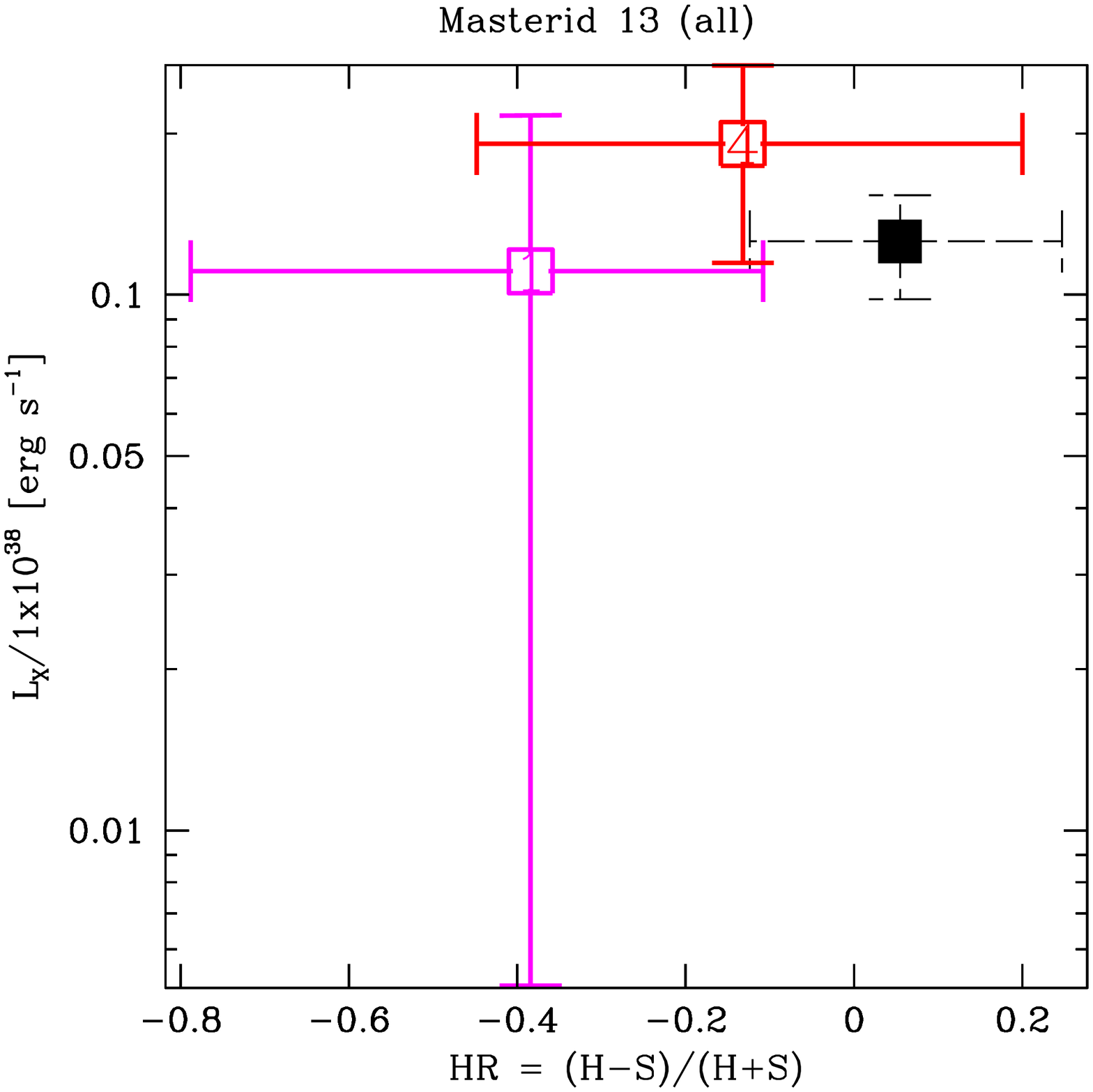}
  
  \end{minipage}
  \begin{minipage}{0.32\linewidth}
  \centering

    \includegraphics[width=\linewidth]{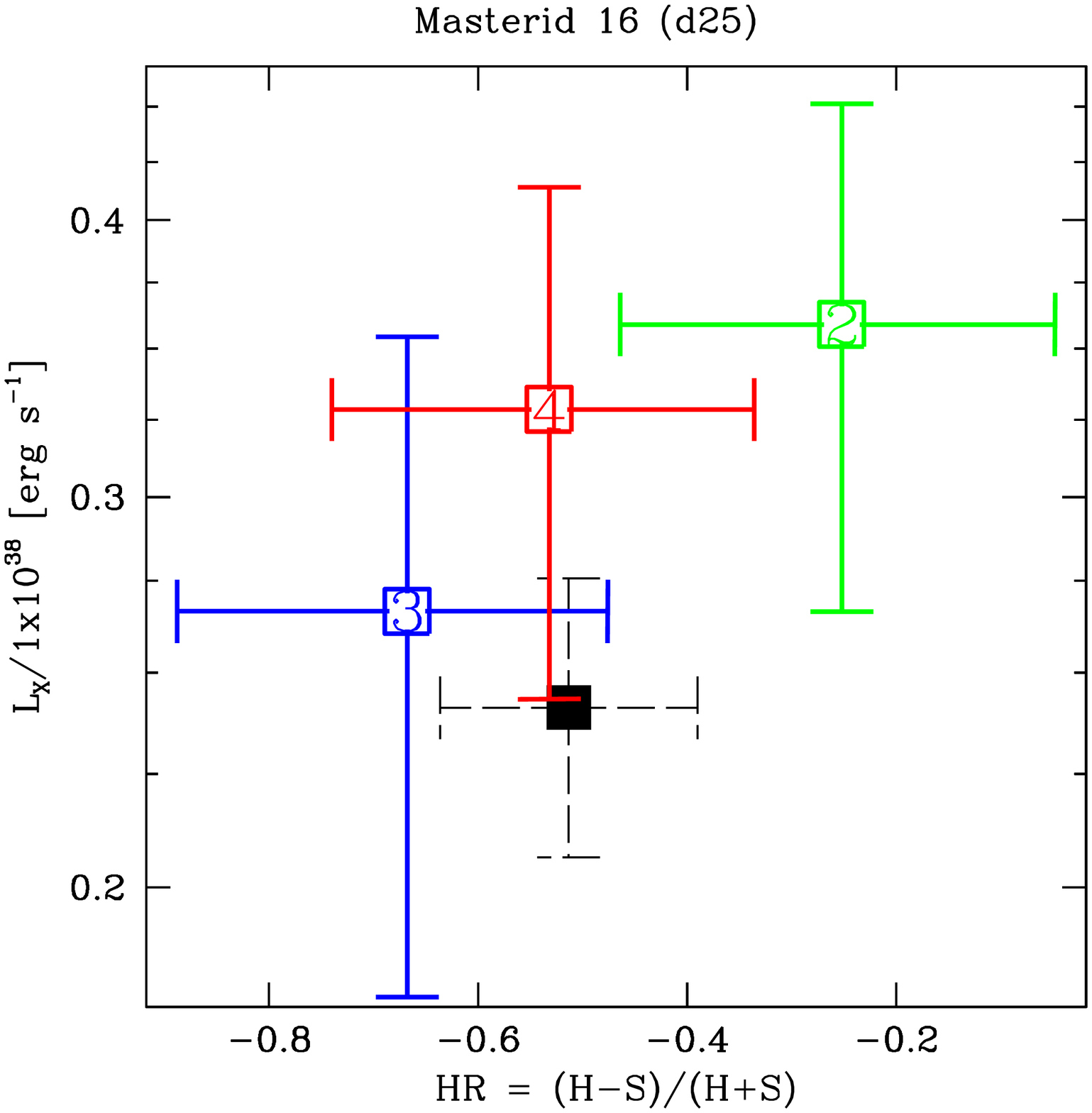}

\end{minipage}
\begin{minipage}{0.32\linewidth}
  \centering

    \includegraphics[width=\linewidth]{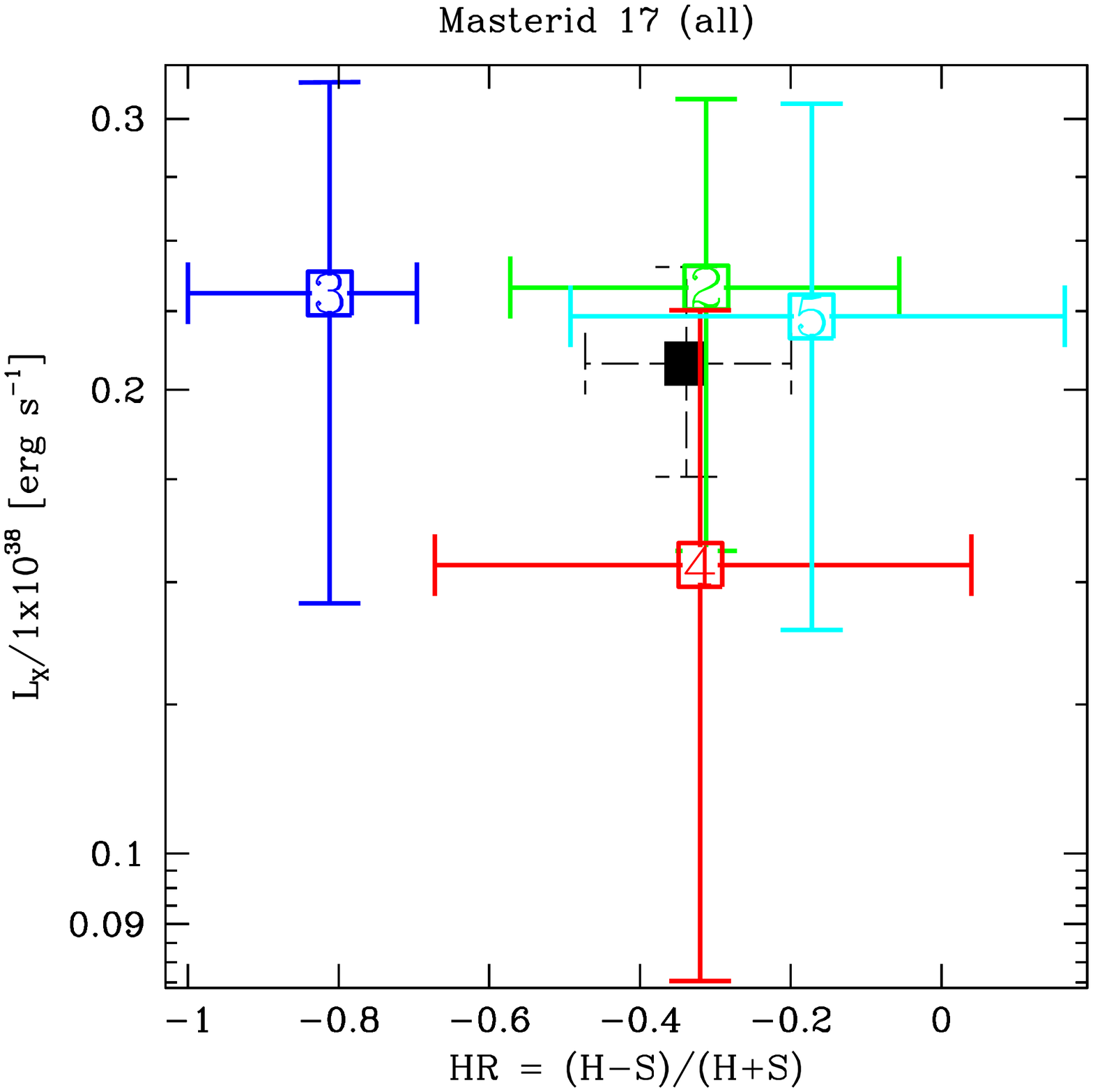}

 \end{minipage}

  \begin{minipage}{0.32\linewidth}
  \centering
  
    \includegraphics[width=\linewidth]{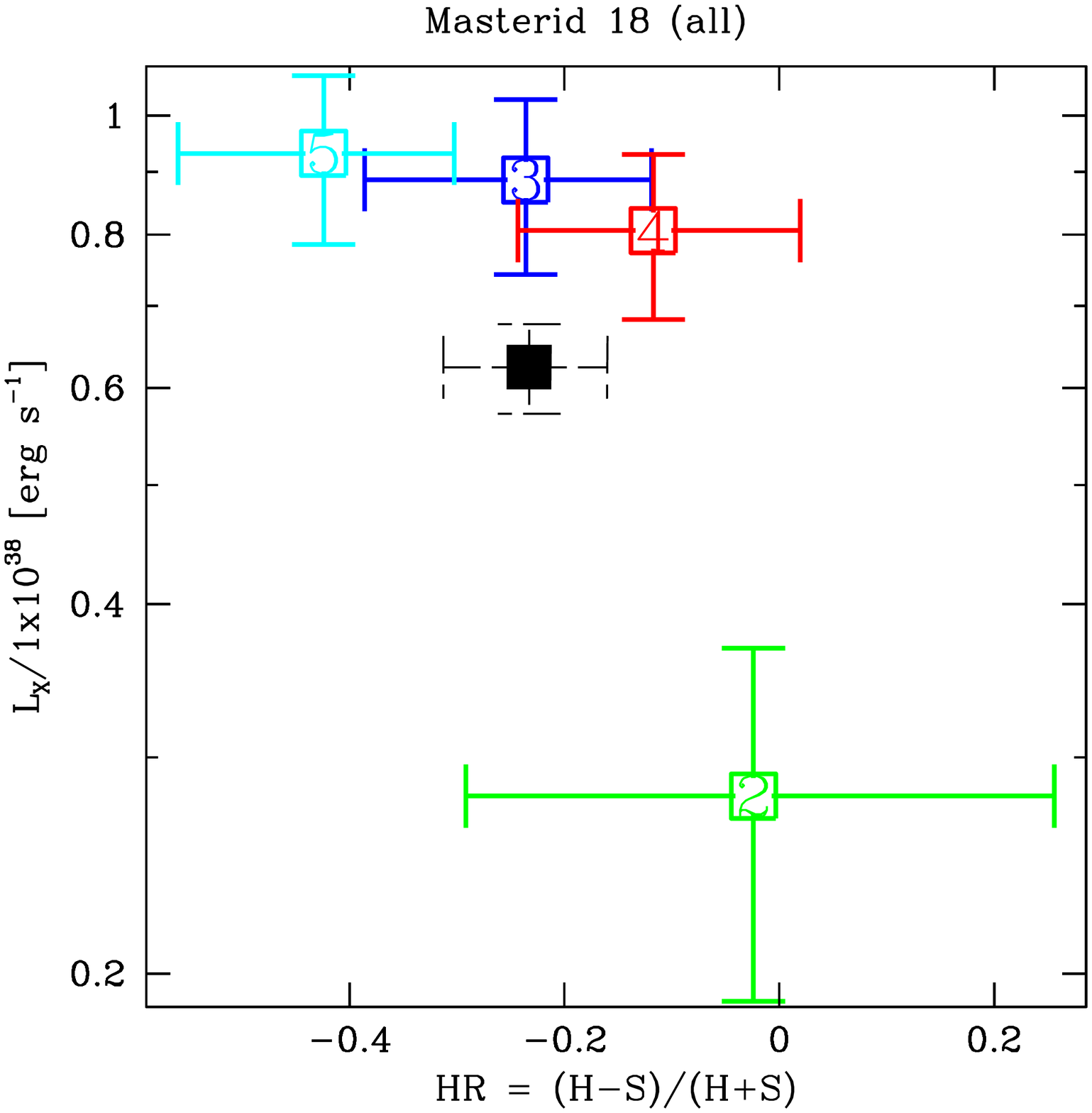}

  \end{minipage}
  \begin{minipage}{0.32\linewidth}
  \centering

    \includegraphics[width=\linewidth]{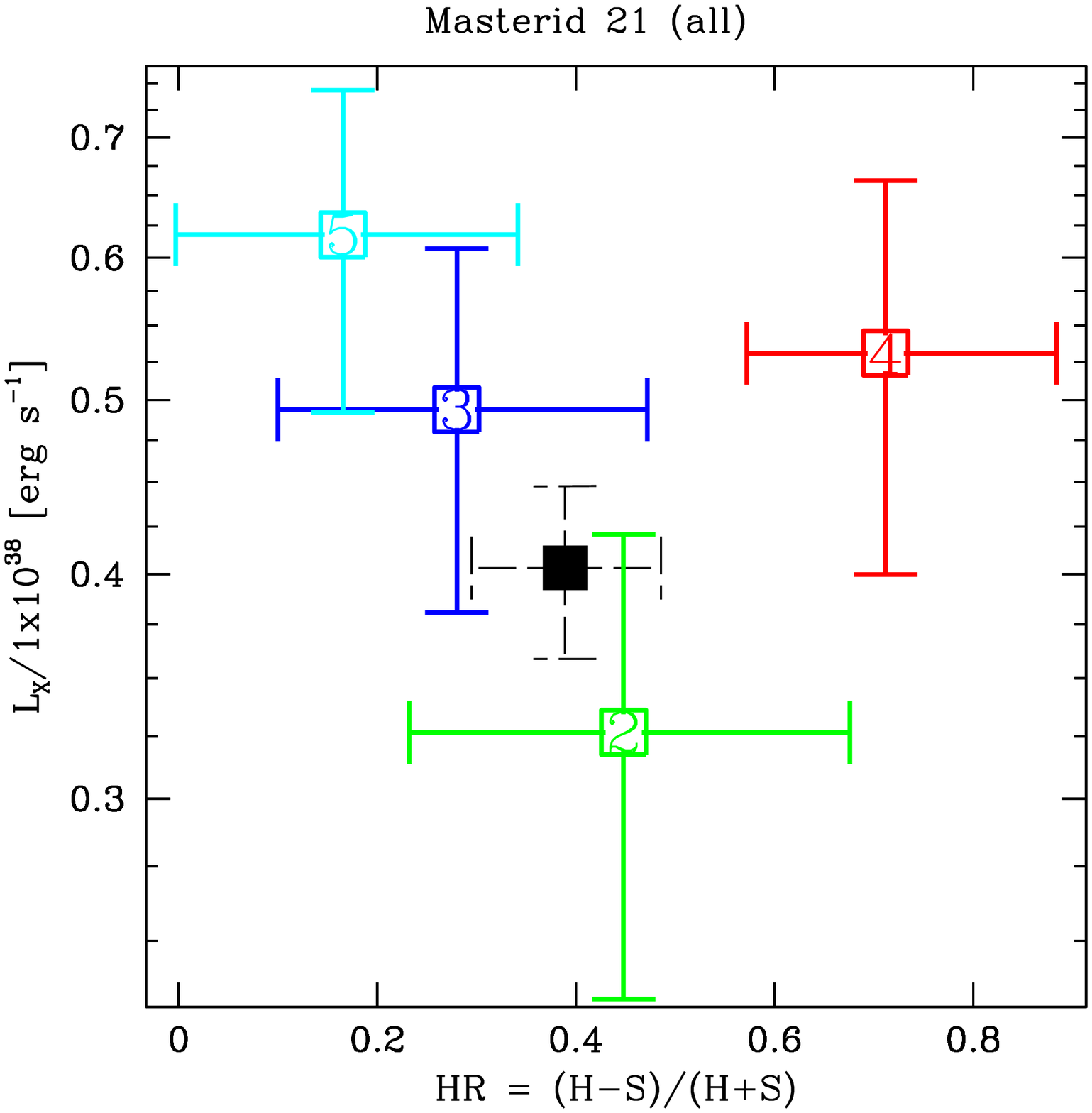}

\end{minipage}
\begin{minipage}{0.32\linewidth}
  \centering

    \includegraphics[width=\linewidth]{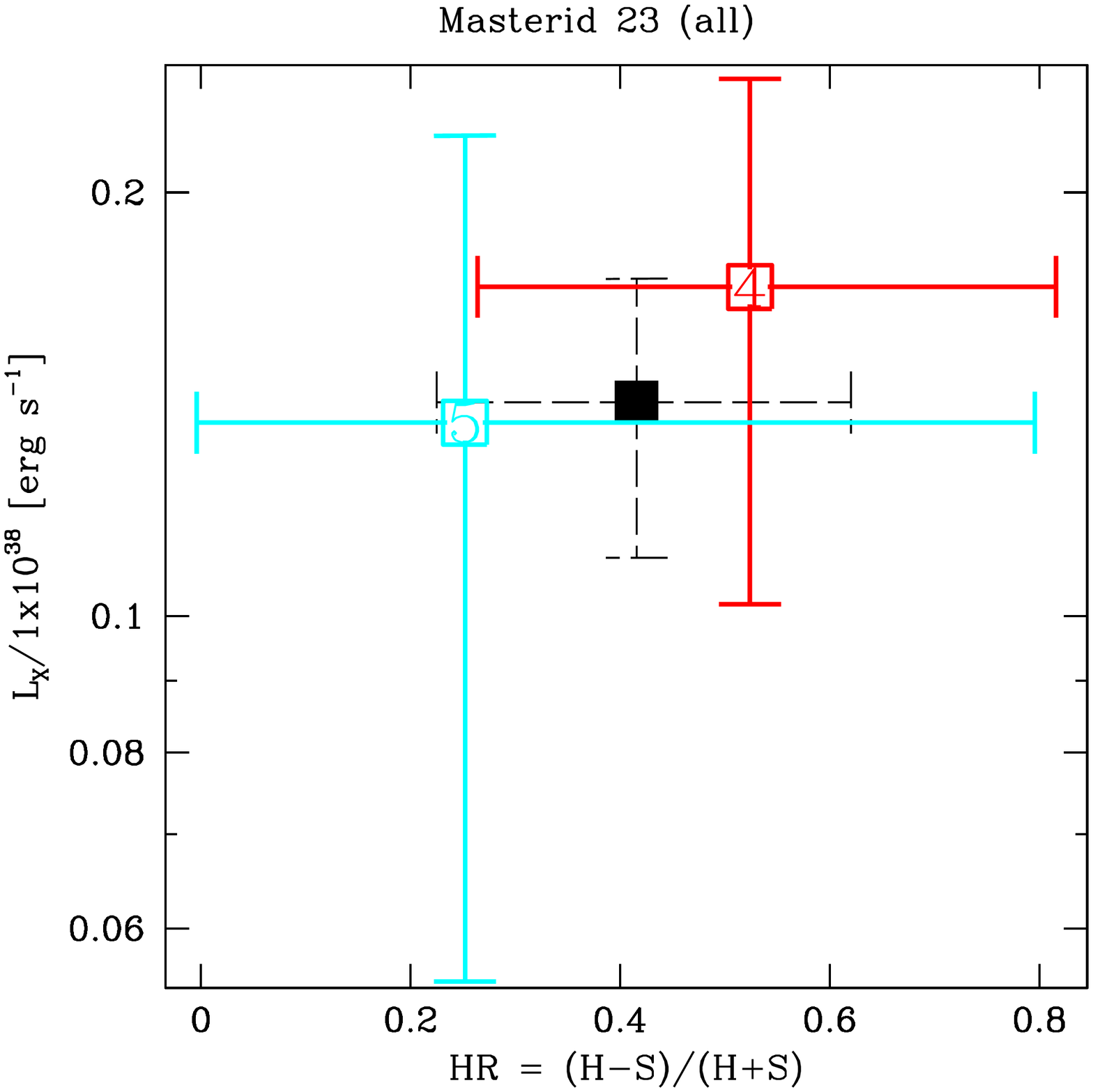}

 \end{minipage}

	\caption{HR$-\ensuremath{L_{\mathrm{X}}}$ plots for each
source detected in more that one individual observation, with each observation plotted in a different color; observation 1
is magenta, observation 2 is green, observation 3 blue, observation 4
red and observation 5 is cyan. The HR and
$\ensuremath{L_{\mathrm{X}}}$ values for the combined observation are
also shown, plotted in black. The hardness ratios are defined to be
HR = H$-$S / H+S, where H is the number of counts in the
hard band (2.0$-$8.0 keV) and S is the number of counts in the soft
band (0.5$-$2.0 keV).  }
\label{fig:lxhrindiv}
\end{figure}

\begin{figure}
  \begin{minipage}{0.32\linewidth}
  \centering
  
    \includegraphics[width=\linewidth]{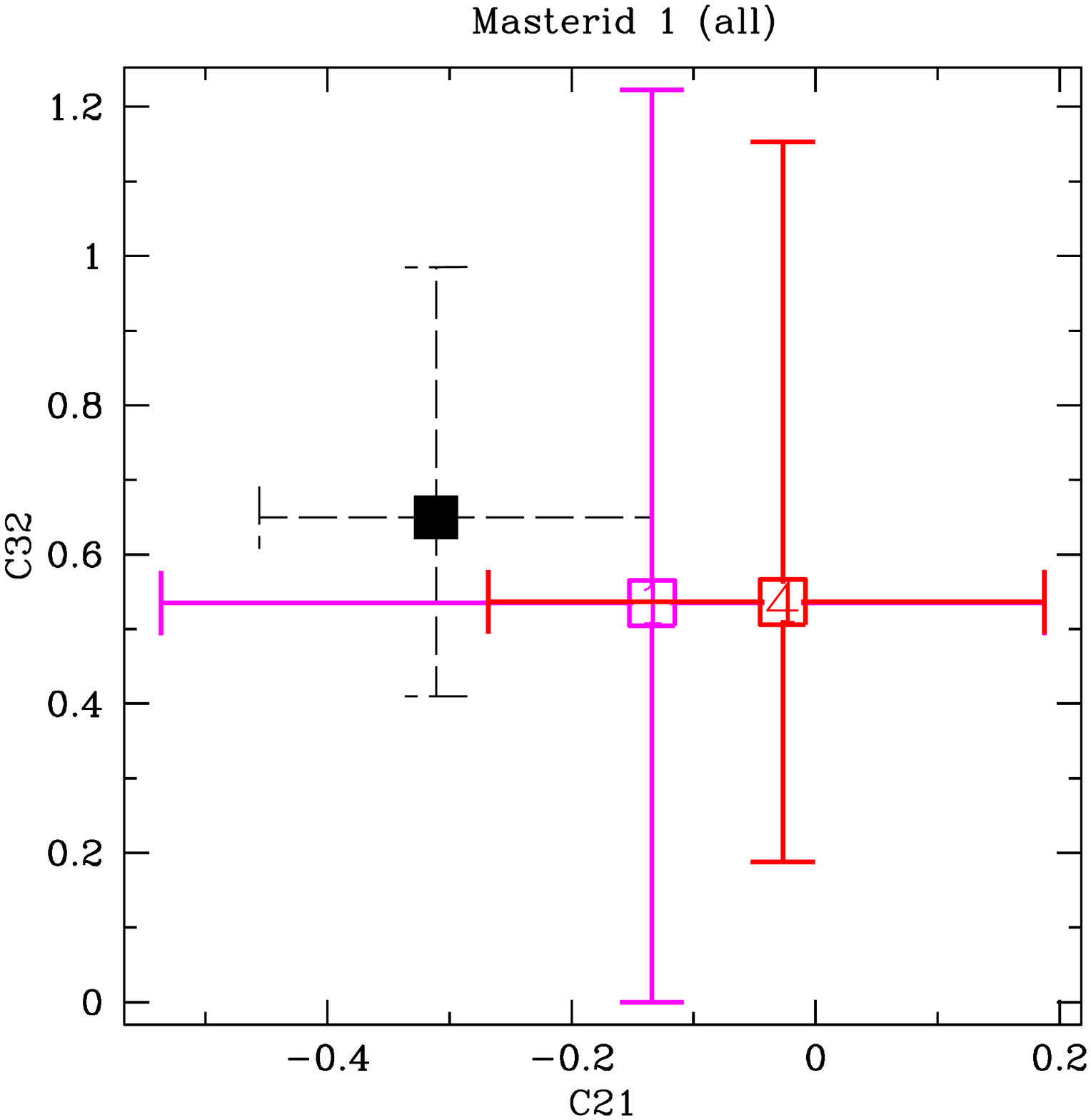}
  
  \end{minipage}
  \begin{minipage}{0.32\linewidth}
  \centering

    \includegraphics[width=\linewidth]{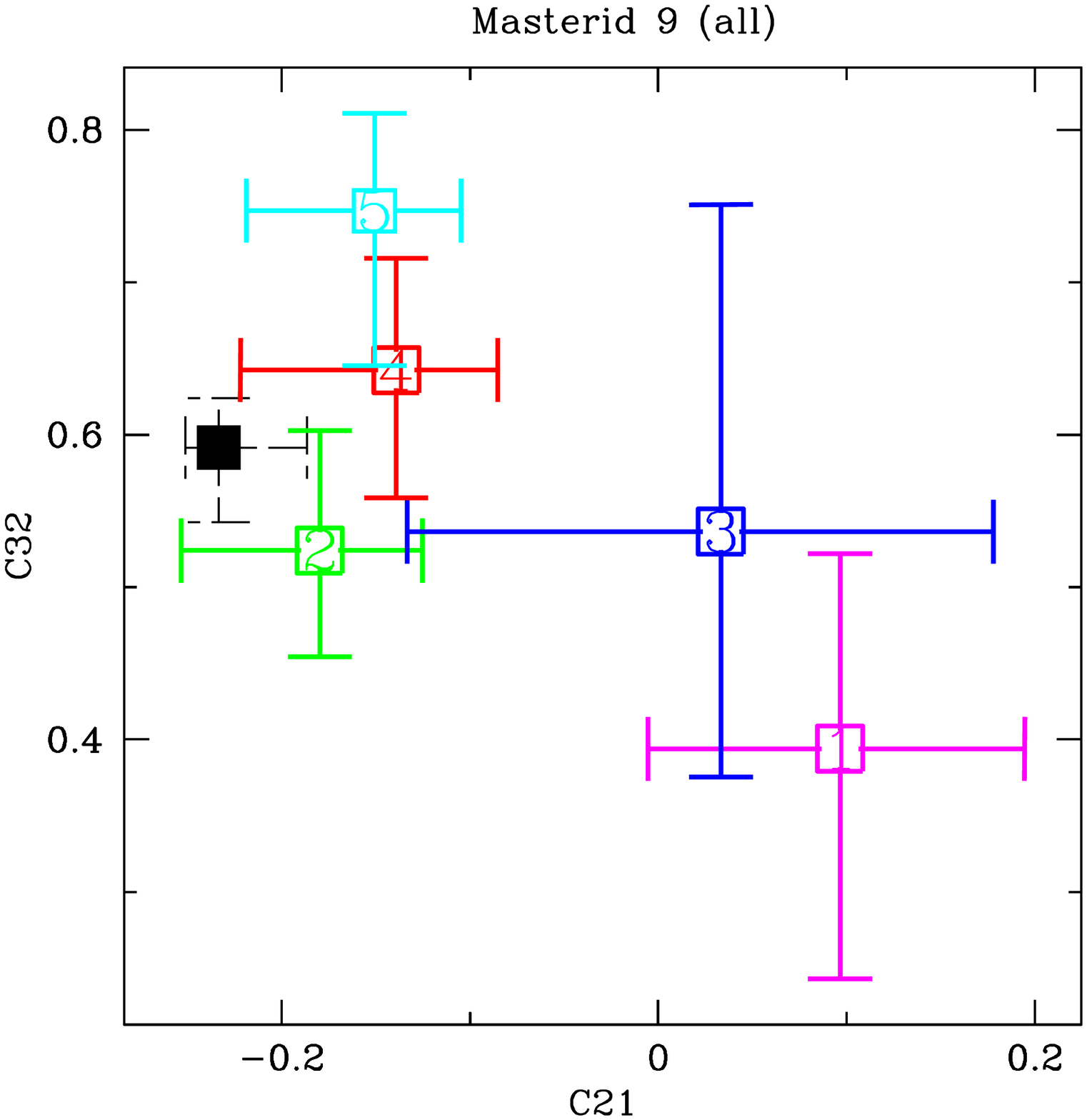}

\end{minipage}
\begin{minipage}{0.32\linewidth}
  \centering

    \includegraphics[width=\linewidth]{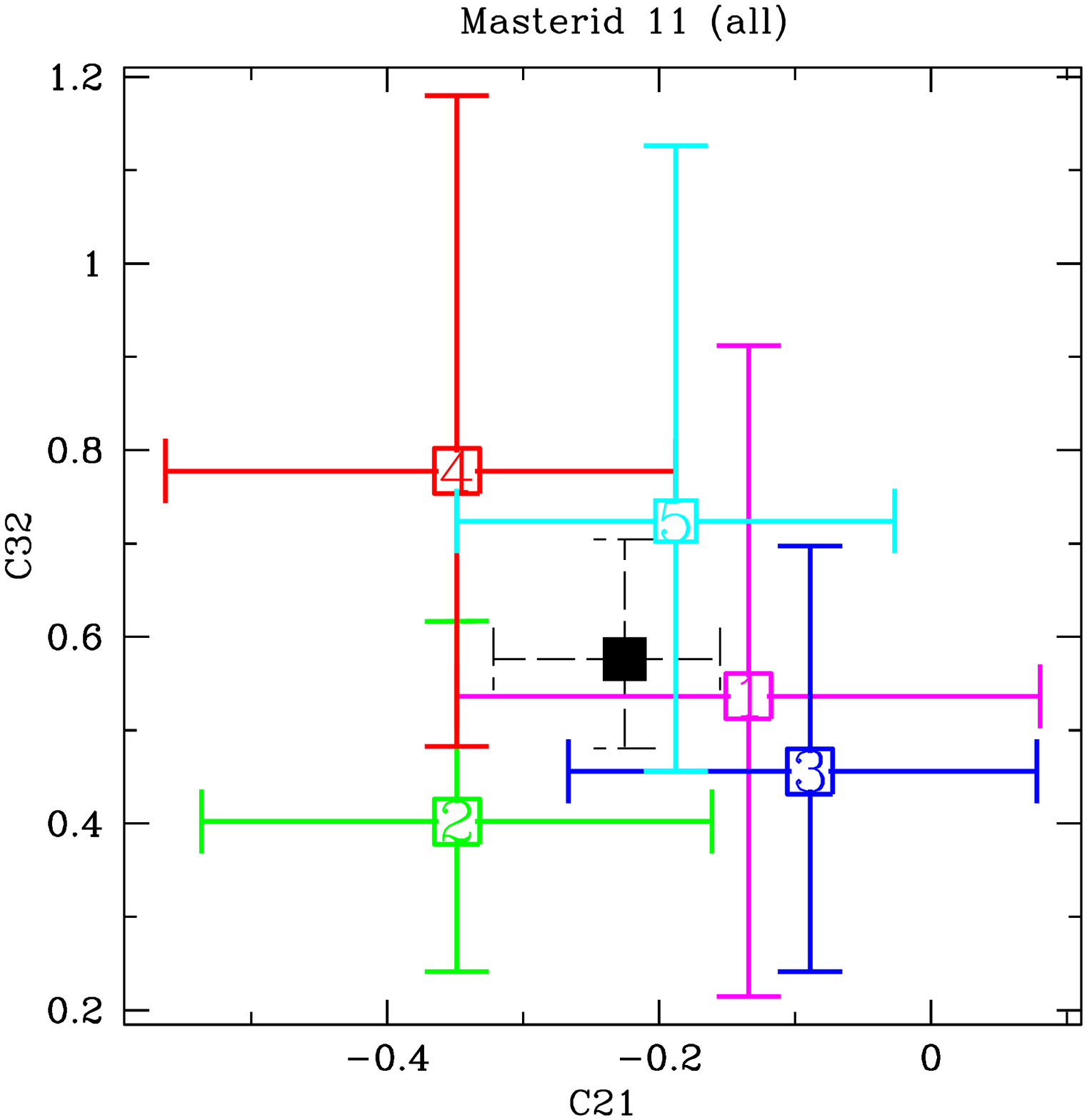}

 \end{minipage}

\begin{minipage}{0.32\linewidth}
  \centering
  
    \includegraphics[width=\linewidth]{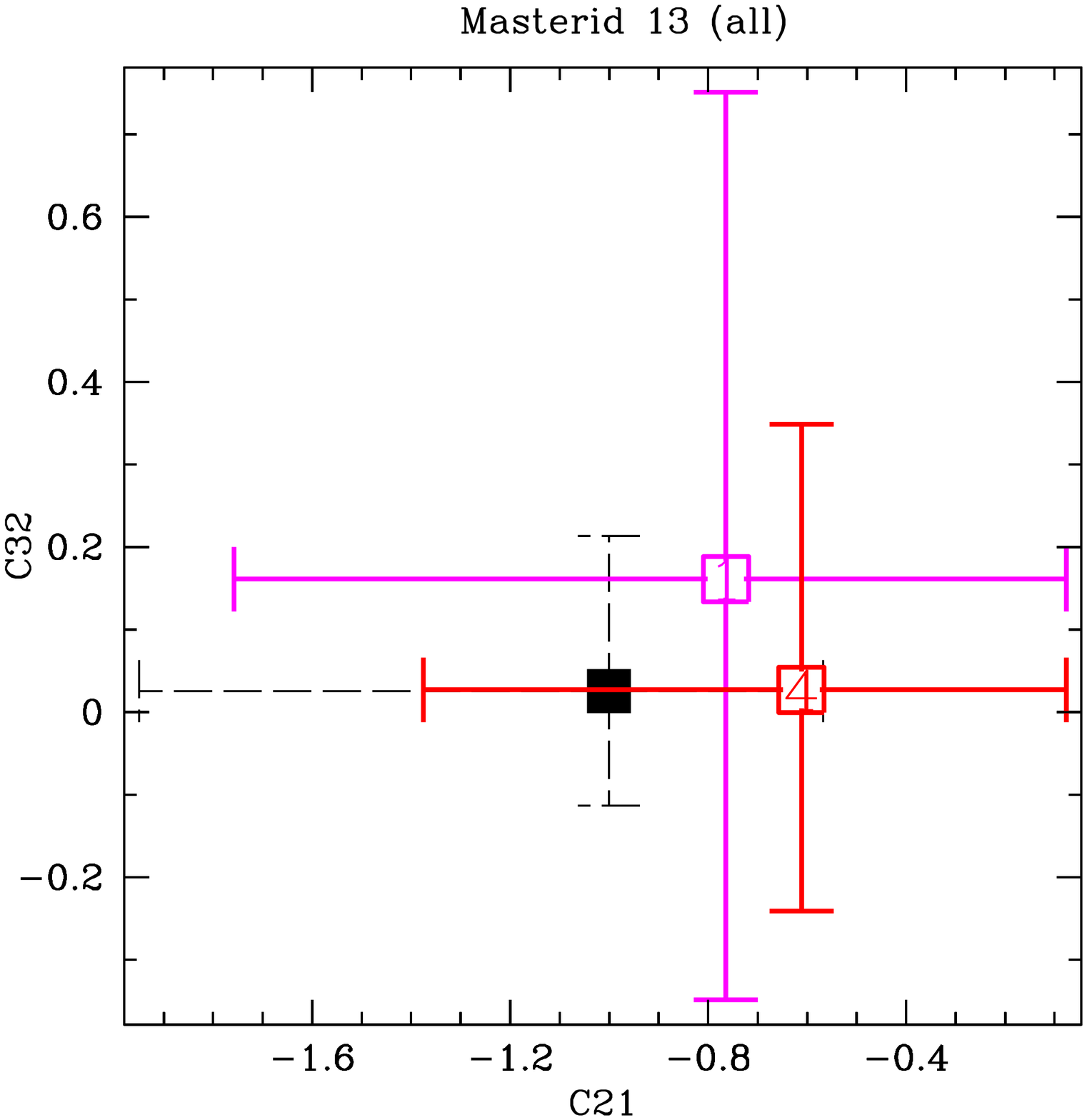}
  
  \end{minipage}
  \begin{minipage}{0.32\linewidth}
  \centering

    \includegraphics[width=\linewidth]{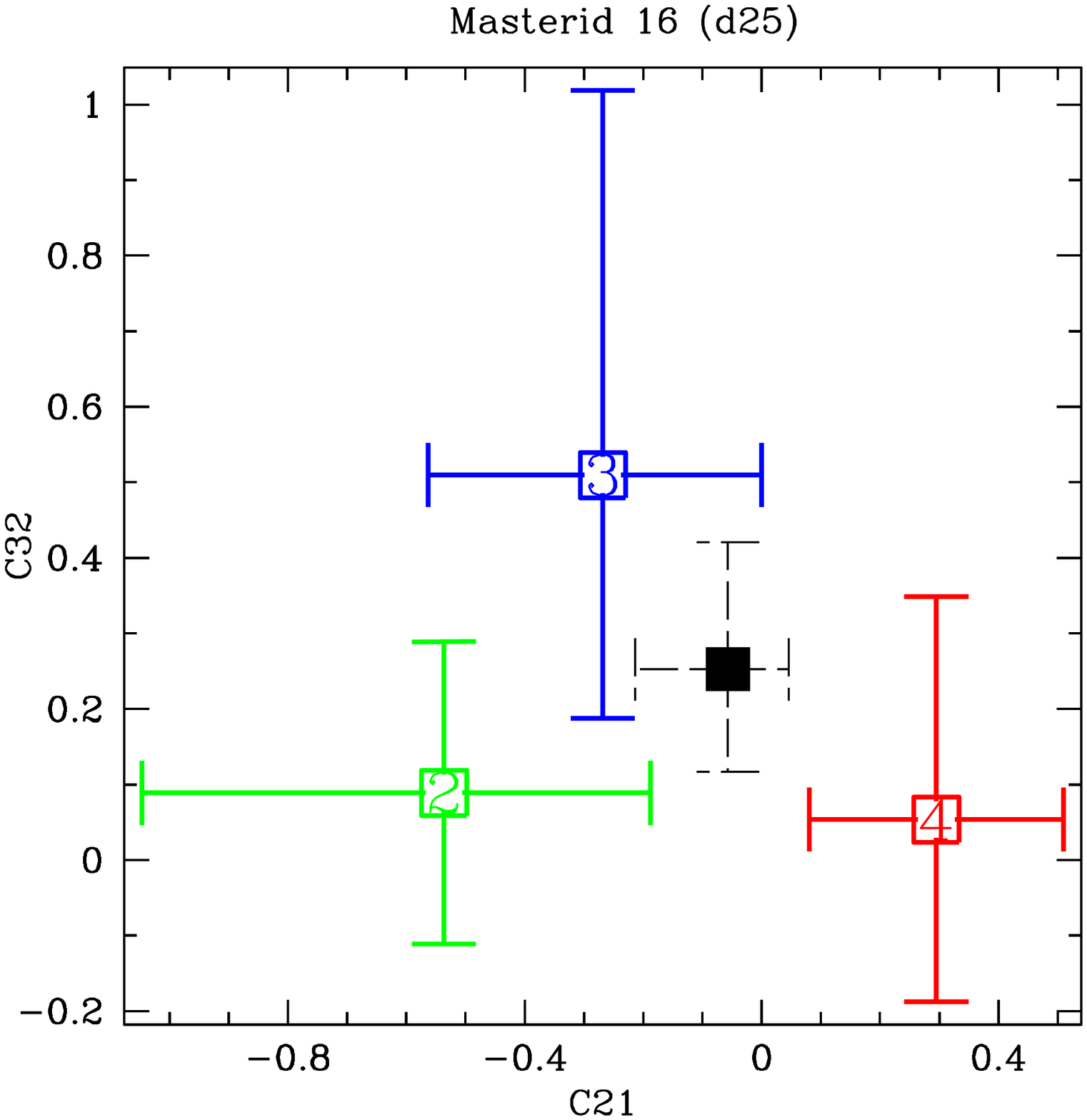}

\end{minipage}
\begin{minipage}{0.32\linewidth}
  \centering

    \includegraphics[width=\linewidth]{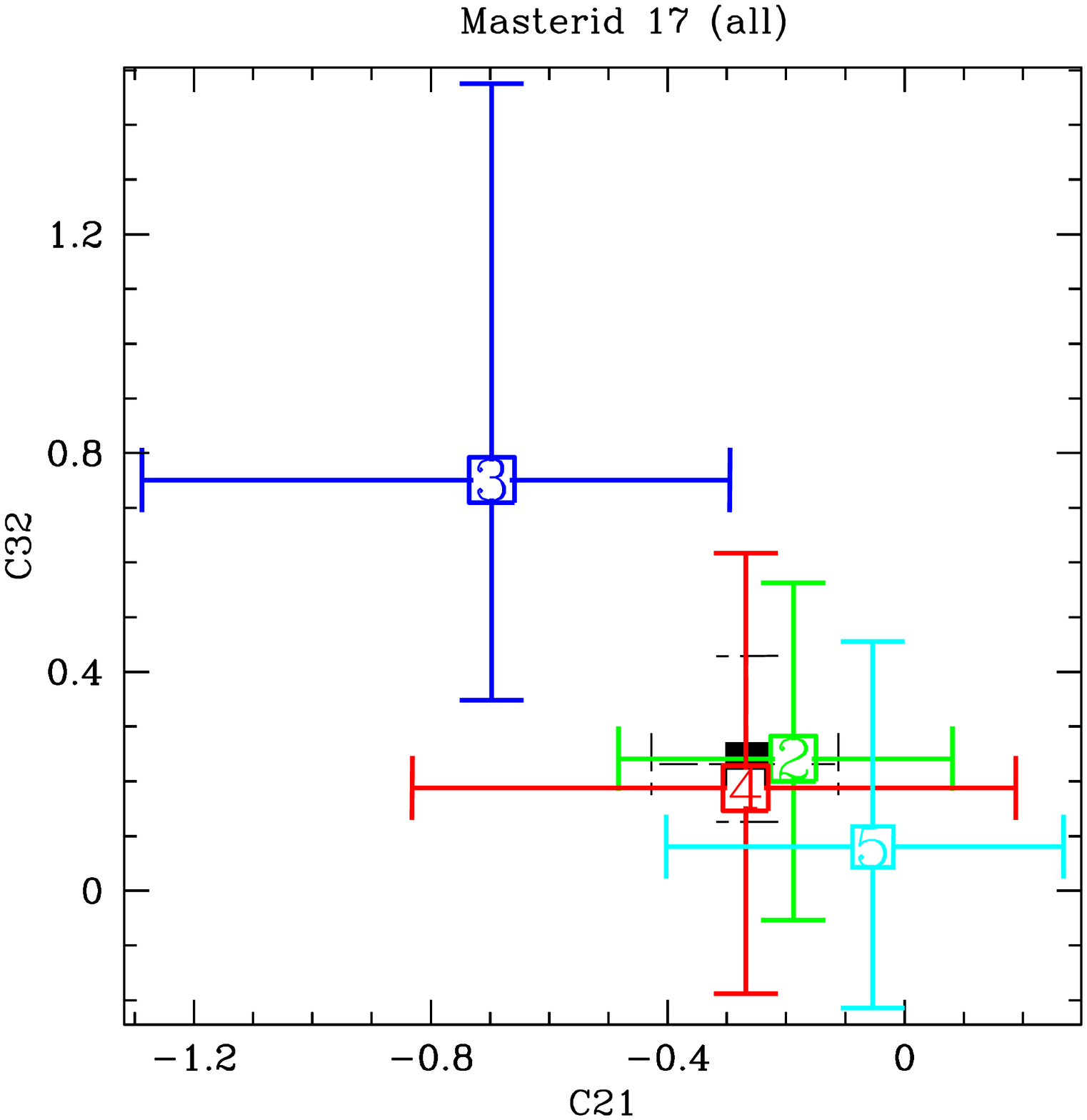}

 \end{minipage}

  \begin{minipage}{0.32\linewidth}
  \centering
  
    \includegraphics[width=\linewidth]{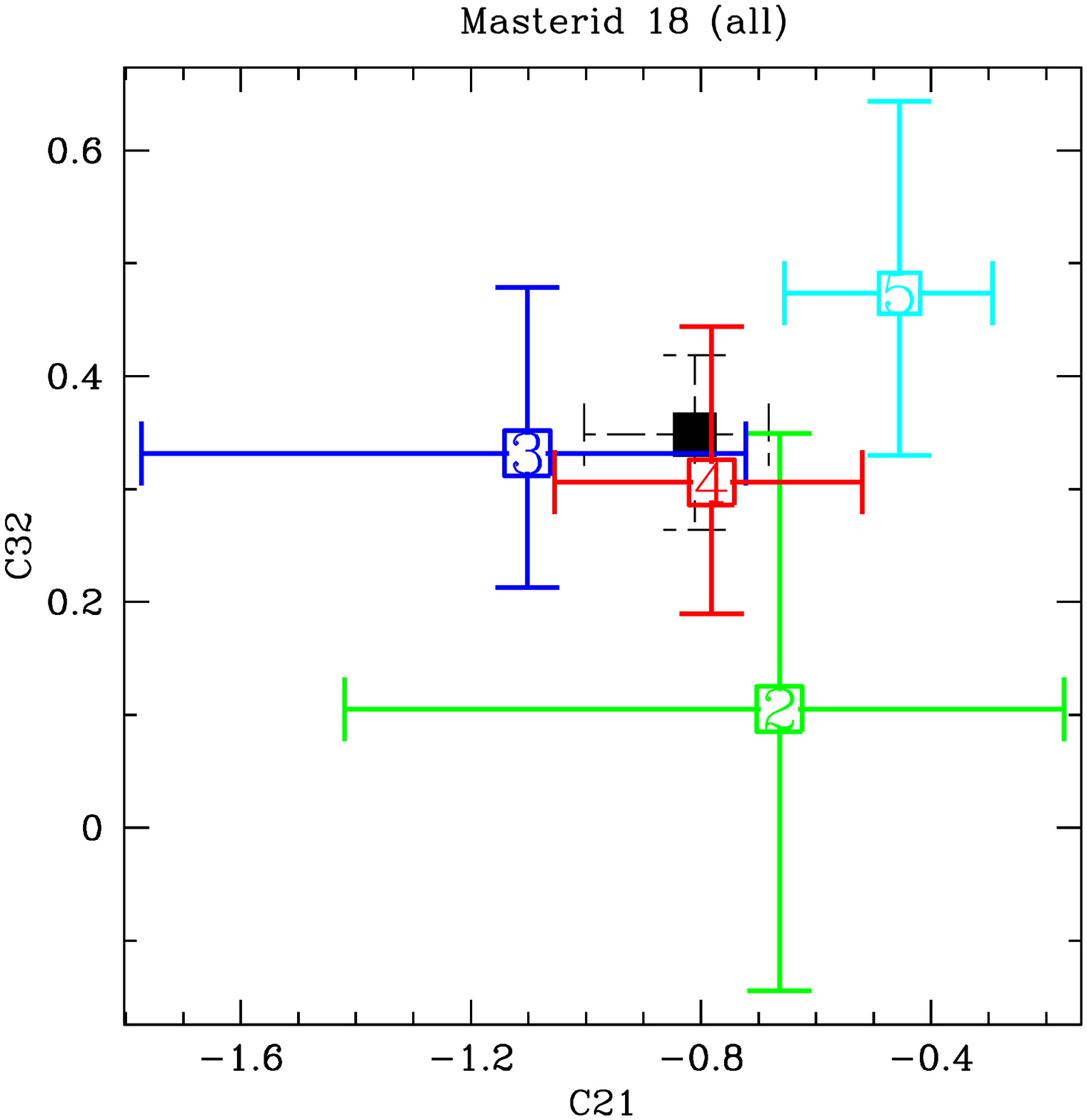}

  \end{minipage}
  \begin{minipage}{0.32\linewidth}
  \centering

    \includegraphics[width=\linewidth]{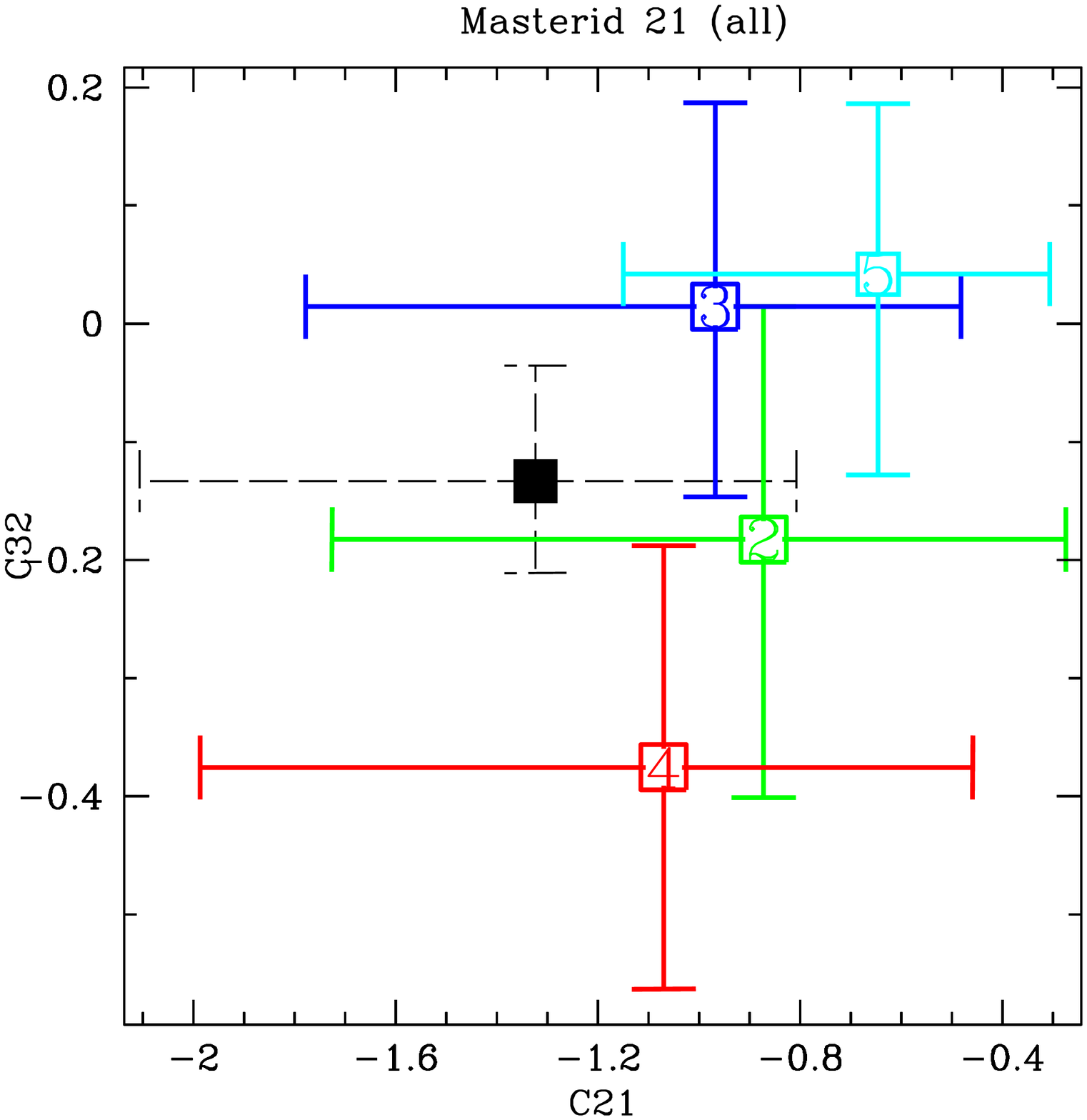}

\end{minipage}
\begin{minipage}{0.32\linewidth}
  \centering

    \includegraphics[width=\linewidth]{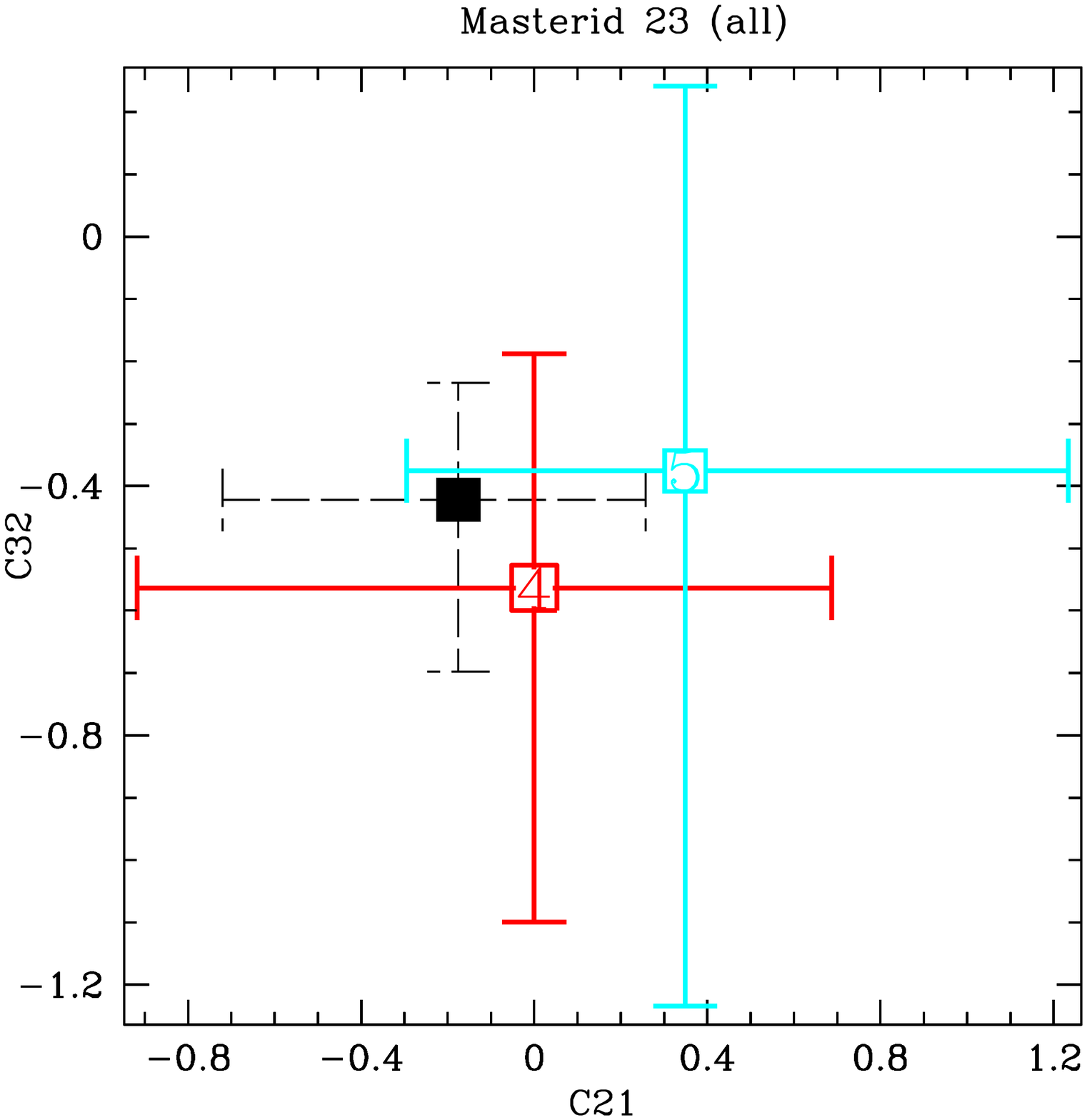}

 \end{minipage}

	\caption{Color-color plots for each
source that has been detected in more
that one individual observation, with each observation plotted in a different color; observation 1
is magenta, observation 2 is green, observation 3 blue, observation 4
red and observation 5 is cyan. The combined observation is also
plotted in black. The color ratios; C21 and C32, are
plotted, where C21={\em log}S2+{\em log}S1 and C32=$-${\em log}H+{\em log}S2. For
the color ratios the bandwidths are defined to be S1=0.3$-$0.9
keV, S2=0.9$-$2.5 keV and H=2.5$-$8.0 keV. }
\label{fig:CCindiv}
\end{figure}

Figure \ref{fig:lxhrindiv} presents the \LX-HR plots for sources with
measured hardness ratios in at least two observations. Each point
shows the X-ray luminosity and hardness ratio value of a source during
each pointing, as well as the values derived from the co-added
observation. Each point is labeled and color coded, where
magenta, green, blue, red and cyan indicate observations 1$-$5 respectively, and
black represents the co-added observation value. Similarly, Figure
\ref{fig:CCindiv} presents the color-color values 
for sources with measured color-color values in at least two
observations, where again individual observations are labeled and
color coded (following the same color scheme as in Figure
\ref{fig:lxhrindiv}), with the co-added observation indicated in
black. 

\section{Discussion}

\subsection{X-ray Source Population}

In the previous sections the data analysis methods, used to determine
the properties of the X-ray binary population of NGC 3379, have been
presented. From the five individual \CHANDRA\ pointings, taken
between February 2001 and January 2007, a co-added observation,
totaling an exposure time of 324-ks, has been produced. From this deep
observation of the galaxy, 132 X-ray point sources have been detected
in the region overlapped by all of the individual pointings, with 98 of these
sources residing within the $D_{25}$ ellipse of the system. These 132
sources are presented in Figure \ref{fig:rgb} where a raw, full band
image from the co-added observation, with the overlap region and
$D_{25}$ ellipse overlaid, is presented in the main image, with source regions
also indicated. The smaller images present the central region of the
galaxy, where the dense population of sources can be
more clearly seen.
Of these 132 sources, based on the hard-band $ChaMP+CDF$ $logN - logS$
relation (Kim \etal\ 2007b), $\sim$36 sources detected in the 
co-added observation are expected to be objects not associated with
NGC 3379. Within the $D_{25}$ ellipse of the galaxy it is expected
that $\sim$17 of these sources are background objects. In Figure
\ref{fig:profile} the number of expected background objects is
indicated in the X-ray source number density profile of the galaxy.

\begin{figure}
\begin{centering}
  \begin{minipage}{0.68\linewidth}
  \centering
  
    \includegraphics[width=\linewidth]{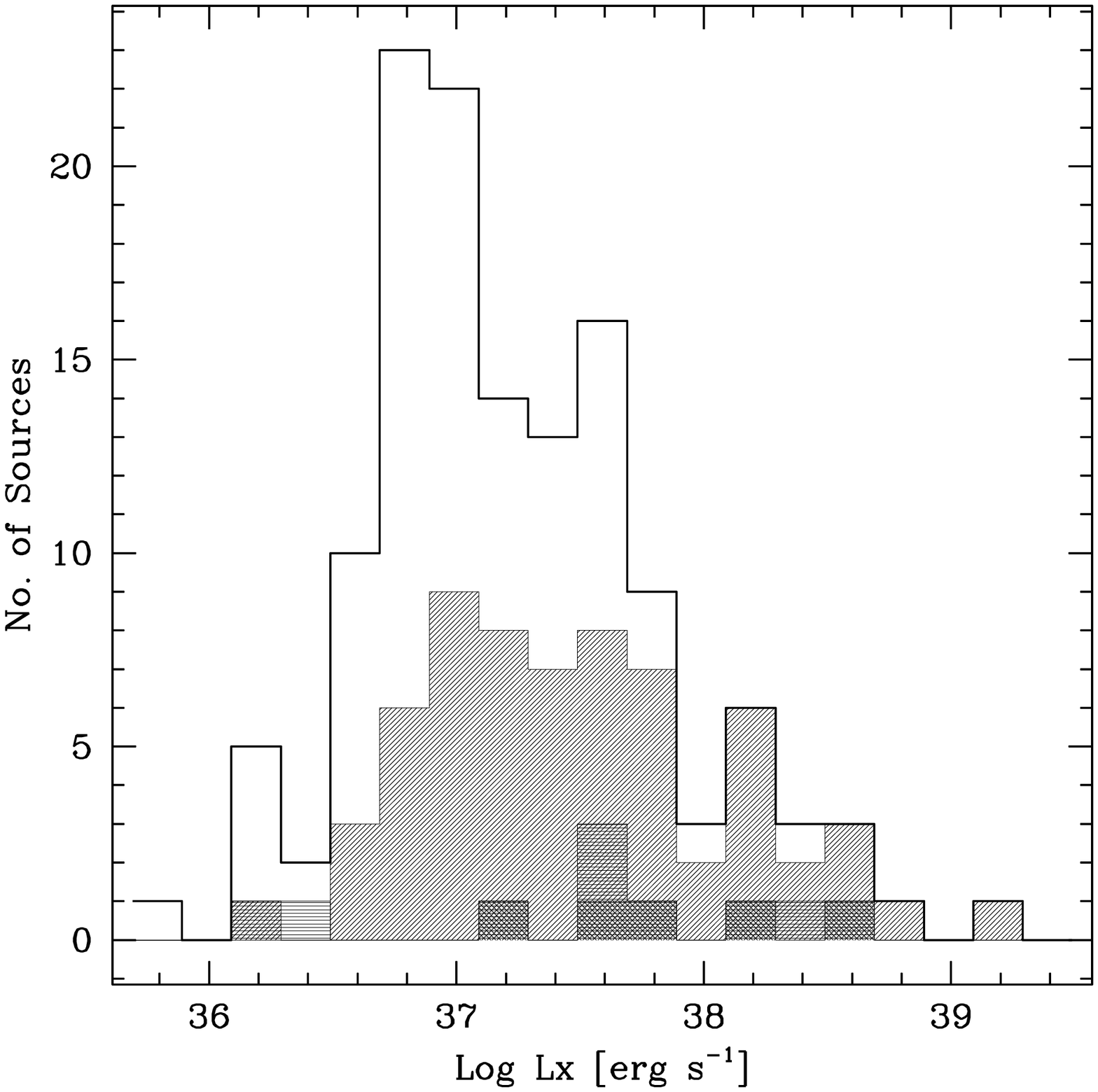}
  
  \end{minipage}
 
\begin{minipage}{0.68\linewidth}
  \centering

    \includegraphics[width=\linewidth]{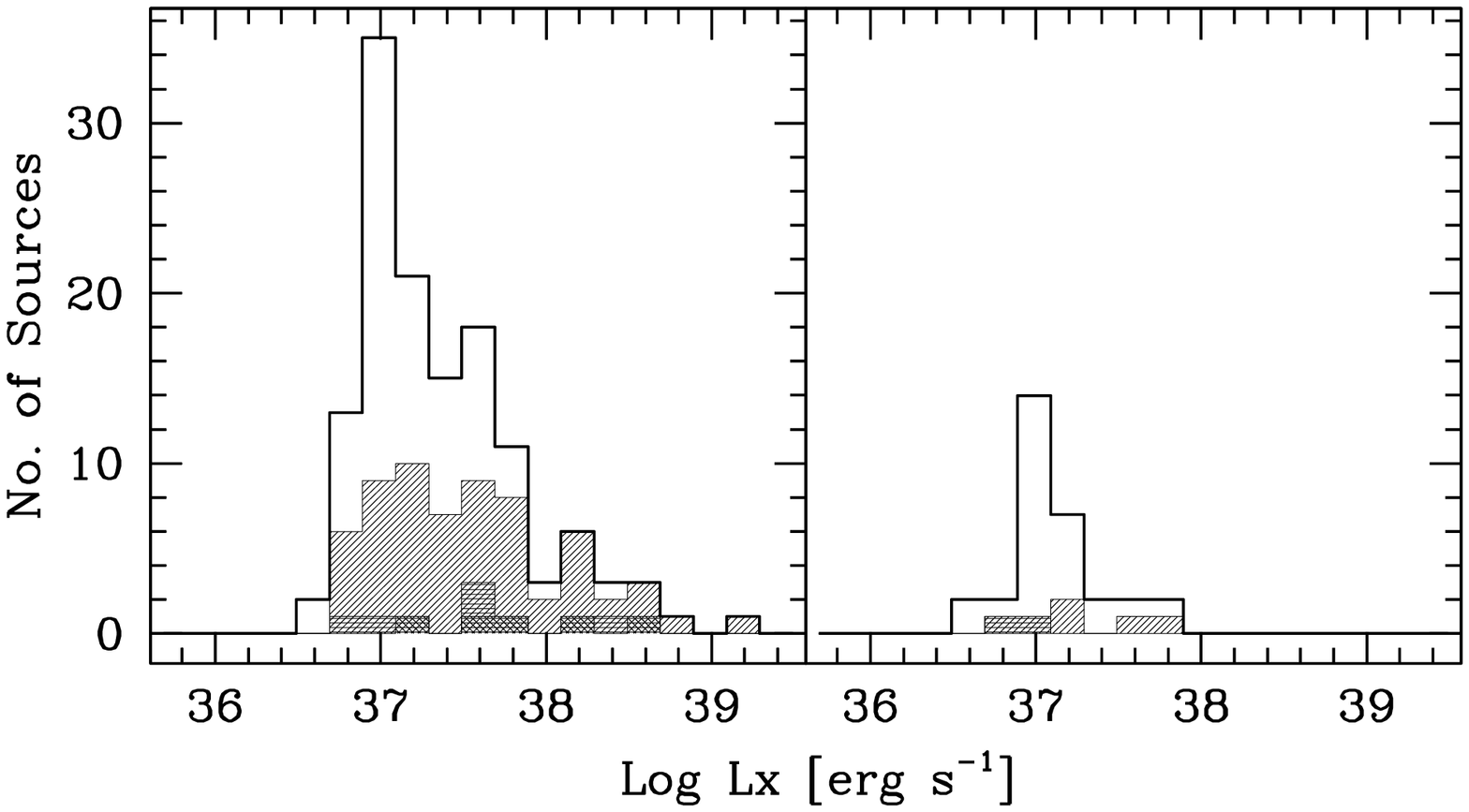}

 \end{minipage}

\caption{The top figure presents the \LX\ distribution of the 132 sources detected within the
overlapping region, covered by all five \CHANDRA\ pointings. The unshaded histogram
indicates sources that have been determined to be non-variable
sources (or we are not able to determine variability), with no GC counterpart. The lightly shaded region shows
variable sources (including both transient classes) that have no GC
counterpart. The darker histogram indicates non-varying sources
associated with a GC and the darkest histogram shows varying sources
that have a confirmed GC counterpart. The bottom left image indicates the same 132 sources, but for those with SNR$<$3, 3$\sigma$ upper limit values have been used in place of \LX. The bottom right image presents these upper limit values only. The shading for these two figures are the same as described for the main histogram.}\label{fig:lxhist}
\end{centering}
\end{figure}

The X-ray luminosity of the sources detected within NGC 3379
ranges from 6$\times~10^{35}$~\ergps\ (with 3$\sigma$ upper limit $\le$4$\times 10^{36}$ \ergps) up to $2\times10^{39}$\ergps,
where the brightest source, 70, is the only ULX detected within this
system. The properties of this source from the first two observations
have been reported in Fabbiano \etal\ (2006), and the full
analysis of the five individual pointings of the ULX will be
presented in the forthcoming paper, Angelini \etal\ (2008, in
prep). The \LX\ distribution of 
all of the detected X-ray sources within NGC 3379 is shown in Figure
\ref{fig:lxhist}, where the GC associations are also
indicated. 

In this figure the main histogram presents the calculated \LX\ values from all sources (with 1$\sigma$ upper limits from the co-added observation provided for sources only detected in a single observation). The bottom left histogram presents these same sources, but for those with SNR$<$3, 3$\sigma$ upper limits are shown), these upper limit values are then presented separately in the bottom right histogram.
From the main figure it can bee seen that the GC-LMXB sources appear to
predominantly lie at the high X-ray luminosity end of this distribution. The properties 
of these 10 GC-LMXB sources and their implications for the
understanding of LMXB evolution in galaxies are presented and
discussed in detail in the companion paper Fabbiano \etal\ (2007). 
Also from this figure it can be seen that the majority of
sources detected from this observation lie in the luminosity range of
$5\times10^{36}$\ergps$-5\times10^{37}$\ergps, with a mode luminosity
of $\sim6\times10^{36}$\ergps\ and with source
incompleteness beginning to affect the source distribution $\sim5\times10^{36}$\ergps. From the histogram including 3$\sigma$ upper limit values, this mode value rises to 1$\times10^{37}$\ergps, with source incompleteness beginning to affect the source distribution $\sim8\times10^{36}$\ergps.

In the forthcoming
paper Kim \etal\ (2008, in prep), the X-ray luminosity function (XLF)
of NGC 3379 will be investigated, and a correction to allow for
source incompleteness will be applied. Some preliminary results, investigating
the XLF of NGC 3379, have been reported in Kim
\etal\ (2006), where sources, down to a 90\% completeness limit of
\LX=1$\times10^{37}$\ergps, from the first two
observations, have been detected. From the even greater sensitivity
afforded to us by combining the five separate pointings, we can
investigate the XLF down to the X-ray luminosity range of normal Galactic
LMXBs. Previously, this has only been possible for the nearby radio galaxy
Centaurus A (NGC 512), where the XLF has been measured down to
$\sim2\times10^{36}$\ergps\ (Kraft \etal\ 2001; Voss \& Gilfanov
2006). With our greater sensitivity we can compare our results to
these studies, allowing us to investigate the shape of the low
luminosity LMXB XLF, although it should be noted that NGC 3379 is a
much more `normal' galaxy than Centaurus A.

In addition the X-ray point sources that have been presented in
this catalog, the optical sources within NGC 3379 have also been
identified. These were detected in a WFPC2 {\em HST} observation,
where 70 confirmed globular clusters have been identified. From these
70 sources, 9 GC-LMXB, with separations $<$ 0.6\arcsec, have been
detected, with one further GC-LMXB connection, found external to
the {\em HST} FOV.

GC-LMXB associations within this galaxy were previously reported
by Kundu, Maccarone \& Zepf (2007), where correlations between the same WFPC2
data used here, and the archival \CHANDRA\ observation, were used
to search for associations. This archival observation is much shorter than
the deep dataset presented here, providing a typical source detection
threshold of $\sim 1-2\times 10^{37}$\ergps. Even so, from their work 7 GC-LMXB
sources were detected. From our study we also detect 7 correlations in
this individual observation, however, due to the better astrometry we
have from our deep observation, used to correct the offsets in
the {\em HST} data, we detect one different optical match to Kundu,
Maccarone \& Zepf (2007), with their additional source determined to be an
`excluded match' in this work.

\subsection{X-ray Colors}
\label{sec:xcol}

\begin{figure}

\begin{centering}
  \begin{minipage}{0.57\linewidth}
  \vspace{-0.5cm}
\includegraphics[width=\linewidth]{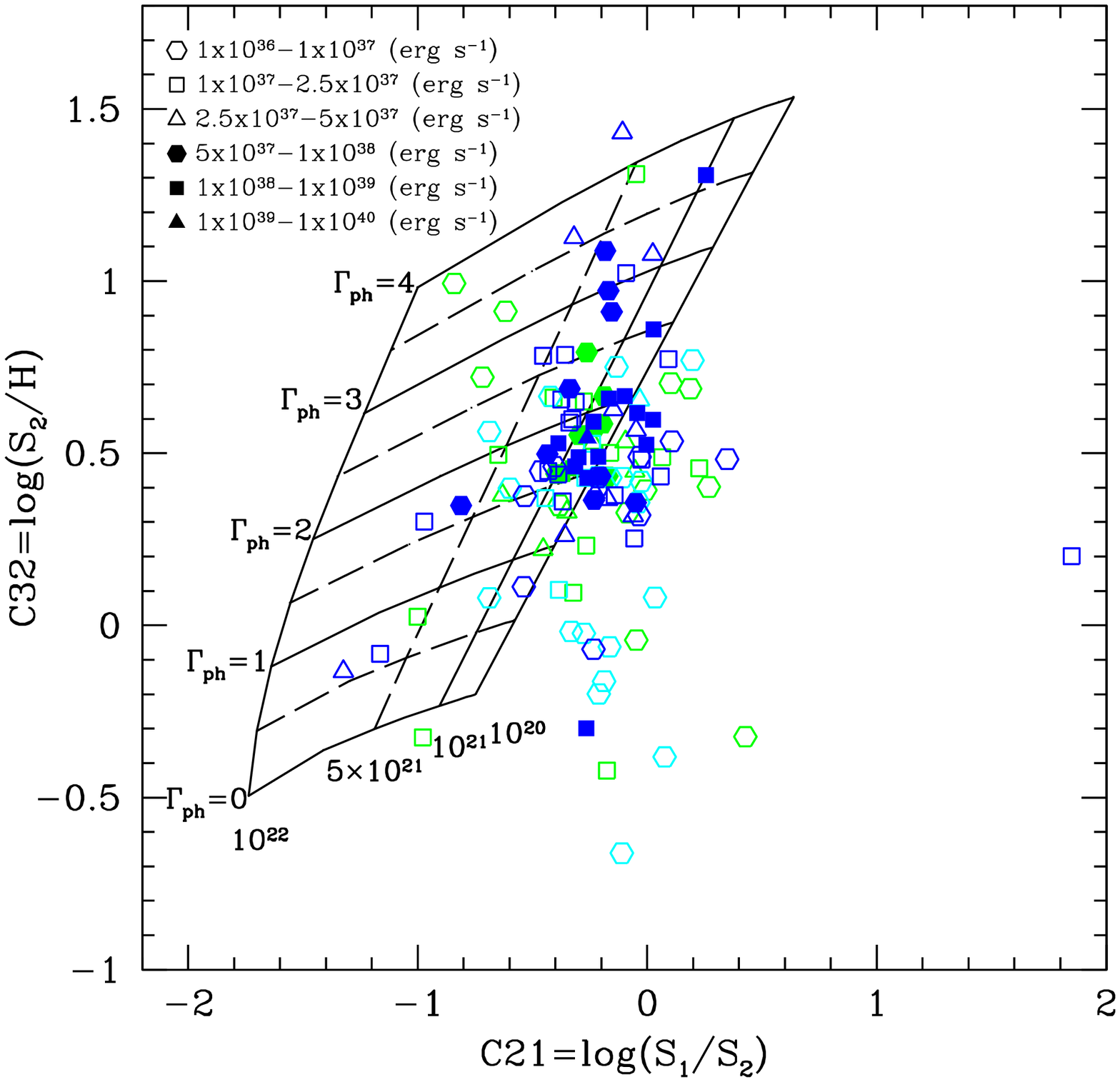}

\end{minipage}

  \begin{minipage}{0.57\linewidth}

\includegraphics[width=\linewidth]{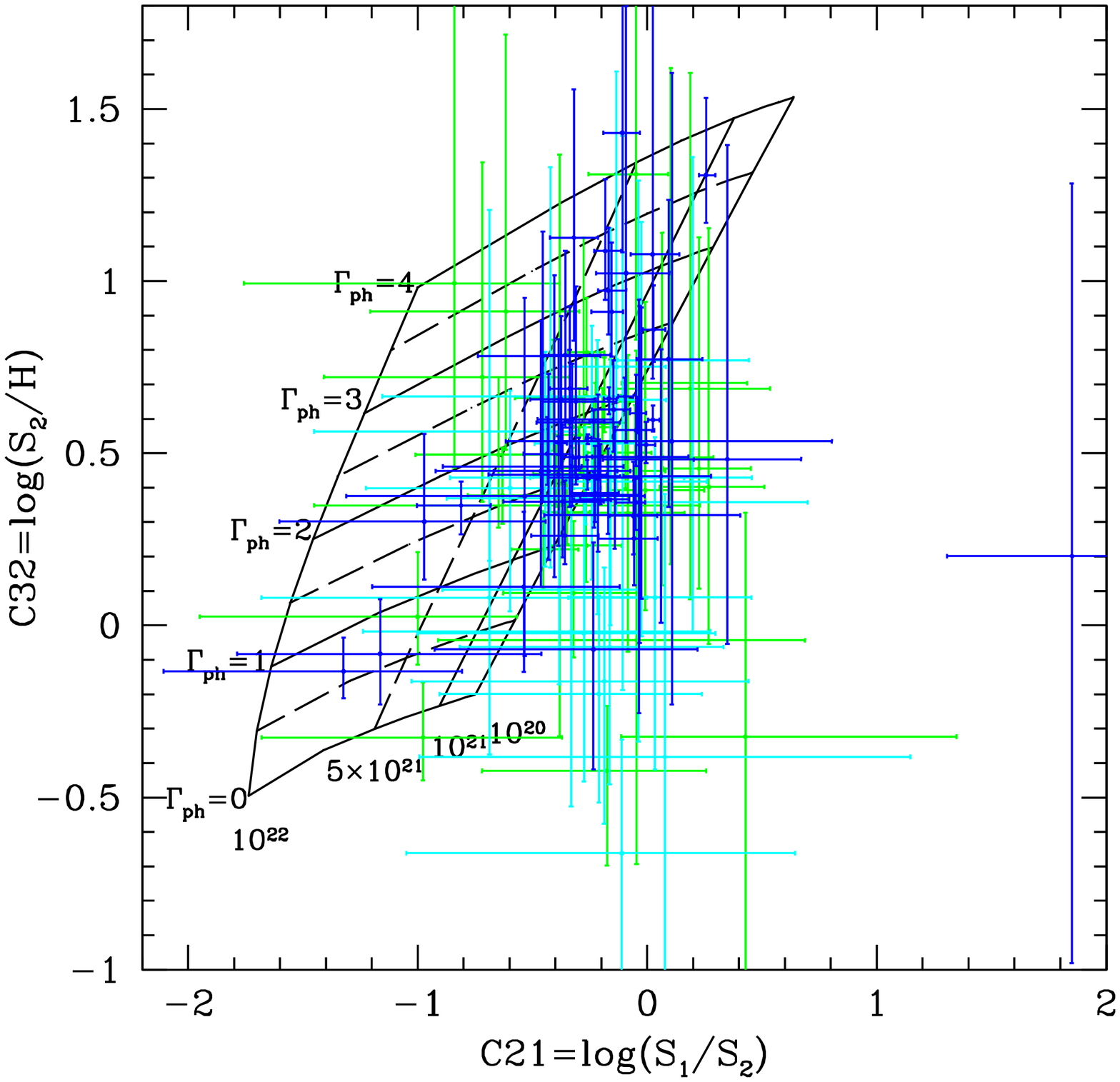}

\end{minipage}
\caption{The color-color diagram of the X-ray point sources detected
in the co-added observation. In the top panel color-color values are
plotted, with the sources divided into
luminosity bins, with symbols of each bin indicated by the labeling in
the panel. Variability is also indicated, where variable sources are
shown in blue, non-variable source are indicated in green and sources that do not have determined variability are shown in cyan. In the
lower panel the error values for each of the sources are presented. In
both of the panels, the grid indicates the predicted locations of the sources at
redshift $z$=0 with various photon indices (0$\le\Gamma_{ph}\le4$,
from top to bottom.) and absorption column densities (10$^{20}\le
$\NH\ $\le10^{22}$ \cmsq, from right to left).}
  \label{fig:cc_pop}
\end{centering}
\end{figure}

\begin{figure}

\begin{centering}
  \begin{minipage}{0.58\linewidth}
  \vspace{-0.5cm}
\includegraphics[width=\linewidth]{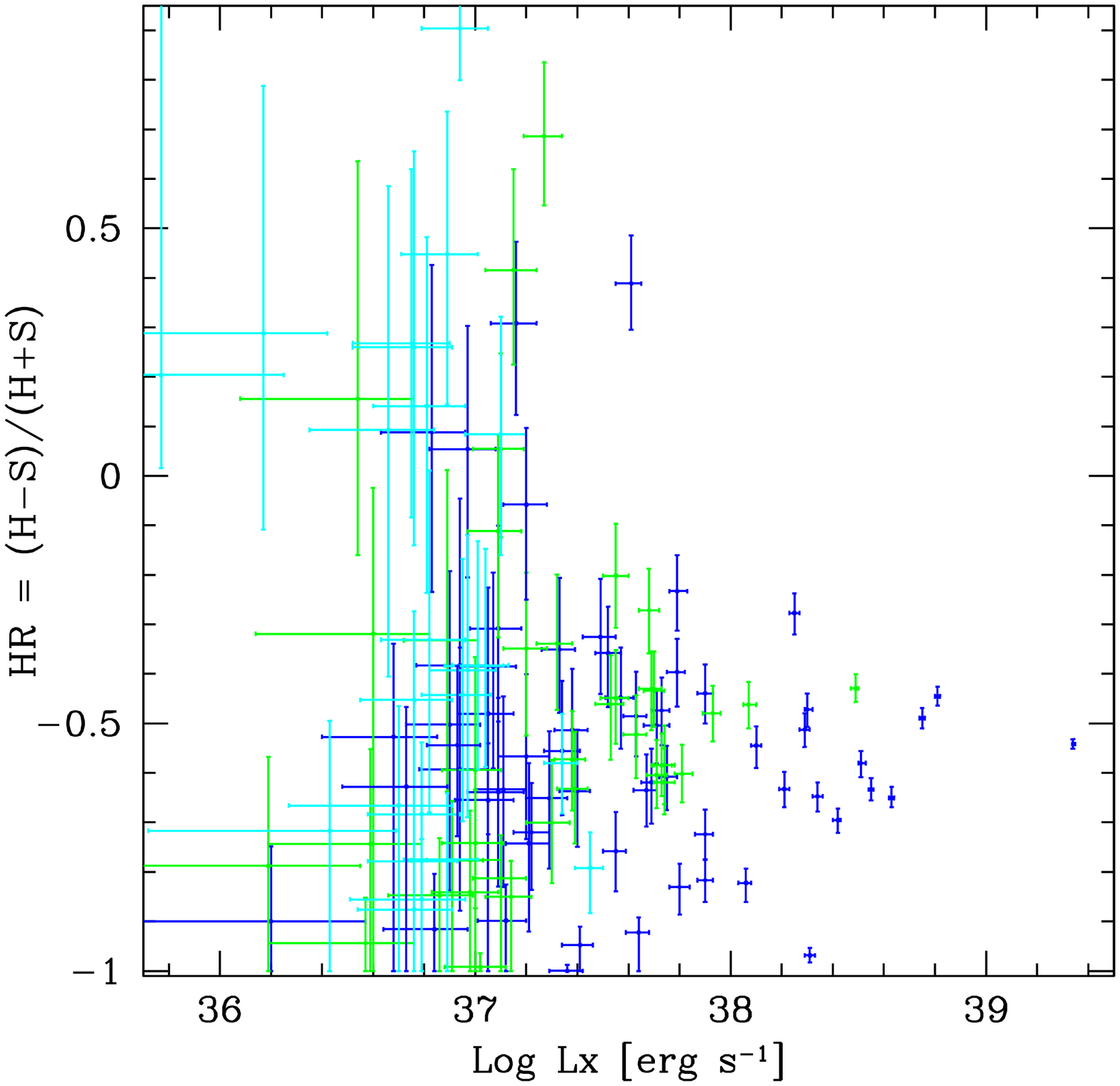}

\end{minipage}

  \begin{minipage}{0.58\linewidth}
\hspace{0.05cm}
\includegraphics[width=\linewidth]{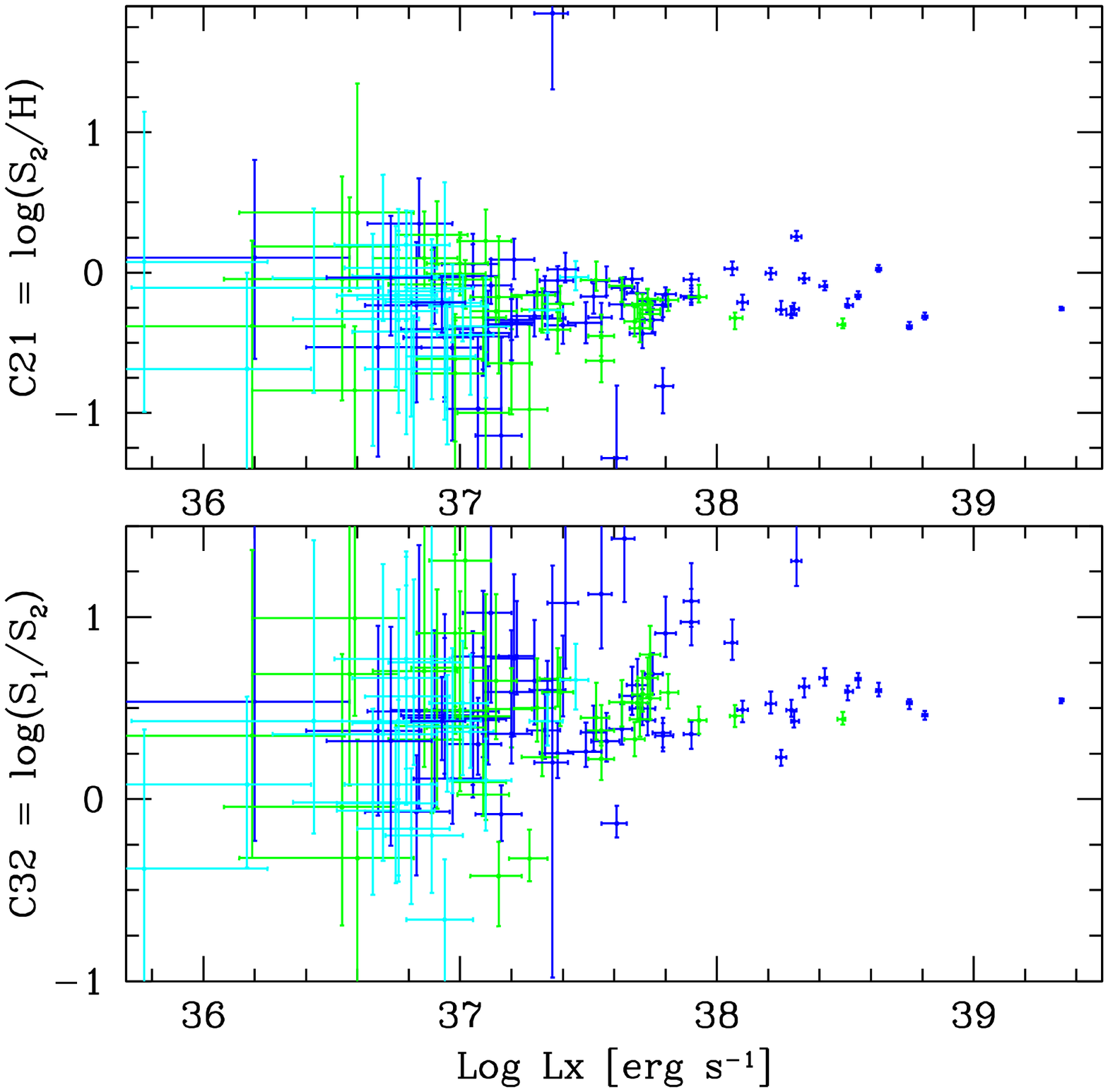}

\end{minipage}
\caption{The top panel presents the \LX-HR diagram of the X-ray point sources detected
in the co-added observation. The second panel shows the \LX-C21 plot
for this population and the bottom panel shows the \LX-C32 values. In
all three panels the variable sources are plotted in blue, non-variable sources are plotted in green and the cyan points indicate sources that do not have enough information to classify their variability.}
  \label{fig:lxhr_pop}
\end{centering}
\end{figure}

In Figure \ref{fig:cc_pop} the LMXB population color-color diagram, based on the
photometry of the co-added observation, is presented. In the top panel
color-color values are plotted, with the sources divided into
luminosity bins, with symbols of each bin indicated by the labeling in
the panel. In the bottom panel, the errorbars for each of these points
are plotted. Also in this figure source variability is indicated,
where variable sources are plotted in blue, non-variable sources
are shown in green and sources with undetermined variability are indicated in cyan. Additionally, in both of the panels a grid has
been overlaid to indicate the predicted locations of the sources at 
redshift $z$=0 for different spectra, described by a power law with
various photon indices (0$\le\Gamma_{ph}\le4$, 
from top to bottom.) and absorption column densities (10$^{20}\le
$\NH\ $\le10^{22}$ \cmsq, from right to left).
In Figure \ref{fig:lxhr_pop} the \LX-HR, \LX-C21 and \LX-C32
population plots are presented, where variability is again indicated
by color, with variable sources shown in blue, non-variable
sources are plotted in green and sources with undetermined variability are shown in cyan.

From the color-color diagram, presented in Figure \ref{fig:cc_pop}, it
can be seen that most of the well defined colors lie within 
the area of a typical LMXB spectrum of $\Gamma=1.5-2.0$, with no
intrinsic absorption (e.g. Irwin, Athey \& Bregman, 2003; Fabbiano
2006). However, there also appears to be a population of sources that
have much harder spectra, again with either no intrinsic absorption,
or sources with a possible soft excess, albeit
with colors that are not as well defined. This sub-population can also
be seen in the \LX-HR population plot presented in the top panel of Figure
\ref{fig:lxhr_pop}, where a significant number of sources have higher
hardness ratios than one would expect from LMXB sources. 

\begin{figure}
\begin{centering}
  \begin{minipage}{0.49\linewidth}
  \includegraphics[width=\linewidth]{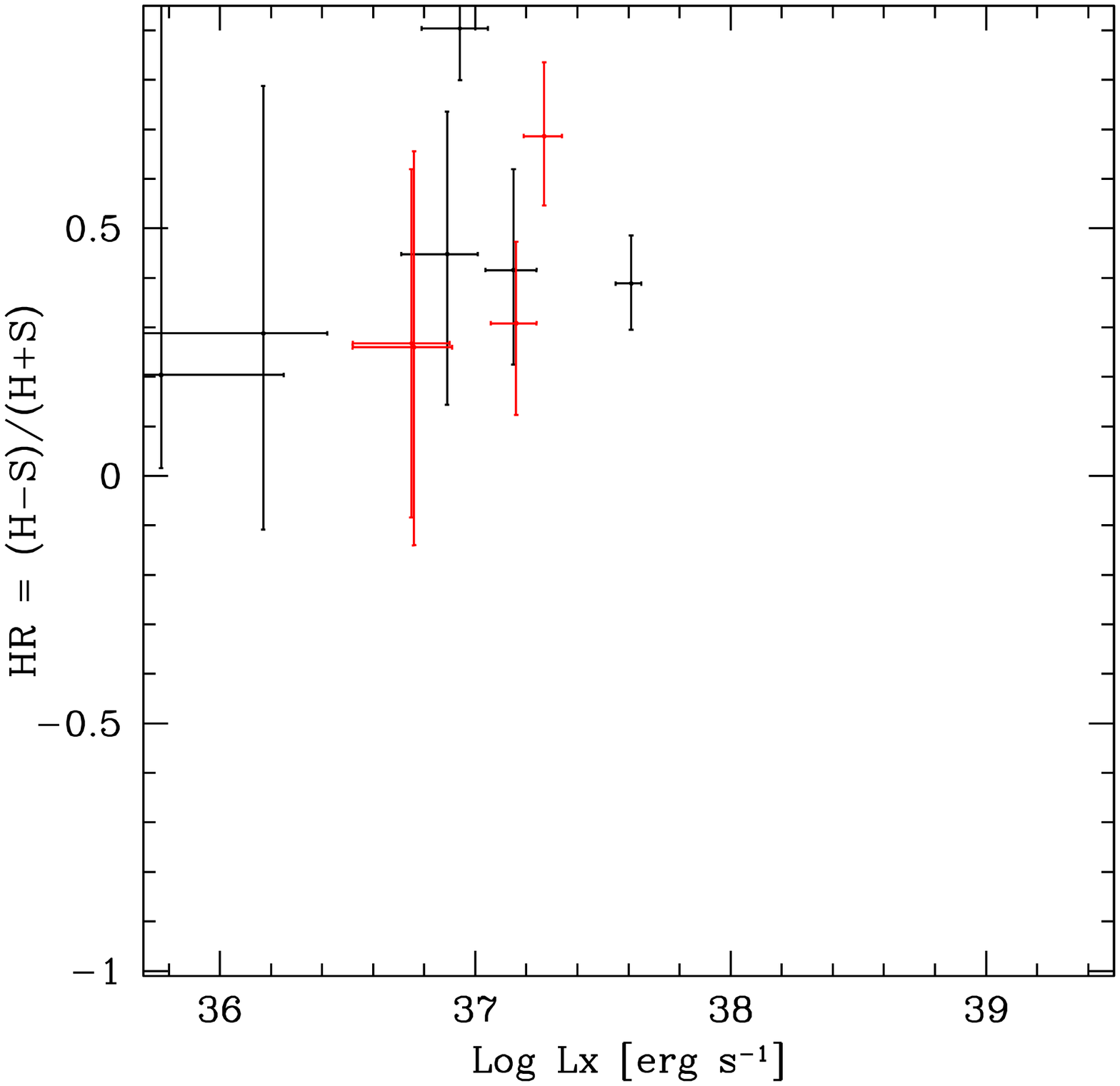}

  \end{minipage}
  \begin{minipage}{0.49\linewidth}

  \includegraphics[width=\linewidth]{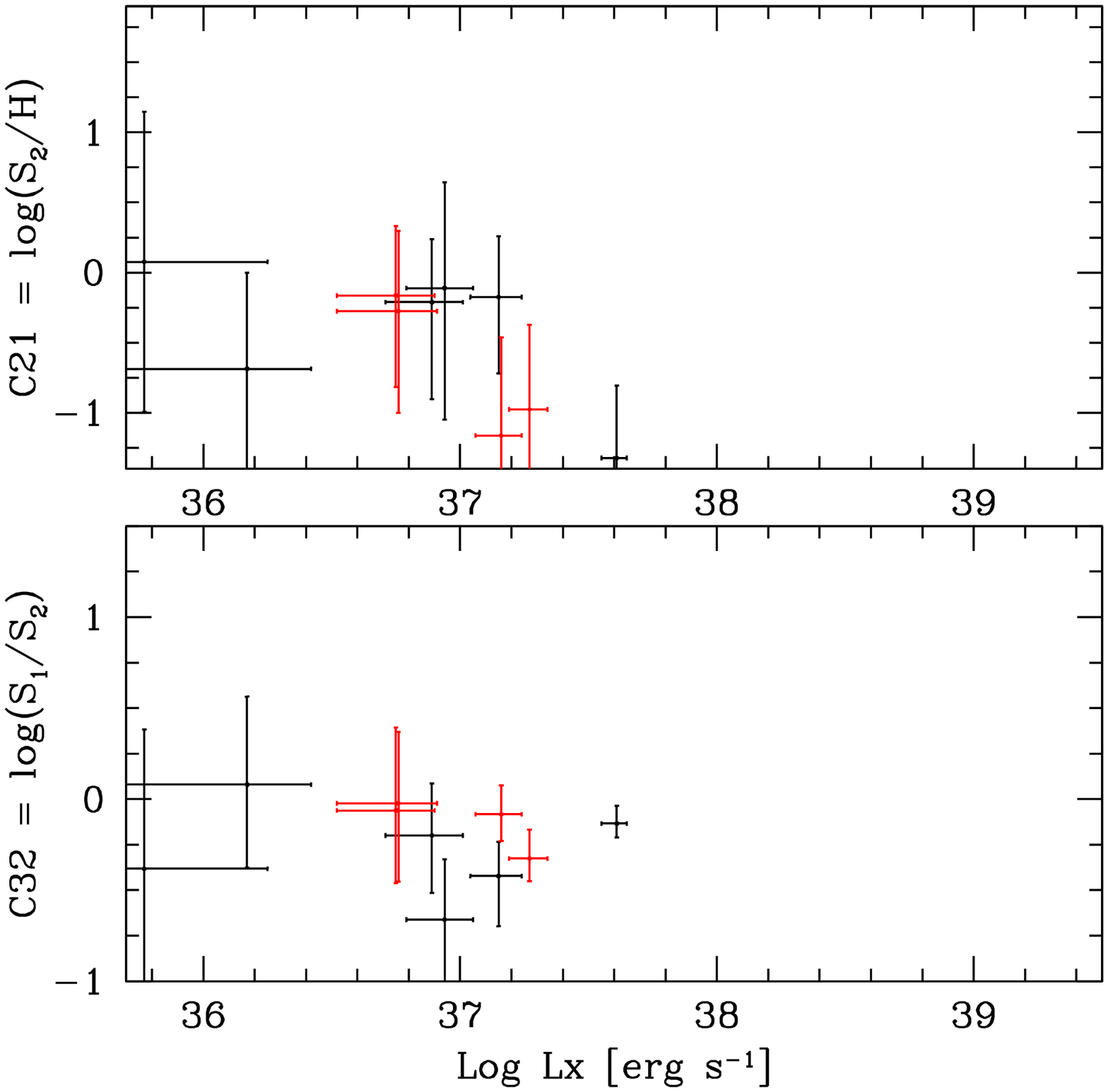}

  \end{minipage}

  \begin{minipage}{0.49\linewidth}

  \includegraphics[width=\linewidth]{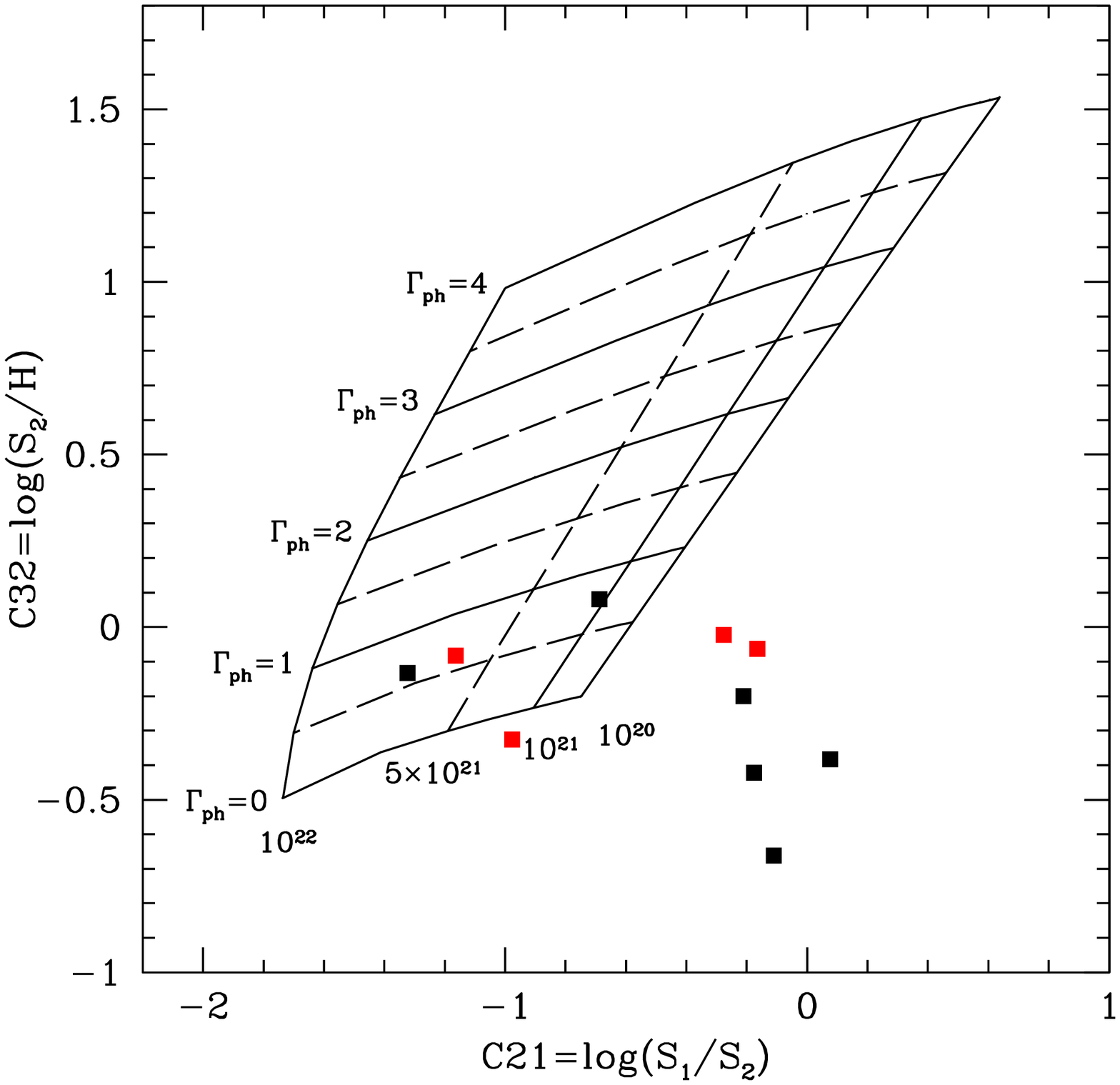}

  \end{minipage}
  \begin{minipage}{0.49\linewidth}
	\vspace{-0.5cm}

  \includegraphics[width=0.95\linewidth]{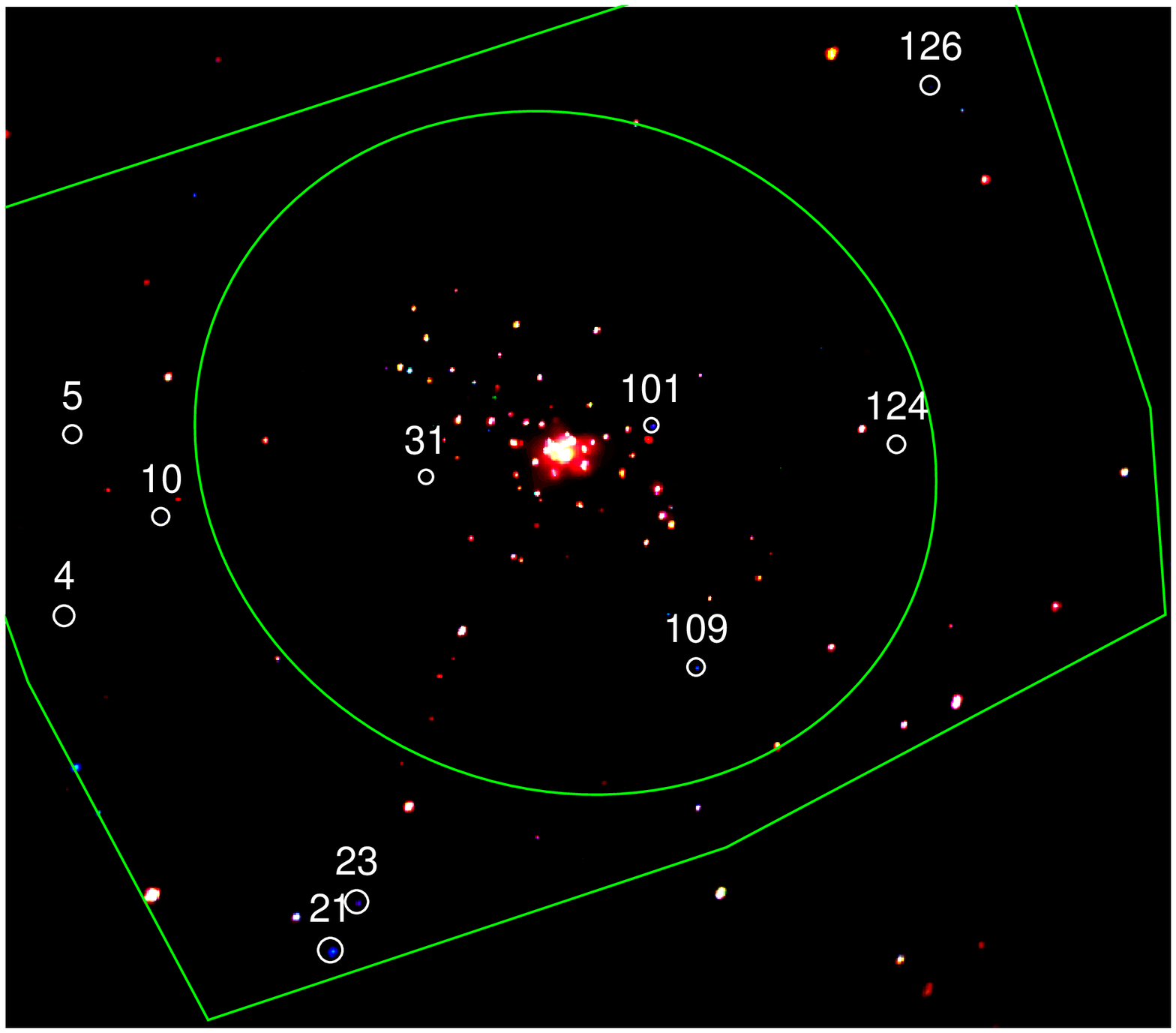}

  \end{minipage}
  \caption{Hardness ratio and color values of the 10 sources that have
been found to exhibit HR values $>$0.2. In the top left panel the
\LX-HR values of these sources are presented. The top right panel
shows the \LX-C21 and \LX-C32 plots for these sources, while the
bottom left panel presents the color-color diagram. In all three of these panels, sources within the
$D_{25}$ ellipse are plotted in red, whilst those external to this
region are shown in black. In the bottom right panel of the figure the
`true color' image of the galaxy is shown, with the $D_{25}$ ellipse
and overlapping region covered in all five pointings overlaid in
green. Also indicated in this image are the 10 X-ray sources, shown in white. }
  \label{fig:hard_pop}
\end{centering}
\end{figure}

This sub-population was investigated by identifying sources with HR
values $>$0.2, resulting in a selection of 10 sources, 4 of which
lie within the $D_{25}$ ellipse of the galaxy. The HR and color-color
values of these objects are presented in Figure \ref{fig:hard_pop}, where
red values indicate sources that lie within the $D_{25}$ ellipse and black points
show those that lie outside of this region. Alongside these plots,
an image indicating the spatial distribution of these objects is also
presented. These plots indicate that these hard sources have similar
C21 values to the majority of the LMXB population but have lower C32
values, indicating that these sources not only have large hardness
ratios but also exhibit spectral hardness in their color values. 

Because many of these objects have
\LX $\le 10^{37}$\ergps, the sources were stacked to ensure that these
harder values are not a consequence of the low counts in the individual
sources. From this stacked photometry, a C21 value of -0.53 (-0.61$-$-0.41) and C32
value of -0.23 (-0.28$-$-0.18) was derived, with a hardness ratio of
0.41 (0.36$-$0.46). These values indicate that these sources are truly
hard objects and, from looking at their spatial distribution
(bottom right panel in Figure \ref{fig:hard_pop}), it is clear
that they are located throughout the galaxy, which suggests that,
coupled with their HR and color values, 
most of these sources are likely to be objects not associated with NGC
3379, possibly absorbed background AGN. This result is consistent with
the number of sources that are expected to be background objects (36)
from the hard-band $ChaMP+CDF$ $logN - logS$ relation. 

However, if we compare the color-color plot in Figure \ref{fig:hard_pop} to the
one presented in Figure \ref{fig:cc_pop}, it can be seen that some of
the harder sources, with no intrinsic absorption (or a soft excess
component) from the whole population plot are not selected with
the HR $>$0.2 cut that has been imposed. From the \LX-HR population
plot in Figure \ref{fig:lxhr_pop} it is clear that there is continuum
of HR values for the sources within NGC 3379, rather than a distinct
separation of different classes. We therefore impose a lower HR value
cut of 0, in an attempt to further identify the population of hard
sources within the galaxy, identified in Figure \ref{fig:cc_pop}. 

Extending this cut down to HR$>$0 increases the sub-population of
harder sources to 17, all of which exhibit colors indicating that they
are sources with hard spectra and no intrinsic absorption. However, 5
of these extra sources lie within the $D_{25}$ ellipse, with only 2
external to this region. This centrally concentrated distribution of
objects indicates that it is likely that some of these additional sources are
associated with the galaxy, and are not AGN. Such an affect is
unsurprising, as it is not unusual for different classes of sources to have
regions of overlap in color-color diagrams (Prestwich \etal\ 2003). 
As a consequence of this confusion, we use the HR$>$0.2 cut to
identify 10 sources that are likely to be absorbed background
AGN, but also note that there are a further seven sources within this
galaxy that, whilst exhibiting lower HR values, also exhibit
spectral hardness in their color values. We suggest here that, due to
their centrally concentrated distribution, it is likely that these sources
are associated with NGC 3379, although, we do not rule out the
possibility that some of these objects could be background AGN.

\subsection{Source Variability}

A characteristic of compact accretion sources such as LMXBs is
variability, and, as a result of the monitoring nature of the
observing campaign, we have been able to search for this variability,
in both the long-term regime, and also over short-term baselines,
where changes over hours and days have been identified. One of the
specific aims of our monitoring campaign has been to identity transient
candidate sources as it has been suggested that field LMXBs are expected
to be transients (Piro \& Bildsten 2002; King 2002) and low luminosity
ultracompact binaries in GCs are also expected to be transient in
nature (Bildsten \& Deloye 2004). In the forthcoming paper Brassington
\etal\ (2008, in prep) we
investigate the sub-population of transient candidates that has been discovered
in NGC 3379. 

Our data represent the most complete variability study for an
extragalactic LMXB population (see Fabbiano 2006; Xu \etal\ 2005),
investigating both long and short term behaviour. In the case of the
long-term variability, sources have been separated into four different
classifications; non-variable and variable sources, and also transient
candidates (TC)
or possible transients candidates (PTC). These two latter definitions have
been applied to sources that either appear or disappear, or are only
detected for a limited amount of `contiguous' time during the
observations, with a lower bound ratio of greater than 10 between the
`on-state' and the `off-state', for TCs, or a lower bound ratio
between 5 and 10 for the PTCs (see \S\ref{sec:var} for a full discussion of this definition). 

The 11 sources that were investigated for transient behavior are presented in Table \ref{tab:trans}, where both the ratio and lower bound ratio, calculated from Bayesian modeling, are presented, along with each source's variability classification. From this table is can be seen that many of these sources appear to be strong TCs from their ratio alone, but when allowing for the uncertainties from their source and background counts, they can only be classified as variable sources. Including the uncertainties when determining TCs is particularly important when dealing with sources with low SNR values, as is the case here for a number of sources in this catalog.

Out of the 132
sources, 56, 42\% of the sources within NGC 3379, have been defined as
variable sources. A further 5 sources are TCs, and 3 are
PTCs, with 44 sources found to be
non-varying in intensity over the five observations. The remaining 24
sources have insufficient data to investigate their long-term variability. These, alongside
the number of sources exhibiting short-term variability, are summarized
in Table \ref{tab:varsum}, where these two variability parameters have
been cross-correlated, to indicate the number of sources exhibiting
both long and short-term variations, although,
the majority of these sources do not have sufficient counts in each
observation to determine their short-term variability. The numbers
within this table indicate the number of sources from the whole observation
and the numbers in brackets represent the sources within the $D_{25}$
ellipse. 

From this table it can be seen that, for the sources with
a defined short-term variability measure, both long-term variable and non-variable
sources have a variety of short-term behavior. For the transient candidates, both classes have few
sources with determined short-term variability, but, for all 
sources that do have short-term measures, all have been found to
also exhibit short-term variability. Also, as an additional point,
nearly all of the TCs, and PTCs found within NGC 3379 reside within
the $D_{25}$ ellipse of the galaxy, with only one confirmed TC,
128, external to this region, indicating that they are likely
LMXBs associated with NGC 3379.

In addition to the \LX\ variability, spectral variations have also
been investigated. These are presented in Figures \ref{fig:lxhrindiv}
and \ref{fig:CCindiv}, where \LX-HR and color-color plots for each
source are shown. From these figures it is clear that the majority of
sources within NGC 3379 are variable, and that their distribution
follows the total source distribution. There is a variety of different
spectral variations within this population, similar to spectral
variability behaviour
discussed in McClintock \& Remillard (2006), with a 
significant number of sources emitting harder spectra as \LX\
increases (e.g sources 25 and 99). Conversely, sources exhibiting
spectral softening with increasing \LX\ are also present within the
galaxy (e.g sources 81 and 119), as well as sources that show little
to no spectral variation with increasing luminosity (e.g. sources 86
and 121), and sources that show no discernible pattern at all (e.g. sources 62
and 98). A more detailed discussion of the spectral variability of
the X-ray sources presented in this catalog will be the subject of a
forthcoming paper.

\section{Conclusions}

We have presented a source catalog and variability atlas resulting from
our monitoring deep observations of the nearby elliptical NGC 3379
with \CHANDRA\ ACIS-S. Our results can be summarized as follows:

\begin{itemize}

\item{132 X-ray point sources have been detected within NGC 3379,
ranging in luminosity from 6$\times 10^{35}$\ergps\  (with 3$\sigma$ upper limit $\le$4$\times 10^{36}$ \ergps) to $\sim2~\times 10^{39}$\ergps, with 98 of these sources residing within the $D_{25}$
ellipse of the galaxy.}

\item{Only one ULX has been identified within this
galaxy, with a galactocentric radius of 6.7\arcsec, and a peak
luminosity of \LX$\sim3\times10^{39}$\ergps.}

\item{Ten globular clusters have been identified to be coincident with
X-ray sources, all of which lie within the $D_{25}$ ellipse of
the galaxy. These GC-LMXB associations tend to have high X-ray
luminosities, with three of these sources exhibiting \LX$>~1\times10^{38}$\ergps.}

\item{From source photometry, it has been determined that the majority
of source with well constrained colors have values that are consistent
with a typical LMXB spectrum of $\Gamma=1.5-2.0$, with no intrinsic absorption.}

\item{A sub-population of 10 sources has been found to exhibit very
hard spectra. These objects are distributed uniformly in the sky and
it is expected that most of these sources are absorbed background
AGN.} 

\item{64 sources, 48\% of the X-ray source population, have been
found to exhibit some type of long-term variability, which clearly
identifies them as accreting compact objects. 5 of these variable
sources have been identified as transient candidates, with a further 3
identified as possible transients.}

\item{Spectral variability analysis has revealed that the sources within NGC
3379 exhibit a range of variability patterns, where both
high/soft$-$low/hard and low/soft$-$high/hard spectral transitions
have been observed, as well as sources that vary in luminosity, but
exhibit no spectral variation, indicating that there are many different
source classes within this galaxy.}

\end{itemize}

In addition to this catalog paper, our companion paper Fabbiano \etal\
(2007) discusses the dearth of low-luminosity GC-LMXBs within this
galaxy. Further highlights from the X-ray binary population of NGC
3379 will also be presented in Brassington \etal\ (2008, in prep),
where the properties of the transient population of NGC 3379 will be presented. 

Forthcoming papers will also present: the properties of the ULX, the
X-ray luminosity function and the diffuse emission of the galaxy, as well
as the properties of the nuclear source and the intensity and spectral variability
of the luminous X-ray binary population.
The results from this deep observation will then be compared to the
X-ray source catalog of the old, GC rich elliptical galaxies NGC 4278,
which has also recently been the subject of a deep \CHANDRA\ observation.


\appendix
\section{Bayesian Estimations of Source Upper-Limits and HR Values}
\label{apen}

 In order to obtain accurate estimates of the source intensities and
their hardness ratios we use a method based on the Bayesian
estimation of the true source intensity in the presence of background,
 and effective area variations  (Park \etal\ 2006).
  This method is based on the posterior predictive probability
distribution of the source intensity, given the  number of
 observed counts (source plus background), and an estimate of the
local background (van Dyk \etal\ 2001). The advantage of this method
is that it
takes into account the Poisson nature of the source {\textit{and}} the
background, allowing a more accurate decomposition  of the
net source intensity. This is particularly important for sources very
close to the detection limit: in these cases the classical method
(e.g. assuming  Gaussian errors, or even the simplified version of the
Gehrels approximation) may give zero or negative counts. However,
 our method overcomes these problems and can provide the full
probability distribution of the source intensities. 
The raw source and background counts (divided by the background to
source area ratio), used to determine the HR and
color values from Bayesian estimations, are presented in Table \ref{tab:raw}.

\begin{figure}
\begin{centering}
\includegraphics[angle=-90,width=\linewidth]{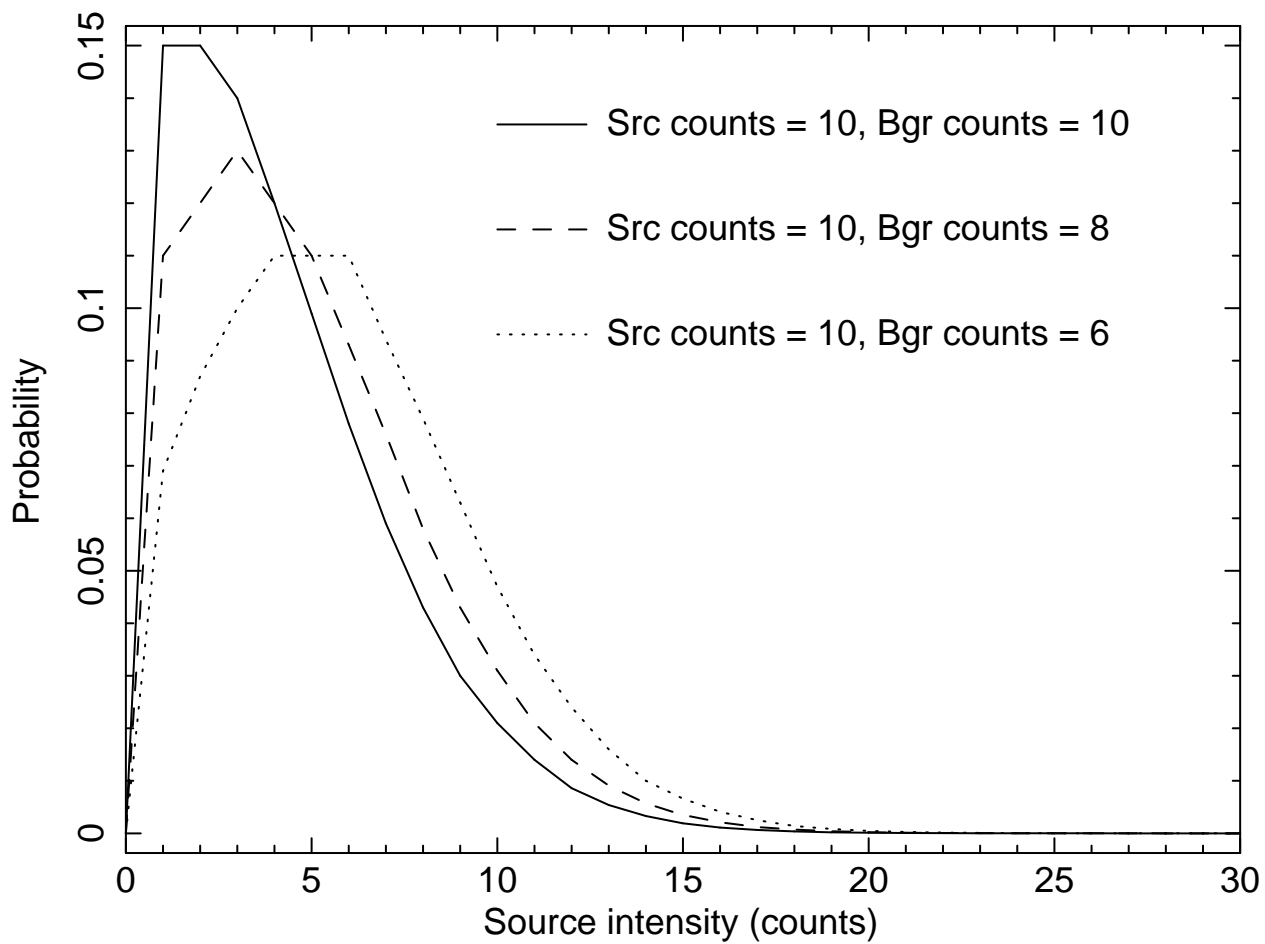}
\caption{The posterior probability distributions for hypothetical
sources with 10 observed counts and background counts of 6, 8 and 10.
Based on Bayesian estimations of the `real' sources intensity. 
 }\label{fig:bayes}
\end{centering}
\end{figure}

For sources very
close to the detection limit we can obtain the mode of the
distribution which, although lower than the average background, might
well be above zero. In those cases we can also estimate the upper 68\%
quantile of the distribution, which would correspond to the 68\%
confidence level on the source intensity (a similar method, which
however does not model the Poisson probability distribution of the
background counts, is presented in Kraft \etal 1991).
 In Figure \ref{fig:bayes} we present the posterior probability distributions for
 hypothetical sources with 10 observed counts and estimated
background of 6, 8, and 10 counts. None of these cases is a formal
`detection' (i.e. flux intensity at least 3$\sigma$ above the background)
in the classical method\footnote{We note however that
significance of the source flux has a different meaning from the
significance of the detection which strongly depends on the detection
process; a more detailed description of these nuanced terms will be
presented in van Dyk \etal\ (2008, in prep).}, however, from these distributions we can
recover the source intensity, and even in the most extreme case (10
background counts) we can estimate the upper confidence bound on the
source intensity.

\acknowledgments

 We thank the CXC DS and SDS teams for their efforts in reducing the data and 
developing the software used for the reduction (SDP) and analysis
(CIAO). We would also like to thank the anonymous referee for helpful
comments which have improved this paper. 
This work was supported by {\em Chandra} G0 grant G06-7079A
(PI:Fabbiano) and subcontract G06-7079B (PI:Kalogera). We acknowledge
partial support from NASA contract NAS8-39073(CXC). A. Zezas
acknowledges support from NASA LTSA grant NAG5-13056. S. Pellegrini
acknowledges partial financial support from the Italian Space Agency
ASI (Agenzia Spaziale Italiana) through grant ASI-INAF I/023/05/0.


{}

\clearpage


\begin{deluxetable}{ccccc}
\tabletypesize{\scriptsize}
\tablecolumns{5}
\tablewidth{0pt}
\tablecaption{Observation Log\label{tab:log}}
\tablehead{\colhead{Obs. Num.} & \colhead{OBSID} & \colhead{Date} &
\colhead{Exposure (sec)} & \colhead{Cleaned Exposure (sec)}  }
\startdata
1  & 1587 & 2001-02-13  &    31519.0  &  28951.9 \\
2  & 7073 & 2006-01-23  &    84099.0  &  80347.1 \\
3  & 7074 & 2006-04-09  &    69066.2  &  66696.6 \\
4  & 7075 & 2006-07-03  &    83110.1  &  79555.7 \\
5  & 7076 & 2007-01-10  &    69249.5  &  68657.1 \\
Total & - & -           &   337043.8  & 324208.4 \\
\enddata
\end{deluxetable}


\begin{deluxetable}{lcccccc}
\tabletypesize{\scriptsize}
\tablecolumns{7}
\tablewidth{0pt}
\tablecaption{Average Right Ascension and Declination offsets of the
twenty brightest sources, detected in all five observations, from the
source position defined in the co-added observation.\label{tab:pointings}}
\tablehead{\colhead{Offset} & \colhead{Obs 1587} & \colhead{Obs 7073}
& \colhead{Obs 7074} & \colhead{Obs 7075}  & \colhead{Original Obs 7076} & \colhead{Corrected Obs 7076} }
\startdata

\multicolumn{3}{l}{Four co-added observation}\\

\hline

RA offset  &    -0.100   &    -0.110   &    0.091   &  0.042  &     -0.025   &   -0.009 \\
DEC offset &    -0.102   &     0.105   &   -0.133   & -0.014  &      0.573   &   -0.008 \\
Separation &     0.196   &     0.168   &    0.181   &  0.100  &      0.598   &    0.151  \\
\hline

\multicolumn{3}{l}{Five co-added observation}\\

\hline

RA offset  &    -0.106  &     -0.114   &    0.082 &  0.036  & - &  -0.039 \\
DEC offset &    -0.081  &      0.114   &   -0.118 & -0.007  & - &  -0.015 \\
Separation &     0.181  &      0.178   &    0.186 &  0.124  & - &   0.129 \\

\enddata
\end{deluxetable}


\begin{deluxetable}{lc}
\tabletypesize{\scriptsize}
\tablecolumns{2}
\tablewidth{0pt}
\tablecaption{Definition of Energy Bands and X-ray Colors \label{tab:bands}}
\tablehead{\colhead{Band} & \colhead{Definition} }
\startdata
Broad (B)	& 0.3$-$8 keV \\
Soft (S)	& 0.3$-$2.5 keV \\
Hard (H)	& 2.5$-$8 keV \\
Soft 1 (S$_1$)	& 0.3$-$0.9 keV \\
Soft 2 (S$_2$)	& 0.9$-$2.5 keV \\
\hline
Conventional Broad (Bc)	& 0.5$-$8 keV \\
Conventional Soft (Sc) & 0.5$-$2 keV \\
Conventional Hard (Hc) & 2$-$8 keV \\
\hline
Hardness Ratio $HR$	& (Hc$-$Sc)/(Hc+Sc) \\
X-ray Color C21 & $-log($S$_2)+log($S$_1)=log($S$_1/$S$_2$) \\
X-ray Color C32 & $-log($H$)+log($S$_2)=log($S$_2/$H) \\
\enddata
\end{deluxetable}

\makeatletter
\def\jnl@aj{AJ}
\ifx\revtex@jnl\jnl@aj\let\tablebreak=\nl\fi
\makeatother
\ptlandscape
\begin{deluxetable}{lcccccc@{}c@{}cccccc}
\tabletypesize{\scriptsize}
\rotate
\tablecolumns{14}
\tablewidth{0pt}
\tablecaption{Master Source List \label{tab:Mainprops} }
\tablehead{  \colhead{Masterid}   & \colhead{CXOU Name} &
\colhead{RA}  & \colhead{Dec}  & \colhead{Radius} &
\colhead{PU} & \colhead{SNR}  & \colhead{Log \LX (0.3$-$8.0 keV)} &
\multicolumn{3}{c}{Variability} & \colhead{Opt Corr}  & \colhead{DG} &
\colhead{Flag} \\
 \colhead{}   &
\colhead{}  & \colhead{(J2000)} & \colhead{(J2000)}  & \colhead{(\arcs)} &
\colhead{(\arcs)} & \colhead{}  & \colhead{(\ergps)} &
\colhead{LT} & \colhead{BB} & \colhead{K-S} & \colhead{}  & \colhead{(\arcs)} &
\colhead{} \\
 \colhead{(1)}   &
\colhead{(2)}  & \colhead{(3)} & \colhead{(4)}  & \colhead{(5)} &
\colhead{(6)} & \colhead{(7)}  & \colhead{(8)} &
\colhead{(9)} & \colhead{(10)}  & \colhead{(11)} &
\colhead{(12)} & \colhead{(13)}  & \colhead{(14)} }
\startdata
1 & J104806.4+123621 & 10:48:06.4 & +12:36:21.0 & 3.13 & 0.54 & 5.8 & 37.3 &  V & - & - & - & 261.1 & - \\
2 & J104806.2+123534 & 10:48:06.2 & +12:35:34.3 & 3.10 & 0.68 & 4.2 & 37.1 &  V & - & - & - & 246.3 & - \\
3 & J104805.5+123423 & 10:48:05.5 & +12:34:22.7 & 3.29 & 1.09 & 3.5 & 37.0 &  - & - & - & - & 235.5 & - \\
7 & J104803.0+123438 & 10:48:03.0 & +12:34:37.8 & 3.00 & 0.68 & 4.4 & 37.1 &  V & - & - & - & 197.5 & - \\
8 & J104801.9+123607 & 10:48:01.9 & +12:36:07.2 & 3.00 & 0.72 & 3.1 & 36.9 &  - & - & - & - & 194.6 & - \\
9 & J104801.7+123145 & 10:48:01.7 & +12:31:44.8 & 4.42 & 0.27 & 30.0 & 38.5 &  V & N & N & - & 259.4 & - \\
10 & J104801.4+123426 & 10:48:01.4 & +12:34:26.4 & 3.00 & 1.28 & 3.4 & 36.9 &  - & - & - & - & 175.5 & BKG? \\
11 & J104801.3+123526 & 10:48:01.3 & +12:35:26.4 & 3.00 & 0.23 & 11.2 & 37.7 &  N & - & - & - & 174.0 & - \\
\enddata
\end{deluxetable}

\makeatletter
\def\jnl@aj{AJ}
\ifx\revtex@jnl\jnl@aj\let\tablebreak=\nl\fi
\makeatother
\ptlandscape
\begin{deluxetable}{l@{}cccccccccc@{}c@{}}
\tabletypesize{\scriptsize}
\rotate
\tablecolumns{12}
\tablewidth{0pt}
\tablecaption{Source counts, hardness ratios and color-color values:
Coadded Observation \label{tab:counts_main} }
\tablehead{  \colhead{MID}   & \multicolumn{7}{c}{Net Counts} &
\colhead{HR} & \colhead{C21}  & \colhead{C32} &
\colhead{Log \LX} \\
\colhead{}   & \colhead{B-band} &
\colhead{S1-band}  & \colhead{S2-band}  & \colhead{S-band} &
\colhead{H-band} & \colhead{Sc-band}  & \colhead{Hc-band} &
\colhead{} & \colhead{}  & \colhead{} &
\colhead{(0.3$-$8.0 keV)} \\
 \colhead{(1)}   &
\colhead{(2)}  & \colhead{(3)} & \colhead{(4)}  & \colhead{(5)} &
\colhead{(6)} & \colhead{(7)}  & \colhead{(8)} &
\colhead{(9)} & \colhead{(10)}  & \colhead{(11)} &
\colhead{(12)} }
\startdata
1 &58.9$\pm$10.2 &14.2$\pm$5.6 &36.4$\pm$7.5 &50.6$\pm$8.9 &8.3$\pm$5.6 &45.7$\pm$8.3 &11.9$\pm$6.1 &-0.65	$_ { -0.14}^ { +0.13}$	&	-0.31$_ { -0.15}^ { +0.18}$	&	0.65$_ { -0.24}^ { +0.34}$	&	37.3\\																	
2 &36.2$\pm$8.7 &1.5$\pm$3.6 &23.0$\pm$6.5 &24.5$\pm$7.0 &11.7$\pm$5.8 &24.8$\pm$6.6 &13.4$\pm$6.1 &-0.39	$_ { -0.20}^ { +0.19}$	&	-0.97$_ { -0.63}^ { +0.53}$	&	0.30$_ { -0.17}^ { +0.26}$	&	37.1\\																	
3 &28.9$\pm$8.3 &7.3$\pm$4.7 &15.1$\pm$5.7 &22.4$\pm$6.9 &6.5$\pm$5.3 &16.5$\pm$5.9 &9.1$\pm$5.8 &-0.39	$_ { -0.30}^ { +0.27}$	&	-0.22$_ { -0.28}^ { +0.28}$	&	0.43$_ { -0.40}^ { +0.40}$	&	37.0\\																	
4 &1.8$\pm$3.6 &-0.1$\pm$2.3 &-0.1$\pm$2.3 &-0.2$\pm$2.7 &2.0$\pm$3.2 &0.8$\pm$2.7 &1.7$\pm$3.2 &0.20	$_ { -0.18}^ { +0.80}$	&	0.08$_ { -1.07}^ { +1.07}$	&	-0.38$_ { -1.07}^ { +0.76}$	&	35.8\\																	
5 &4.6$\pm$3.6 &-0.3$\pm$1.9 &2.6$\pm$2.9 &2.3$\pm$2.9 &2.3$\pm$2.9 &1.6$\pm$2.7 &3.2$\pm$3.2 &0.29	$_ { -0.40}^ { +0.50}$	&	-0.69$_ { -0.99}^ { +0.69}$	&	0.08$_ { -0.46}^ { +0.48}$	&	36.2\\																	
6 &$\le$10.6 &- &- &- &- &- &- &-		&	-	&	-	&	$\le$36.6\\																	
7 &38.7$\pm$8.8 &7.0$\pm$4.6 &23.9$\pm$6.5 &30.8$\pm$7.5 &7.8$\pm$5.3 &23.1$\pm$6.5 &14.5$\pm$6.2 &-0.31	$_ { -0.19}^ { +0.21}$	&	-0.39$_ { -0.30}^ { +0.21}$	&	0.44$_ { -0.21}^ { +0.39}$	&	37.1\\																	
8 &24.7$\pm$8.0 &8.6$\pm$4.7 &15.4$\pm$5.7 &24.0$\pm$6.9 &0.7$\pm$4.7 &17.2$\pm$5.9 &3.0$\pm$5.2 &-0.78	$_ { -0.22}^ { +0.14}$	&	-0.13$_ { -0.25}^ { +0.21}$	&	0.75$_ { -0.43}^ { +0.86}$	&	36.9\\
\enddata
\tablecomments{Col. (1): Master ID, cols. (2)$-$(8): net counts, in each of
the 7 energy bands (see Table \ref{tab:bands} for definitions of these
bands), col. (9): hardness ratio, cols. (10) and (11) color values,
col. (12): log \LX\ (0.3$-$8.0 keV). Upper limit \LX\ values are at
the 68\% confidence level. }
\end{deluxetable}

\clearpage
\pagestyle{empty}
\makeatletter
\def\jnl@aj{AJ}
\ifx\revtex@jnl\jnl@aj\let\tablebreak=\nl\fi
\makeatother
\ptlandscape
\begin{deluxetable}{l@{}cccccccc@{}c@{}c@{}c@{}c@{}c@{}c@{}}
\tabletypesize{\scriptsize}
\rotate
\tablecolumns{15}
\tablewidth{0pt}
\tablecaption{Source counts, hardness ratios, color-color values and variability:
Observation 1 \label{tab:counts_ob1} }
\tablehead{  \colhead{MID}   & \multicolumn{7}{c}{Net Counts}
& \colhead{HR} & \colhead{C21}  & \colhead{C32} &
  \multicolumn{3}{c}{Variability} & \colhead{Log \LX } \\
\colhead{}   & \colhead{B-band} &
\colhead{S1-band}  & \colhead{S2-band}  & \colhead{S-band} &
\colhead{H-band} & \colhead{Sc-band}  & \colhead{Hc-band} &
\colhead{} & \colhead{}  & \colhead{} &
\colhead{BB} & \colhead{k-S} & \colhead{signif.} &\colhead{(\ergps)} \\
 \colhead{(1)}   &
\colhead{(2)}  & \colhead{(3)} & \colhead{(4)}  & \colhead{(5)} &
\colhead{(6)} & \colhead{(7)}  & \colhead{(8)} &
\colhead{(9)} & \colhead{(10)}  & \colhead{(11)} &
\colhead{(12)} & \colhead{(13)}  & \colhead{(14)}& \colhead{(15)} }
\startdata
1 &8.3$\pm$4.3 &2.6$\pm$2.9 &4.7$\pm$3.4 &7.3$\pm$4.0 &0.9$\pm$2.7 &6.6$\pm$3.8 &1.9$\pm$2.9 &-0.66	$_ { -0.23}^ { +0.17}$	&	-0.13$_ { -0.41}^ { +0.32}$	&	0.53$_ { -0.53}^ { +0.69}$	&	-	&	-	&	-	&	37.5\\														
2 &$\le$1.3 &- &- &- &- &- &- &-		&	-	&	-	&	-	&	-	&	-	&	$\le$36.7\\														
3 &$\le$3.0 &- &- &- &- &- &- &-		&	-	&	-	&	-	&	-	&	-	&	$\le$37.0\\														
4 &$\le$2.7 &- &- &- &- &- &- &-		&	-	&	-	&	-	&	-	&	-	&	$\le$37.0\\														
5 &$\le$0.5 &- &- &- &- &- &- &-		&	-	&	-	&	-	&	-	&	-	&	$\le$36.2\\														
6 &$\le$2.1 &- &- &- &- &- &- &-		&	-	&	-	&	-	&	-	&	-	&	$\le$36.9\\														
7 &14.9$\pm$5.1 &4.7$\pm$3.4 &8.6$\pm$4.1 &13.3$\pm$4.8 &1.6$\pm$2.7 &12.4$\pm$4.7 &2.6$\pm$2.9 &-0.68	$_ { -0.22}^ { +0.16}$	&	-0.21$_ { -0.27}^ { +0.24}$	&	0.64$_ { -0.37}^ { +0.46}$	&	-	&	-	&	-	&	37.7\\														
8 &$\le$3.3 &- &- &- &- &- &- &-		&	-	&	-	&	-	&	-	&	-	&	$\le$37.0\\
\enddata
\tablecomments{Col. (1): Master ID, cols. (2)$-$(8): net counts, in each of
the 7 energy bands (see Table \ref{tab:bands} for definitions of these
bands), col. (9): hardness ratio, cols. (10) and (11) color values, errors are given as 1$\sigma$,
cols. (12) and (13): short-term variability, where (BB) indicate
Bayesian block analysis and (K-S) indicates the Kolmogorov-Smirnov test, in both columns symbols
indicate - (N) non-variable in all observations, (V) variable in at
least one observation, (P) possible variability in at least one
observation, col. (14): the significance of the change in
\LX\ between the previous observation and the current observation
respectively (equation \ref{equ:sig}), col. (15): log \LX\ (0.3$-$8.0
keV).  Upper limit values of net B and \LX\ are at
the 68\% confidence level.}
\end{deluxetable}

\makeatletter
\def\jnl@aj{AJ}
\ifx\revtex@jnl\jnl@aj\let\tablebreak=\nl\fi
\makeatother
\ptlandscape
\begin{deluxetable}{l@{}cccccccc@{}c@{}c@{}c@{}c@{}c@{}c@{}}
\tabletypesize{\scriptsize}
\rotate
\tablecolumns{15}
\tablewidth{0pt}
\tablecaption{Source counts, hardness ratios, color-color values and variability:
Observation 2 \label{tab:counts_ob2} }
\tablehead{  \colhead{MID}   & \multicolumn{7}{c}{Net Counts}
& \colhead{HR} & \colhead{C21}  & \colhead{C32} &
  \multicolumn{3}{c}{Variability} & \colhead{Log \LX } \\
\colhead{}   & \colhead{B-band} &
\colhead{S1-band}  & \colhead{S2-band}  & \colhead{S-band} &
\colhead{H-band} & \colhead{Sc-band}  & \colhead{Hc-band} &
\colhead{} & \colhead{}  & \colhead{} &
\colhead{BB} & \colhead{k-S} & \colhead{signif.} &\colhead{(\ergps)} \\
 \colhead{(1)}   &
\colhead{(2)}  & \colhead{(3)} & \colhead{(4)}  & \colhead{(5)} &
\colhead{(6)} & \colhead{(7)}  & \colhead{(8)} &
\colhead{(9)} & \colhead{(10)}  & \colhead{(11)} &
\colhead{(12)} & \colhead{(13)}  & \colhead{(14)}& \colhead{(15)}}
\startdata
1 &$\le$16.6 &- &- &- &- &- &- &-		&	-	&	-	&	-	&	-	&	0.3	&	$\le$37.4\\														
2 &$\le$2.2 &- &- &- &- &- &- &-		&	-	&	-	&	-	&	-	&	0.3	&	$\le$36.5\\														
3 &$\le$9.9 &- &- &- &- &- &- &-		&	-	&	-	&	-	&	-	&	0.3	&	$\le$37.1\\														
4 &$\le$1.0 &- &- &- &- &- &- &-		&	-	&	-	&	-	&	-	&	0.9	&	$\le$36.1\\														
5 &$\le$1.4 &- &- &- &- &- &- &-		&	-	&	-	&	-	&	-	&	0.1	&	$\le$36.3\\														
6 &3.7$\pm$4.3 &2.4$\pm$3.2 &2.4$\pm$3.2 &4.8$\pm$4.0 &-1.0$\pm$2.7 &1.3$\pm$2.9 &0.4$\pm$3.2 &-0.27	$_ { -0.73}^ { +0.17}$	&	0.16$_ { -0.64}^ { +0.67}$	&	0.46$_ { -0.69}^ { +1.14}$	&	-	&	-	&	0.2	&	36.7\\														
7 &$\le$4.7 &- &- &- &- &- &- &-		&	-	&	-	&	-	&	-	&	2.4	&	$\le$36.8\\														
8 &$\le$10.6 &- &- &- &- &- &- &-		&	-	&	-	&	-	&	-	&	0.3	&	$\le$37.1\\
\enddata
\tablecomments{Col. (1): Master ID, cols. (2)$-$(8): net counts, in each of
the 7 energy bands (see Table \ref{tab:bands} for definitions of these
bands), col. (9): hardness ratio, cols. (10) and (11) color values, errors are given as 1$\sigma$,
cols. (12) and (13): short-term variability, where (BB) indicate
Bayesian block analysis and (K-S) indicates the Kolmogorov-Smirnov test, in both columns symbols
indicate - (N) non-variable in all observations, (V) variable in at
least one observation, (P) possible variability in at least one
observation, col. (14): the significance of the change in
\LX\ between the previous observation and the current observation
respectively (equation \ref{equ:sig}), col. (15): log \LX\ (0.3$-$8.0
keV).  Upper limit values of net B and \LX\ are at
the 68\% confidence level.}
\end{deluxetable}

\makeatletter
\def\jnl@aj{AJ}
\ifx\revtex@jnl\jnl@aj\let\tablebreak=\nl\fi
\makeatother
\ptlandscape
\begin{deluxetable}{l@{}cccccccc@{}c@{}c@{}c@{}c@{}c@{}c@{}}
\tabletypesize{\scriptsize}
\rotate
\tablecolumns{15}
\tablewidth{0pt}
\tablecaption{Source counts, hardness ratios, color-color values and variability:
Observation 3 \label{tab:counts_ob3} }
\tablehead{  \colhead{MID}   & \multicolumn{7}{c}{Net Counts}
& \colhead{HR} & \colhead{C21}  & \colhead{C32} &
  \multicolumn{3}{c}{Variability} & \colhead{Log \LX } \\
\colhead{}   & \colhead{B-band} &
\colhead{S1-band}  & \colhead{S2-band}  & \colhead{S-band} &
\colhead{H-band} & \colhead{Sc-band}  & \colhead{Hc-band} &
\colhead{} & \colhead{}  & \colhead{} &
\colhead{BB} & \colhead{k-S} & \colhead{signif.} &\colhead{(\ergps)} \\
 \colhead{(1)}   &
\colhead{(2)}  & \colhead{(3)} & \colhead{(4)}  & \colhead{(5)} &
\colhead{(6)} & \colhead{(7)}  & \colhead{(8)} &
\colhead{(9)} & \colhead{(10)}  & \colhead{(11)} &
\colhead{(12)} & \colhead{(13)}  & \colhead{(14)}& \colhead{(15)} }
\startdata
1 &$\le$2.7 &- &- &- &- &- &- &-		&	-	&	-	&	-	&	-	&	2.6	&	$\le$36.6\\														
2 &$\le$16.2 &- &- &- &- &- &- &-		&	-	&	-	&	-	&	-	&	3.1	&	$\le$37.4\\														
3 &$\le$4.3 &- &- &- &- &- &- &-		&	-	&	-	&	-	&	-	&	0.5	&	$\le$36.9\\														
4 &$\le$1.1 &- &- &- &- &- &- &-		&	-	&	-	&	-	&	-	&	0.3	&	$\le$36.3\\														
5 &$\le$0.6 &- &- &- &- &- &- &-		&	-	&	-	&	-	&	-	&	0.4	&	$\le$36.0\\														
6 &$\le$1.1 &- &- &- &- &- &- &-		&	-	&	-	&	-	&	-	&	0.5	&	$\le$36.3\\														
7 &$\le$10.9 &- &- &- &- &- &- &-		&	-	&	-	&	-	&	-	&	1.5	&	$\le$37.3\\														
8 &$\le$8.6 &- &- &- &- &- &- &-		&	-	&	-	&	-	&	-	&	0.0	&	$\le$37.1\\
\enddata
\tablecomments{Col. (1): Master ID, cols. (2)$-$(8): net counts, in each of
the 7 energy bands (see Table \ref{tab:bands} for definitions of these
bands), col. (9): hardness ratio, cols. (10) and (11) color values, errors are given as 1$\sigma$,
cols. (12) and (13): short-term variability, where (BB) indicate
Bayesian block analysis and (K-S) indicates the Kolmogorov-Smirnov test, in both columns symbols
indicate - (N) non-variable in all observations, (V) variable in at
least one observation, (P) possible variability in at least one
observation, col. (14): the significance of the change in
\LX\ between the previous observation and the current observation
respectively (equation \ref{equ:sig}), col. (15): log \LX\ (0.3$-$8.0
keV).  Upper limit values of net B and \LX\ are at
the 68\% confidence level.}
\end{deluxetable}

\makeatletter
\def\jnl@aj{AJ}
\ifx\revtex@jnl\jnl@aj\let\tablebreak=\nl\fi
\makeatother
\ptlandscape
\begin{deluxetable}{l@{}cccccccc@{}c@{}c@{}c@{}c@{}c@{}c@{}}
\tabletypesize{\scriptsize}
\rotate
\tablecolumns{15}
\tablewidth{0pt}
\tablecaption{Source counts, hardness ratios, color-color values and variability:
Observation 4 \label{tab:counts_ob4} }
\tablehead{  \colhead{MID}   & \multicolumn{7}{c}{Net Counts}
& \colhead{HR} & \colhead{C21}  & \colhead{C32} &
  \multicolumn{3}{c}{Variability} & \colhead{Log \LX } \\
\colhead{}   & \colhead{B-band} &
\colhead{S1-band}  & \colhead{S2-band}  & \colhead{S-band} &
\colhead{H-band} & \colhead{Sc-band}  & \colhead{Hc-band} &
\colhead{} & \colhead{}  & \colhead{} &
\colhead{BB} & \colhead{k-S} & \colhead{signif.} &\colhead{(\ergps)} \\
 \colhead{(1)}   &
\colhead{(2)}  & \colhead{(3)} & \colhead{(4)}  & \colhead{(5)} &
\colhead{(6)} & \colhead{(7)}  & \colhead{(8)} &
\colhead{(9)} & \colhead{(10)}  & \colhead{(11)} &
\colhead{(12)} & \colhead{(13)}  & \colhead{(14)}& \colhead{(15)} }
\startdata
1 &24.3$\pm$7.2 &7.7$\pm$4.3 &13.3$\pm$5.2 &21.0$\pm$6.3 &3.3$\pm$4.3 &16.0$\pm$5.6 &5.5$\pm$4.7 &-0.65	$_ { -0.21}^ { +0.21}$	&	-0.03$_ { -0.24}^ { +0.22}$	&	0.54$_ { -0.35}^ { +0.61}$	&	N	&	N	&	2.7	&	37.6\\														
2 &$\le$16.1 &- &- &- &- &- &- &-		&	-	&	-	&	-	&	-	&	0.3	&	$\le$37.4\\														
3 &$\le$10.8 &- &- &- &- &- &- &-		&	-	&	-	&	-	&	-	&	0.8	&	$\le$37.2\\														
4 &$\le$0.8 &- &- &- &- &- &- &-		&	-	&	-	&	-	&	-	&	0.3	&	$\le$36.1\\														
5 &$\le$2.4 &- &- &- &- &- &- &-		&	-	&	-	&	-	&	-	&	0.7	&	$\le$36.6\\														
6 &$\le$2.5 &- &- &- &- &- &- &-		&	-	&	-	&	-	&	-	&	0.4	&	$\le$36.6\\														
7 &$\le$8.3 &- &- &- &- &- &- &-		&	-	&	-	&	-	&	-	&	0.9	&	$\le$37.1\\														
8 &$\le$0.7 &- &- &- &- &- &- &-		&	-	&	-	&	-	&	-	&	2.1	&	$\le$36.0\\
\enddata
\tablecomments{Col. (1): Master ID, cols. (2)$-$(8): net counts, in each of
the 7 energy bands (see Table \ref{tab:bands} for definitions of these
bands), col. (9): hardness ratio, cols. (10) and (11) color values, errors are given as 1$\sigma$,
cols. (12) and (13): short-term variability, where (BB) indicate
Bayesian block analysis and (K-S) indicates the Kolmogorov-Smirnov test, in both columns symbols
indicate - (N) non-variable in all observations, (V) variable in at
least one observation, (P) possible variability in at least one
observation, col. (14): the significance of the change in
\LX\ between the previous observation and the current observation
respectively (equation \ref{equ:sig}), col. (15): log \LX\ (0.3$-$8.0
keV).  Upper limit values of net B and \LX\ are at
the 68\% confidence level.}
\end{deluxetable}

\makeatletter
\def\jnl@aj{AJ}
\ifx\revtex@jnl\jnl@aj\let\tablebreak=\nl\fi
\makeatother
\ptlandscape
\begin{deluxetable}{l@{}cccccccc@{}c@{}c@{}c@{}c@{}c@{}c@{}}
\tabletypesize{\scriptsize}
\rotate
\tablecolumns{15}
\tablewidth{0pt}
\tablecaption{Source counts, hardness ratios, color-color values and variability:
Observation 5 \label{tab:counts_ob5} }
\tablehead{  \colhead{MID}   & \multicolumn{7}{c}{Net Counts}
& \colhead{HR} & \colhead{C21}  & \colhead{C32} &
  \multicolumn{3}{c}{Variability} & \colhead{Log \LX } \\
\colhead{}   & \colhead{B-band} &
\colhead{S1-band}  & \colhead{S2-band}  & \colhead{S-band} &
\colhead{H-band} & \colhead{Sc-band}  & \colhead{Hc-band} &
\colhead{} & \colhead{}  & \colhead{} &
\colhead{BB} & \colhead{k-S} & \colhead{signif.} &\colhead{(\ergps)} \\
 \colhead{(1)}   &
\colhead{(2)}  & \colhead{(3)} & \colhead{(4)}  & \colhead{(5)} &
\colhead{(6)} & \colhead{(7)}  & \colhead{(8)} &
\colhead{(9)} & \colhead{(10)}  & \colhead{(11)} &
\colhead{(12)} & \colhead{(13)}  & \colhead{(14)}& \colhead{(15)} }
\startdata
1 &$\le$16.3 &- &- &- &- &- &- &-		&	-	&	-	&	-	&	-	&	0.3	&	$\le$37.5\\														
2 &11.7$\pm$5.2 &0.9$\pm$2.7 &8.3$\pm$4.3 &9.2$\pm$4.6 &2.5$\pm$3.4 &9.6$\pm$4.4 &3.0$\pm$3.6 &-0.66	$_ { -0.24}^ { +0.15}$	&	-0.53$_ { -0.77}^ { +0.38}$	&	0.46$_ { -0.38}^ { +0.56}$	&	-	&	-	&	0.5	&	37.3\\														
3 &$\le$10.1 &- &- &- &- &- &- &-		&	-	&	-	&	-	&	-	&	0.0	&	$\le$37.2\\														
4 &$\le$2.0 &- &- &- &- &- &- &-		&	-	&	-	&	-	&	-	&	0.6	&	$\le$36.5\\														
5 &$\le$1.7 &- &- &- &- &- &- &-		&	-	&	-	&	-	&	-	&	0.2	&	$\le$36.4\\														
6 &$\le$5.4 &- &- &- &- &- &- &-		&	-	&	-	&	-	&	-	&	0.5	&	$\le$36.9\\														
7 &$\le$8.0 &- &- &- &- &- &- &-		&	-	&	-	&	-	&	-	&	0.2	&	$\le$37.1\\														
8 &$\le$14.7 &- &- &- &- &- &- &-		&	-	&	-	&	-	&	-	&	3.2	&	$\le$37.3\\
\enddata
\tablecomments{Col. (1): Master ID, cols. (2)$-$(8): net counts, in each of
the 7 energy bands (see Table \ref{tab:bands} for definitions of these
bands), col. (9): hardness ratio, cols. (10) and (11) color values, errors are given as 1$\sigma$,
cols. (12) and (13): short-term variability, where (BB) indicate
Bayesian block analysis and (K-S) indicates the Kolmogorov-Smirnov test, in both columns symbols
indicate - (N) non-variable in all observations, (V) variable in at
least one observation, (P) possible variability in at least one
observation, col. (14): the significance of the change in
\LX\ between the previous observation and the current observation
respectively (equation \ref{equ:sig}), col. (15): log \LX\ (0.3$-$8.0
keV).  Upper limit values of net B and \LX\ are at
the 68\% confidence level.}
\end{deluxetable}

\clearpage
\pagestyle{plaintop}

\makeatletter
\def\jnl@aj{AJ}
\ifx\revtex@jnl\jnl@aj\let\tablebreak=\nl\fi
\makeatother
\ptlandscape
\begin{deluxetable}{lccccccccccccccccc}
\tabletypesize{\scriptsize}
\rotate
\tablecolumns{11}
\tablewidth{0pt}
\tablecaption{Properties of optical sources that are correlated with
an X-ray point source\label{tab:GCprops_corr} }
\tablehead{  \colhead{Masterid}   &
\colhead{V}  & \colhead{I} &
\colhead{V$-$I} & \colhead{B$-$V}  & \colhead{V$-$R} &
\colhead{B$-$R} & \colhead{Radial Velocity}  & \colhead{GC separation} &
\colhead{Ratio} &
\colhead{Reference} \\
 \colhead{}   &
\colhead{}  & \colhead{}  & \colhead{} &
\colhead{} & \colhead{}  & \colhead{} &
\colhead{} & \colhead{(arcsec)}  & \colhead{} &
\colhead{}  \\
\colhead{(1)}   &
\colhead{(2)}  & \colhead{(3)}  & \colhead{(4)} &
\colhead{(5)} & \colhead{(6)}  & \colhead{(7)} &
\colhead{(8)} & \colhead{(9)}  & \colhead{(10)} & \colhead{(11)}  }

\startdata
41 & 21.148$\pm$0.008 & 20.297$\pm$0.010 & 0.851$\pm$0.013 & - & - & - & - & 0.307 & 1.31 & 4\\
42 & 21.880$\pm$0.011 & 20.753$\pm$0.012 & 1.127$\pm$0.016 & 1.074 & 0.519 & 1.590 & 889$\pm$7 & 0.448 & 1.84 & 1,4,6\\
50 & 22.569$\pm$0.022 & 21.307$\pm$0.021 & 1.262$\pm$0.030 & - & - & - & - & 0.410 & 1.23 & 4\\
55 & 20.059$\pm$0.004 & 19.007$\pm$0.004 & 1.052$\pm$0.006 & - & - & 1.240 & 776$\pm$4 & 0.380 & 1.59 & 1,4\\
61 & 22.923$\pm$0.032 & 22.094$\pm$0.043 & 0.829$\pm$0.054 & - & - & - & - & 0.219 & 0.32 & 4\\
62 & 21.006$\pm$0.010 & 19.877$\pm$0.012 & 1.129$\pm$0.016 & - & - & 1.260 & 1071$\pm$18 & 0.146 & 0.54 & 3,4\\
66 & 21.517$\pm$0.013 & 20.334$\pm$0.013 & 1.183$\pm$0.018 & - & - & 1.310 & 1258$\pm$38 & 0.174 & 0.67 & 2,4\\
67 & 21.567$\pm$0.025 & 20.702$\pm$0.038 & 0.865$\pm$0.045 & - & - & - & - & 0.141 & 0.61 & 4\\
\enddata
\tablecomments{The sources in the top section of the table, denoted by
the horizontal line, have been
confirmed as globular clusters, while those in the bottom section of
the table have all been identified as background objects.}
\tablerefs{1. Bergond et al. 2006, 2. Puzia et al. 2004, 3. Pierce et
al. 2006, 4. Kundu \& Whitmore (2001), confirmed GCs, 5. Kundu \&
Whitmore (2001), background objects,
6. Rhode \& Zepf, 2004.}

\end{deluxetable}

\clearpage

\makeatletter
\def\jnl@aj{AJ}
\ifx\revtex@jnl\jnl@aj\let\tablebreak=\nl\fi
\makeatother
\ptlandscape
\begin{deluxetable}{lccccccccccccccccc}
\tabletypesize{\scriptsize}
\rotate
\tablecolumns{11}
\tablewidth{0pt}
\tablecaption{Properties of optical sources that have been classified
as `excluded matches'; sources detected between 0.6\arcs\ and 3\arcs\
of an X-ray point source\label{tab:GCprops_near} }
\tablehead{  \colhead{Masterid}   &
\colhead{V}  & \colhead{I} &
\colhead{V$-$I} & \colhead{B$-$V}  & \colhead{V$-$R} &
\colhead{B$-$R} & \colhead{Radial Velocity}  & \colhead{GC separation} &
\colhead{Ratio} &
\colhead{Reference} \\
 \colhead{}   &
\colhead{}  & \colhead{}  & \colhead{} &
\colhead{} & \colhead{}  & \colhead{} &
\colhead{} & \colhead{(arcsec)}  & \colhead{} &
\colhead{} \\
 \colhead{(1)}   &
\colhead{(2)}  & \colhead{(3)}  & \colhead{(4)} &
\colhead{(5)} & \colhead{(6)}  & \colhead{(7)} &
\colhead{(8)} & \colhead{(9)}  & \colhead{(10)} &
\colhead{(11)} }
\startdata
28 & 23.213$\pm$0.030 & 21.984$\pm$6 & 1.229$\pm$0.042 & - & - & - & - & 1.555 & 5.55 & 4\\
34 & 22.041$\pm$0.014 & 20.931$\pm$6 & 1.110$\pm$0.020 & - & - & 1.220 & 790$\pm$41 & 1.088 & 2.05 & 3,4\\
40 & 25.379$\pm$0.450 & 24.040$\pm$6 & 1.339$\pm$0.567 & - & - & - & - & 1.341 & 2.95 & 4\\
46 & 21.573$\pm$0.012 & 20.474$\pm$6 & 1.099$\pm$0.019 & - & - & 1.280 & 1080$\pm$32 & 2.197 & 4.53 & 2,4\\
49 & 23.782$\pm$0.063 & 22.629$\pm$6 & 1.153$\pm$0.099 & - & - & - & - & 2.971 & 6.66 & 4\\
51 & 23.226$\pm$0.050 & 22.105$\pm$6 & 1.121$\pm$0.076 & - & - & - & - & 2.101 & 5.27 & 4\\
58 & 24.061$\pm$0.068 & 23.545$\pm$6 & 0.516$\pm$0.130 & - & - & - & - & 2.405 & 5.97 & 4\\
70 & 22.319$\pm$0.077 & 21.253$\pm$6 & 1.066$\pm$0.118 & - & - & - & - & 1.352 & 6.30 & 4\\
\enddata
\tablecomments{The sources in the top section of the table, denoted by
the horizontal line, have been
confirmed as globular clusters, while those in the bottom section of
the table have all been identified as background objects.}
\tablerefs{1. Bergond et al. 2006, 2. Puzia et al. 2004, 3. Pierce et
al. 2006, 4. Kundu \& Whitmore (2001), confirmed GCs, 5. Kundu \&
Whitmore (2001), background objects, 6. Rhode \& Zepf, 2004.}

\end{deluxetable}
\clearpage

\clearpage

\begin{deluxetable}{cccc}
\tabletypesize{\scriptsize}
\tablecolumns{4}
\tablewidth{0pt}
\tablecaption{Ratio Values of Potential Transient Candidates \label{tab:trans}}
\tablehead{\colhead{Masterid} & \colhead{Mode Ratio} &\colhead{Lower Bound Ratio}& \colhead{Variability}}
\startdata
25 & 161.39 & 43.48 & TC \\
50 & 6.83 & 5.87 & PTC \\
53 & 11.40 & 4.20 & V \\
85 & 30.19 & 9.00 & PTC \\
89 & 18.39 & 16.42 & TC \\
91 & 8.49 & 4.75 & V \\
92 & 7.44 & 3.15 & V \\
94 & 19.21 & 13.18 & TC \\
100 & 20.21 & 14.05 & TC \\
115 & 12.88 & 6.84 & PTC \\
128 & 85.09 & 50.69 & TC \\
\enddata
\tablecomments{The 11 sources that were identified as potential transients candidates. Mode ratios and lower bound ratios were derived from Bayesian modeling, see \S\ref{sec:var} for more details. Sources were determined to be transient candidates (TC) if the lower bound $>$10, possible transient candidates (PTC) if the lower bound $>$5 and variable (V), if lower than 5.}
\end{deluxetable}

\begin{deluxetable}{ccccccc}
\tabletypesize{\scriptsize}
\tablecolumns{6}
\tablewidth{0pt}
\tablecaption{Summary of Long-Term and Short-Term Source Variability \label{tab:varsum}}
\tablehead{\colhead{Short-Term} & \multicolumn{5}{c}{Long-Term
Variability} &\colhead{Total} \\
\colhead{Variability} & \colhead{N} & \colhead{V} & \colhead{TC} &
\colhead{PTC} &\colhead{-} &\colhead{}}
\startdata
N  & 2 {\bf (2)} & 8 {\bf (7)} &  0 {\bf (0)} & 0 {\bf (0)}  & 0 {\bf (0)} & 10 {\bf (9)} \\ 
V  & 2 {\bf (2)} & 3 {\bf (2)} &  2 {\bf (1)} & 0 {\bf (0)}  & 0 {\bf (0)} & 7 {\bf (5)} \\
P  & 3 {\bf (3)} & 4 {\bf (4)} &  0 {\bf (0)} & 0 {\bf (0)}  & 0 {\bf (0)} & 7 {\bf (7)} \\
-  & 37 {\bf (27)} & 41 {\bf (30)} &  3 {\bf (3)} & 3 {\bf (3)}  & 24 {\bf (14)} & 108 {\bf (77)} \\
\hline
Total & 44 {\bf (34)} &  56 {\bf (43)} &  5 {\bf (4)}  & 3 {\bf (3)}  & 24 {\bf (14)} & 132 {\bf (98)} \\
\enddata
\tablecomments{The long-term variability definitions: N-non-variable,
V-variable, TC-transient candidate, PTC-possible transient
candidate and - unable to determine variability as all individual
observations were upper limits. Short-term variability definitions: N-non-variable in all
five observation, V-variable in a least one observation, P-possible
variability in at least one observation (see \S \ref{sec:var} for
full definition), - too few counts in all five observations to
determine variability. Bold values in brackets indicate the number of
sources within the $D_{25}$ ellipse.}

\end{deluxetable}

\makeatletter
\def\jnl@aj{AJ}
\ifx\revtex@jnl\jnl@aj\let\tablebreak=\nl\fi
\makeatother
\ptlandscape
\begin{deluxetable}{lccccccccccccccc}
\tabletypesize{\scriptsize}
\rotate
\tablecolumns{16}
\tablewidth{0pt}
\tablecaption{Raw source and background counts from the co-added observation \label{tab:raw} }
\tablehead{ \colhead{Masterid}  & \colhead{Net B Counts} &
\multicolumn{14}{c}{Raw Counts} \\
\colhead{}   & \colhead{} &
\multicolumn{2}{c}{B-band}
& \multicolumn{2}{c}{S1-band} & \multicolumn{2}{c}{S2-band} &
\multicolumn{2}{c}{H-band} & \multicolumn{2}{c}{Bc-band} & \multicolumn{2}{c}{Sc-band} & 
\multicolumn{2}{c}{Hc-band} \\
 \colhead{}   &
\colhead{}  & \colhead{(Src )} & \colhead{(Bkg )}  &
\colhead{(Src )} &
\colhead{(Bkg )} & \colhead{(Src )}  & \colhead{(Bkg )} &
\colhead{(Src )} & \colhead{(Bkg )}  & \colhead{(Src )} &
\colhead{(Bkg )} & \colhead{(Src )}  & \colhead{(Bkg )}  & \colhead{(Src )} &
\colhead{(Bkg )} \\
 \colhead{(1)}   &
\colhead{(2)}  & \colhead{(3)} & \colhead{(4)}  & \colhead{(5)} &
\colhead{(6)} & \colhead{(7)}  & \colhead{(8)} &
\colhead{(9)} & \colhead{(10)}  & \colhead{(11)} &
\colhead{(12)} & \colhead{(13)}  & \colhead{(14)}& \colhead{(15)} & \colhead{(16)} }
\startdata
1 & 58.9 & 82 & 23.1 & 20 & 5.8 & 42 & 5.6 & 20 & 11.7 & 77 & 19.4 & 52 & 6.3 & 25 & 13.1 \\
2 & 36.2 & 57 & 20.8 & 6 & 4.5 & 29 & 6.0 & 22 & 10.3 & 56 & 17.9 & 31 & 6.2 & 25 & 11.6 \\
3 & 28.9 & 52 & 23.1 & 13 & 5.7 & 21 & 5.9 & 18 & 11.5 & 45 & 19.3 & 23 & 6.5 & 22 & 12.9 \\
4 & 1.8 & 6 & 4.2 & 1 & 1.1 & 1 & 1.1 & 4 & 2.0 & 6 & 3.5 & 2 & 1.2 & 4 & 2.3 \\
5 & 4.6 & 6 & 1.4 & 0 & 0.3 & 3 & 0.4 & 3 & 0.7 & 6 & 1.2 & 2 & 0.4 & 4 & 0.8 \\
6 & - & - & - & - & - & - & - & - & - & - & - & - & - & - & - \\
7 & 38.7 & 59 & 20.3 & 12 & 5.0 & 29 & 5.1 & 18 & 10.2 & 55 & 17.4 & 29 & 5.9 & 26 & 11.5 \\
8 & 24.7 & 47 & 22.3 & 13 & 4.4 & 21 & 5.6 & 13 & 12.3 & 40 & 19.9 & 23 & 5.8 & 17 & 14.0 \\
\enddata
\tablecomments{Col. (1): Master ID, cols. (2): net broad-band
counts. Cols. (3), (5), (7), (9), (11), (13) and (15): raw source counts in each of
the 7 energy bands (see Table \ref{tab:bands} for definitions of these
bands), cols. (4), (6), (8), (10), (12), (14) and (16): background
counts in each of the 7 energy bands.
In some instances background counts are very low. For these sources
standard aperture photometry results in negative net counts, so,
instead, the source cell determined by {\em wavdetect} has been used. This
results in a large area ratio between the background and source
regions and therefore a low background count value is derived.}
\end{deluxetable}

\clearpage

\clearpage
\begin{centering}
  \begin{minipage}{0.95\linewidth}

  \includegraphics[angle=-90,width=\linewidth]{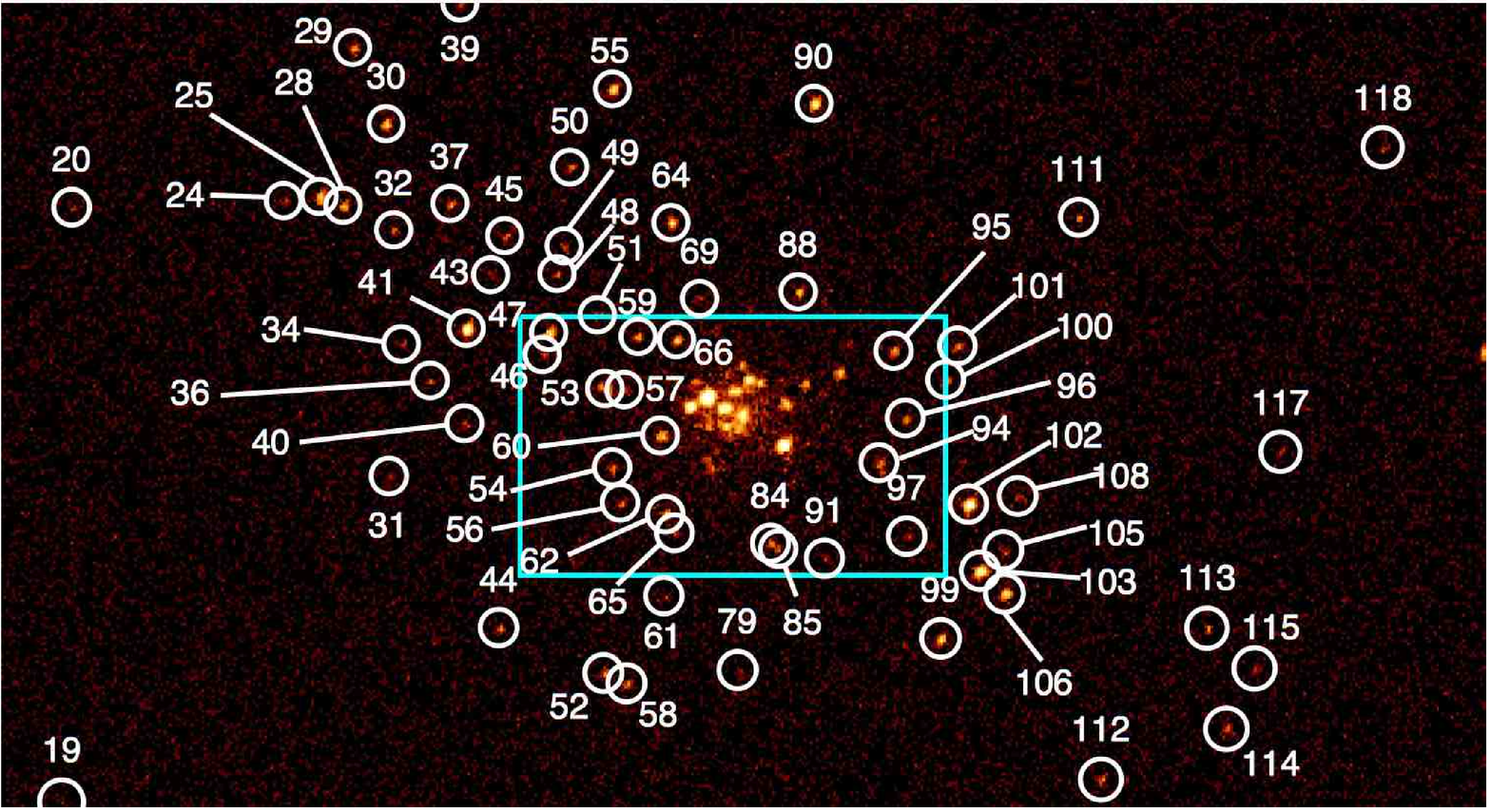}

  \end{minipage}

\vspace{1cm}

  \begin{minipage}{0.95\linewidth}

  \includegraphics[angle=-90,width=\linewidth]{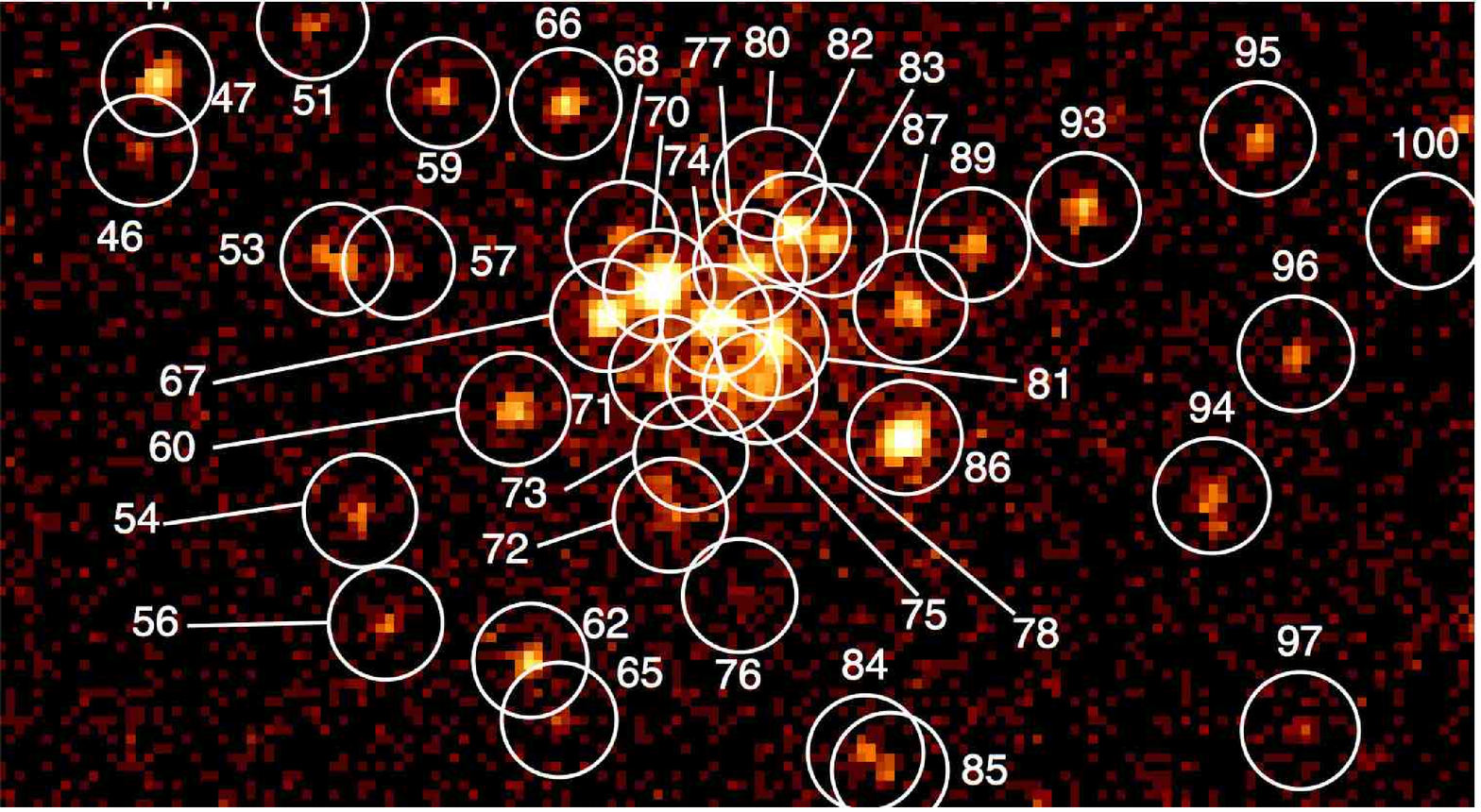}

  \end{minipage}

\end{centering}
\clearpage

\begin{figure}

  \begin{minipage}{0.485\linewidth}
  \centering

    \includegraphics[width=\linewidth]{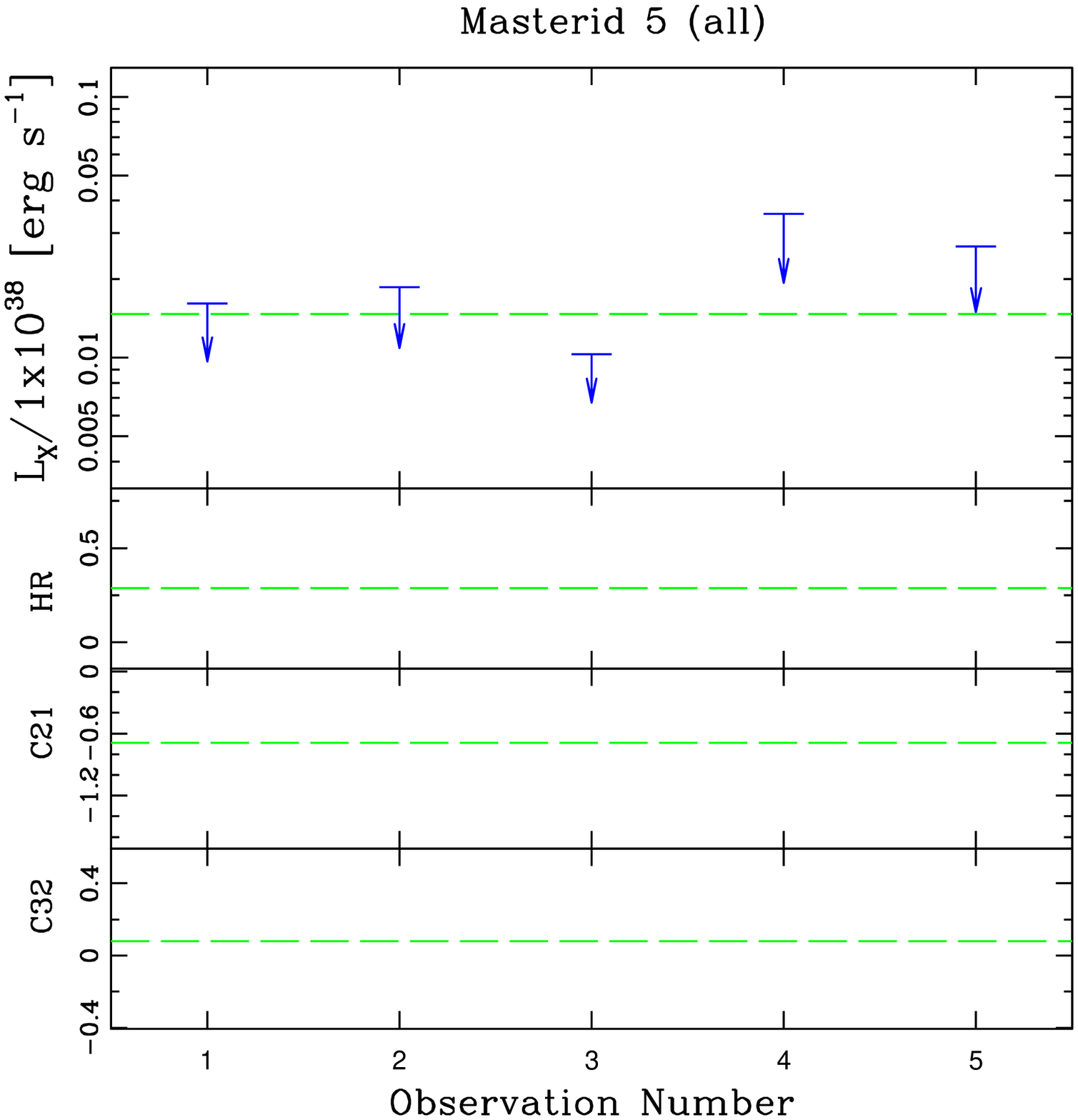}

\end{minipage}\hspace{0.02\linewidth}
\begin{minipage}{0.485\linewidth}
  \centering

    \includegraphics[width=\linewidth]{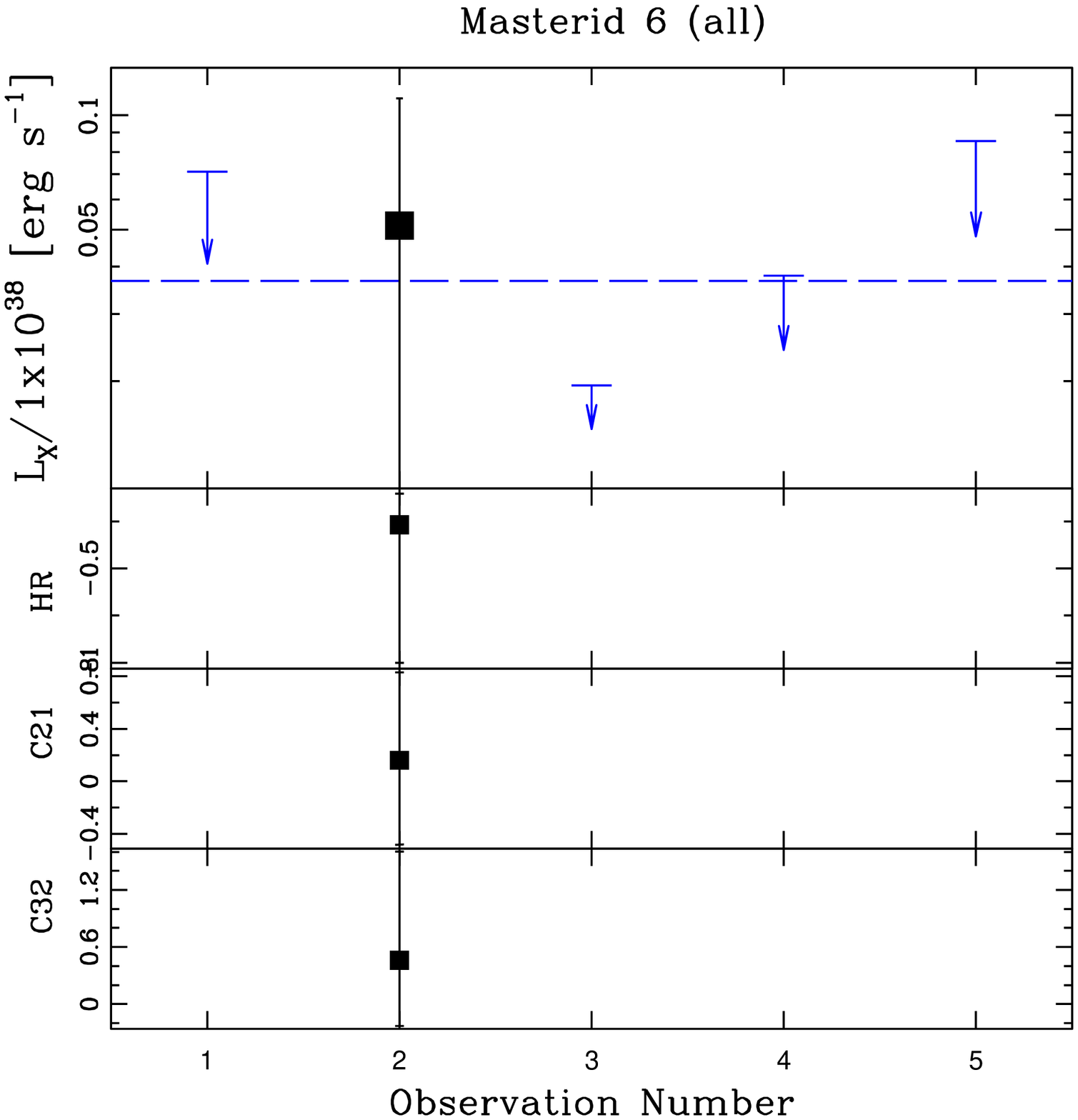}

 \end{minipage}\hspace{0.02\linewidth}

  \begin{minipage}{0.485\linewidth}
  \centering
  
    \includegraphics[width=\linewidth]{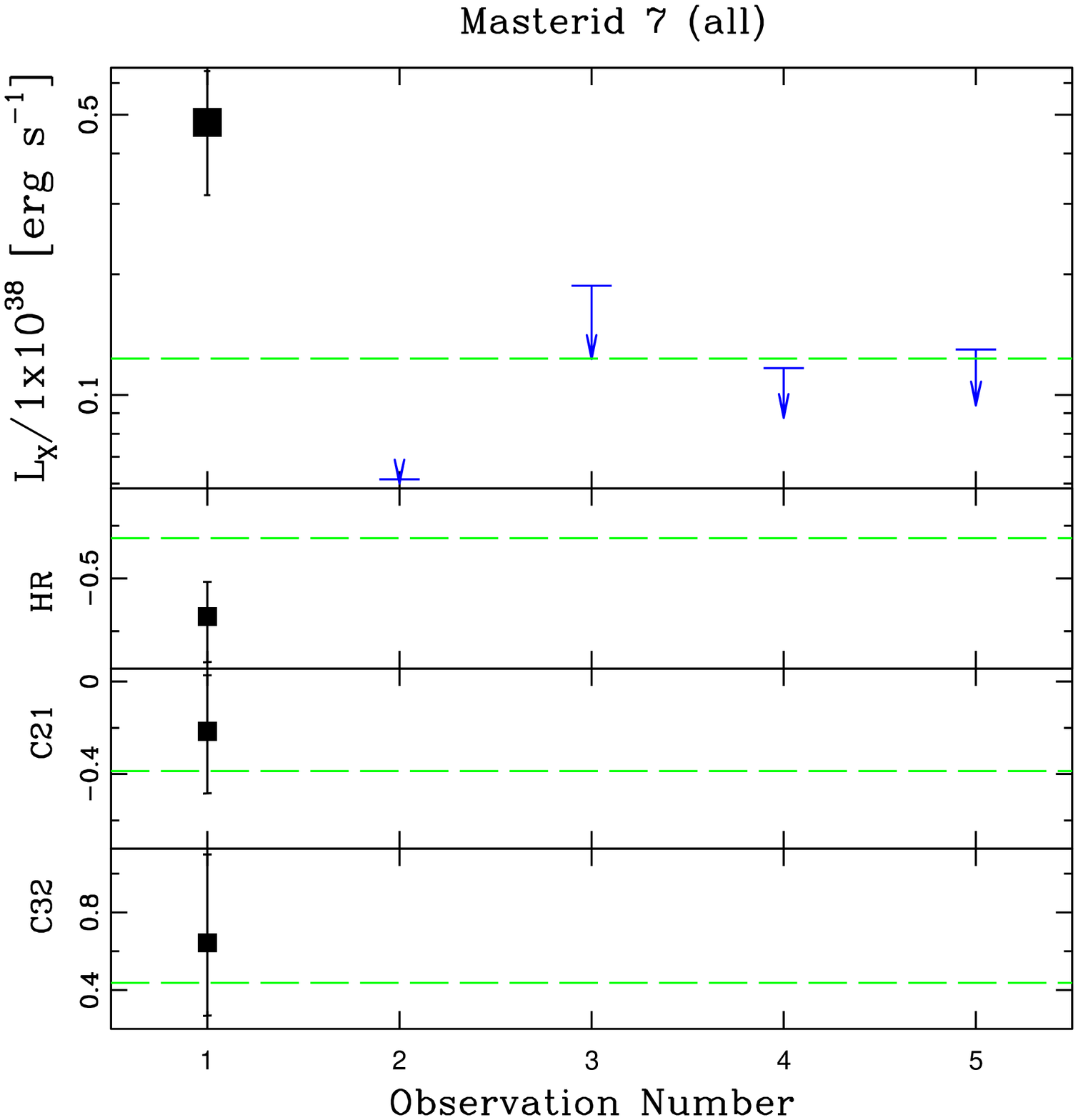}

  \end{minipage}\hspace{0.02\linewidth}
  \begin{minipage}{0.485\linewidth}
  \centering

    \includegraphics[width=\linewidth]{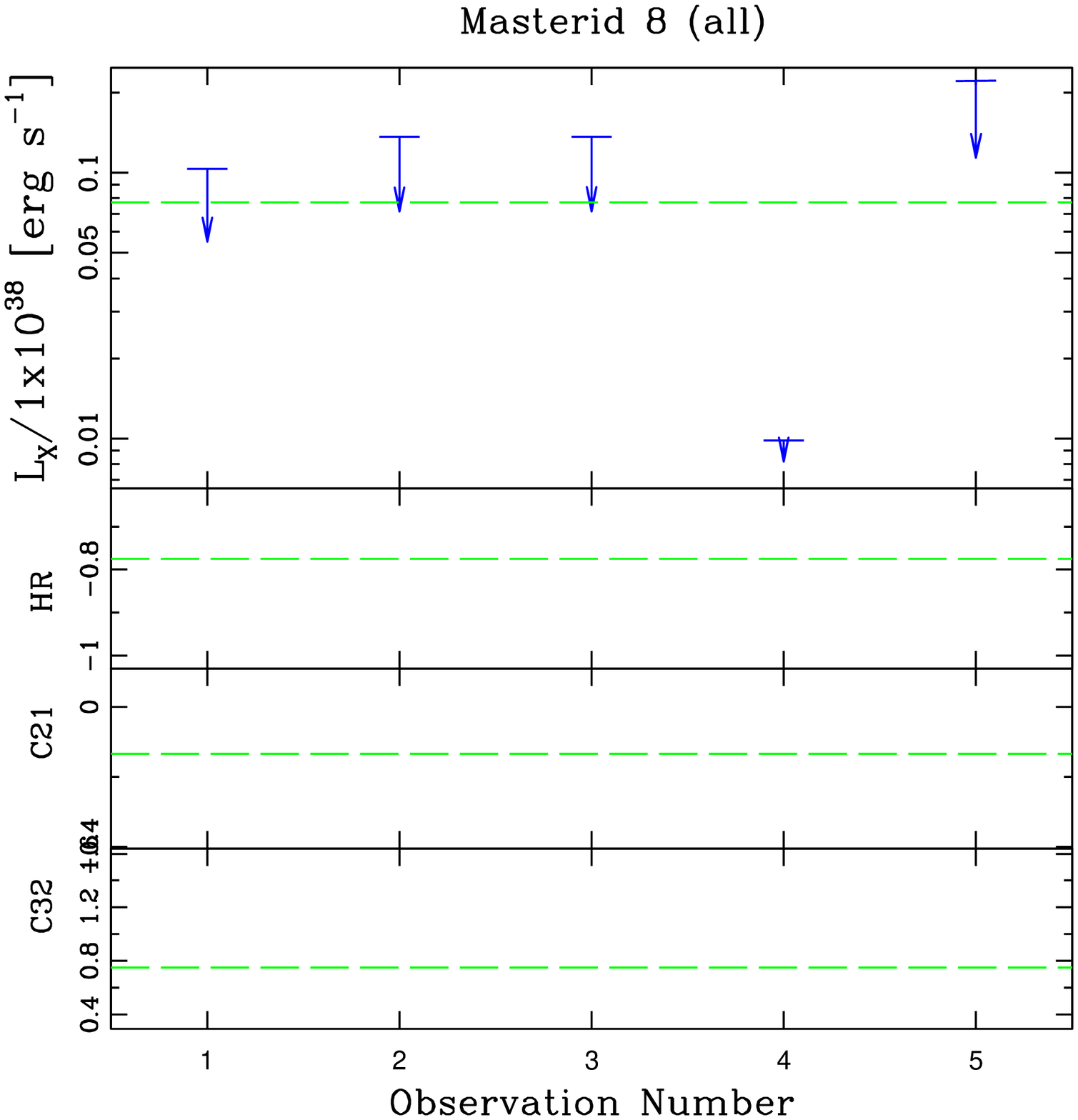}

\end{minipage}
\begin{minipage}{0.485\linewidth}
  \centering

    \includegraphics[width=\linewidth]{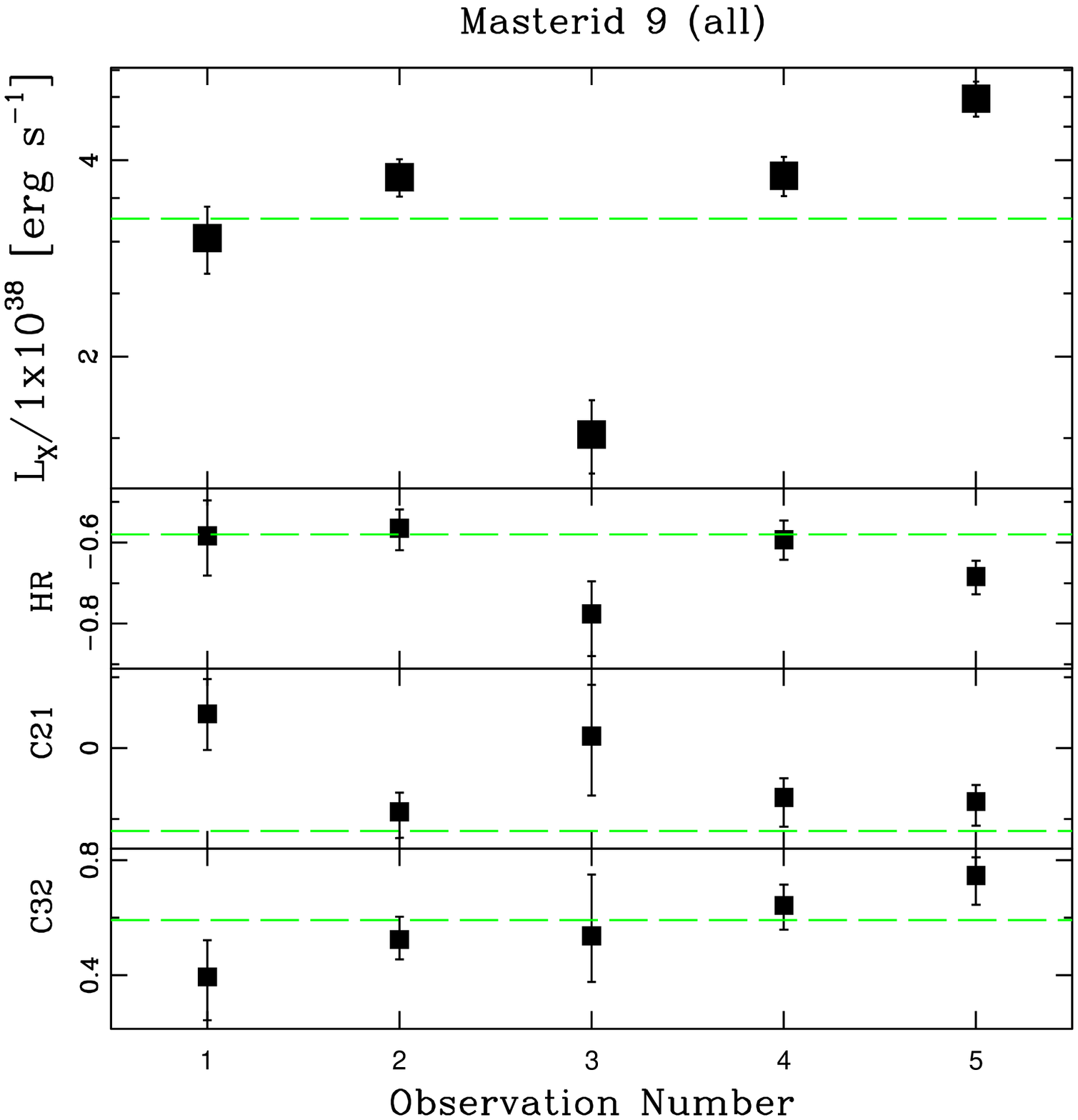}

 \end{minipage}\hspace{0.02\linewidth}
\begin{minipage}{0.485\linewidth}
  \centering
  
    \includegraphics[width=\linewidth]{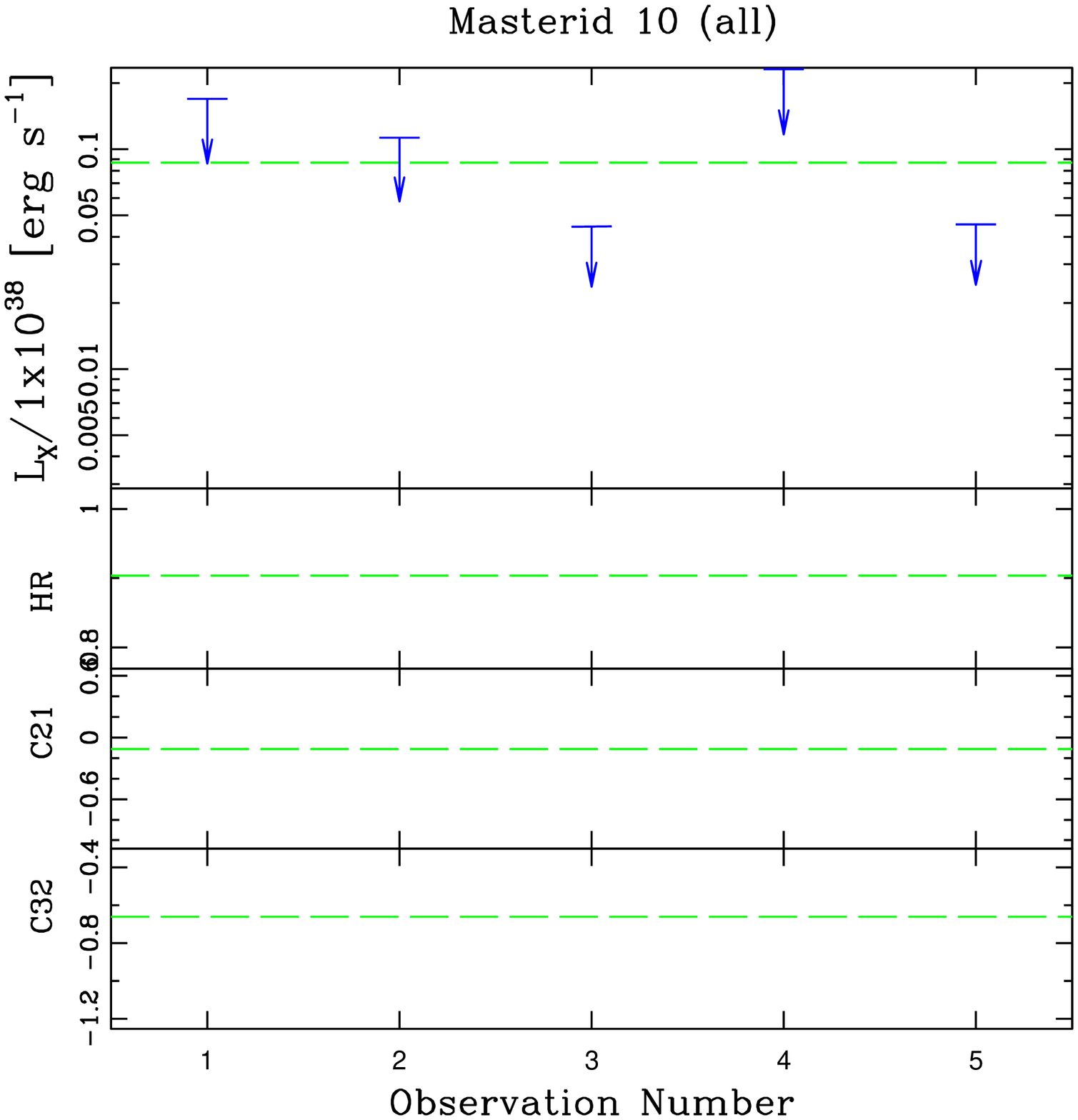}

  \end{minipage}\hspace{0.02\linewidth}

\end{figure}

\begin{figure}

  \begin{minipage}{0.485\linewidth}
  \centering

    \includegraphics[width=\linewidth]{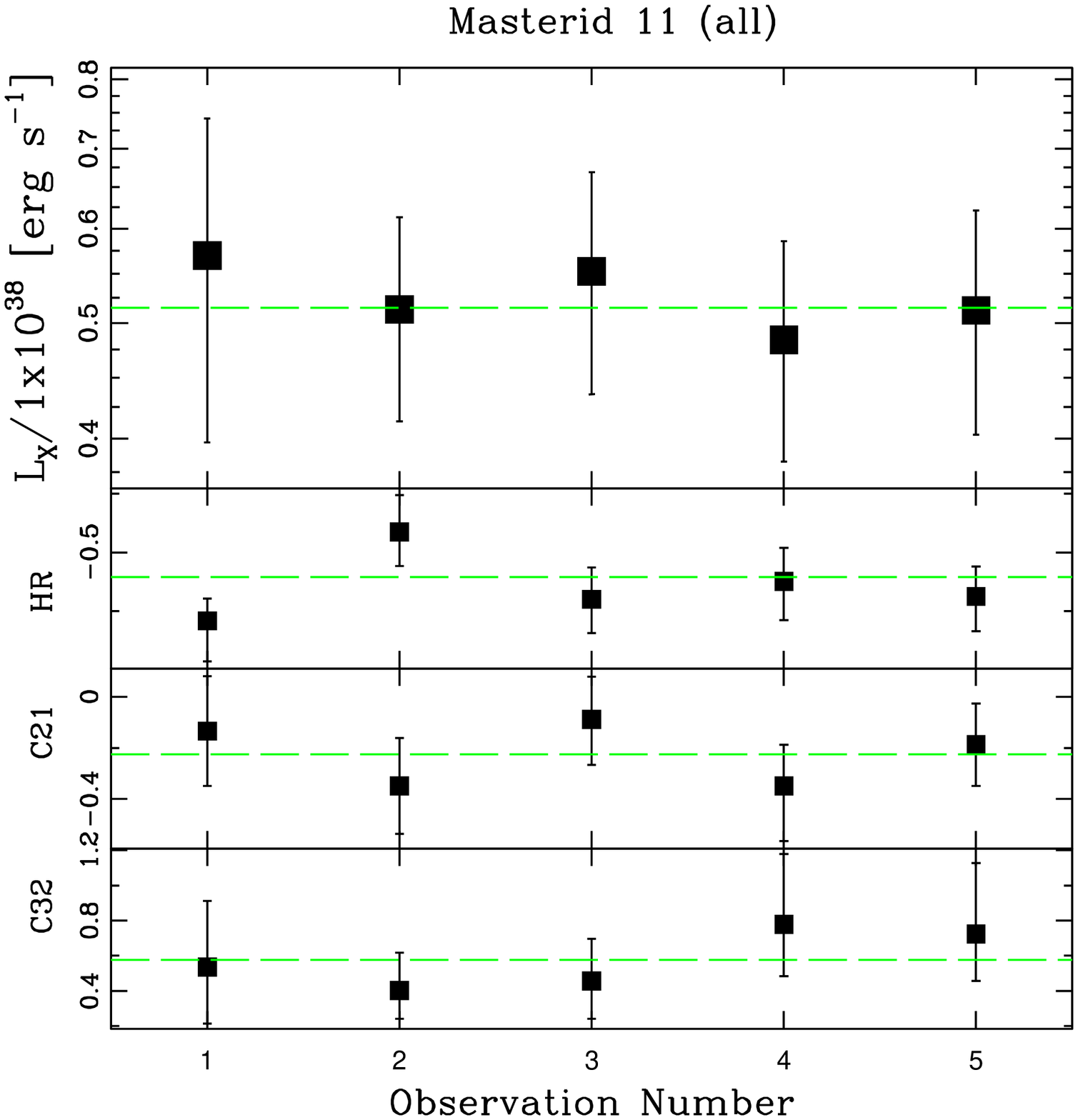}

\end{minipage}\hspace{0.02\linewidth}
\begin{minipage}{0.485\linewidth}
  \centering

    \includegraphics[width=\linewidth]{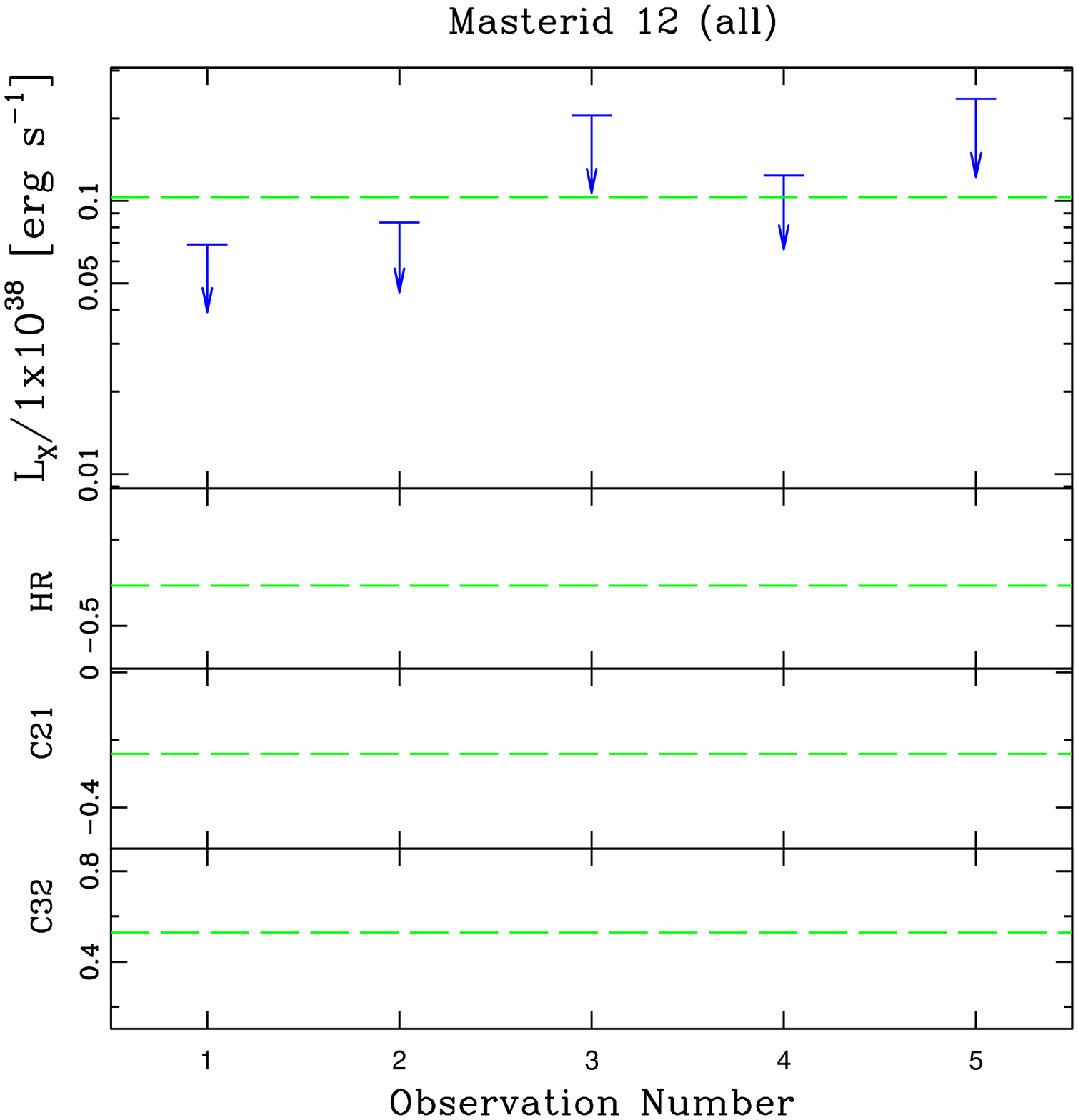}

 \end{minipage}\hspace{0.02\linewidth}

  \begin{minipage}{0.485\linewidth}
  \centering
  
    \includegraphics[width=\linewidth]{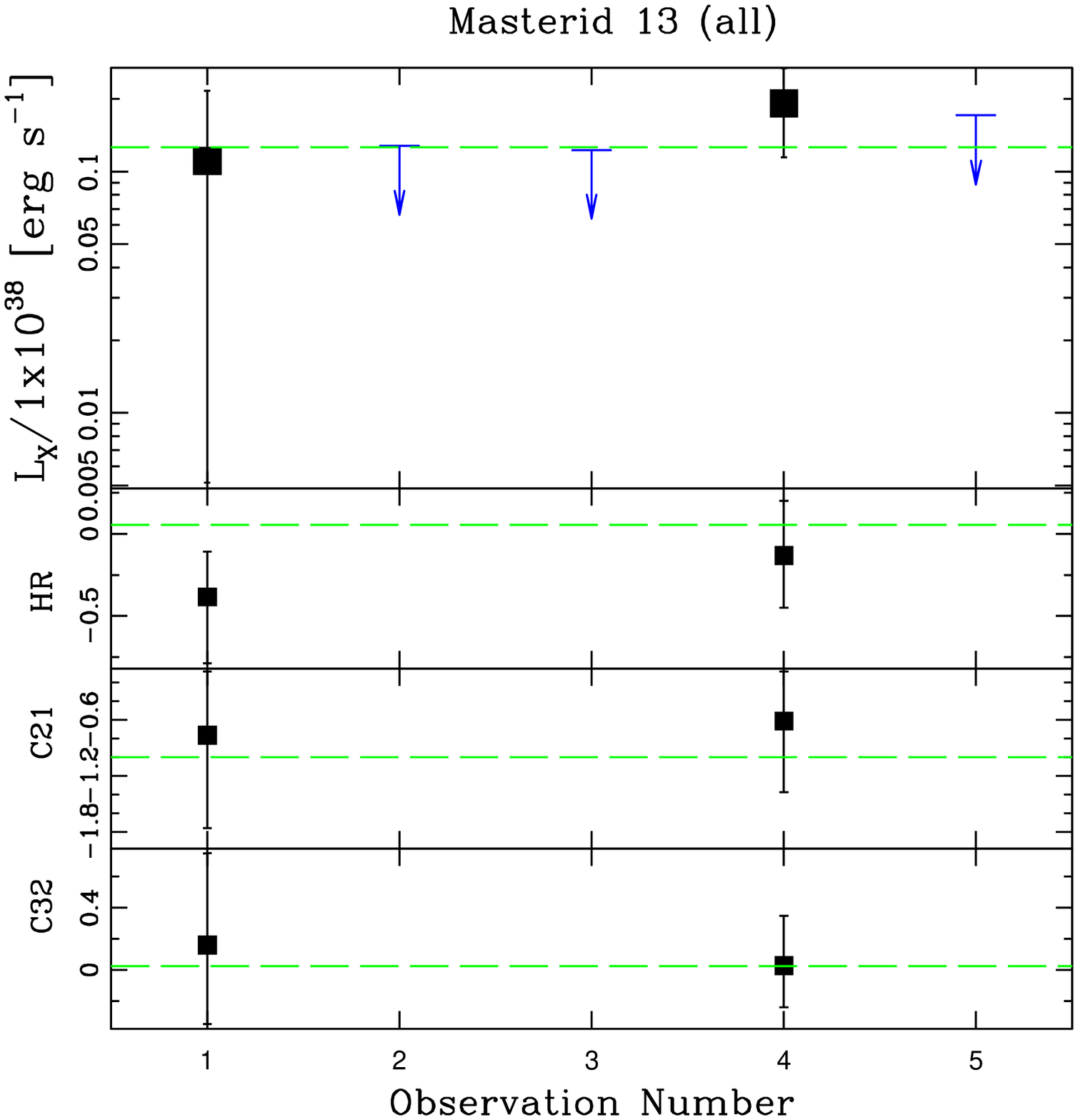}

  \end{minipage}\hspace{0.02\linewidth}
  \begin{minipage}{0.485\linewidth}
  \centering

    \includegraphics[width=\linewidth]{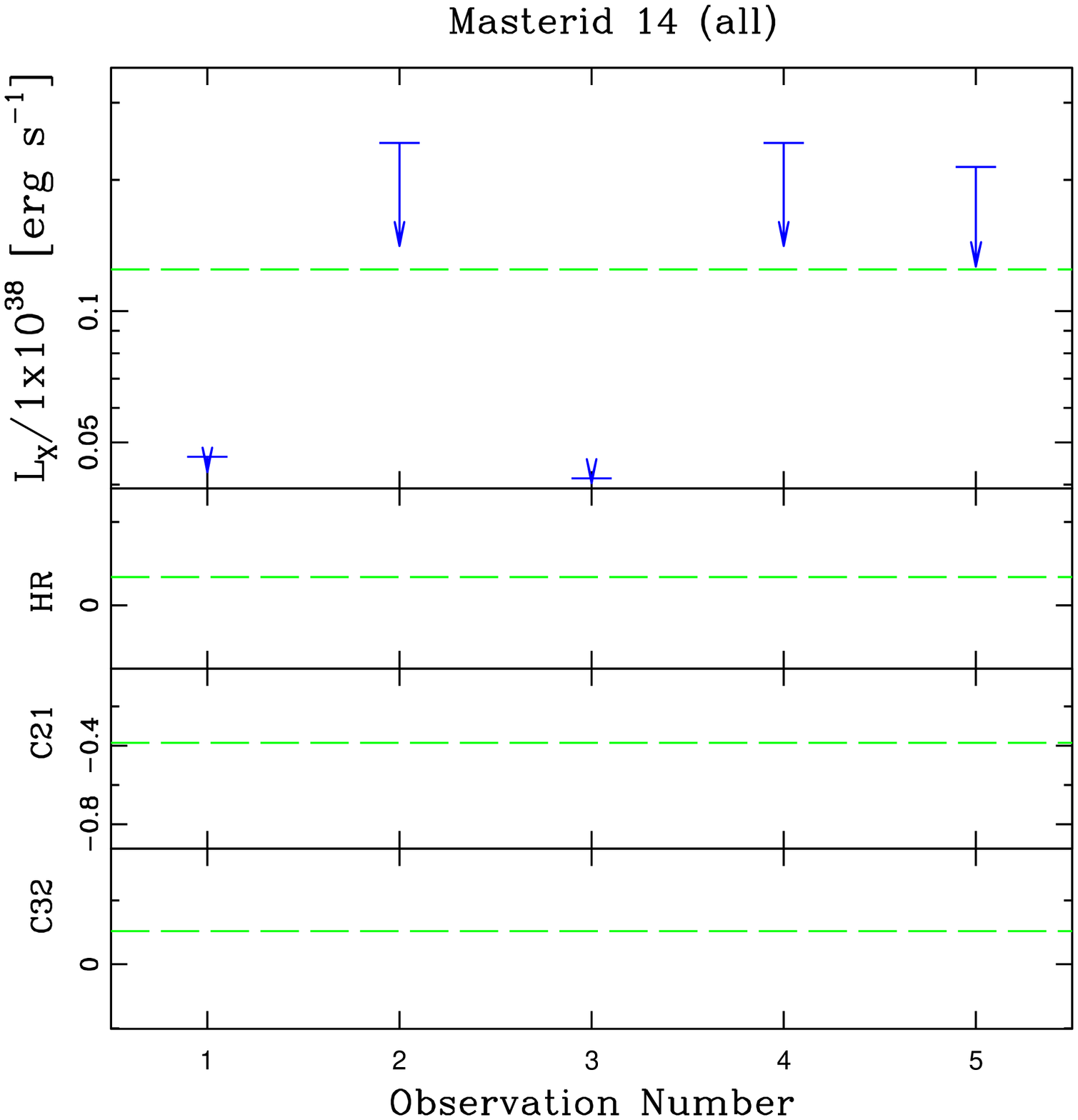}

\end{minipage}\hspace{0.02\linewidth}

\begin{minipage}{0.485\linewidth}
  \centering

    \includegraphics[width=\linewidth]{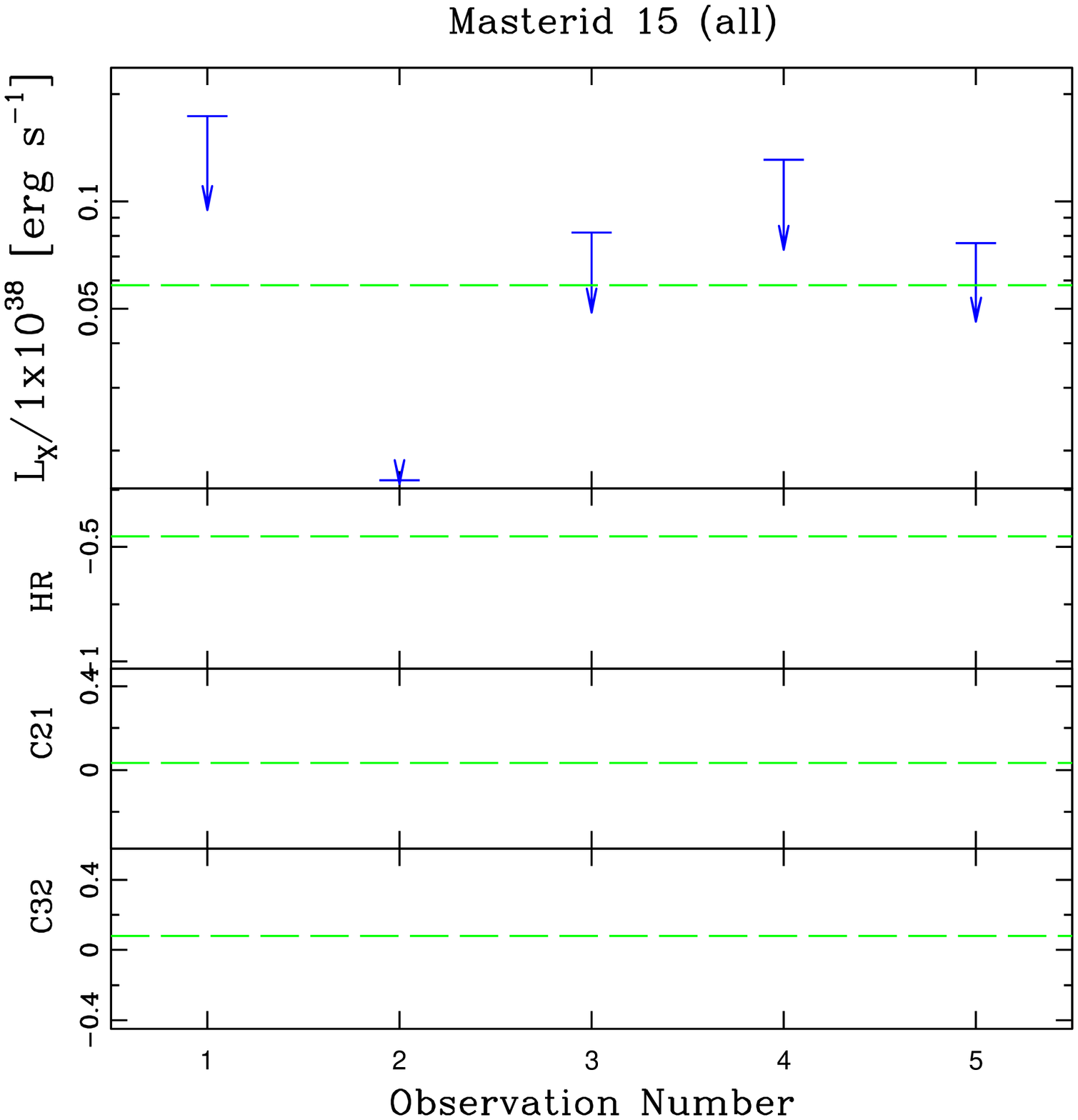}

 \end{minipage}\hspace{0.02\linewidth}
\begin{minipage}{0.485\linewidth}
  \centering
  
    \includegraphics[width=\linewidth]{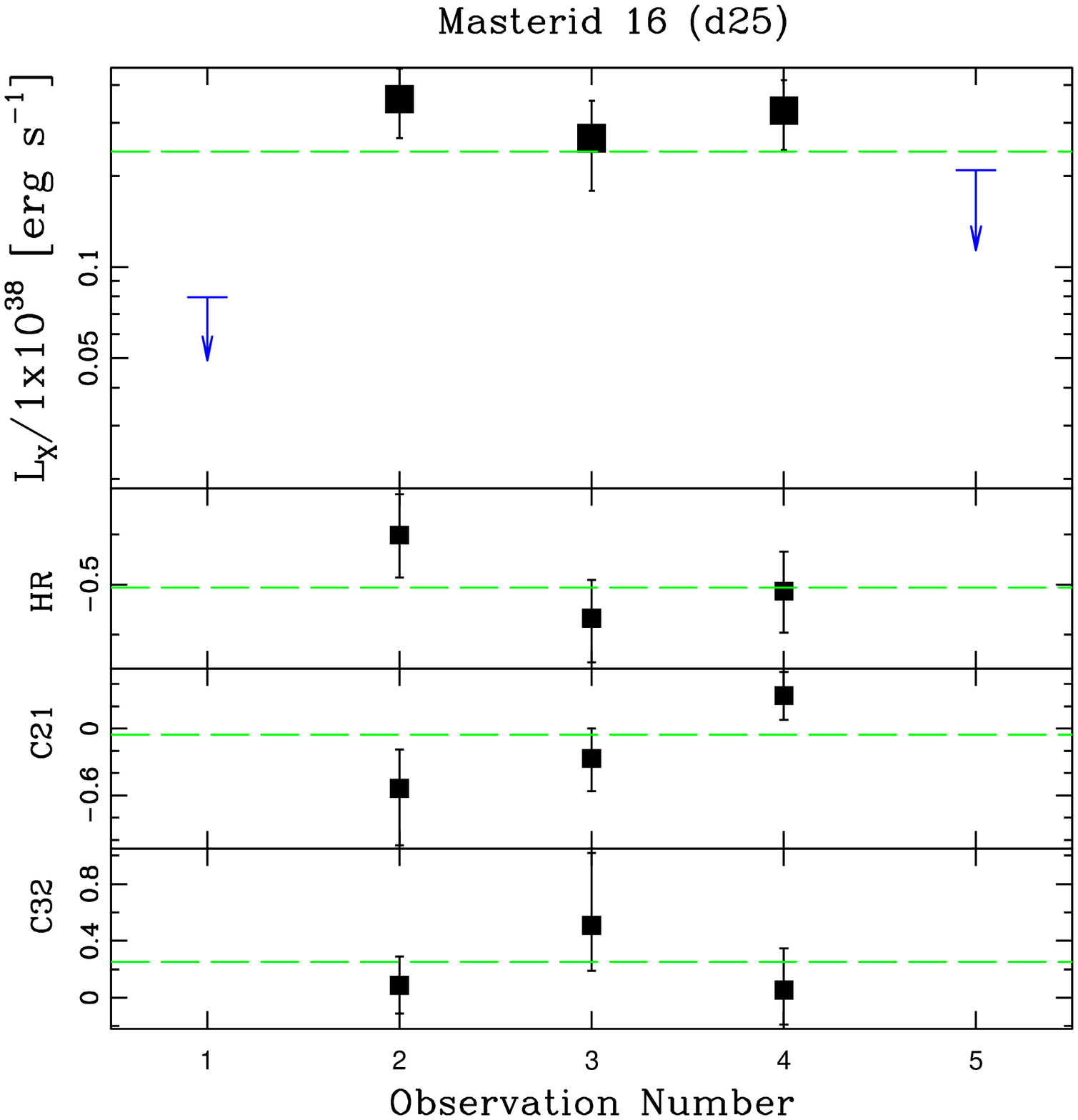}

  \end{minipage}\hspace{0.02\linewidth}

\end{figure}

\begin{figure}

  \begin{minipage}{0.485\linewidth}
  \centering

    \includegraphics[width=\linewidth]{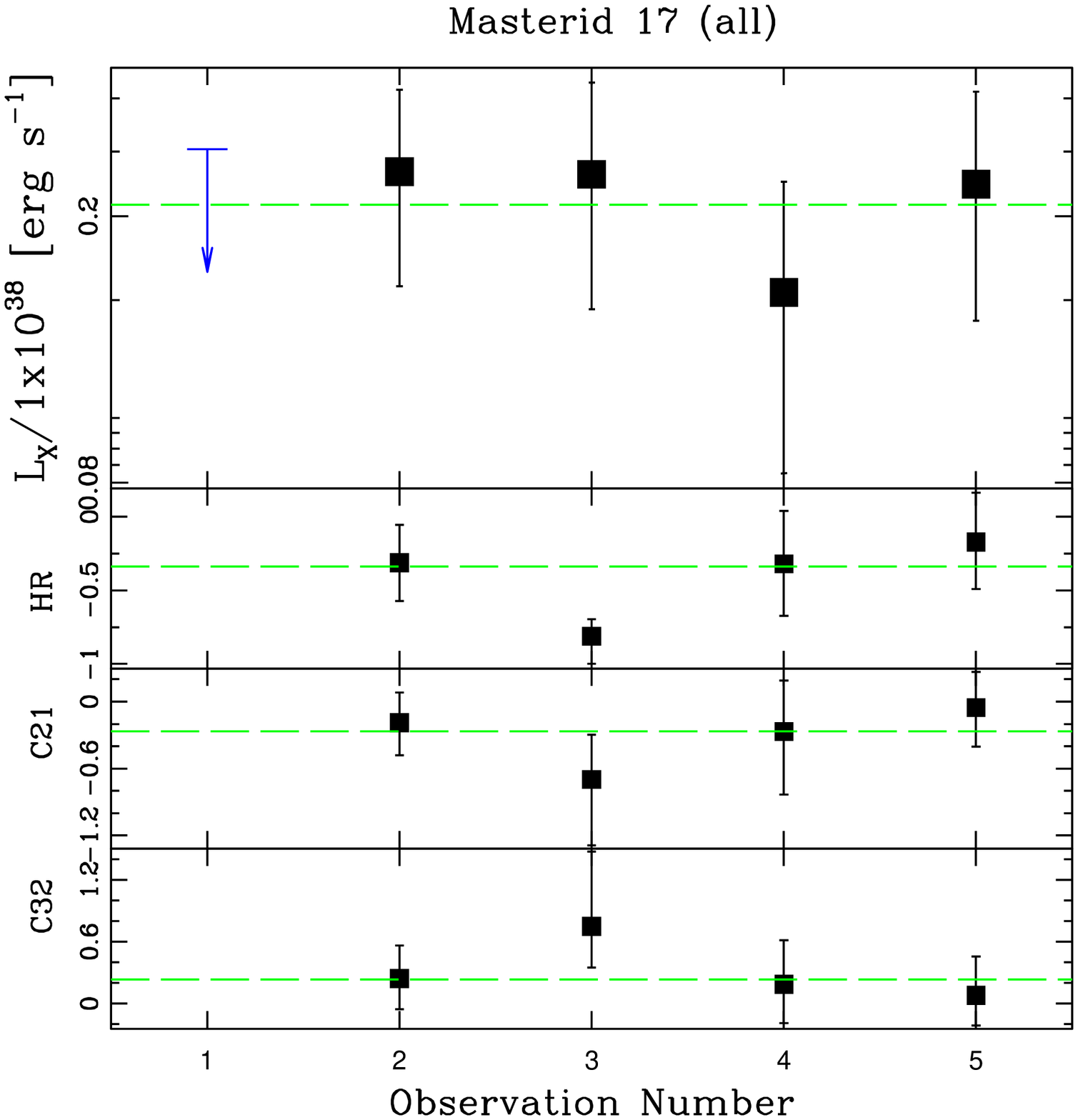}

\end{minipage}\hspace{0.02\linewidth}
\begin{minipage}{0.485\linewidth}
  \centering

    \includegraphics[width=\linewidth]{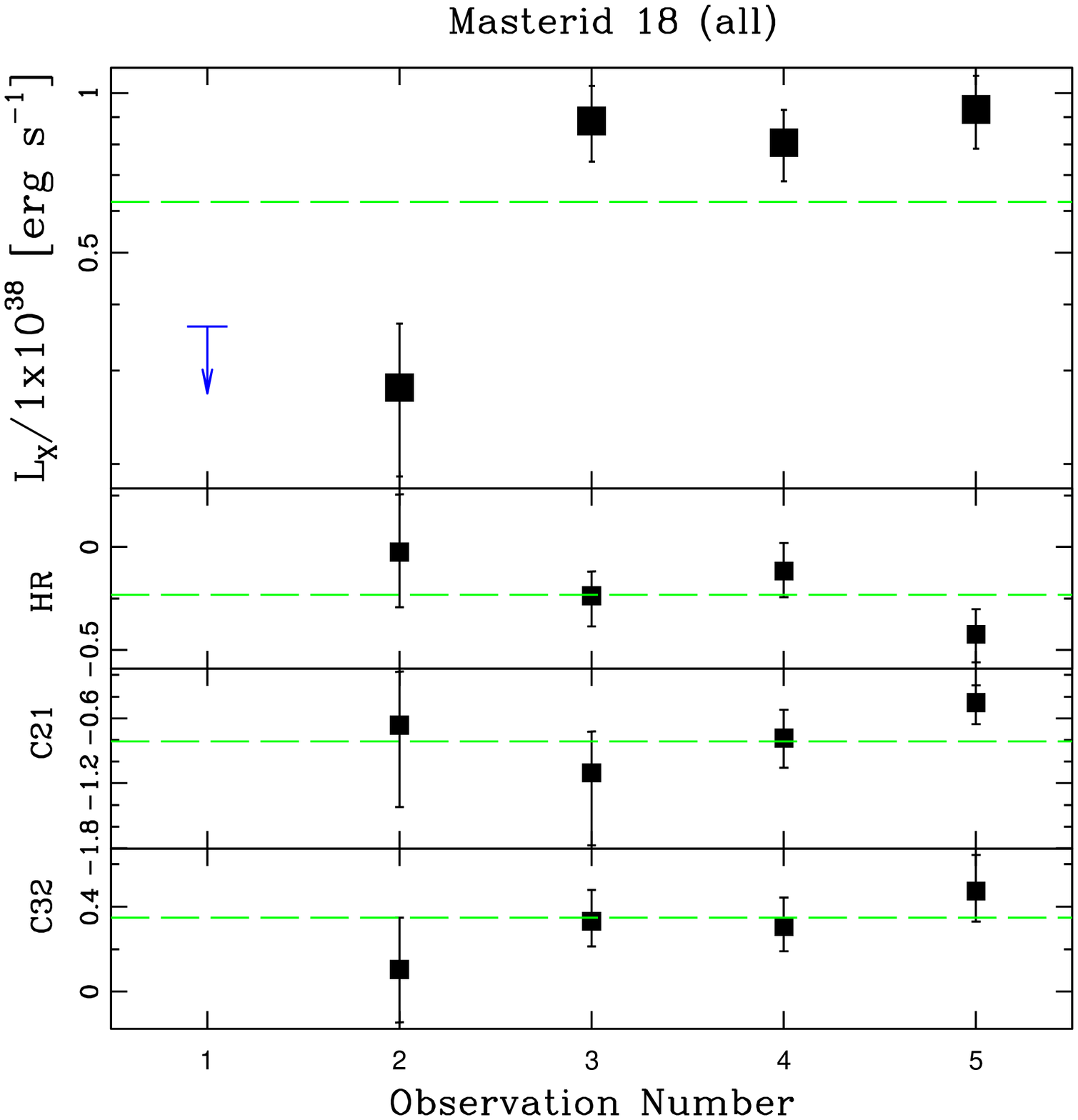}

 \end{minipage}\hspace{0.02\linewidth}
  
  \begin{minipage}{0.485\linewidth}
  \centering
  
    \includegraphics[width=\linewidth]{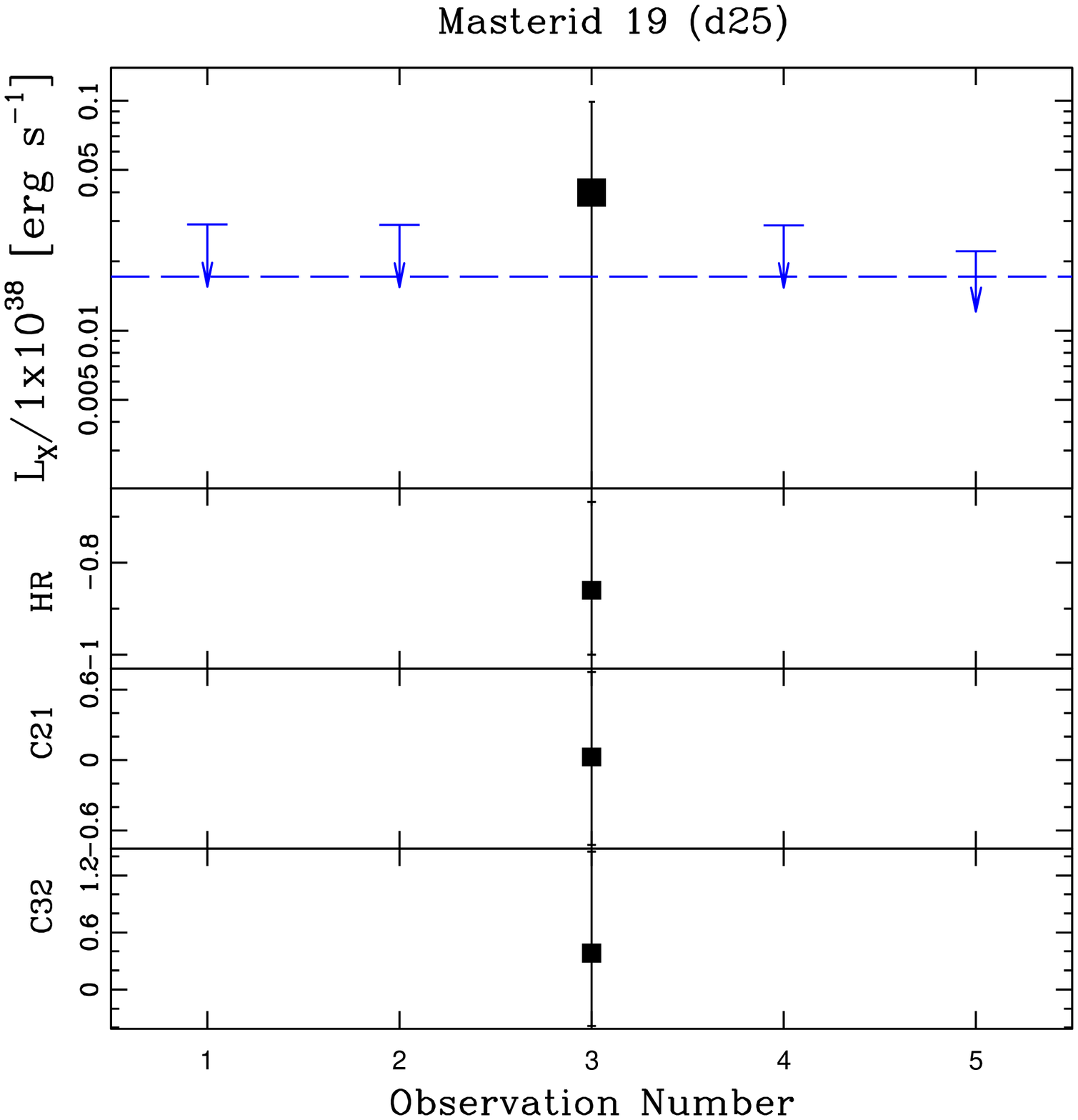}

  \end{minipage}\hspace{0.02\linewidth}
  \begin{minipage}{0.485\linewidth}
  \centering

    \includegraphics[width=\linewidth]{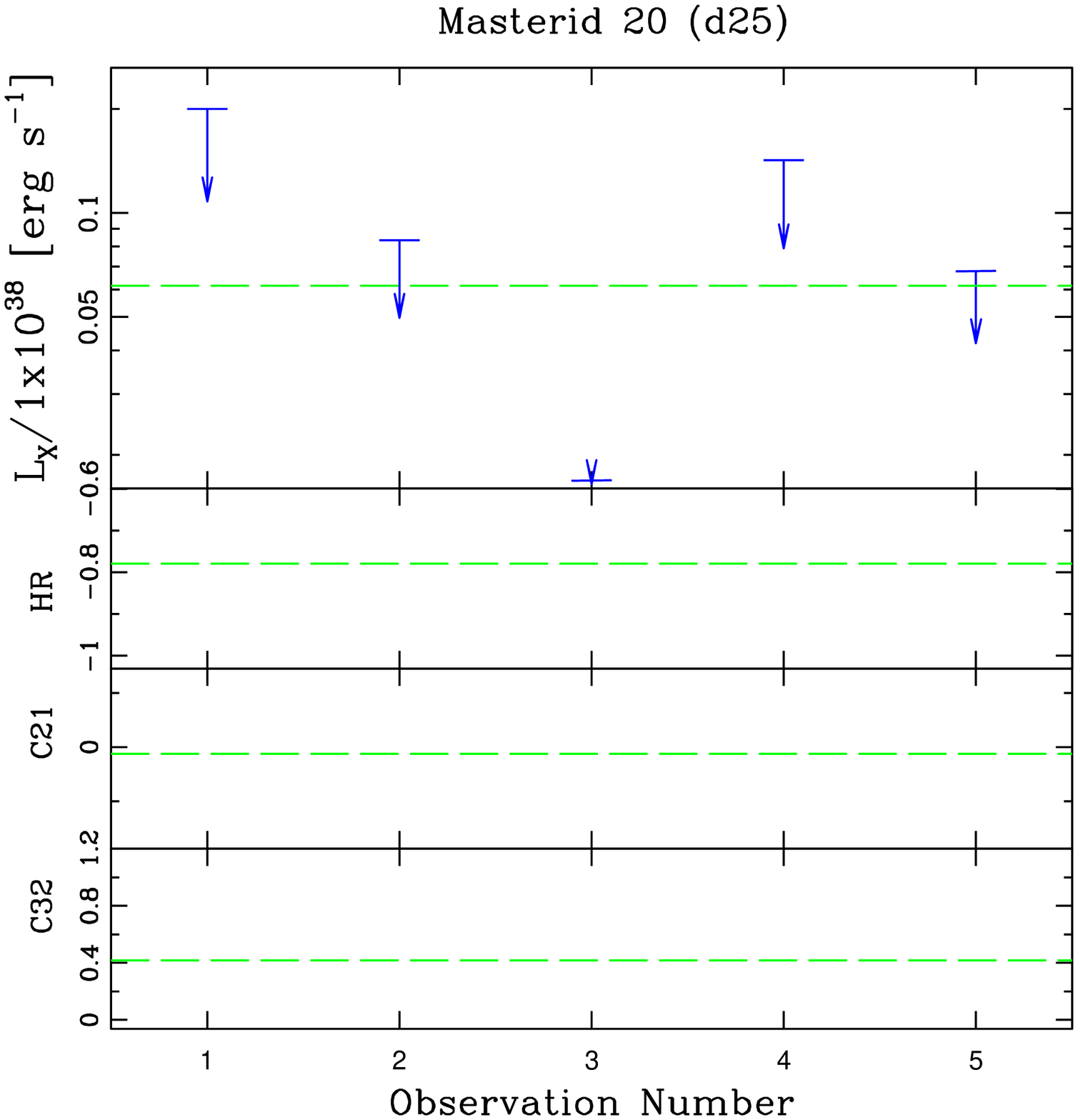}

\end{minipage}\hspace{0.02\linewidth}

\begin{minipage}{0.485\linewidth}
  \centering

    \includegraphics[width=\linewidth]{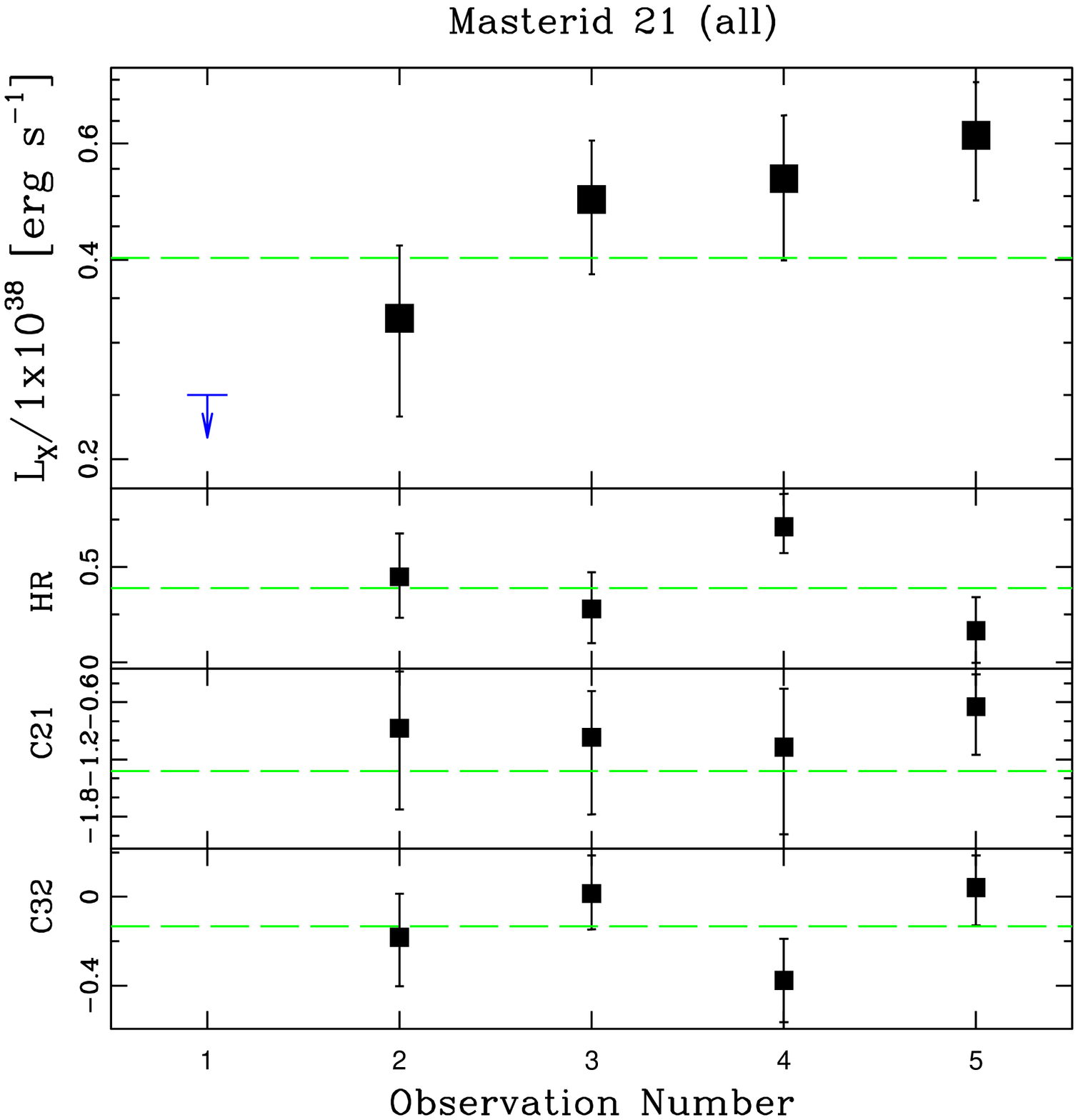}

 \end{minipage}\hspace{0.02\linewidth}
\begin{minipage}{0.485\linewidth}
  \centering
  
    \includegraphics[width=\linewidth]{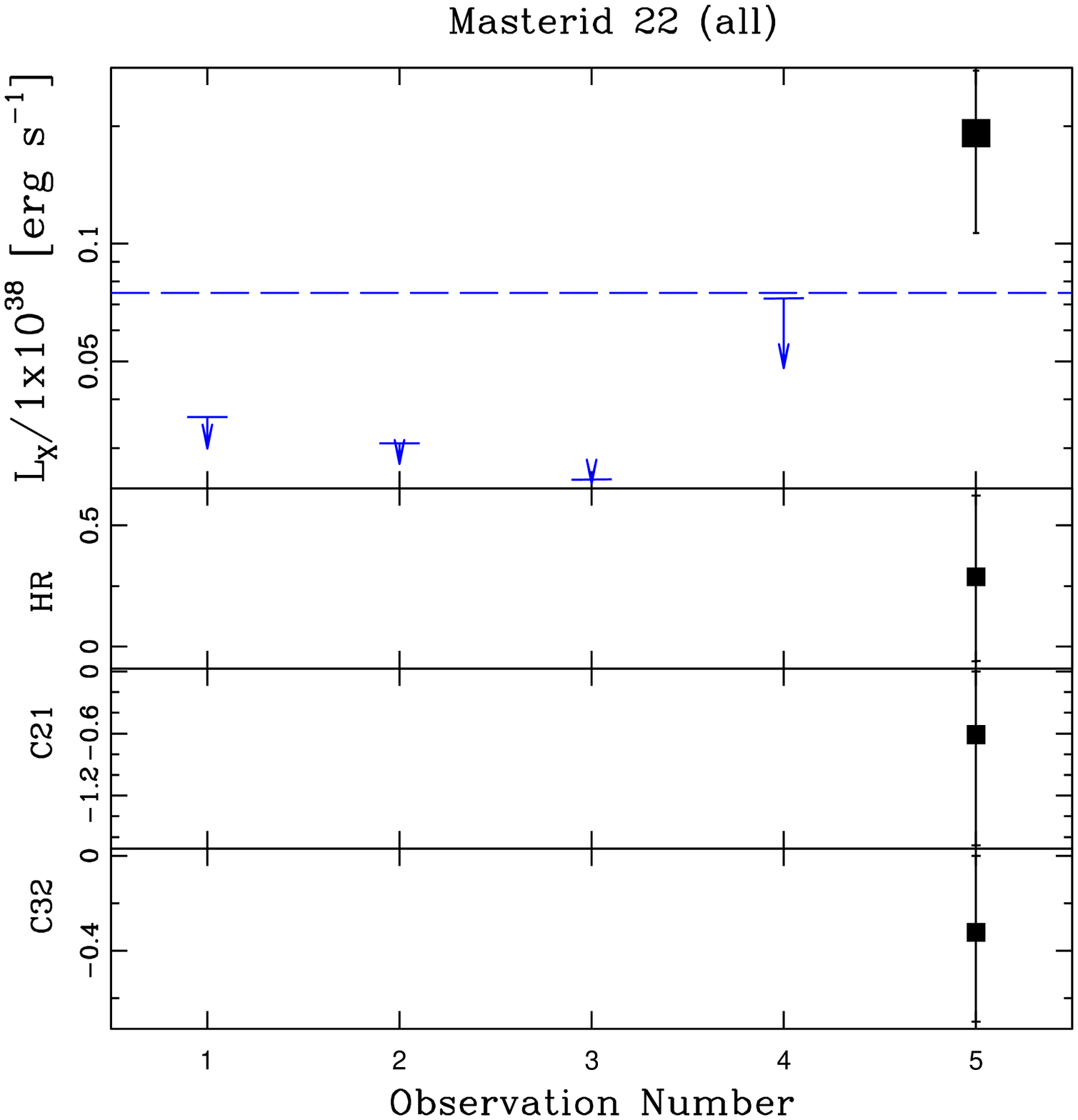}

  \end{minipage}\hspace{0.02\linewidth}

\end{figure}

\begin{figure}

  \begin{minipage}{0.485\linewidth}
  \centering

    \includegraphics[width=\linewidth]{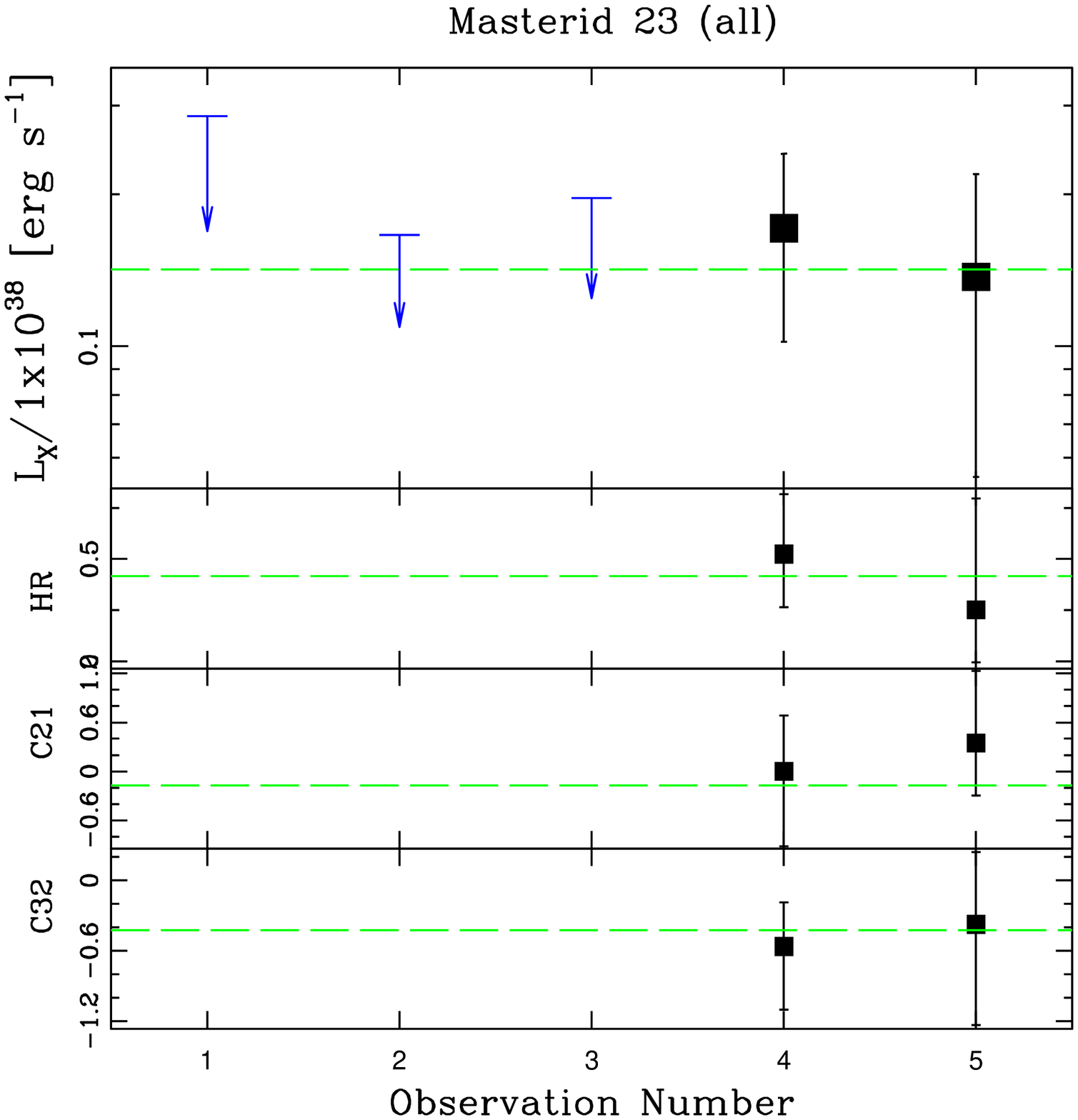}

\end{minipage}\hspace{0.02\linewidth}
\begin{minipage}{0.485\linewidth}
  \centering

    \includegraphics[width=\linewidth]{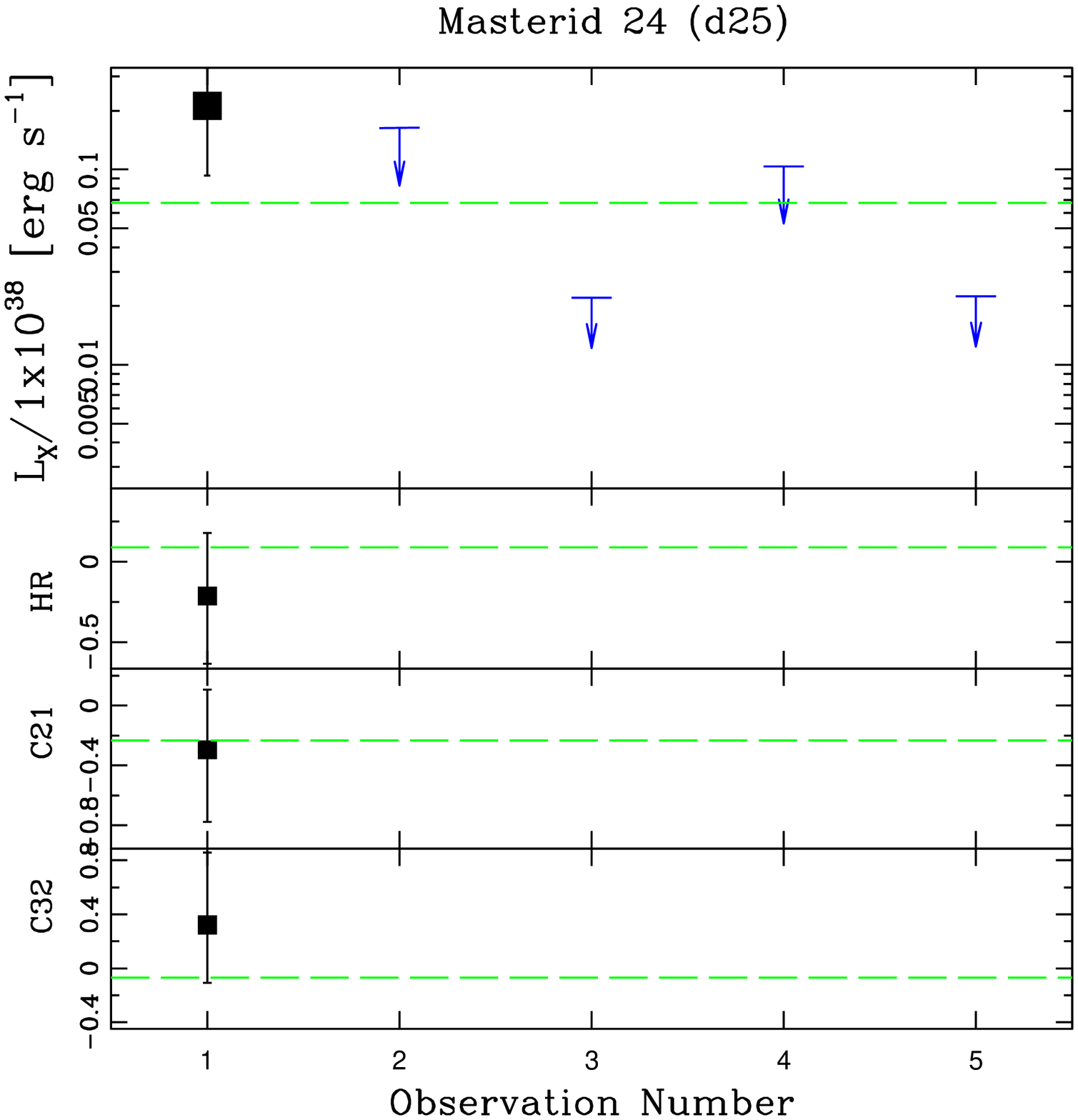}

 \end{minipage}\hspace{0.02\linewidth}

  \begin{minipage}{0.485\linewidth}
  \centering
  
    \includegraphics[width=\linewidth]{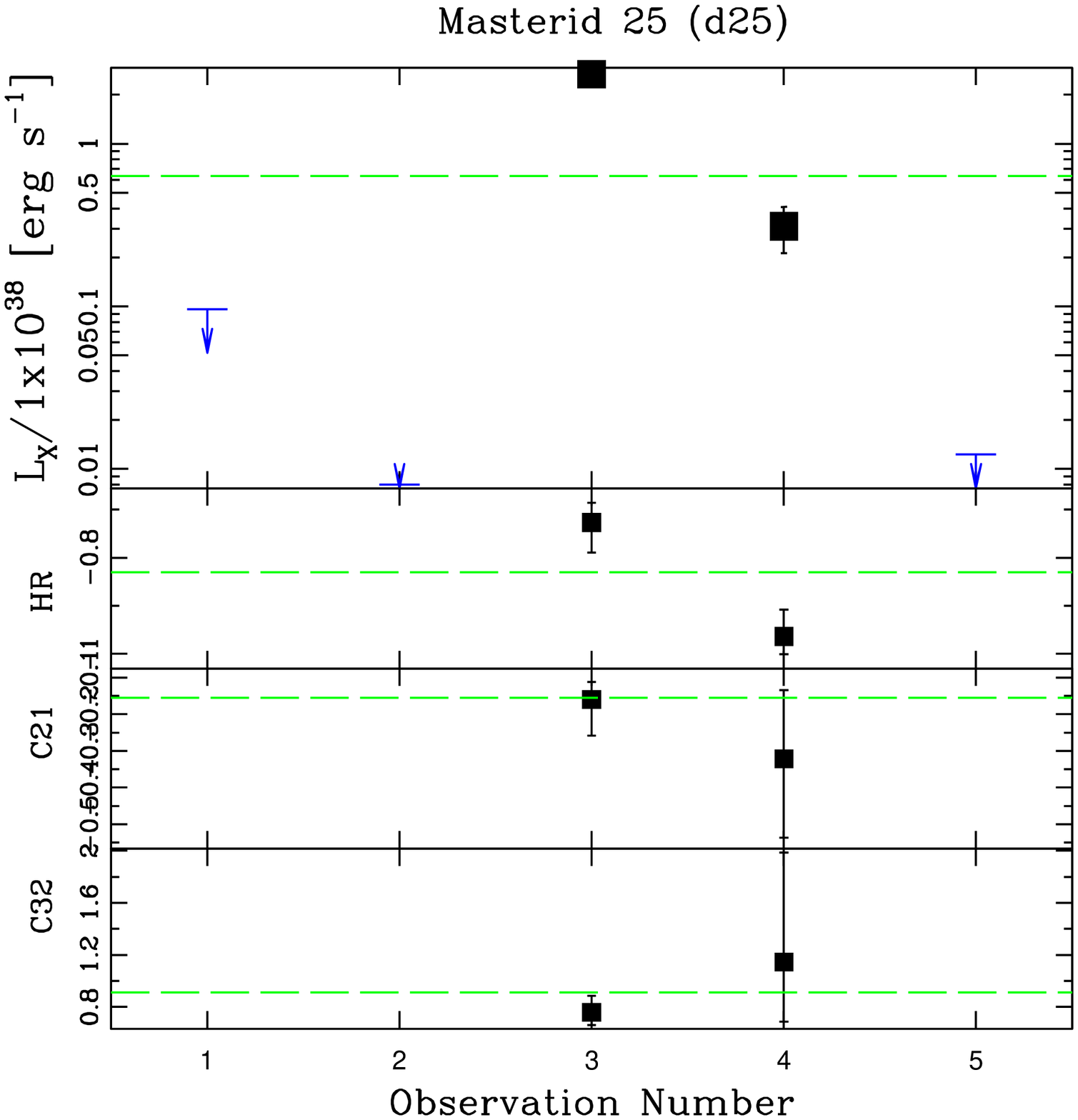}

  \end{minipage}\hspace{0.02\linewidth}
  \begin{minipage}{0.485\linewidth}
  \centering

    \includegraphics[width=\linewidth]{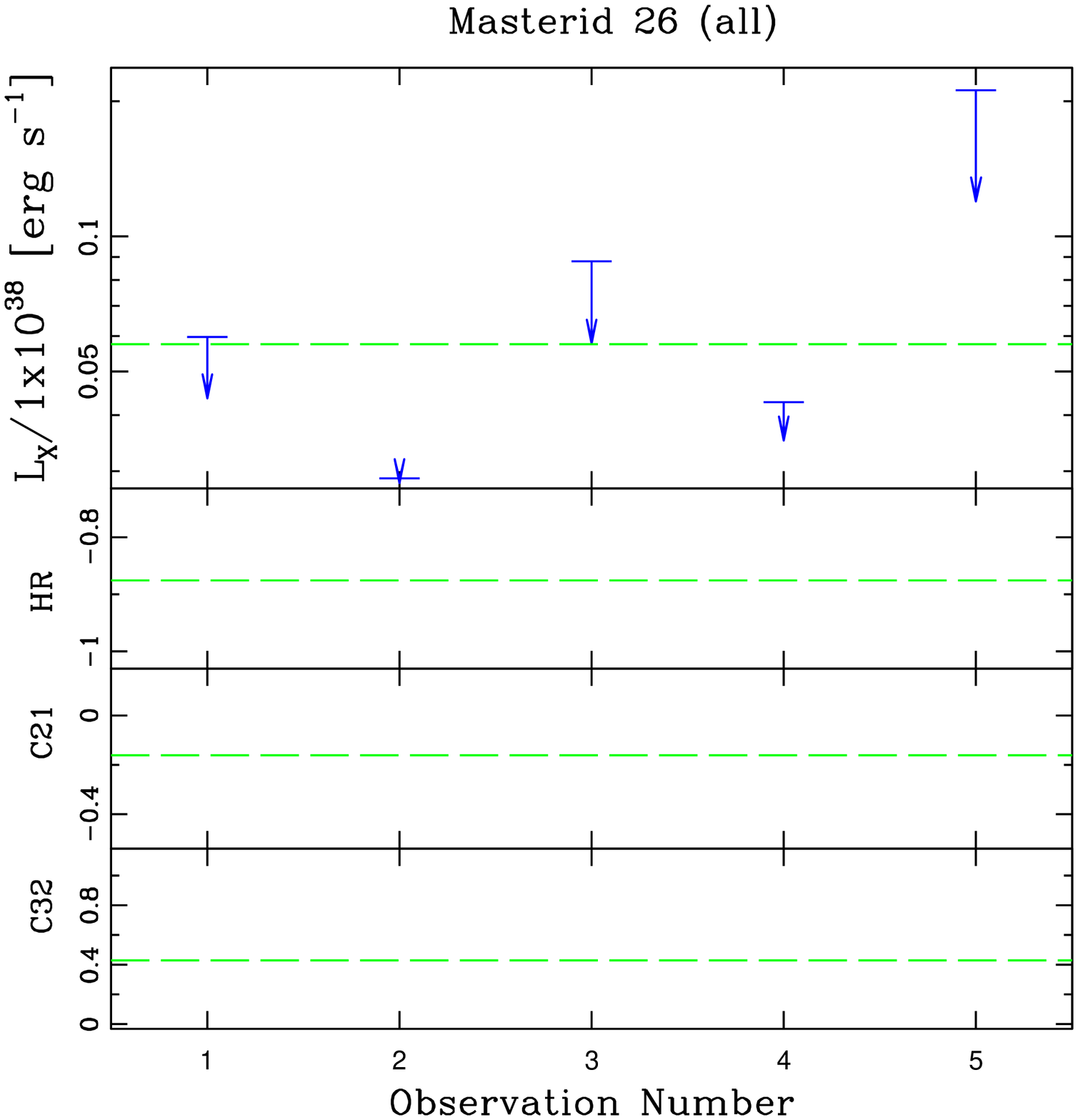}

\end{minipage}\hspace{0.02\linewidth}

\begin{minipage}{0.485\linewidth}
  \centering

    \includegraphics[width=\linewidth]{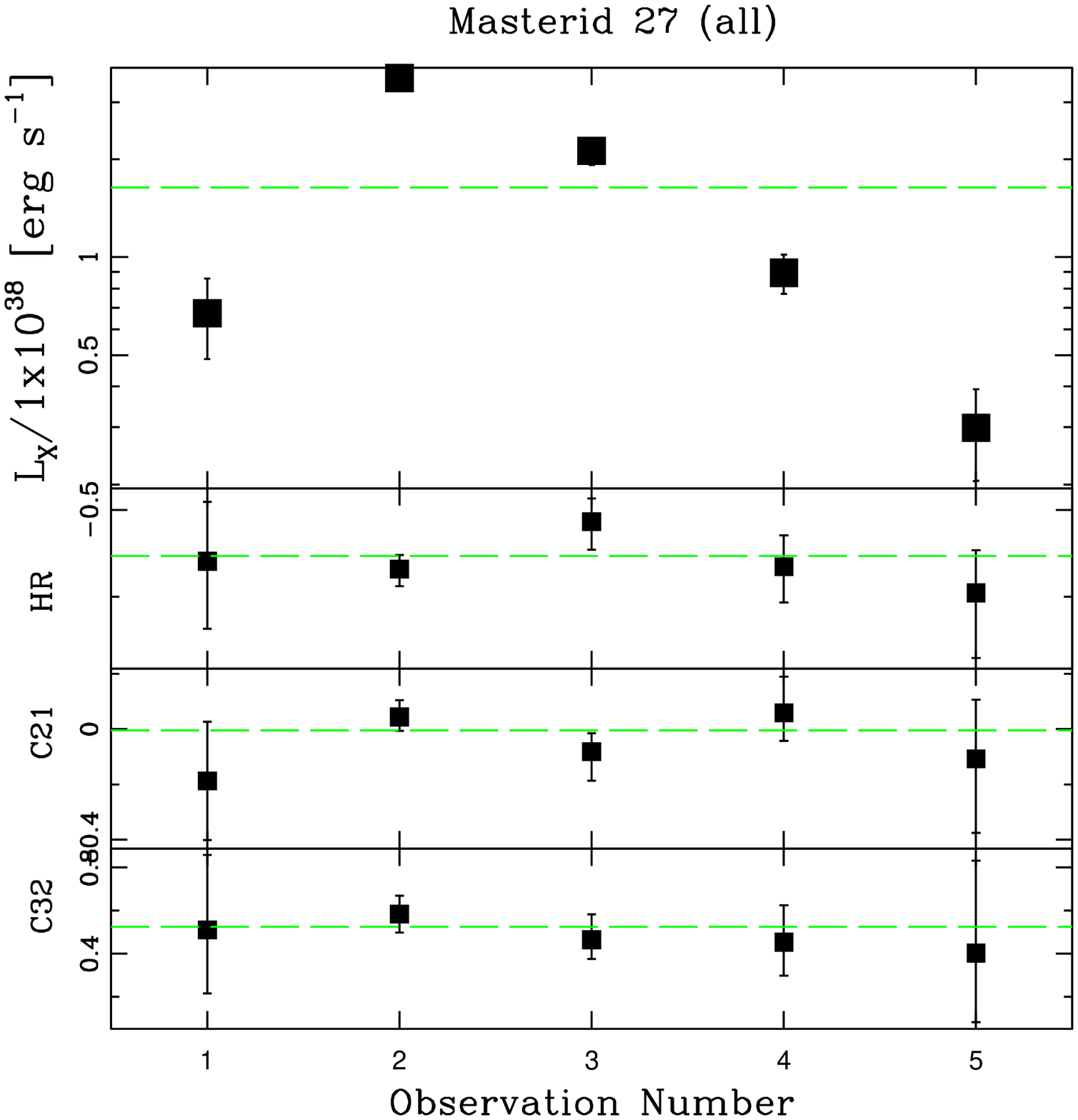}

 \end{minipage}\hspace{0.02\linewidth}
\begin{minipage}{0.485\linewidth}
  \centering
  
    \includegraphics[width=\linewidth]{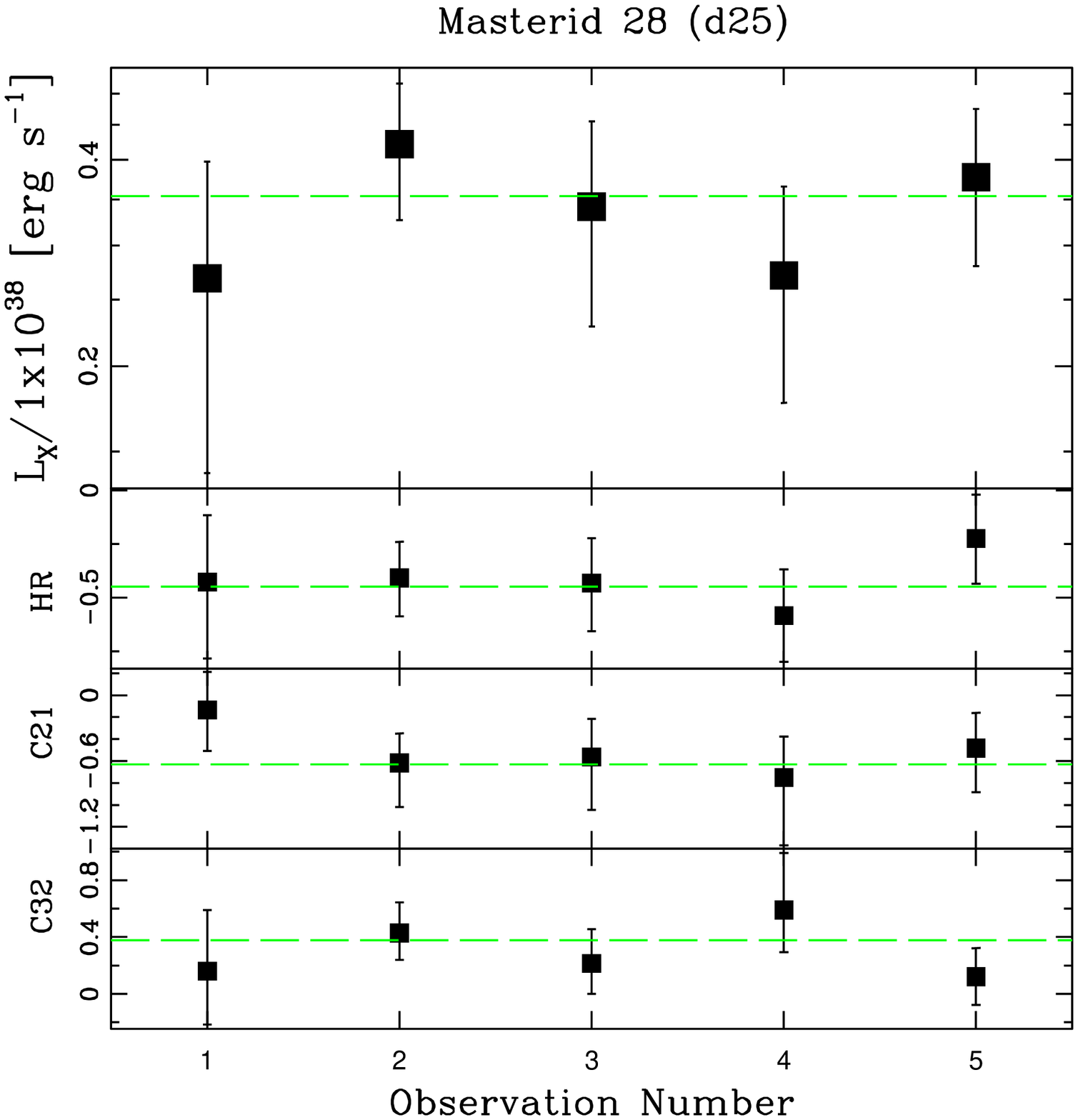}

  \end{minipage}\hspace{0.02\linewidth}
  
\end{figure}

\clearpage

\begin{figure}

  \begin{minipage}{0.485\linewidth}
  \centering

    \includegraphics[width=\linewidth]{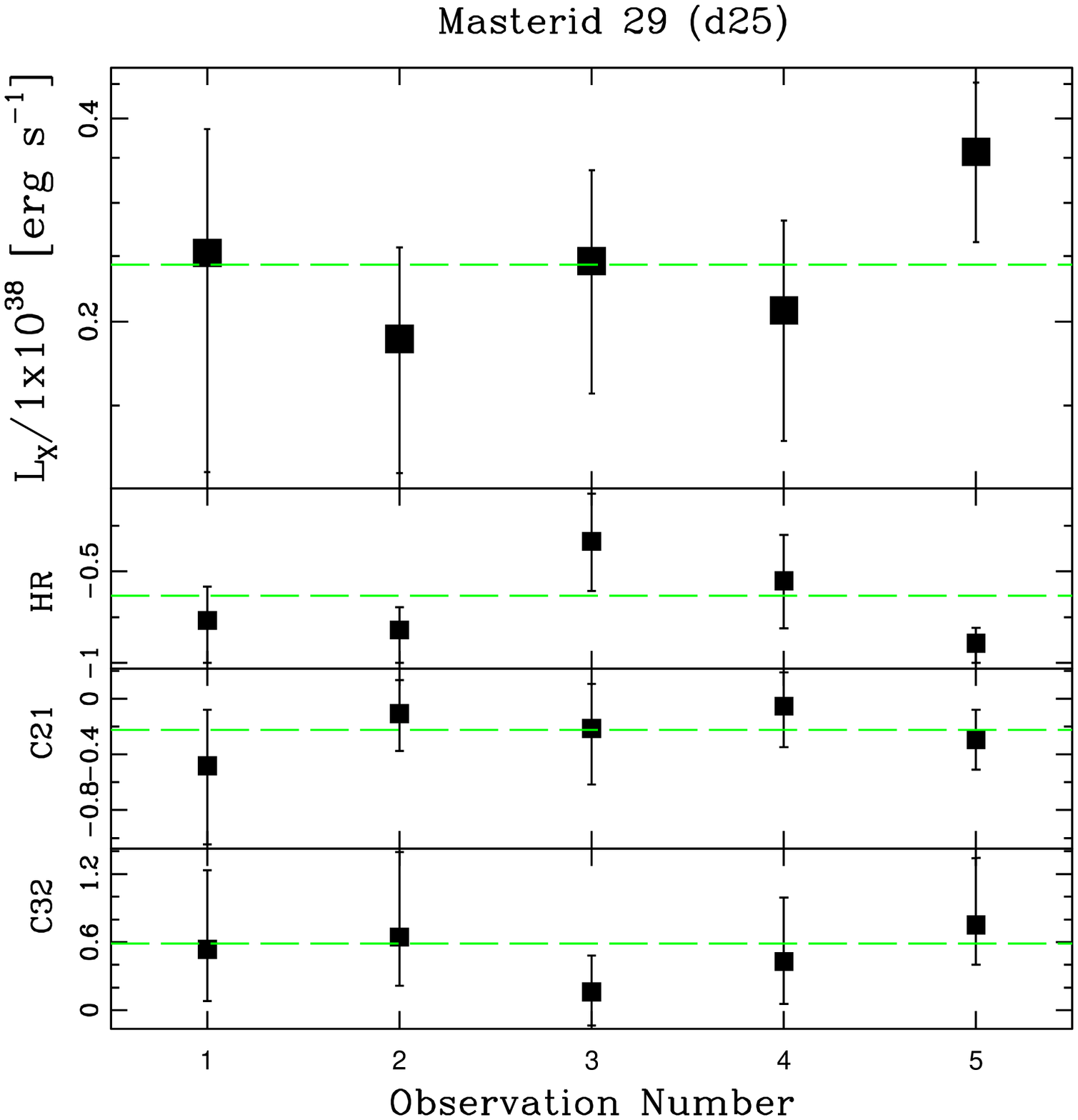}

\end{minipage}\hspace{0.02\linewidth}
\begin{minipage}{0.485\linewidth}
  \centering

    \includegraphics[width=\linewidth]{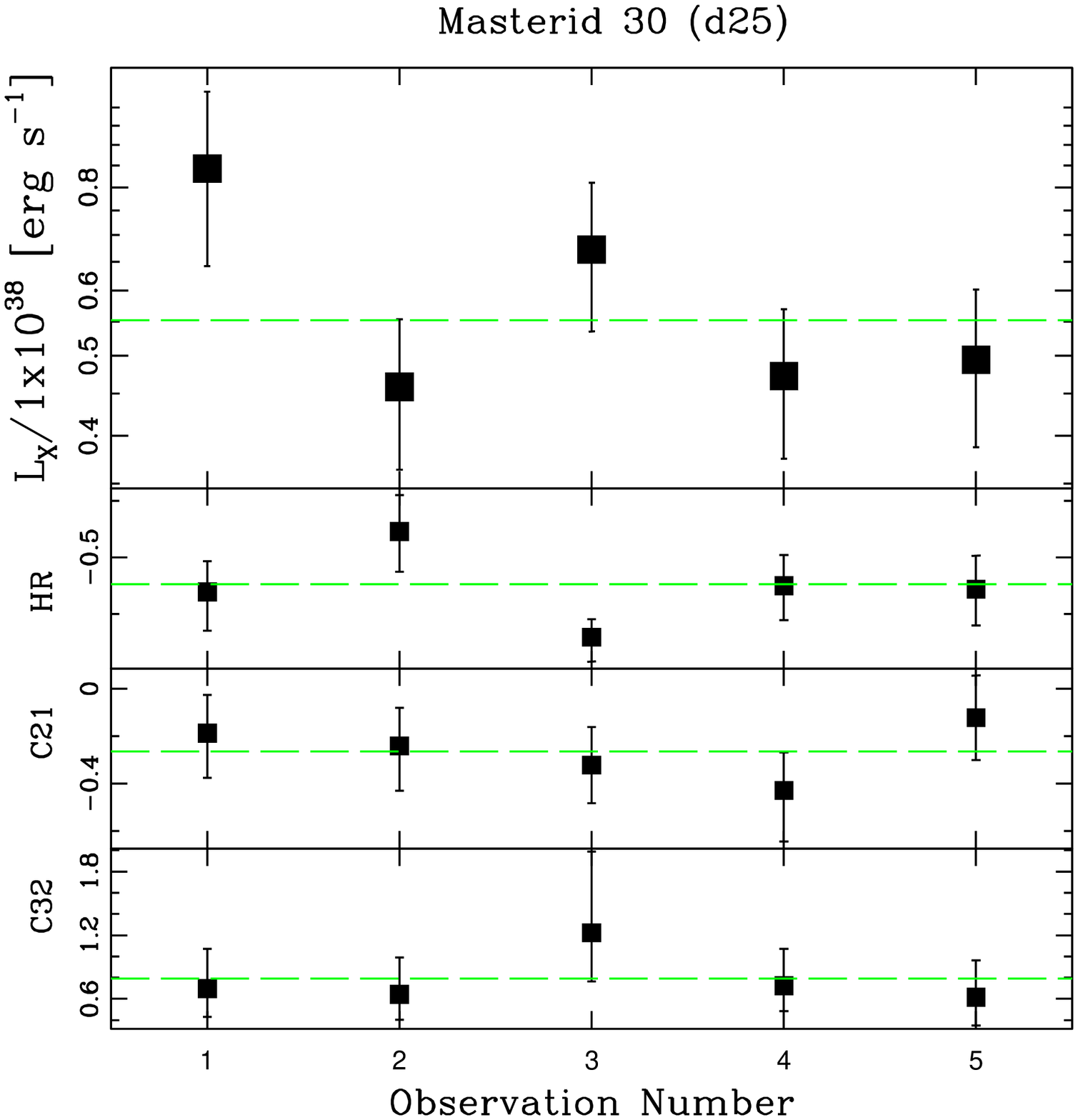}

 \end{minipage}\hspace{0.02\linewidth}

  \begin{minipage}{0.485\linewidth}
  \centering
  
    \includegraphics[width=\linewidth]{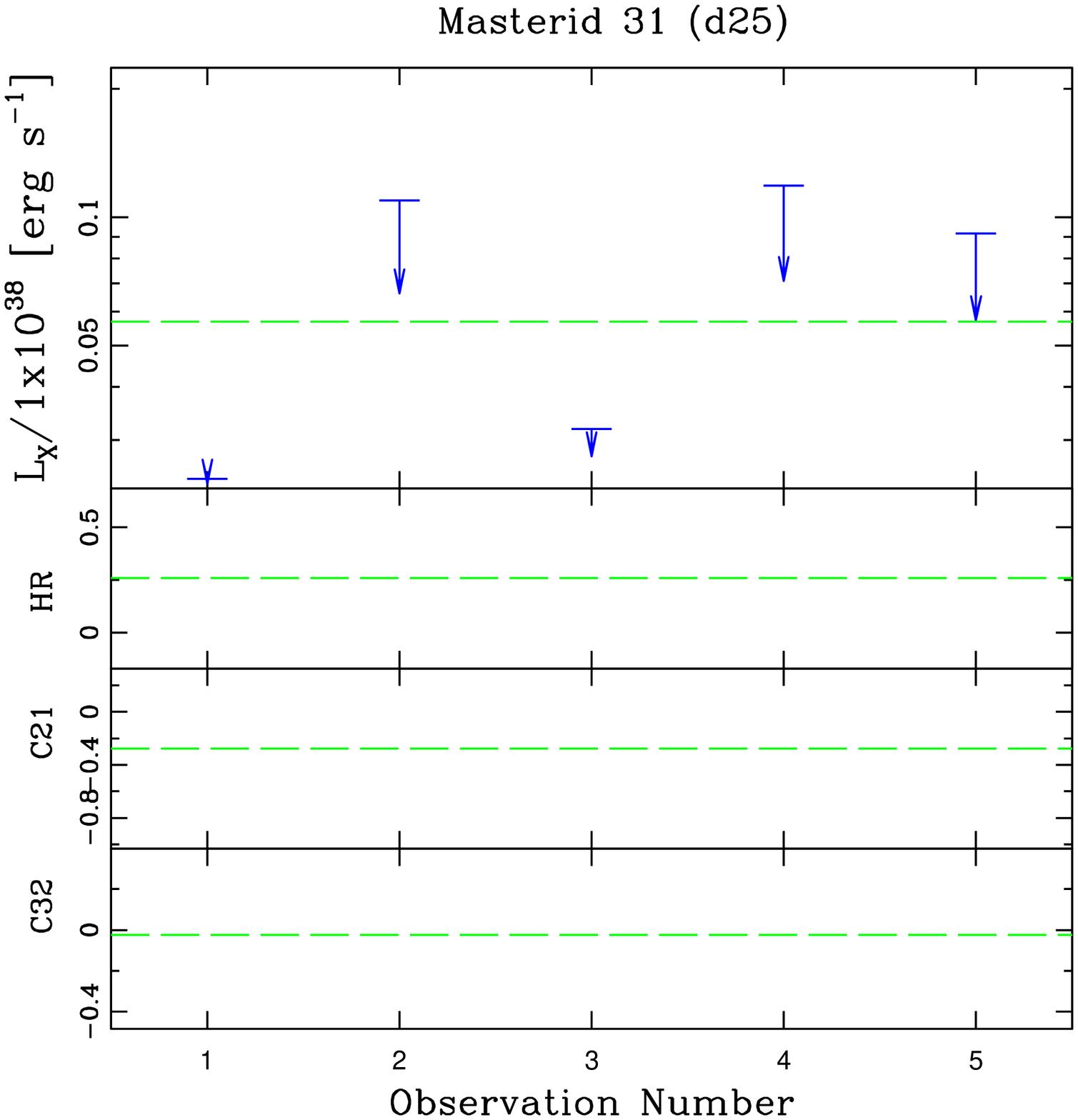}

  \end{minipage}\hspace{0.02\linewidth}
  \begin{minipage}{0.485\linewidth}
  \centering

    \includegraphics[width=\linewidth]{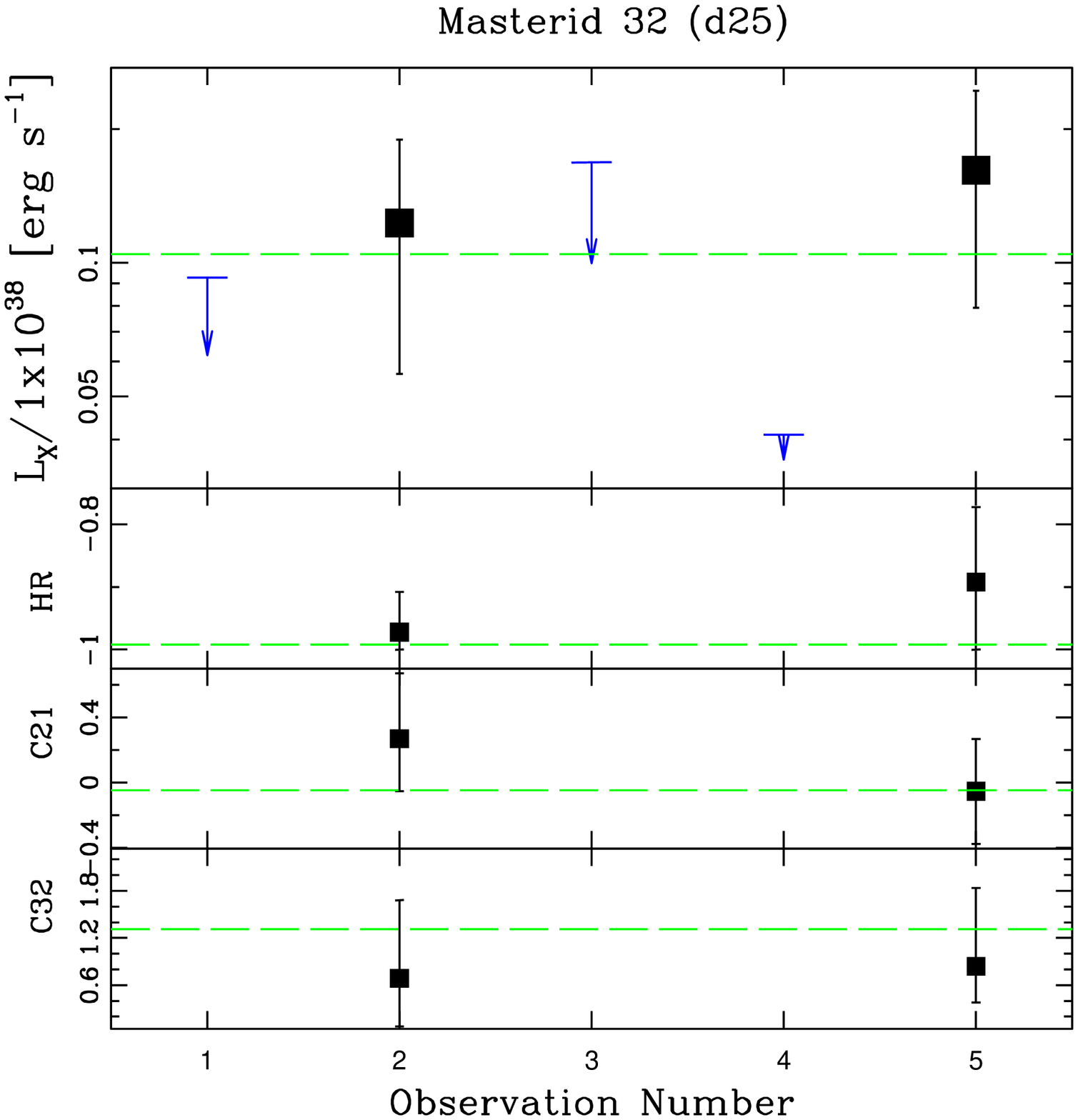}

\end{minipage}\hspace{0.02\linewidth}

\begin{minipage}{0.485\linewidth}
  \centering

    \includegraphics[width=\linewidth]{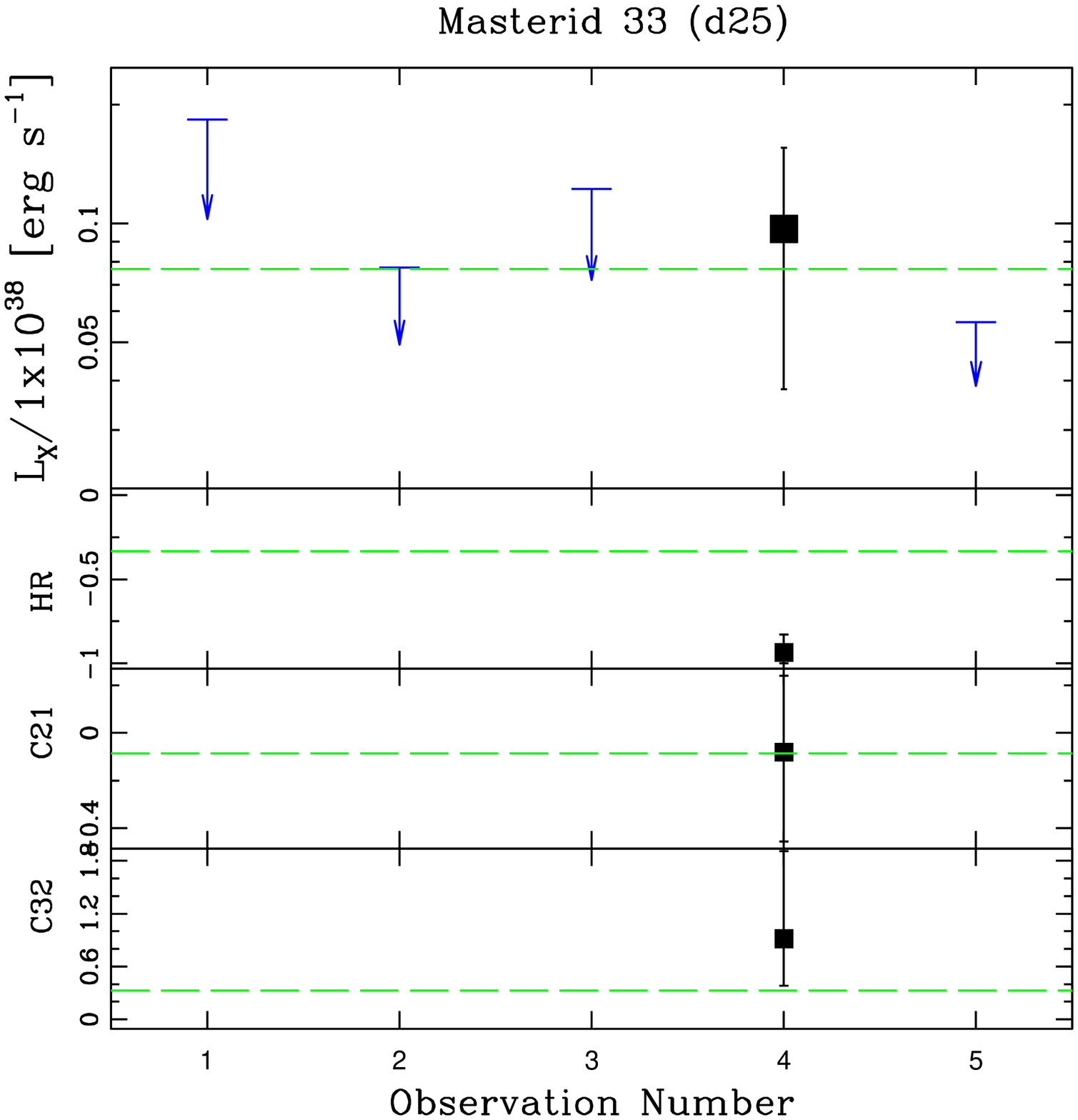}

 \end{minipage}\hspace{0.02\linewidth}
\begin{minipage}{0.485\linewidth}
  \centering
  
    \includegraphics[width=\linewidth]{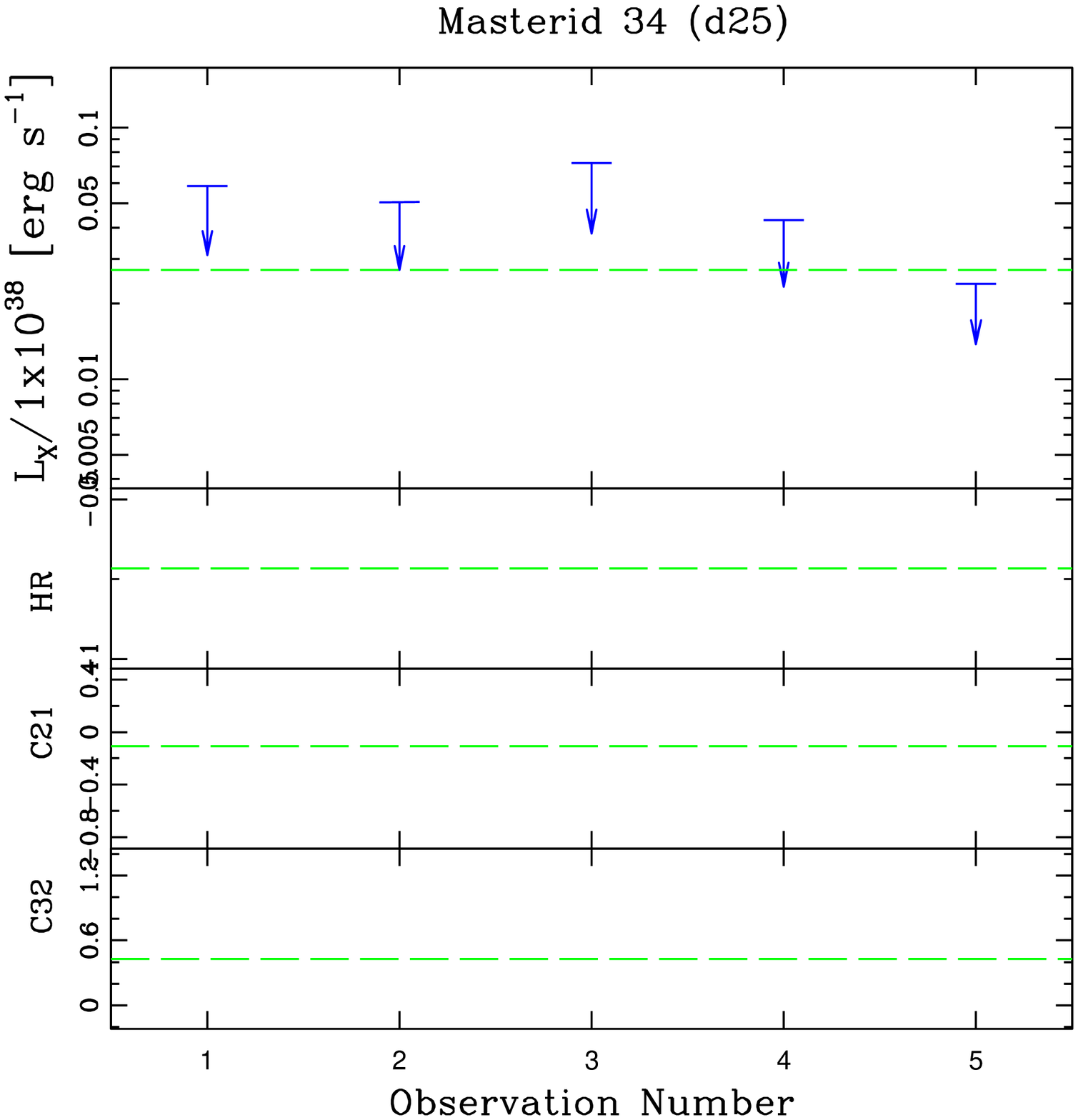}

  \end{minipage}\hspace{0.02\linewidth}

\end{figure}

\begin{figure}

  \begin{minipage}{0.485\linewidth}
  \centering

    \includegraphics[width=\linewidth]{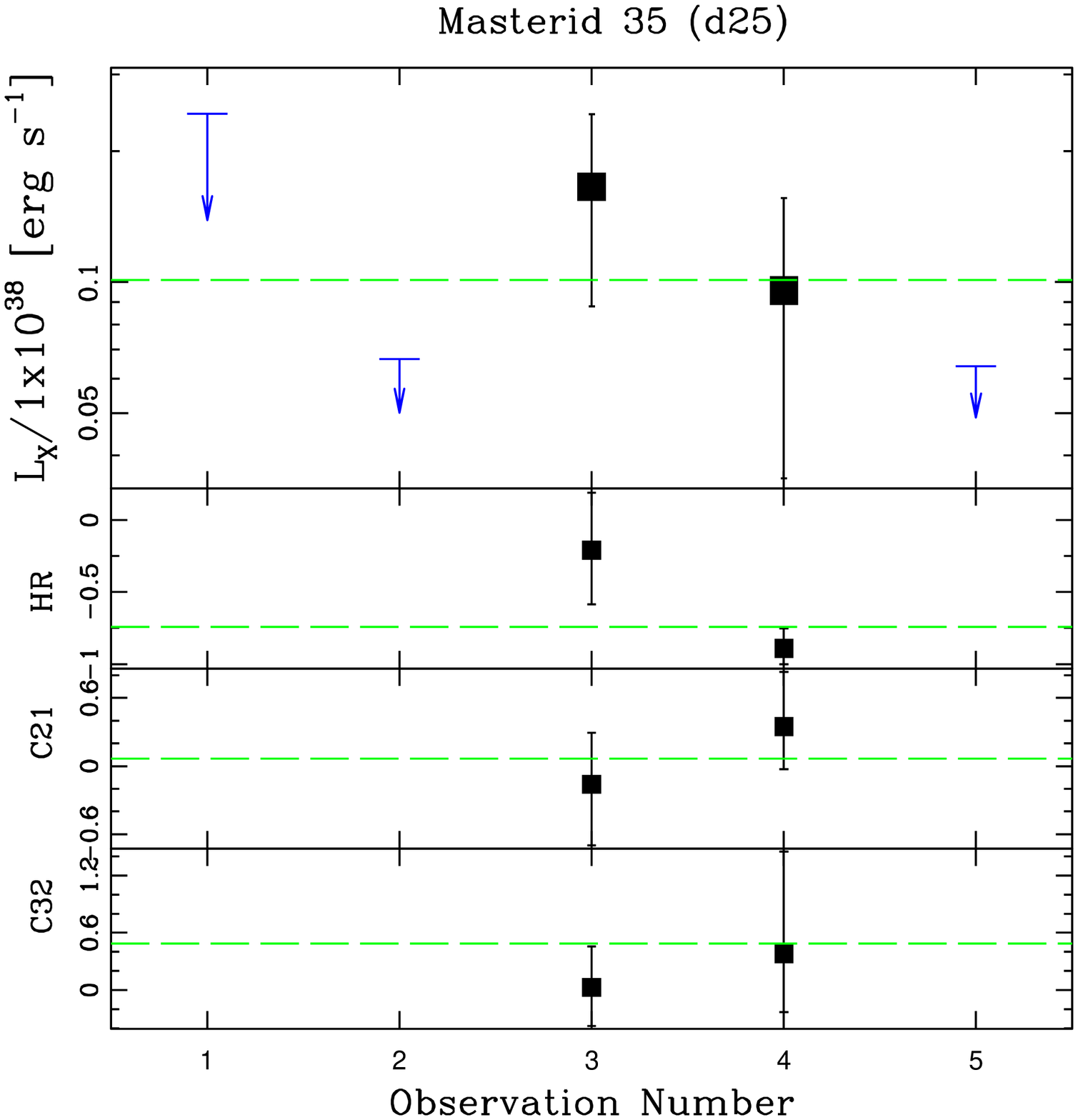}

\end{minipage}\hspace{0.02\linewidth}
\begin{minipage}{0.485\linewidth}
  \centering

    \includegraphics[width=\linewidth]{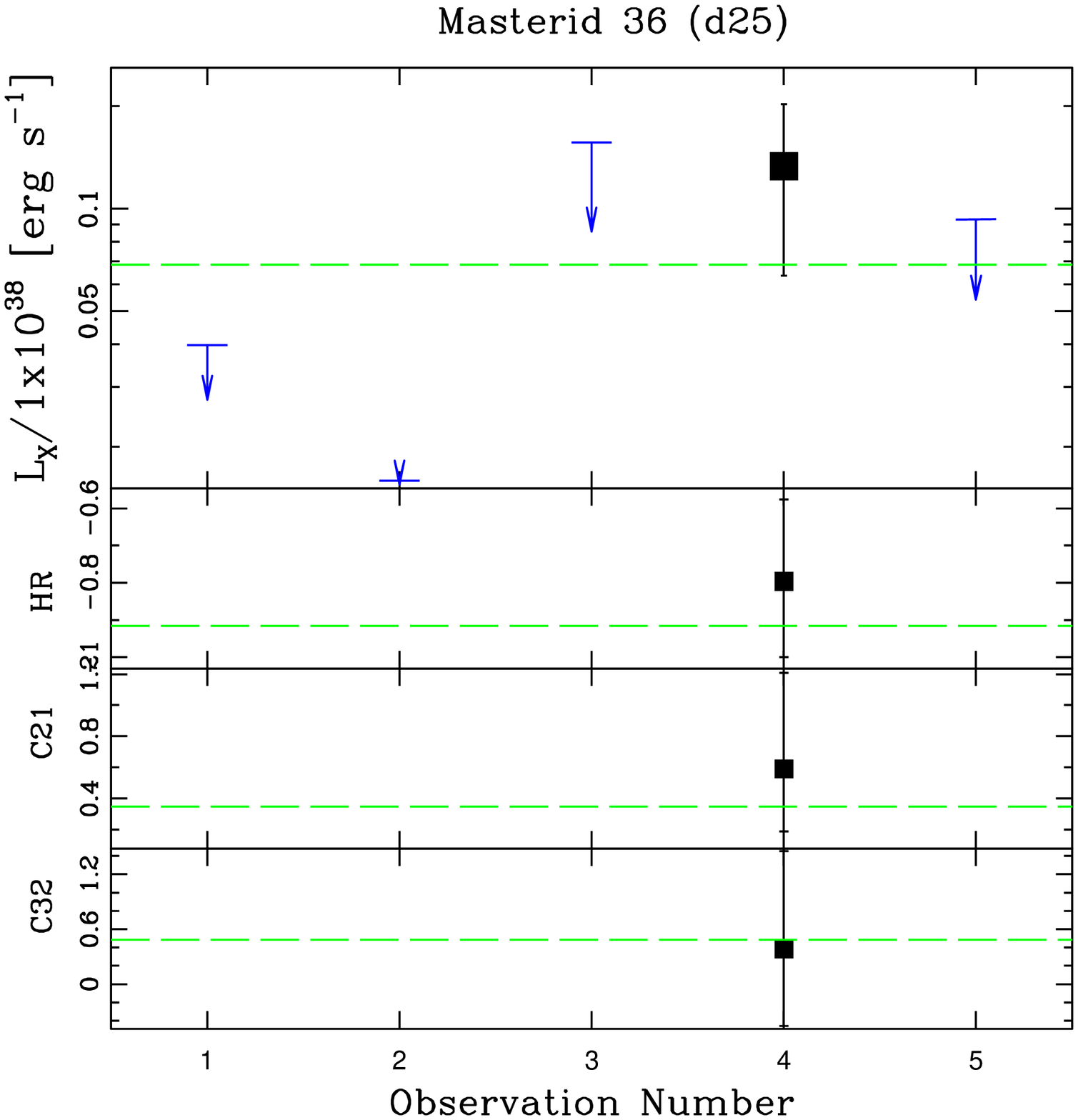}

 \end{minipage}\hspace{0.02\linewidth}

  \begin{minipage}{0.485\linewidth}
  \centering
  
    \includegraphics[width=\linewidth]{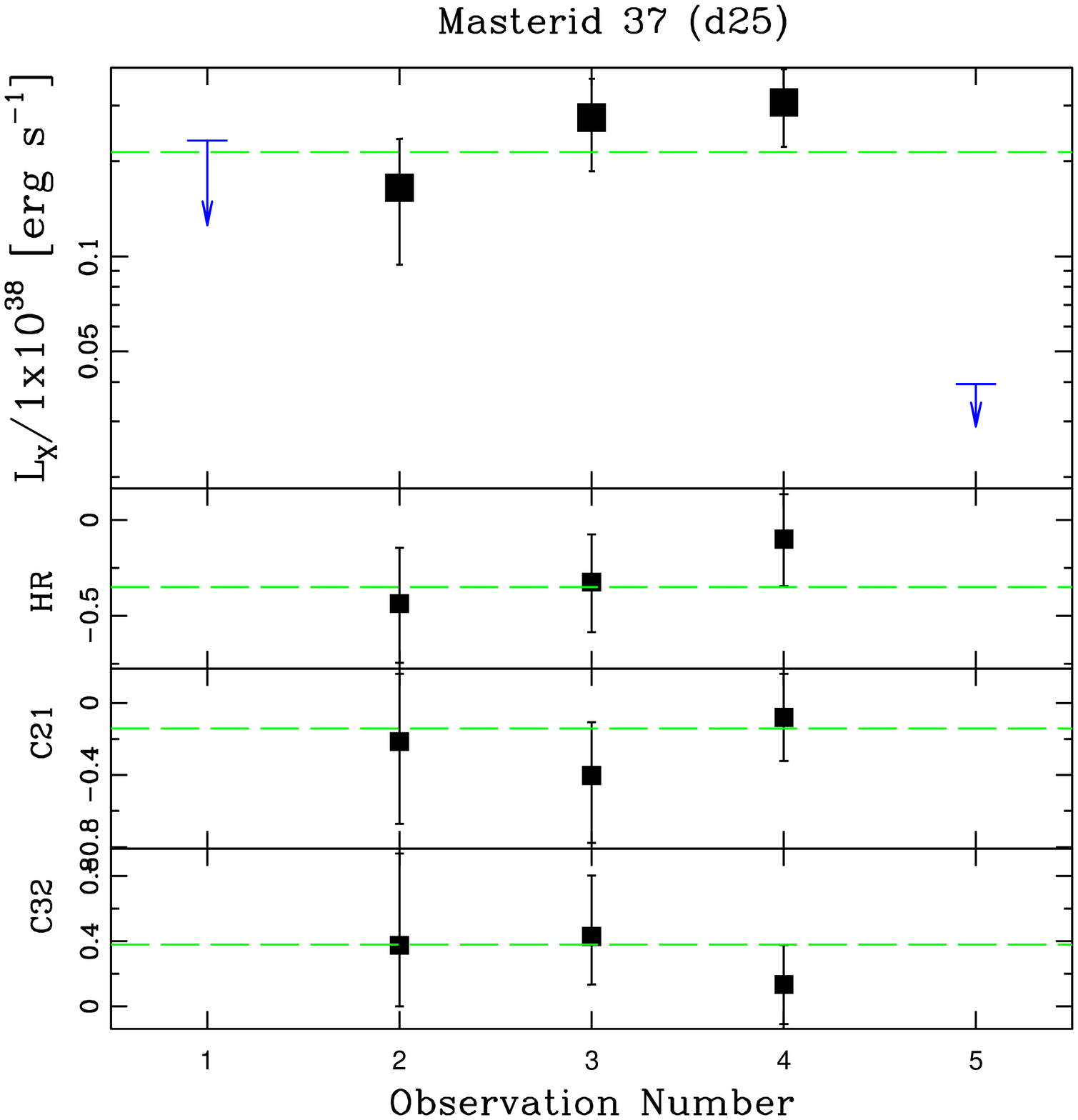}

  \end{minipage}\hspace{0.02\linewidth}
  \begin{minipage}{0.485\linewidth}
  \centering

    \includegraphics[width=\linewidth]{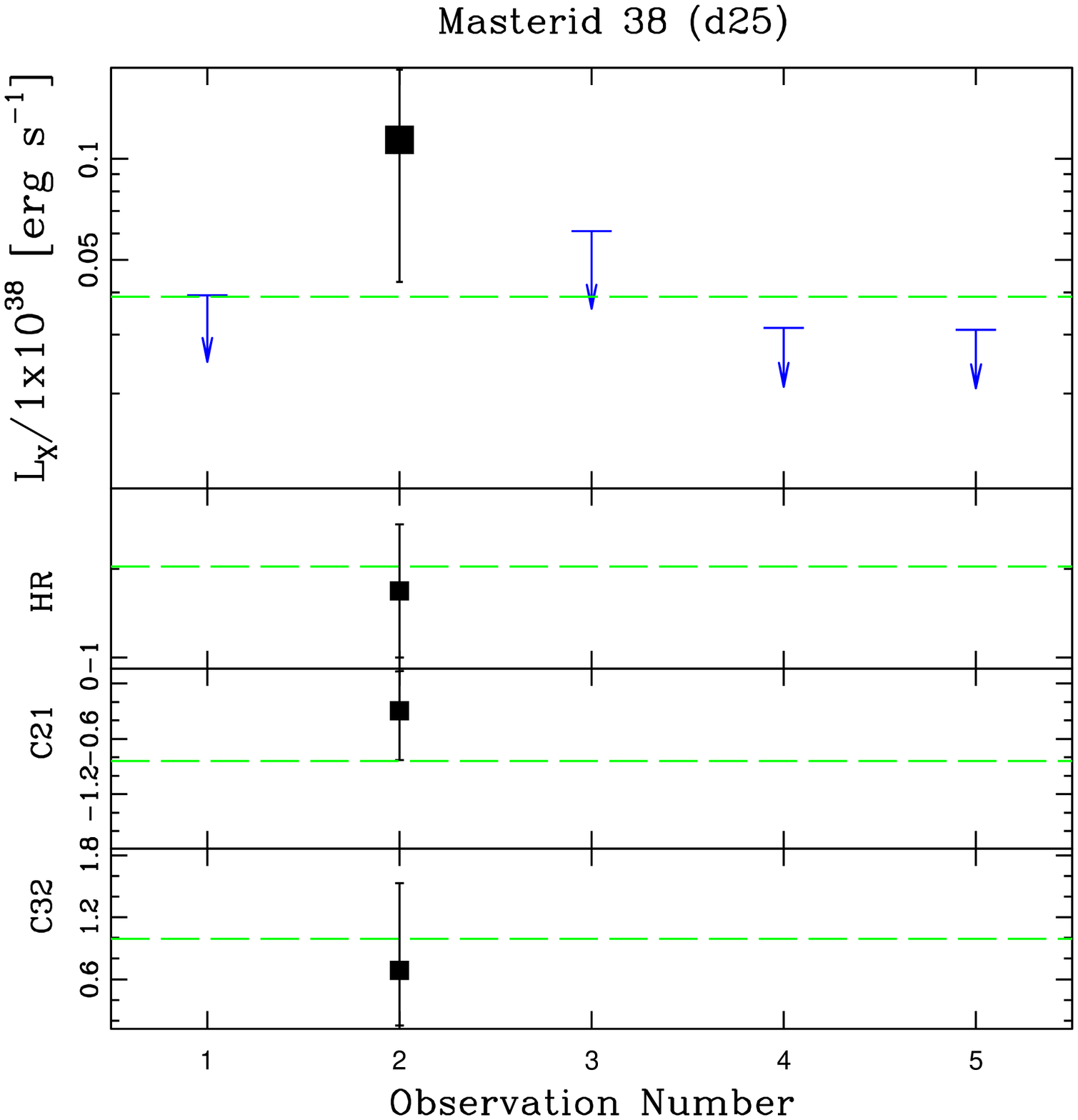}

\end{minipage}\hspace{0.02\linewidth}

\begin{minipage}{0.485\linewidth}
  \centering

    \includegraphics[width=\linewidth]{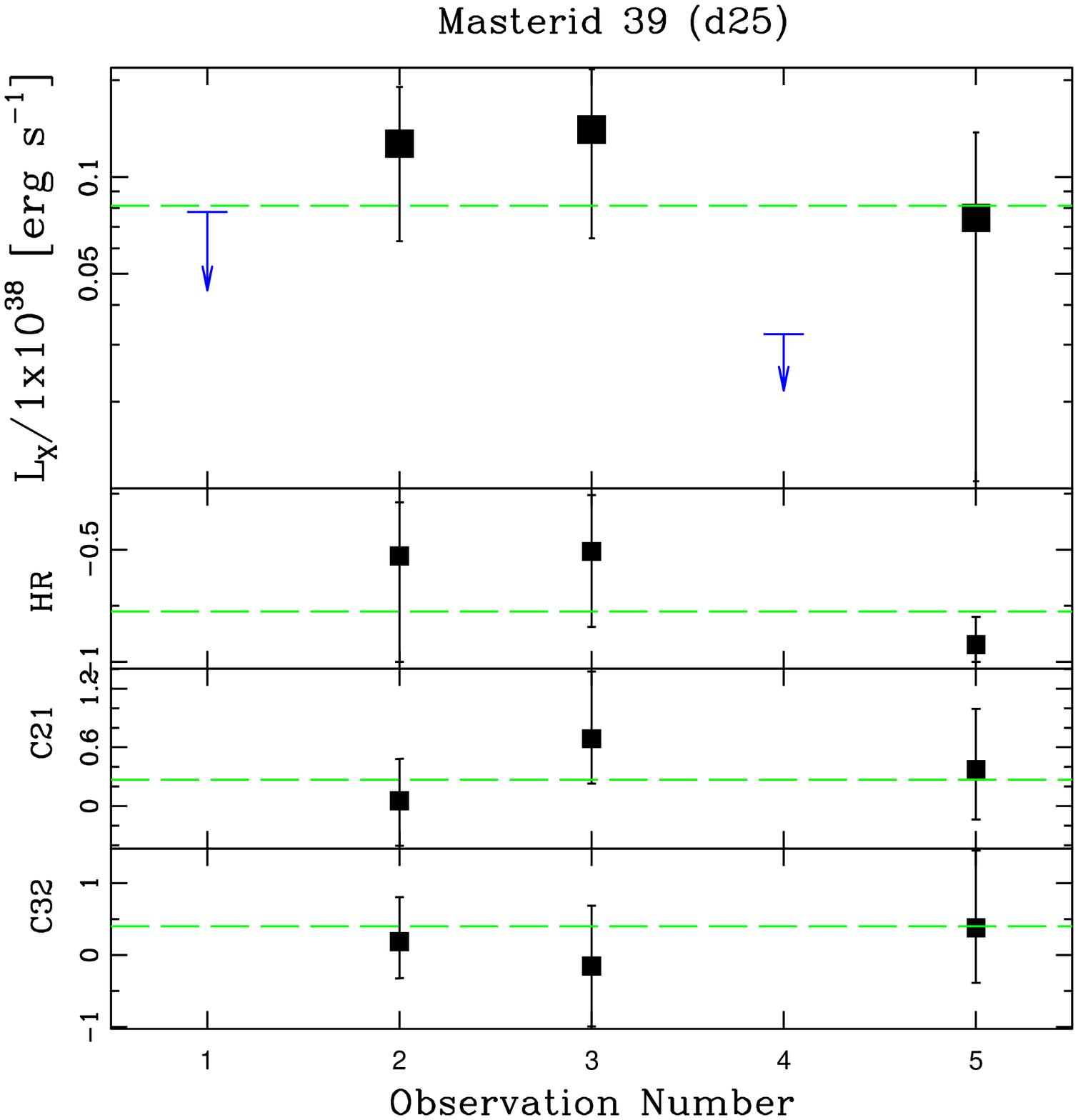}

 \end{minipage}\hspace{0.02\linewidth}
\begin{minipage}{0.485\linewidth}
  \centering
  
    \includegraphics[width=\linewidth]{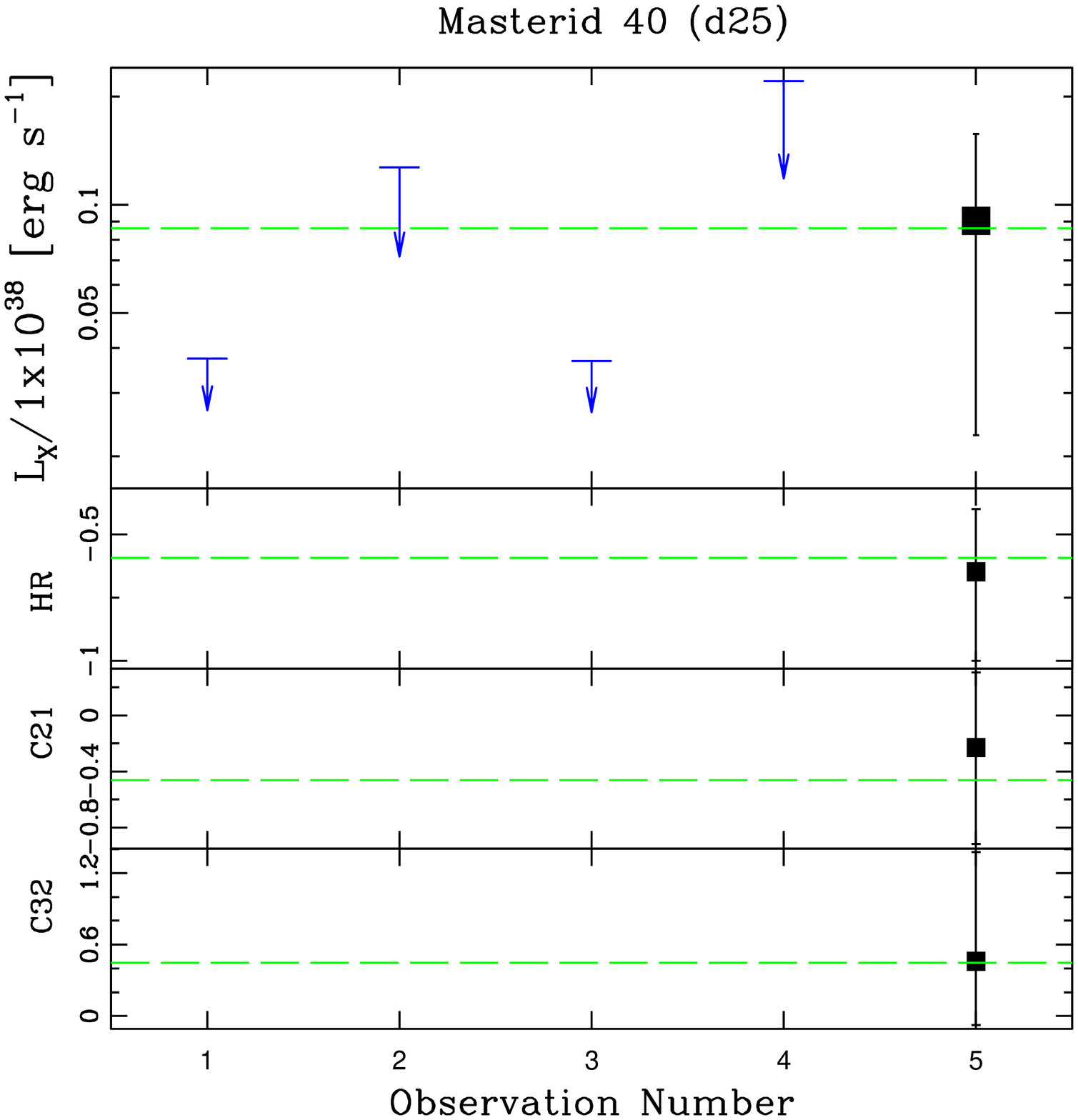}

  \end{minipage}\hspace{0.02\linewidth}

\end{figure}

\begin{figure}

  \begin{minipage}{0.485\linewidth}
  \centering

    \includegraphics[width=\linewidth]{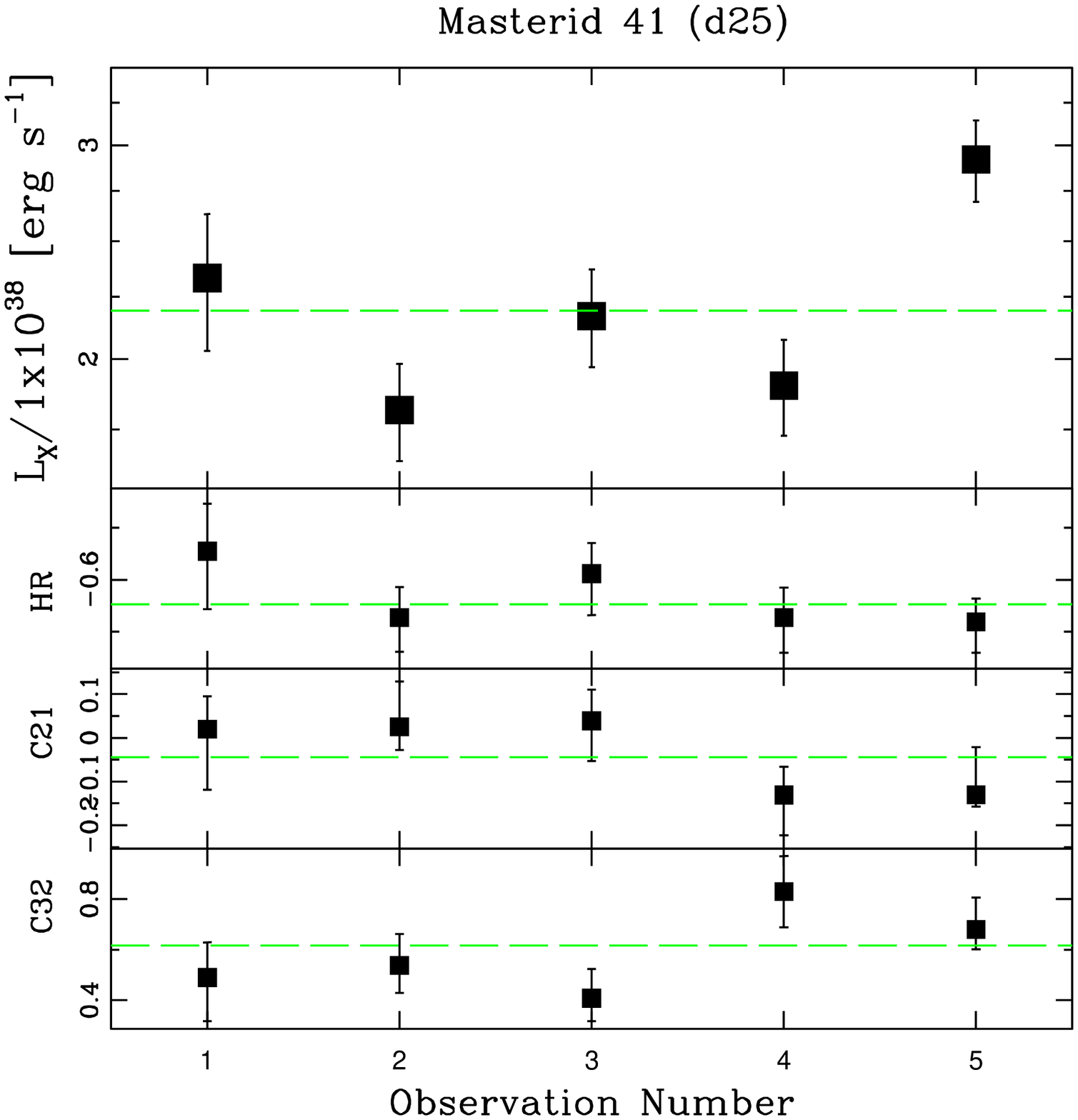}

\end{minipage}\hspace{0.02\linewidth}
\begin{minipage}{0.485\linewidth}
  \centering

    \includegraphics[width=\linewidth]{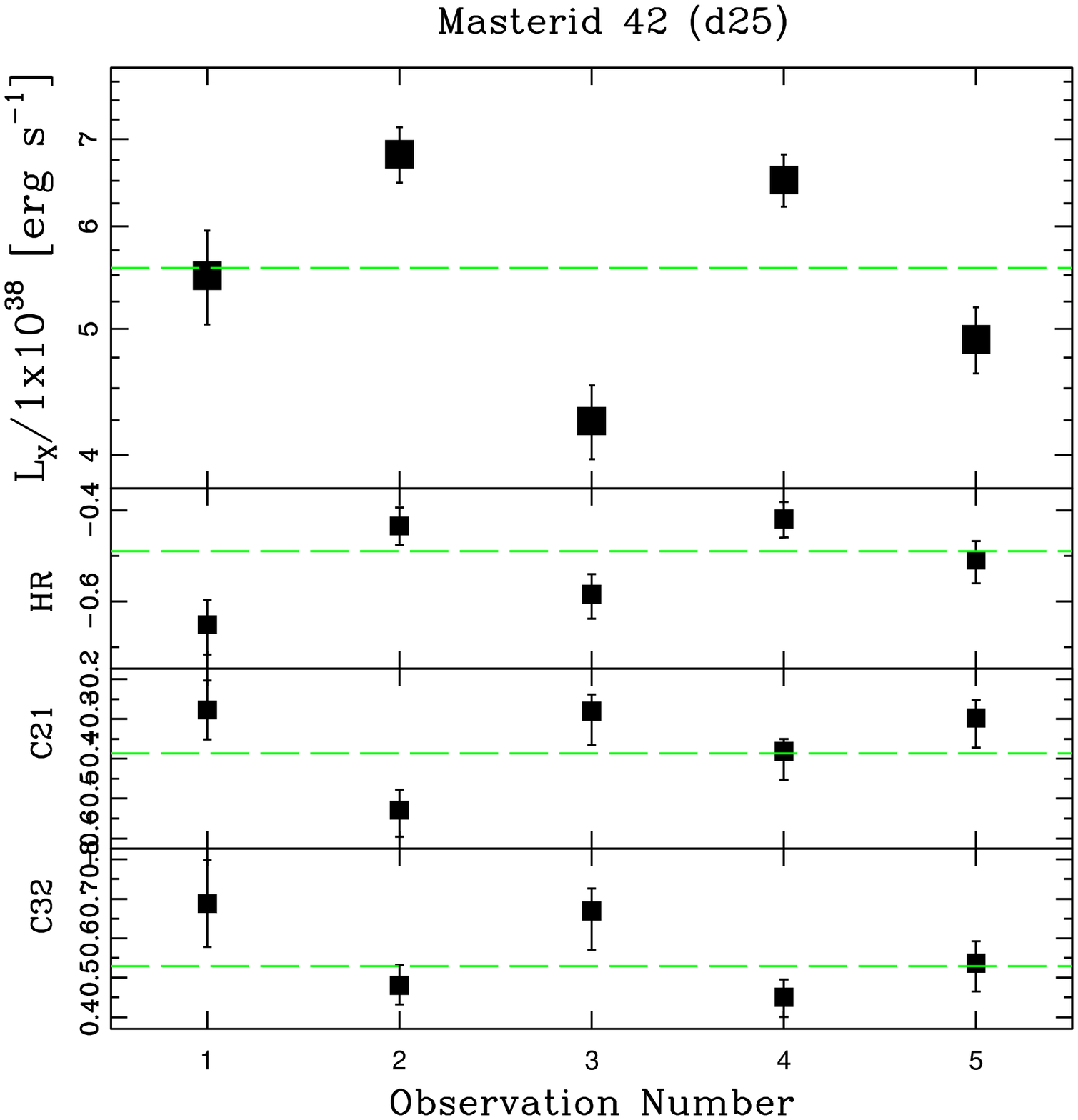}

 \end{minipage}\hspace{0.02\linewidth}

  \begin{minipage}{0.485\linewidth}
  \centering
  
    \includegraphics[width=\linewidth]{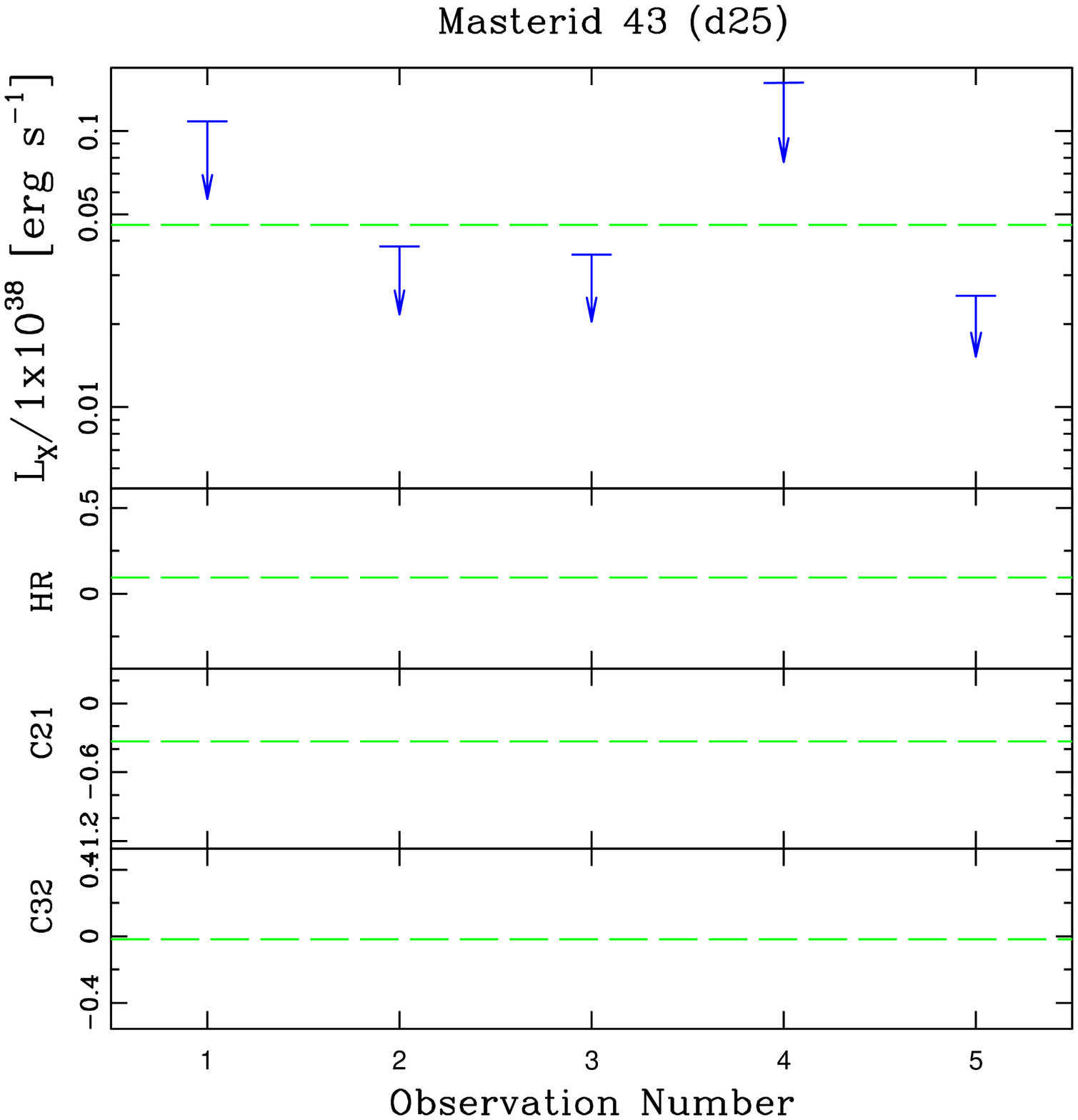}

  \end{minipage}\hspace{0.02\linewidth}
  \begin{minipage}{0.485\linewidth}
  \centering

    \includegraphics[width=\linewidth]{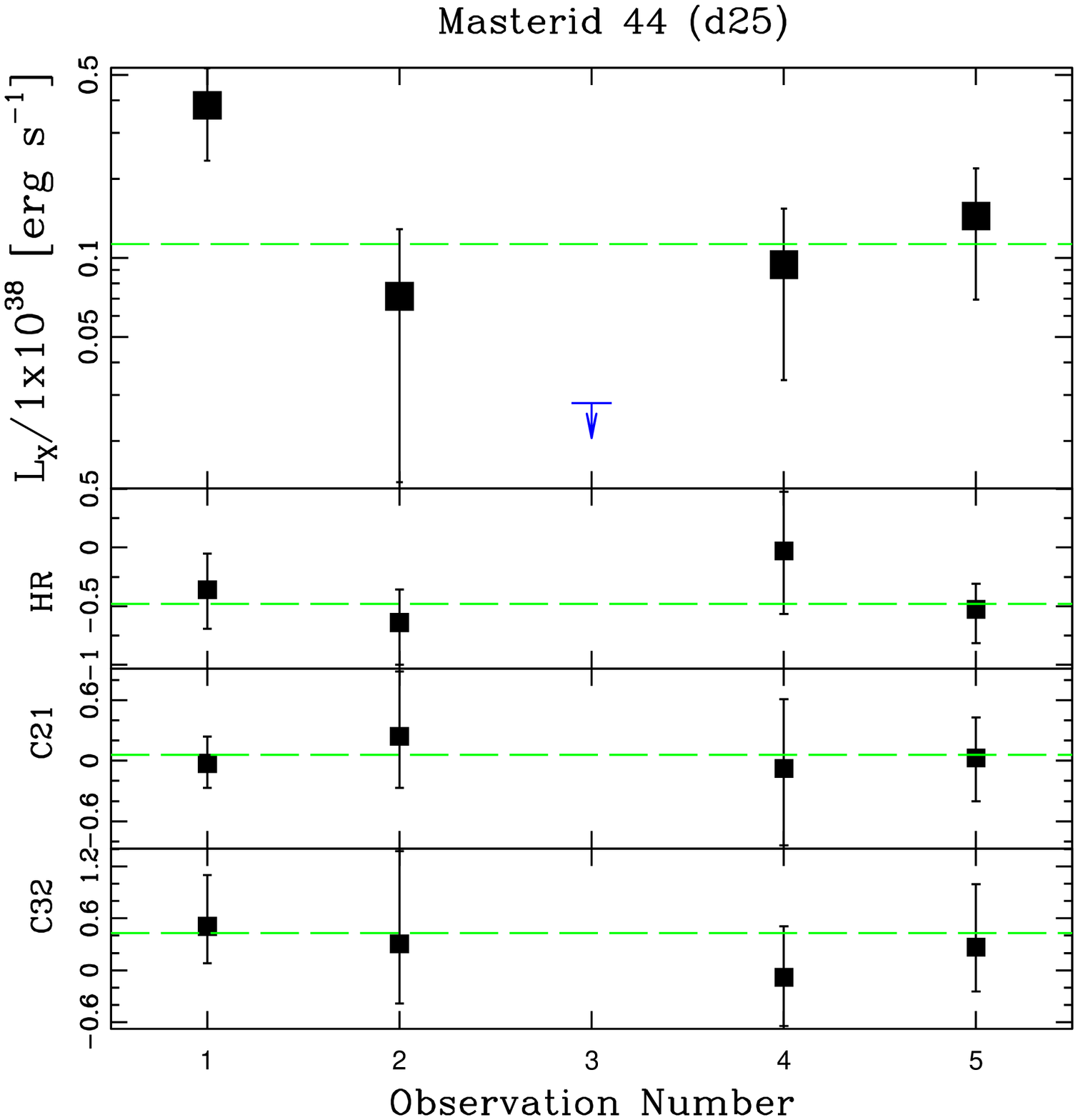}

\end{minipage}\hspace{0.02\linewidth}

\begin{minipage}{0.485\linewidth}
  \centering

    \includegraphics[width=\linewidth]{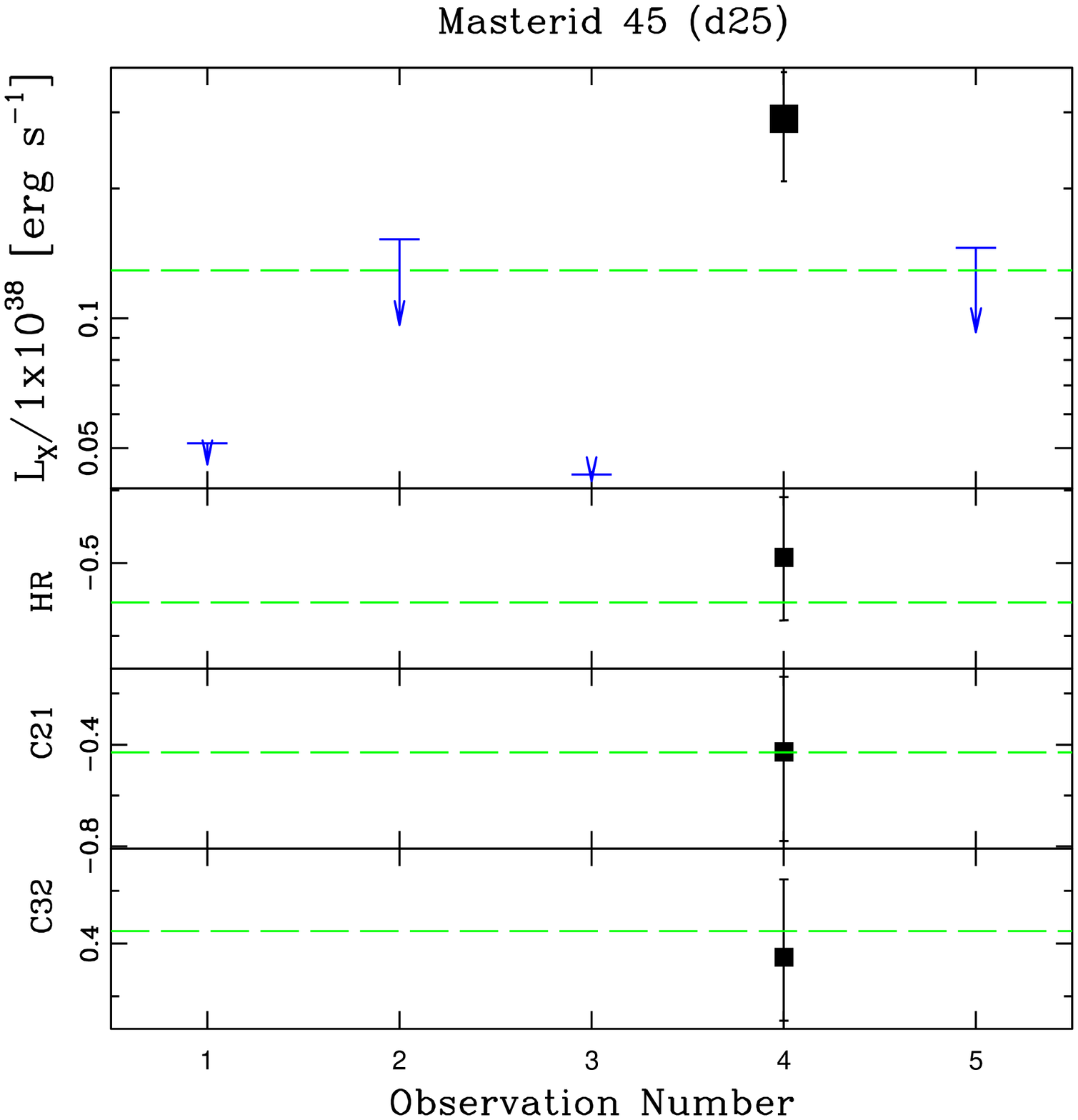}

 \end{minipage}\hspace{0.02\linewidth}
\begin{minipage}{0.485\linewidth}
  \centering
  
    \includegraphics[width=\linewidth]{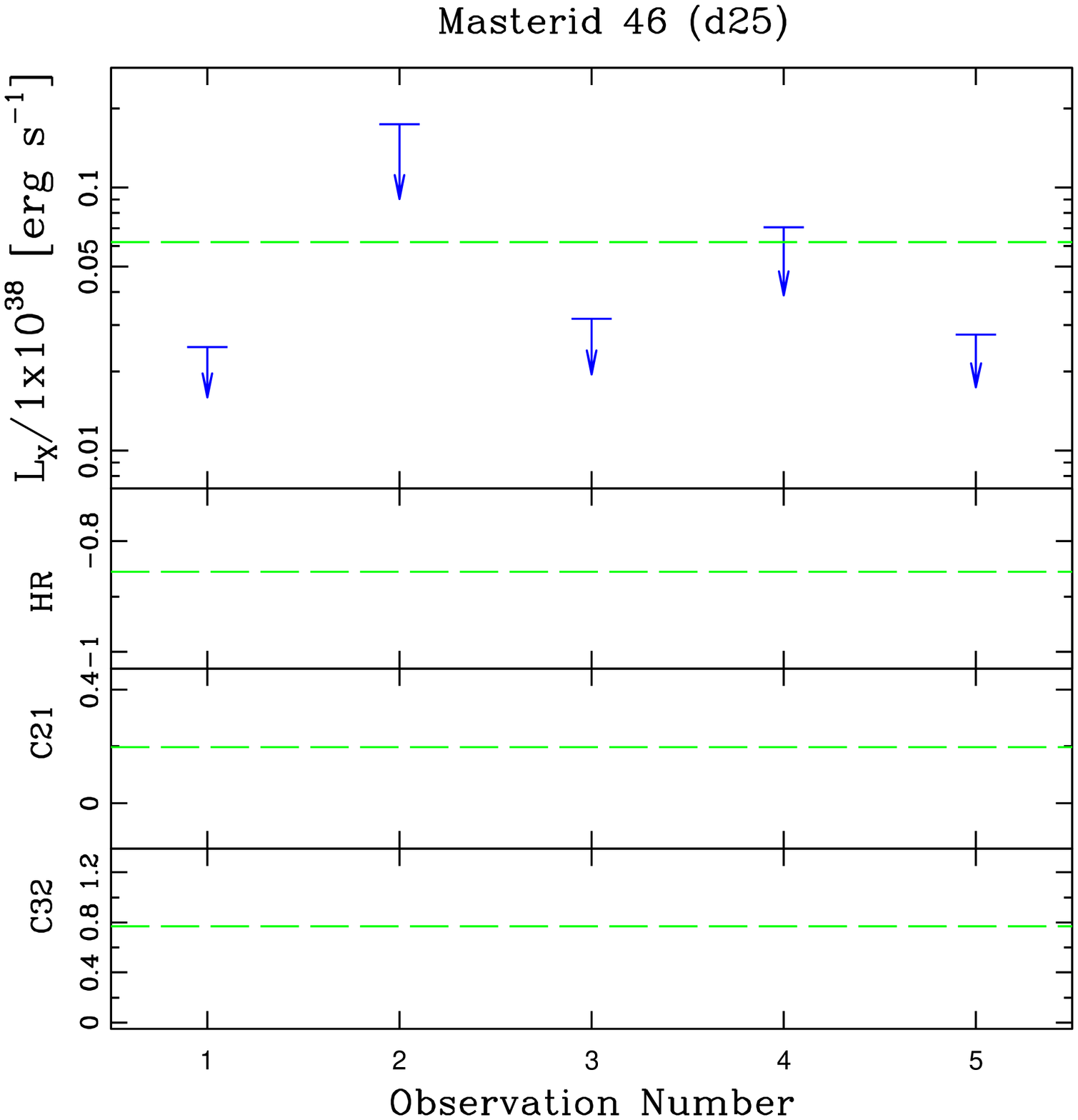}

  \end{minipage}\hspace{0.02\linewidth}
  
\end{figure}

\begin{figure}

  \begin{minipage}{0.485\linewidth}
  \centering

    \includegraphics[width=\linewidth]{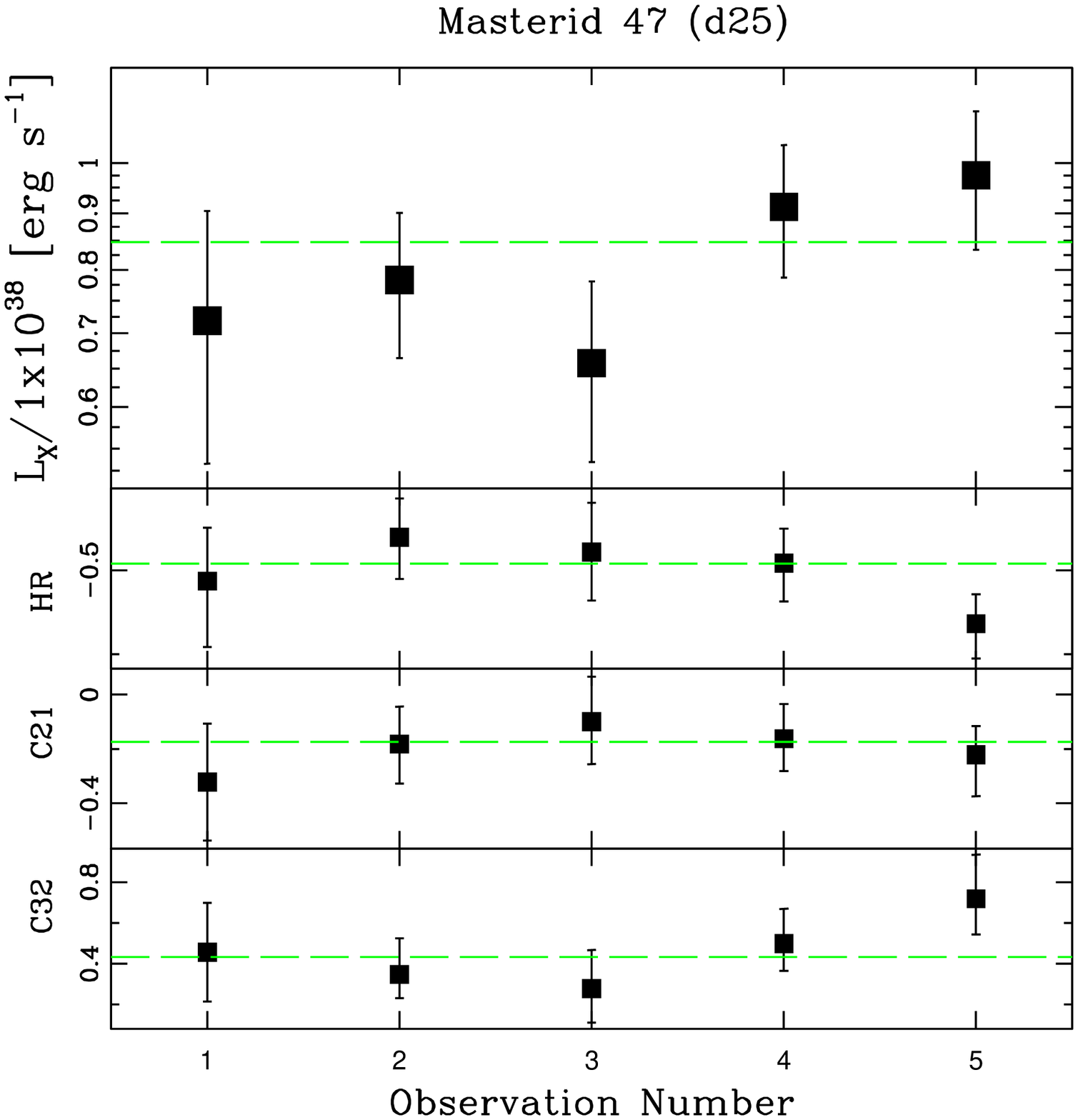}

\end{minipage}\hspace{0.02\linewidth}
\begin{minipage}{0.485\linewidth}
  \centering

    \includegraphics[width=\linewidth]{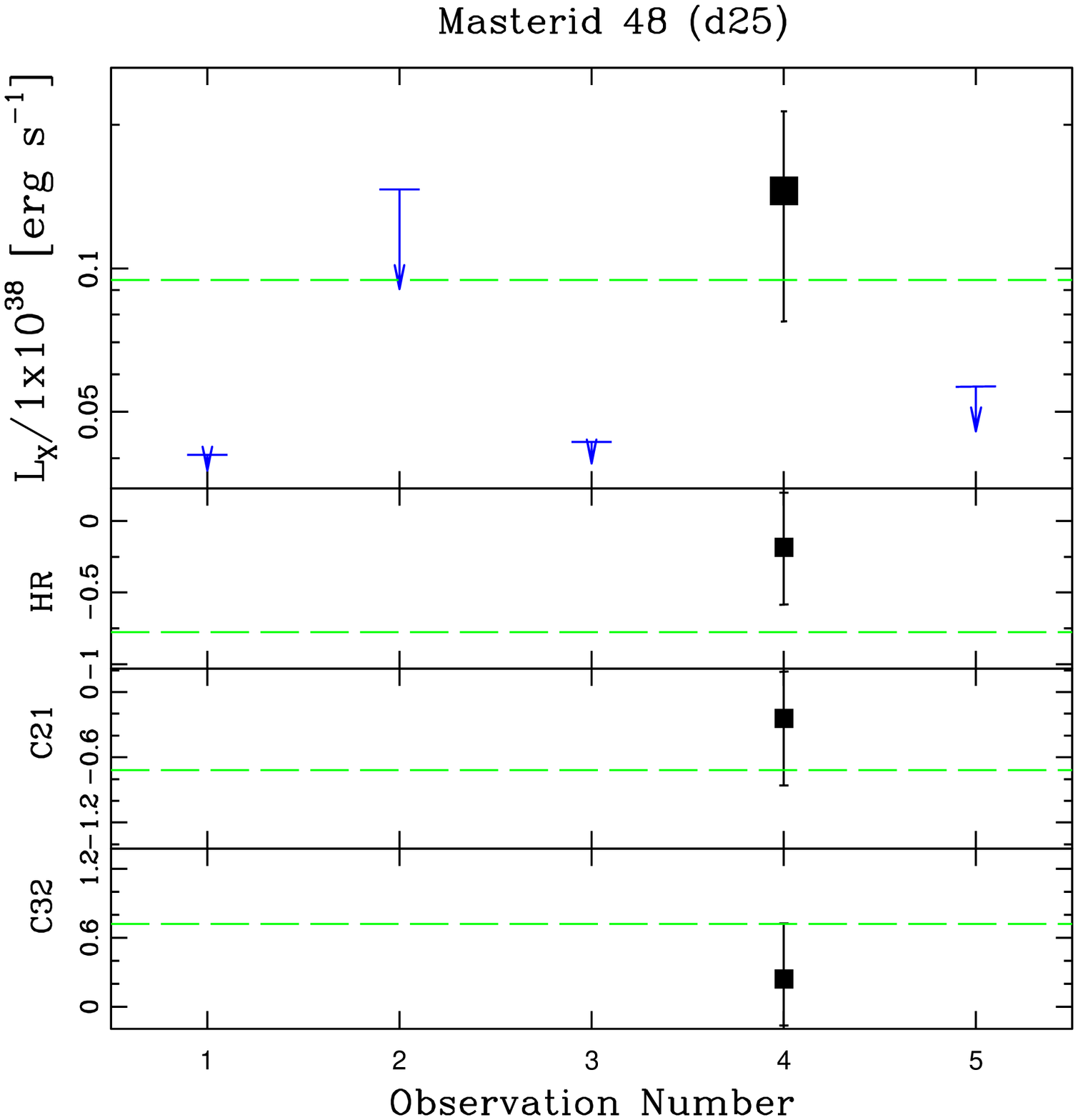}

 \end{minipage}\hspace{0.02\linewidth}

  \begin{minipage}{0.485\linewidth}
  \centering
  
    \includegraphics[width=\linewidth]{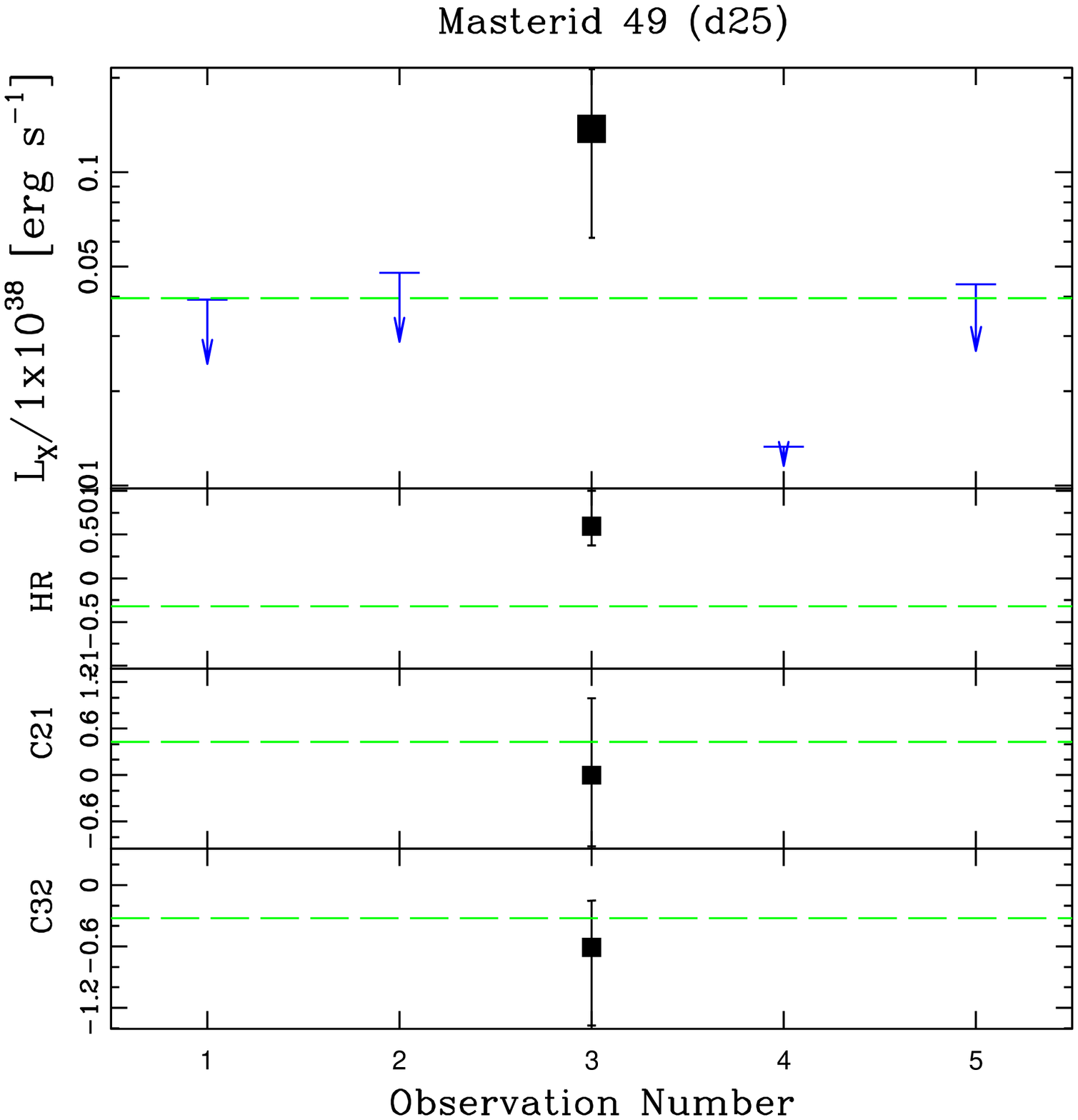}

  \end{minipage}\hspace{0.02\linewidth}
  \begin{minipage}{0.485\linewidth}
  \centering

    \includegraphics[width=\linewidth]{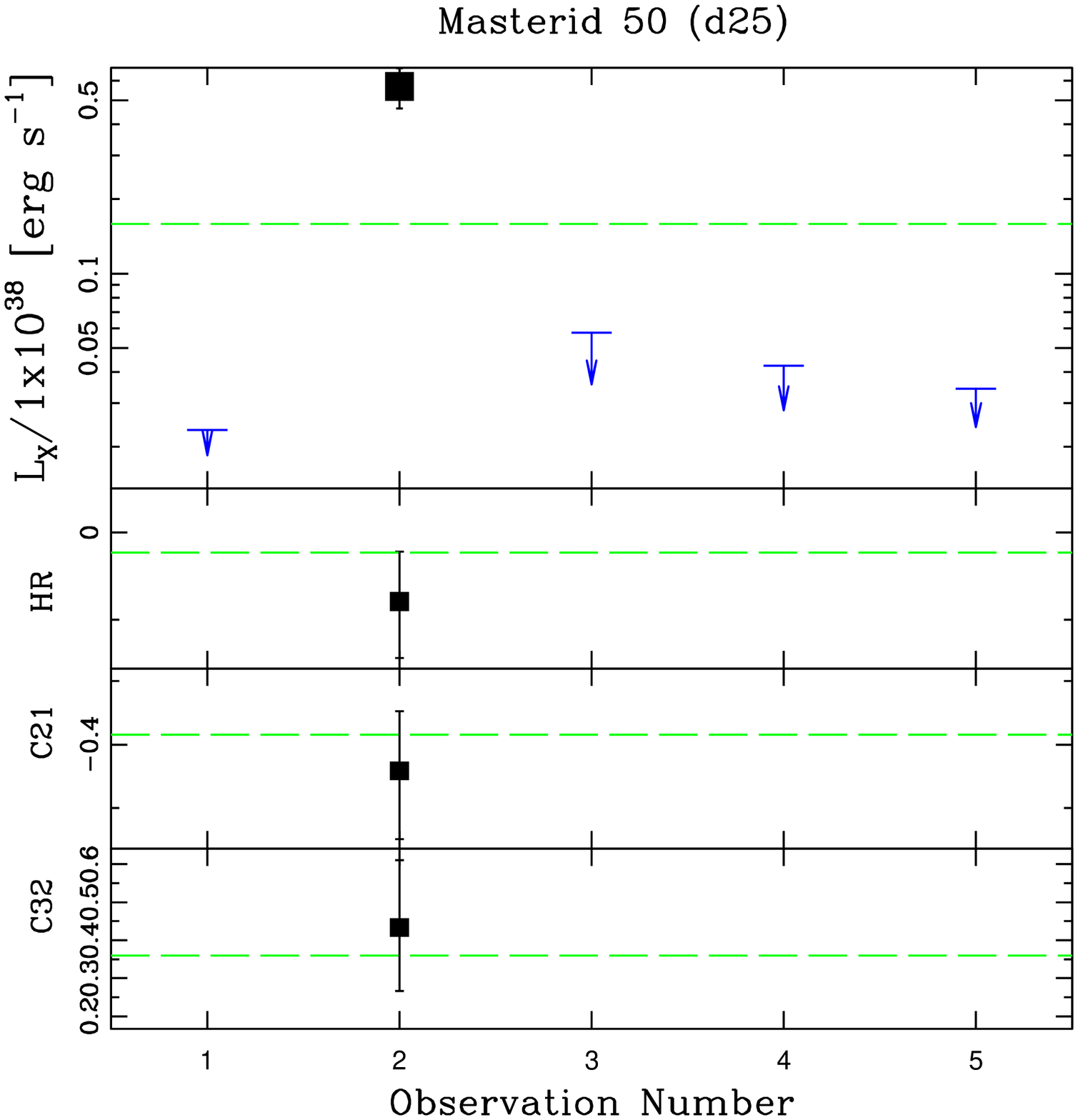}

\end{minipage}\hspace{0.02\linewidth}

\begin{minipage}{0.485\linewidth}
  \centering

    \includegraphics[width=\linewidth]{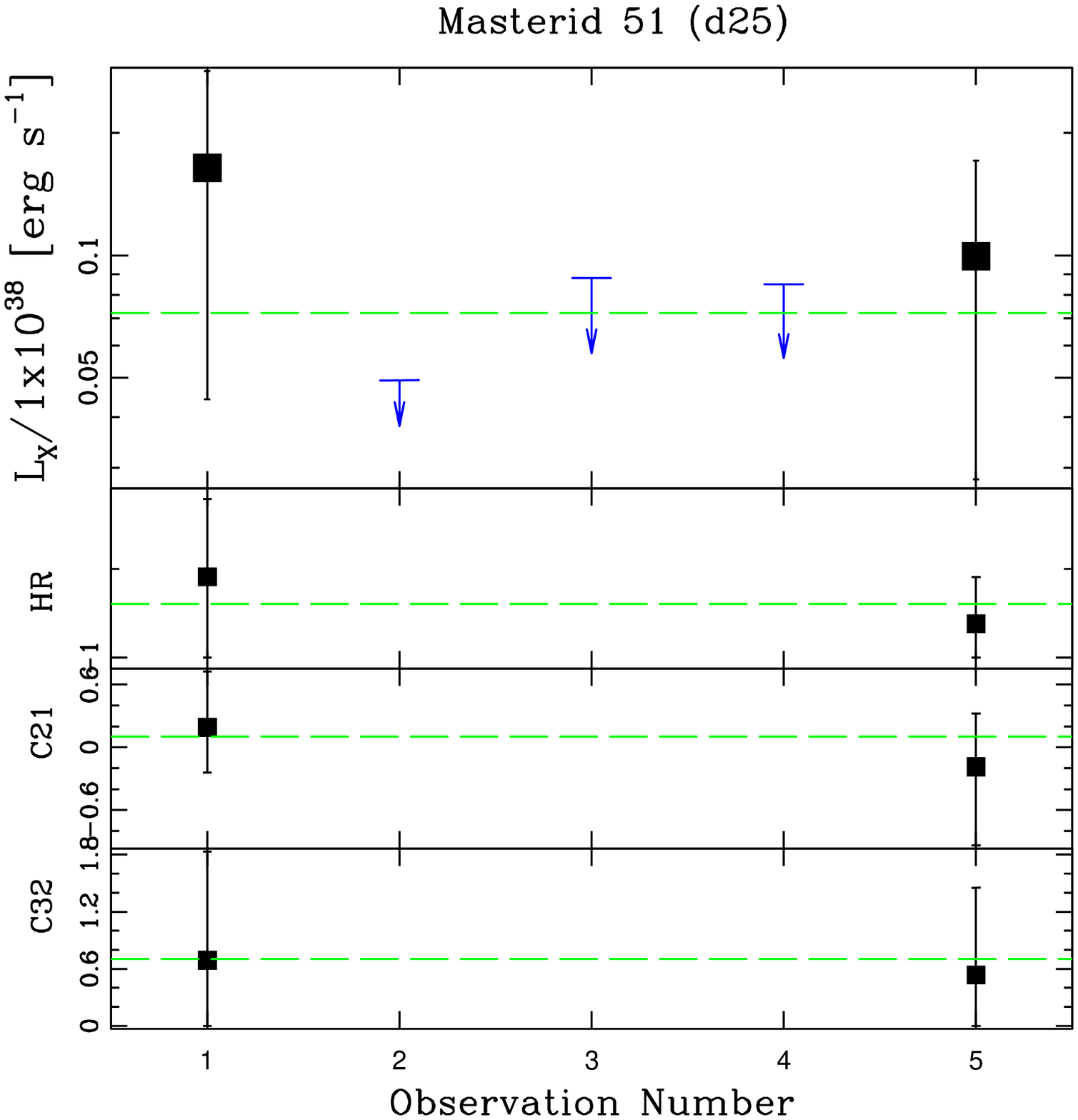}

 \end{minipage}\hspace{0.02\linewidth}
\begin{minipage}{0.485\linewidth}
  \centering
  
    \includegraphics[width=\linewidth]{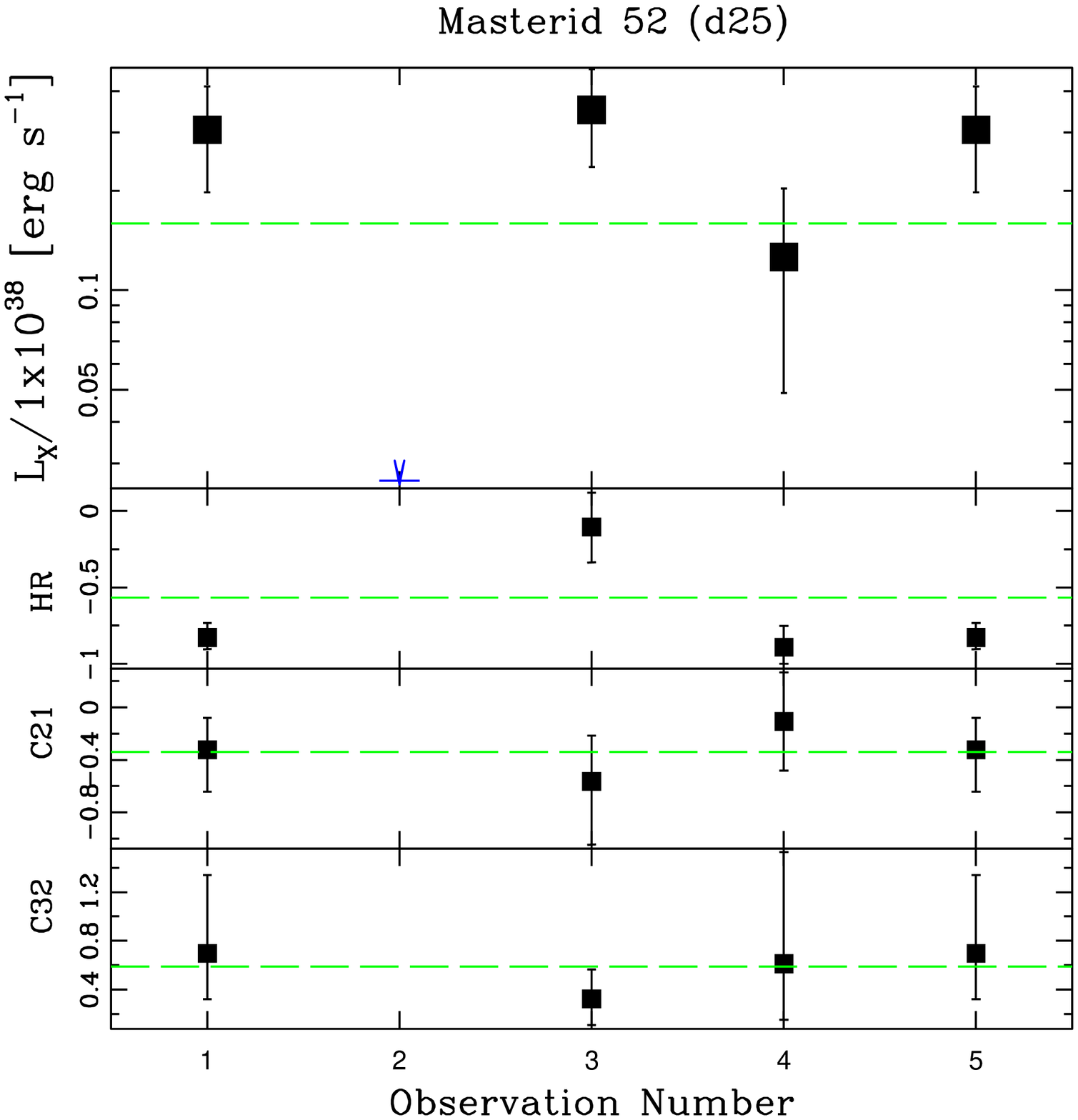}

  \end{minipage}\hspace{0.02\linewidth}

\end{figure}

\begin{figure}

  \begin{minipage}{0.485\linewidth}
  \centering

    \includegraphics[width=\linewidth]{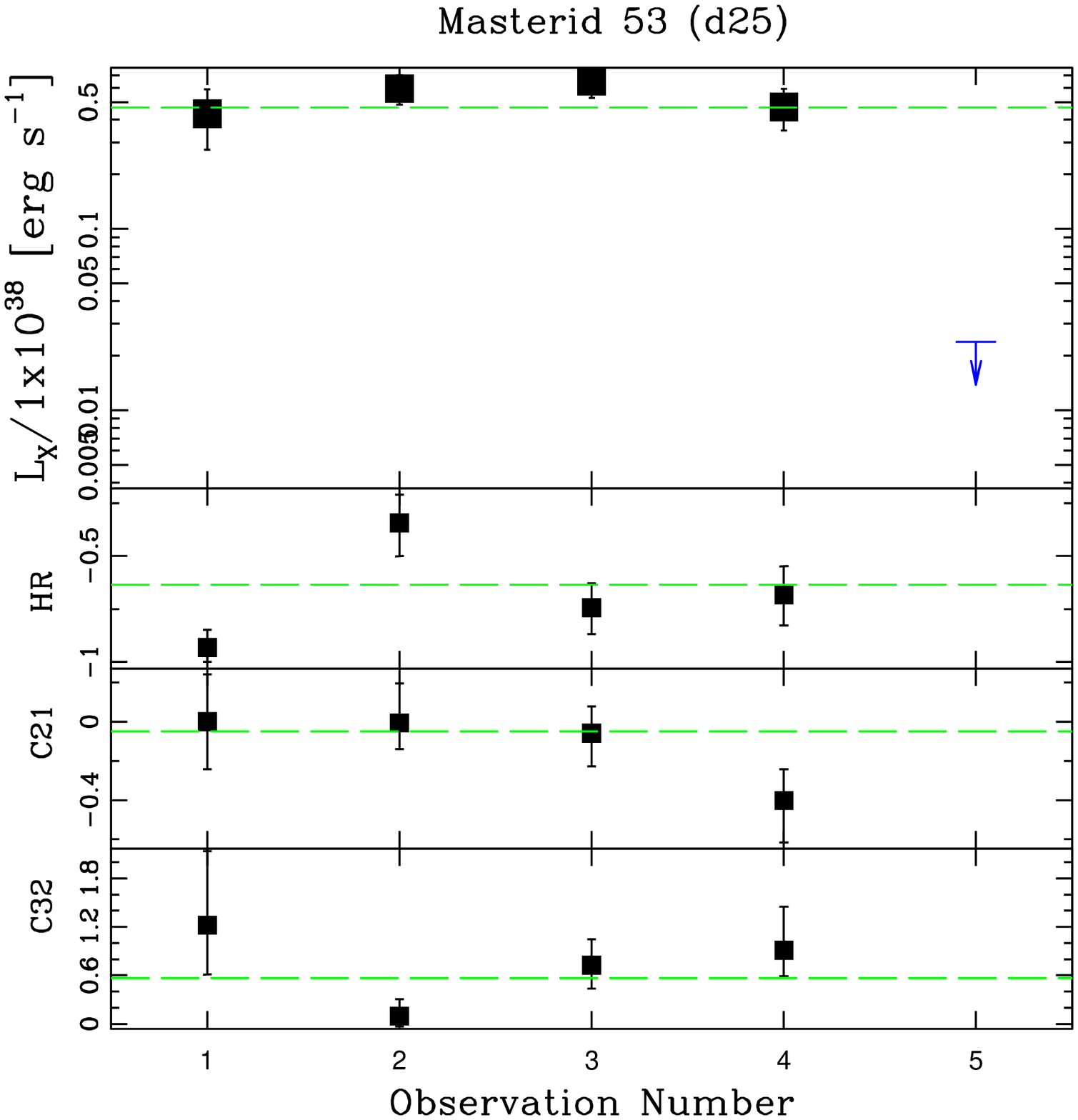}

\end{minipage}\hspace{0.02\linewidth}
\begin{minipage}{0.485\linewidth}
  \centering

    \includegraphics[width=\linewidth]{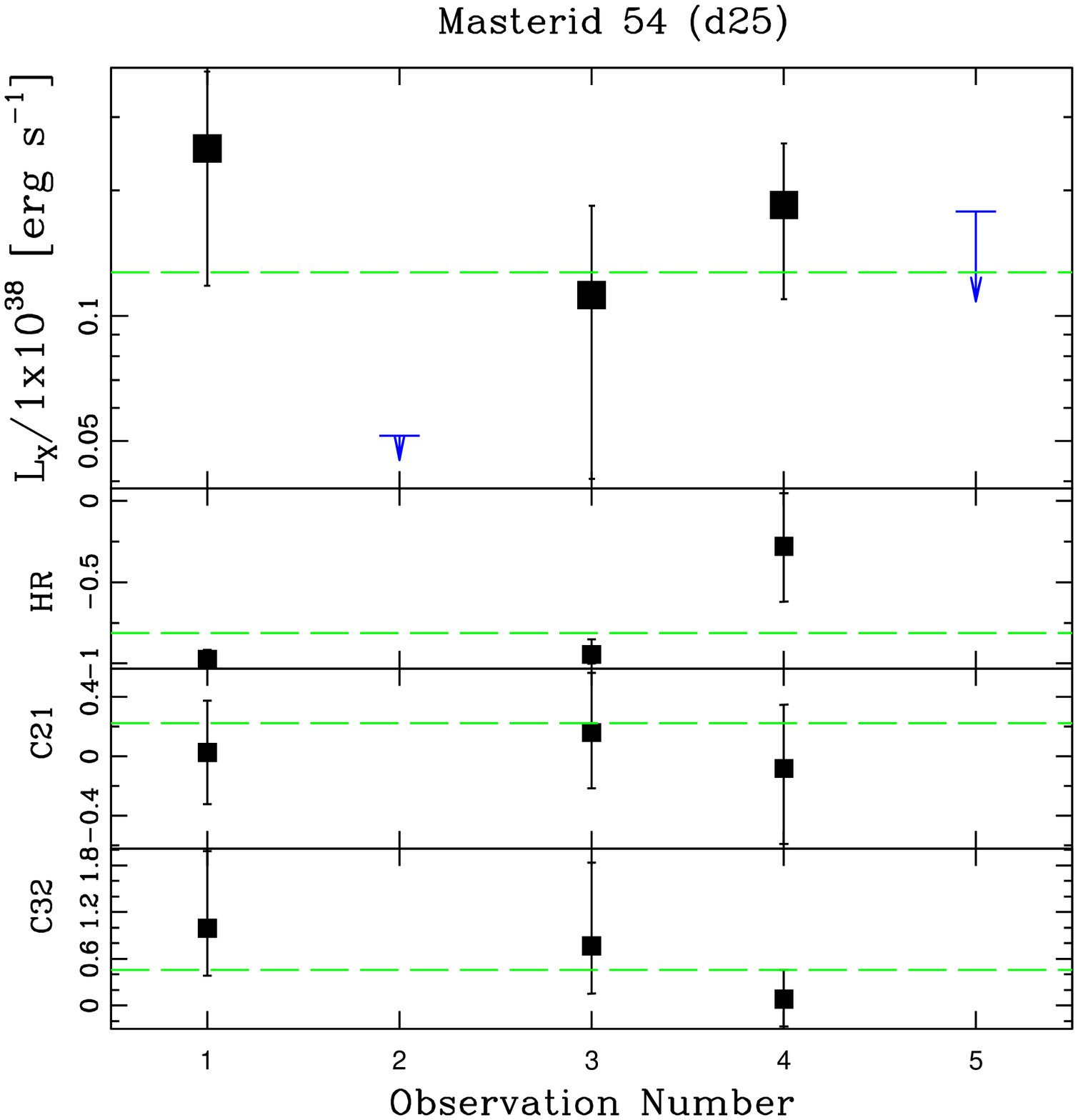}

 \end{minipage}\hspace{0.02\linewidth}
  
  \begin{minipage}{0.485\linewidth}
  \centering
  
    \includegraphics[width=\linewidth]{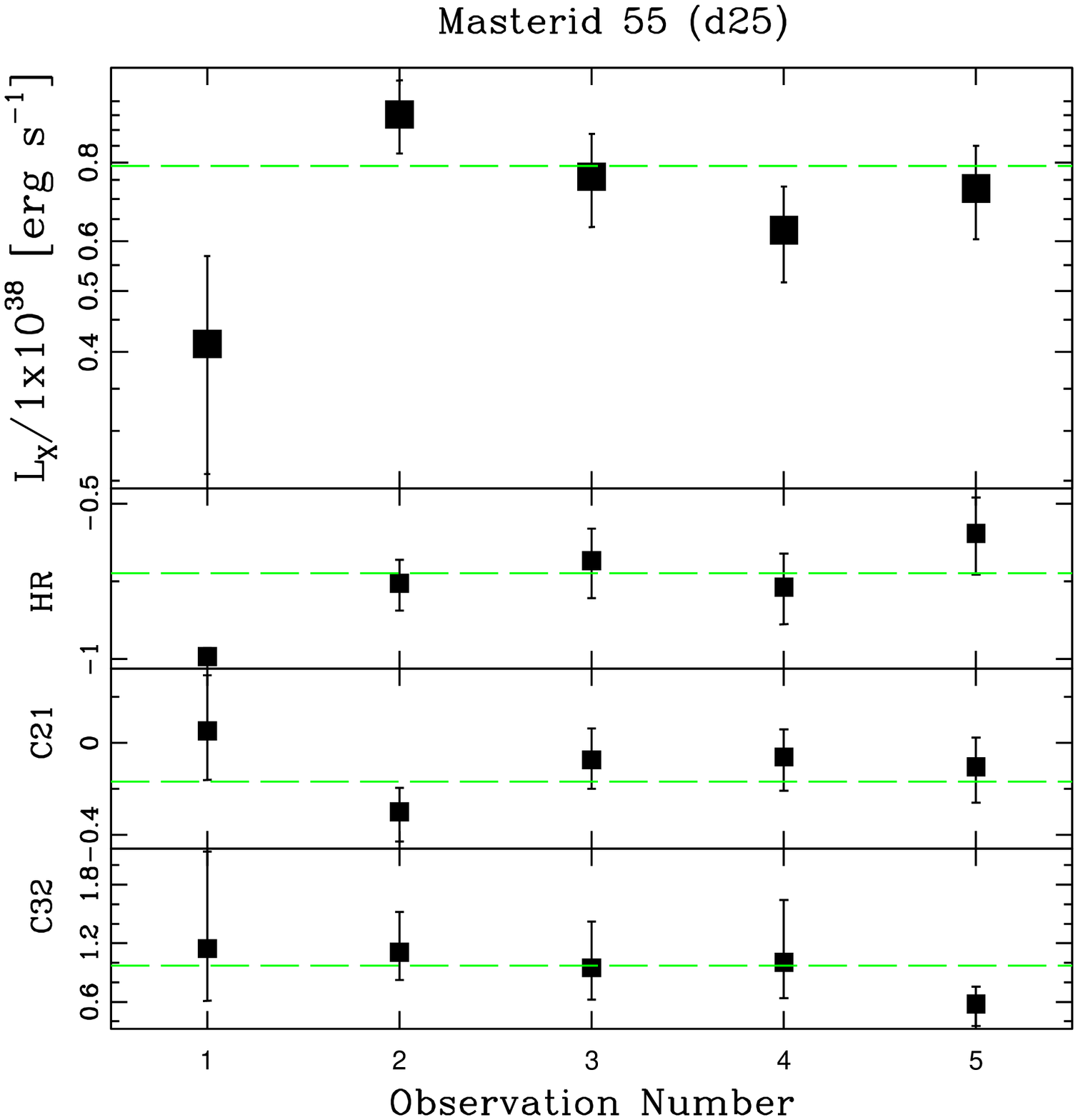}

  \end{minipage}\hspace{0.02\linewidth}
  \begin{minipage}{0.485\linewidth}
  \centering

    \includegraphics[width=\linewidth]{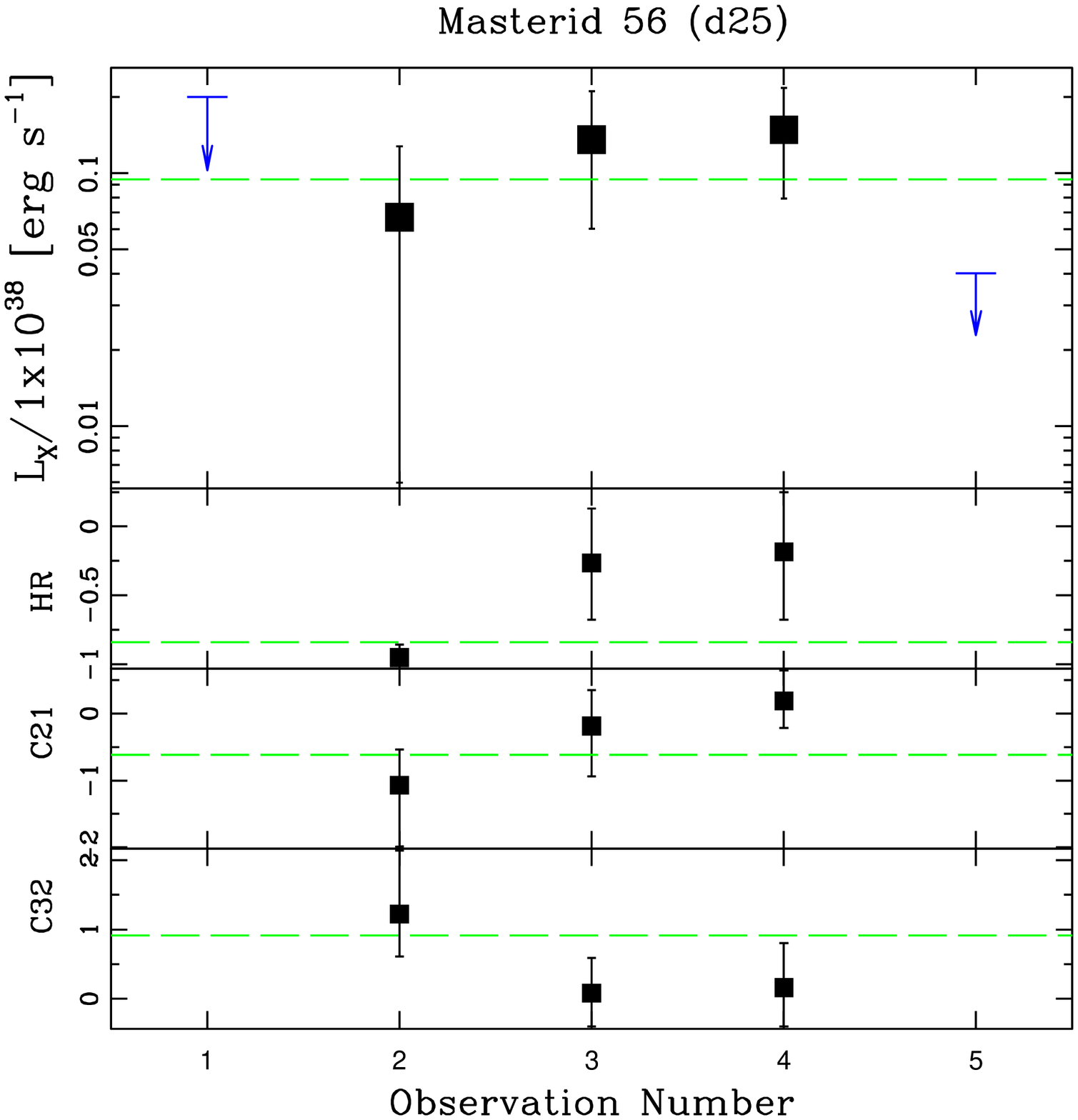}

\end{minipage}\hspace{0.02\linewidth}

\begin{minipage}{0.485\linewidth}
  \centering

    \includegraphics[width=\linewidth]{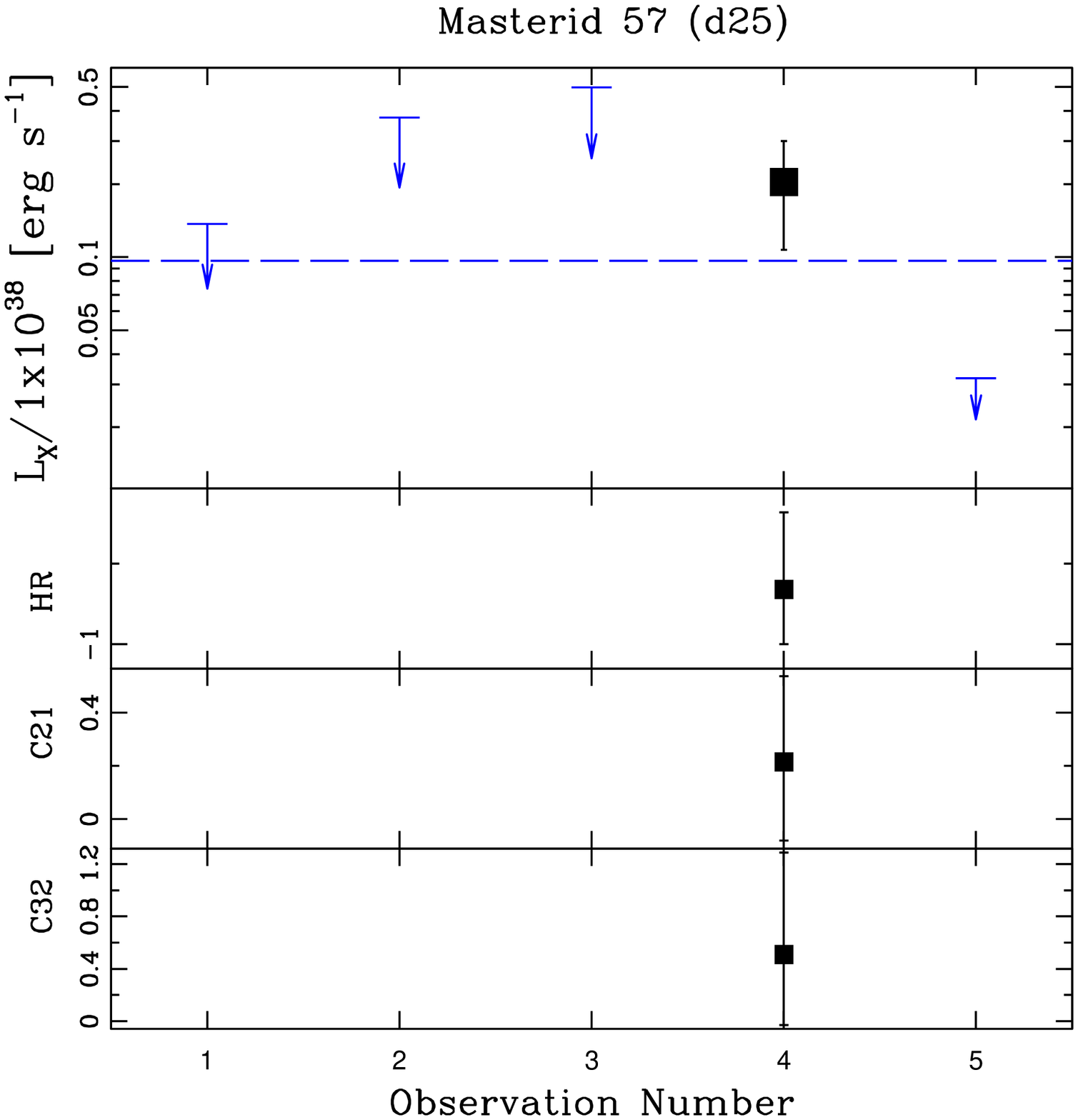}

 \end{minipage}\hspace{0.02\linewidth}
\begin{minipage}{0.485\linewidth}
  \centering
  
    \includegraphics[width=\linewidth]{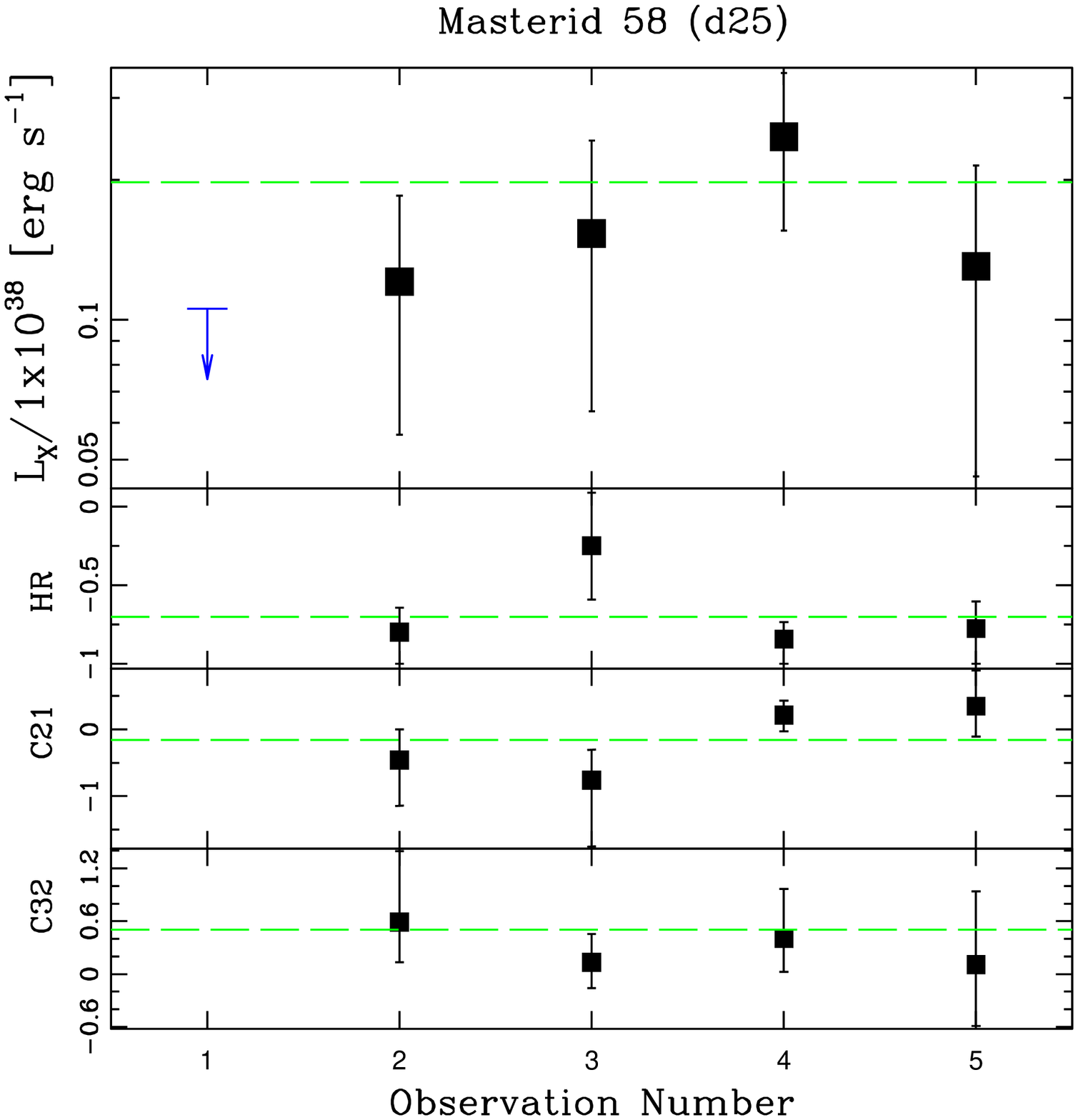}

  \end{minipage}\hspace{0.02\linewidth}

\end{figure}

\begin{figure}

  \begin{minipage}{0.485\linewidth}
  \centering

    \includegraphics[width=\linewidth]{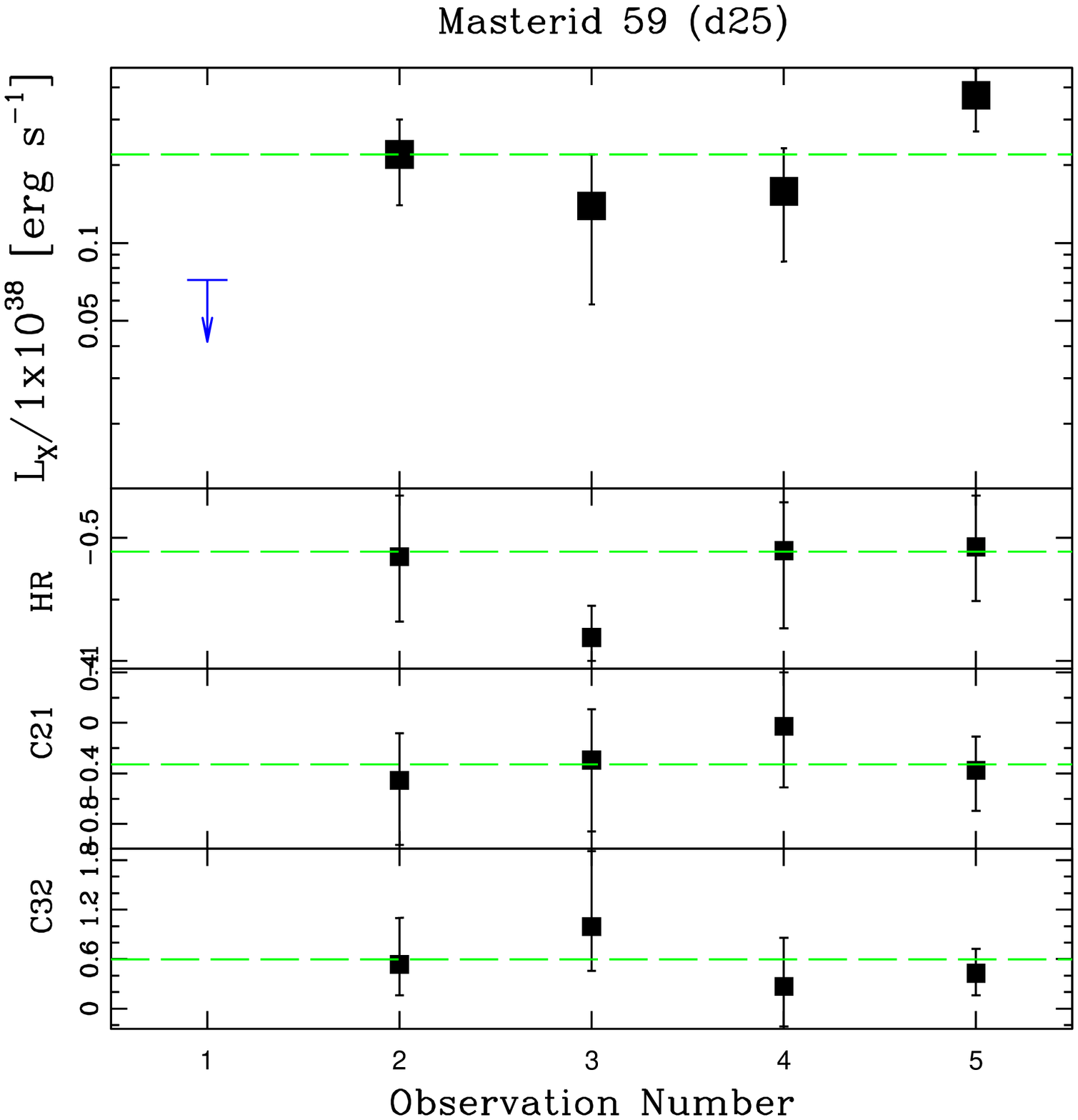}

\end{minipage}\hspace{0.02\linewidth}
\begin{minipage}{0.485\linewidth}
  \centering

    \includegraphics[width=\linewidth]{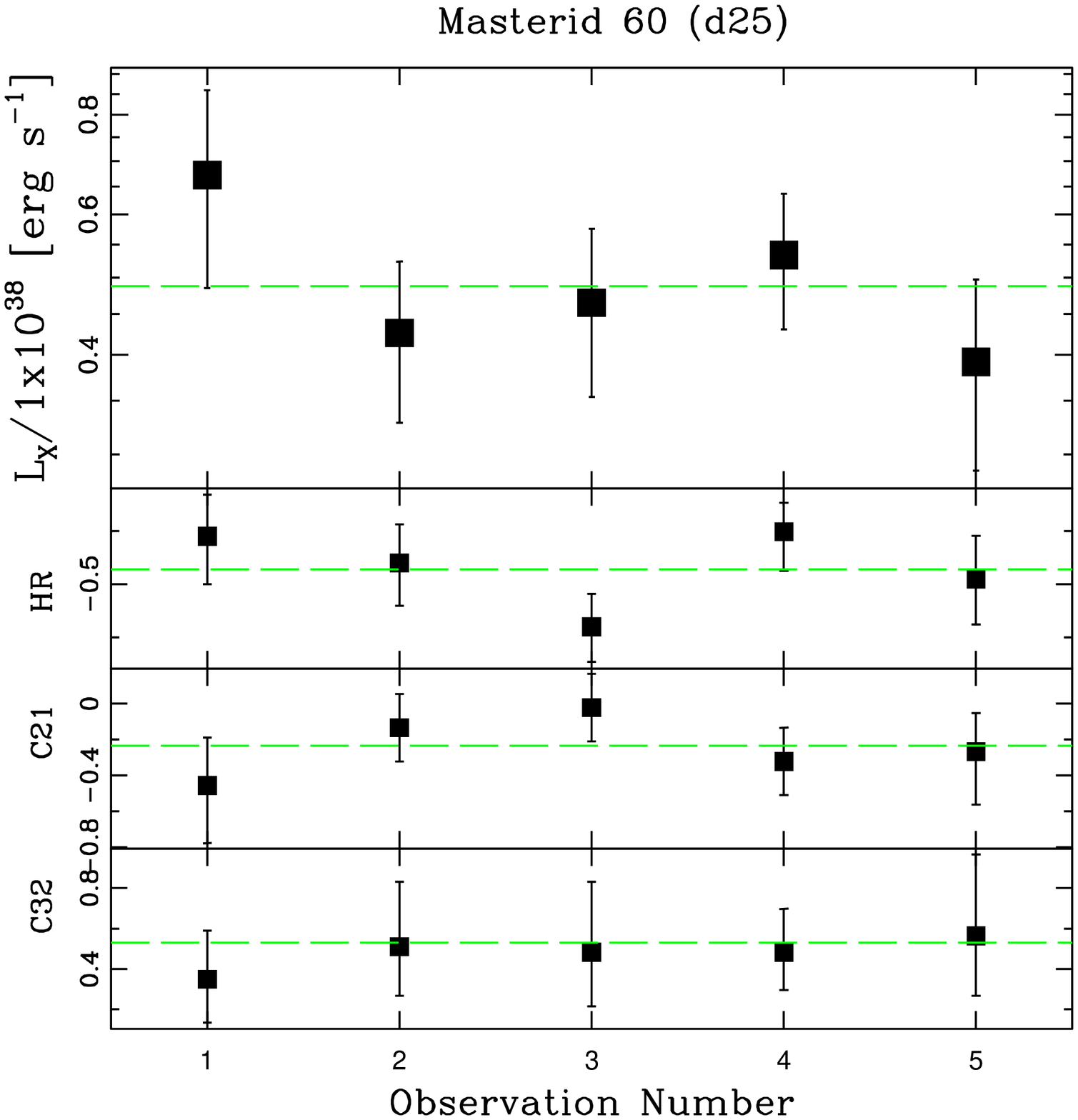}

 \end{minipage}\hspace{0.02\linewidth}

  \begin{minipage}{0.485\linewidth}
  \centering
  
    \includegraphics[width=\linewidth]{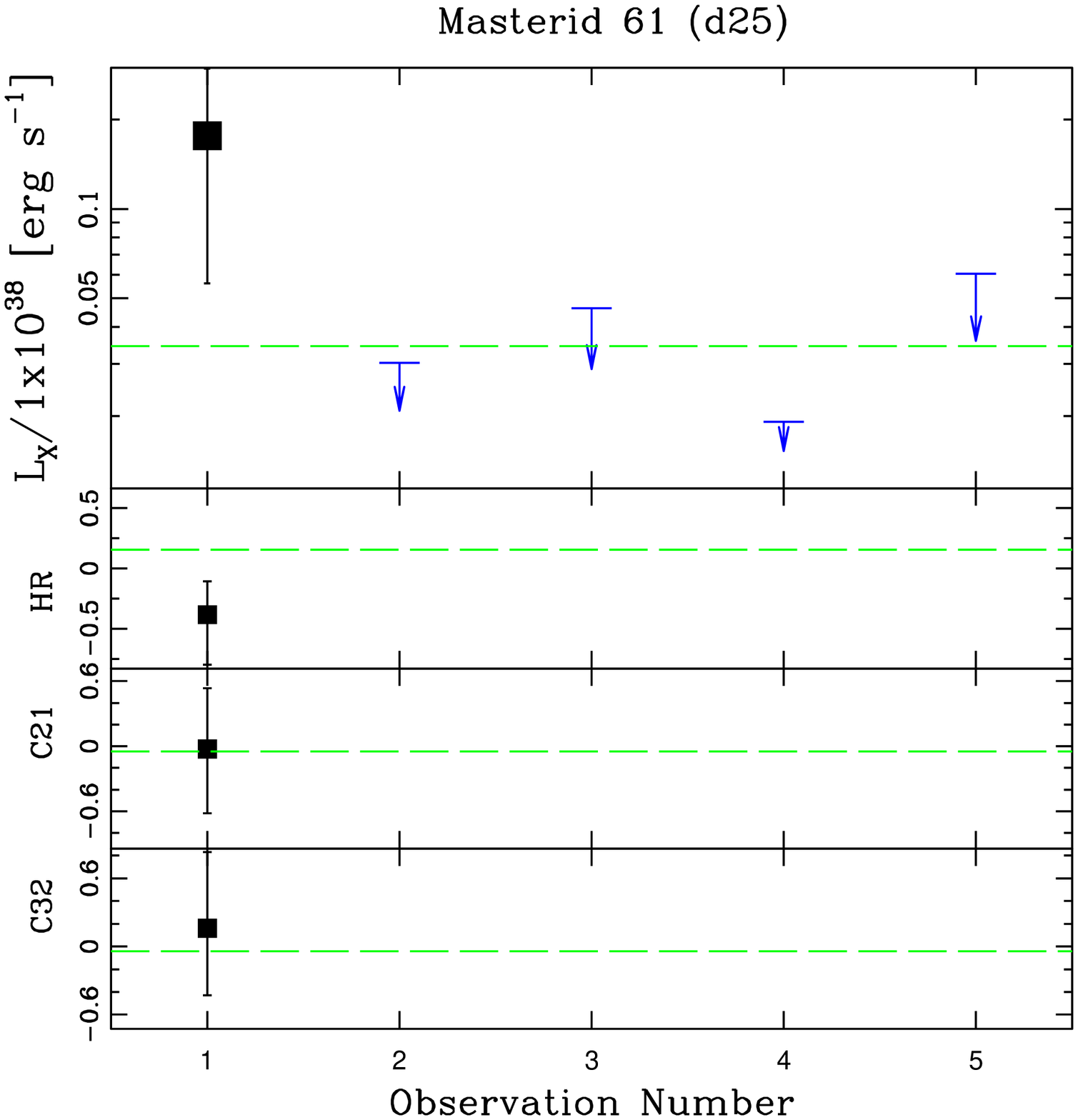}

  \end{minipage}\hspace{0.02\linewidth}
  \begin{minipage}{0.485\linewidth}
  \centering

    \includegraphics[width=\linewidth]{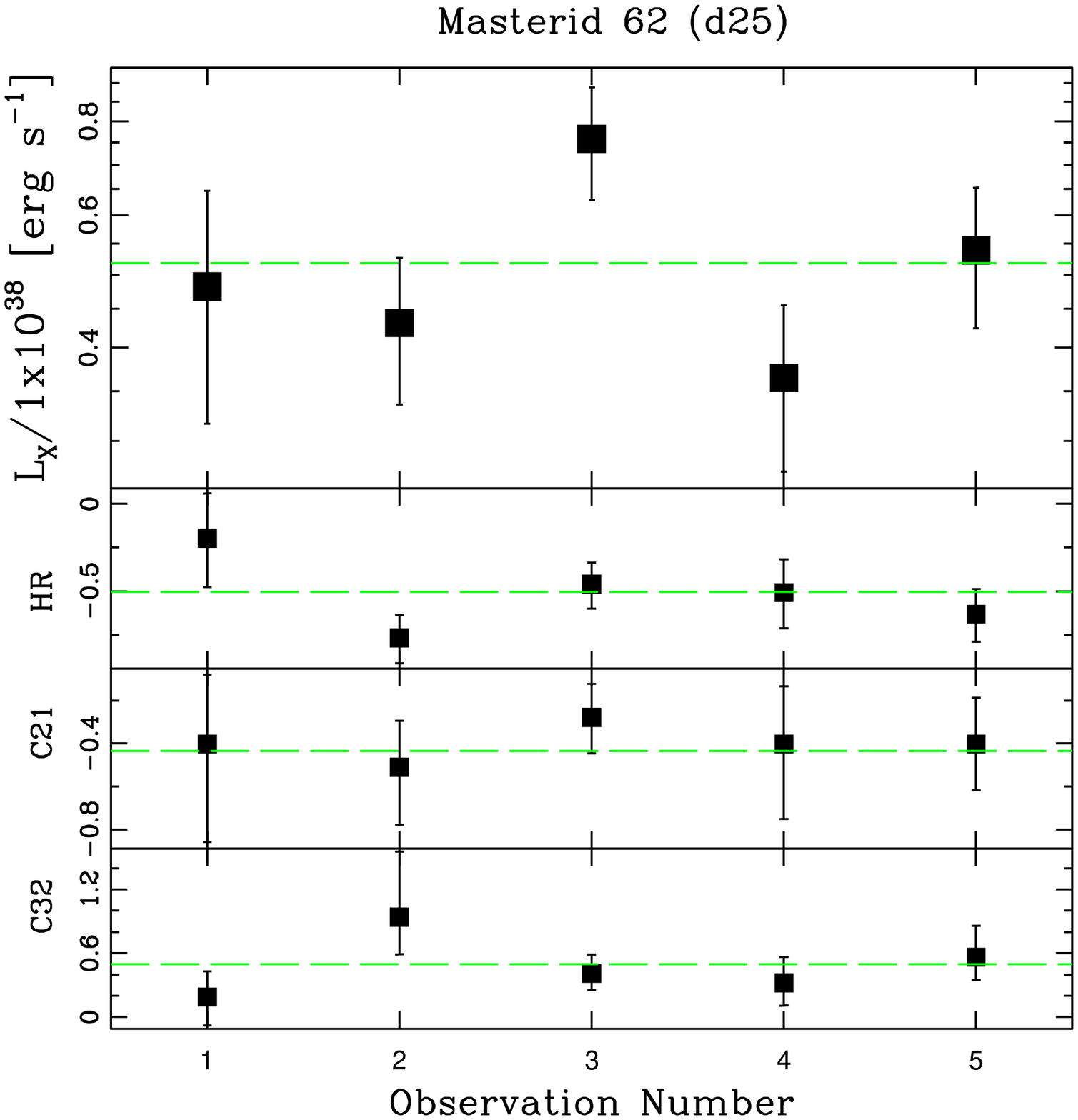}

\end{minipage}\hspace{0.02\linewidth}

\begin{minipage}{0.485\linewidth}
  \centering

    \includegraphics[width=\linewidth]{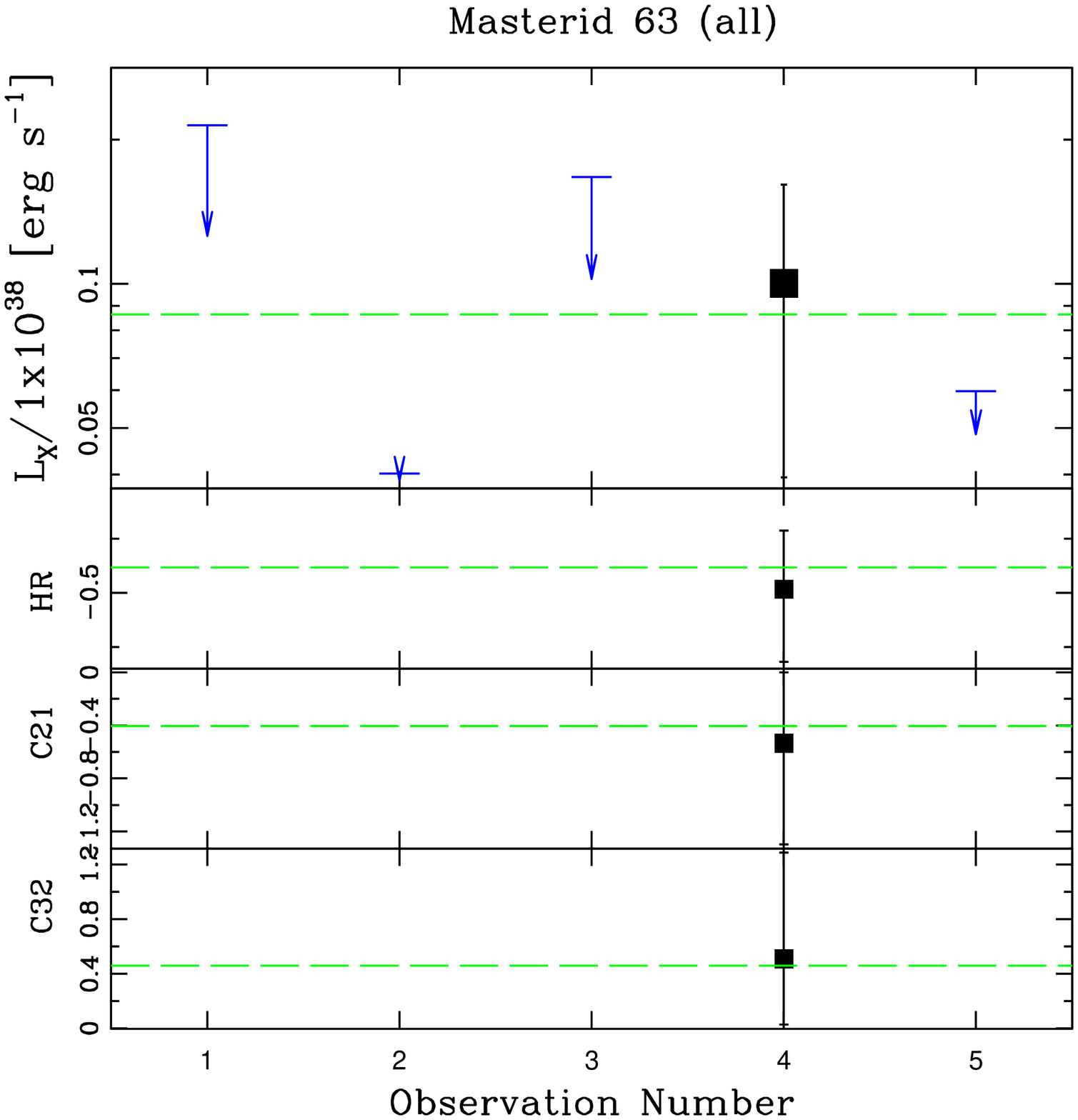}

\end{minipage}\hspace{0.02\linewidth}
\begin{minipage}{0.485\linewidth}
  \centering
  
    \includegraphics[width=\linewidth]{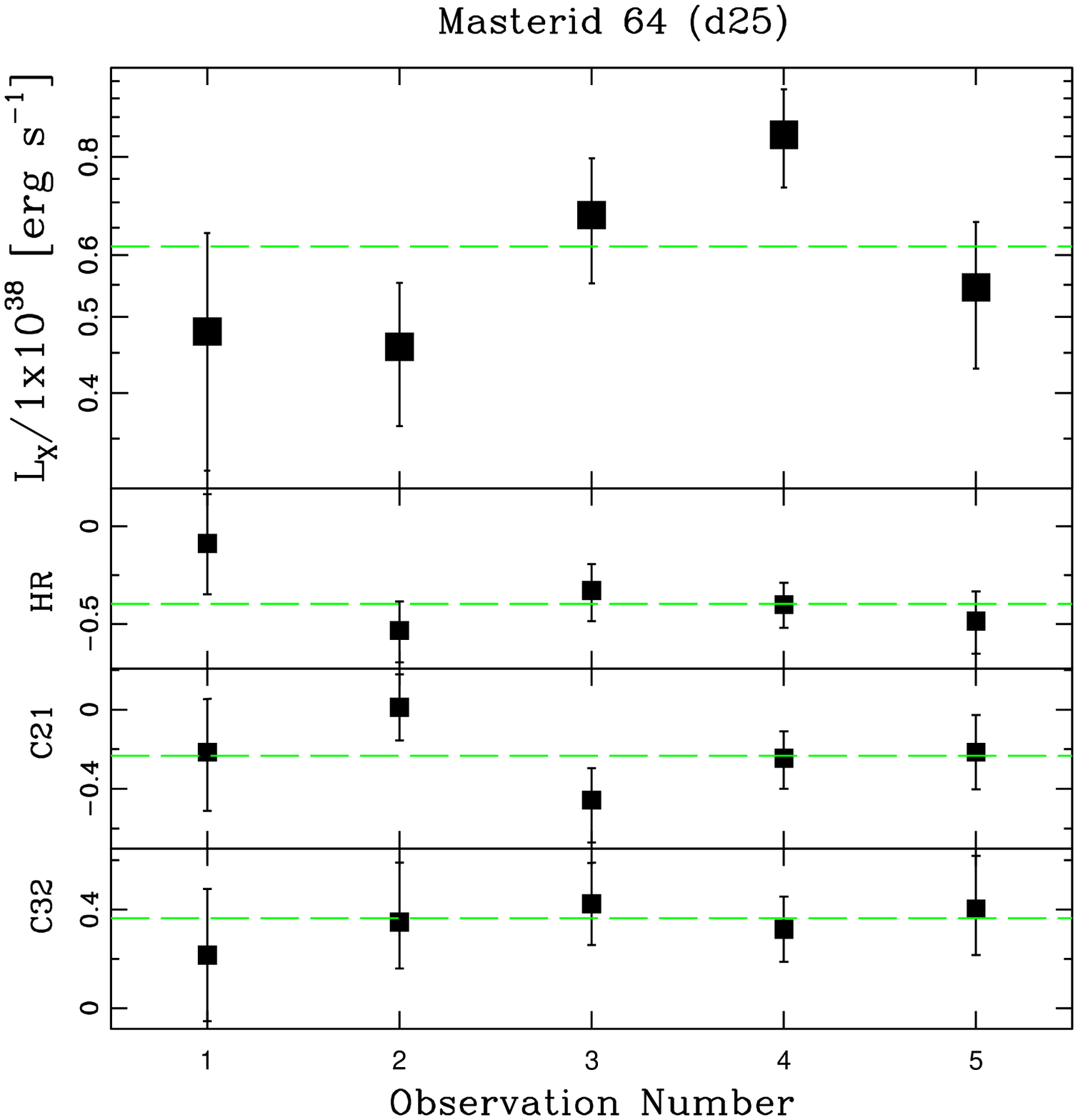}

  \end{minipage}\hspace{0.02\linewidth}

\end{figure}

\begin{figure}
  \begin{minipage}{0.485\linewidth}
  \centering

    \includegraphics[width=\linewidth]{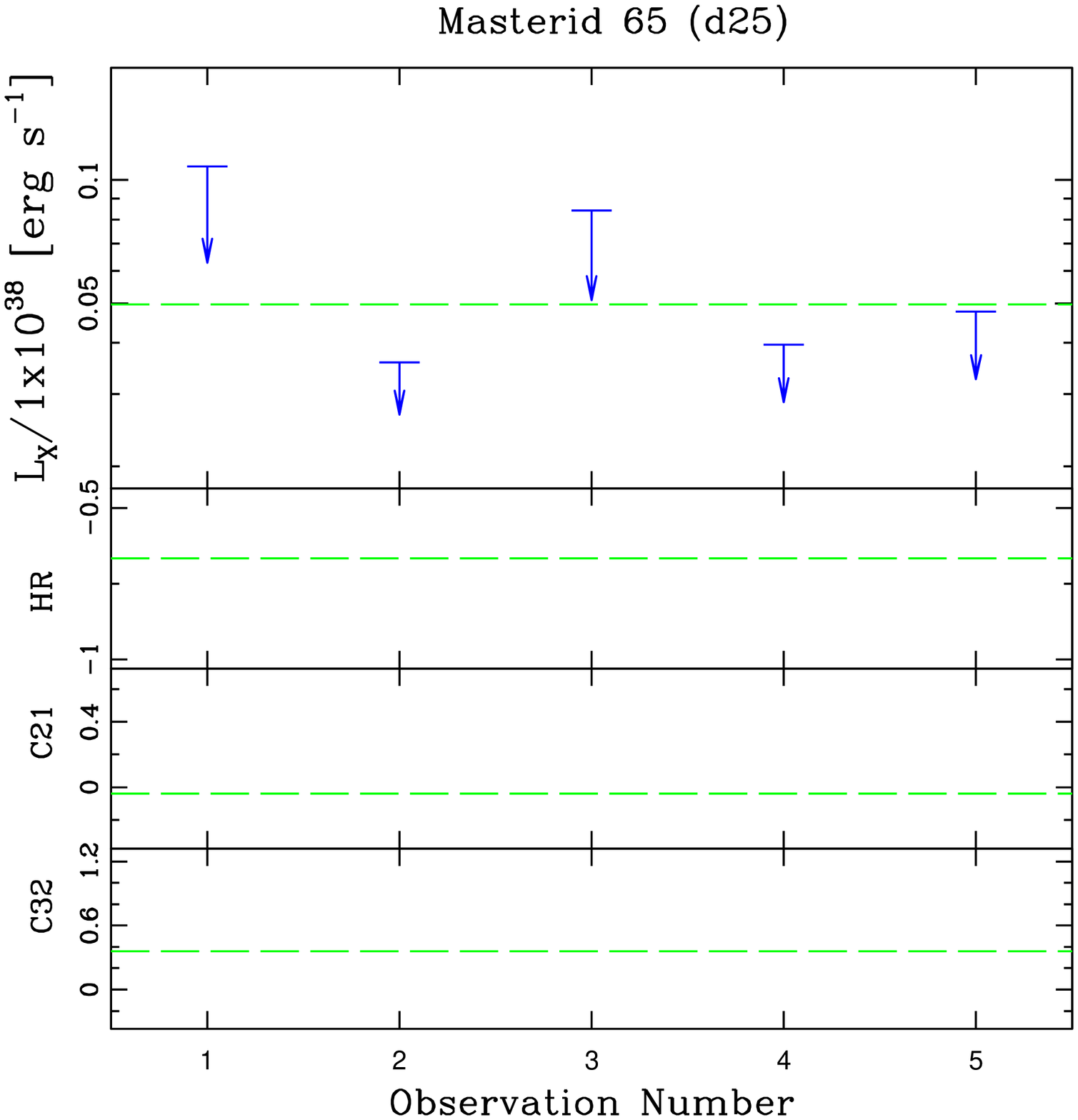}

\end{minipage}\hspace{0.02\linewidth}
\begin{minipage}{0.485\linewidth}
  \centering

    \includegraphics[width=\linewidth]{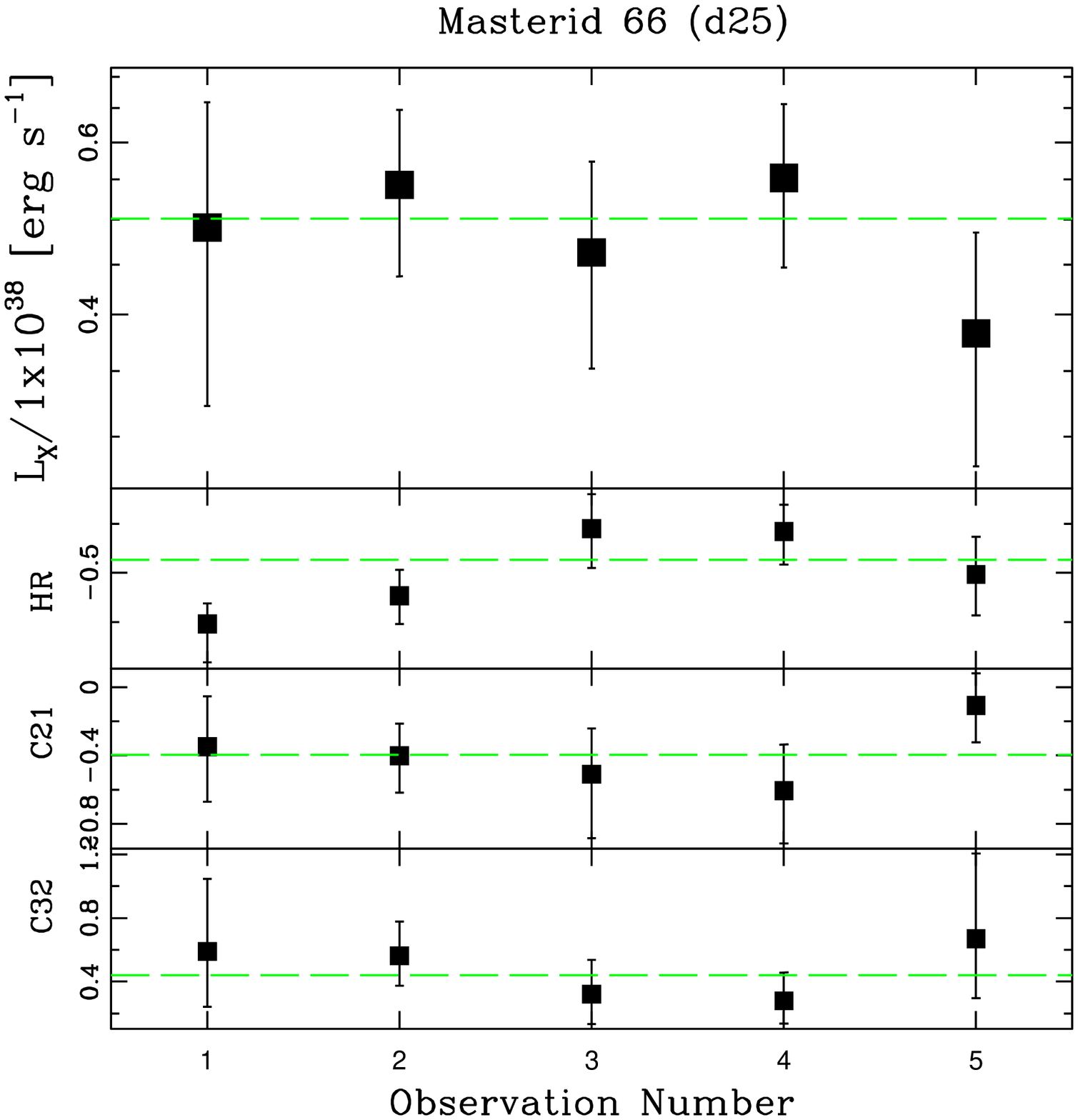}

\end{minipage}\hspace{0.02\linewidth}

  \begin{minipage}{0.485\linewidth}
  \centering
  
    \includegraphics[width=\linewidth]{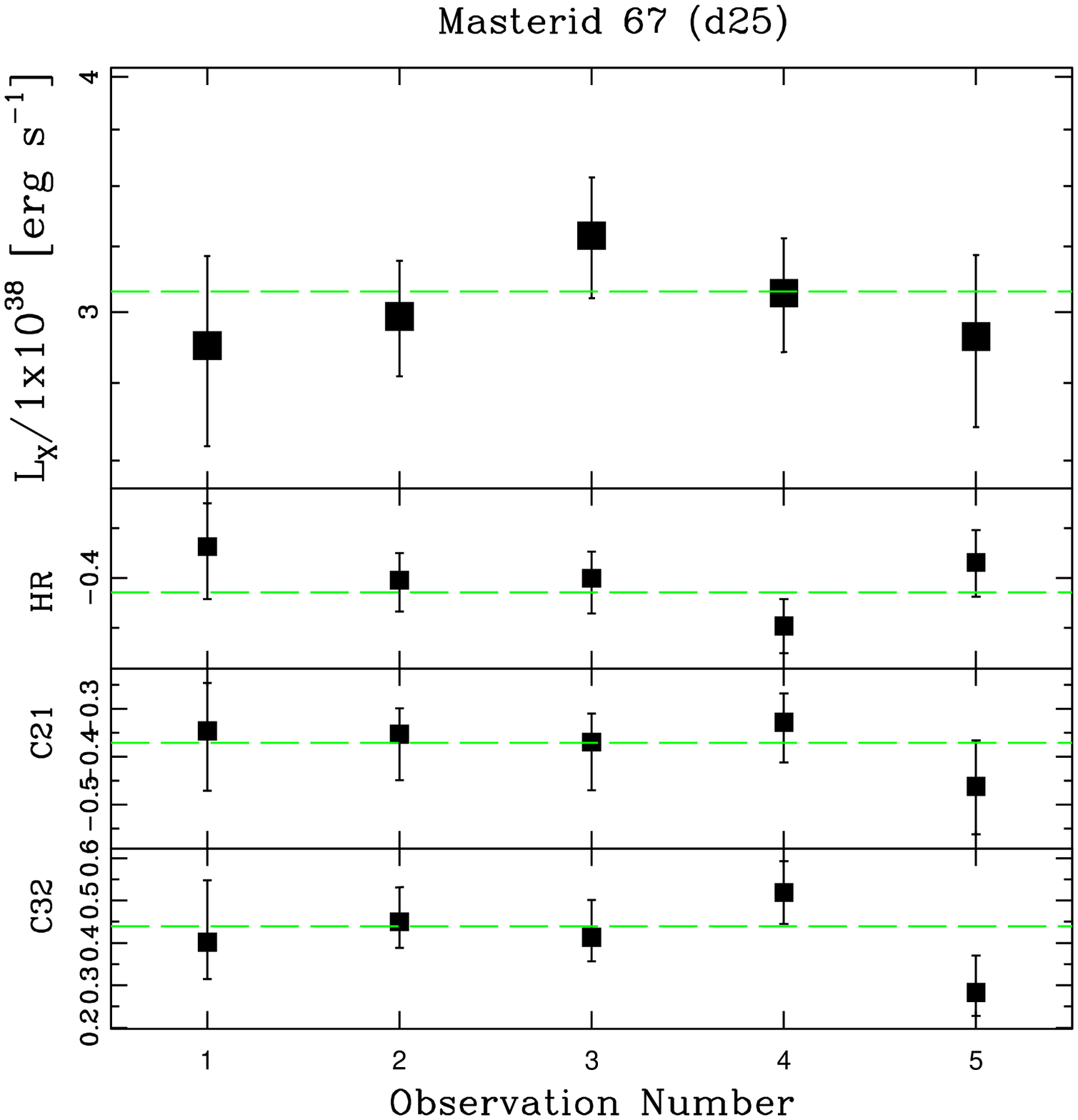}

  \end{minipage}\hspace{0.02\linewidth}
  \begin{minipage}{0.485\linewidth}
  \centering

    \includegraphics[width=\linewidth]{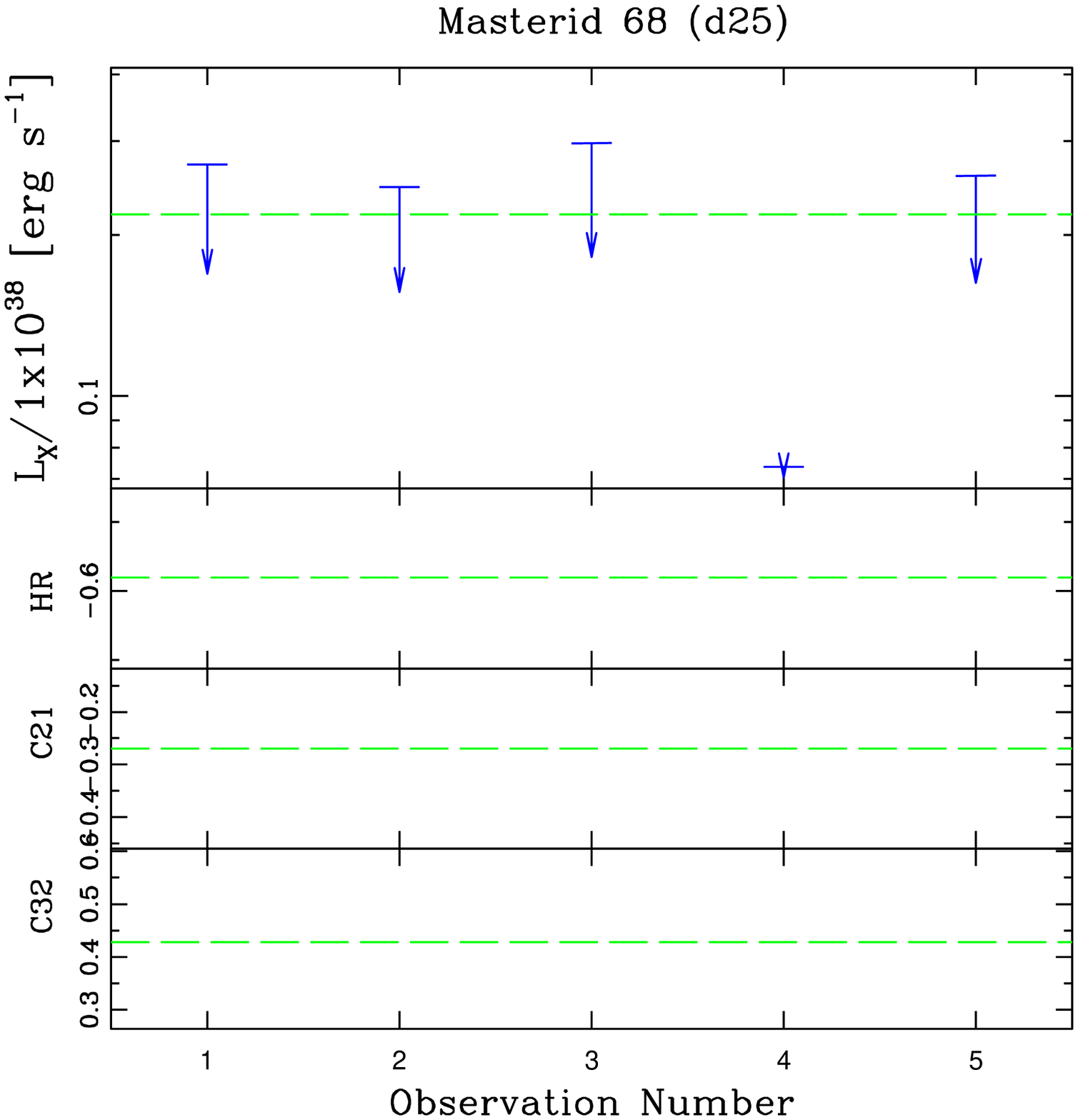}

\end{minipage}\hspace{0.02\linewidth}

\begin{minipage}{0.485\linewidth}
  \centering

    \includegraphics[width=\linewidth]{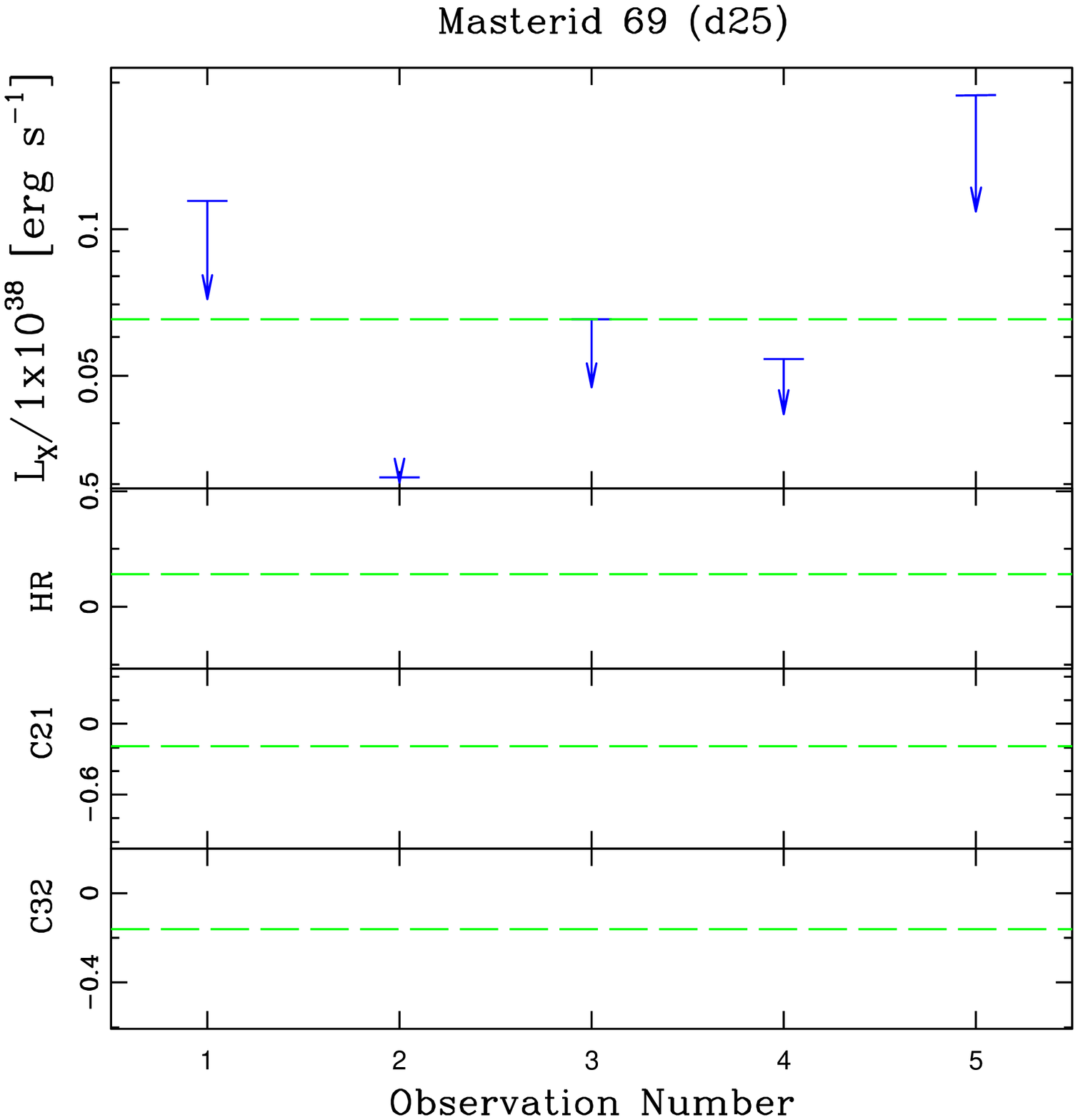}

 \end{minipage}\hspace{0.02\linewidth}
\begin{minipage}{0.485\linewidth}
  \centering
  
    \includegraphics[width=\linewidth]{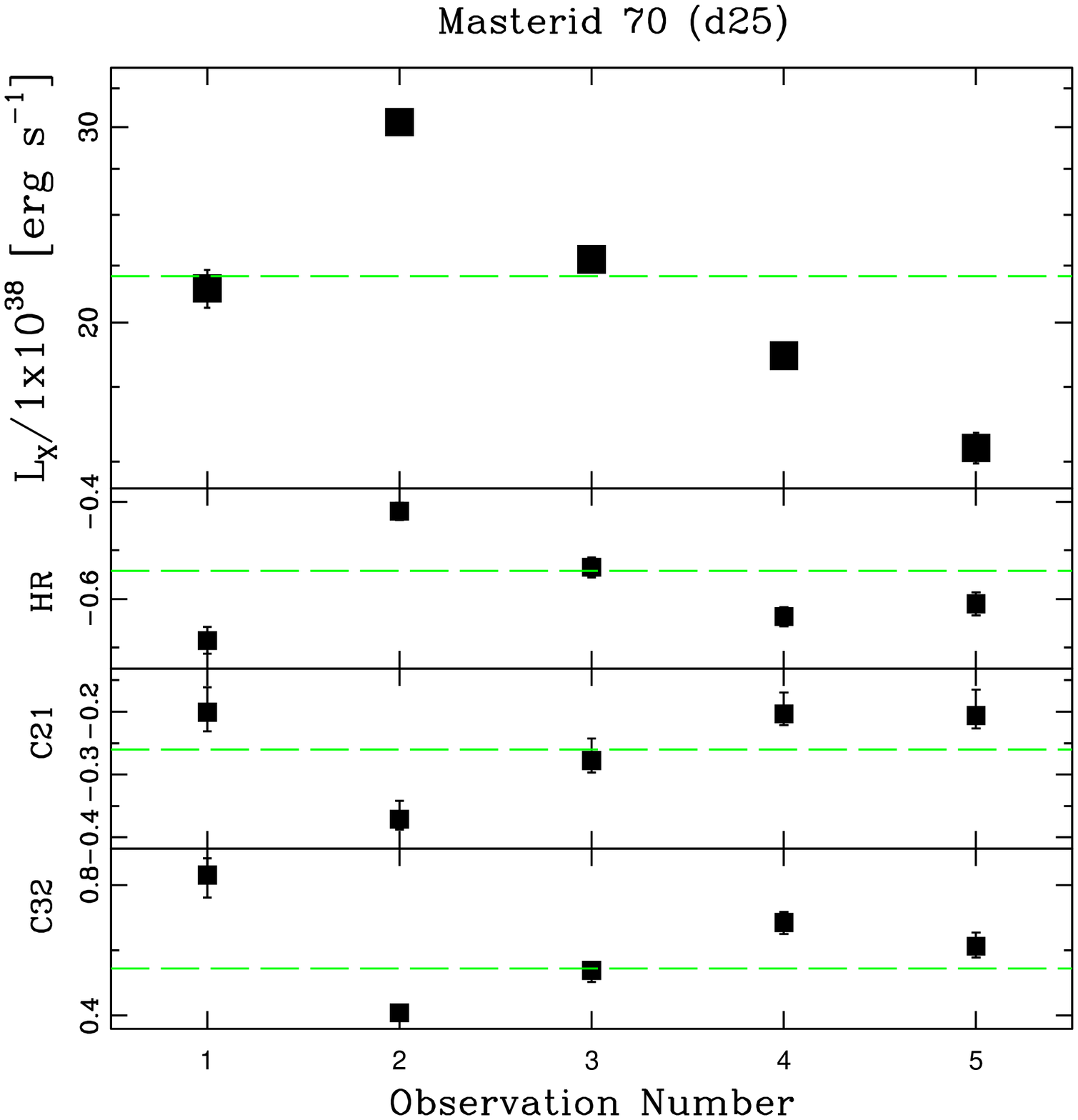}

  \end{minipage}\hspace{0.02\linewidth}
\end{figure}

\begin{figure}
  \begin{minipage}{0.485\linewidth}
  \centering

    \includegraphics[width=\linewidth]{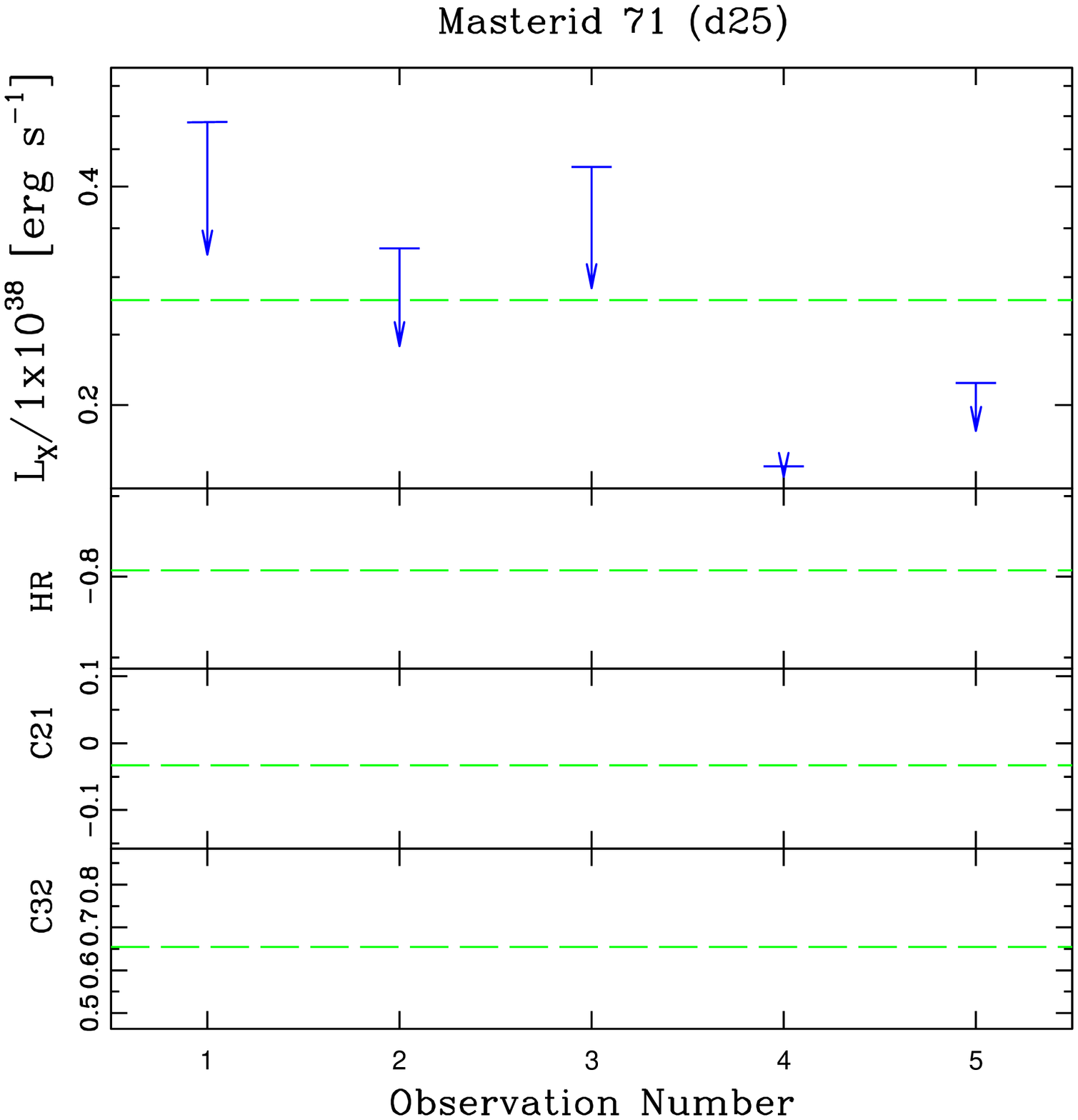}

\end{minipage}\hspace{0.02\linewidth}
\begin{minipage}{0.485\linewidth}
  \centering

    \includegraphics[width=\linewidth]{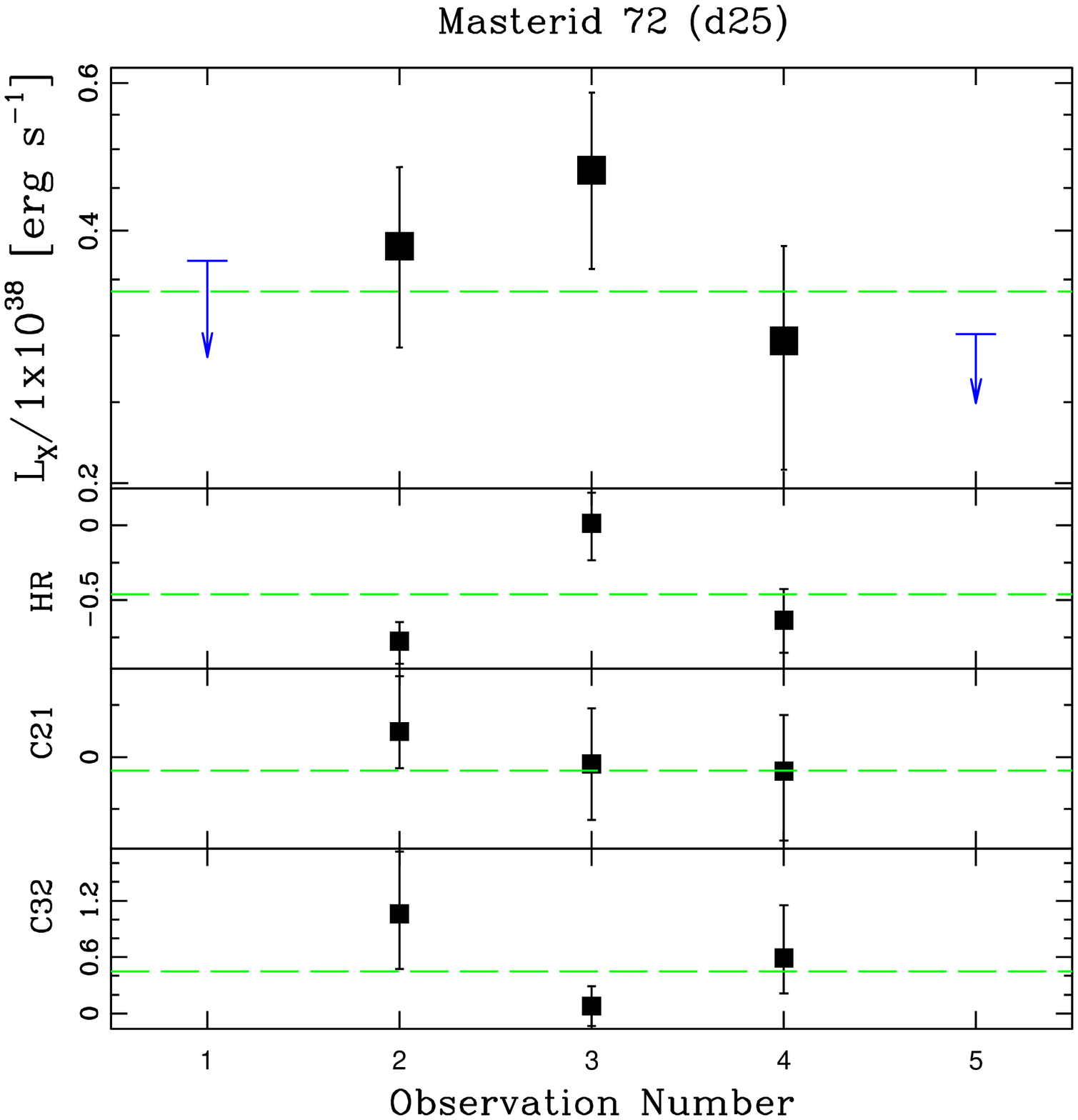}

 \end{minipage}\hspace{0.02\linewidth}

  \begin{minipage}{0.485\linewidth}
  \centering
  
    \includegraphics[width=\linewidth]{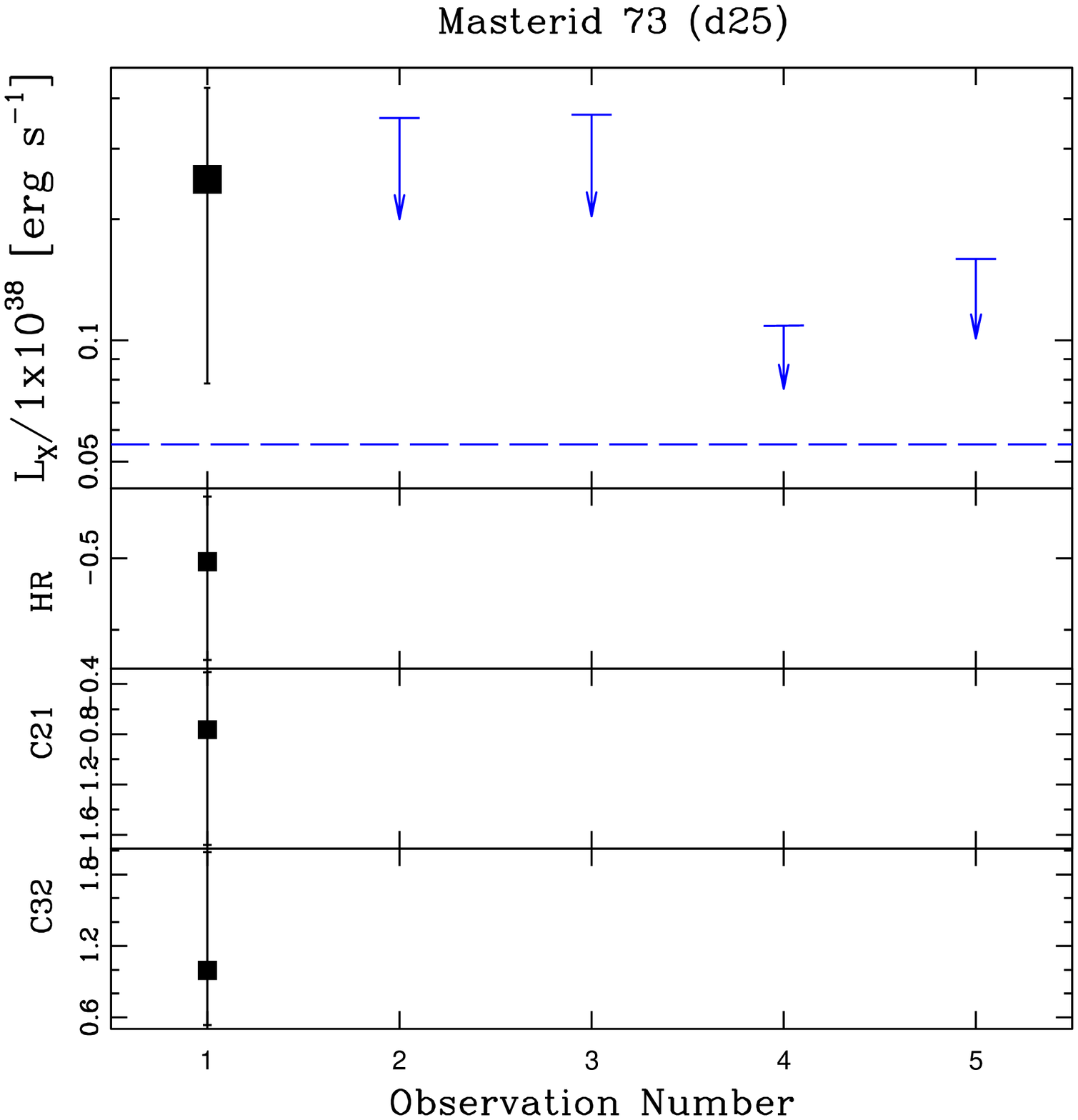}

  \end{minipage}\hspace{0.02\linewidth}
  \begin{minipage}{0.485\linewidth}
  \centering

    \includegraphics[width=\linewidth]{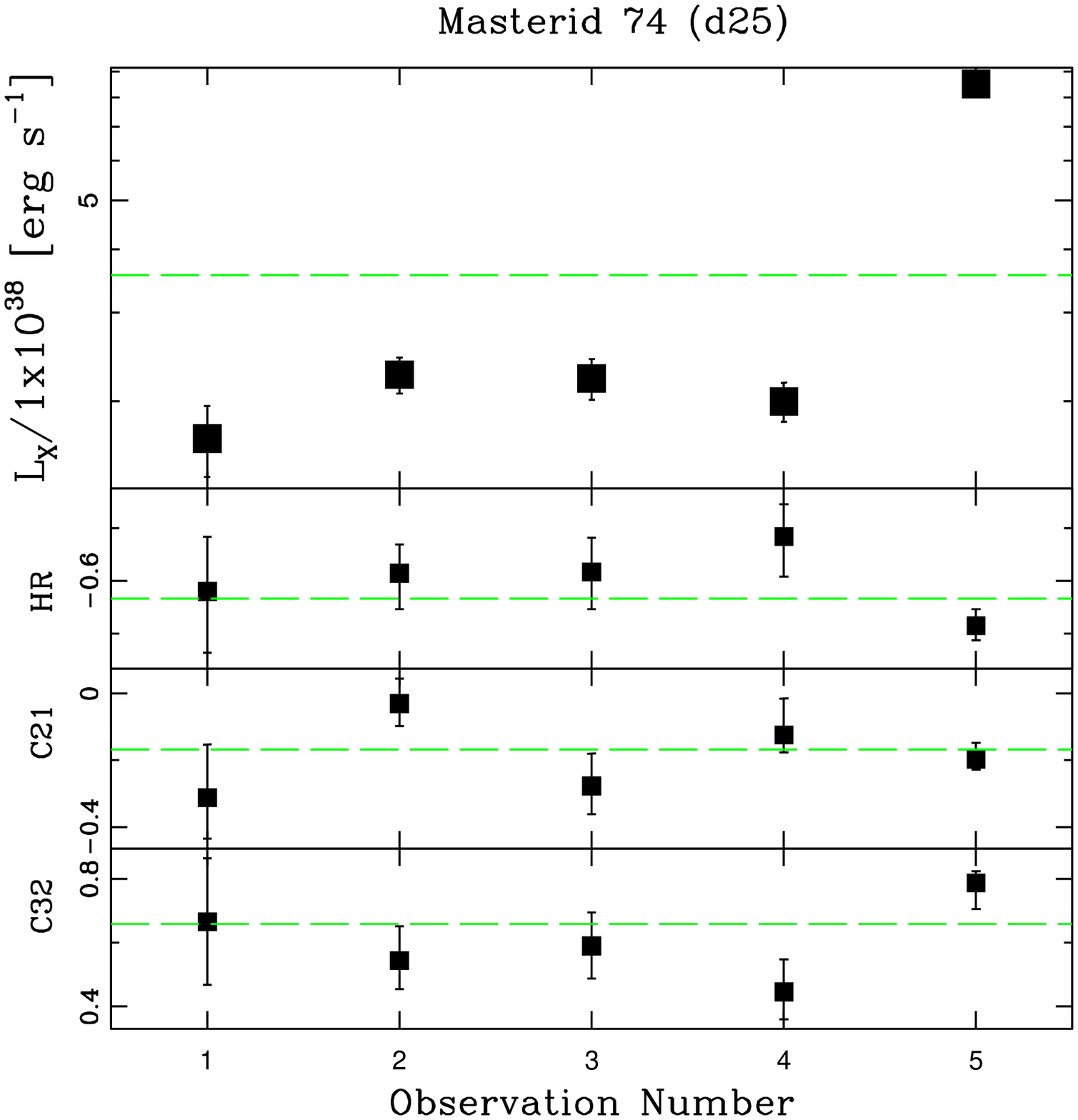}

\end{minipage}\hspace{0.02\linewidth}

\begin{minipage}{0.485\linewidth}
  \centering

    \includegraphics[width=\linewidth]{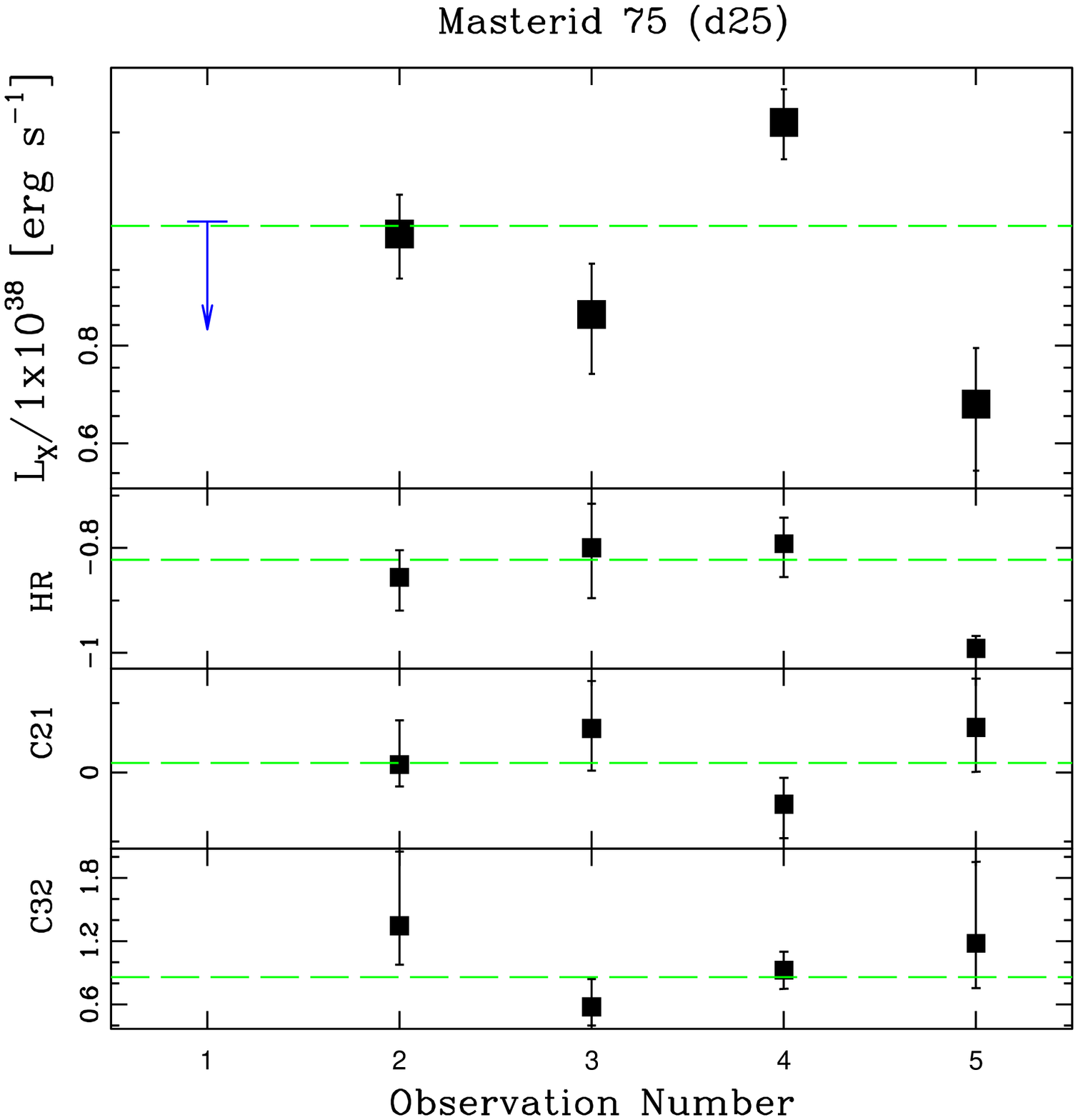}

\end{minipage}\hspace{0.02\linewidth}
\begin{minipage}{0.485\linewidth}
  \centering
  
    \includegraphics[width=\linewidth]{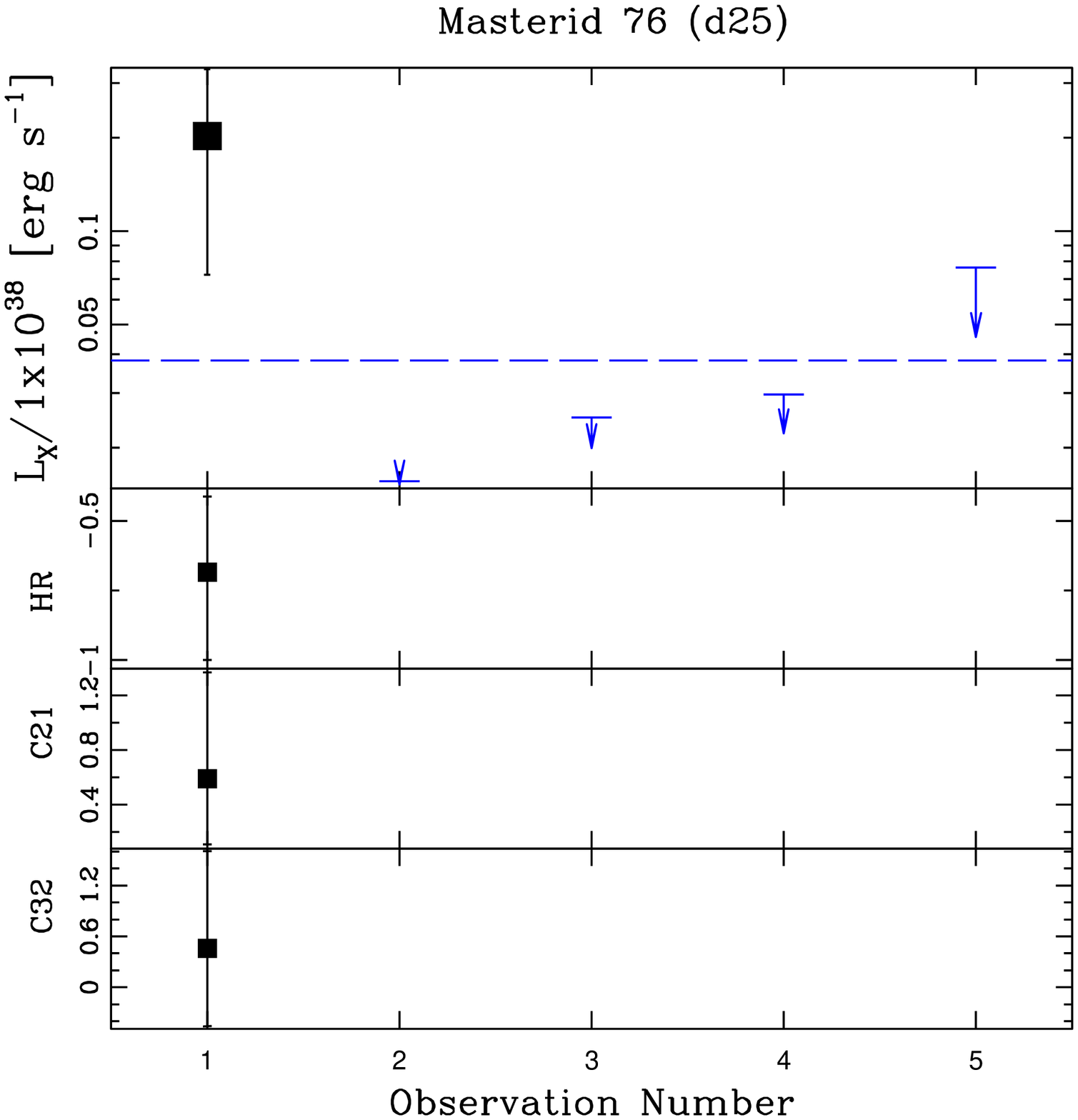}

  \end{minipage}\hspace{0.02\linewidth}
\end{figure}

\clearpage

\begin{figure}
  \begin{minipage}{0.485\linewidth}
  \centering

    \includegraphics[width=\linewidth]{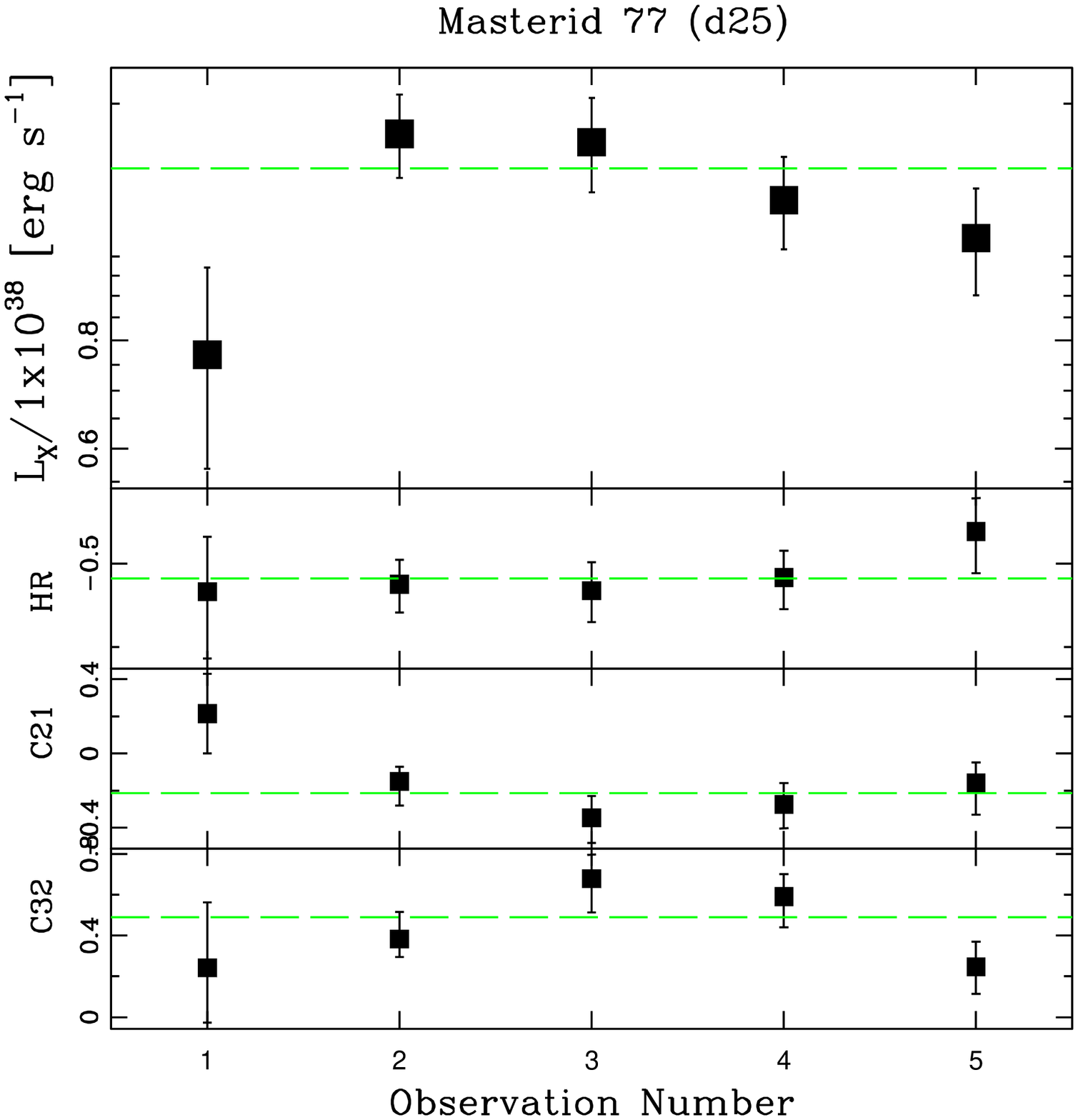}

\end{minipage}\hspace{0.02\linewidth}
\begin{minipage}{0.485\linewidth}
  \centering

    \includegraphics[width=\linewidth]{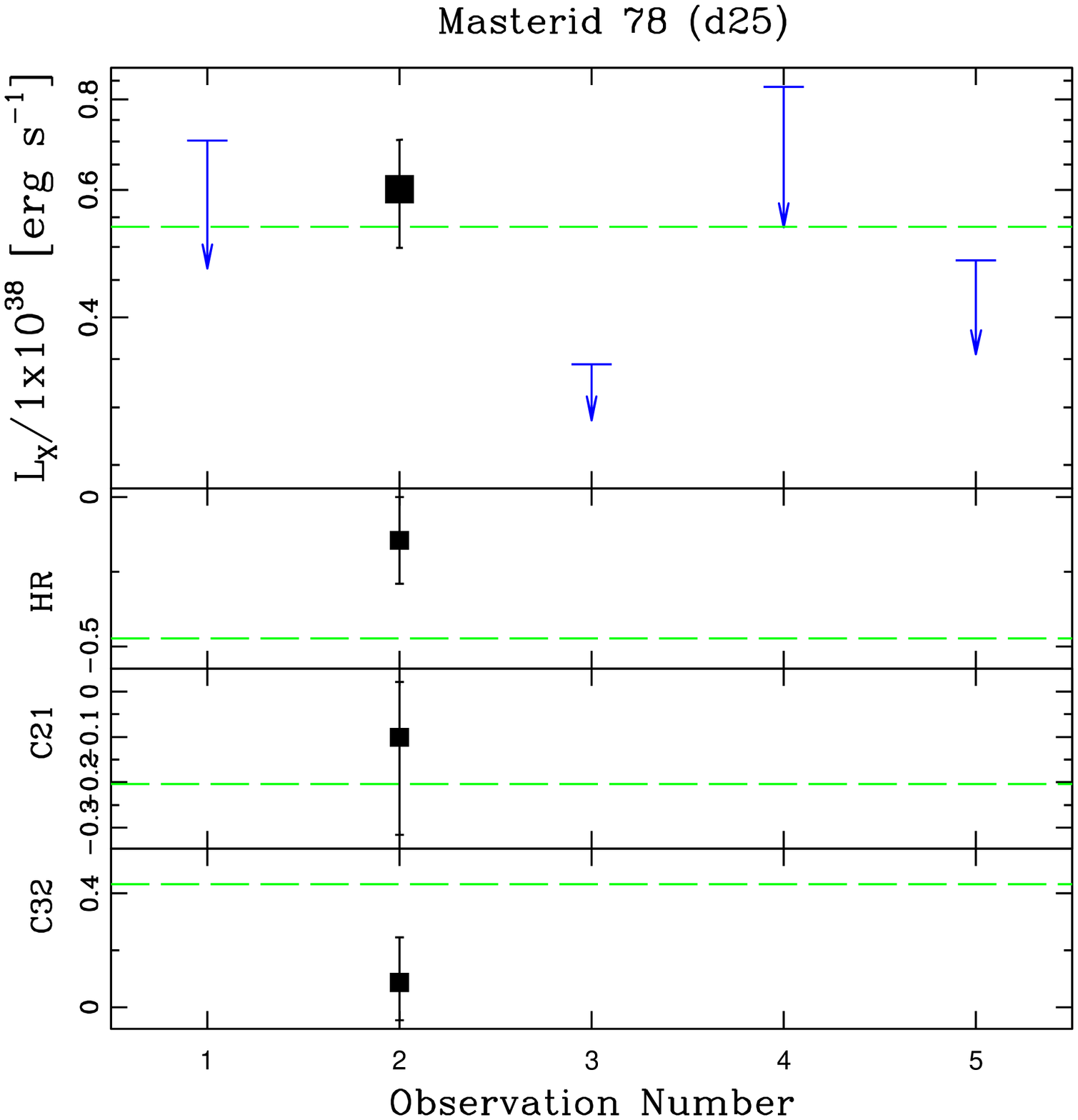}

\end{minipage}\hspace{0.02\linewidth}

  \begin{minipage}{0.485\linewidth}
  \centering
  
    \includegraphics[width=\linewidth]{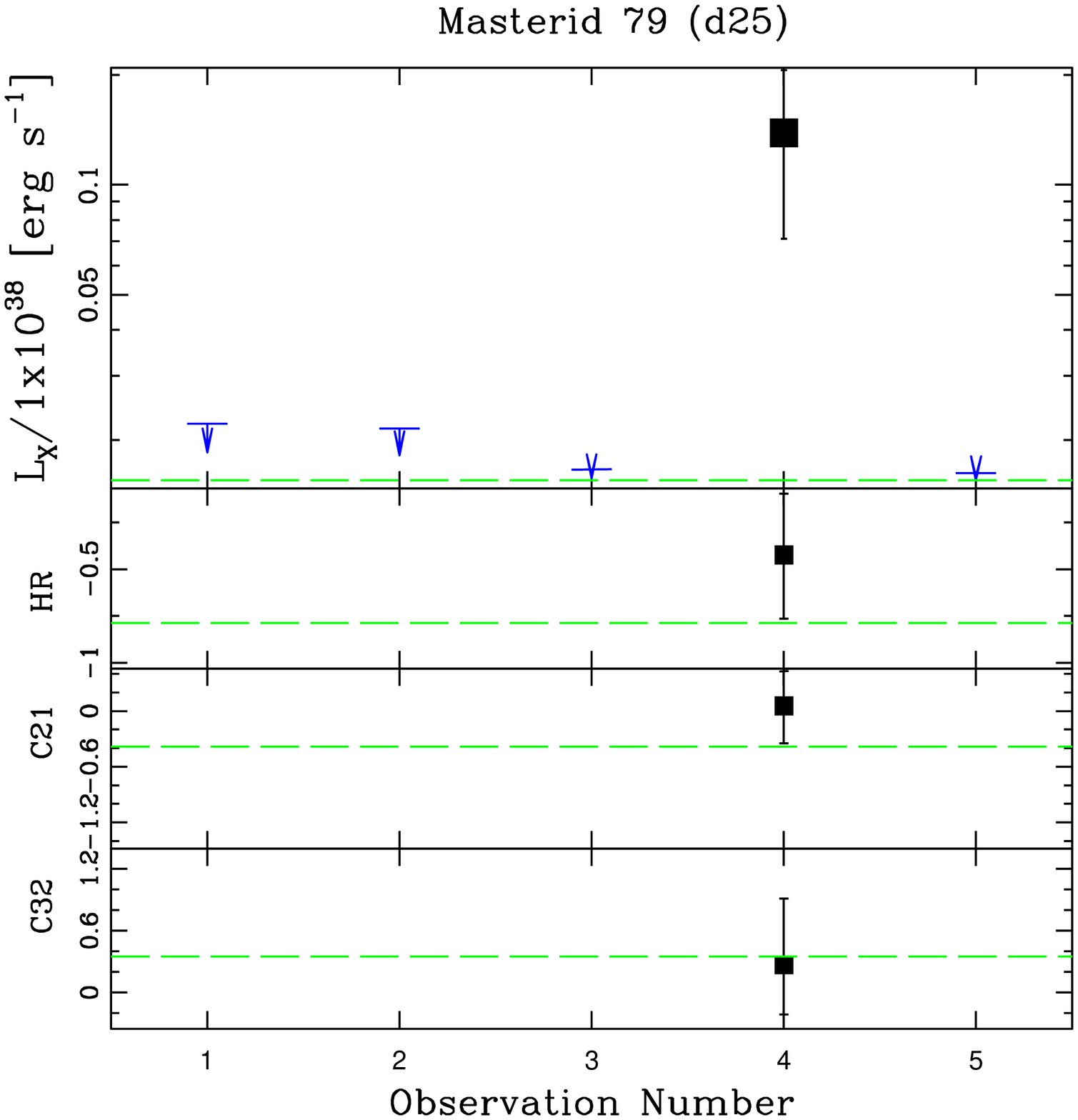}

  \end{minipage}\hspace{0.02\linewidth}
  \begin{minipage}{0.485\linewidth}
  \centering

    \includegraphics[width=\linewidth]{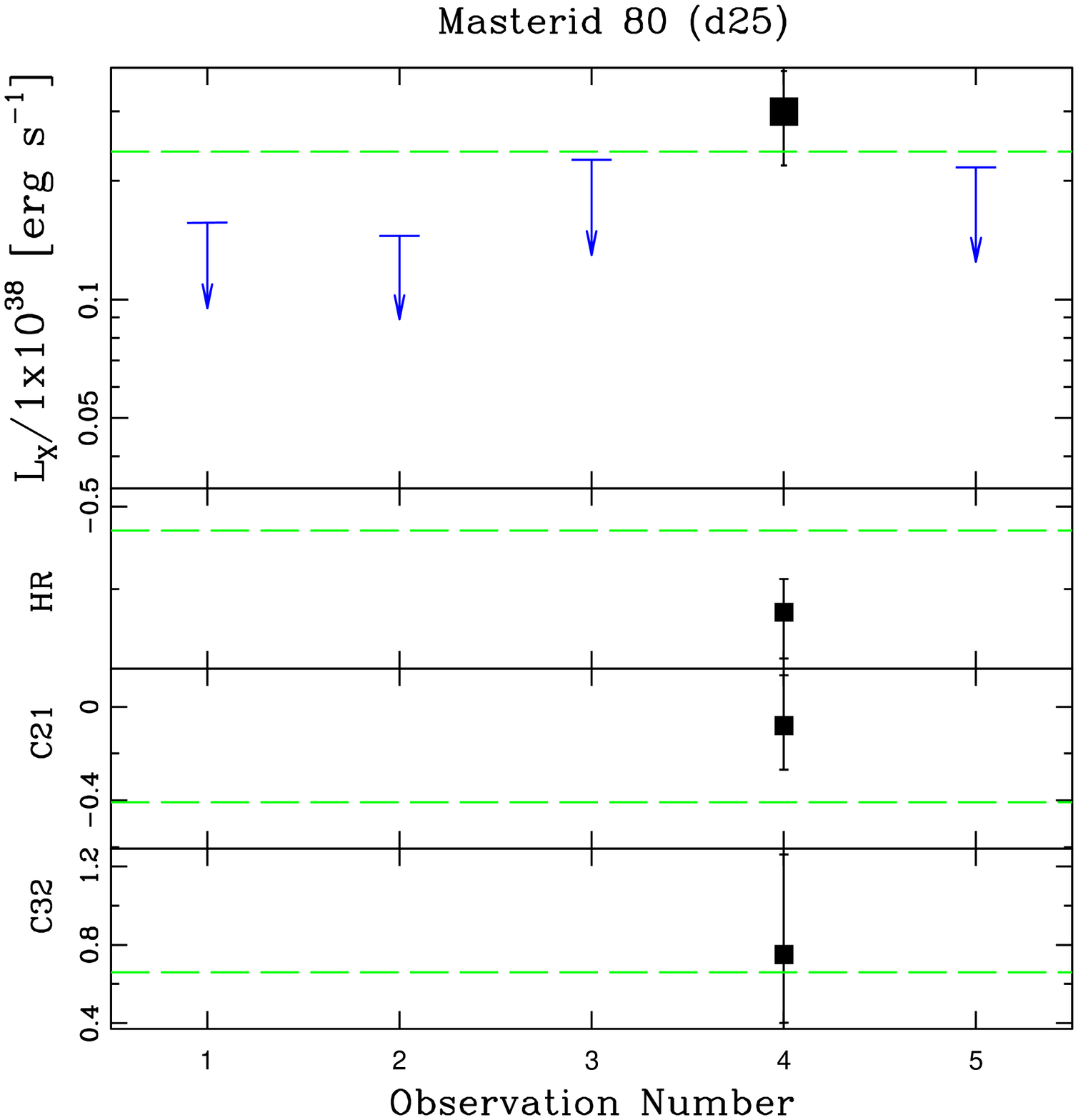}

\end{minipage}\hspace{0.02\linewidth}

\begin{minipage}{0.485\linewidth}
  \centering

    \includegraphics[width=\linewidth]{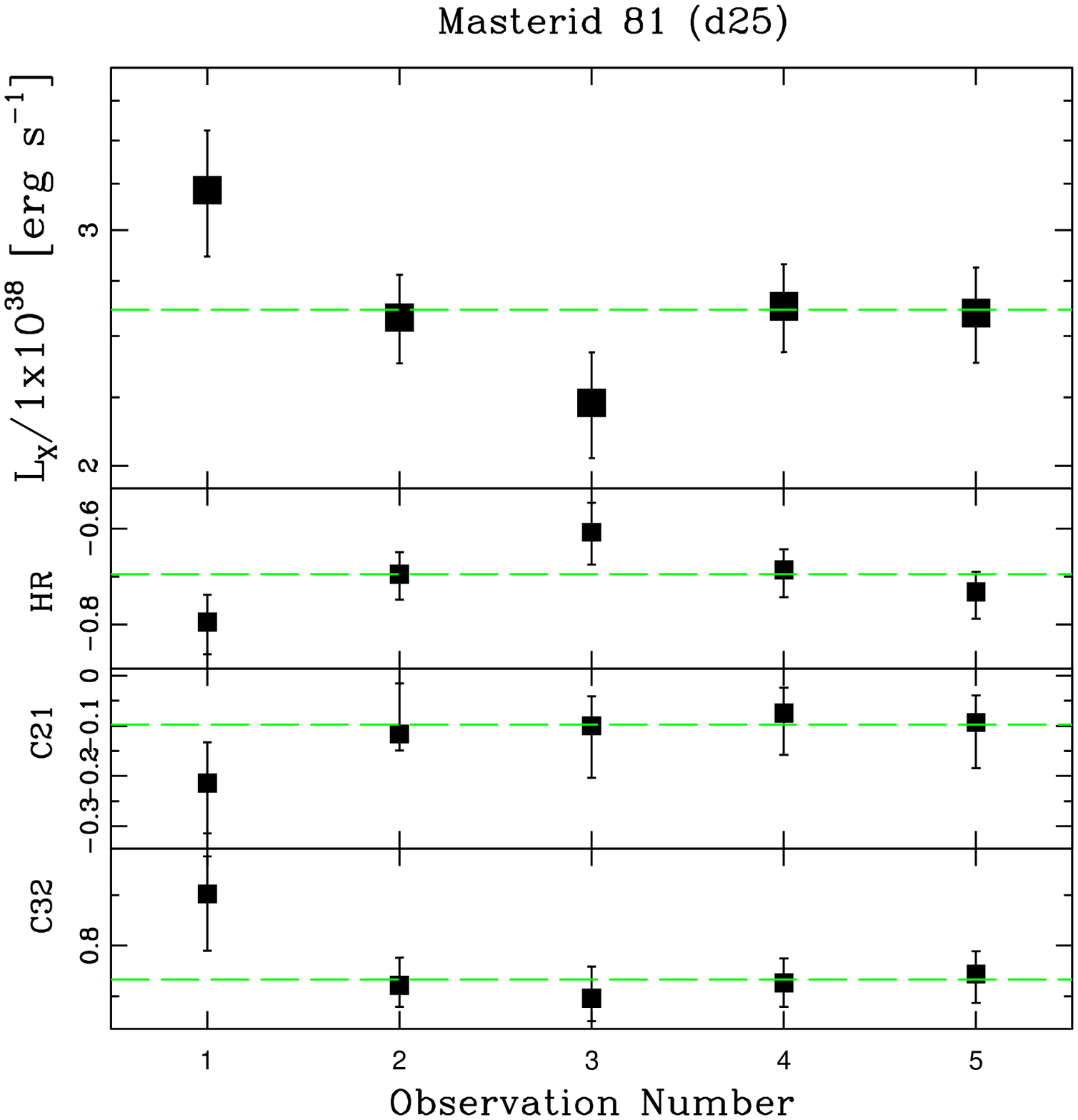}

 \end{minipage}\hspace{0.02\linewidth}
\begin{minipage}{0.485\linewidth}
  \centering
  
    \includegraphics[width=\linewidth]{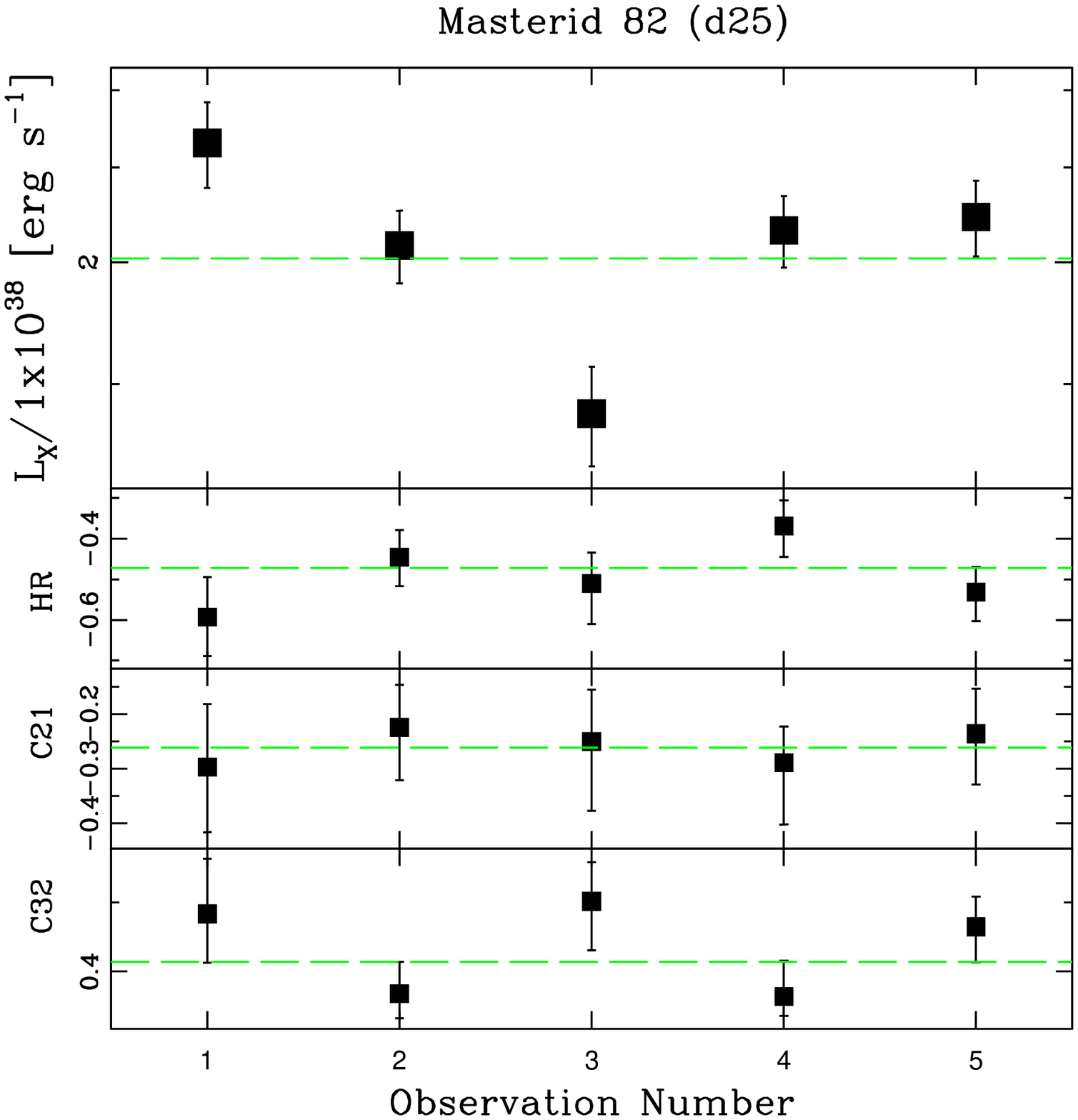}

  \end{minipage}\hspace{0.02\linewidth}
\end{figure}

\begin{figure}
  \begin{minipage}{0.485\linewidth}
  \centering

    \includegraphics[width=\linewidth]{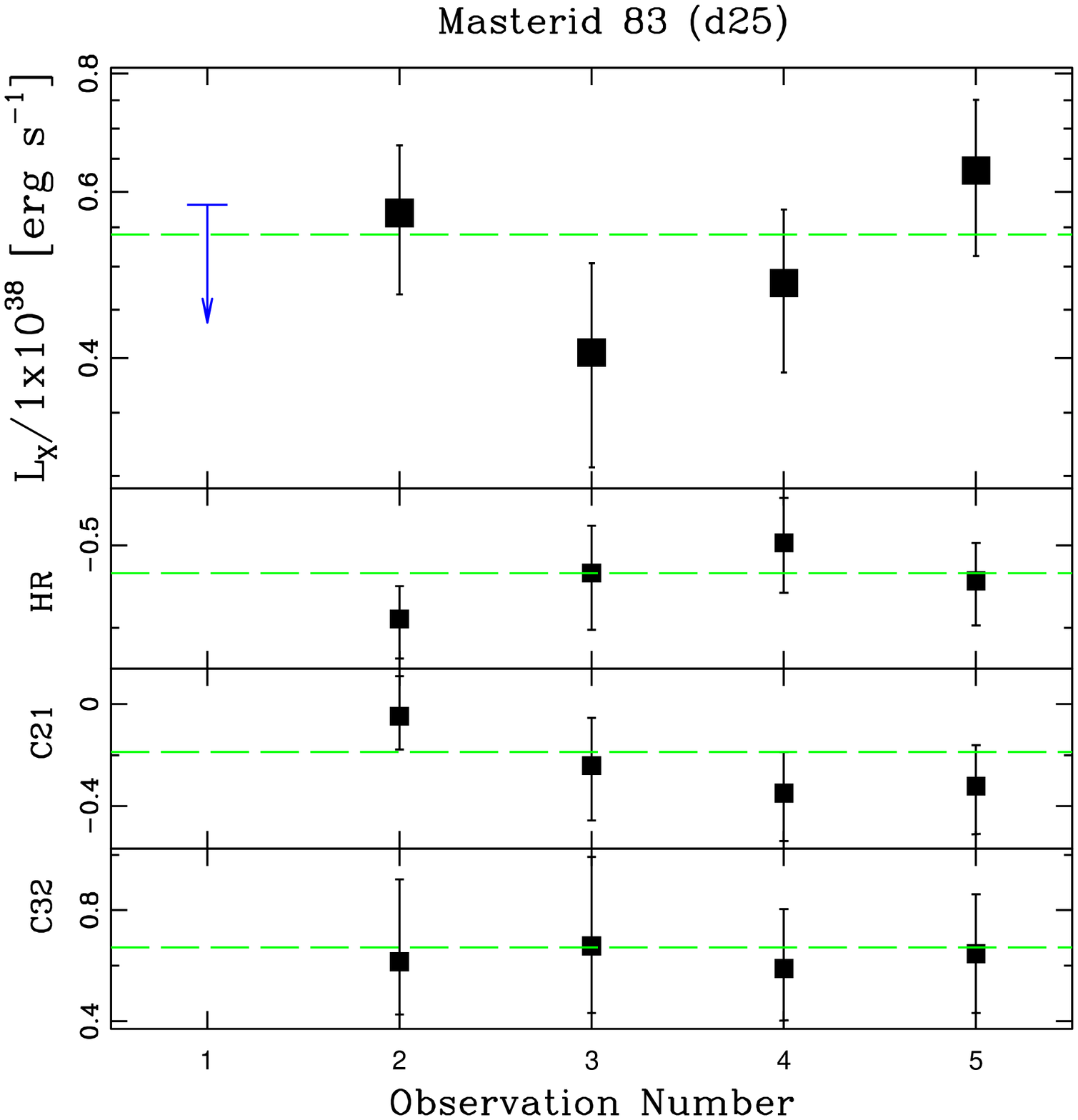}

\end{minipage}\hspace{0.02\linewidth}
\begin{minipage}{0.485\linewidth}
  \centering

    \includegraphics[width=\linewidth]{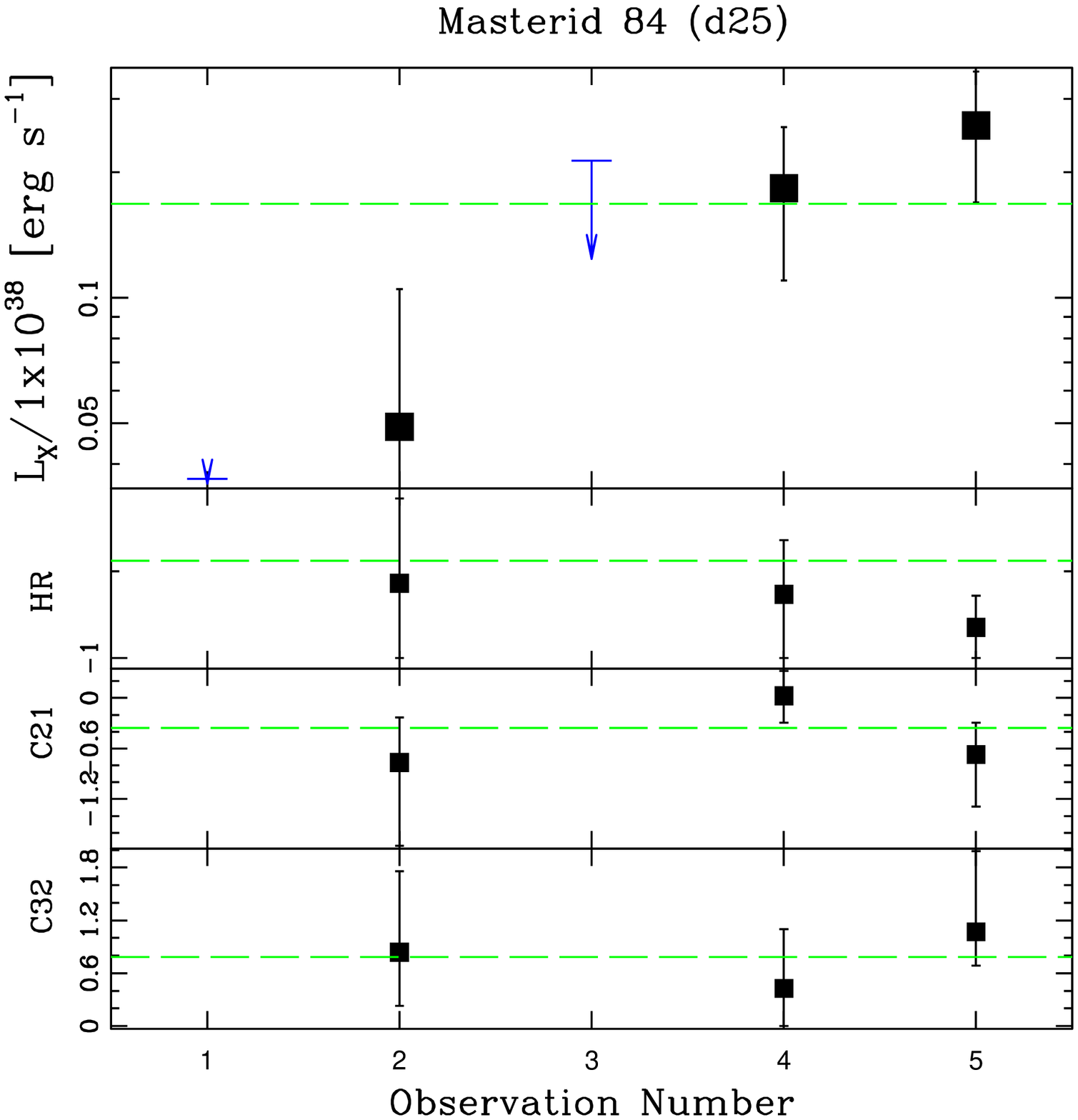}

 \end{minipage}\hspace{0.02\linewidth}

  \begin{minipage}{0.485\linewidth}
  \centering
  
    \includegraphics[width=\linewidth]{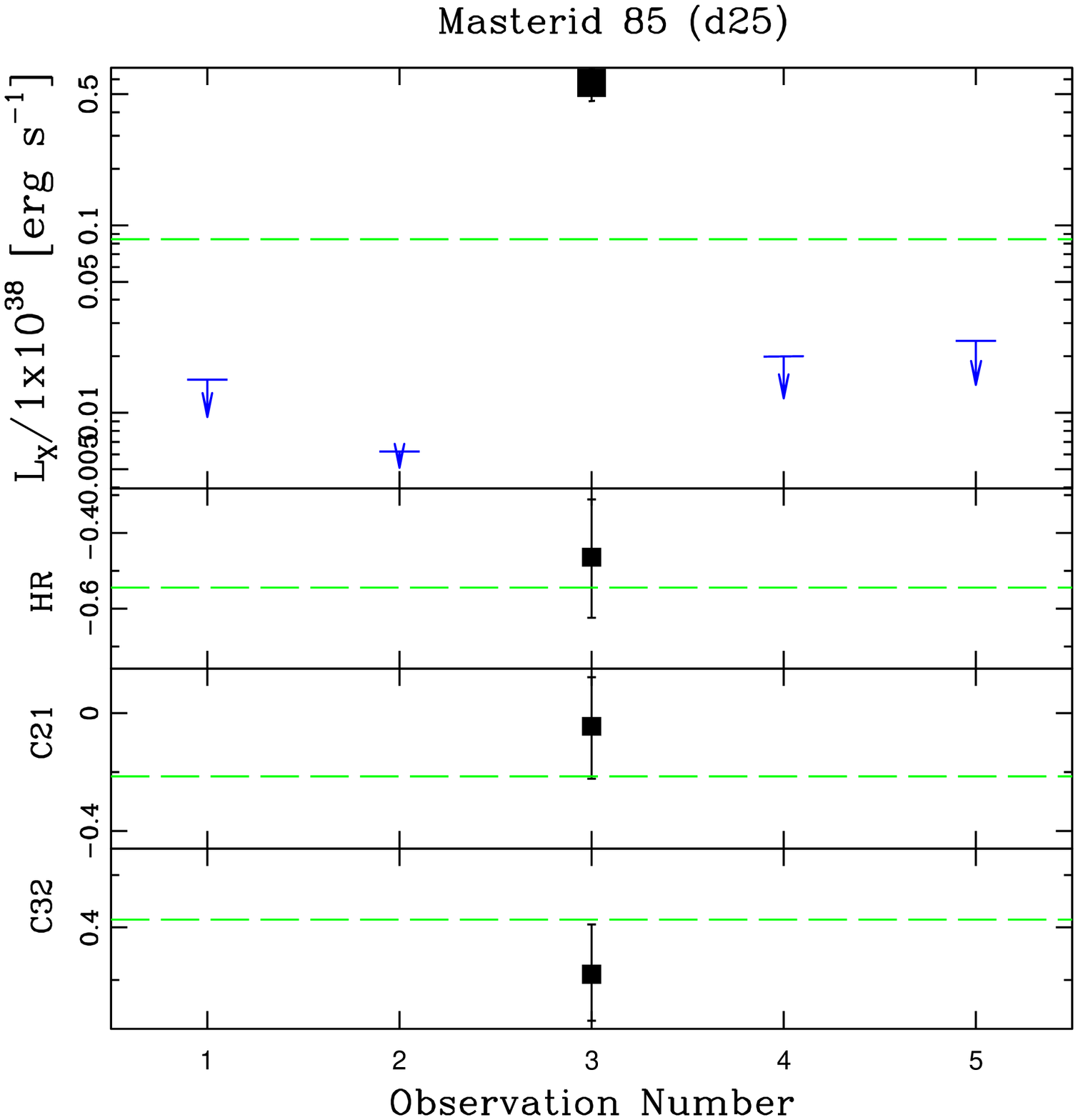}

  \end{minipage}\hspace{0.02\linewidth}
  \begin{minipage}{0.485\linewidth}
  \centering

    \includegraphics[width=\linewidth]{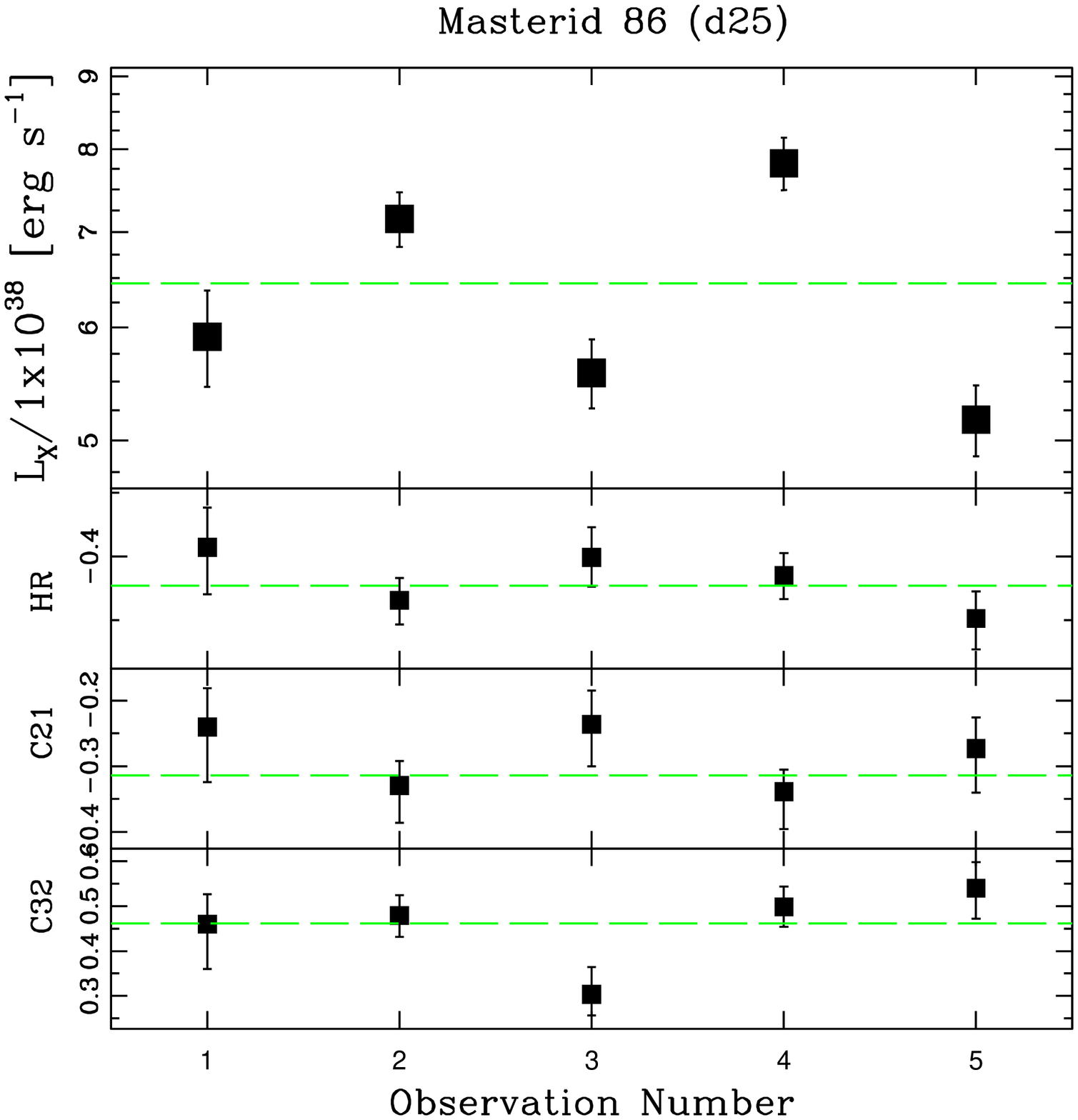}

\end{minipage}\hspace{0.02\linewidth}

\begin{minipage}{0.485\linewidth}
  \centering

    \includegraphics[width=\linewidth]{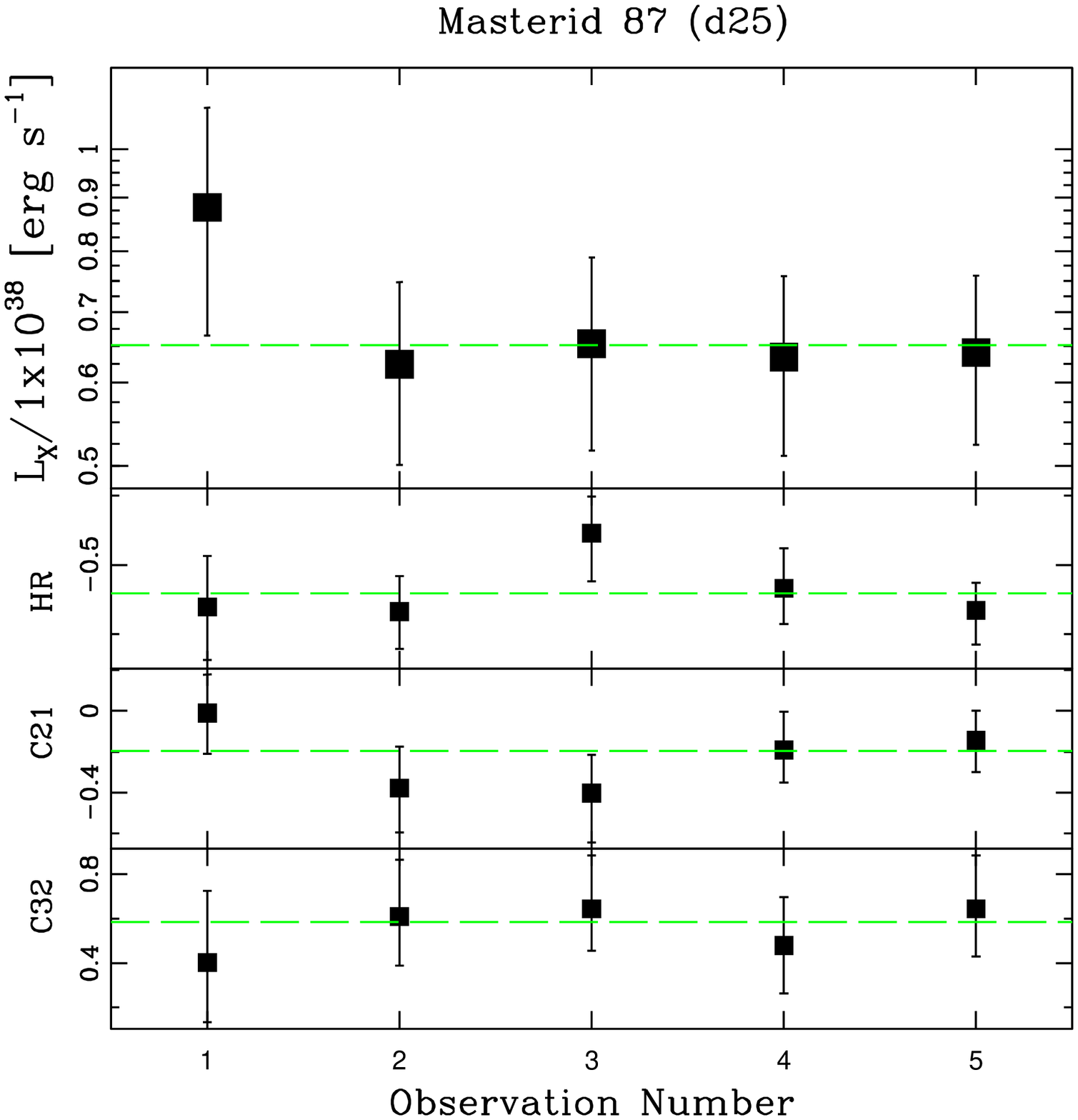}

\end{minipage}\hspace{0.02\linewidth}
\begin{minipage}{0.485\linewidth}
  \centering
  
    \includegraphics[width=\linewidth]{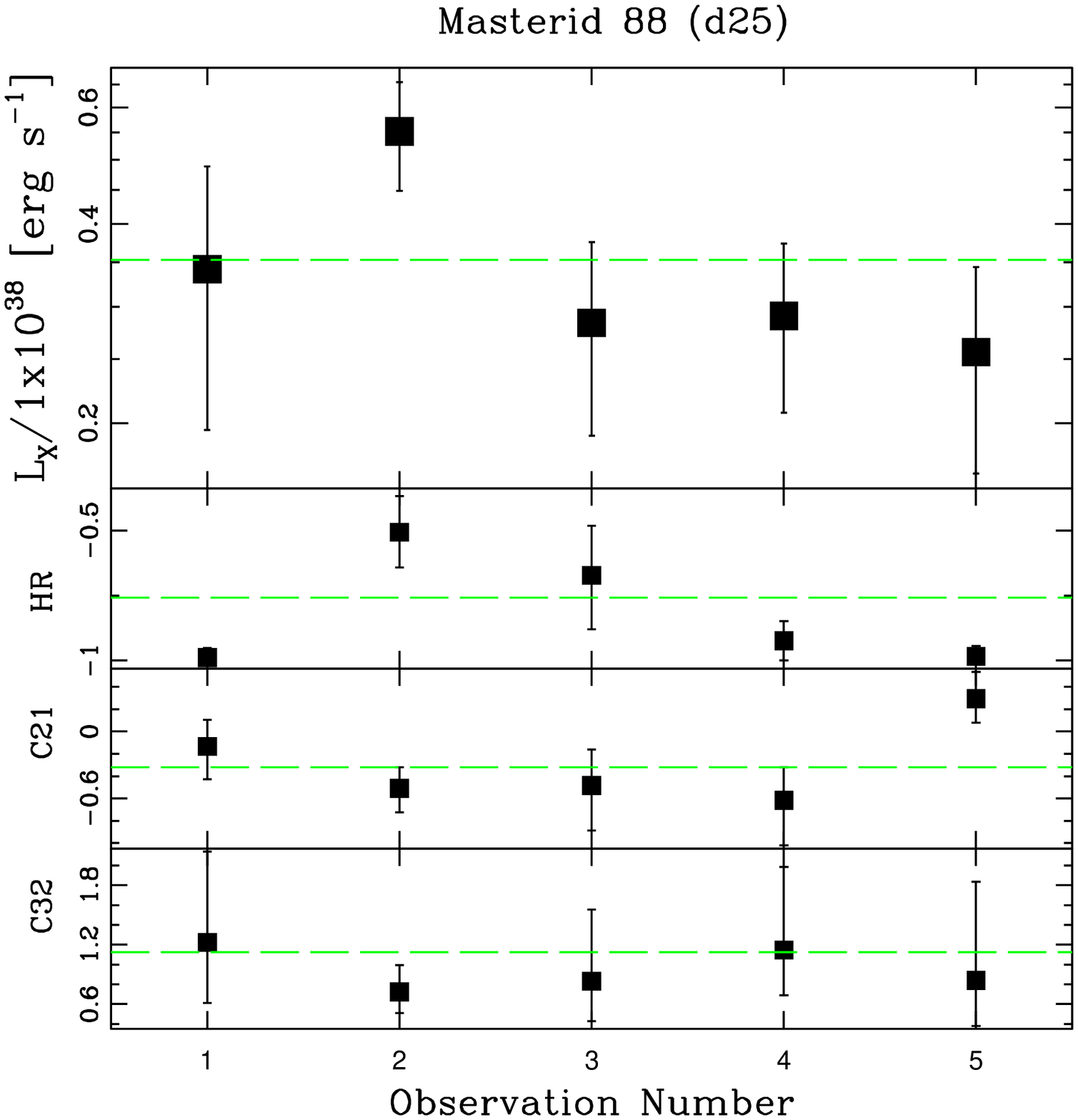}

  \end{minipage}\hspace{0.02\linewidth}

\end{figure}

\begin{figure}
  \begin{minipage}{0.485\linewidth}
  \centering

    \includegraphics[width=\linewidth]{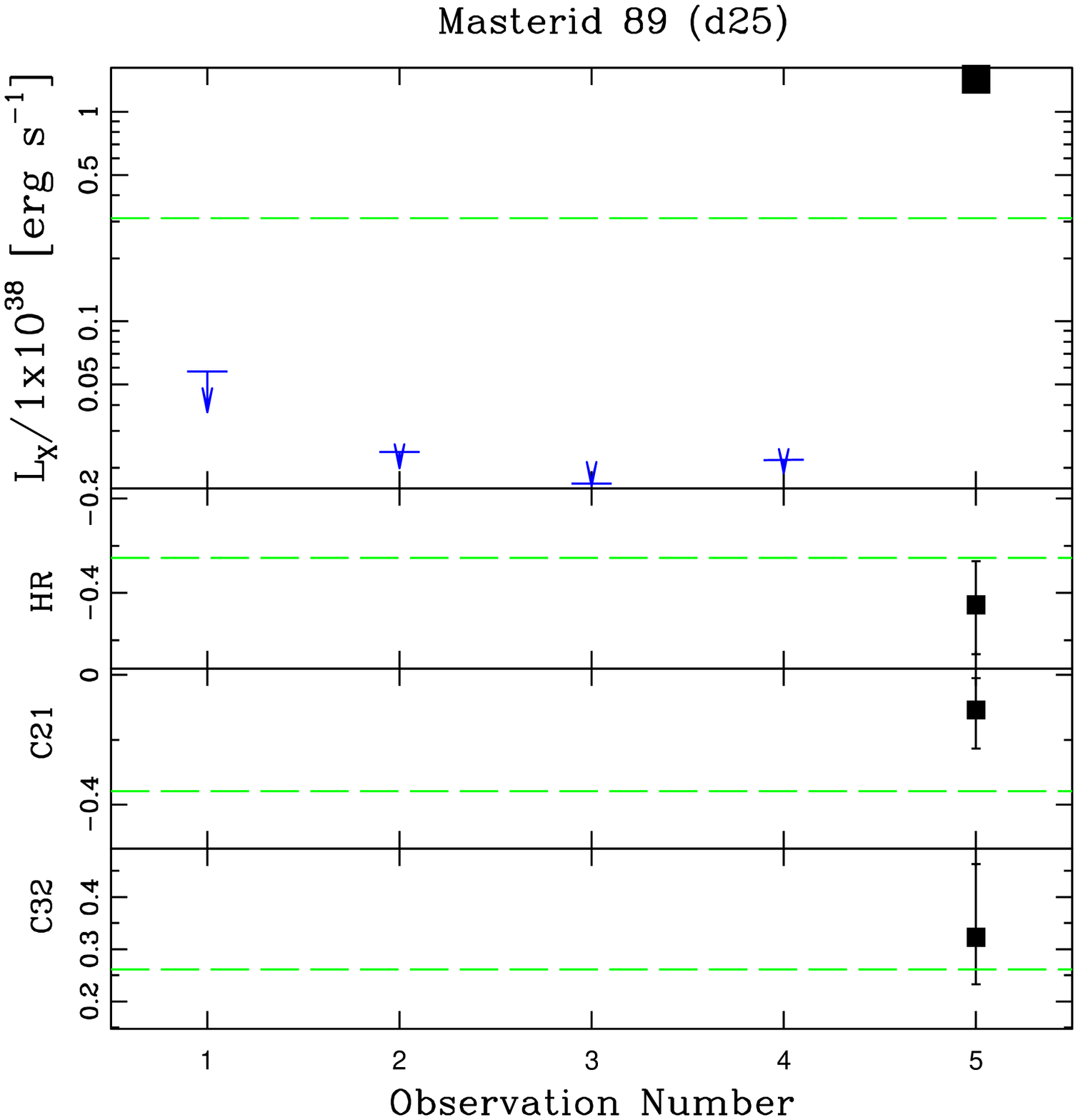}

\end{minipage}\hspace{0.02\linewidth}
\begin{minipage}{0.485\linewidth}
  \centering

    \includegraphics[width=\linewidth]{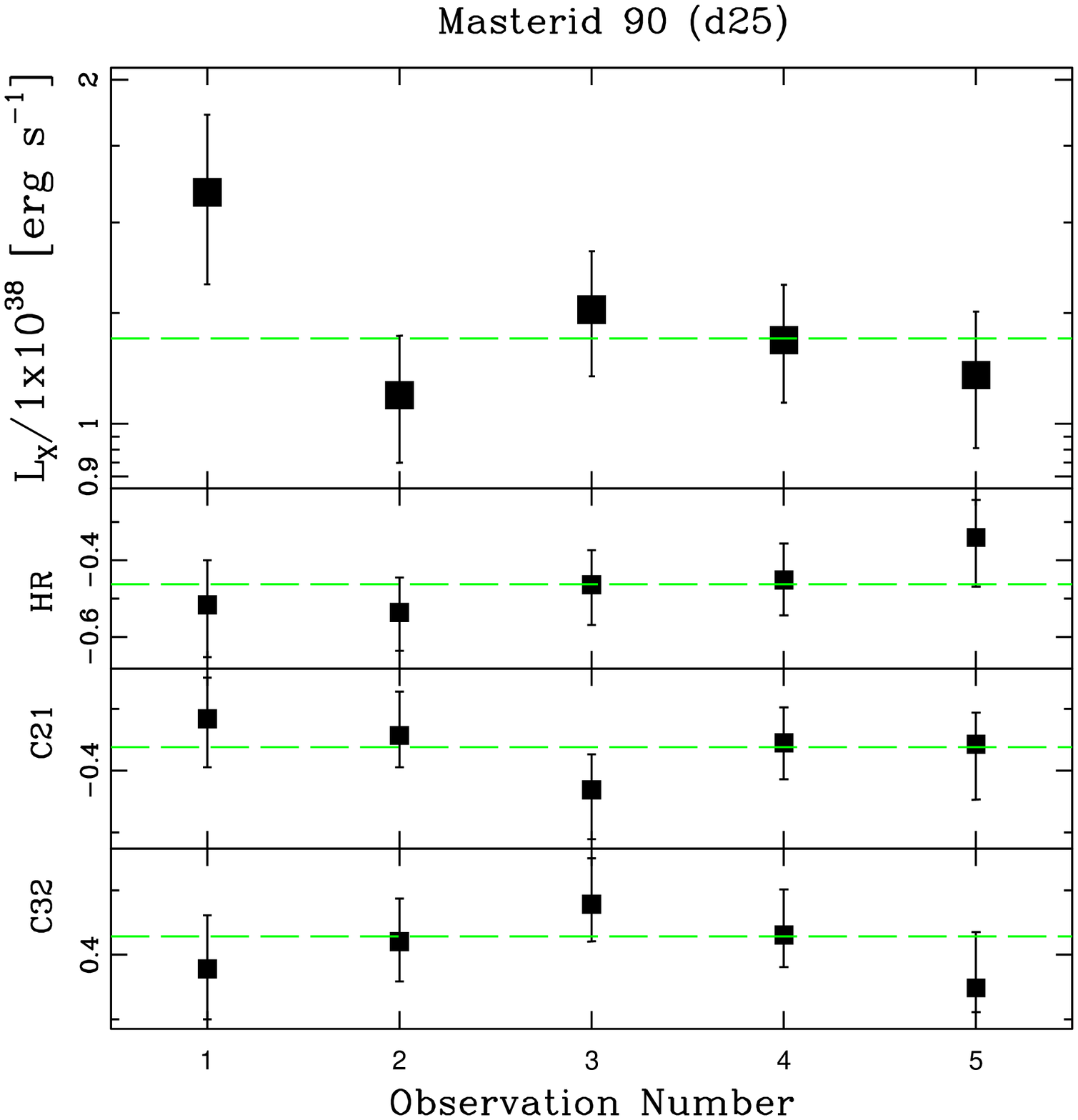}

\end{minipage}\hspace{0.02\linewidth}

  \begin{minipage}{0.485\linewidth}
  \centering
  
    \includegraphics[width=\linewidth]{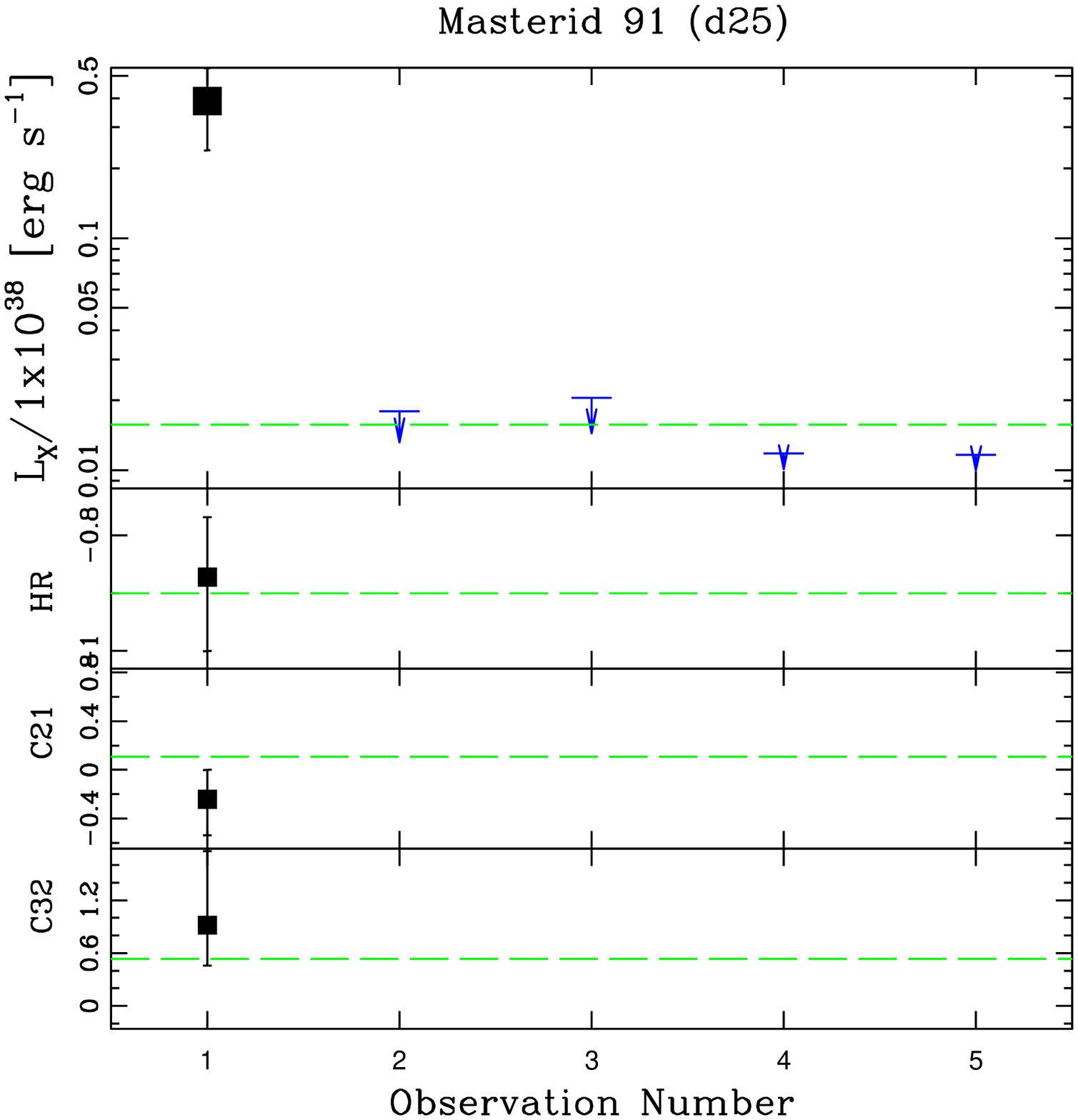}

  \end{minipage}\hspace{0.02\linewidth}
  \begin{minipage}{0.485\linewidth}
  \centering

    \includegraphics[width=\linewidth]{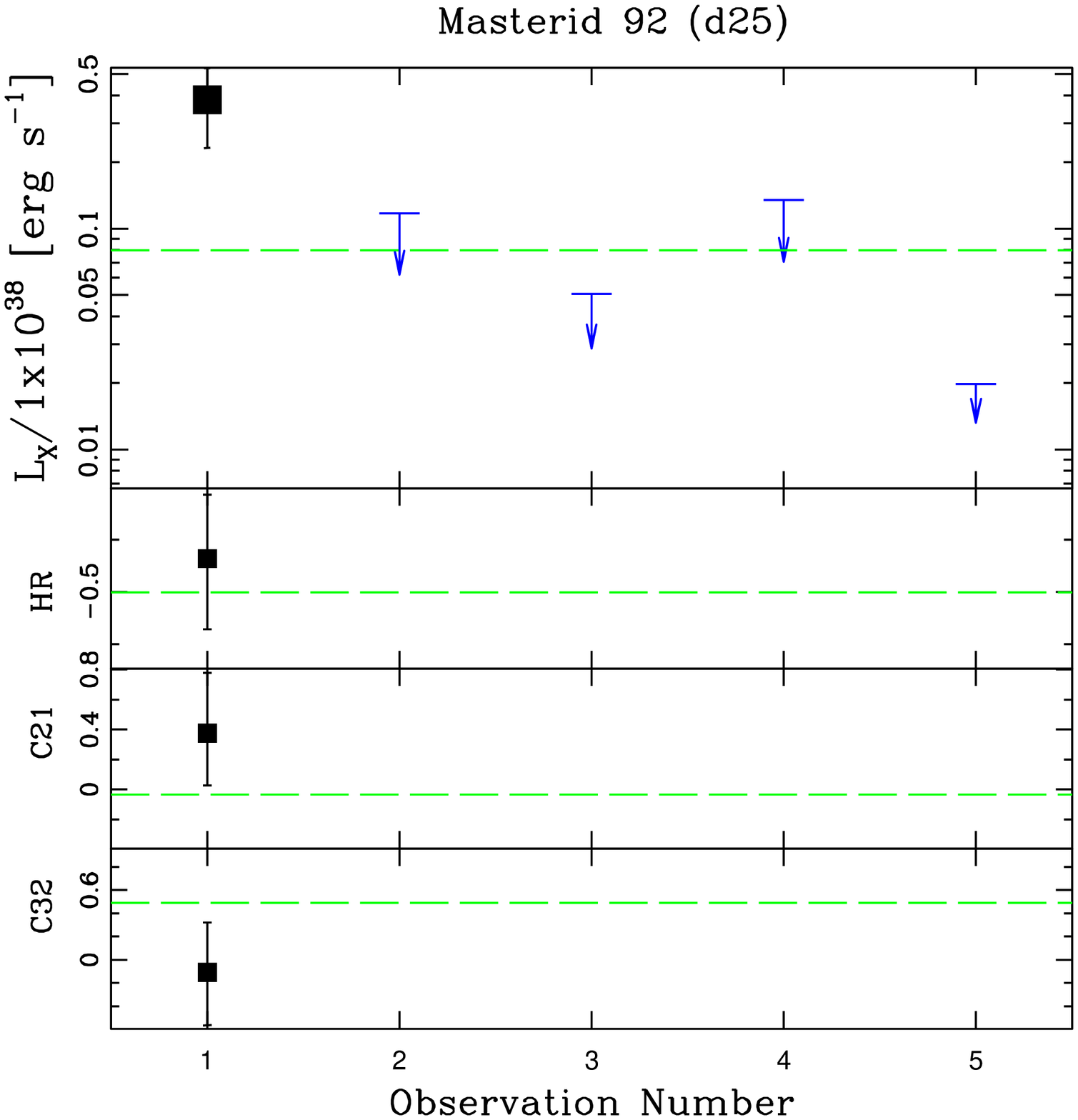}

\end{minipage}\hspace{0.02\linewidth}

\begin{minipage}{0.485\linewidth}
  \centering

    \includegraphics[width=\linewidth]{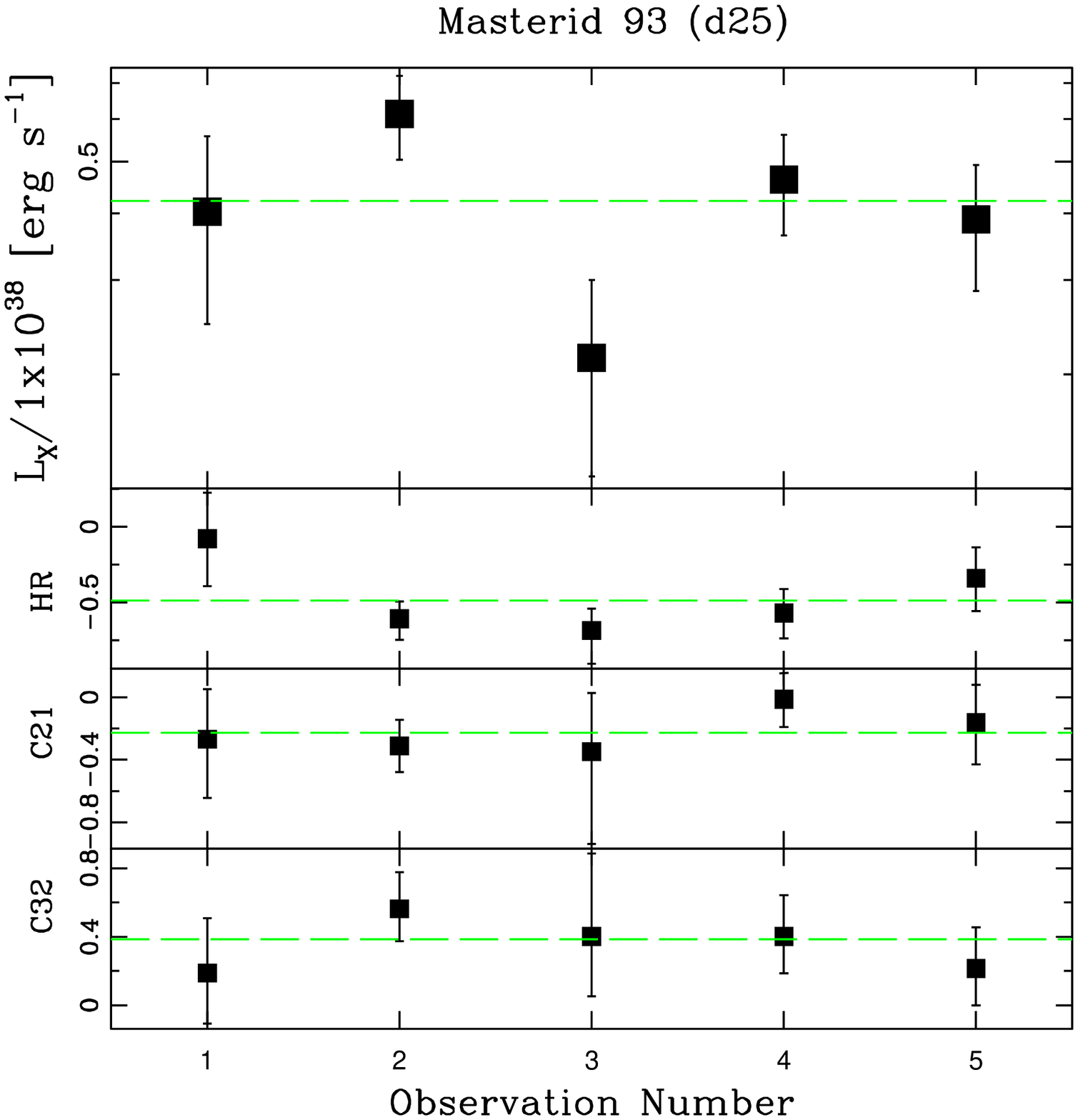}

 \end{minipage}\hspace{0.02\linewidth}
\begin{minipage}{0.485\linewidth}
  \centering
  
    \includegraphics[width=\linewidth]{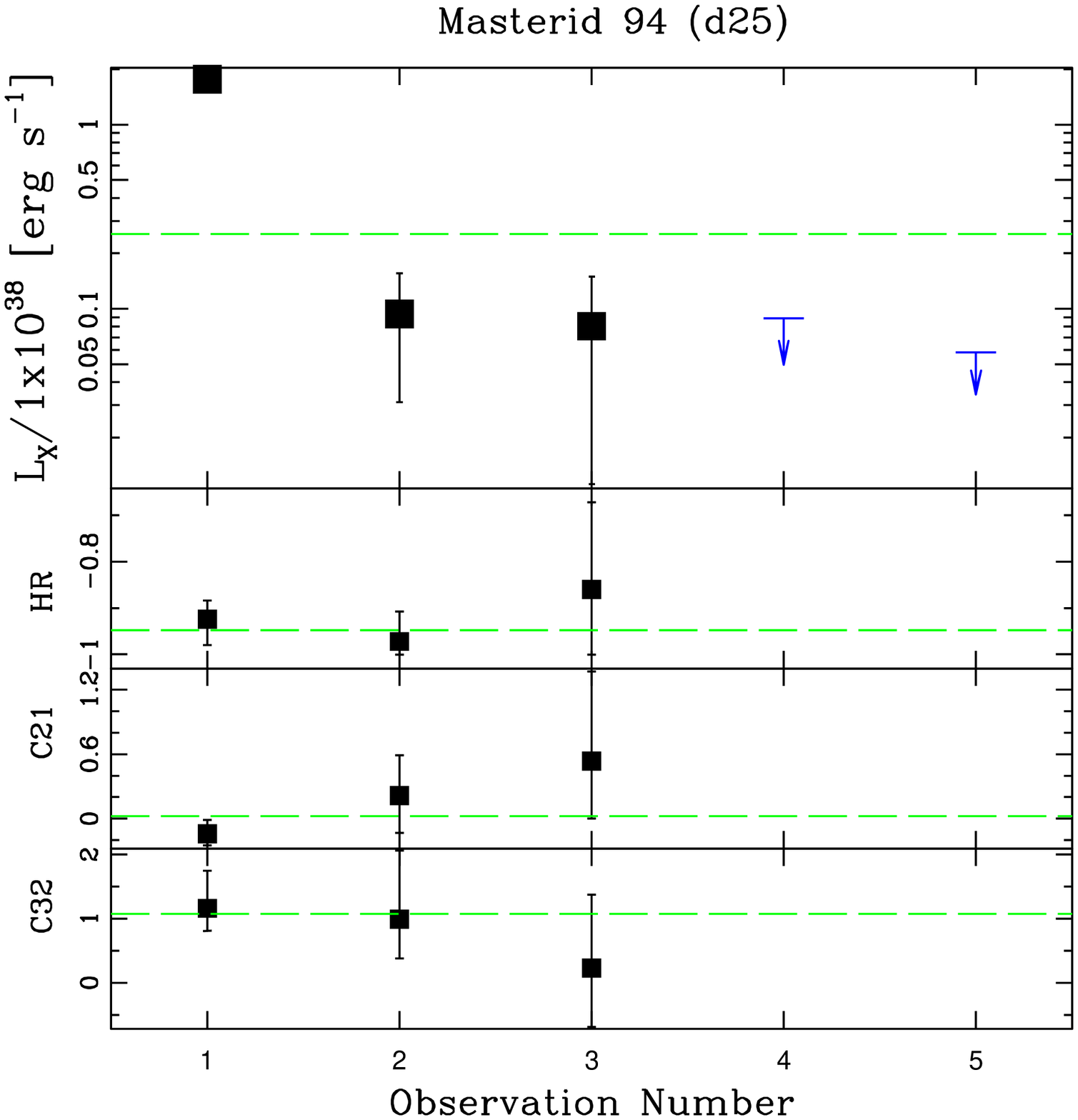}

  \end{minipage}\hspace{0.02\linewidth}
\end{figure}

\begin{figure}
  \begin{minipage}{0.485\linewidth}
  \centering

    \includegraphics[width=\linewidth]{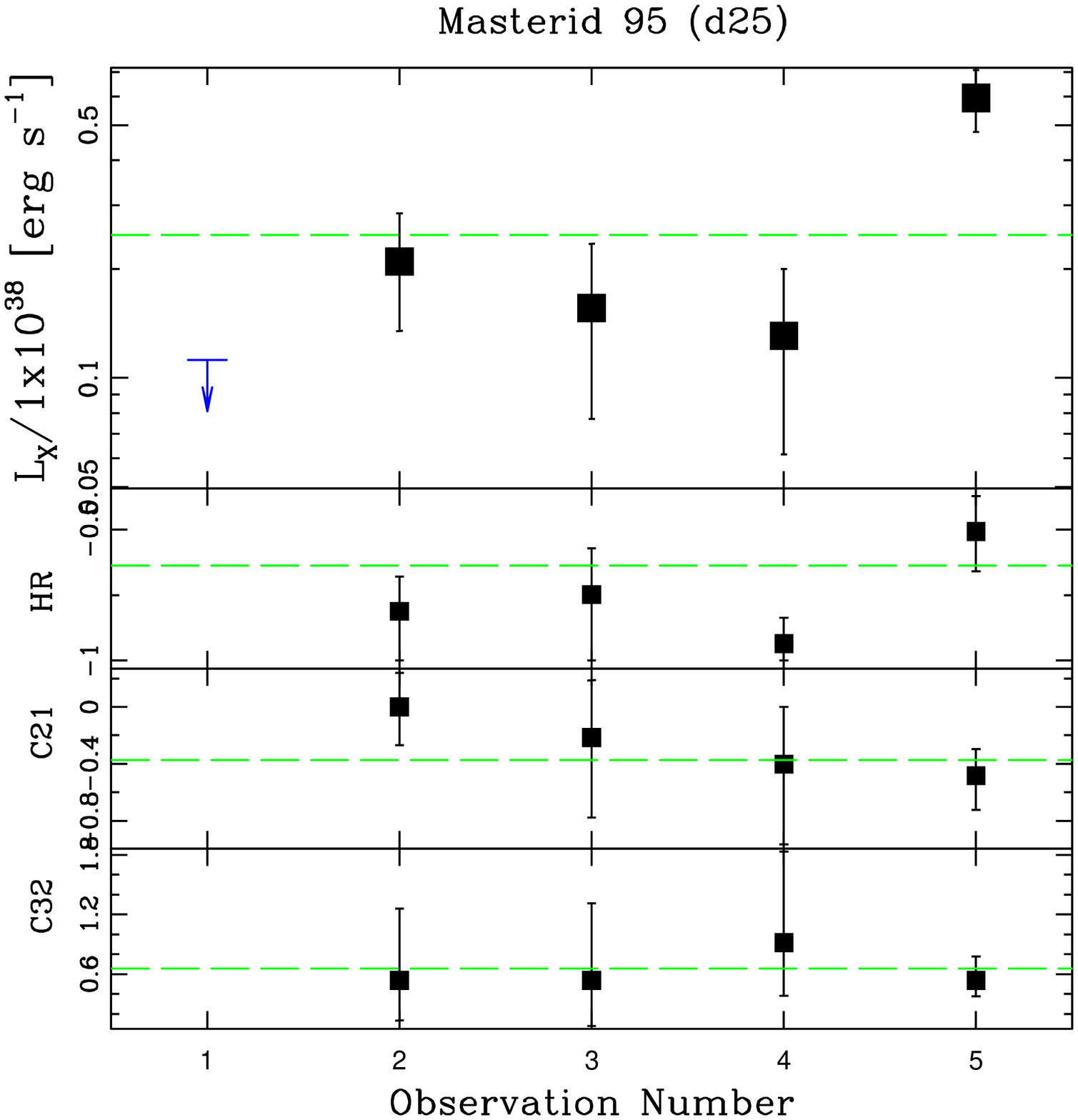}

\end{minipage}\hspace{0.02\linewidth}
\begin{minipage}{0.485\linewidth}
  \centering

    \includegraphics[width=\linewidth]{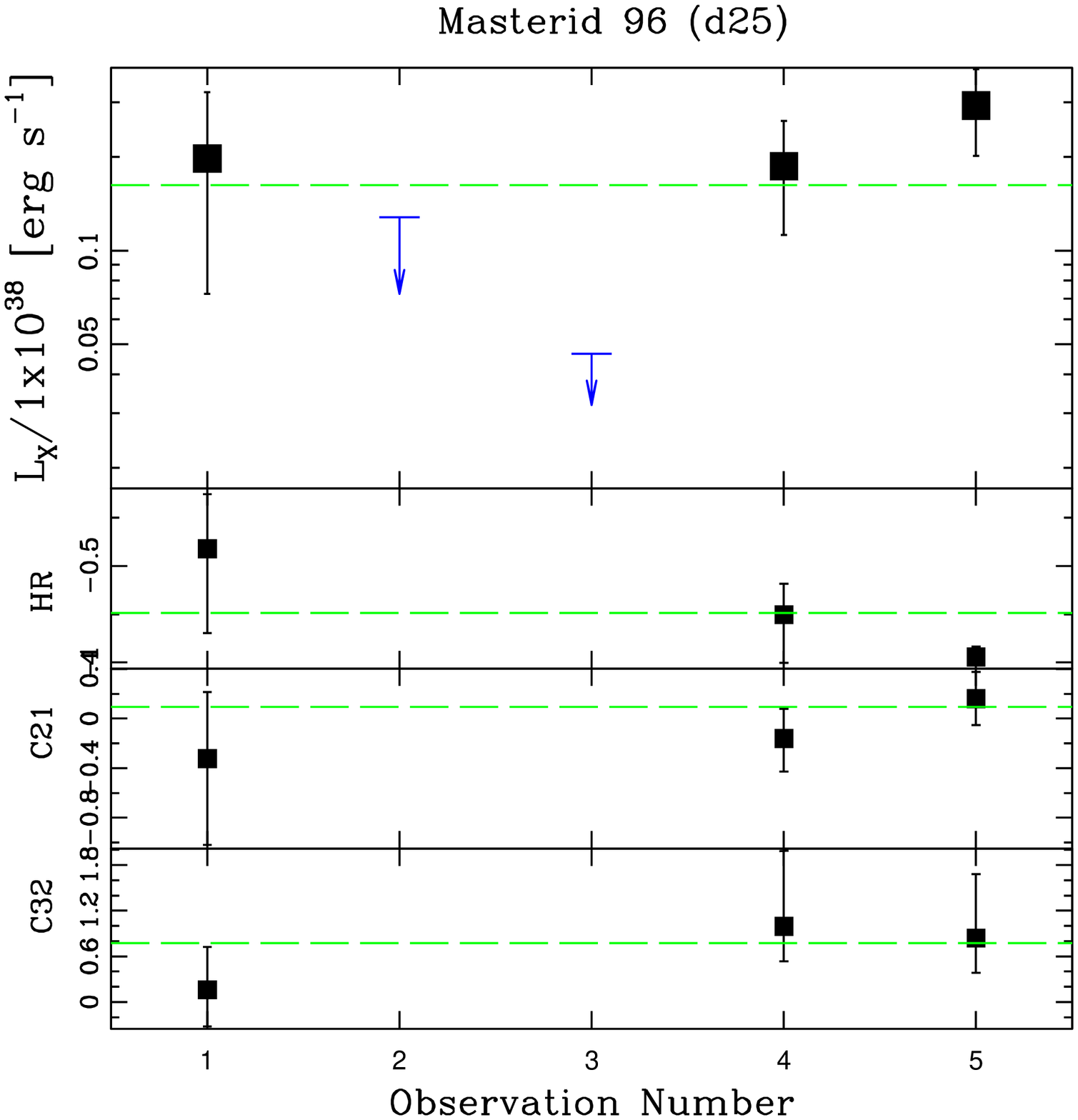}

 \end{minipage}\hspace{0.02\linewidth}

  \begin{minipage}{0.485\linewidth}
  \centering
  
    \includegraphics[width=\linewidth]{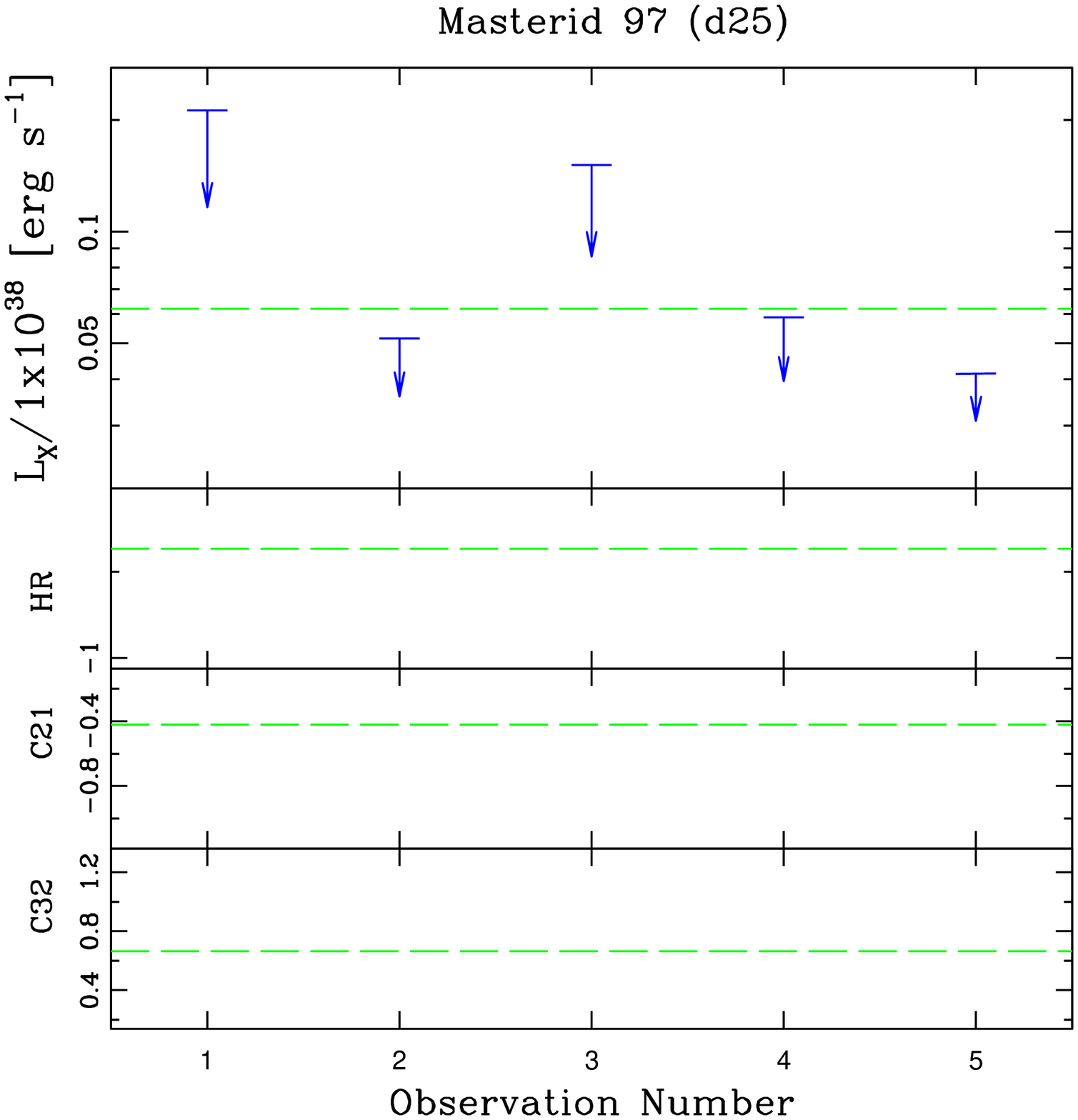}

  \end{minipage}\hspace{0.02\linewidth}
  \begin{minipage}{0.485\linewidth}
  \centering

    \includegraphics[width=\linewidth]{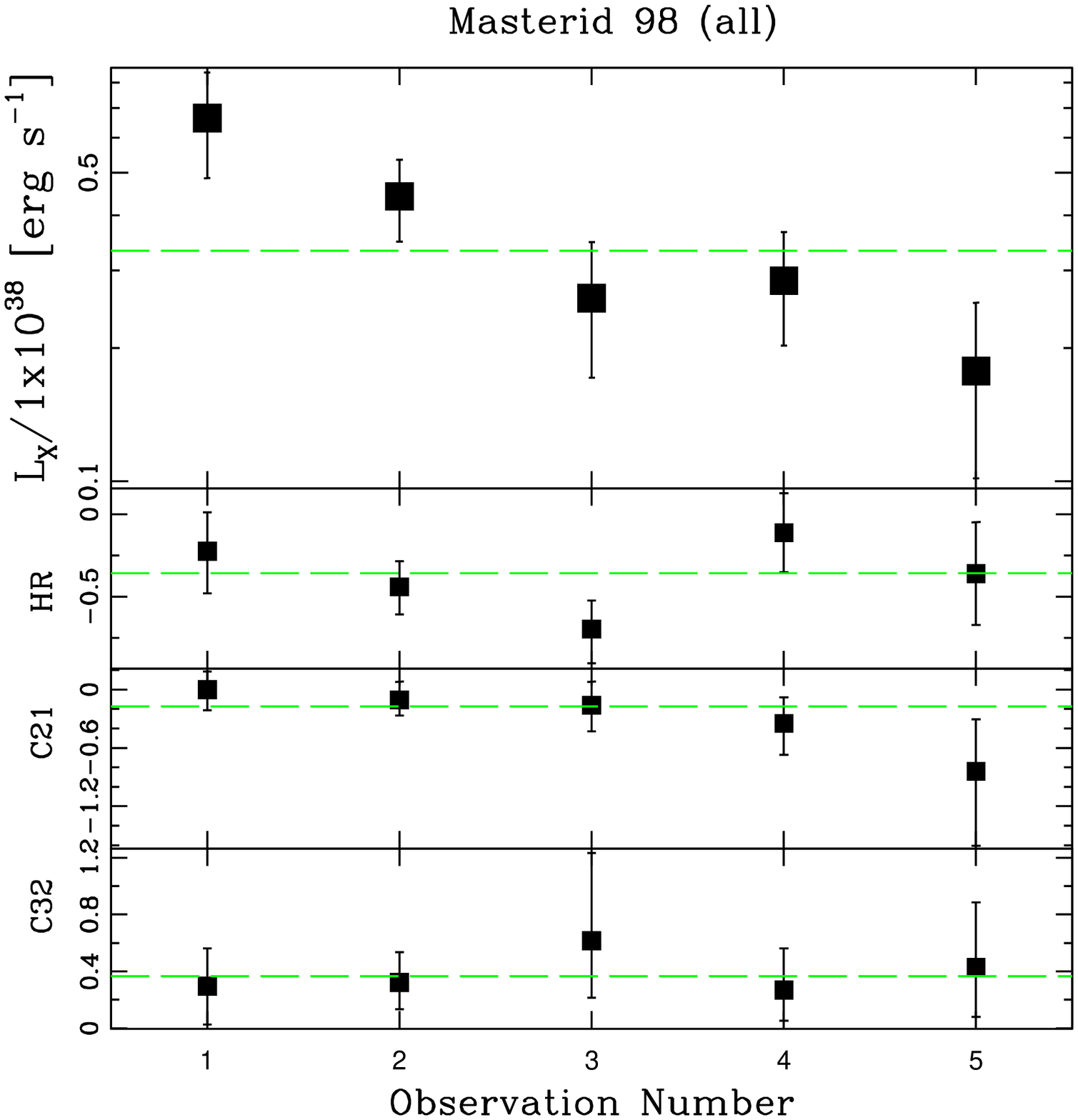}

\end{minipage}\hspace{0.02\linewidth}

\begin{minipage}{0.485\linewidth}
  \centering

    \includegraphics[width=\linewidth]{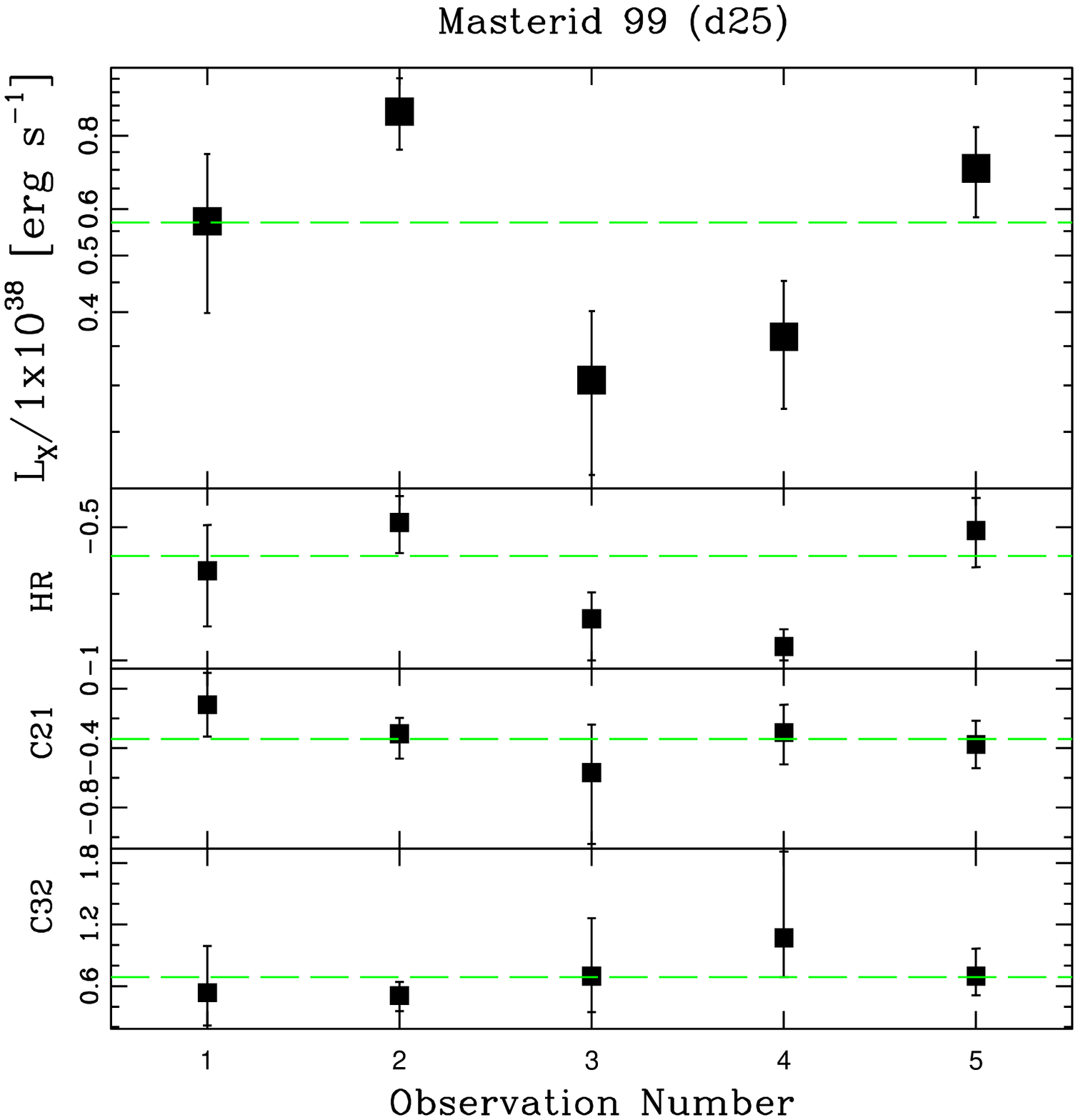}

\end{minipage}\hspace{0.02\linewidth}
\begin{minipage}{0.485\linewidth}
  \centering
  
    \includegraphics[width=\linewidth]{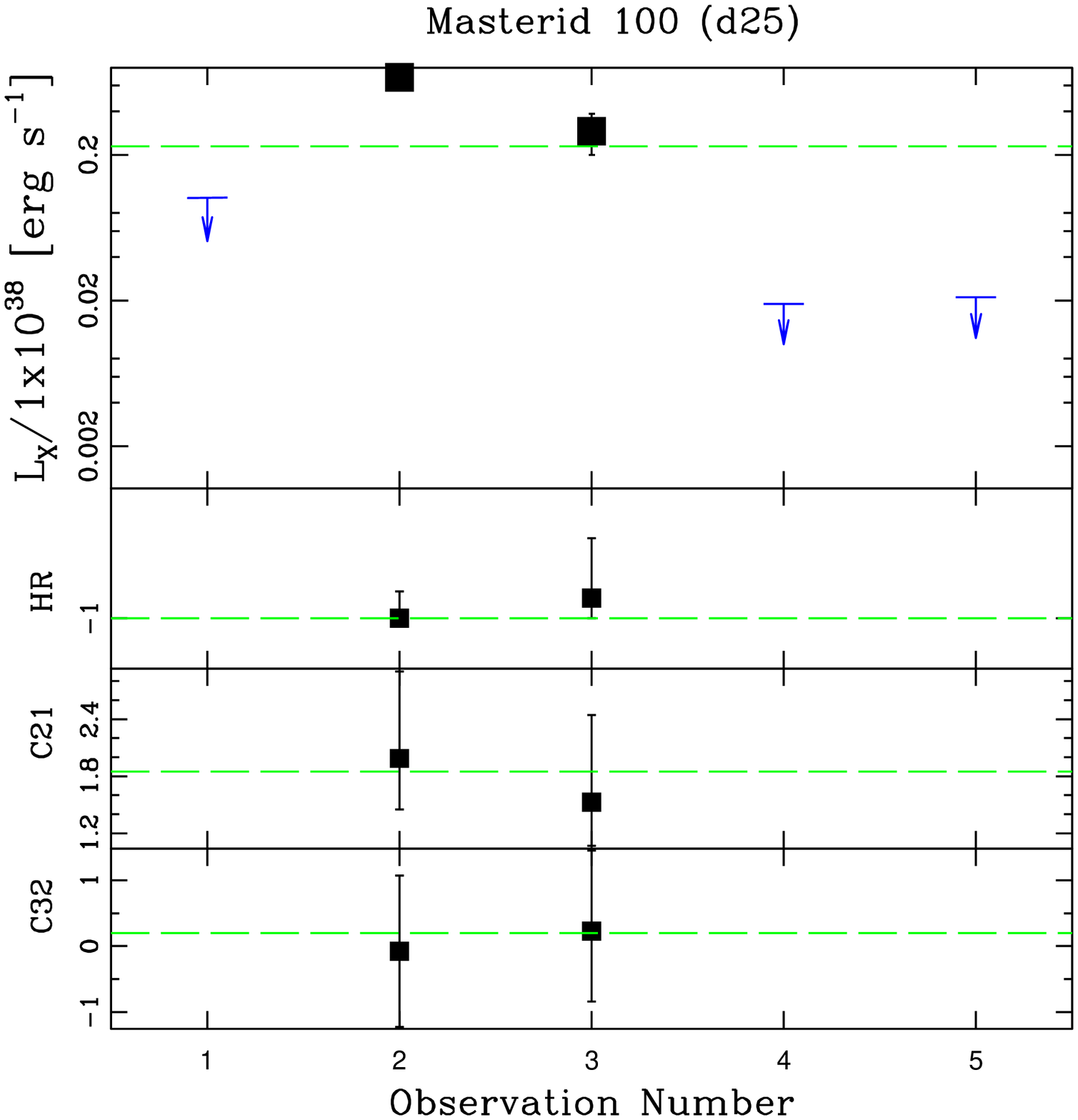}

  \end{minipage}\hspace{0.02\linewidth}
\end{figure}

\clearpage

\begin{figure}
  \begin{minipage}{0.485\linewidth}
  \centering

    \includegraphics[width=\linewidth]{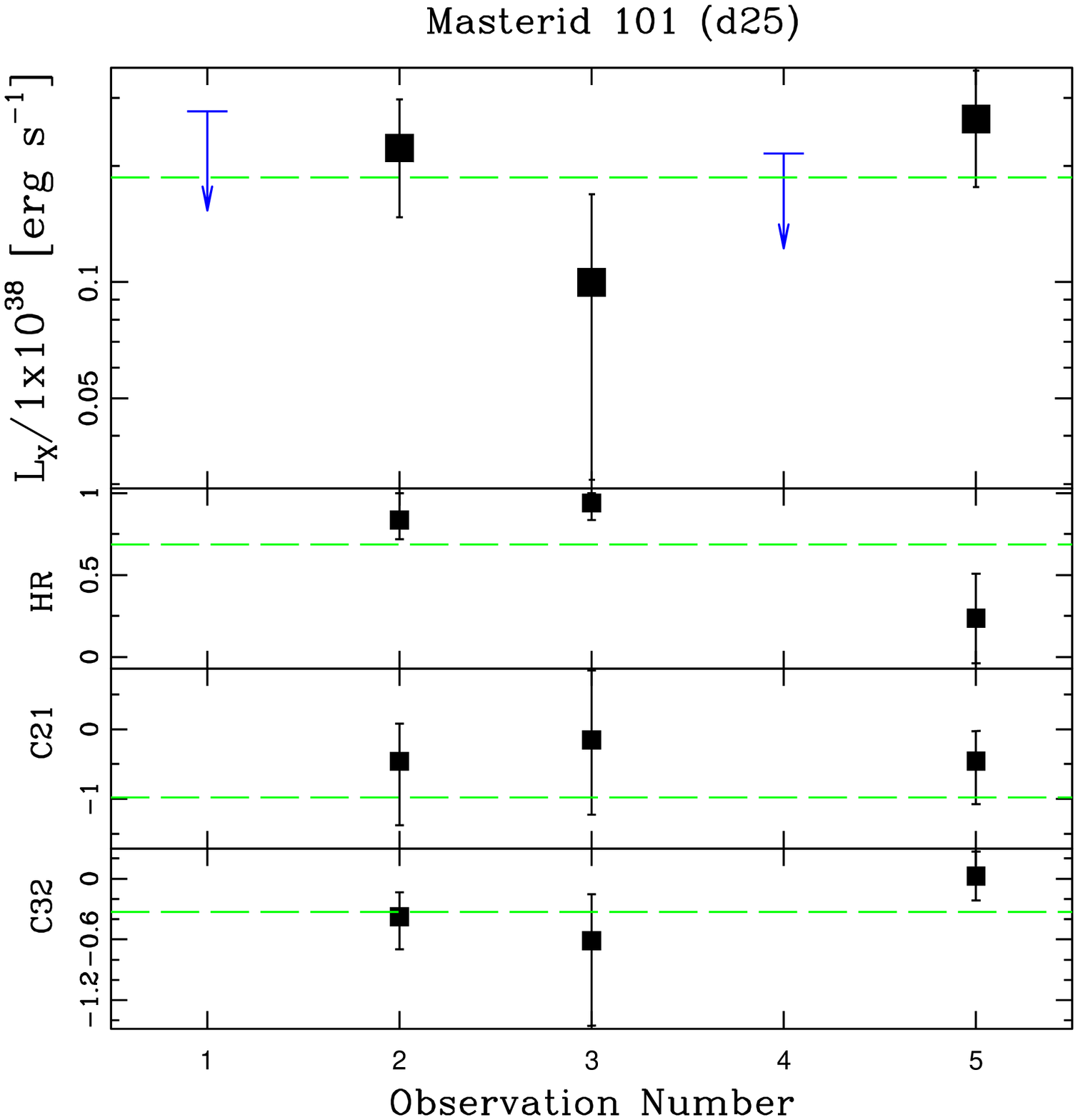}

\end{minipage}\hspace{0.02\linewidth}
\begin{minipage}{0.485\linewidth}
  \centering

    \includegraphics[width=\linewidth]{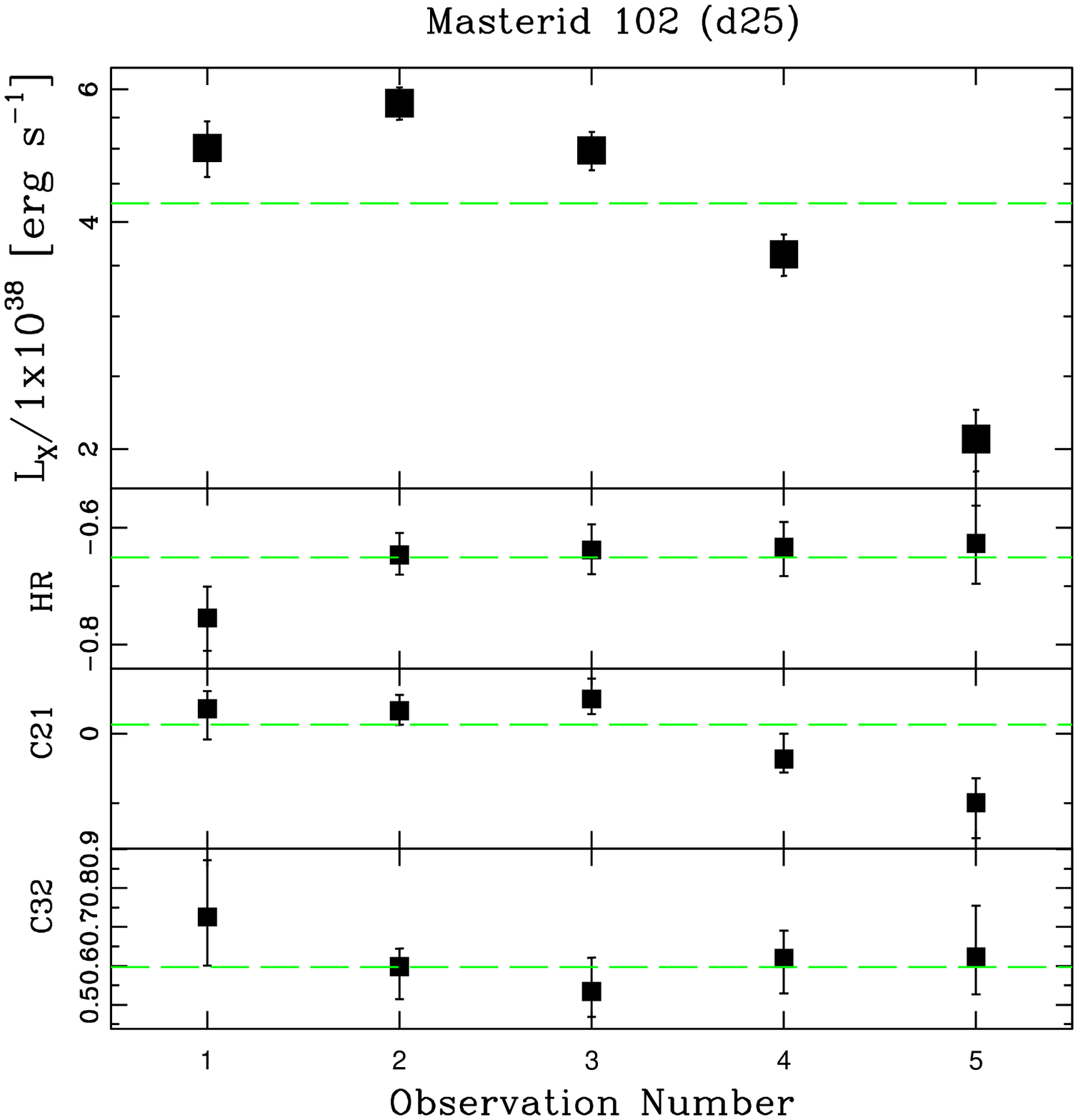}

\end{minipage}\hspace{0.02\linewidth}

  \begin{minipage}{0.485\linewidth}
  \centering
  
    \includegraphics[width=\linewidth]{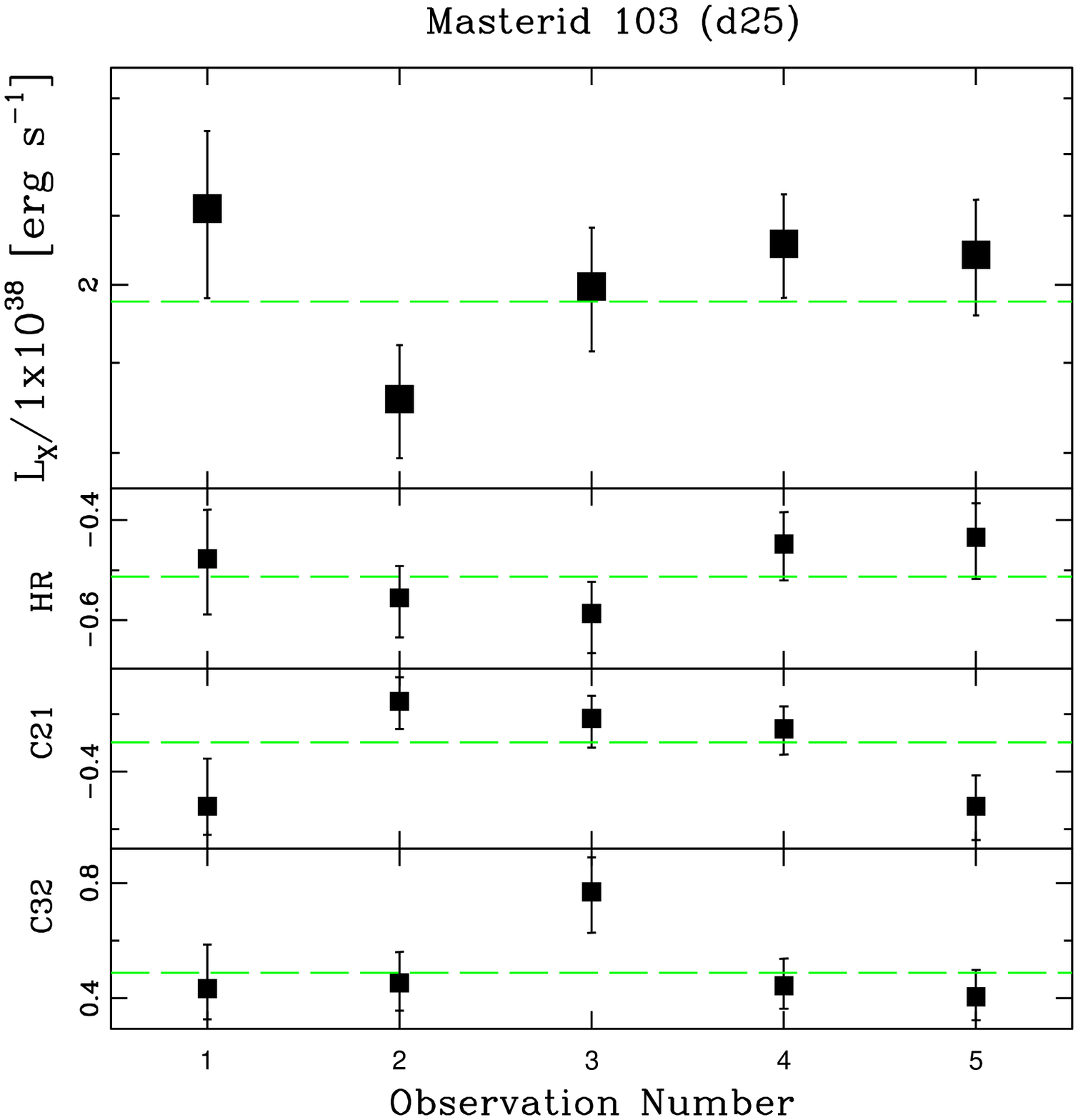}

  \end{minipage}\hspace{0.02\linewidth}
  \begin{minipage}{0.485\linewidth}
  \centering

    \includegraphics[width=\linewidth]{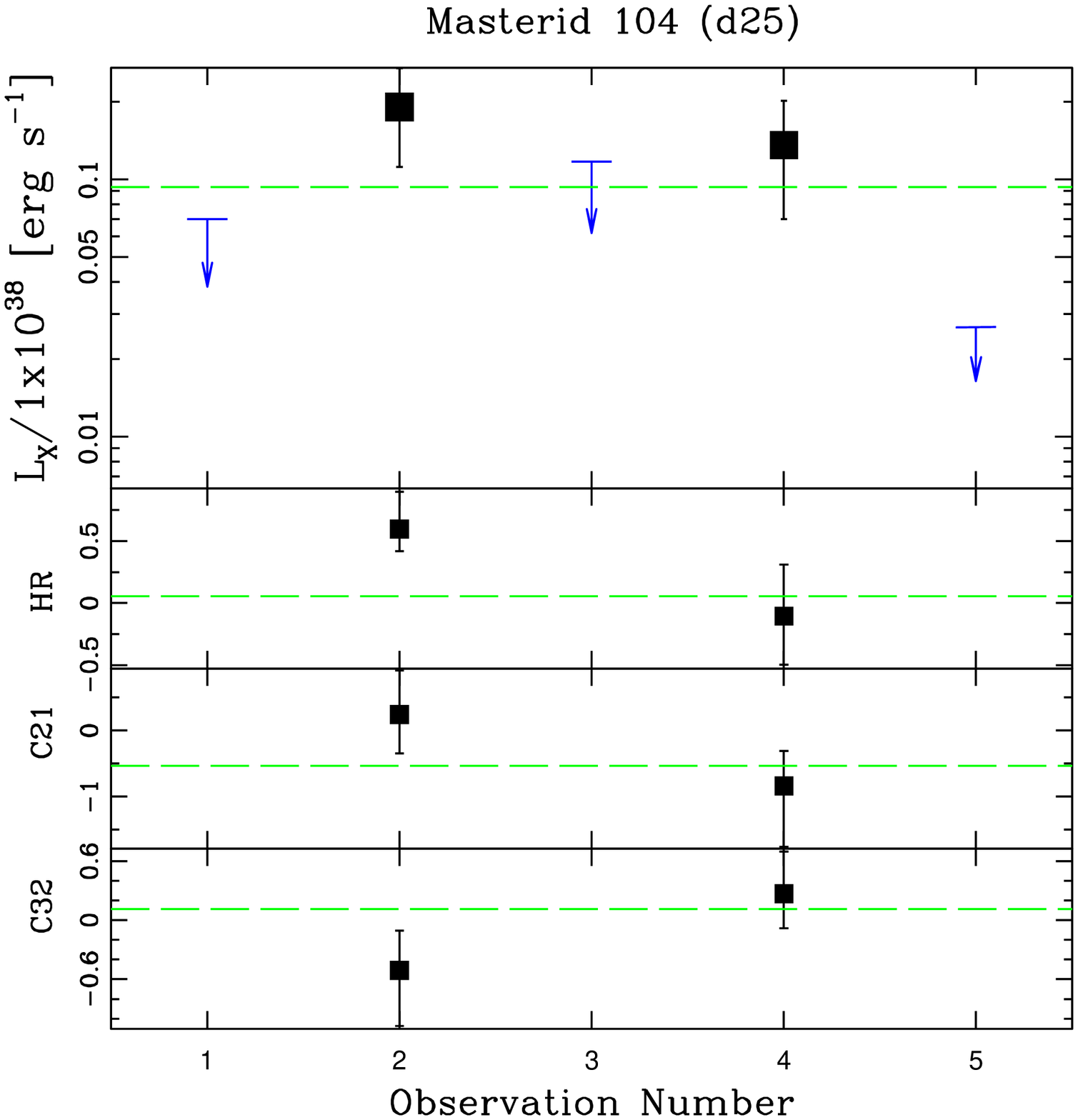}

\end{minipage}\hspace{0.02\linewidth}

\begin{minipage}{0.485\linewidth}
  \centering

    \includegraphics[width=\linewidth]{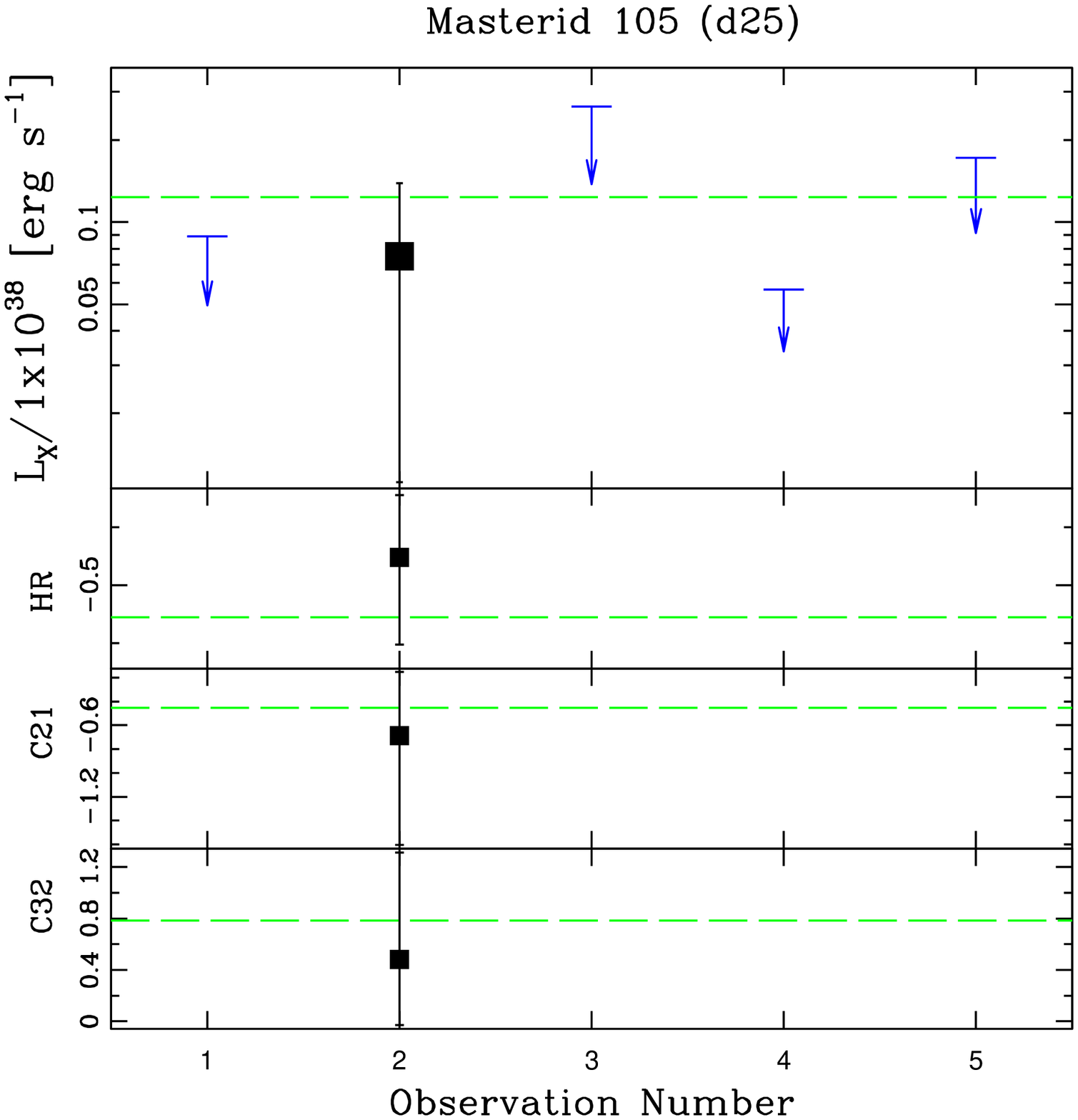}

 \end{minipage}\hspace{0.02\linewidth}
\begin{minipage}{0.485\linewidth}
  \centering
  
    \includegraphics[width=\linewidth]{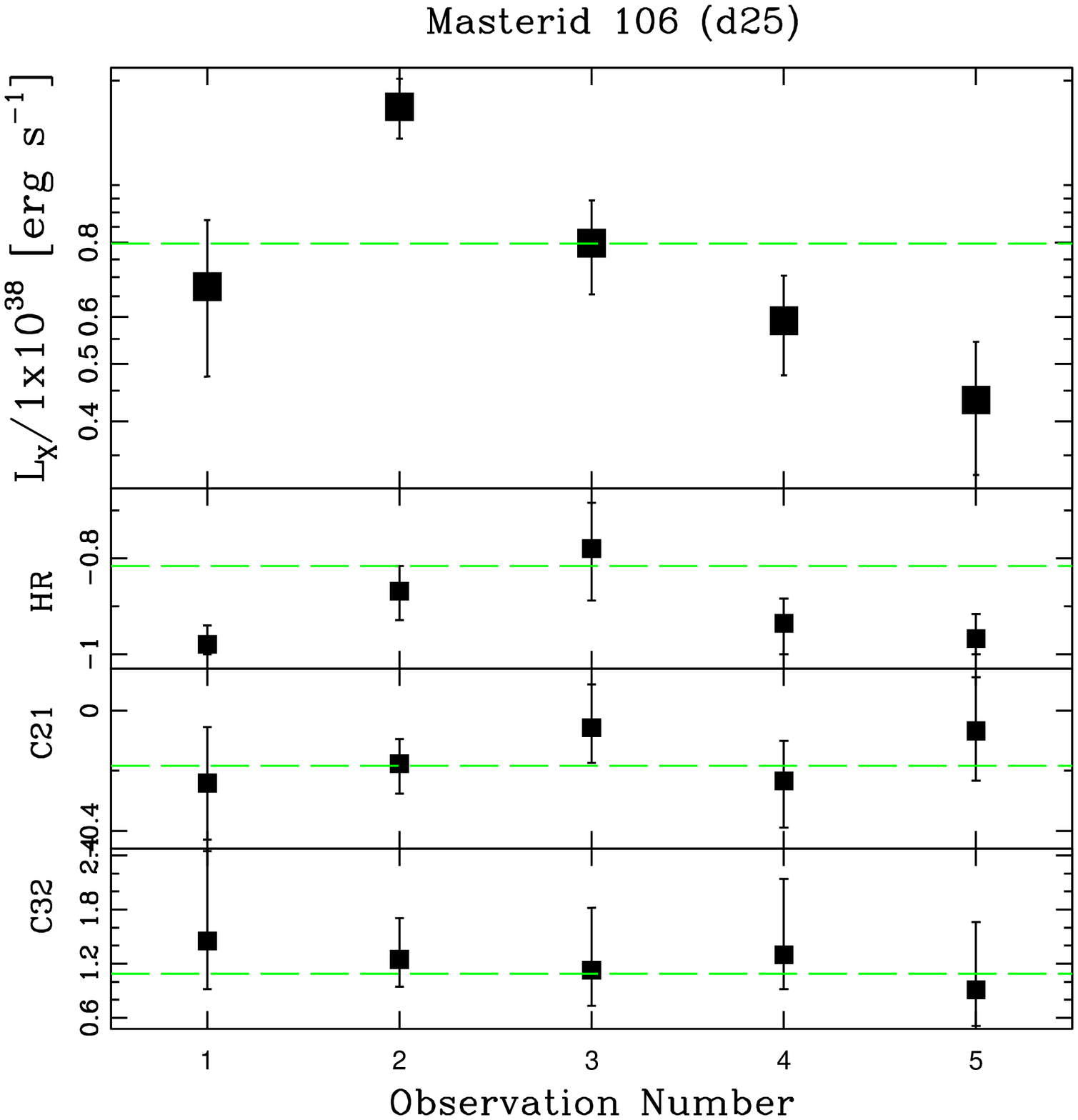}

  \end{minipage}\hspace{0.02\linewidth}
\end{figure}

\begin{figure}
  \begin{minipage}{0.485\linewidth}
  \centering

    \includegraphics[width=\linewidth]{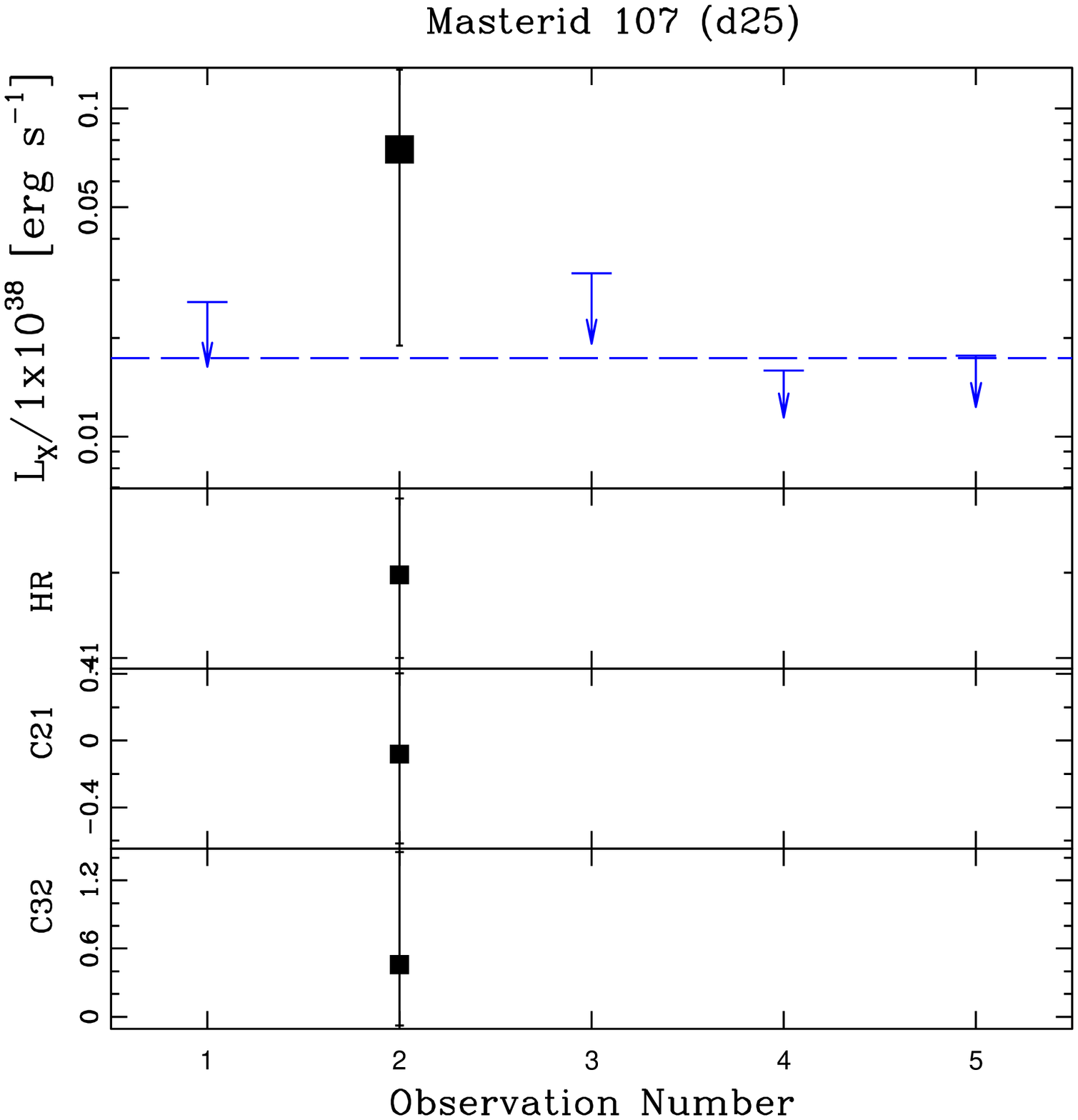}

\end{minipage}\hspace{0.02\linewidth}
\begin{minipage}{0.485\linewidth}
  \centering

    \includegraphics[width=\linewidth]{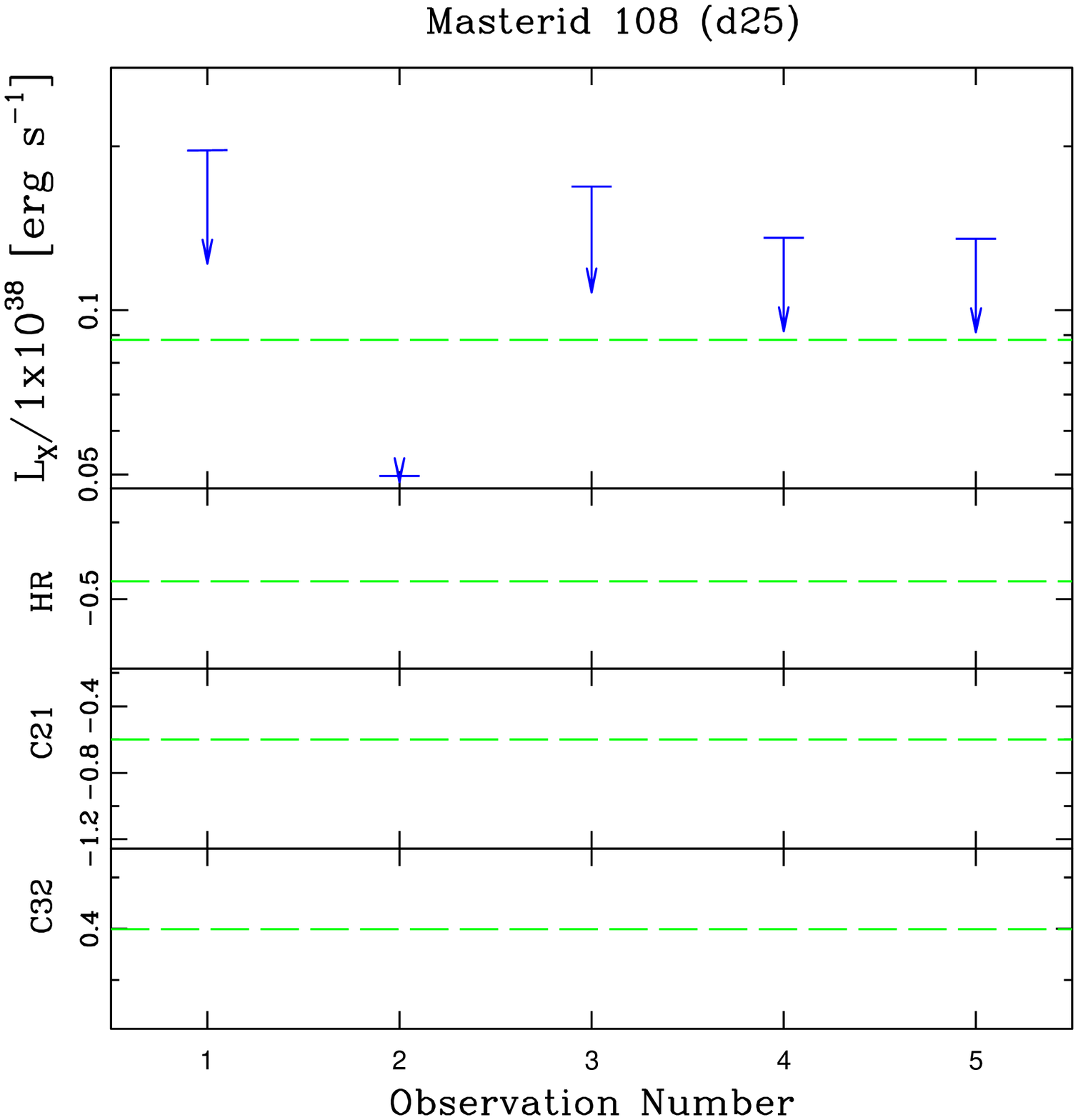}

 \end{minipage}\hspace{0.02\linewidth}

  \begin{minipage}{0.485\linewidth}
  \centering
  
    \includegraphics[width=\linewidth]{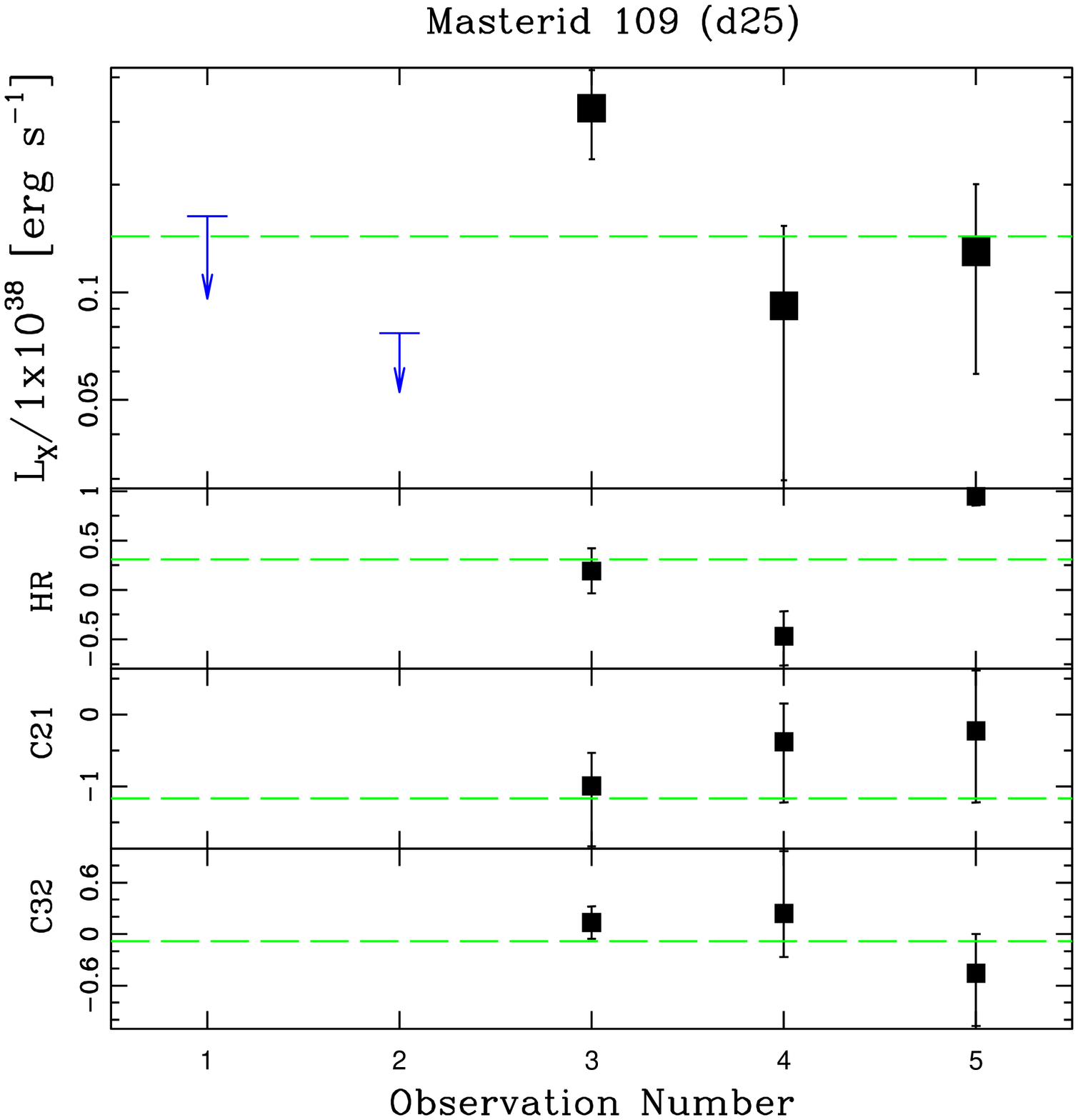}

  \end{minipage}\hspace{0.02\linewidth}
  \begin{minipage}{0.485\linewidth}
  \centering

    \includegraphics[width=\linewidth]{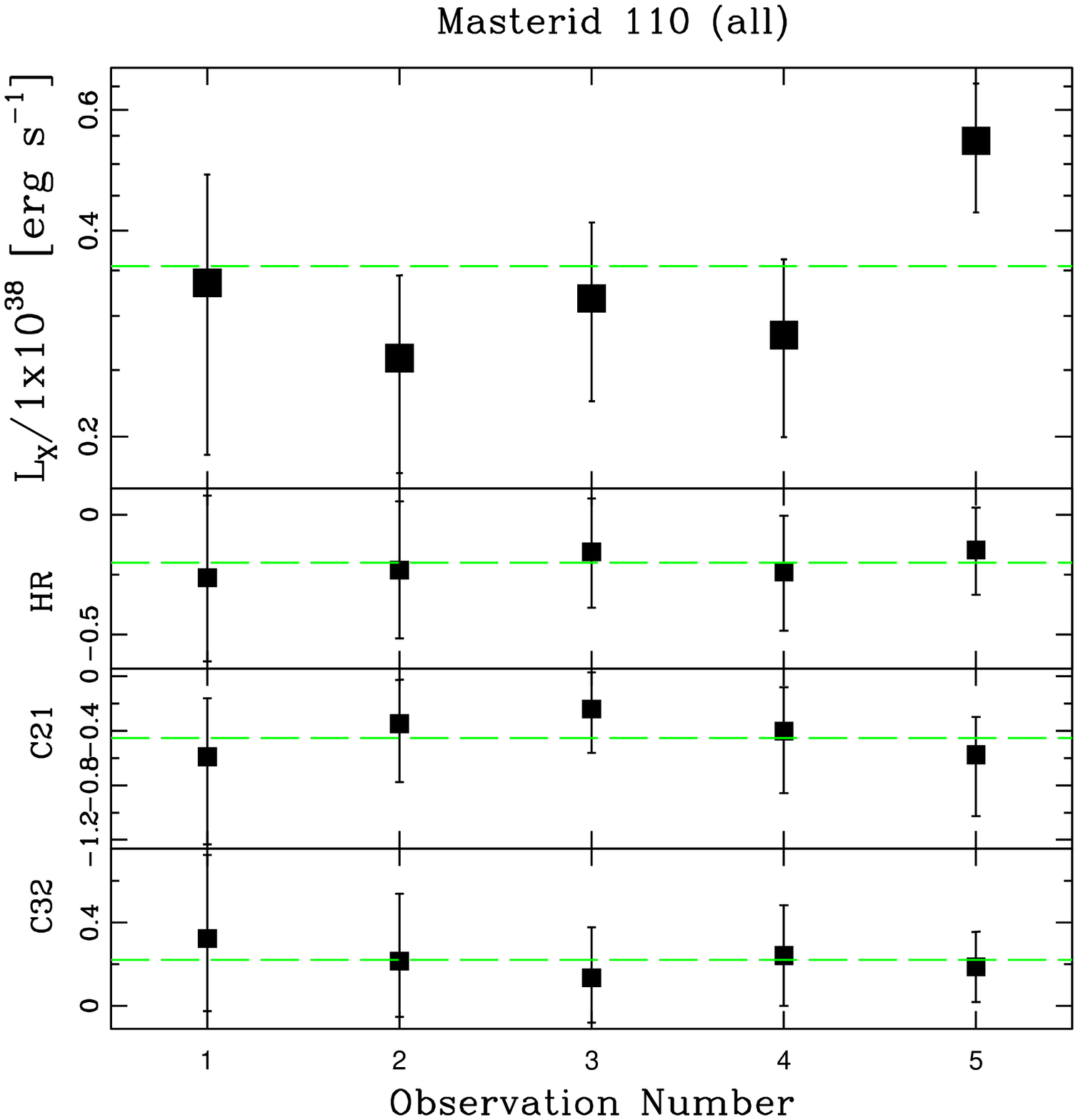}

\end{minipage}\hspace{0.02\linewidth}

\begin{minipage}{0.485\linewidth}
  \centering

    \includegraphics[width=\linewidth]{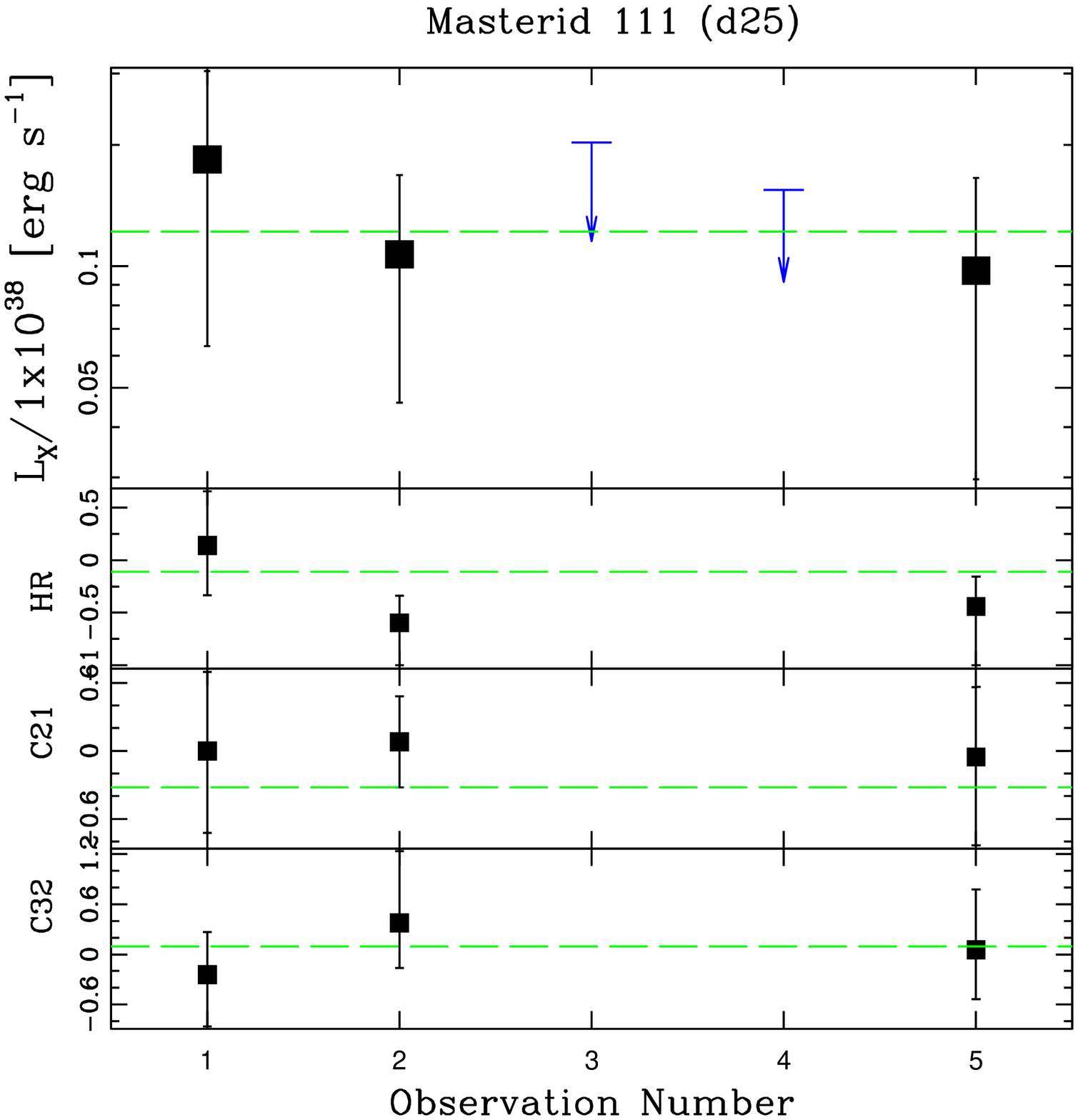}

\end{minipage}\hspace{0.02\linewidth}
\begin{minipage}{0.485\linewidth}
  \centering
  
    \includegraphics[width=\linewidth]{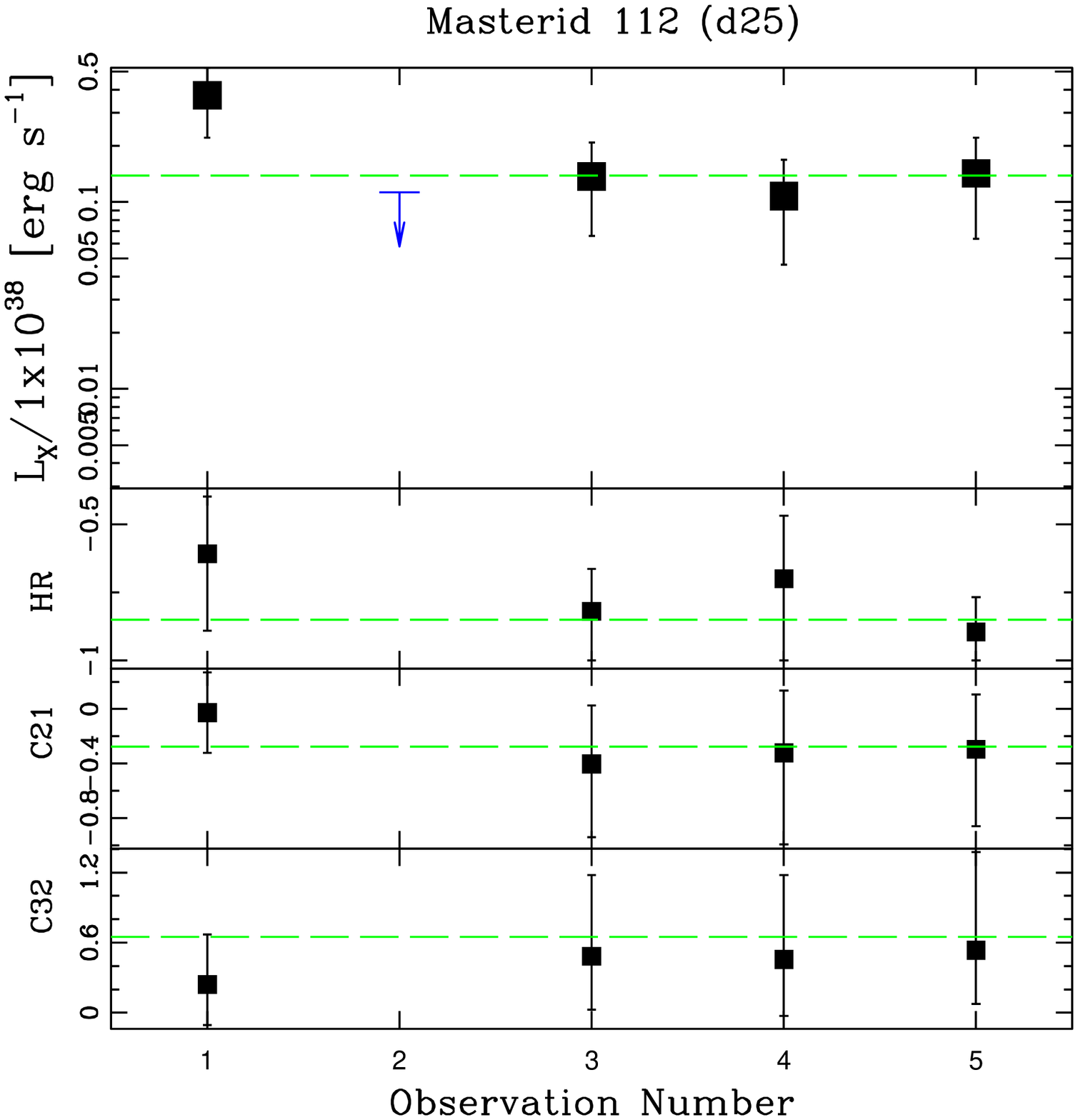}

  \end{minipage}\hspace{0.02\linewidth}

\end{figure}

\begin{figure}
  \begin{minipage}{0.485\linewidth}
  \centering

    \includegraphics[width=\linewidth]{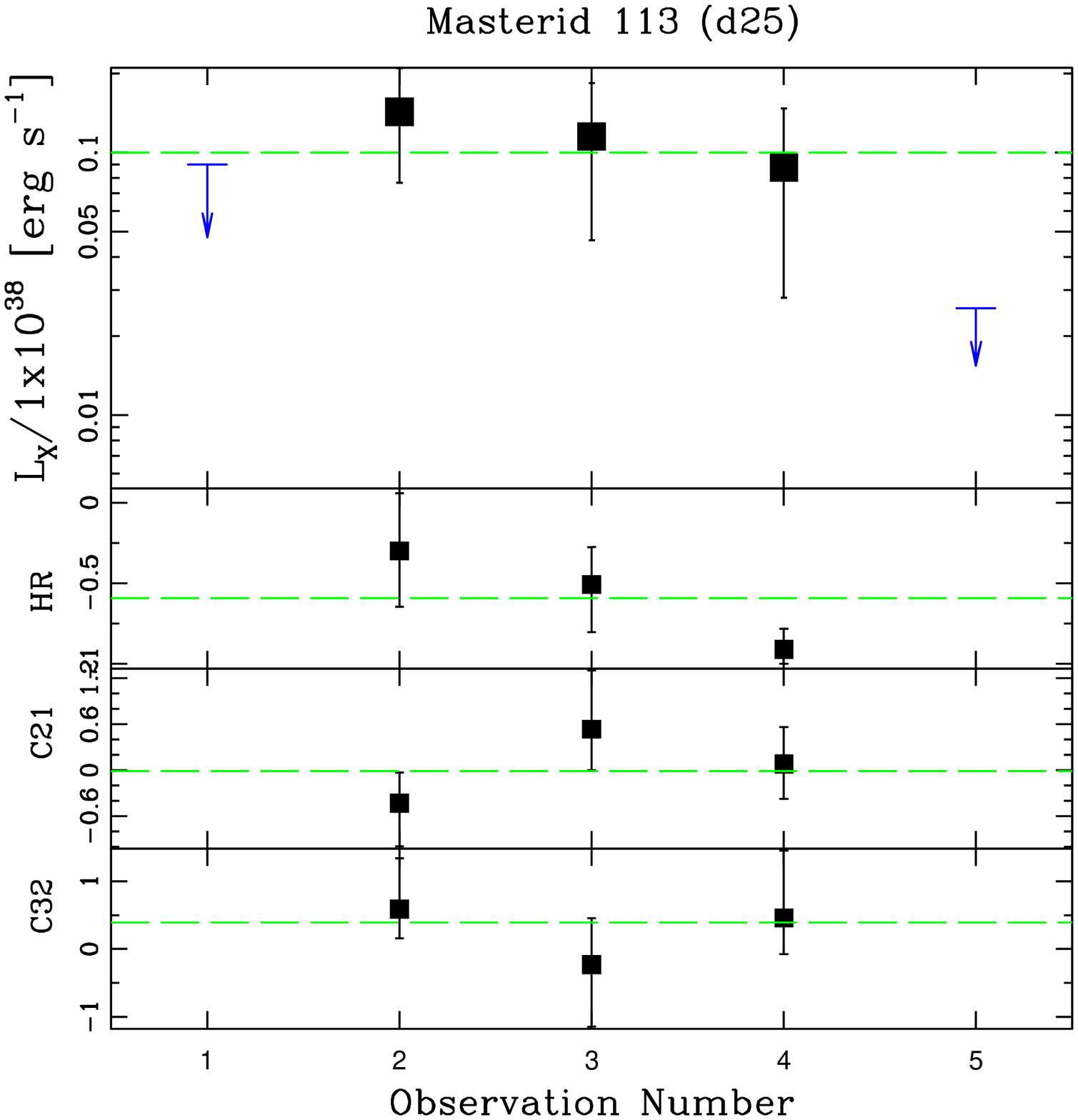}

\end{minipage}\hspace{0.02\linewidth}
\begin{minipage}{0.485\linewidth}
  \centering

    \includegraphics[width=\linewidth]{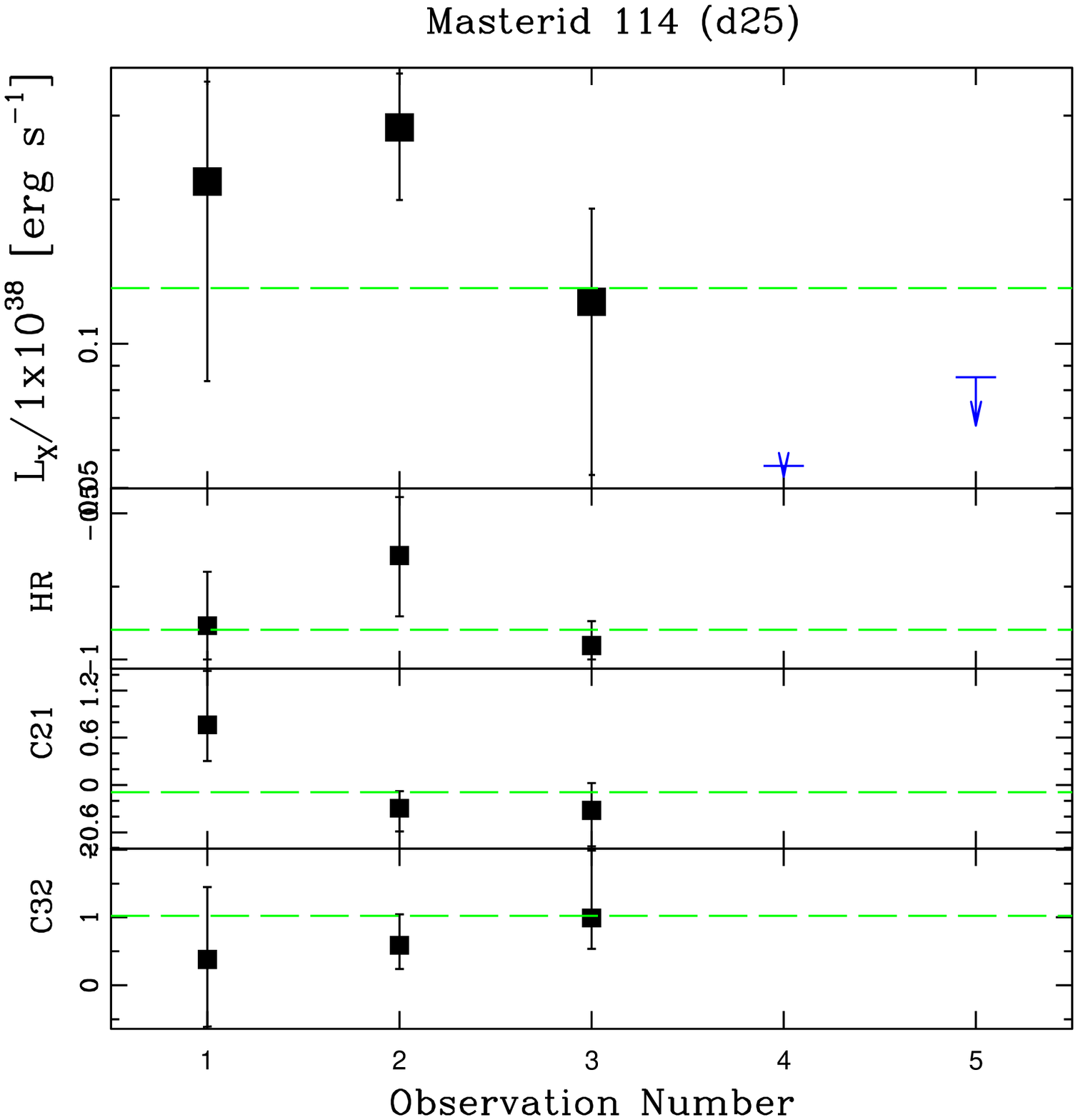}

\end{minipage}\hspace{0.02\linewidth}

  \begin{minipage}{0.485\linewidth}
  \centering
  
    \includegraphics[width=\linewidth]{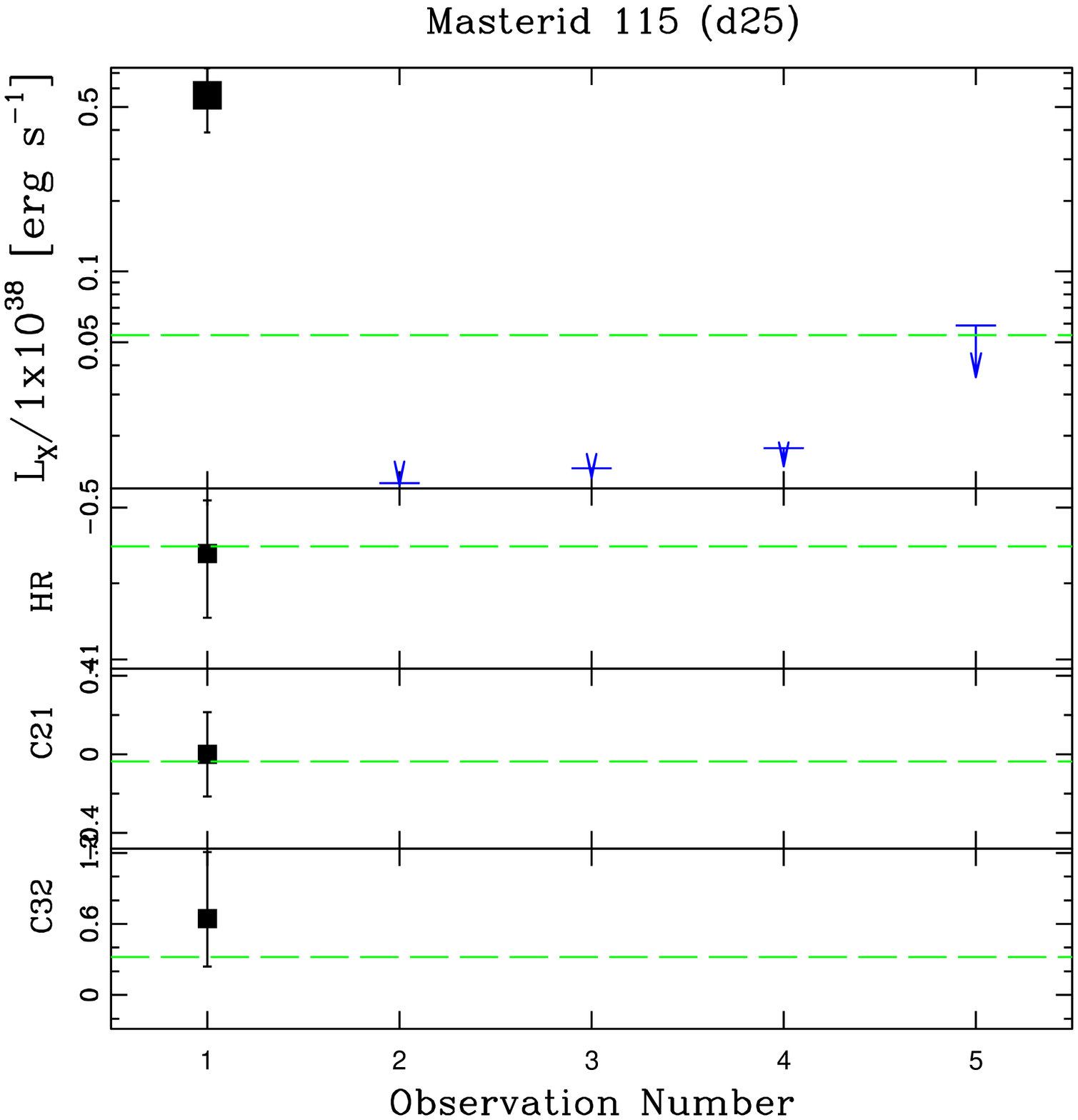}

  \end{minipage}\hspace{0.02\linewidth}
  \begin{minipage}{0.485\linewidth}
  \centering

    \includegraphics[width=\linewidth]{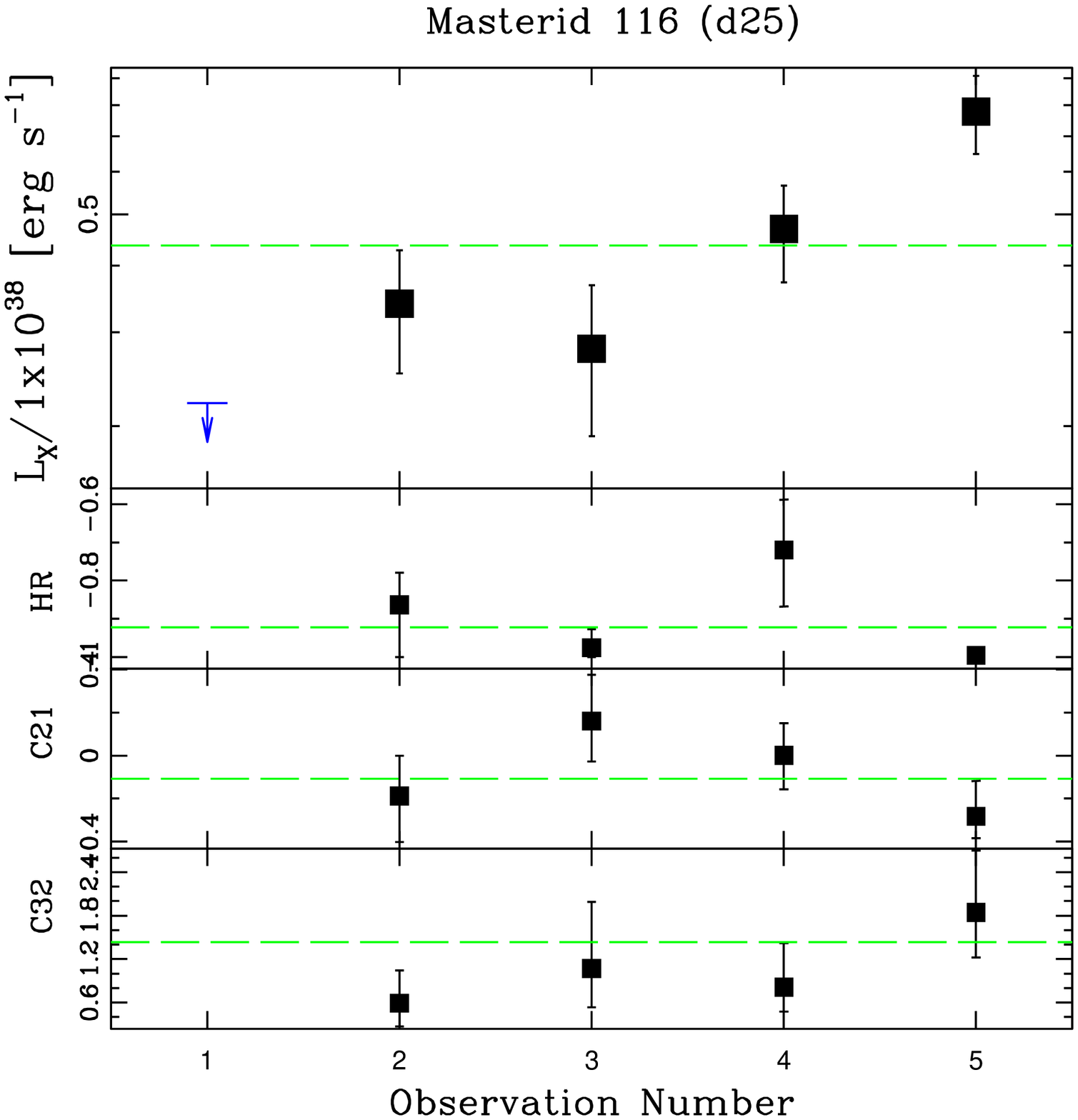}

\end{minipage}

\begin{minipage}{0.485\linewidth}
  \centering

    \includegraphics[width=\linewidth]{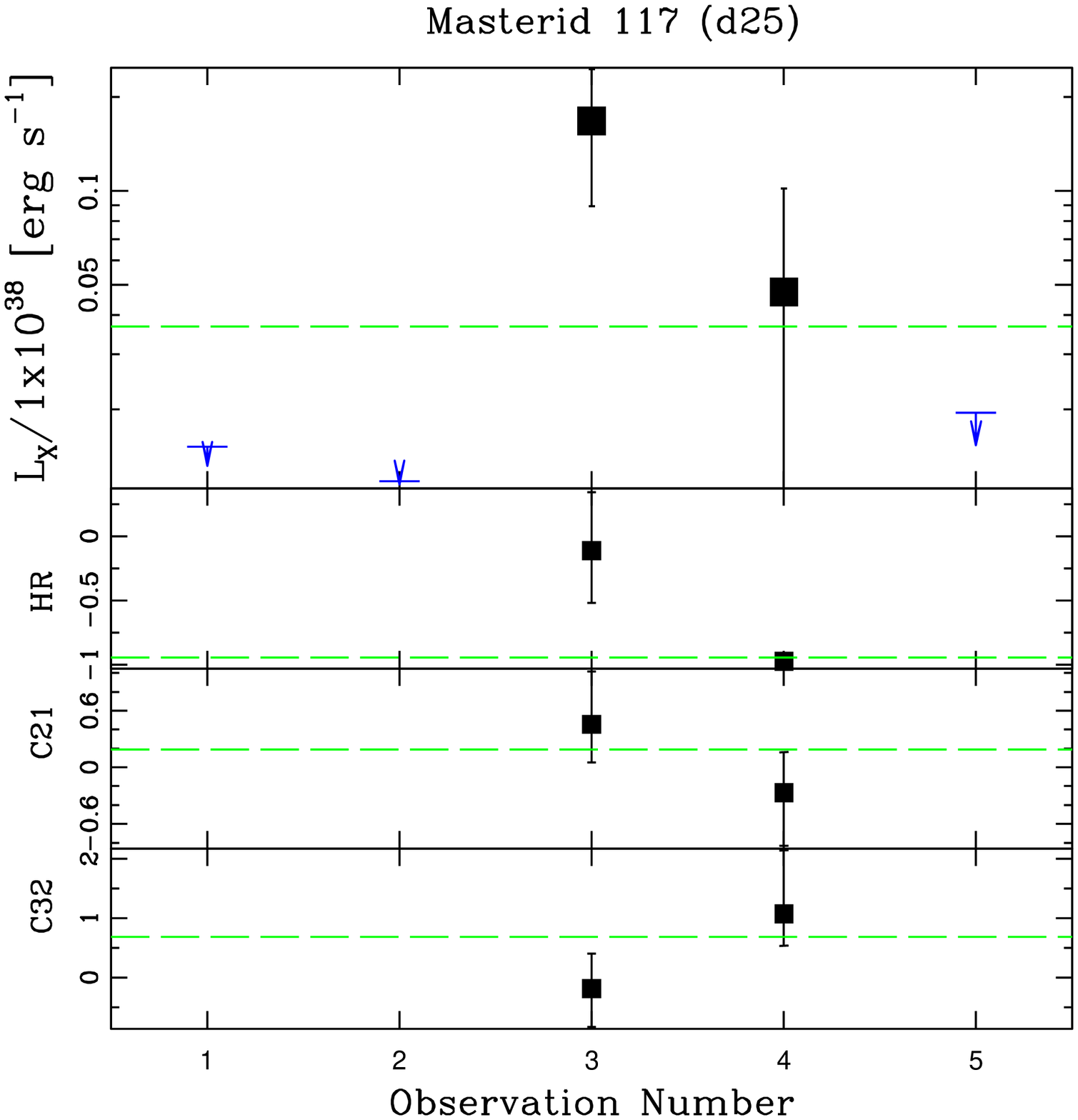}

 \end{minipage}\hspace{0.02\linewidth}
\begin{minipage}{0.485\linewidth}
  \centering
  
    \includegraphics[width=\linewidth]{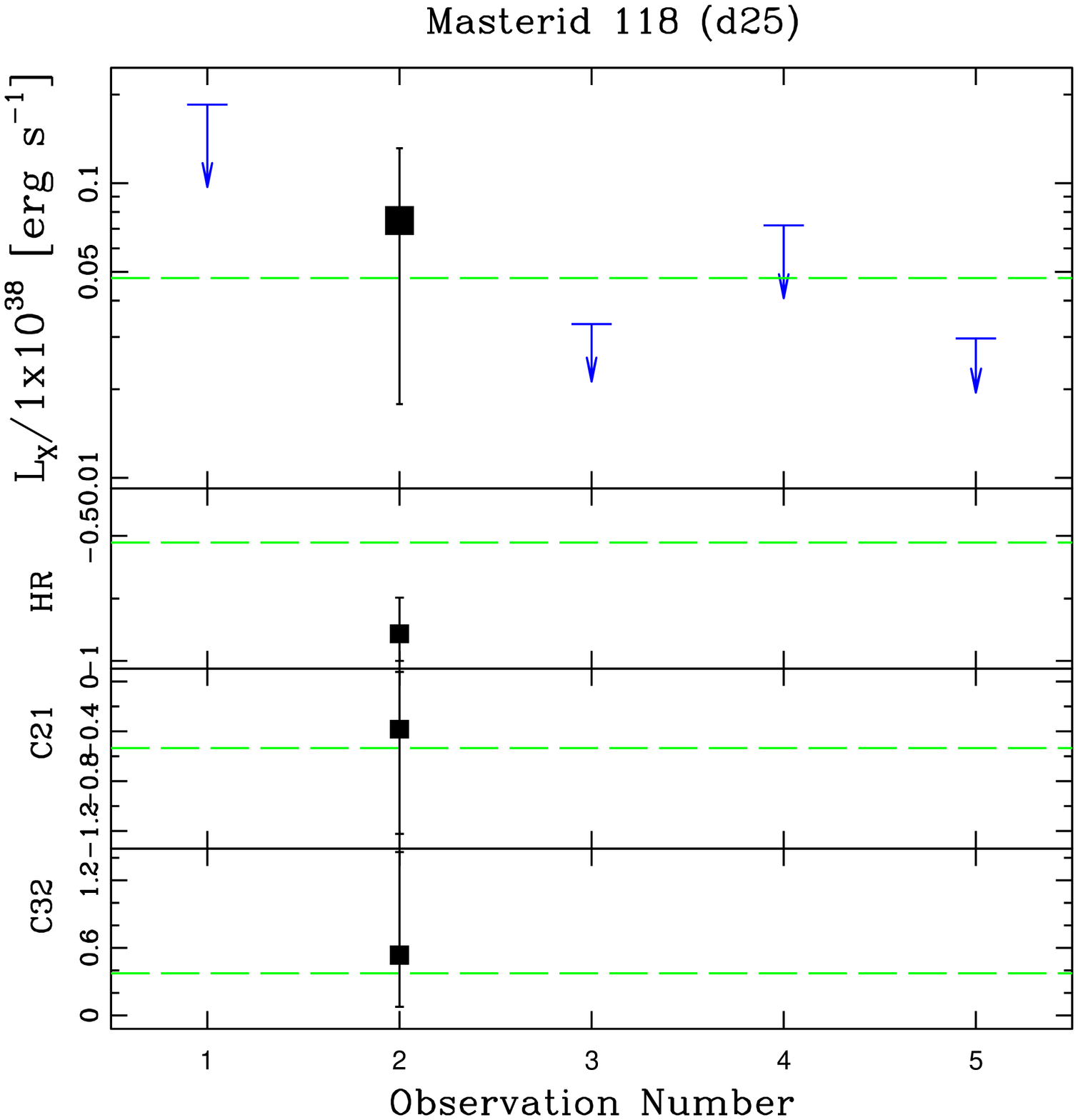}

  \end{minipage}\hspace{0.02\linewidth}
\end{figure}

\clearpage

\begin{figure}
  \begin{minipage}{0.485\linewidth}
  \centering

    \includegraphics[width=\linewidth]{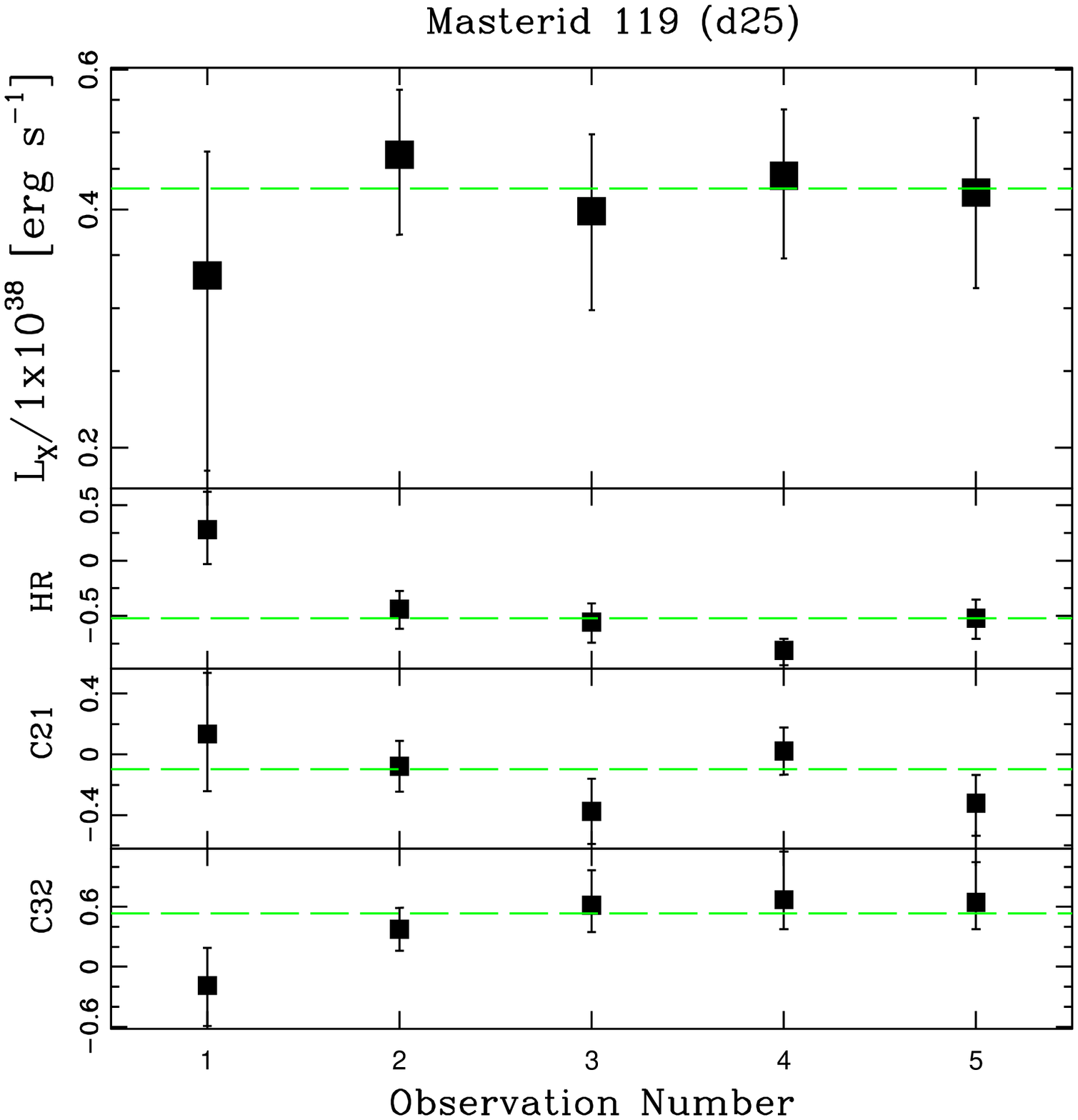}

\end{minipage}\hspace{0.02\linewidth}
\begin{minipage}{0.485\linewidth}
  \centering

    \includegraphics[width=\linewidth]{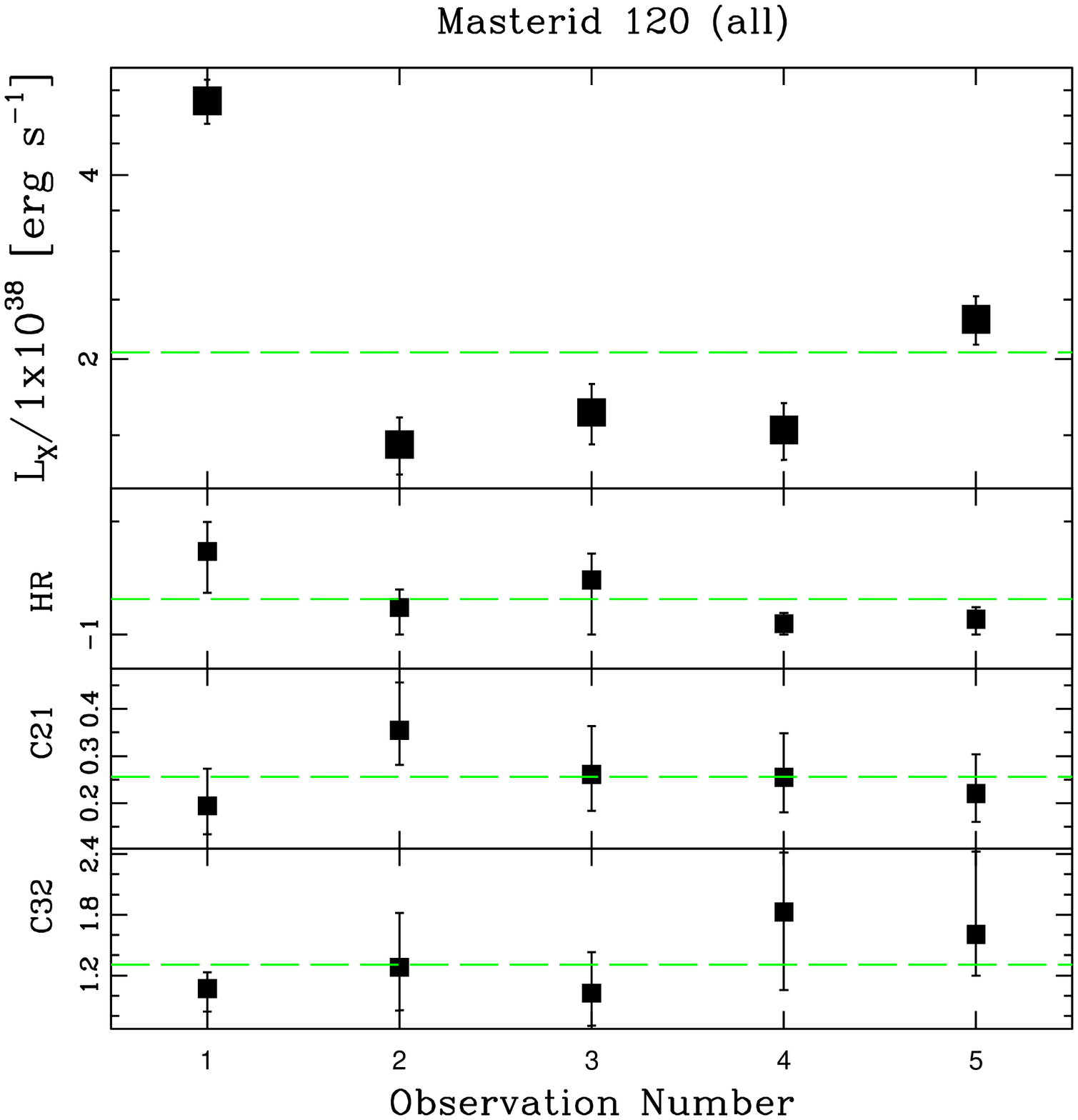}

 \end{minipage}\hspace{0.02\linewidth}

  \begin{minipage}{0.485\linewidth}
  \centering
  
    \includegraphics[width=\linewidth]{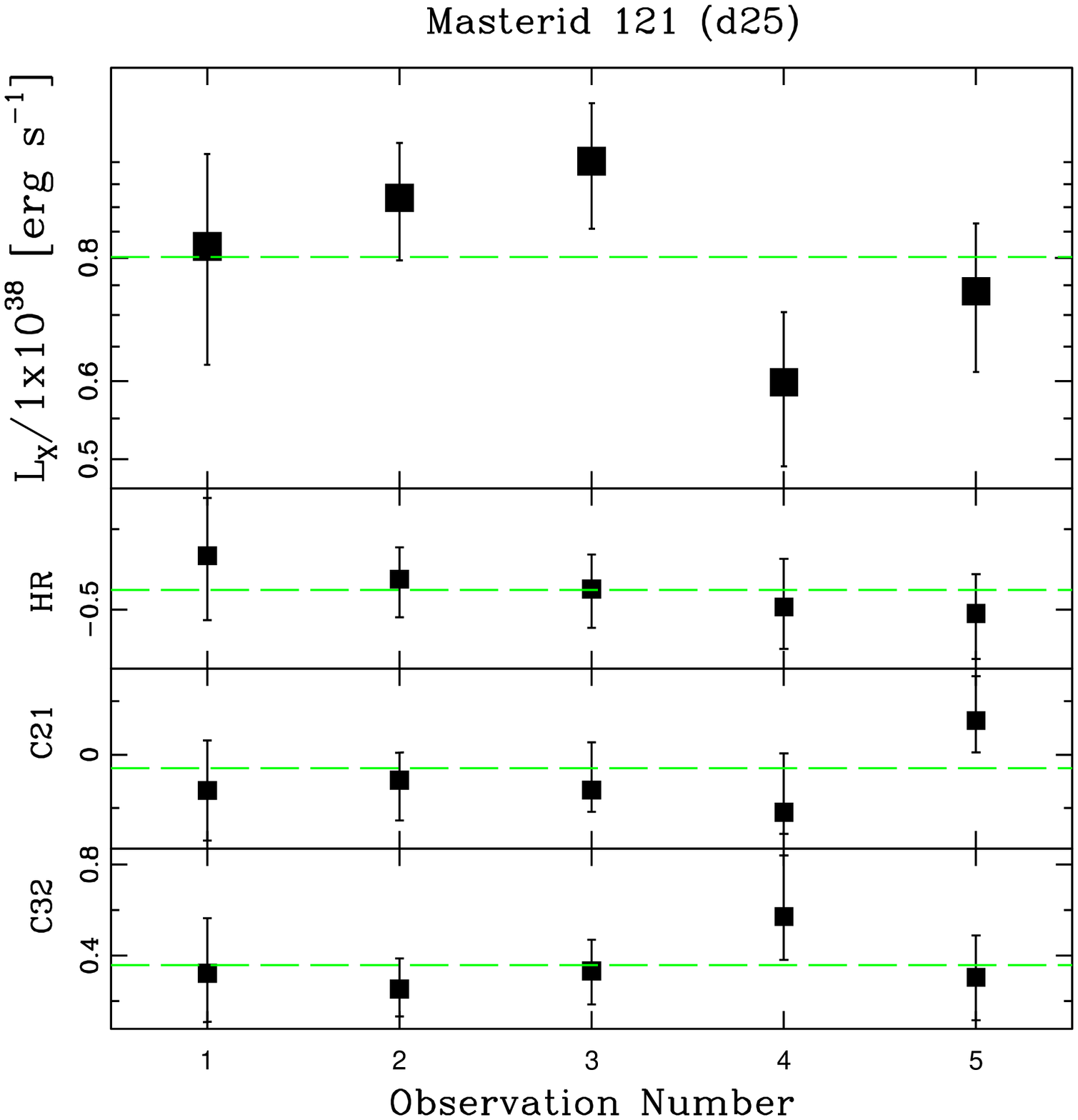}

  \end{minipage}\hspace{0.02\linewidth}
  \begin{minipage}{0.485\linewidth}
  \centering

    \includegraphics[width=\linewidth]{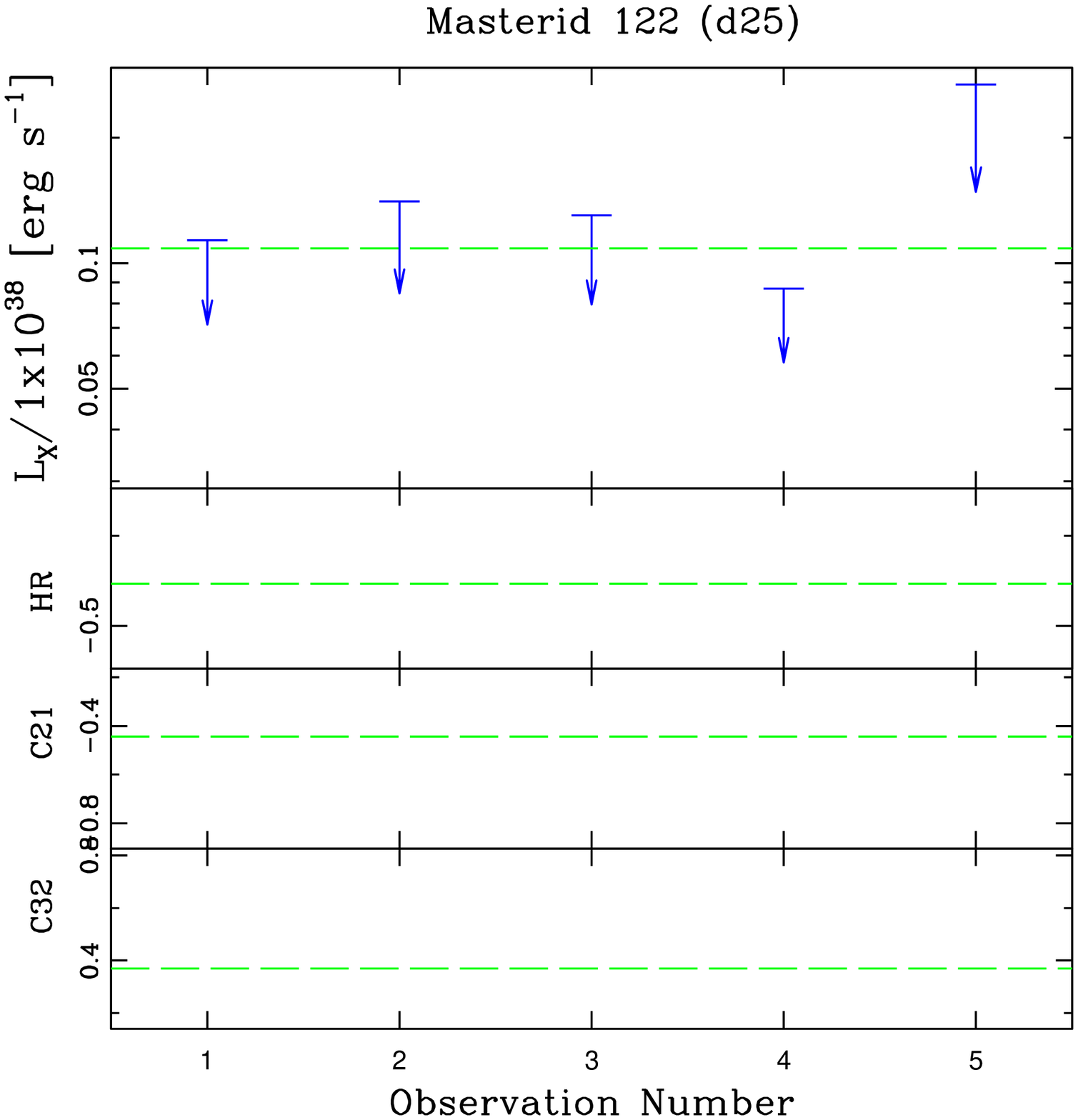}

\end{minipage}\hspace{0.02\linewidth}

\begin{minipage}{0.485\linewidth}
  \centering

    \includegraphics[width=\linewidth]{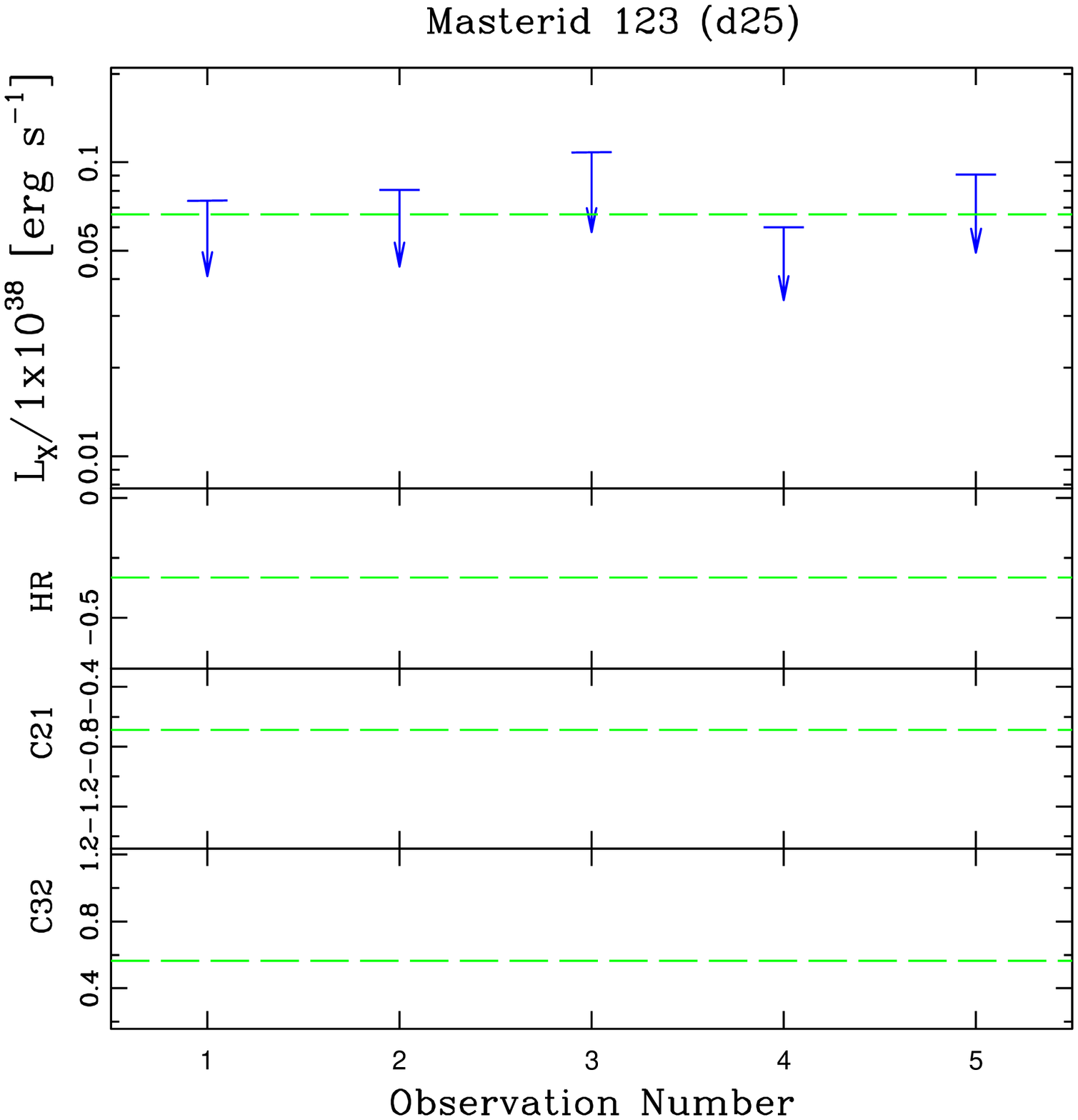}

\end{minipage}\hspace{0.02\linewidth}
\begin{minipage}{0.485\linewidth}
  \centering
  
    \includegraphics[width=\linewidth]{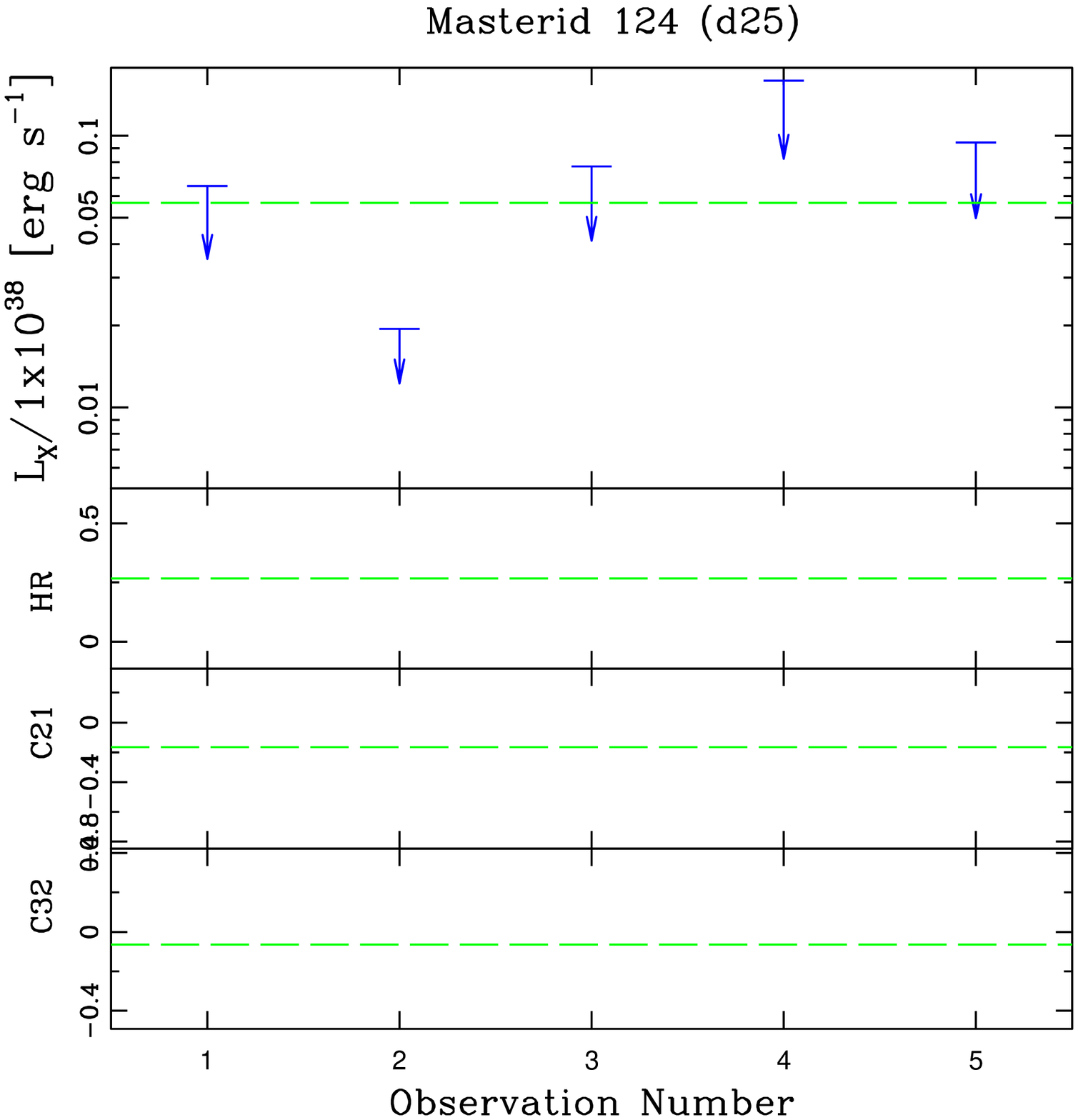}

  \end{minipage}\hspace{0.02\linewidth}

\end{figure}

\begin{figure}
  \begin{minipage}{0.485\linewidth}
  \centering

    \includegraphics[width=\linewidth]{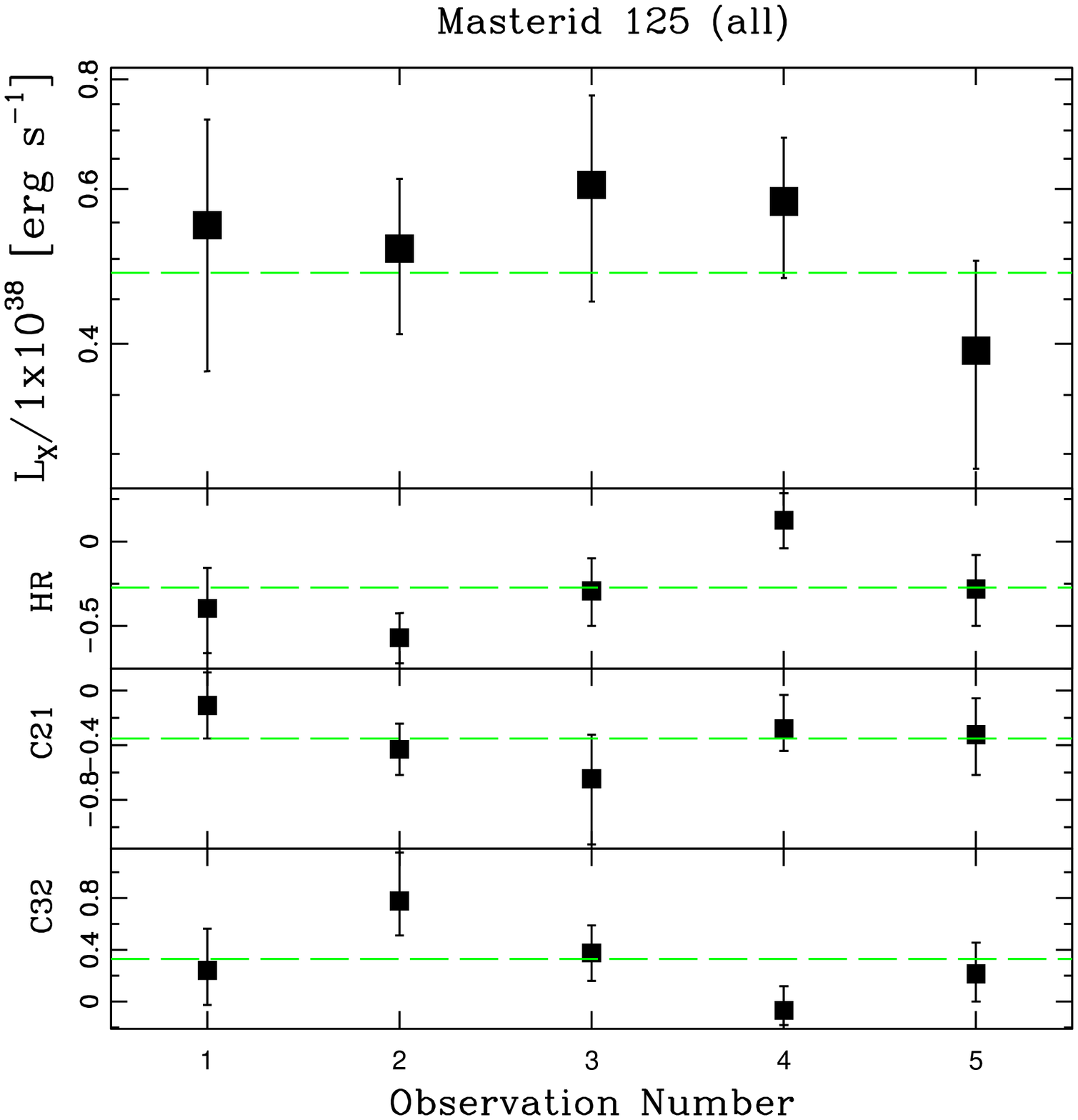}

\end{minipage}\hspace{0.02\linewidth}
\begin{minipage}{0.485\linewidth}
  \centering

    \includegraphics[width=\linewidth]{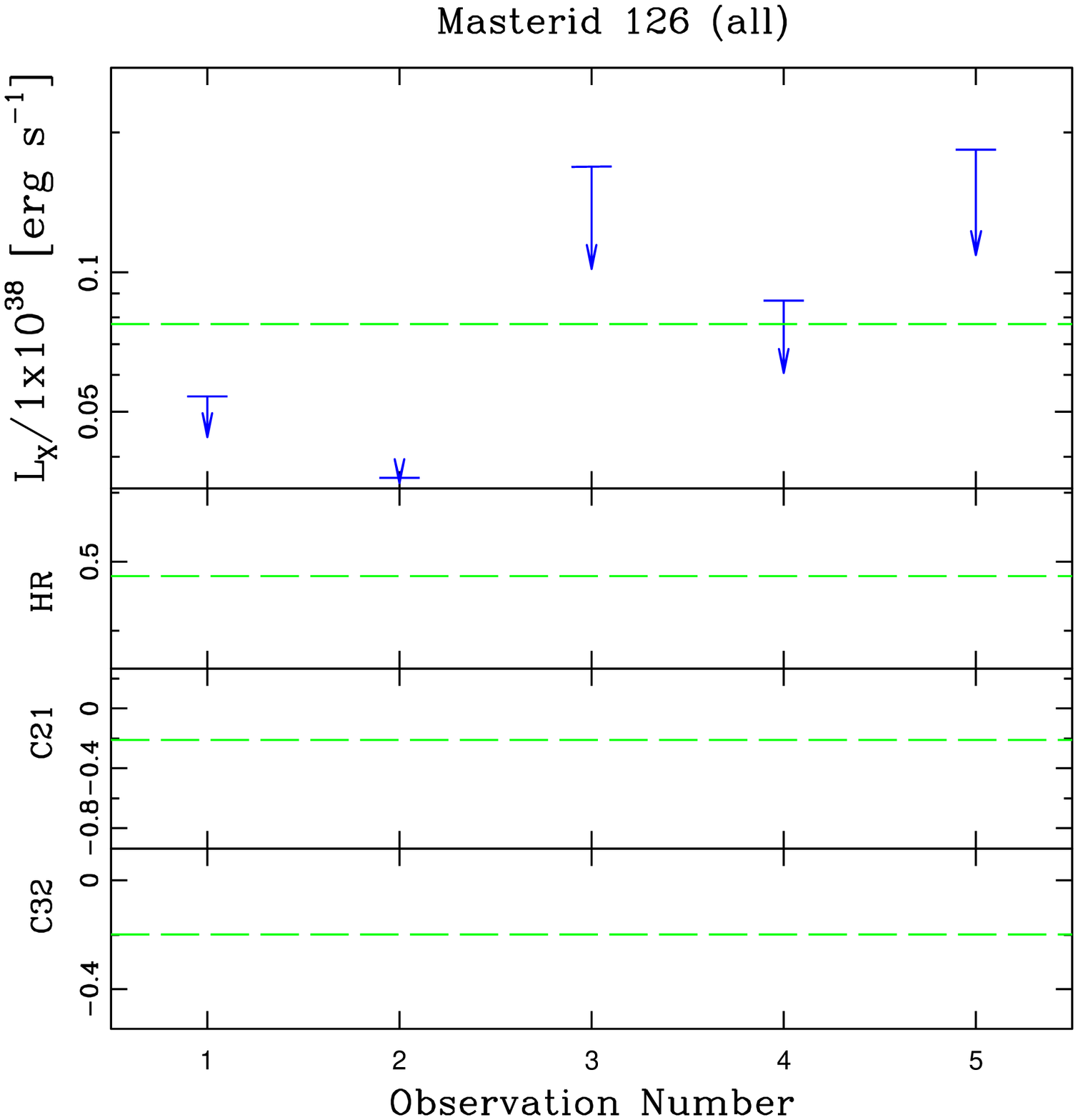}

\end{minipage}\hspace{0.02\linewidth}

  \begin{minipage}{0.485\linewidth}
  \centering
  
    \includegraphics[width=\linewidth]{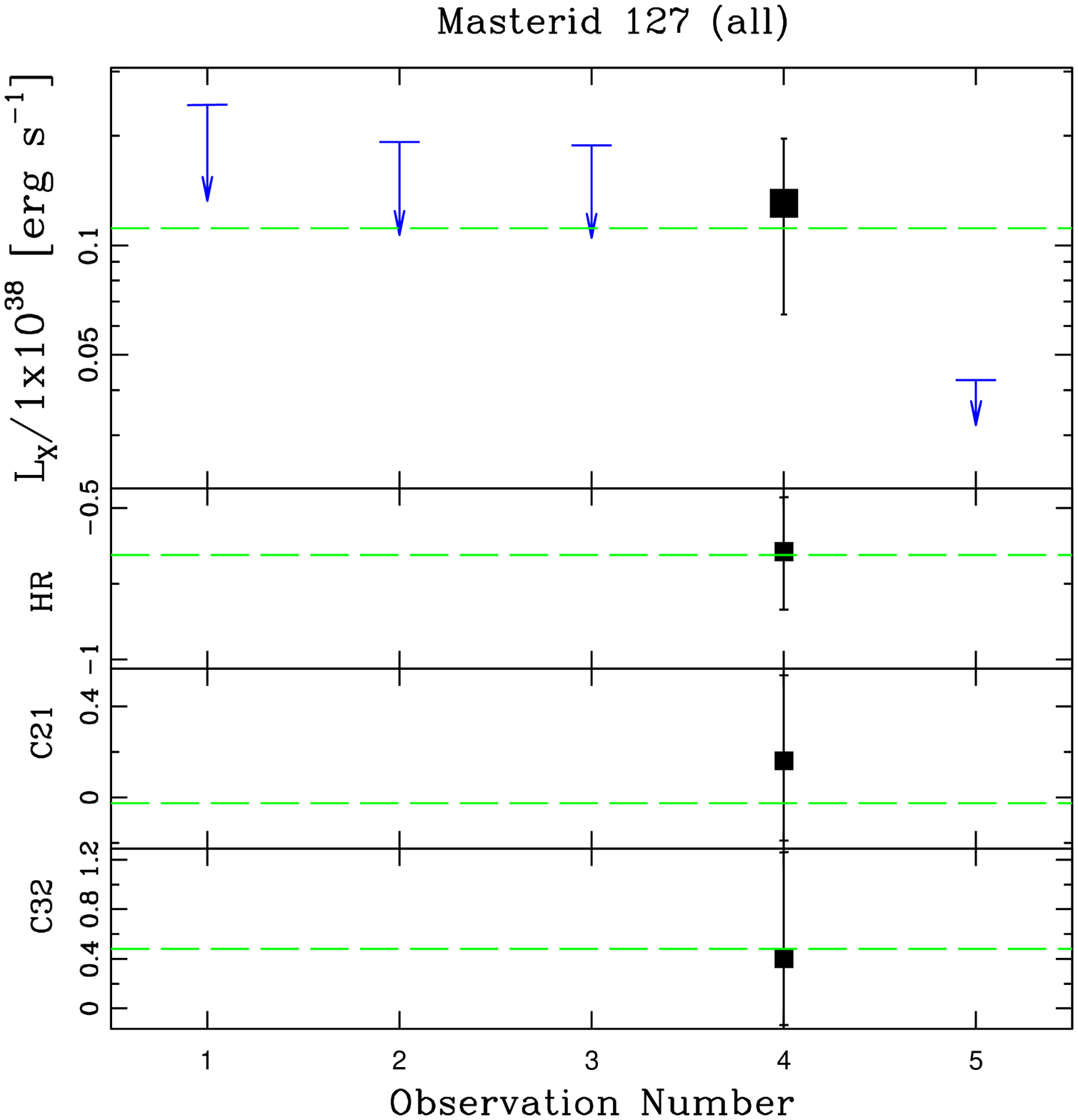}

  \end{minipage}\hspace{0.02\linewidth}
  \begin{minipage}{0.485\linewidth}
  \centering

    \includegraphics[width=\linewidth]{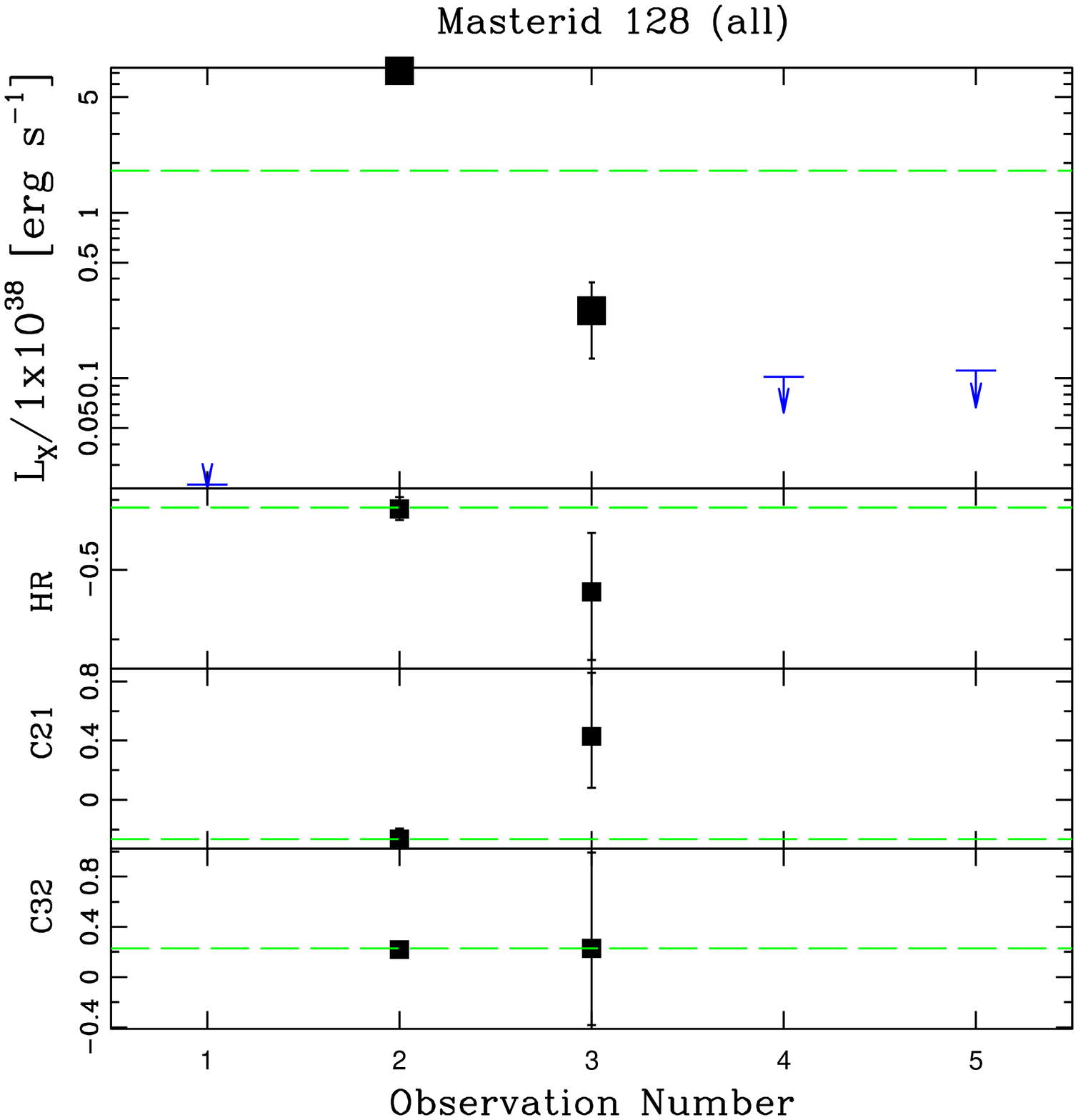}

\end{minipage}\hspace{0.02\linewidth}

\begin{minipage}{0.485\linewidth}
  \centering

    \includegraphics[width=\linewidth]{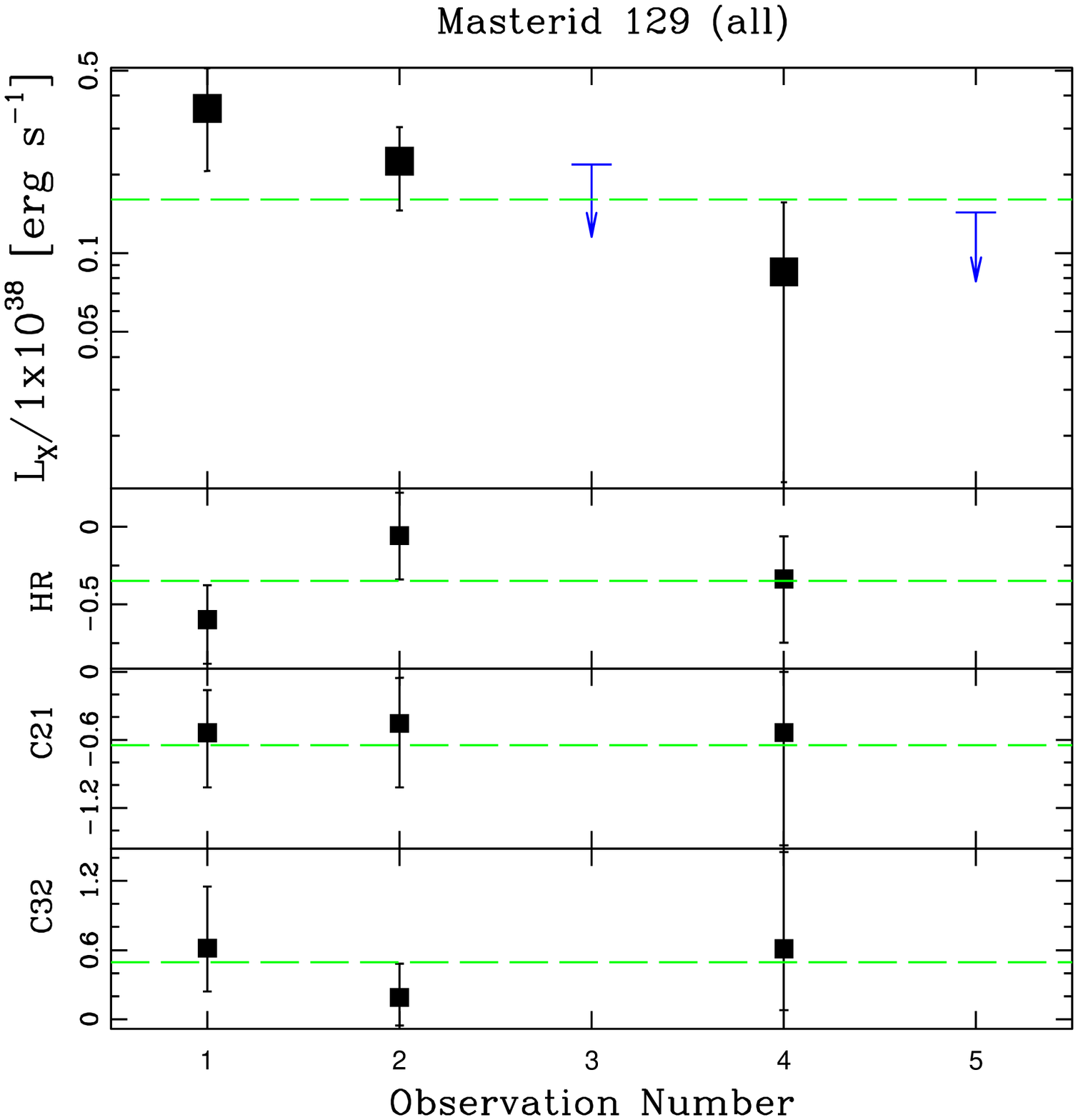}

 \end{minipage}\hspace{0.02\linewidth}
\begin{minipage}{0.485\linewidth}
  \centering
  
    \includegraphics[width=\linewidth]{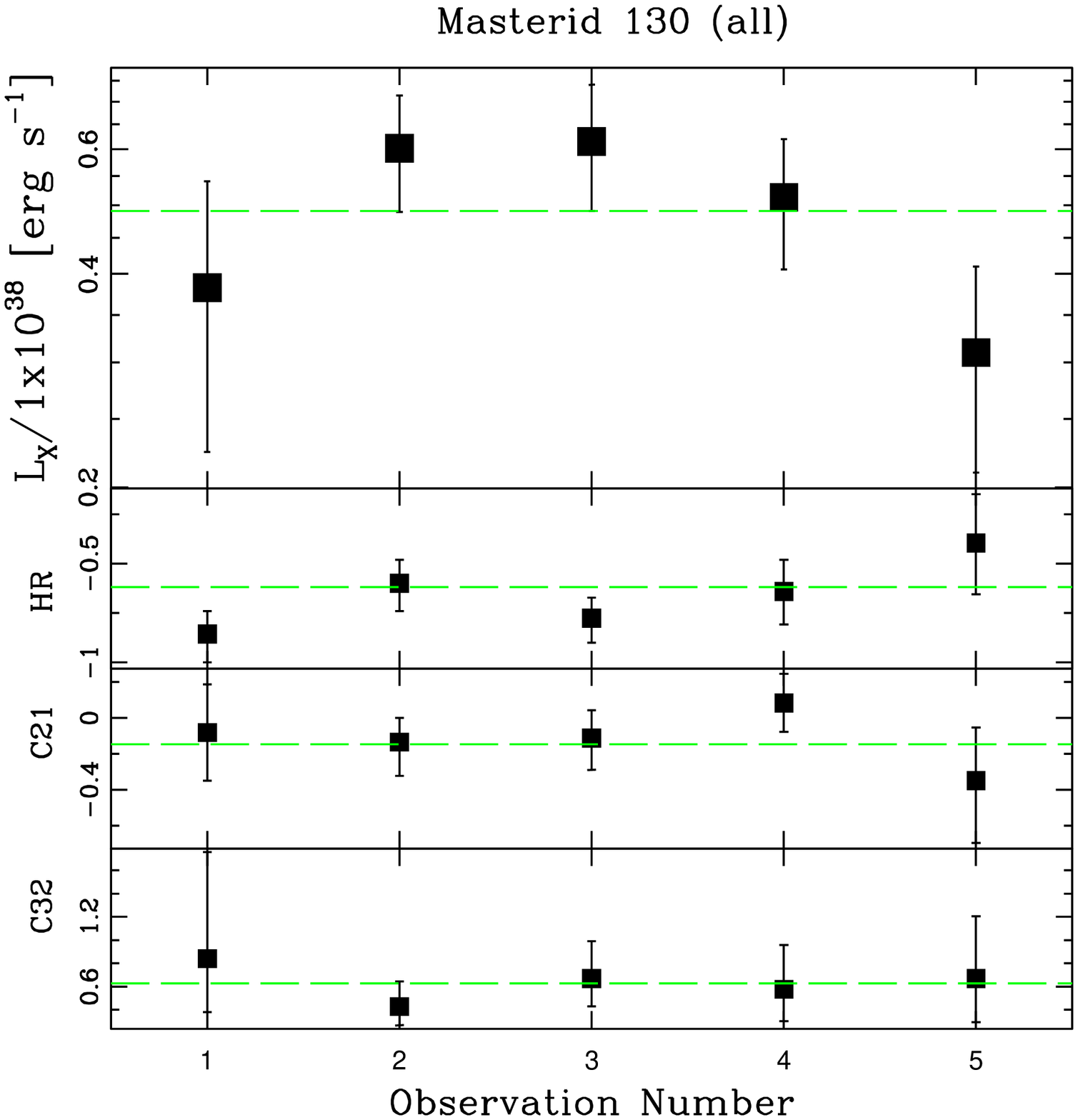}

  \end{minipage}\hspace{0.02\linewidth}
\end{figure}

\begin{figure}
  \begin{minipage}{0.485\linewidth}
  \centering

    \includegraphics[width=\linewidth]{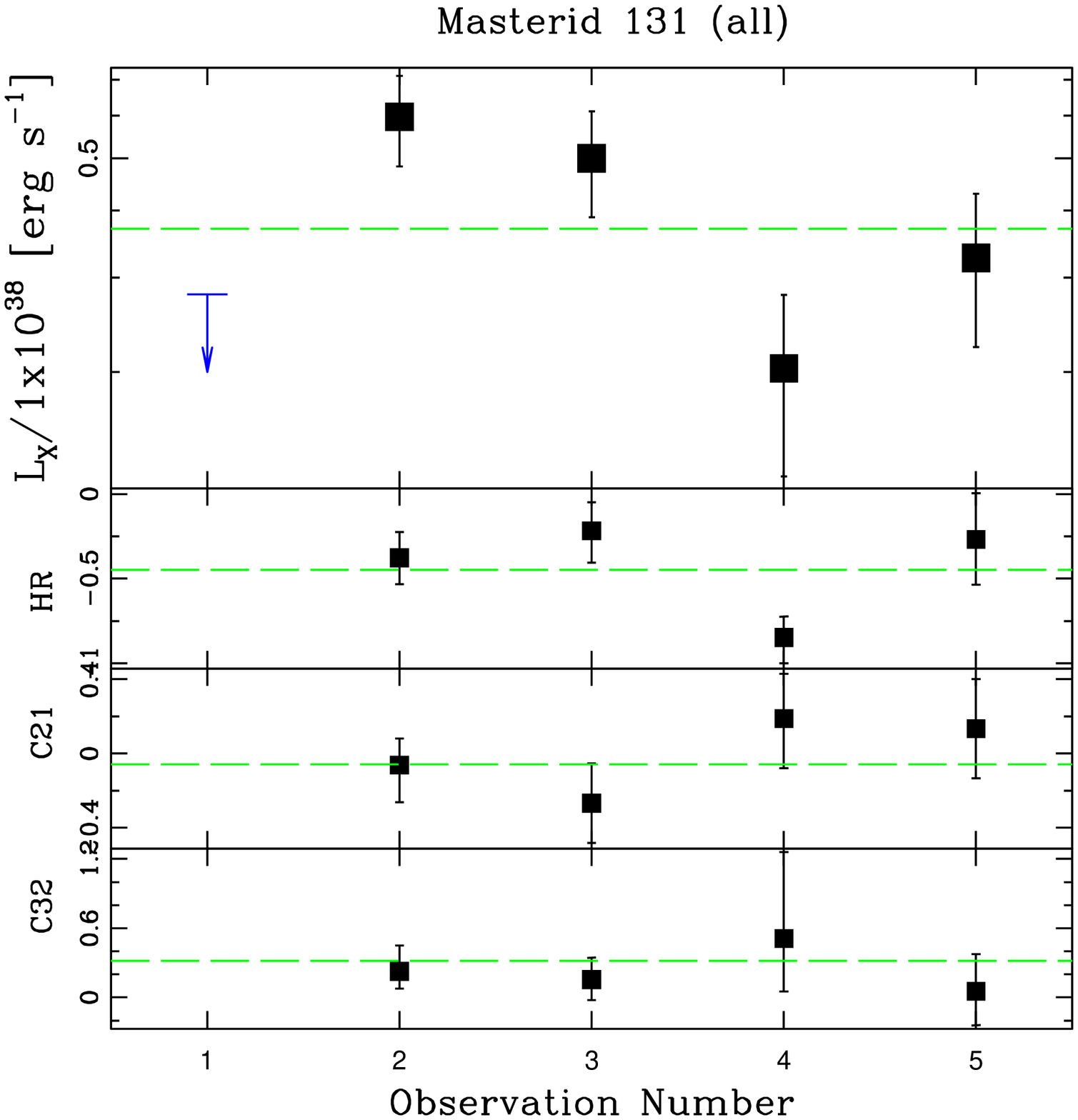}

\end{minipage}\hspace{0.02\linewidth}
\begin{minipage}{0.485\linewidth}
  \centering

    \includegraphics[width=\linewidth]{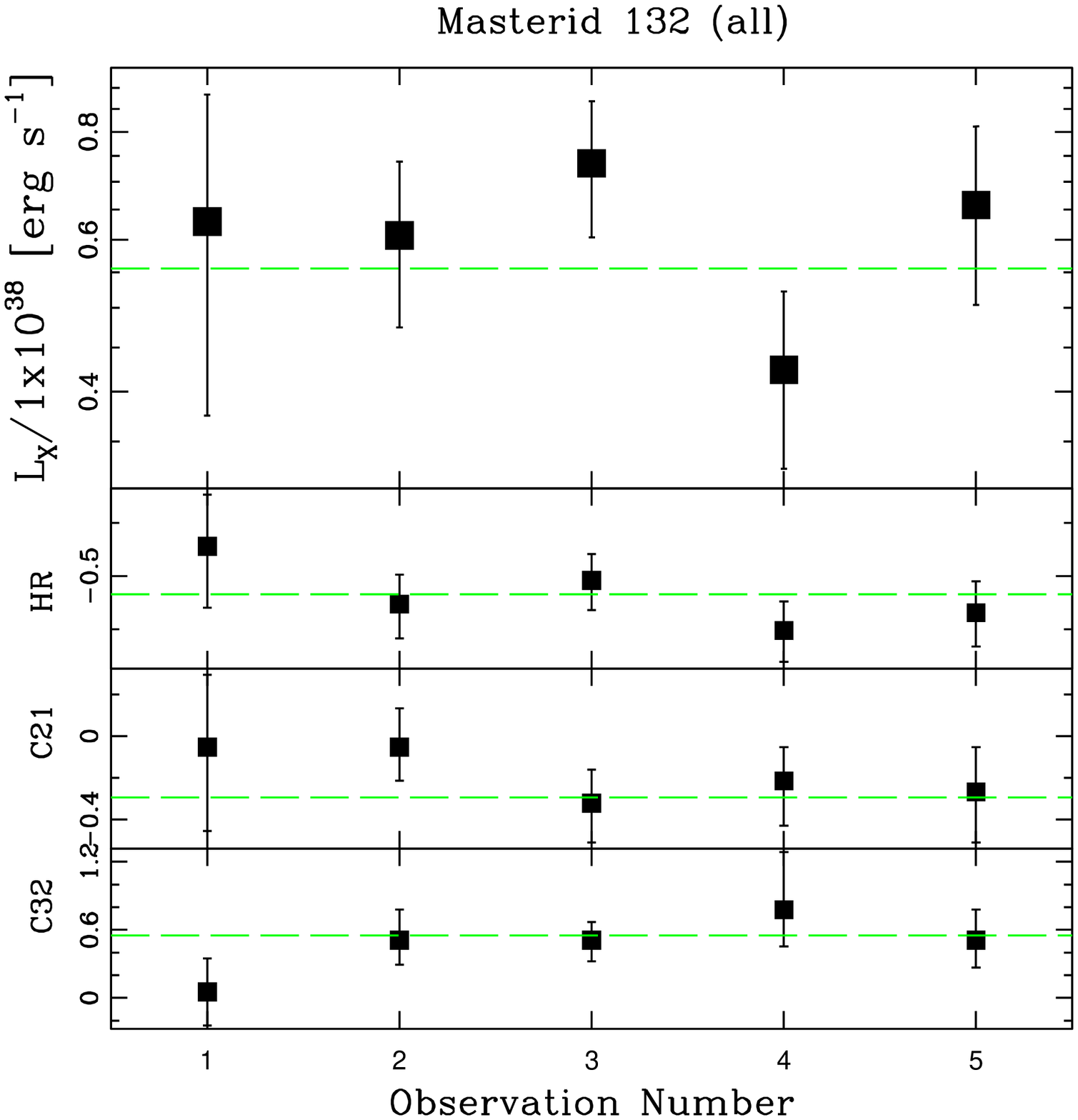}

  \end{minipage}\hspace{0.02\linewidth}
\end{figure}

\begin{figure}
  \begin{minipage}{0.32\linewidth}
  \centering
  
    \includegraphics[width=\linewidth]{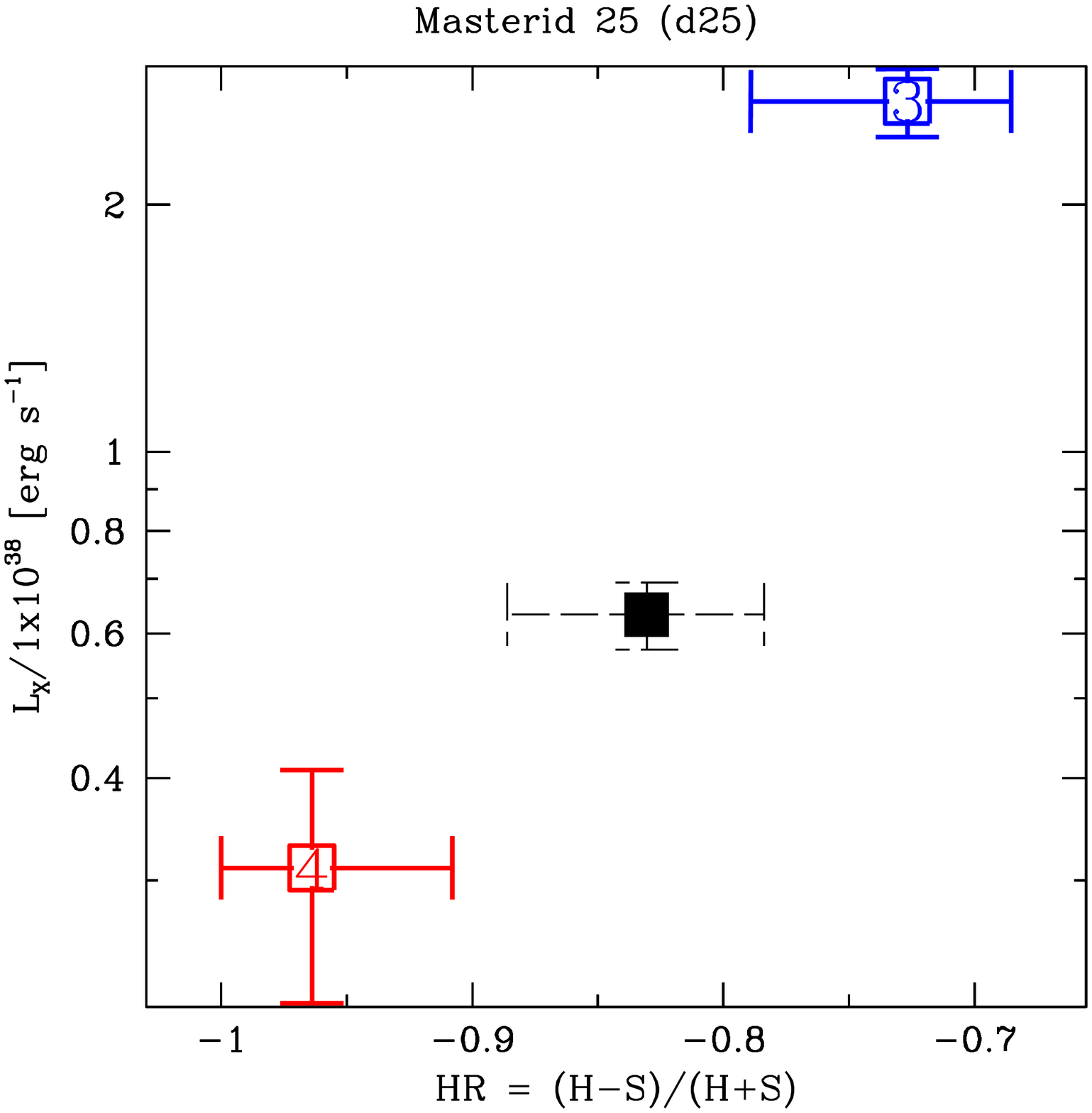}
  
  \end{minipage}
  \begin{minipage}{0.32\linewidth}
  \centering

    \includegraphics[width=\linewidth]{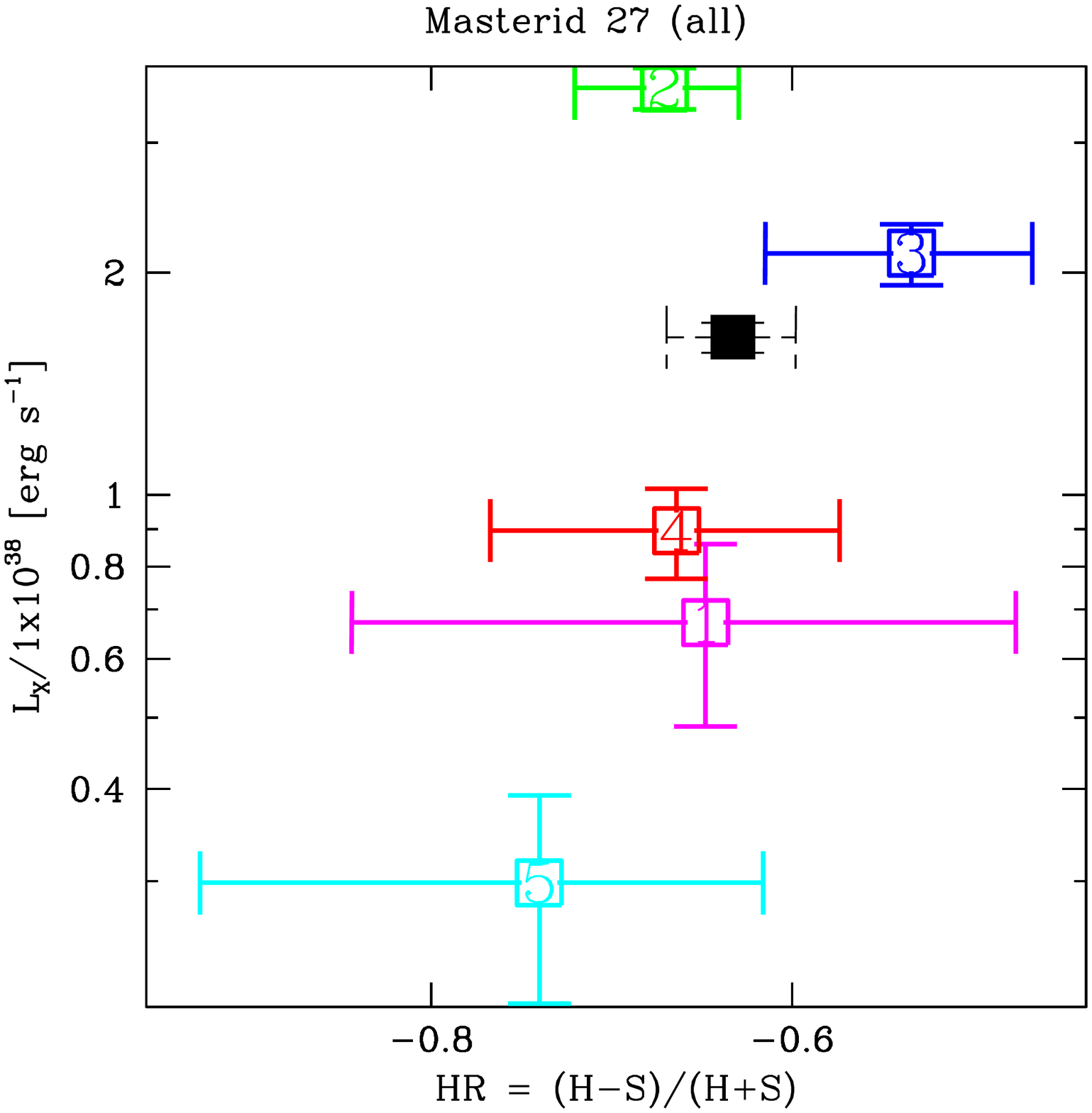}

\end{minipage}
\begin{minipage}{0.32\linewidth}
  \centering

    \includegraphics[width=\linewidth]{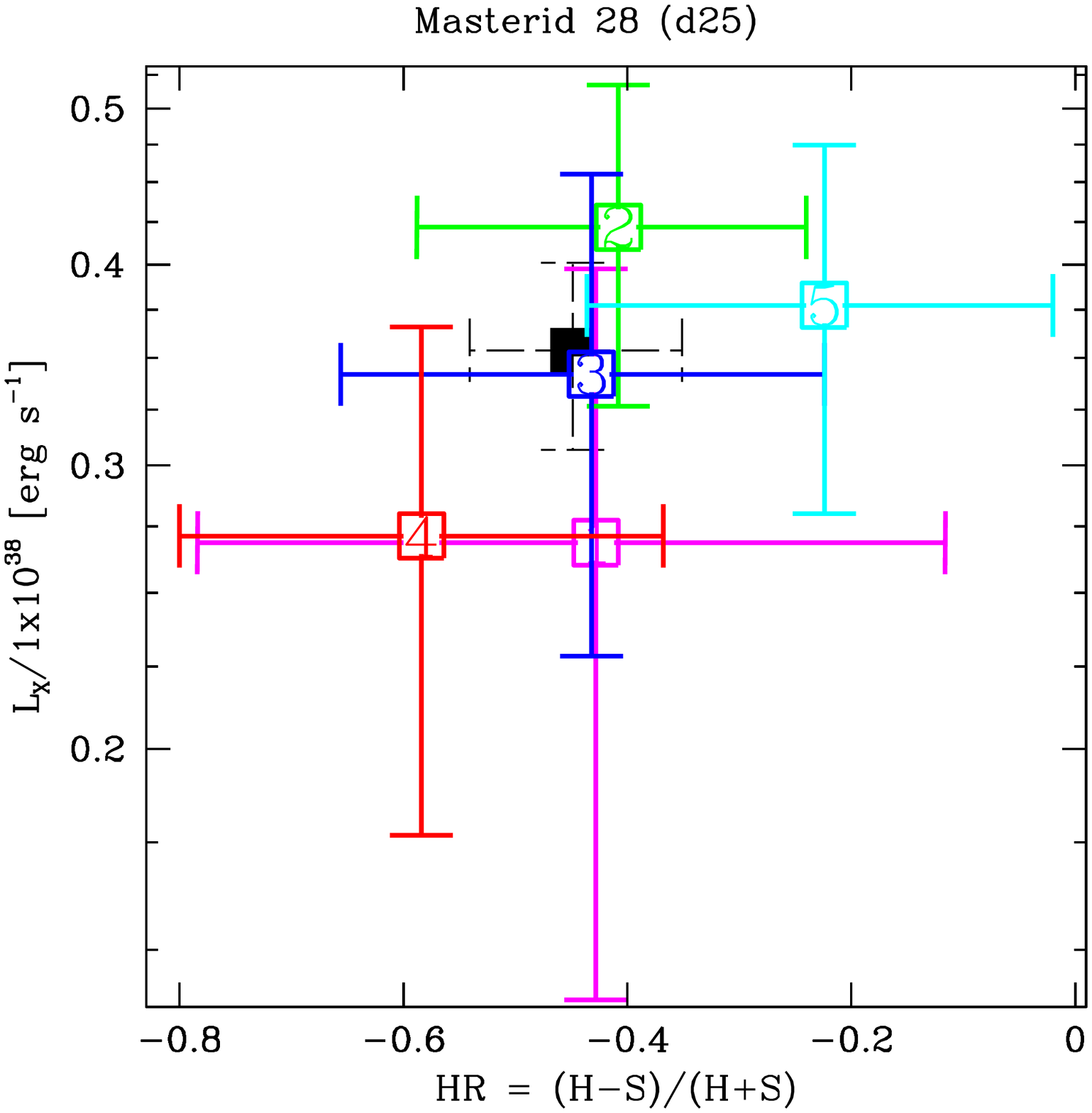}

 \end{minipage}

\begin{minipage}{0.32\linewidth}
  \centering
  
    \includegraphics[width=\linewidth]{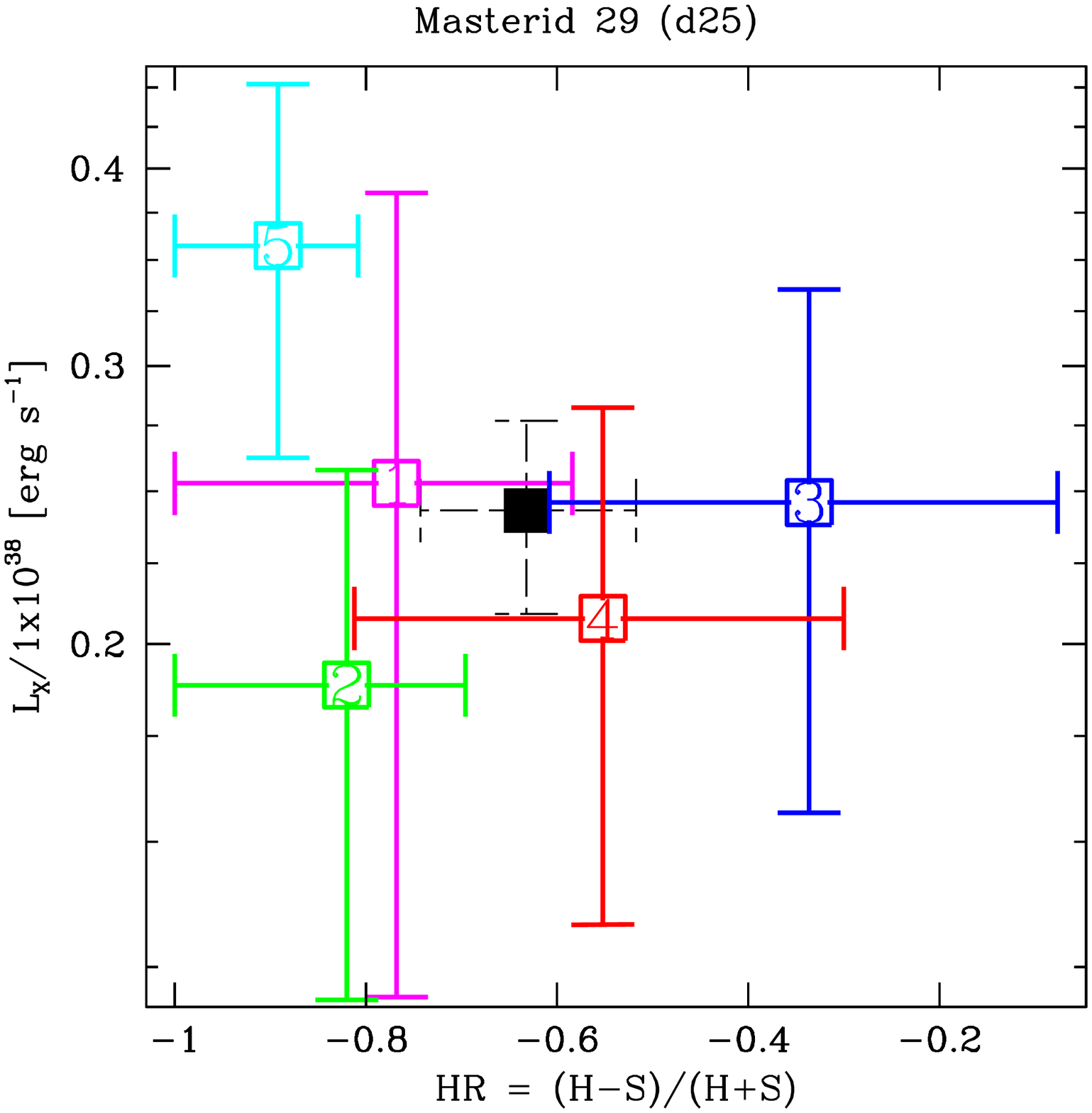}
  
  \end{minipage}
  \begin{minipage}{0.32\linewidth}
  \centering

    \includegraphics[width=\linewidth]{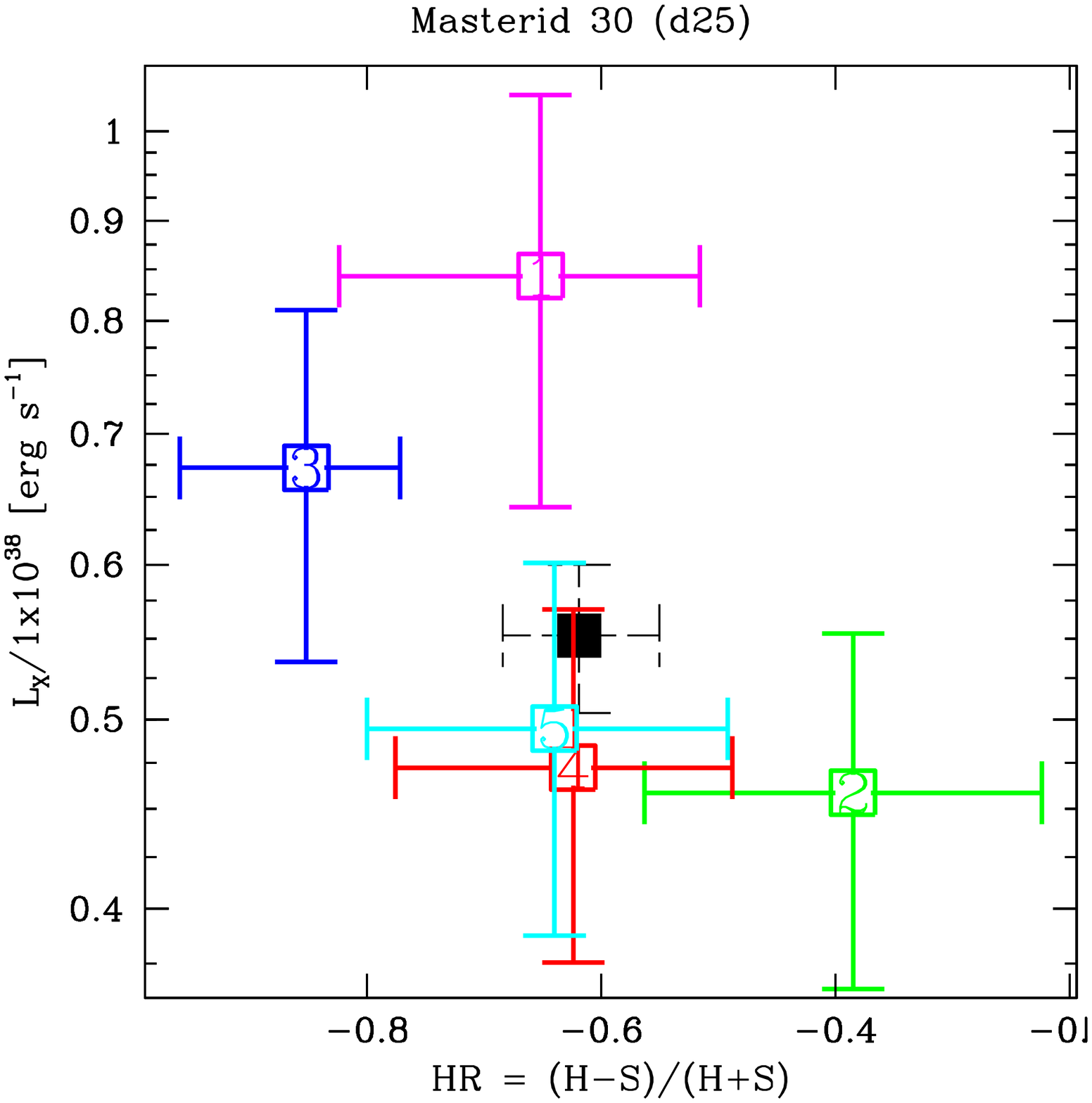}

\end{minipage}
\begin{minipage}{0.32\linewidth}
  \centering

    \includegraphics[width=\linewidth]{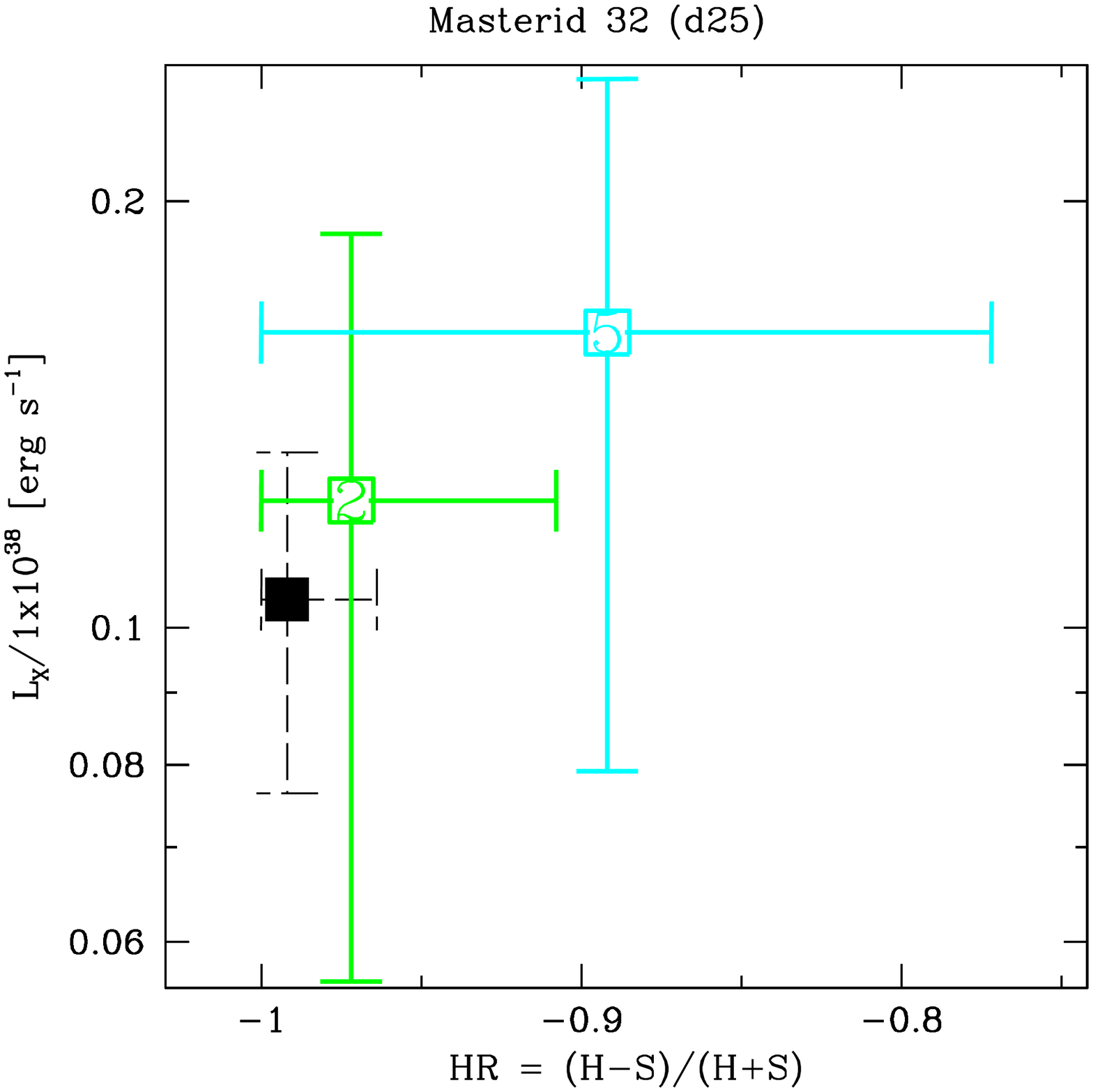}

 \end{minipage}

  \begin{minipage}{0.32\linewidth}
  \centering
  
    \includegraphics[width=\linewidth]{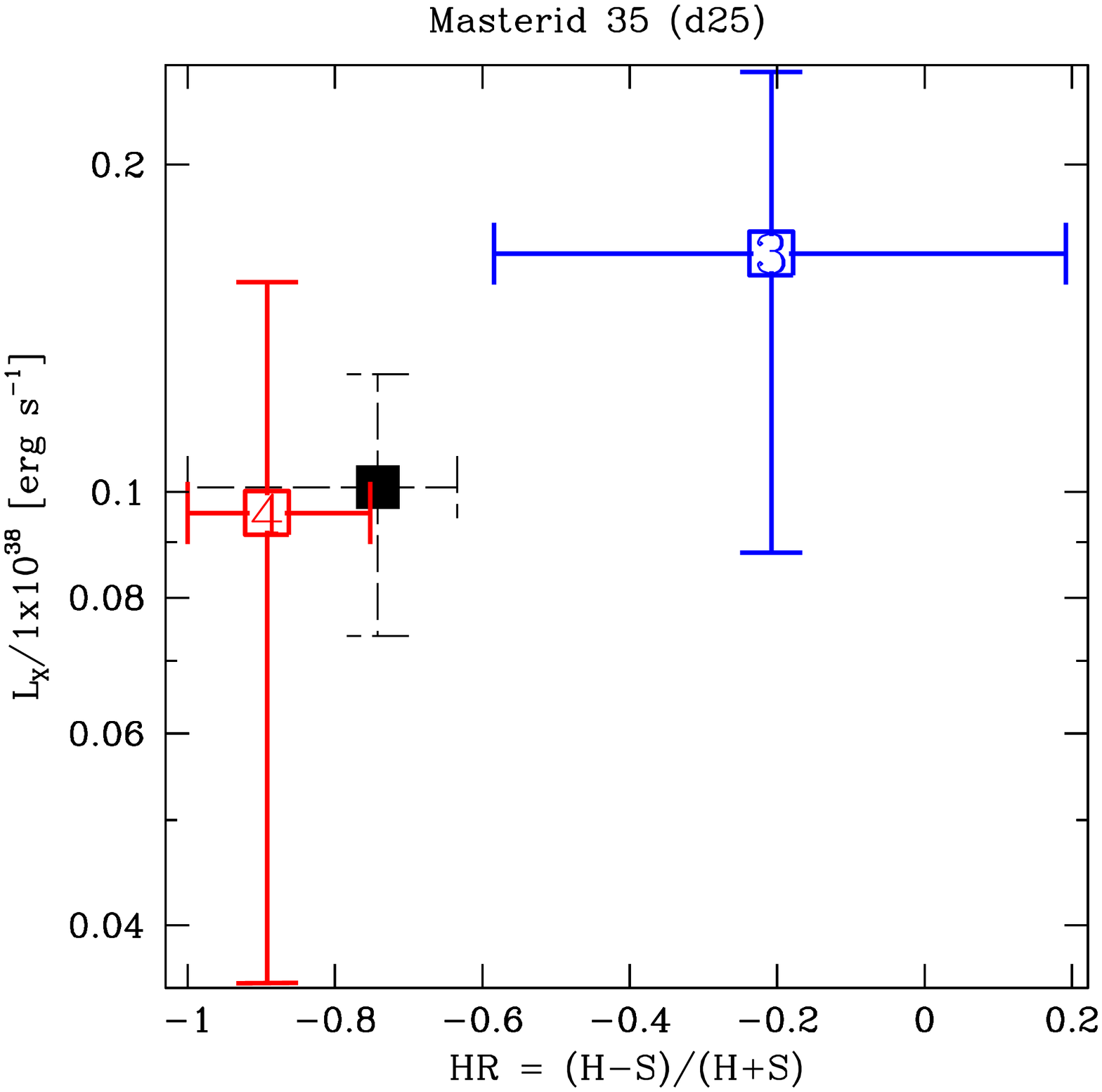}
  
  \end{minipage}
  \begin{minipage}{0.32\linewidth}
  \centering

    \includegraphics[width=\linewidth]{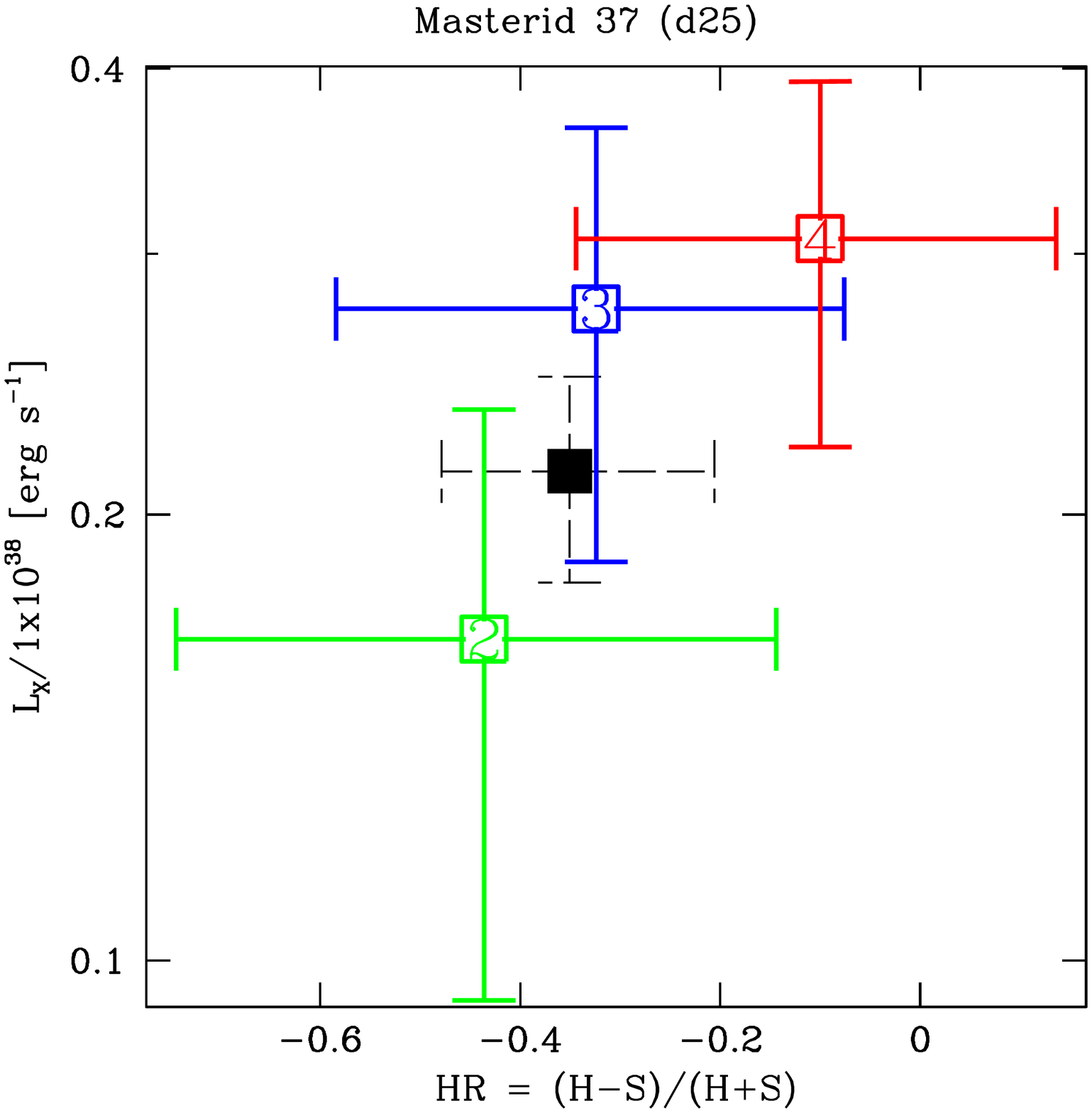}

\end{minipage}
\begin{minipage}{0.32\linewidth}
  \centering

    \includegraphics[width=\linewidth]{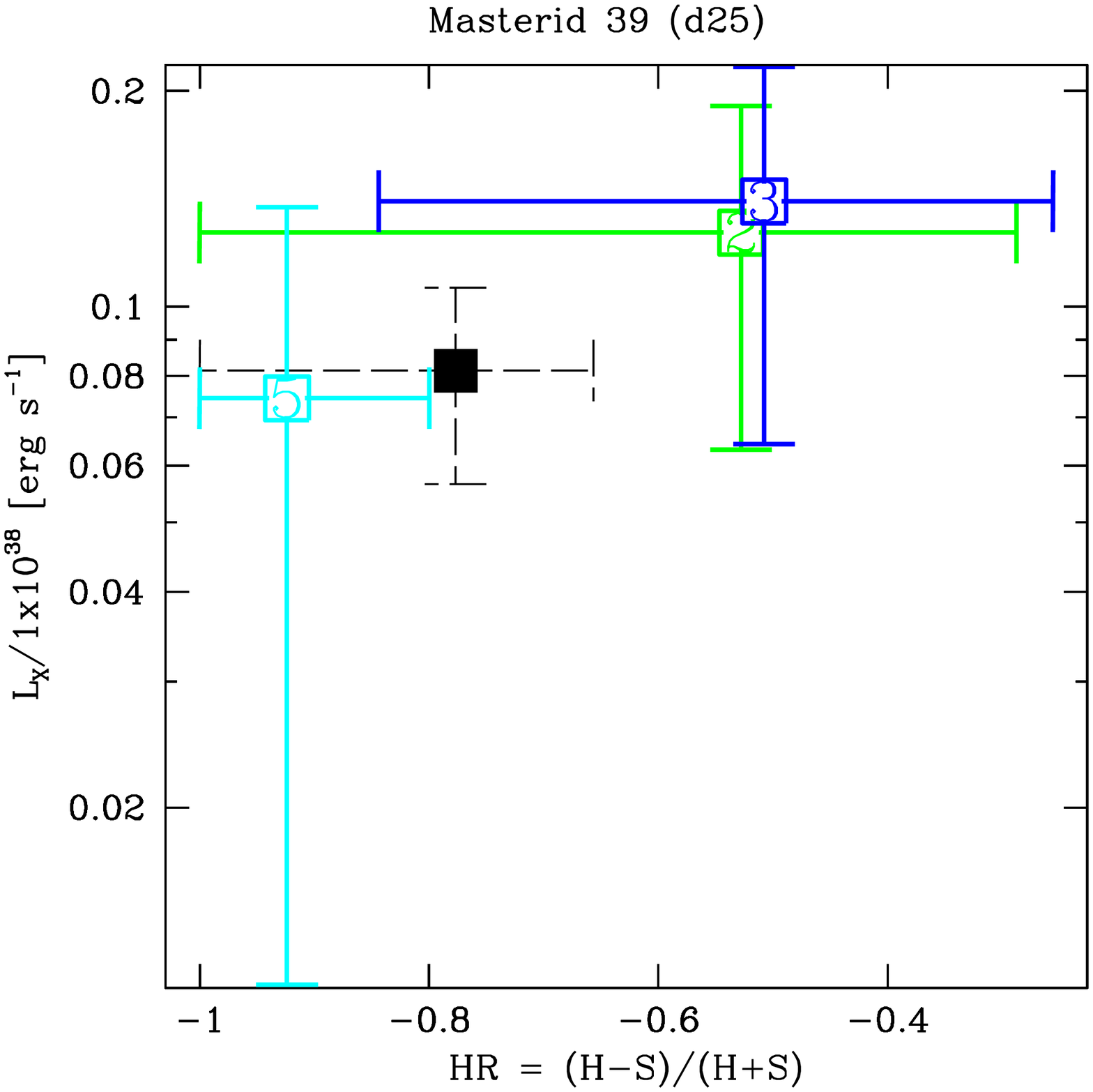}

 \end{minipage}

\begin{minipage}{0.32\linewidth}
  \centering
  
    \includegraphics[width=\linewidth]{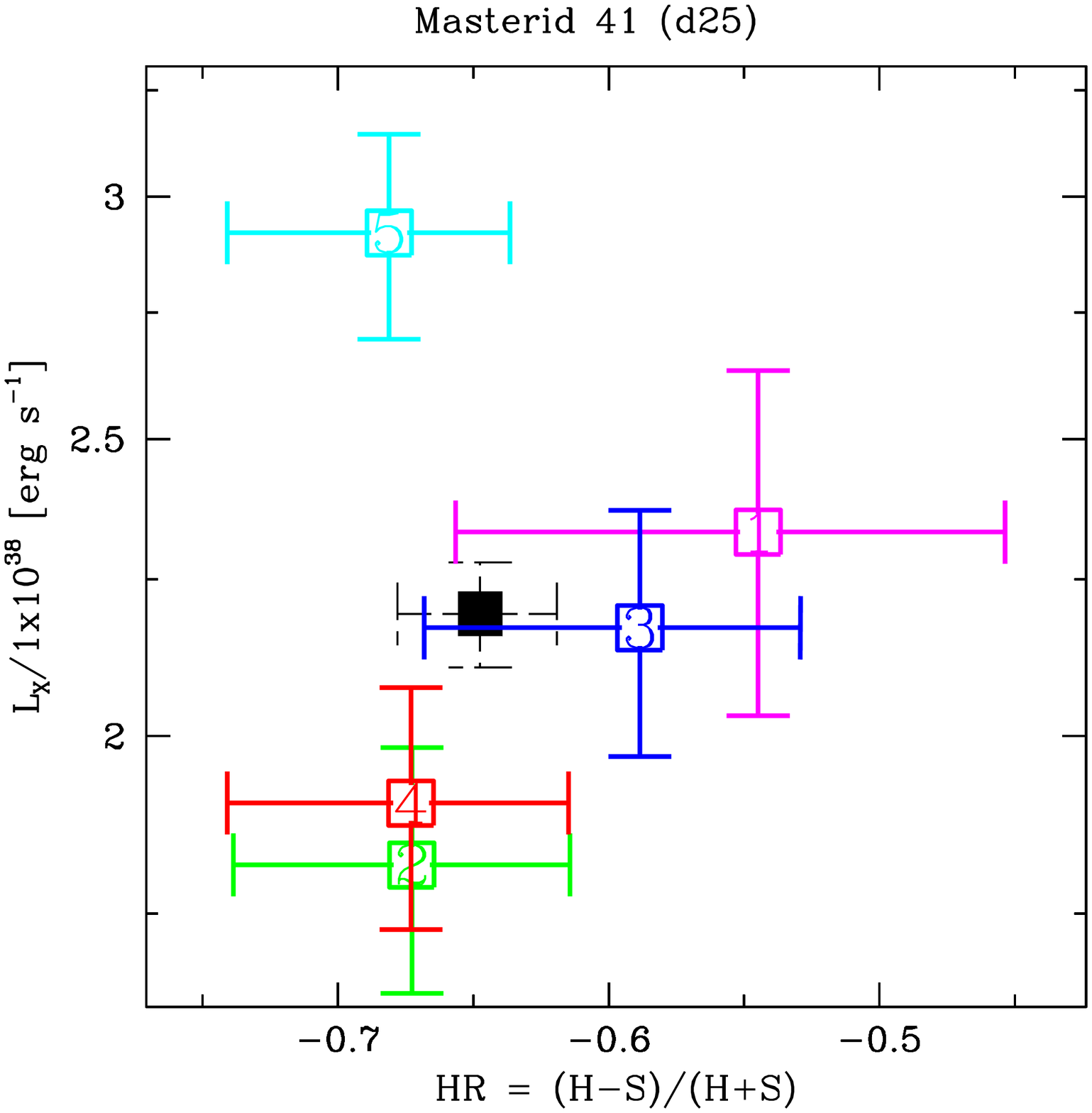}
  
  \end{minipage}
  \begin{minipage}{0.32\linewidth}
  \centering

    \includegraphics[width=\linewidth]{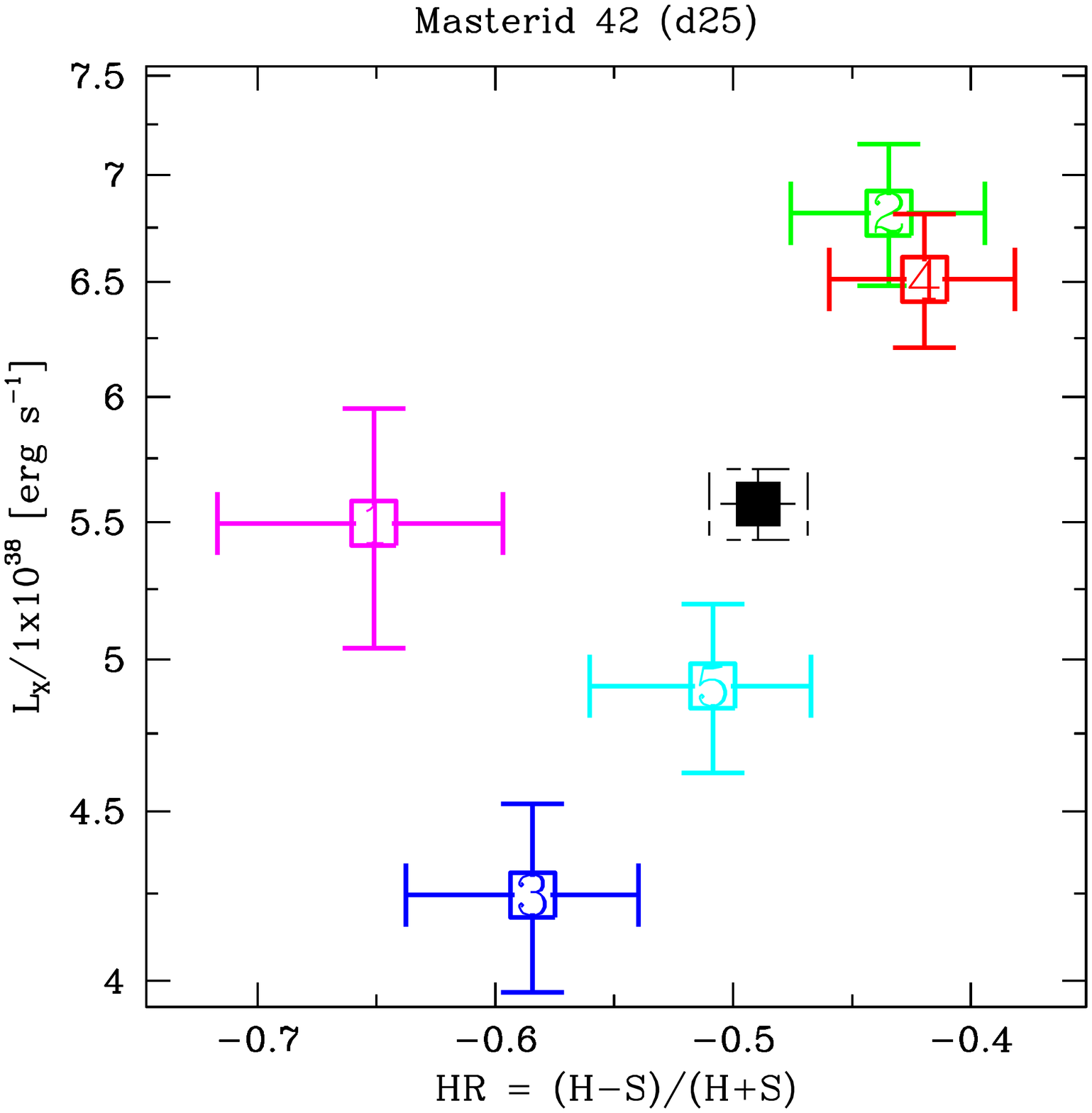}

\end{minipage}
\begin{minipage}{0.32\linewidth}
  \centering

    \includegraphics[width=\linewidth]{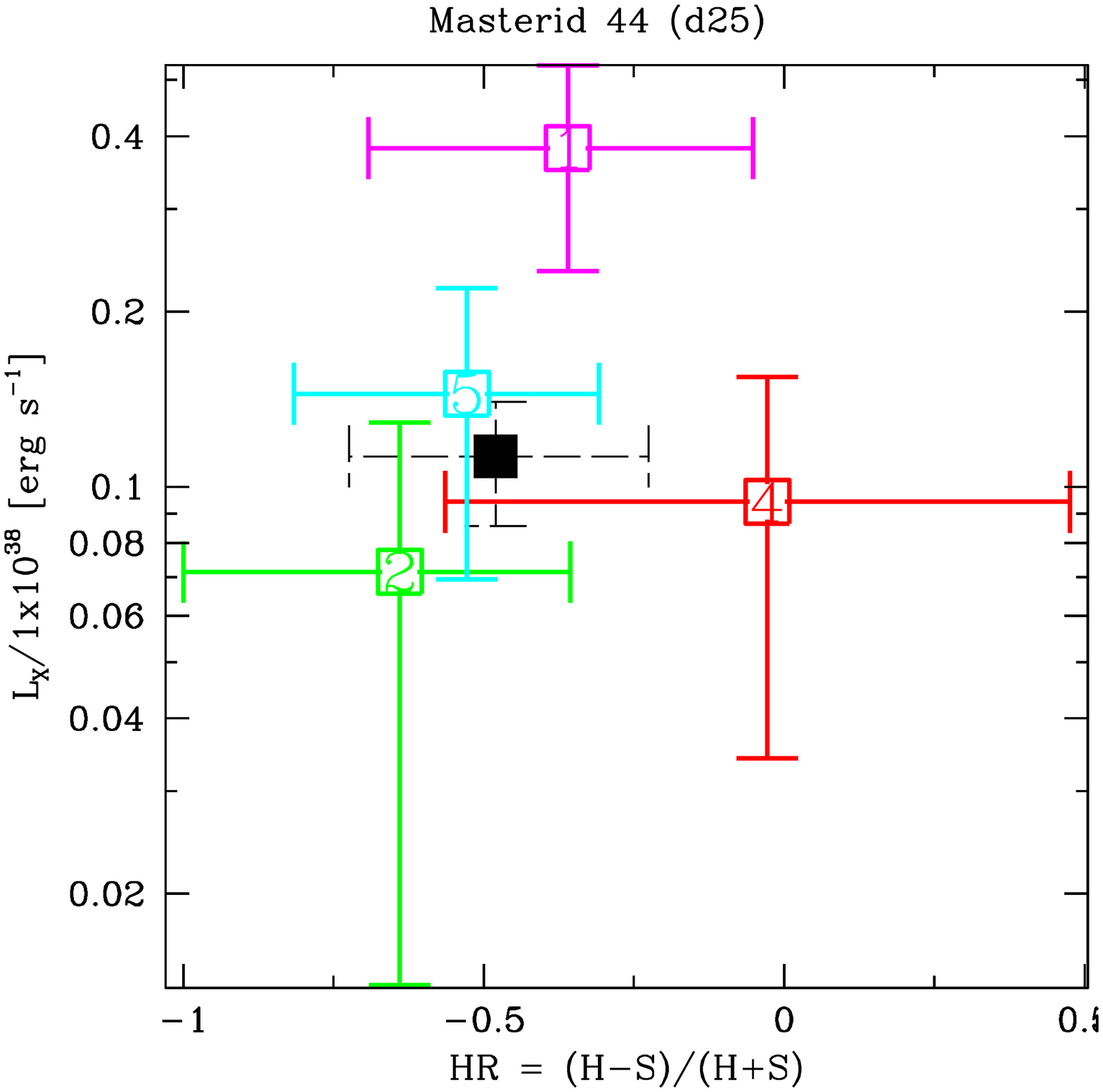}

 \end{minipage}
  
\end{figure}

\clearpage

\begin{figure}
  \begin{minipage}{0.32\linewidth}
  \centering
  
    \includegraphics[width=\linewidth]{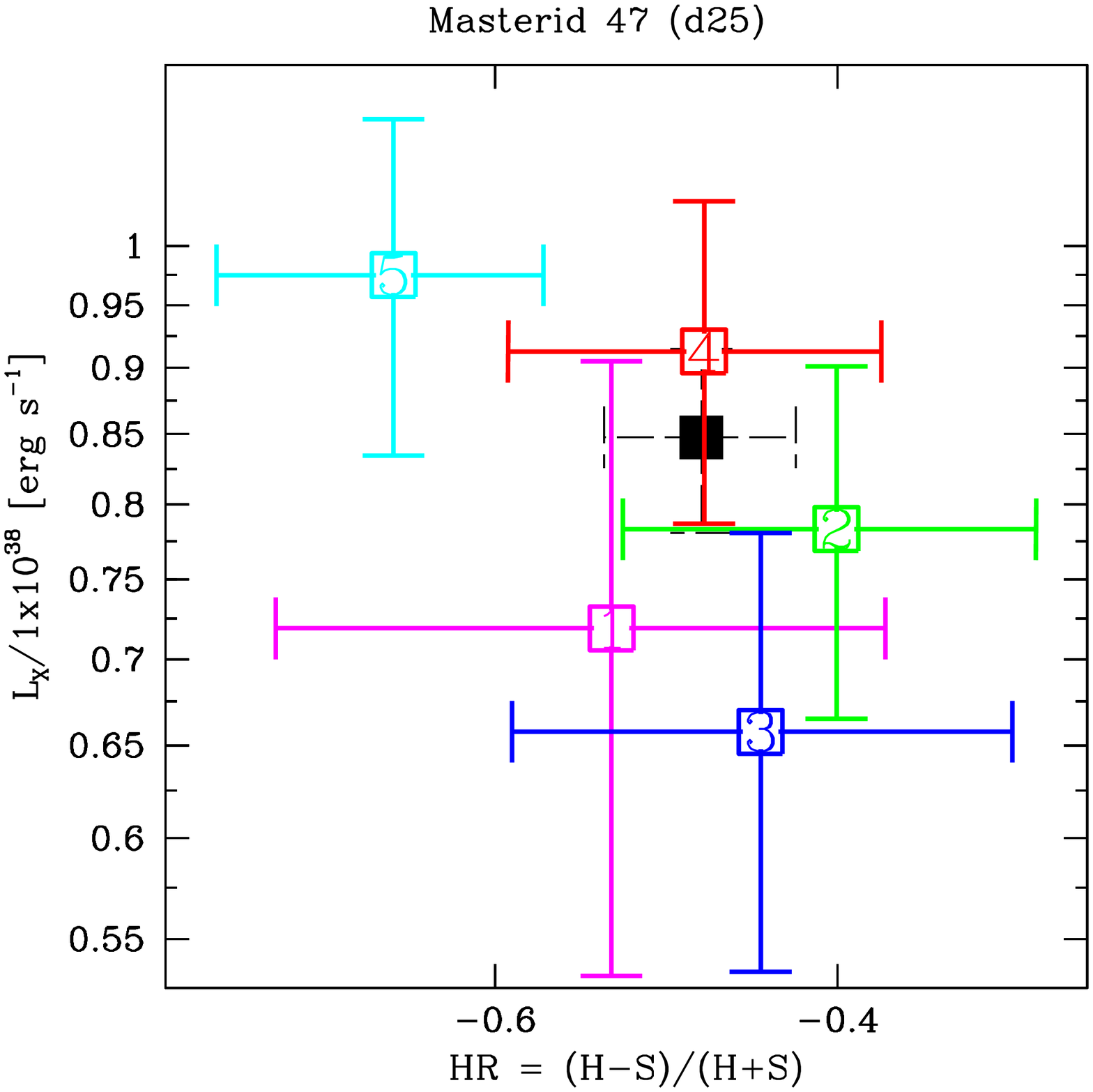}
  
  \end{minipage}
  \begin{minipage}{0.32\linewidth}
  \centering

    \includegraphics[width=\linewidth]{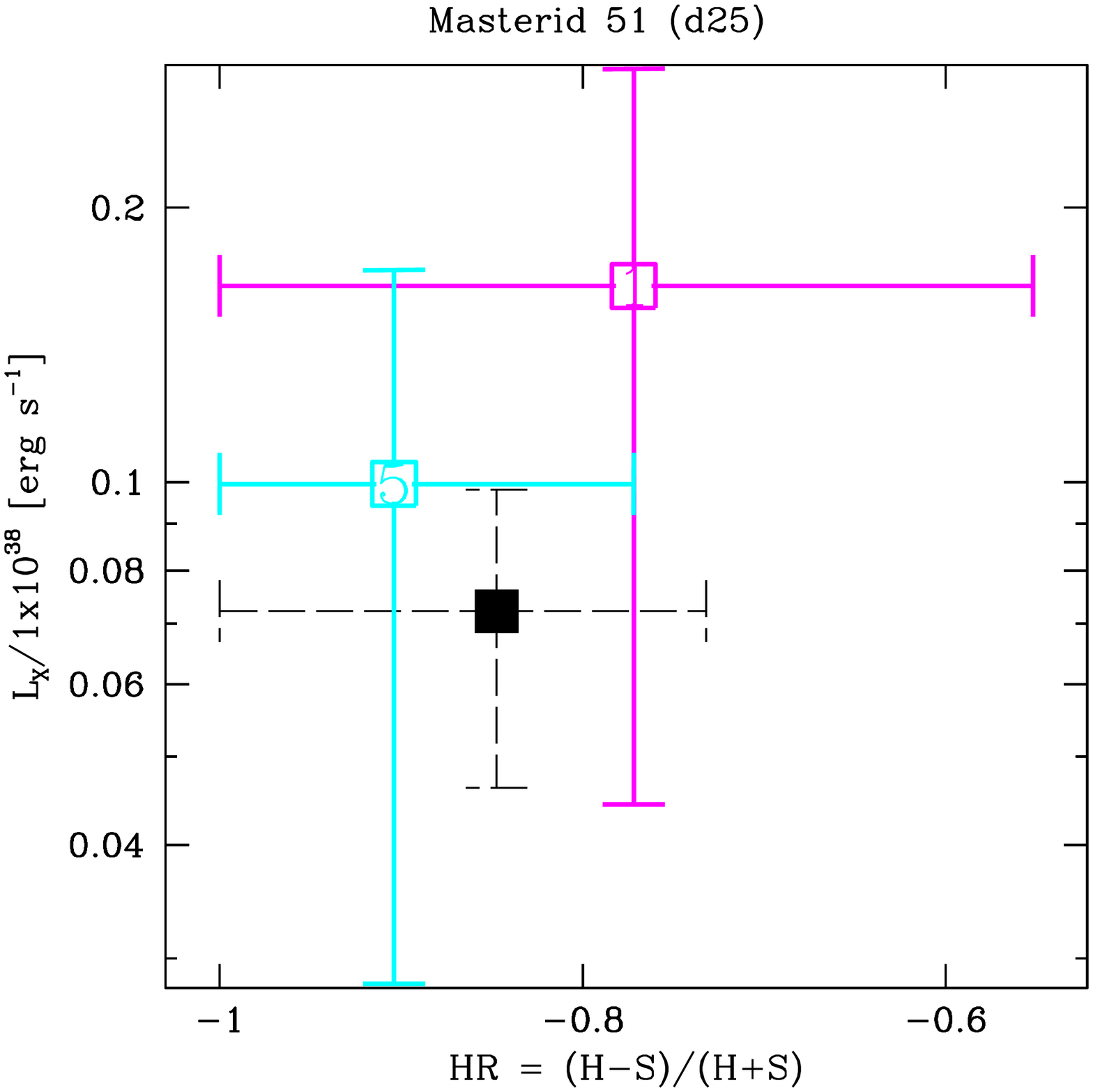}

\end{minipage}
\begin{minipage}{0.32\linewidth}
  \centering

    \includegraphics[width=\linewidth]{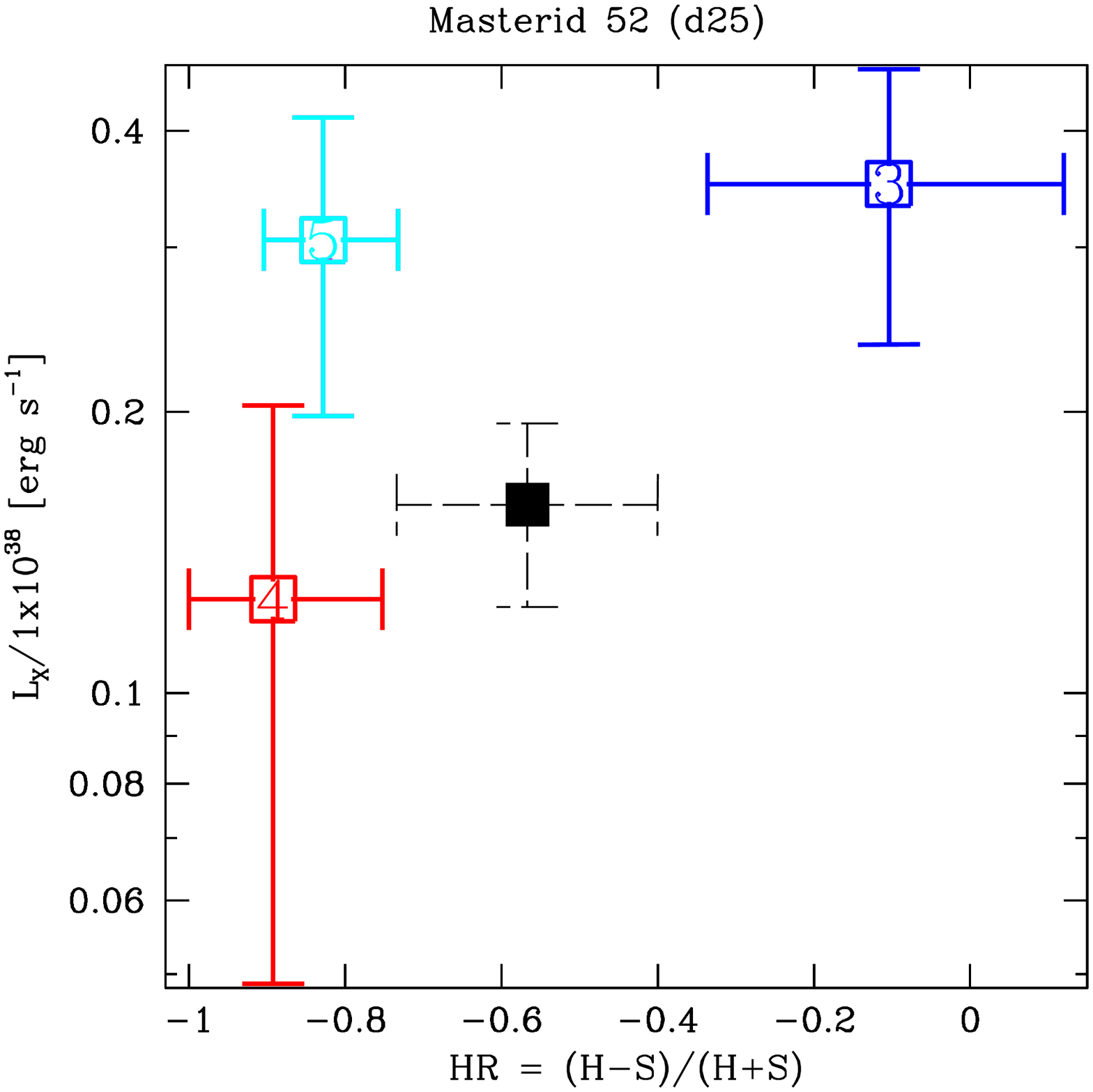}

 \end{minipage}

\begin{minipage}{0.32\linewidth}
  \centering
  
    \includegraphics[width=\linewidth]{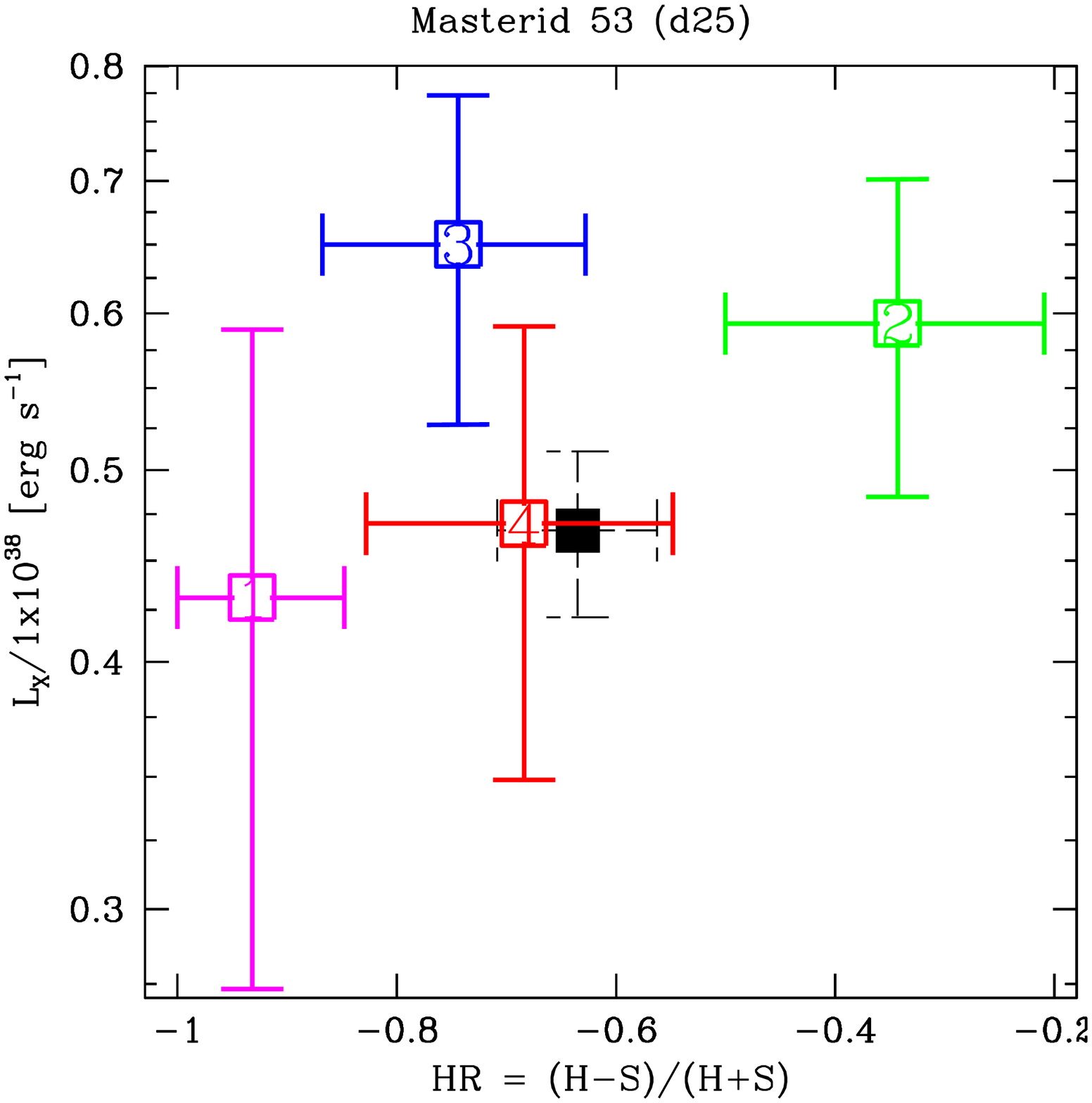}

  \end{minipage}
  \begin{minipage}{0.32\linewidth}
  \centering

    \includegraphics[width=\linewidth]{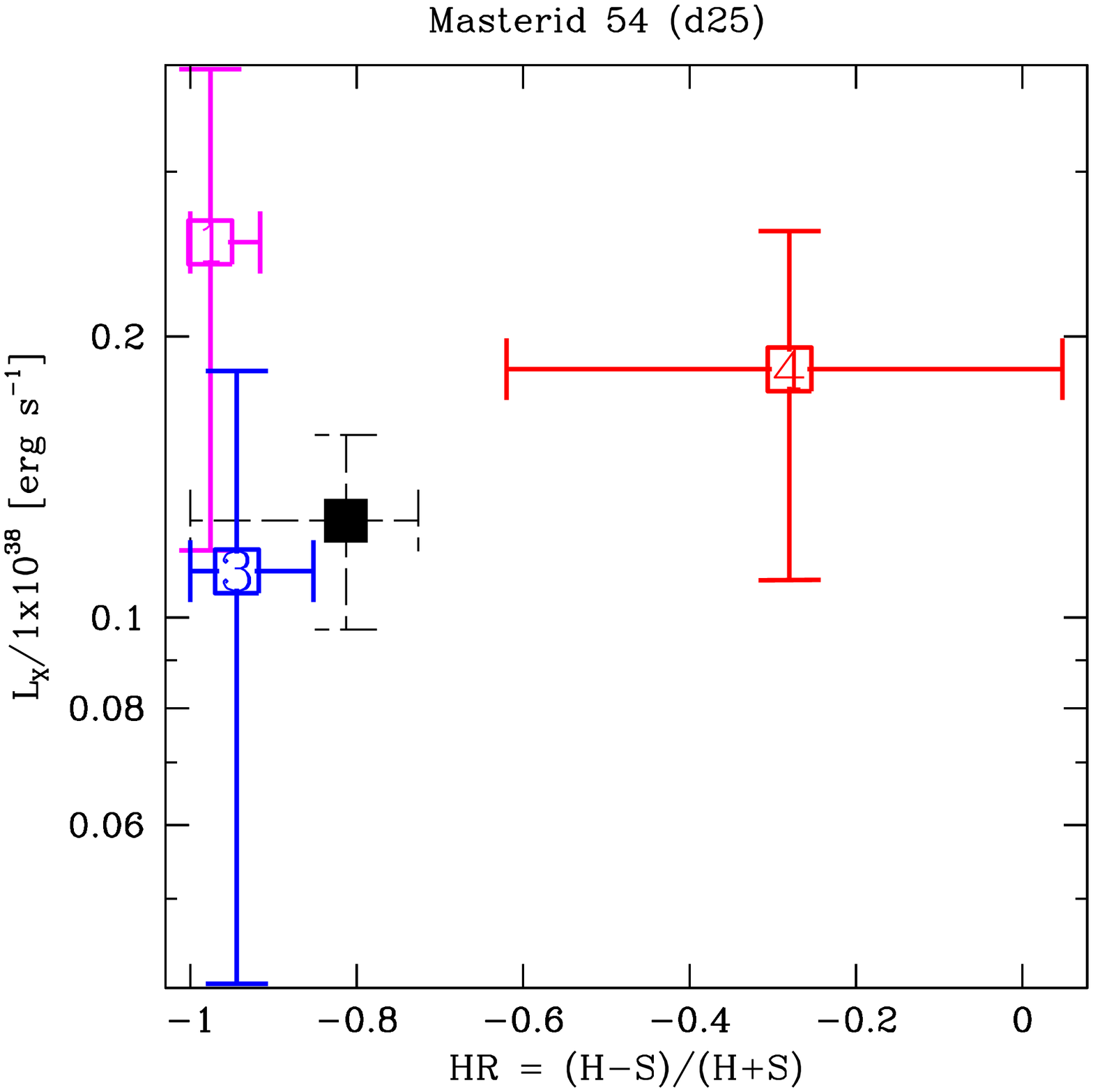}

\end{minipage}
\begin{minipage}{0.32\linewidth}
  \centering

    \includegraphics[width=\linewidth]{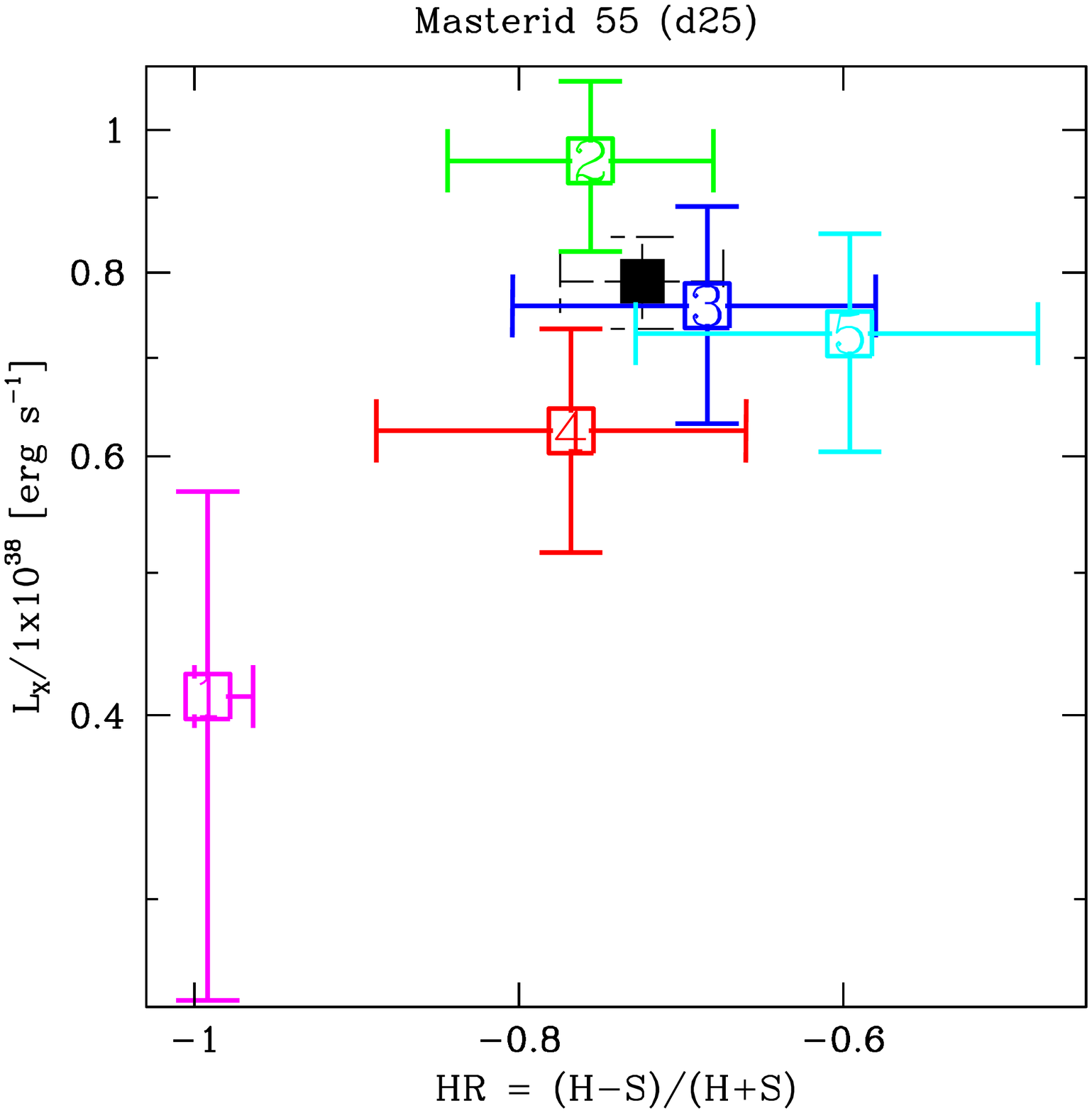}

 \end{minipage}

  \begin{minipage}{0.32\linewidth}
  \centering
  
    \includegraphics[width=\linewidth]{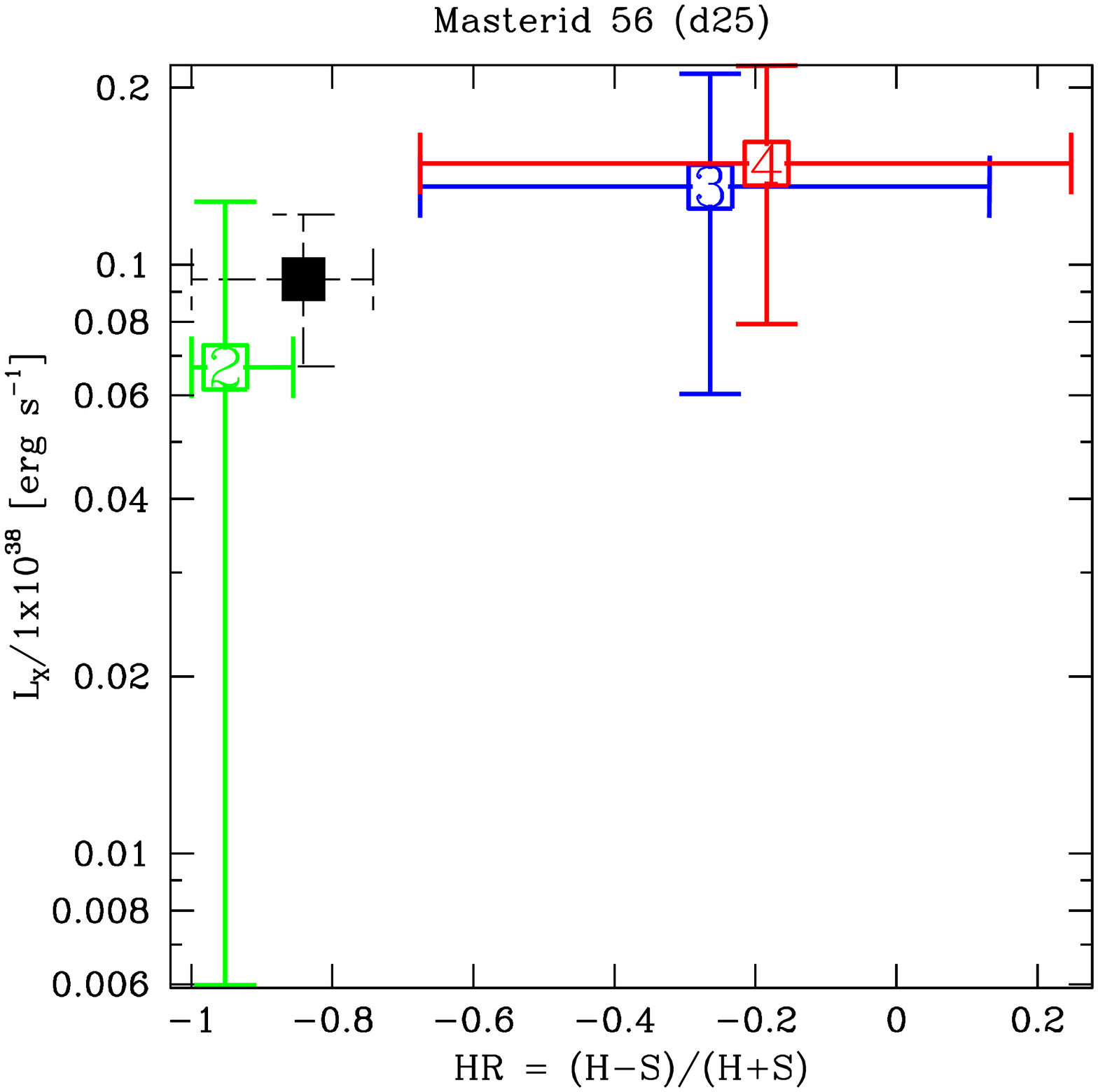}

  \end{minipage}
  \begin{minipage}{0.32\linewidth}
  \centering

    \includegraphics[width=\linewidth]{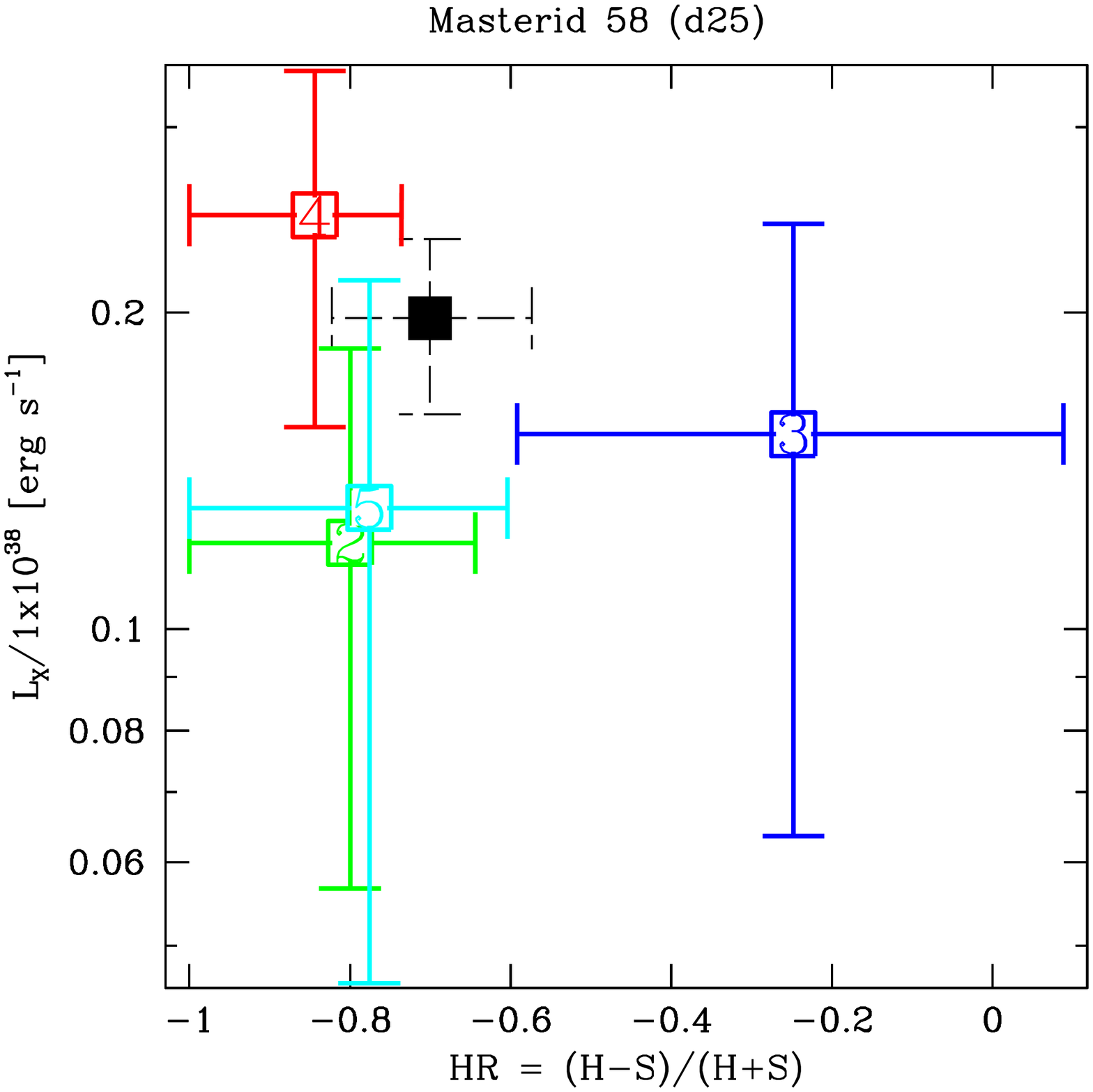}

\end{minipage}
\begin{minipage}{0.32\linewidth}
  \centering

    \includegraphics[width=\linewidth]{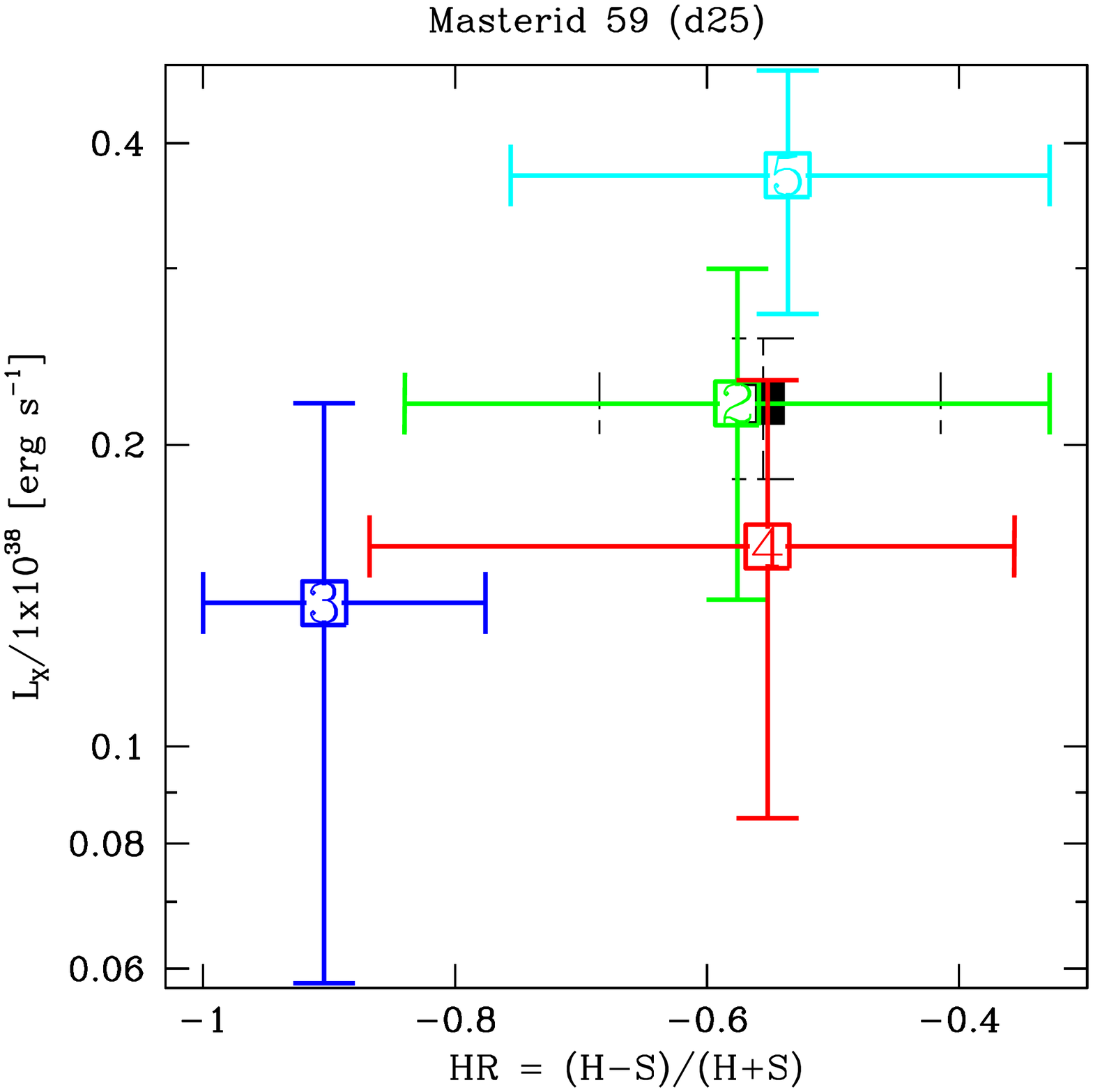}

 \end{minipage}

\begin{minipage}{0.32\linewidth}
  \centering
  
    \includegraphics[width=\linewidth]{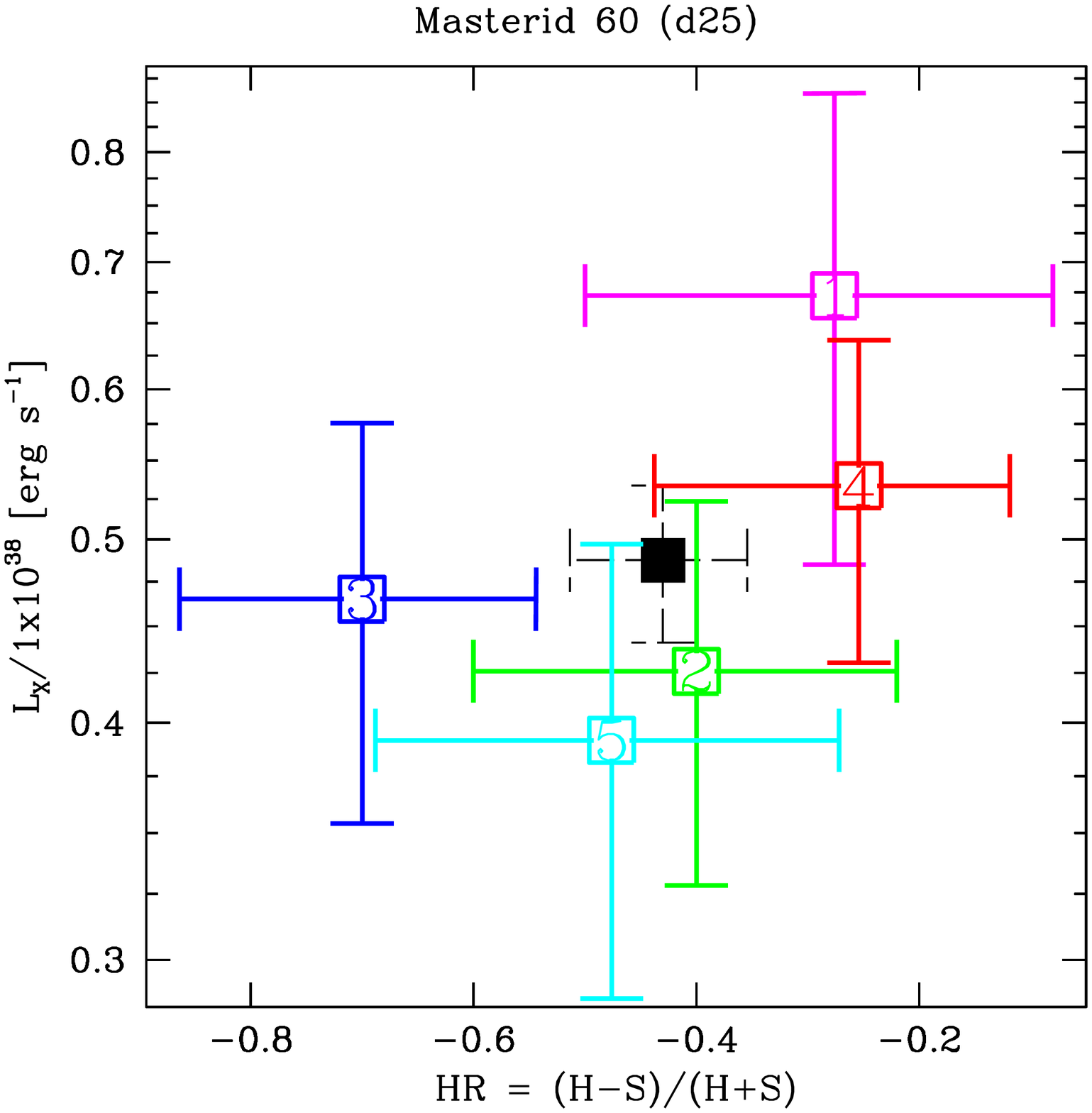}
  
  \end{minipage}
  \begin{minipage}{0.32\linewidth}
  \centering

    \includegraphics[width=\linewidth]{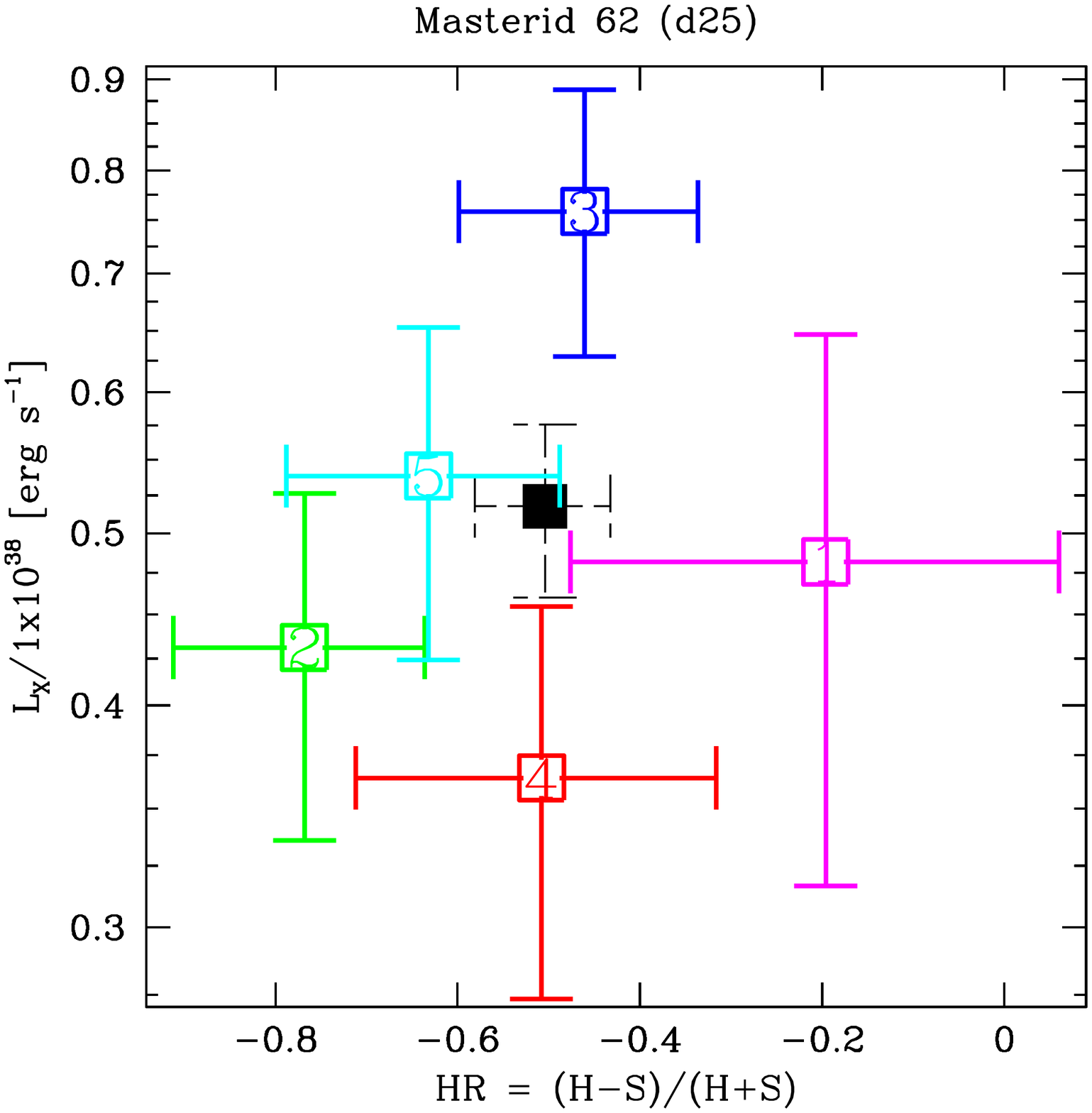}

\end{minipage}
\begin{minipage}{0.32\linewidth}
  \centering

    \includegraphics[width=\linewidth]{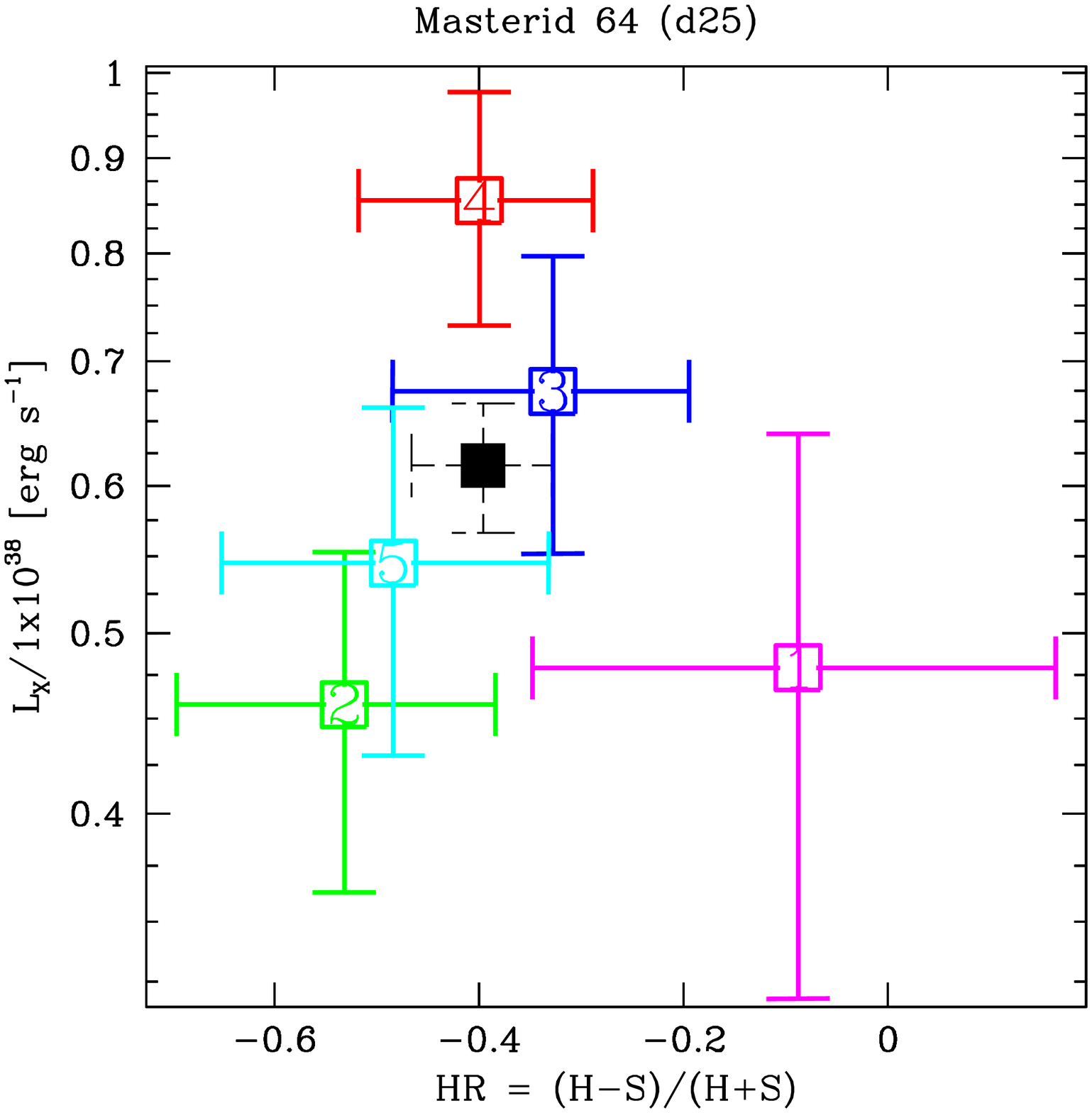}

 \end{minipage}
  
\end{figure}

\begin{figure}
  \begin{minipage}{0.32\linewidth}
  \centering
  
    \includegraphics[width=\linewidth]{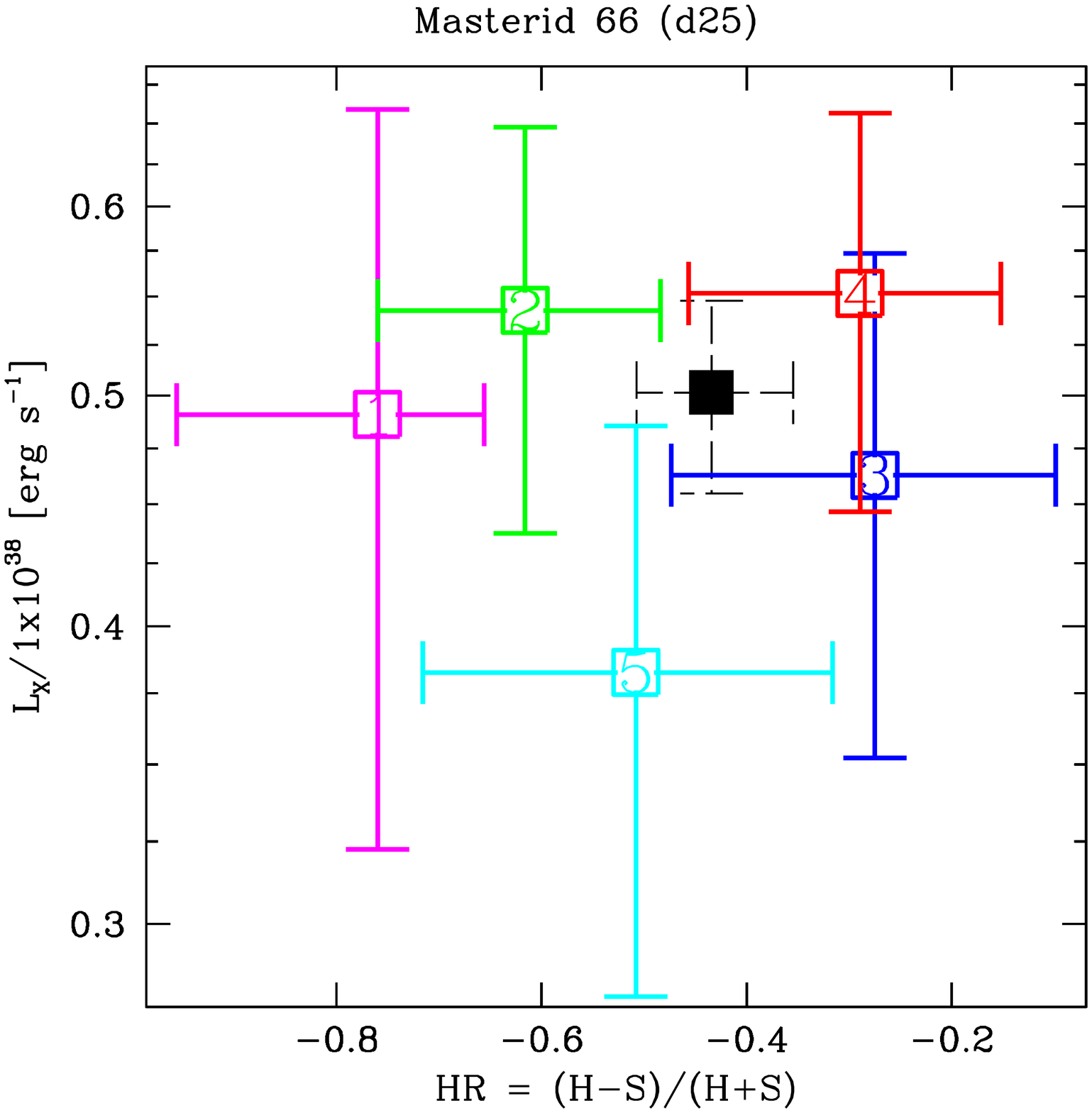}

  \end{minipage}
  \begin{minipage}{0.32\linewidth}
  \centering

    \includegraphics[width=\linewidth]{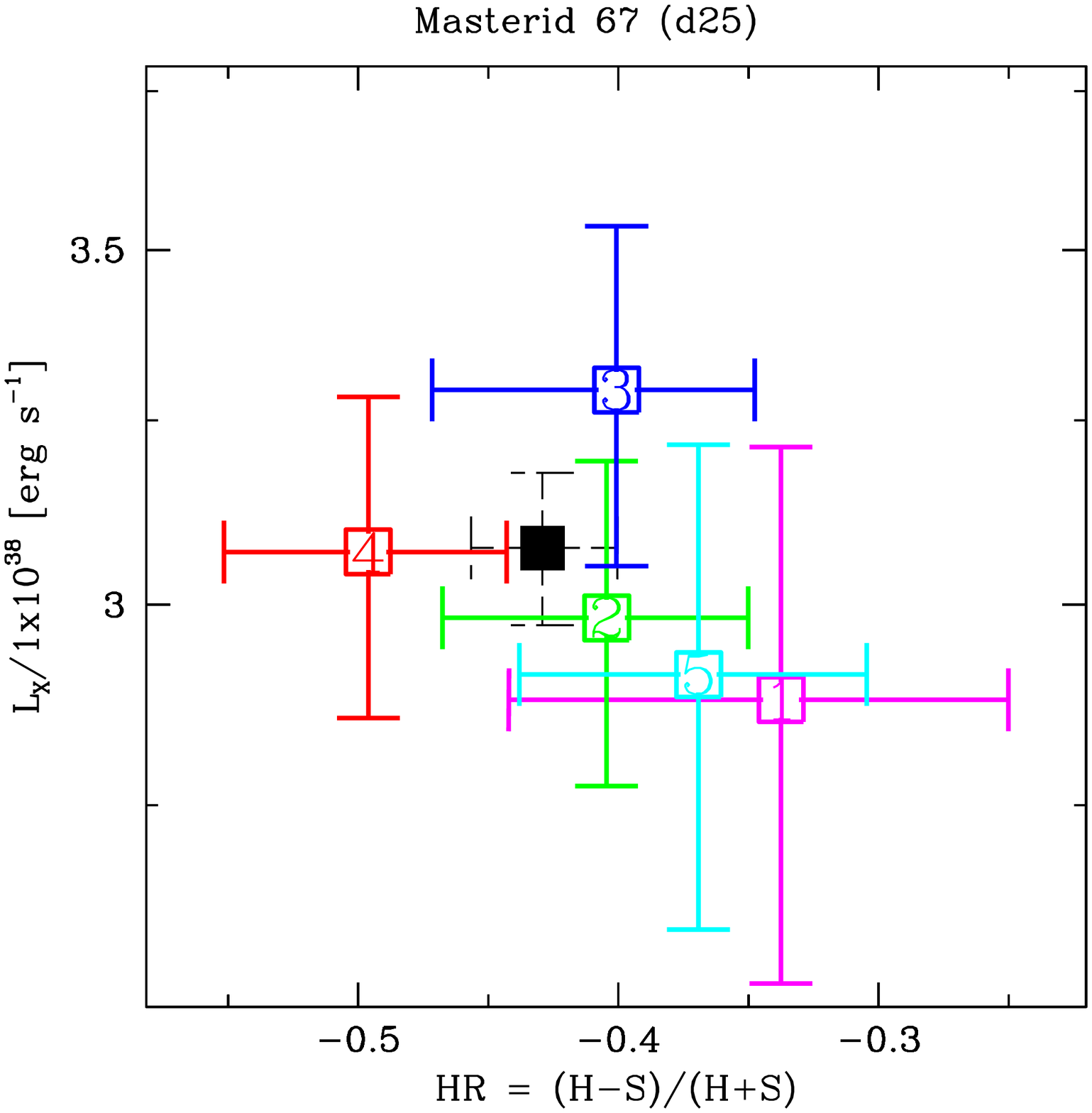}
 
\end{minipage}
\begin{minipage}{0.32\linewidth}
  \centering

    \includegraphics[width=\linewidth]{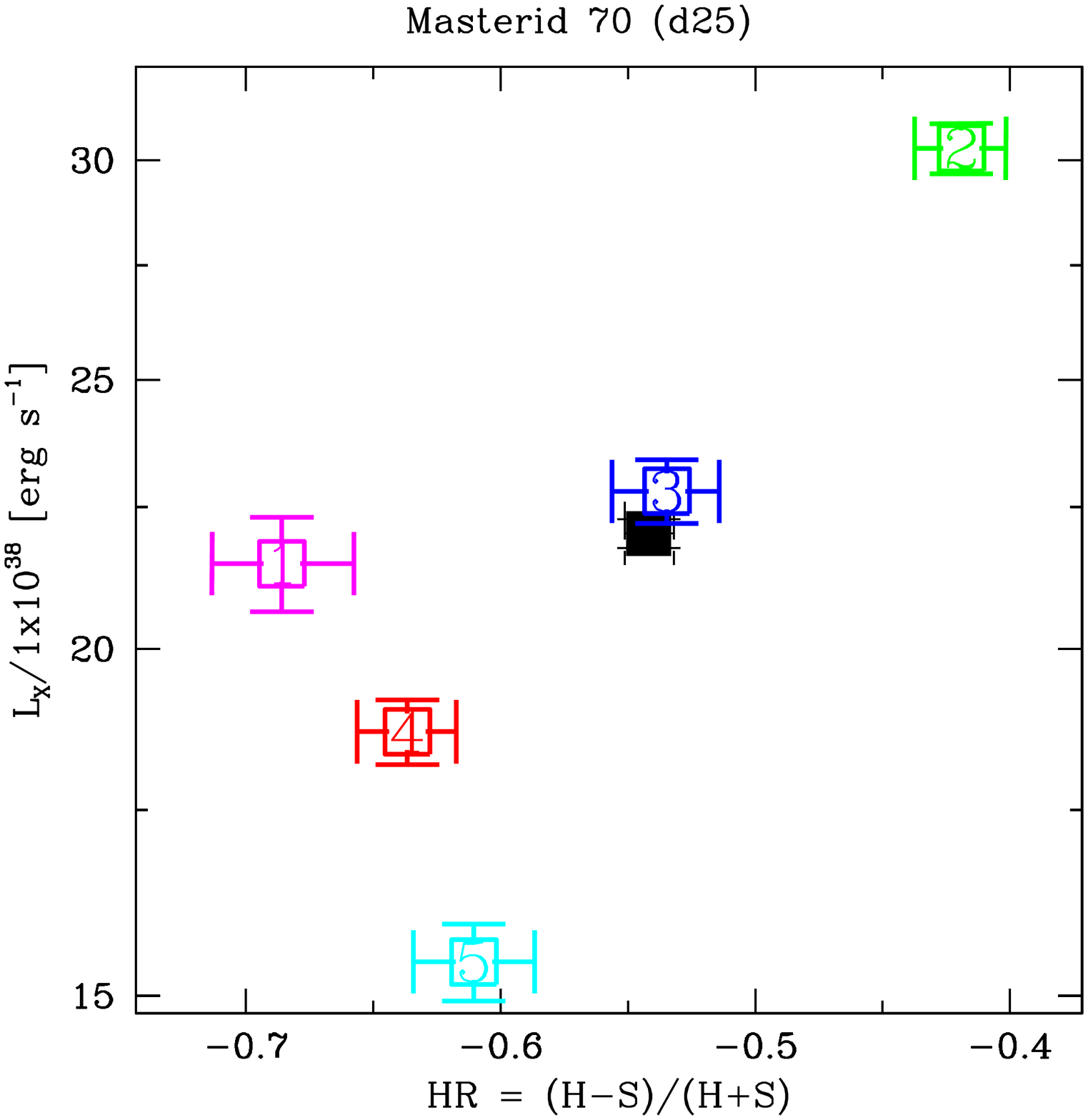}

 \end{minipage}

\begin{minipage}{0.32\linewidth}
  \centering
  
    \includegraphics[width=\linewidth]{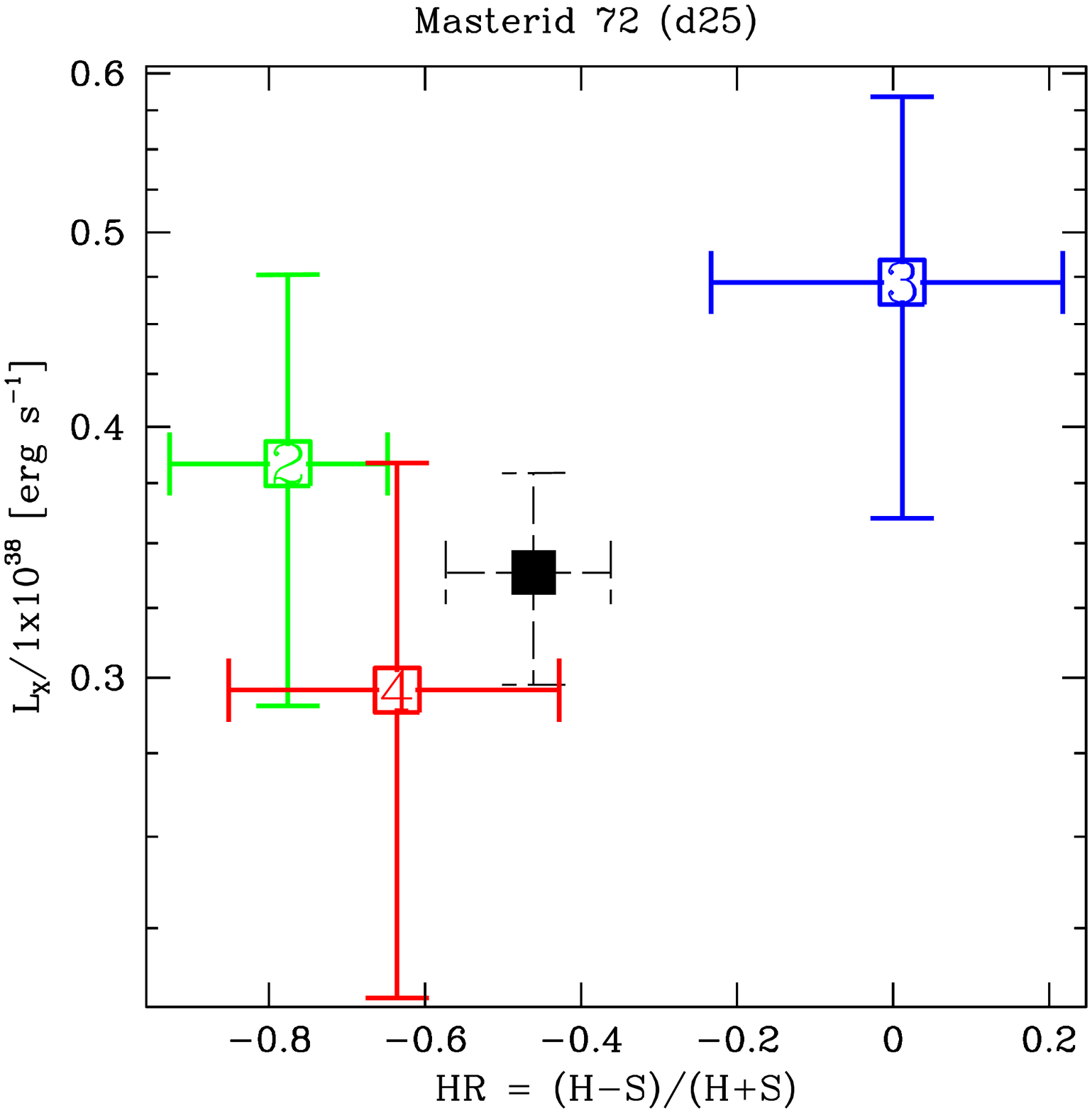}
  
  \end{minipage}
  \begin{minipage}{0.32\linewidth}
  \centering

    \includegraphics[width=\linewidth]{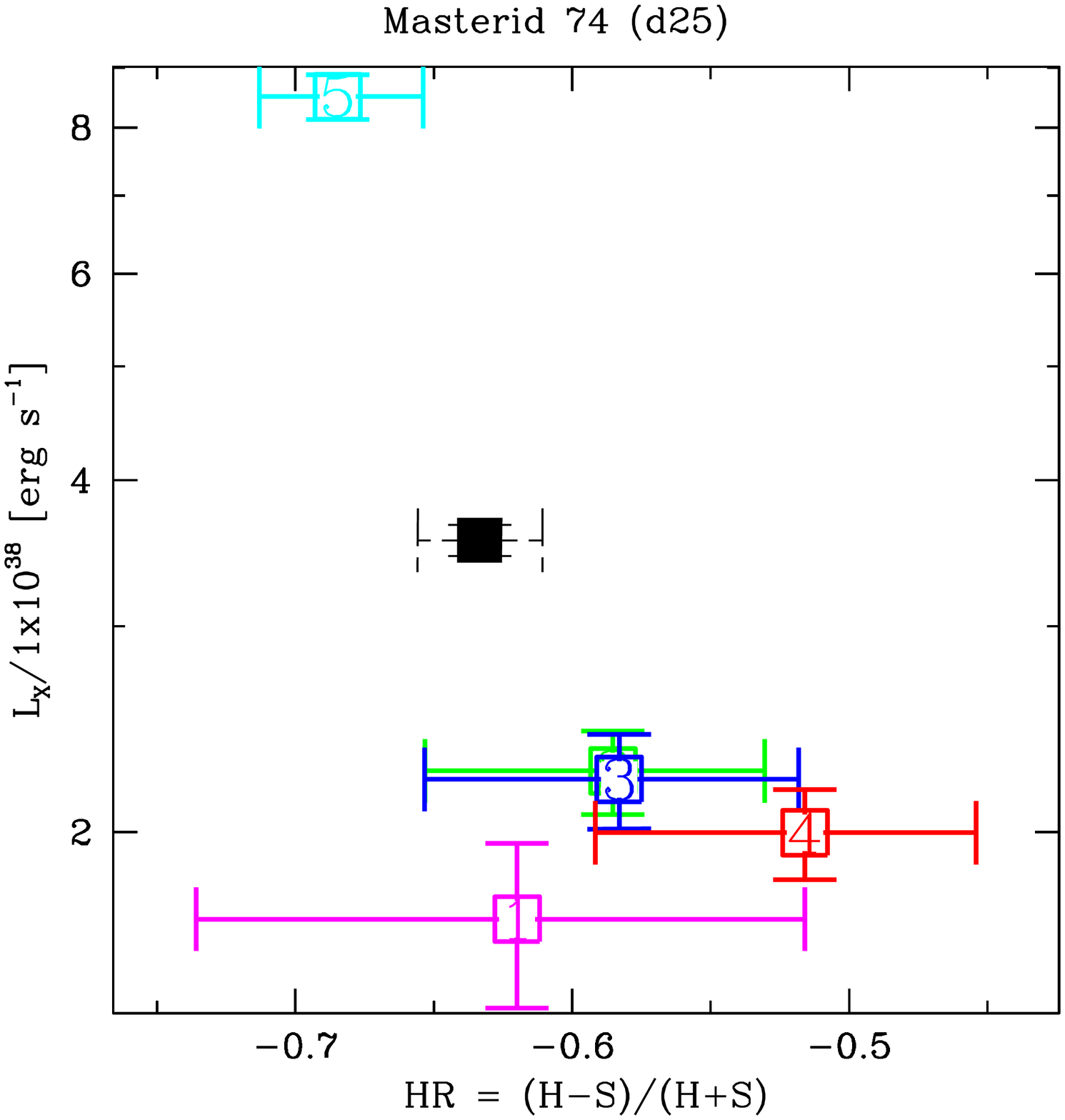}

\end{minipage}
\begin{minipage}{0.32\linewidth}
  \centering

    \includegraphics[width=\linewidth]{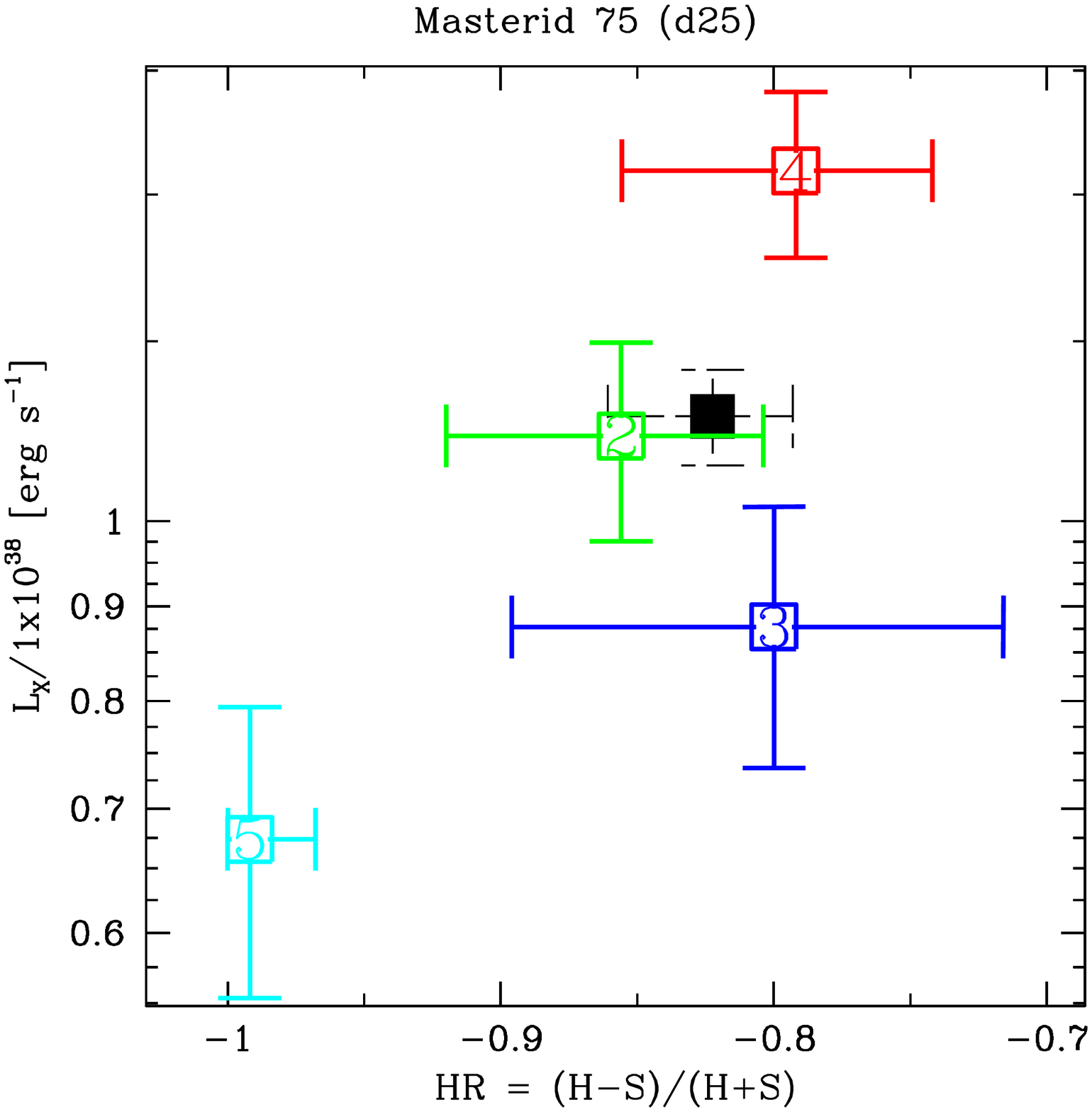}

 \end{minipage}

  \begin{minipage}{0.32\linewidth}
  \centering
  
    \includegraphics[width=\linewidth]{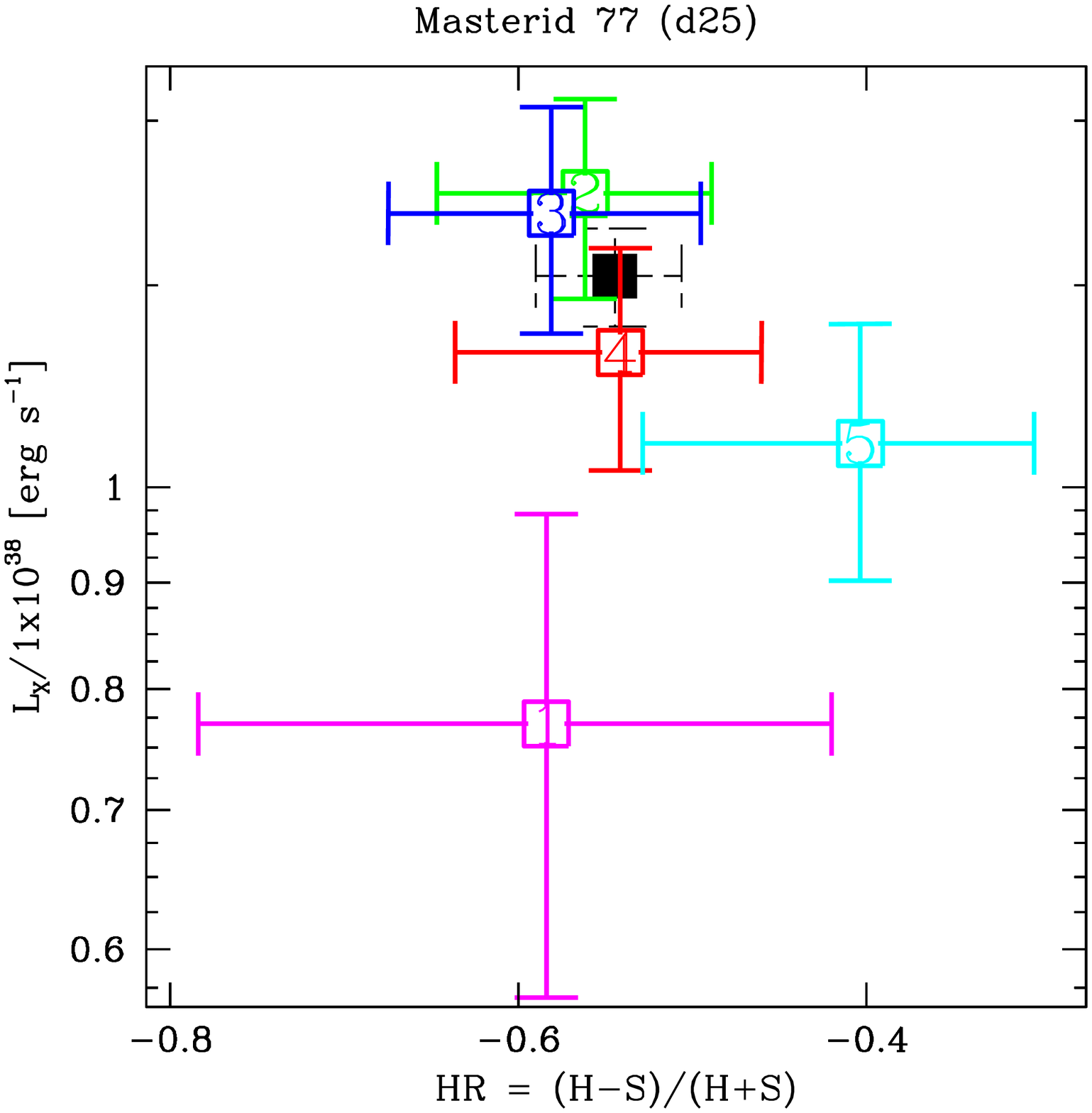}
  
  \end{minipage}
  \begin{minipage}{0.32\linewidth}
  \centering

    \includegraphics[width=\linewidth]{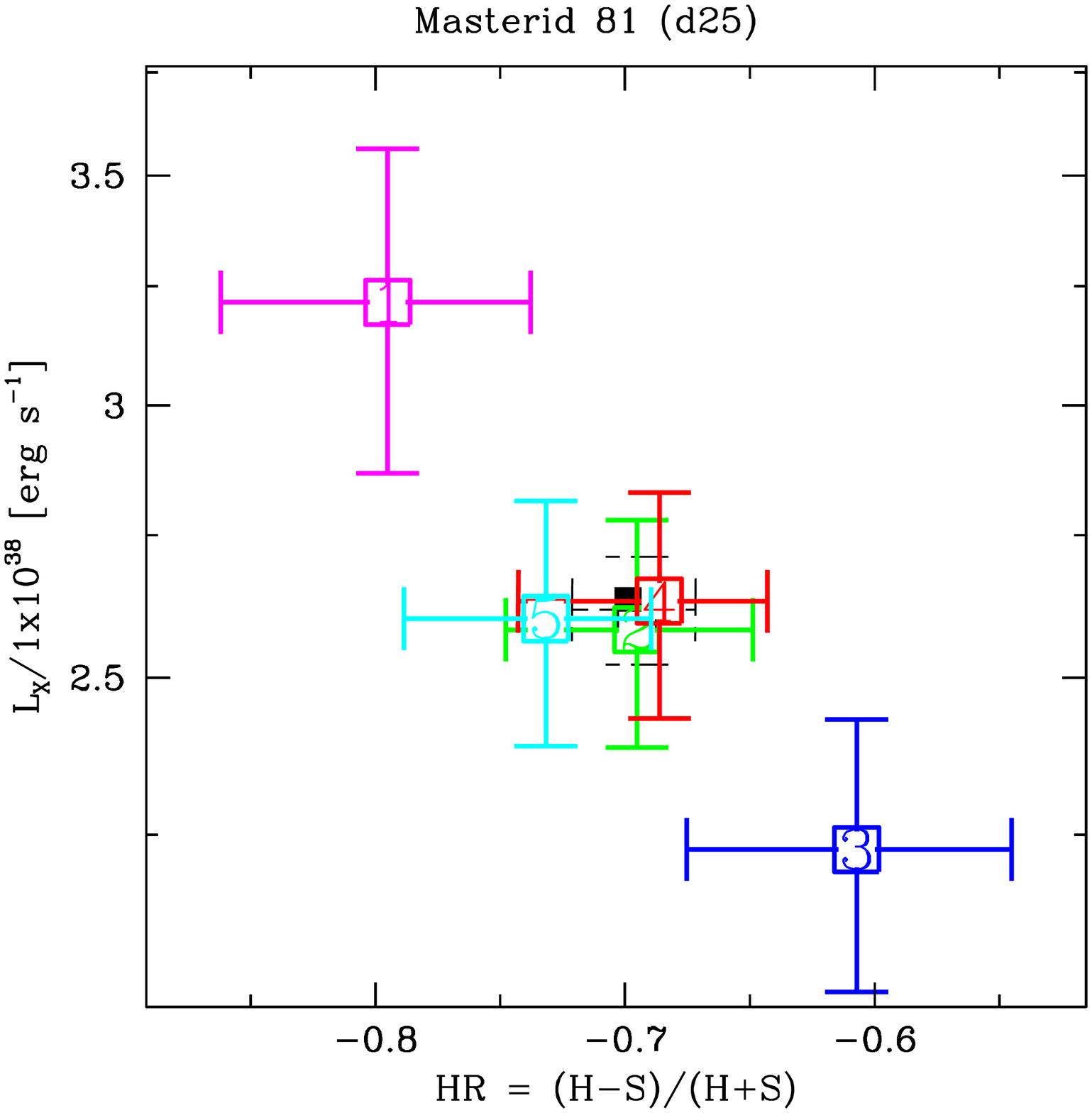}

\end{minipage}
\begin{minipage}{0.32\linewidth}
  \centering

    \includegraphics[width=\linewidth]{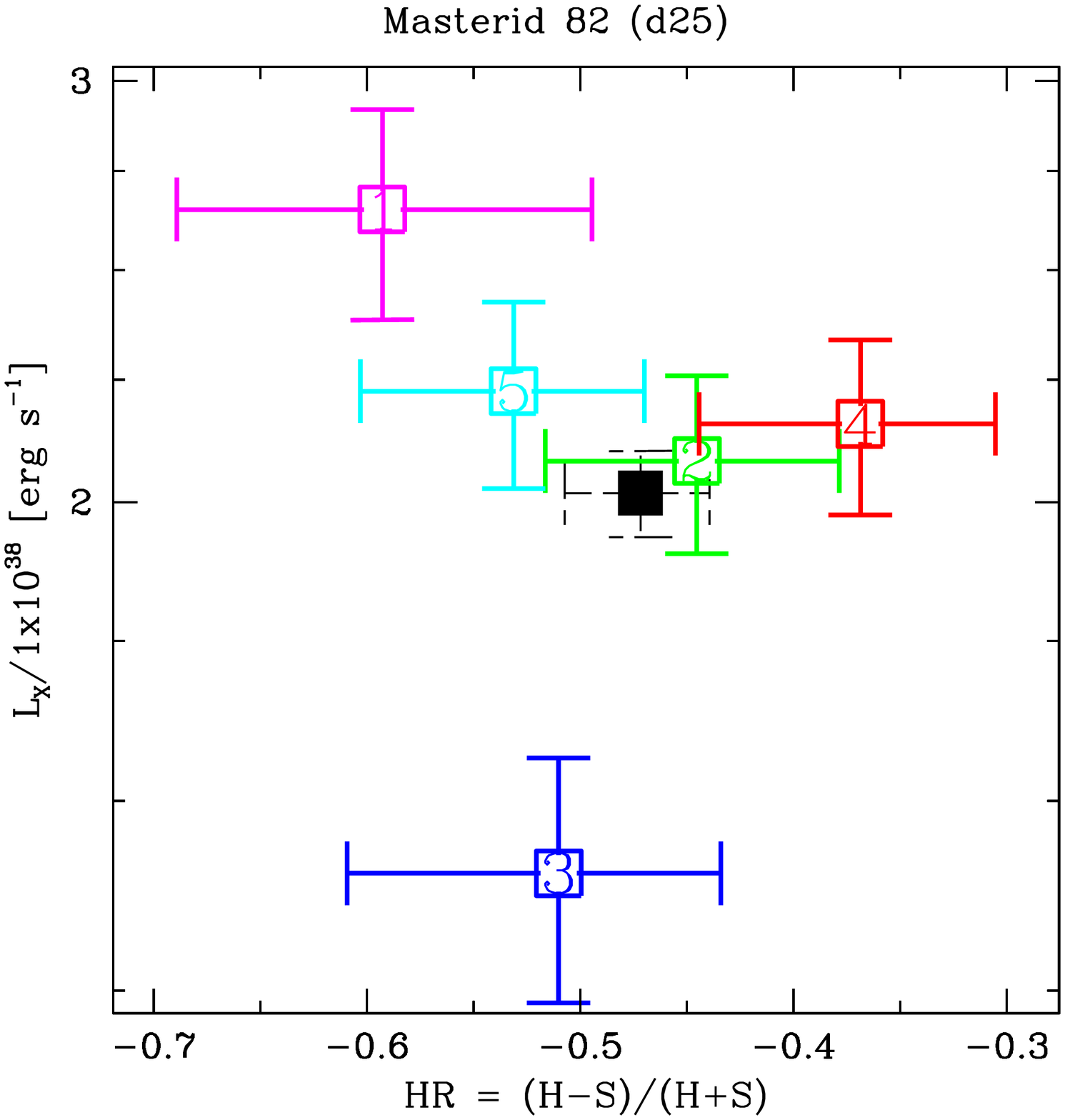}

 \end{minipage}

\begin{minipage}{0.32\linewidth}
  \centering
  
    \includegraphics[width=\linewidth]{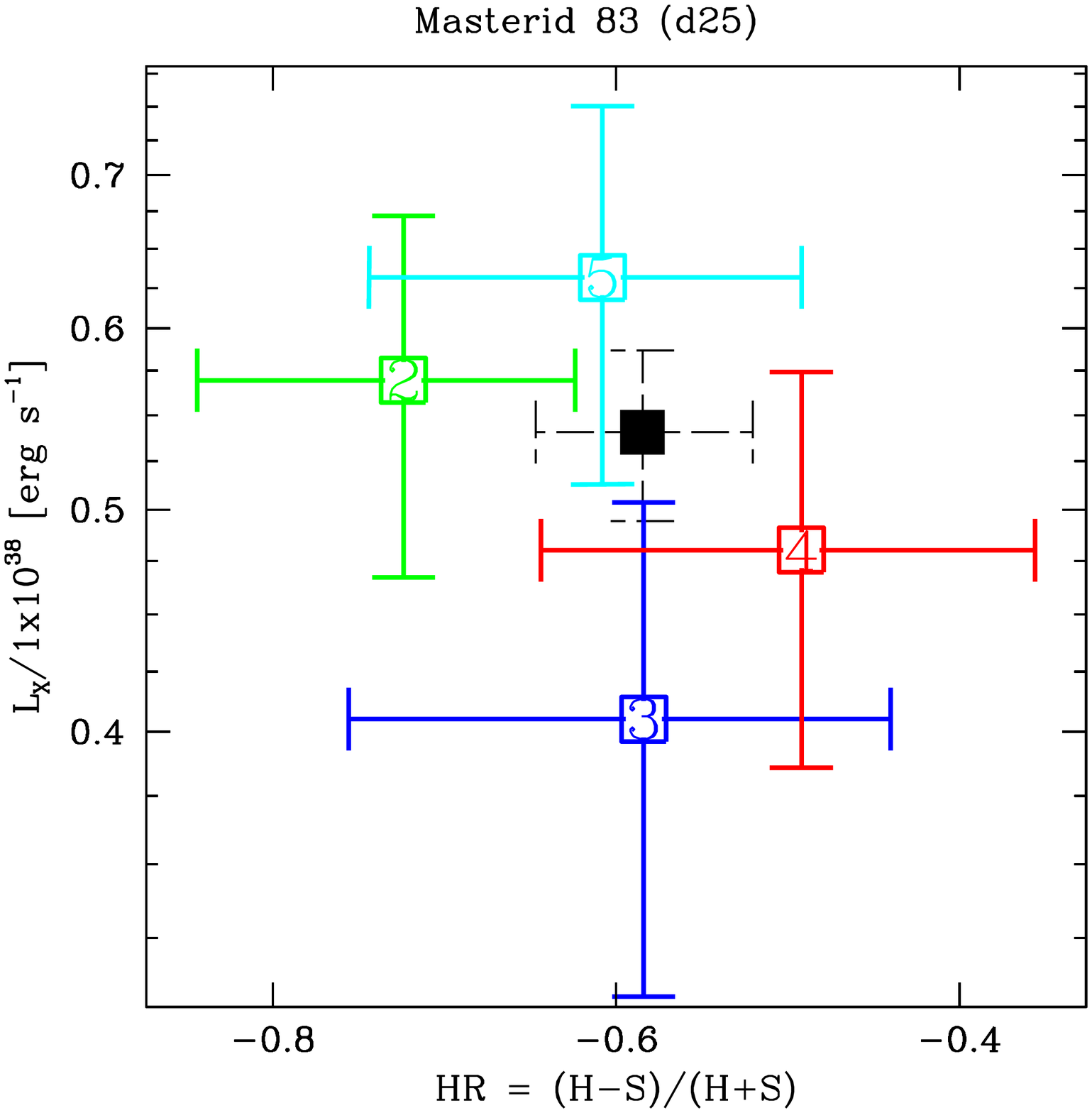}
  
  \end{minipage}
  \begin{minipage}{0.32\linewidth}
  \centering

    \includegraphics[width=\linewidth]{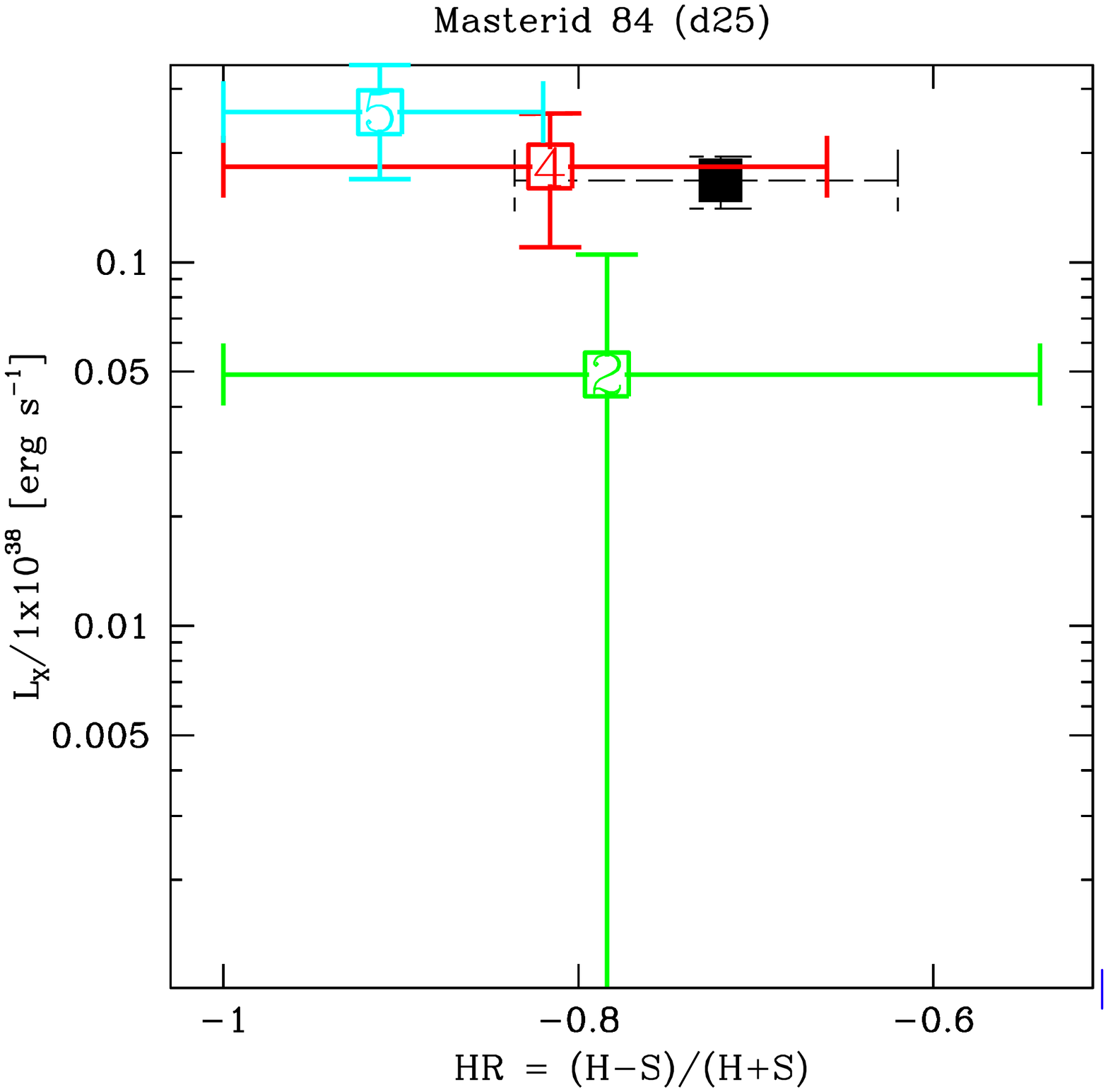}

\end{minipage}
\begin{minipage}{0.32\linewidth}
  \centering

    \includegraphics[width=\linewidth]{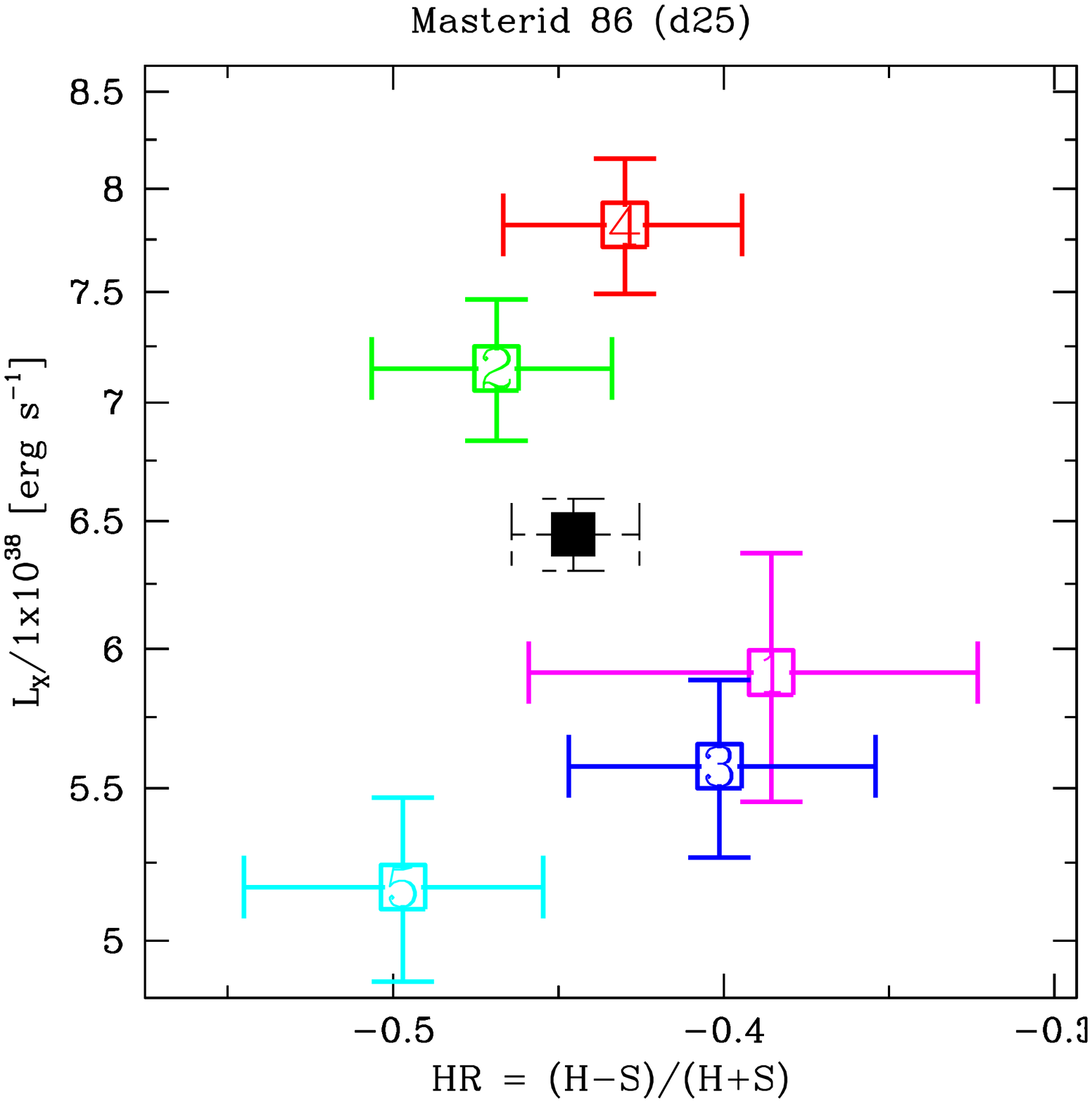}

 \end{minipage}
  
\end{figure}

\clearpage

\begin{figure}
  \begin{minipage}{0.32\linewidth}
  \centering
  
    \includegraphics[width=\linewidth]{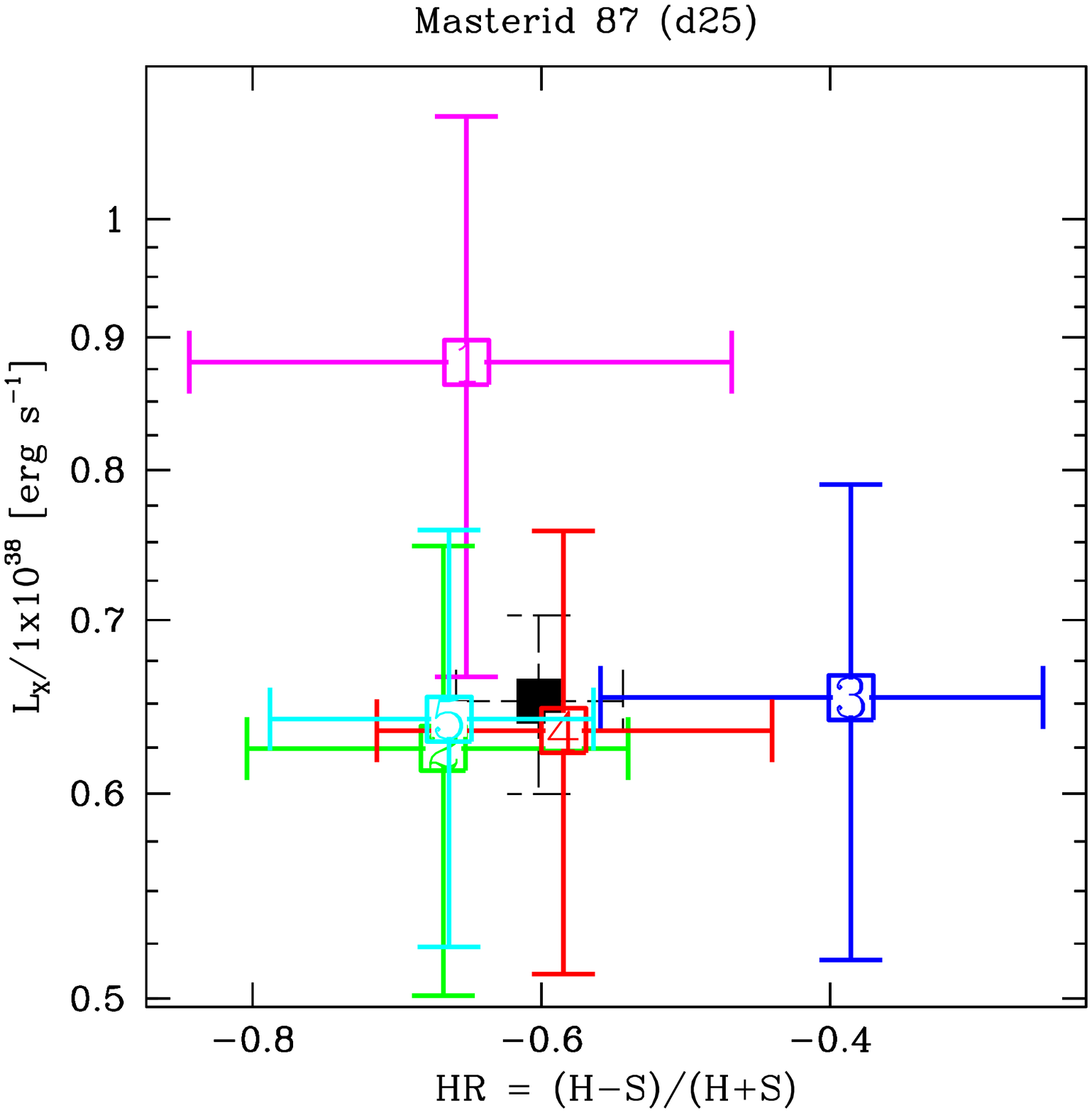}

  \end{minipage}
  \begin{minipage}{0.32\linewidth}
  \centering

    \includegraphics[width=\linewidth]{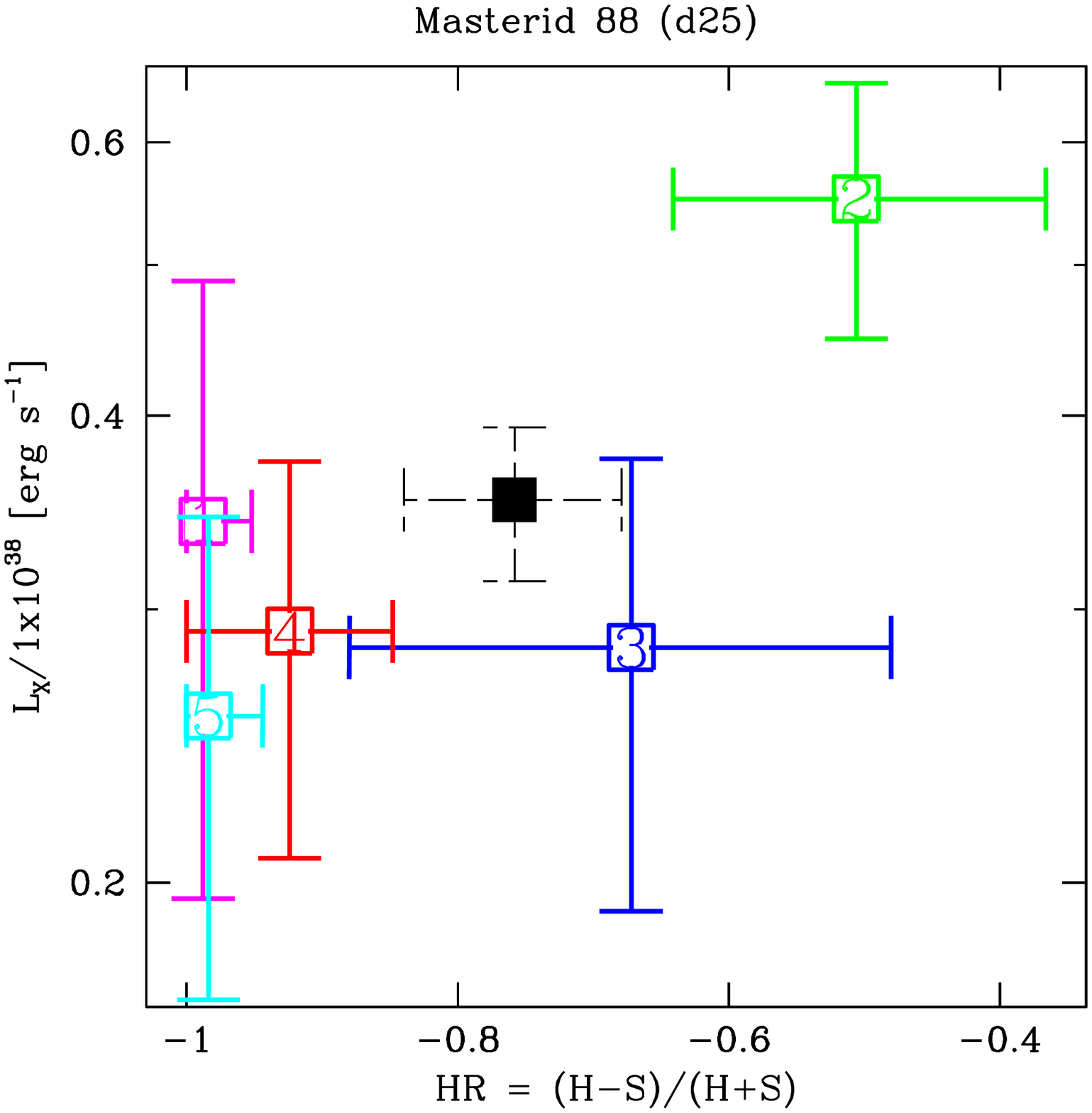}

\end{minipage}
\begin{minipage}{0.32\linewidth}
  \centering

    \includegraphics[width=\linewidth]{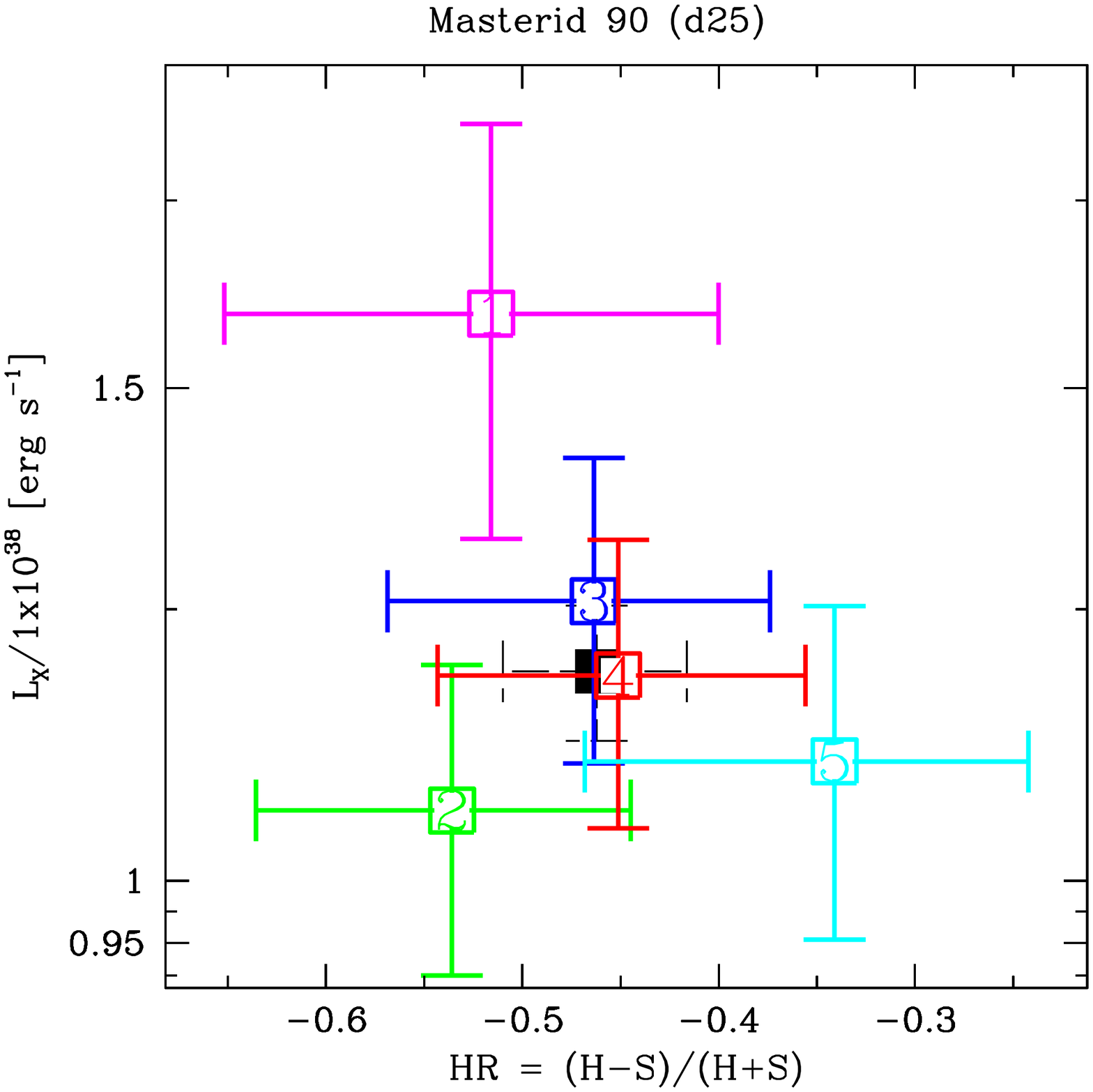}

 \end{minipage}

\begin{minipage}{0.32\linewidth}
  \centering
  
    \includegraphics[width=\linewidth]{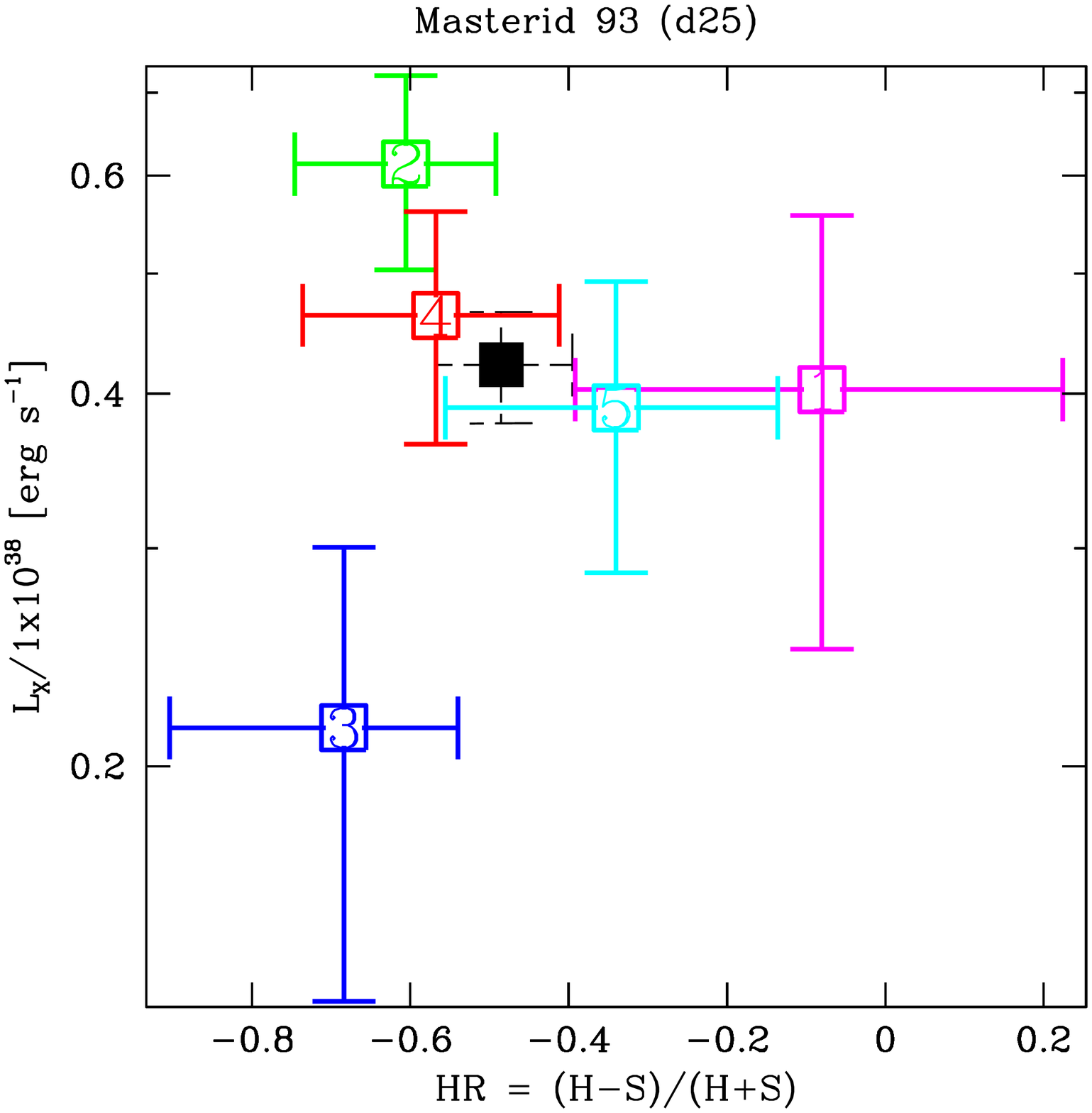}

  \end{minipage}
  \begin{minipage}{0.32\linewidth}
  \centering

    \includegraphics[width=\linewidth]{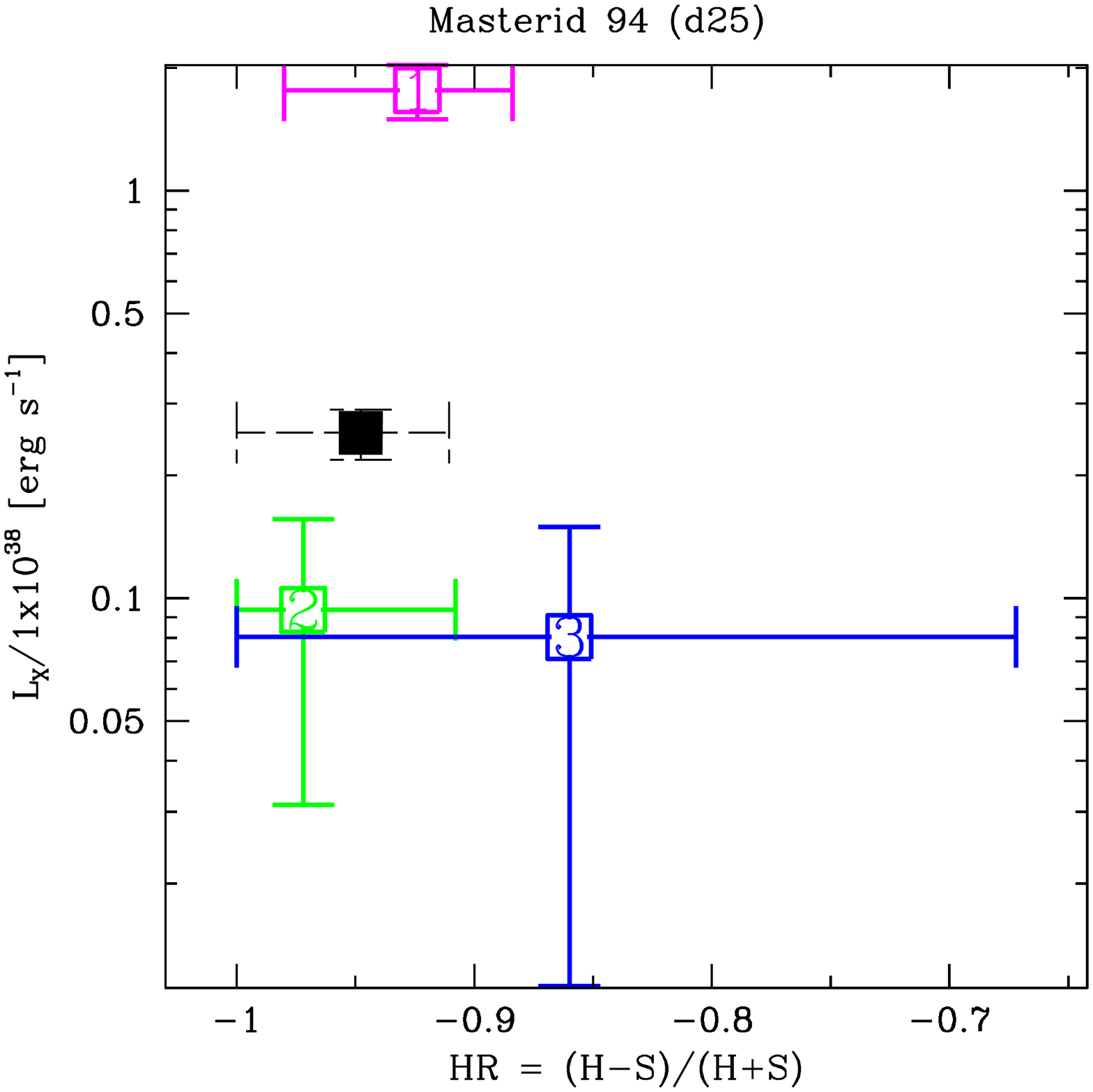}

\end{minipage}
\begin{minipage}{0.32\linewidth}
  \centering

    \includegraphics[width=\linewidth]{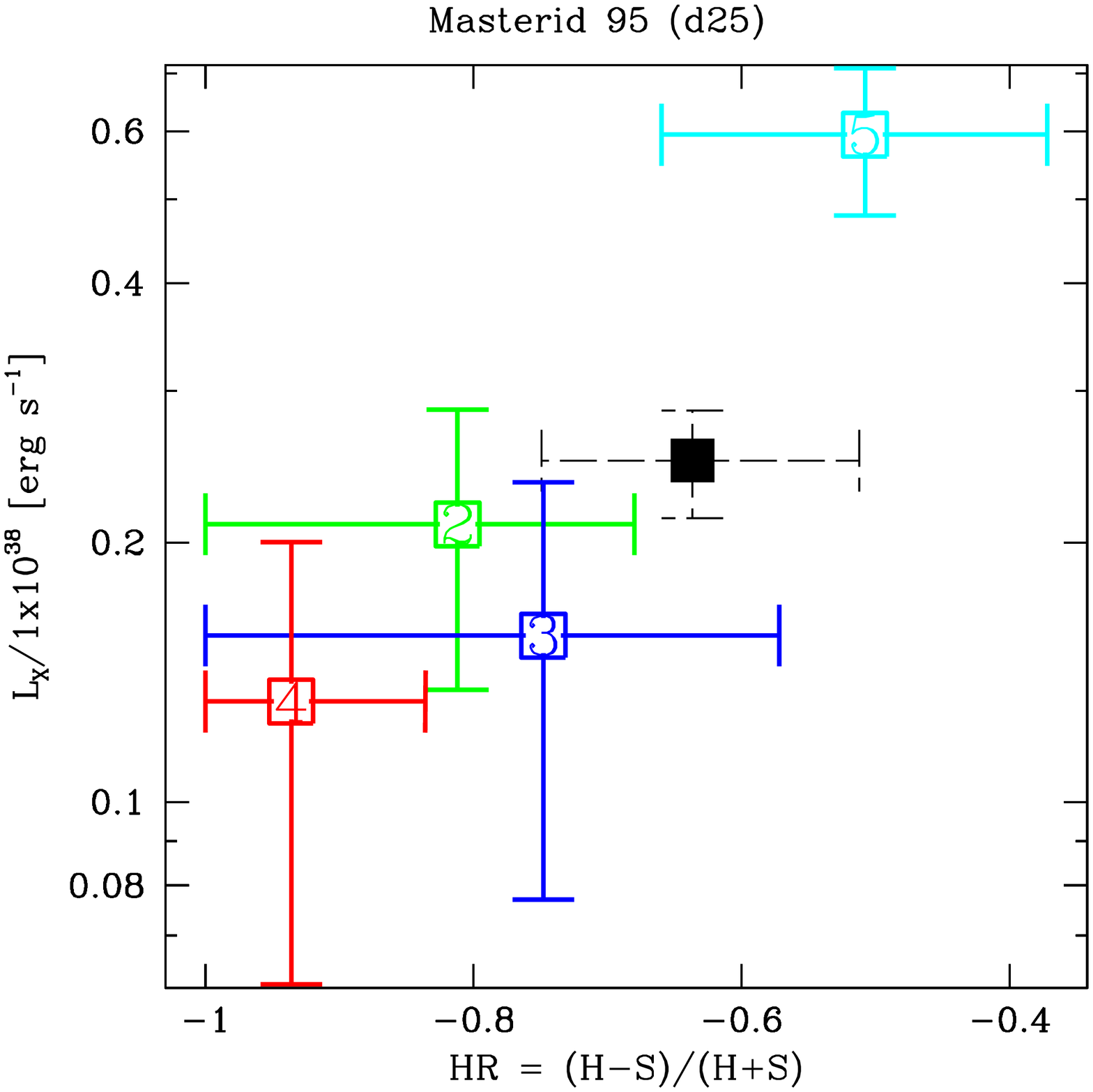}

 \end{minipage}

  \begin{minipage}{0.32\linewidth}
  \centering
  
    \includegraphics[width=\linewidth]{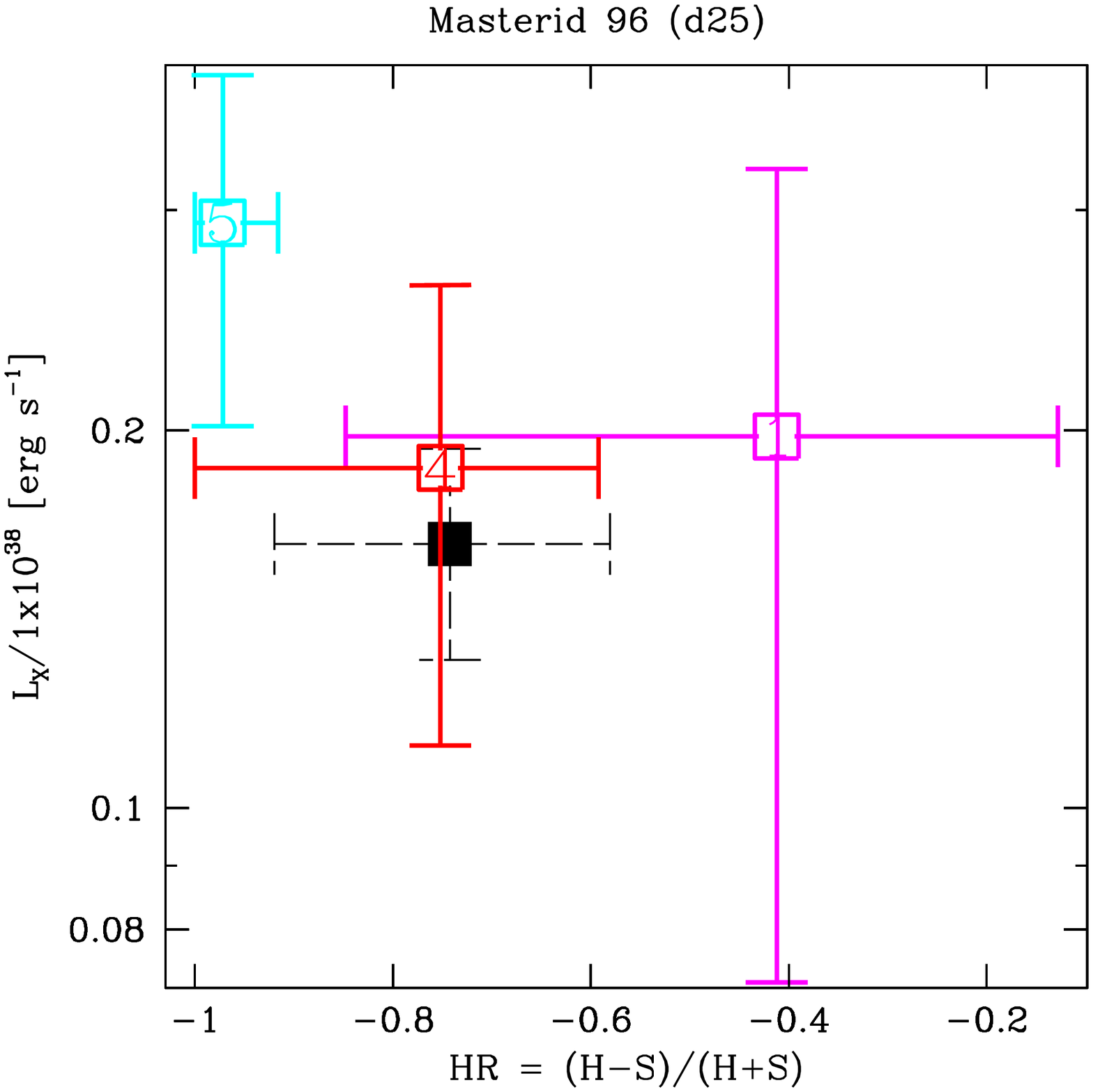}
  
  \end{minipage}
  \begin{minipage}{0.32\linewidth}
  \centering

    \includegraphics[width=\linewidth]{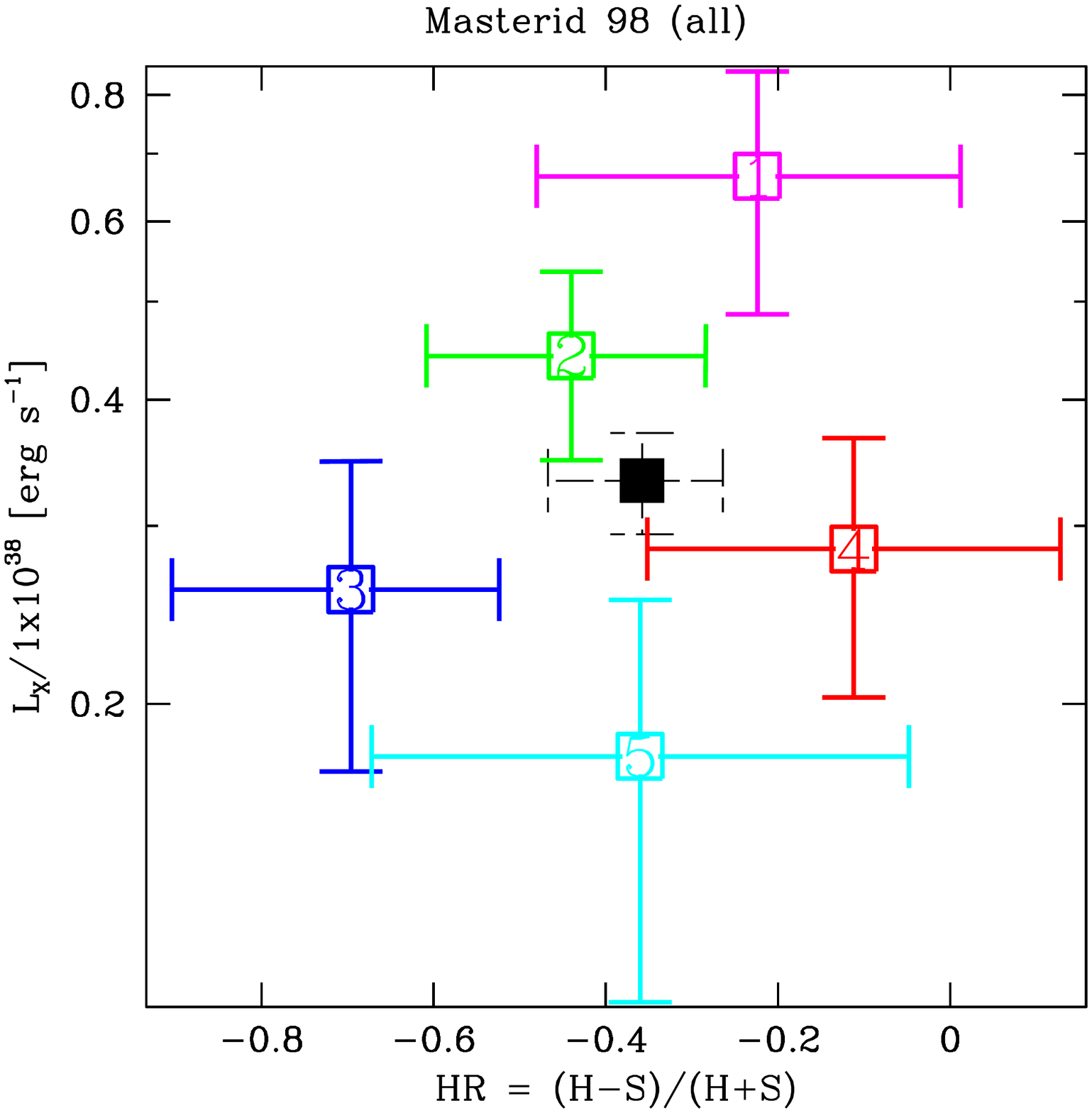}

\end{minipage}
\begin{minipage}{0.32\linewidth}
  \centering

    \includegraphics[width=\linewidth]{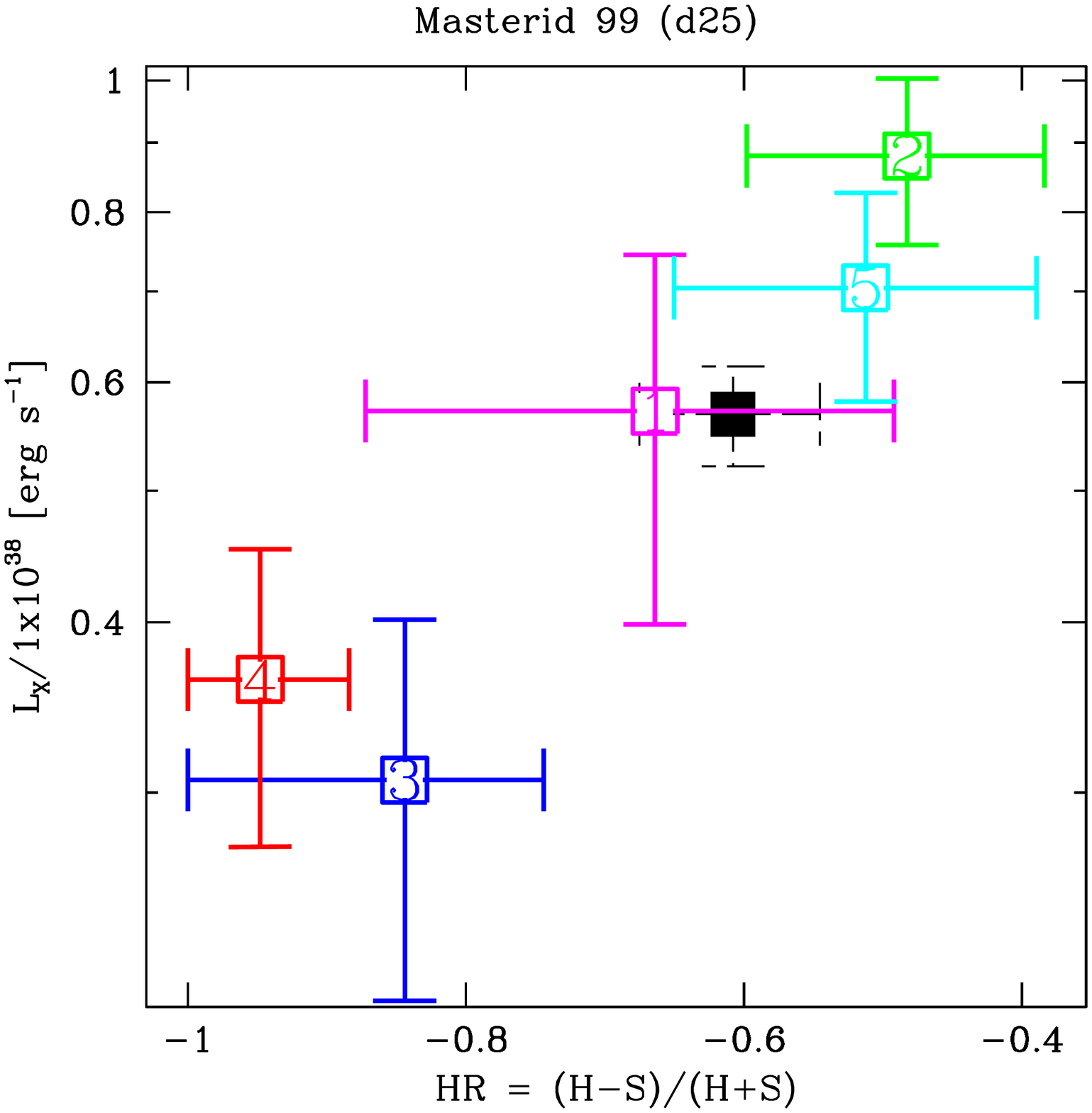}

 \end{minipage}

\begin{minipage}{0.32\linewidth}
  \centering
  
    \includegraphics[width=\linewidth]{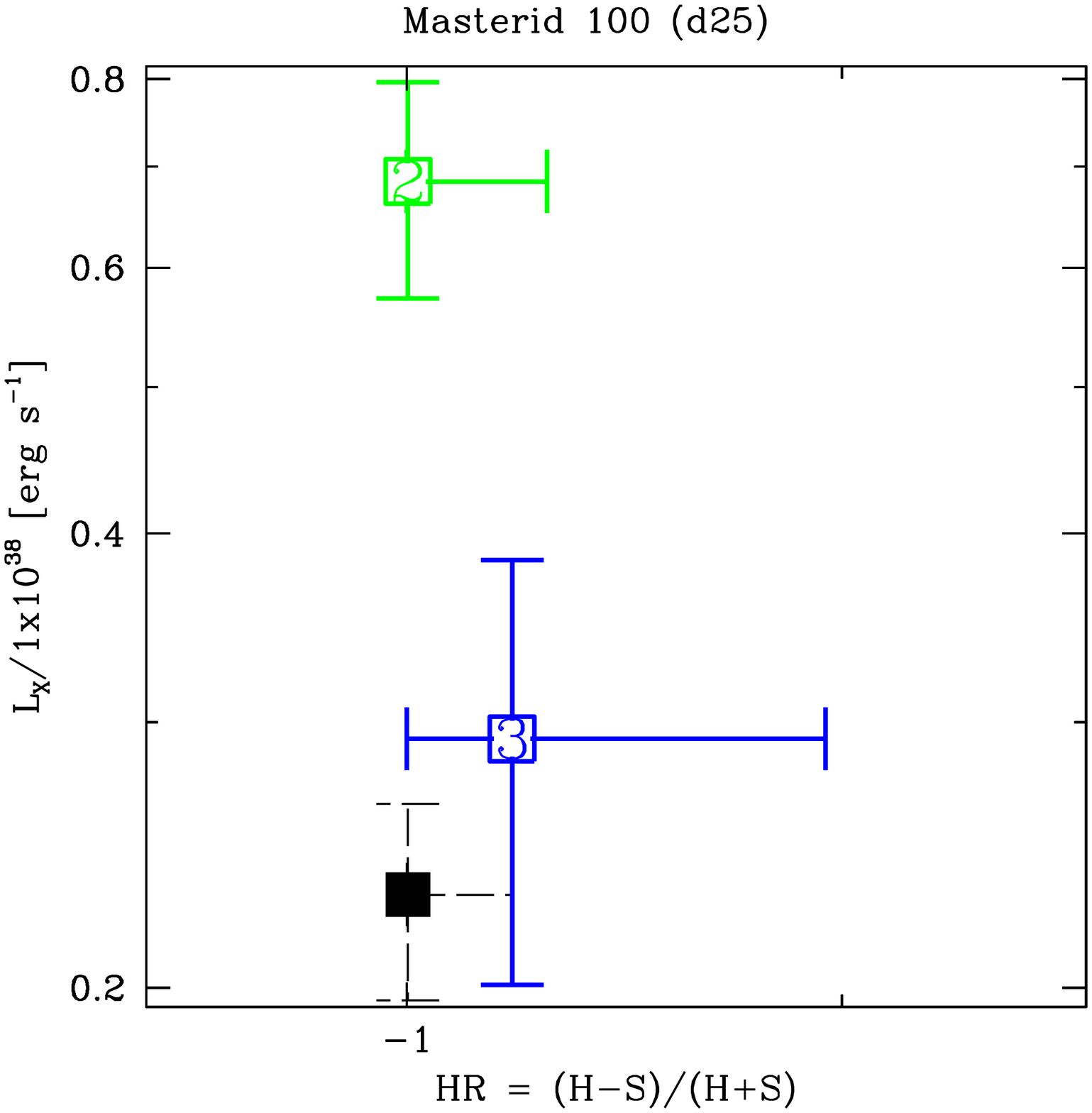}
  
  \end{minipage}
  \begin{minipage}{0.32\linewidth}
  \centering

    \includegraphics[width=\linewidth]{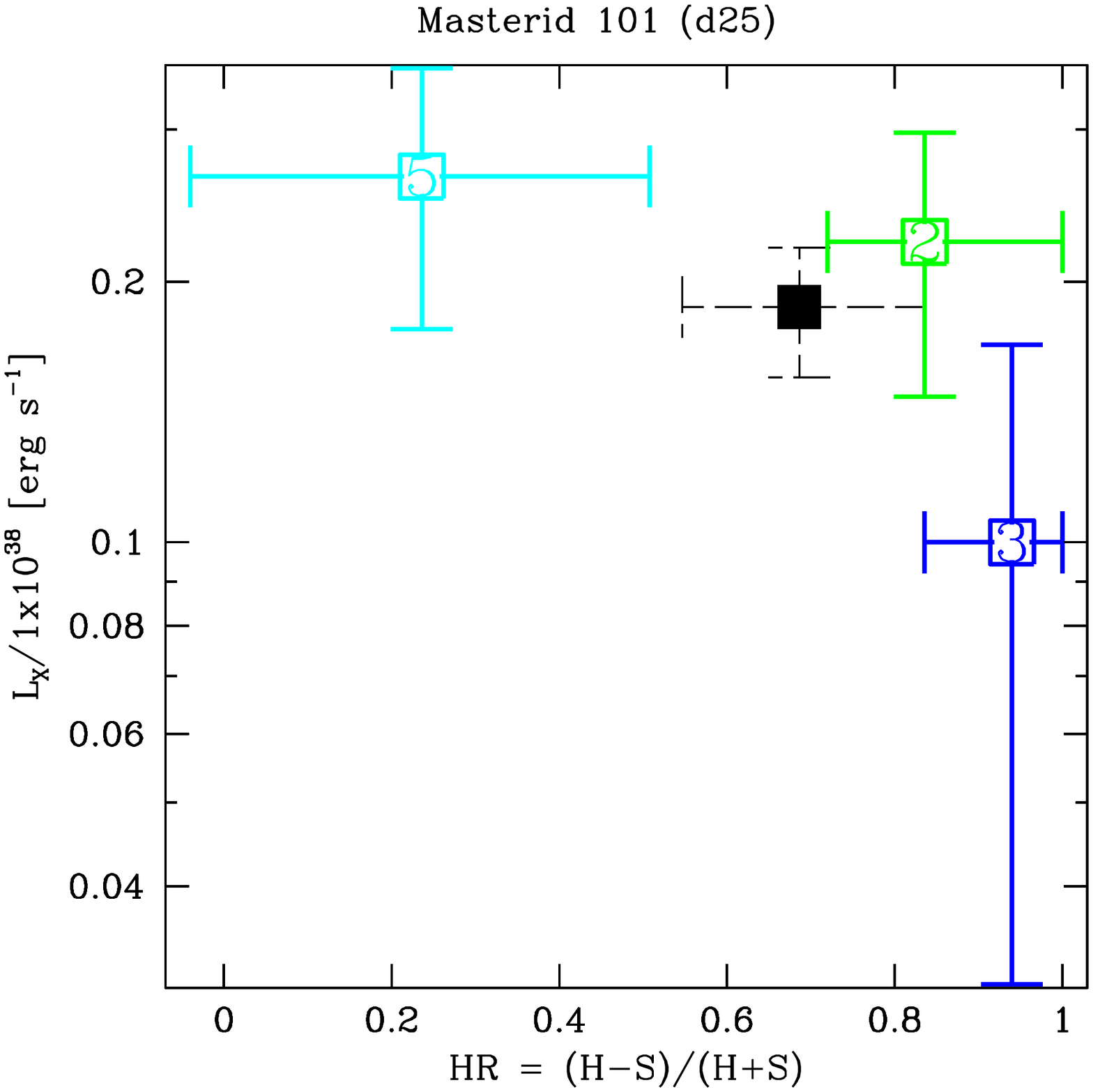}

\end{minipage}
\begin{minipage}{0.32\linewidth}
  \centering

    \includegraphics[width=\linewidth]{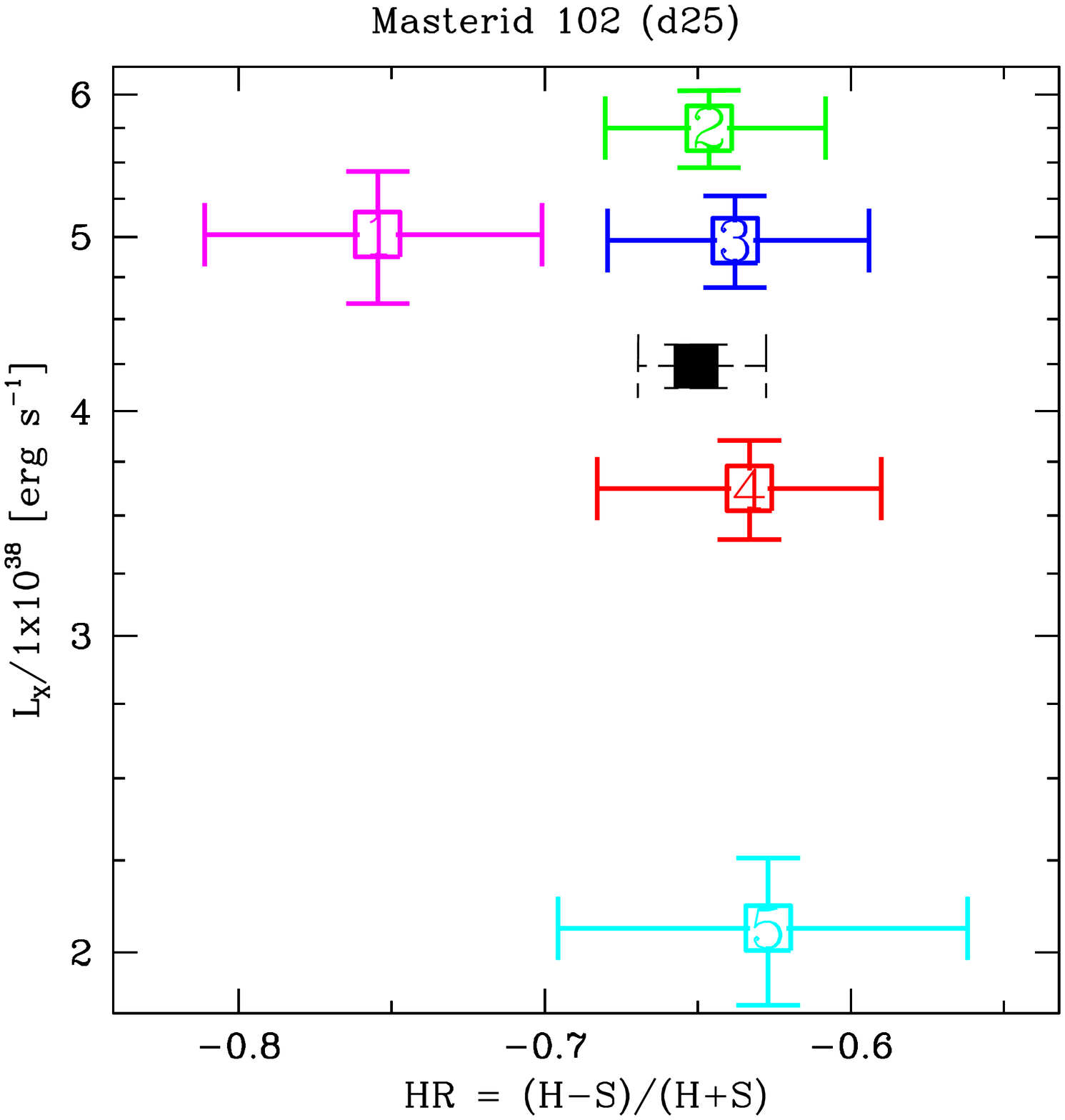}

 \end{minipage}
  
\end{figure}

\begin{figure}
  \begin{minipage}{0.32\linewidth}
  \centering
  
    \includegraphics[width=\linewidth]{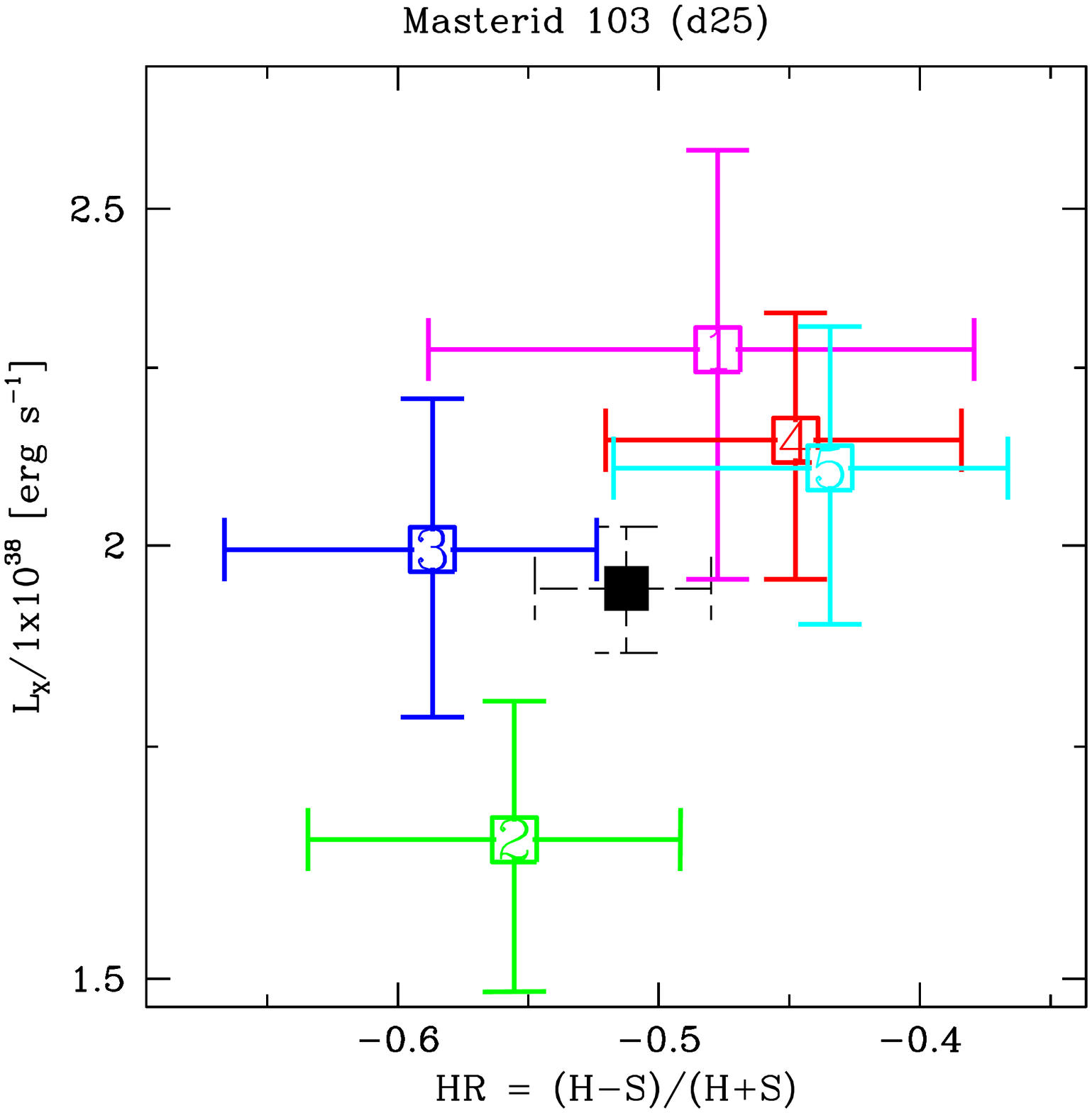}
  
  \end{minipage}
  \begin{minipage}{0.32\linewidth}
  \centering

    \includegraphics[width=\linewidth]{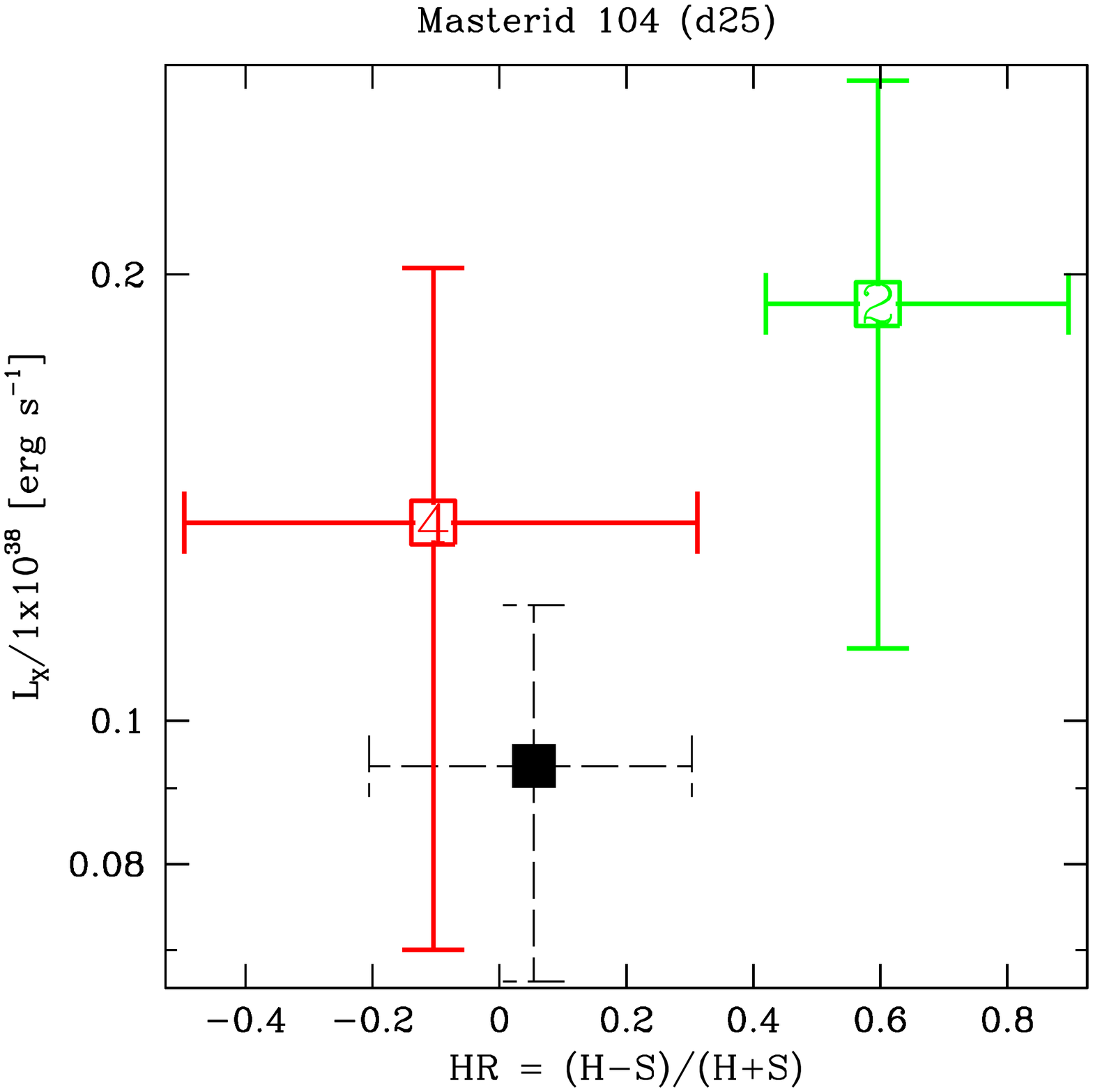}

\end{minipage}
\begin{minipage}{0.32\linewidth}
  \centering

    \includegraphics[width=\linewidth]{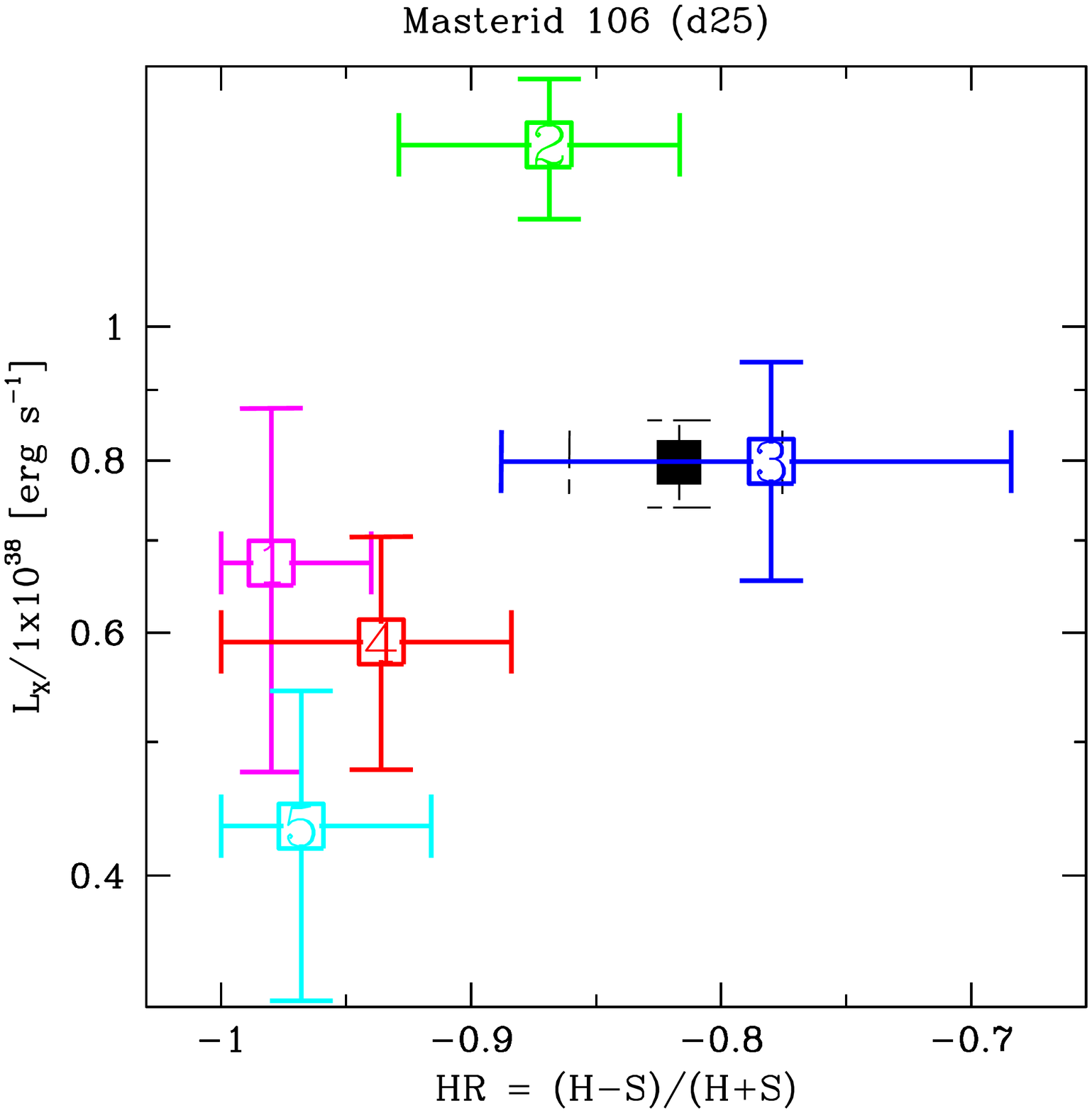}

 \end{minipage}

\begin{minipage}{0.32\linewidth}
  \centering
  
    \includegraphics[width=\linewidth]{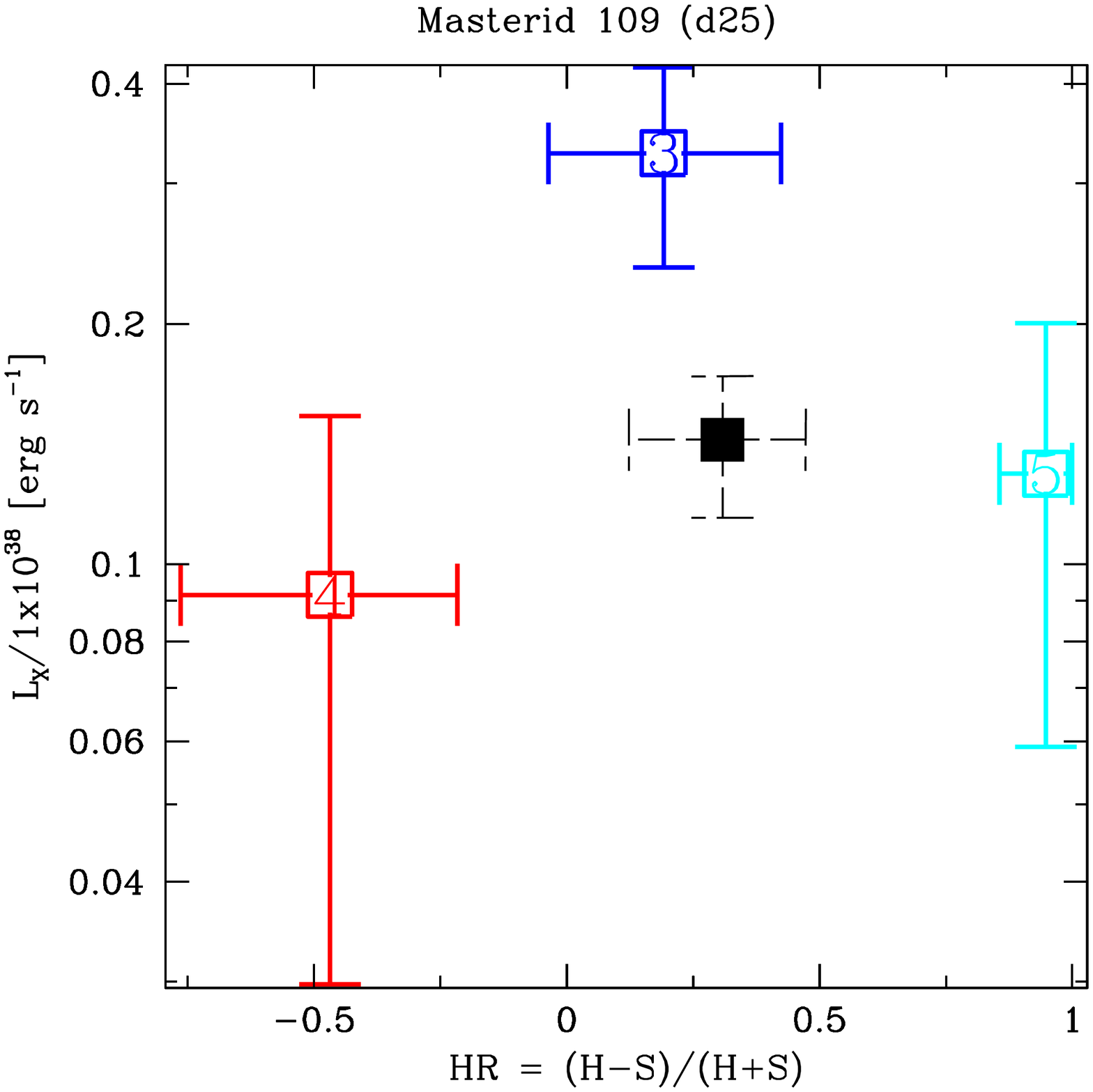}
  
  \end{minipage}
  \begin{minipage}{0.32\linewidth}
  \centering

    \includegraphics[width=\linewidth]{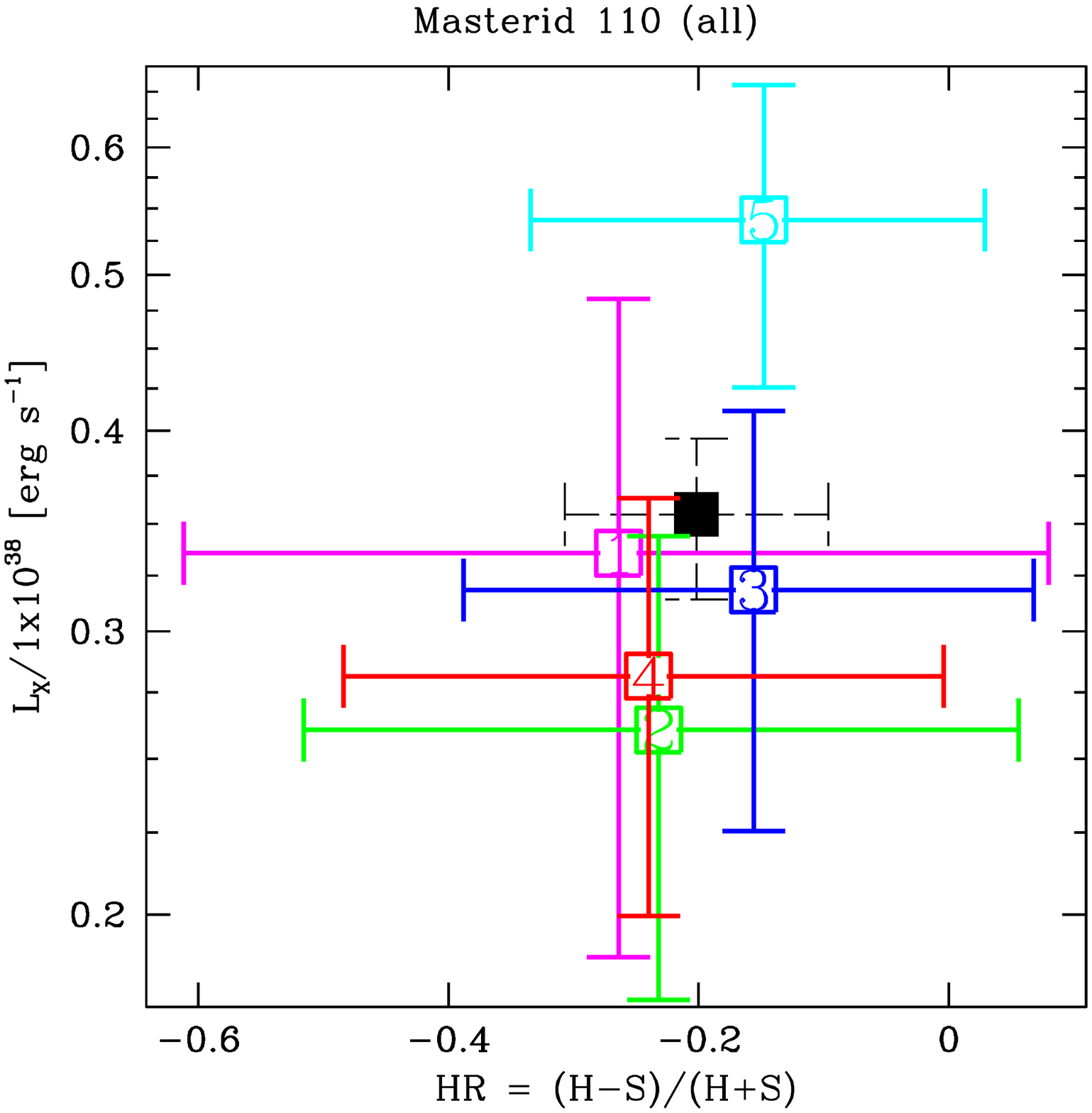}

\end{minipage}
\begin{minipage}{0.32\linewidth}
  \centering

    \includegraphics[width=\linewidth]{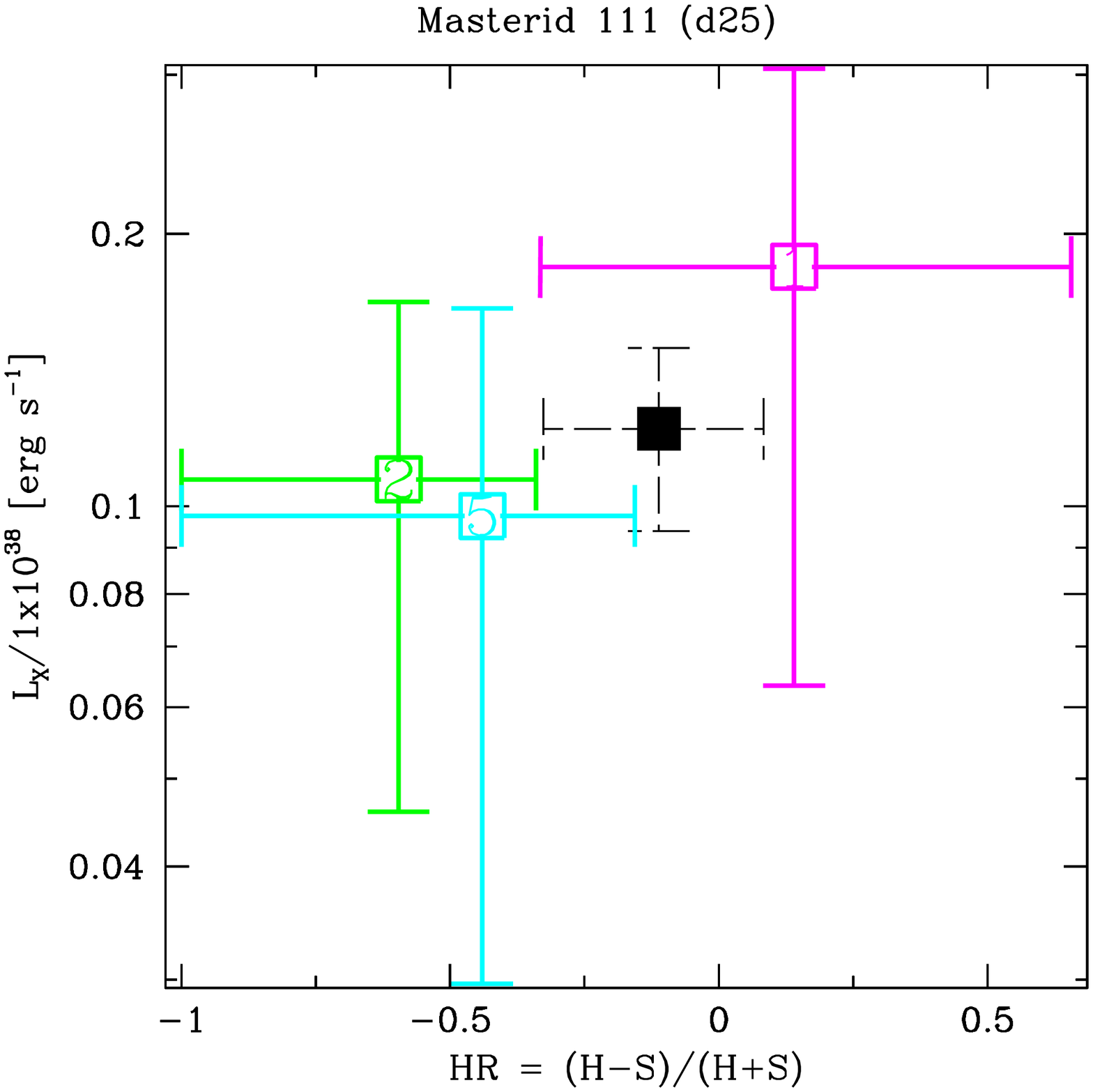}

 \end{minipage}

  \begin{minipage}{0.32\linewidth}
  \centering
  
    \includegraphics[width=\linewidth]{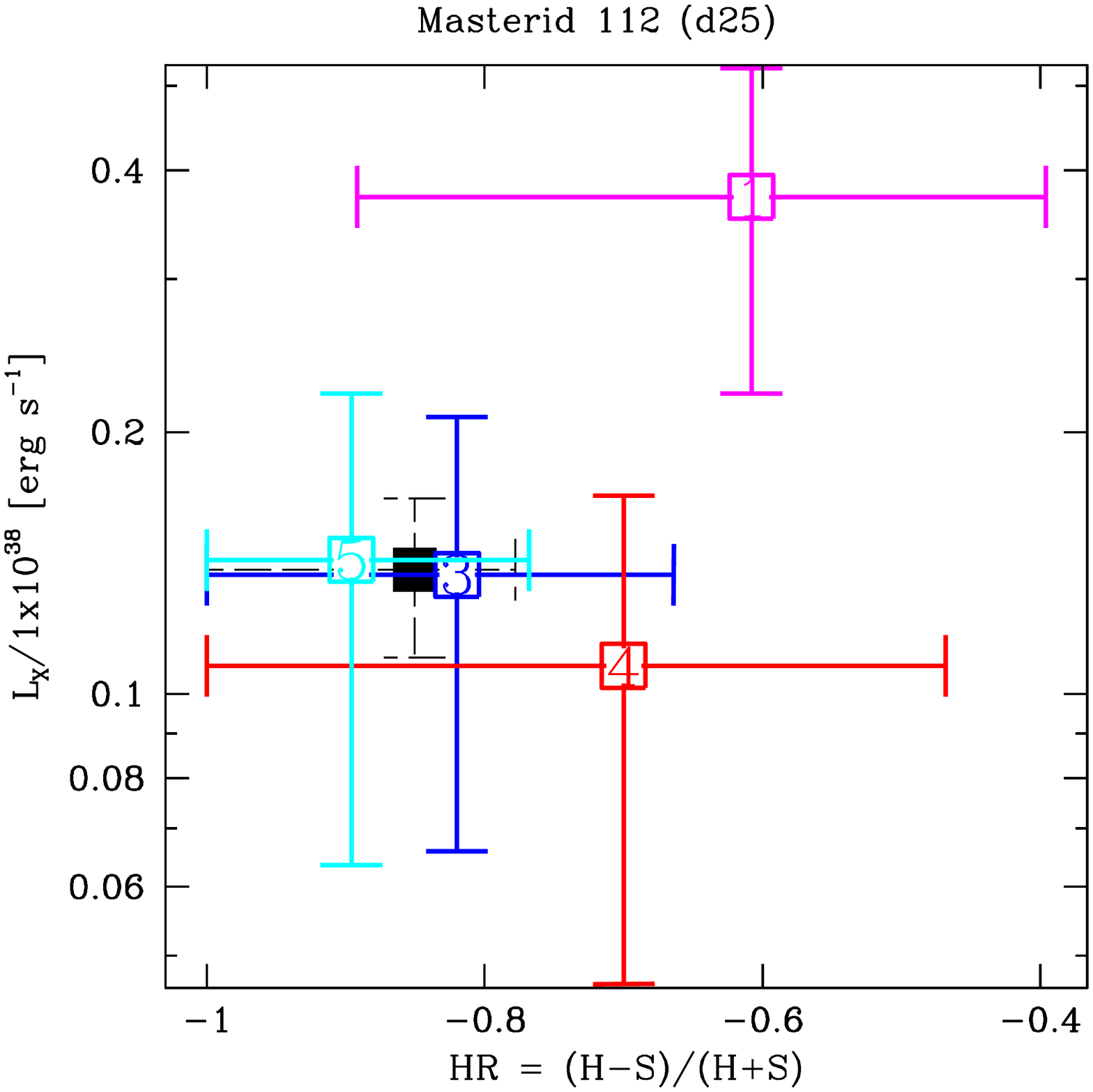}
  
  \end{minipage}
  \begin{minipage}{0.32\linewidth}
  \centering

    \includegraphics[width=\linewidth]{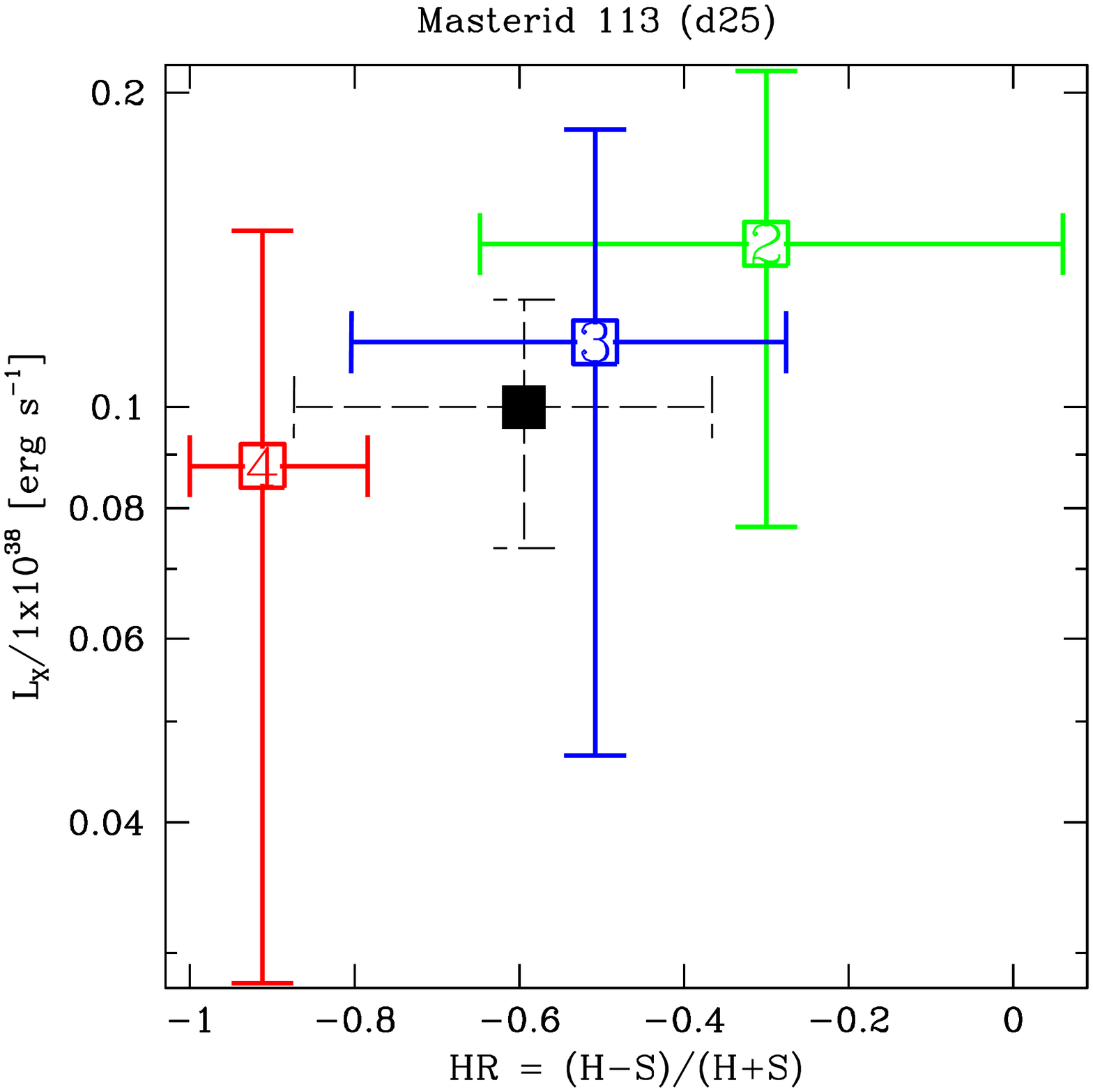}

\end{minipage}
\begin{minipage}{0.32\linewidth}
  \centering

    \includegraphics[width=\linewidth]{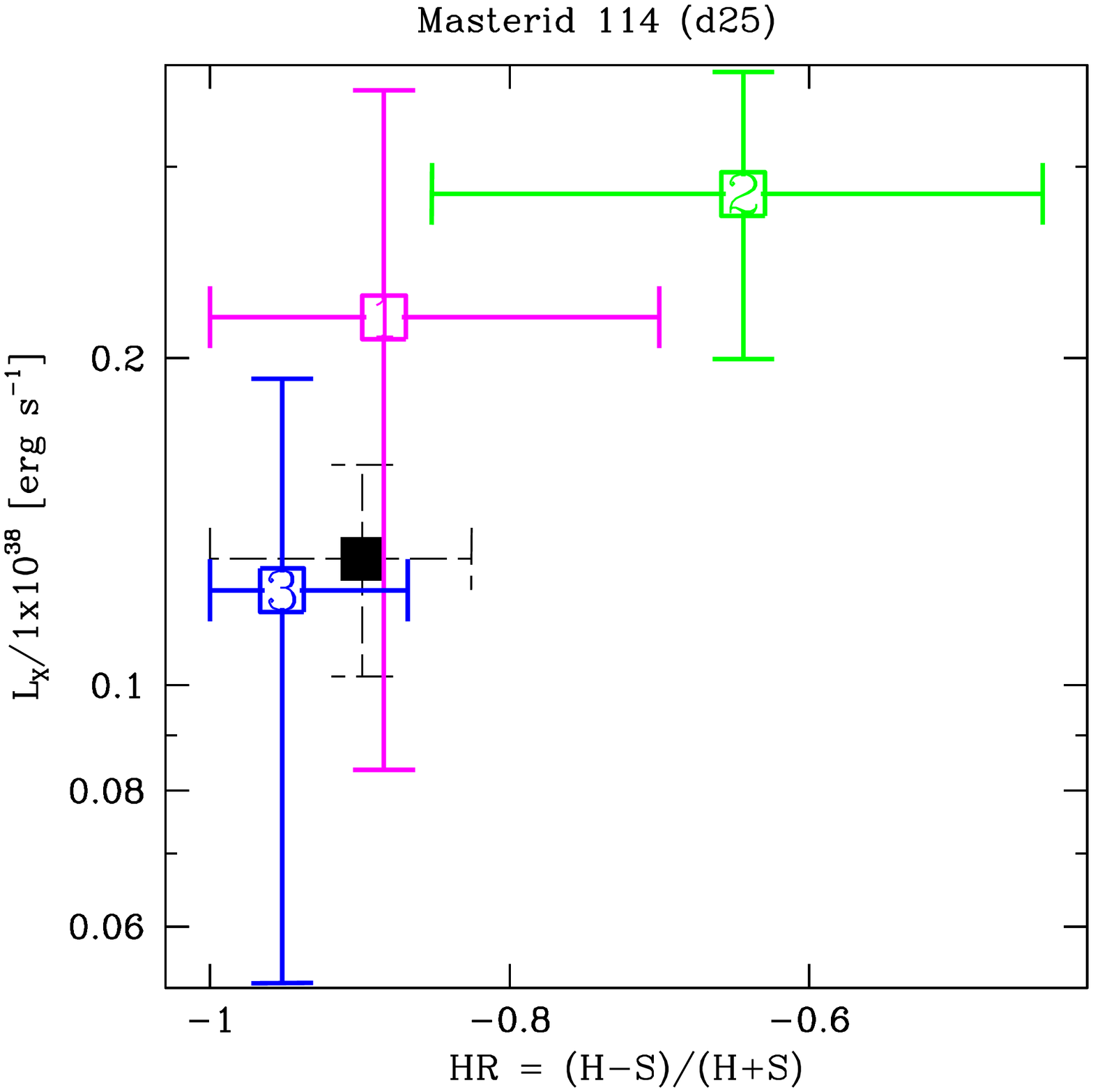}

 \end{minipage}

\begin{minipage}{0.32\linewidth}
  \centering
  
    \includegraphics[width=\linewidth]{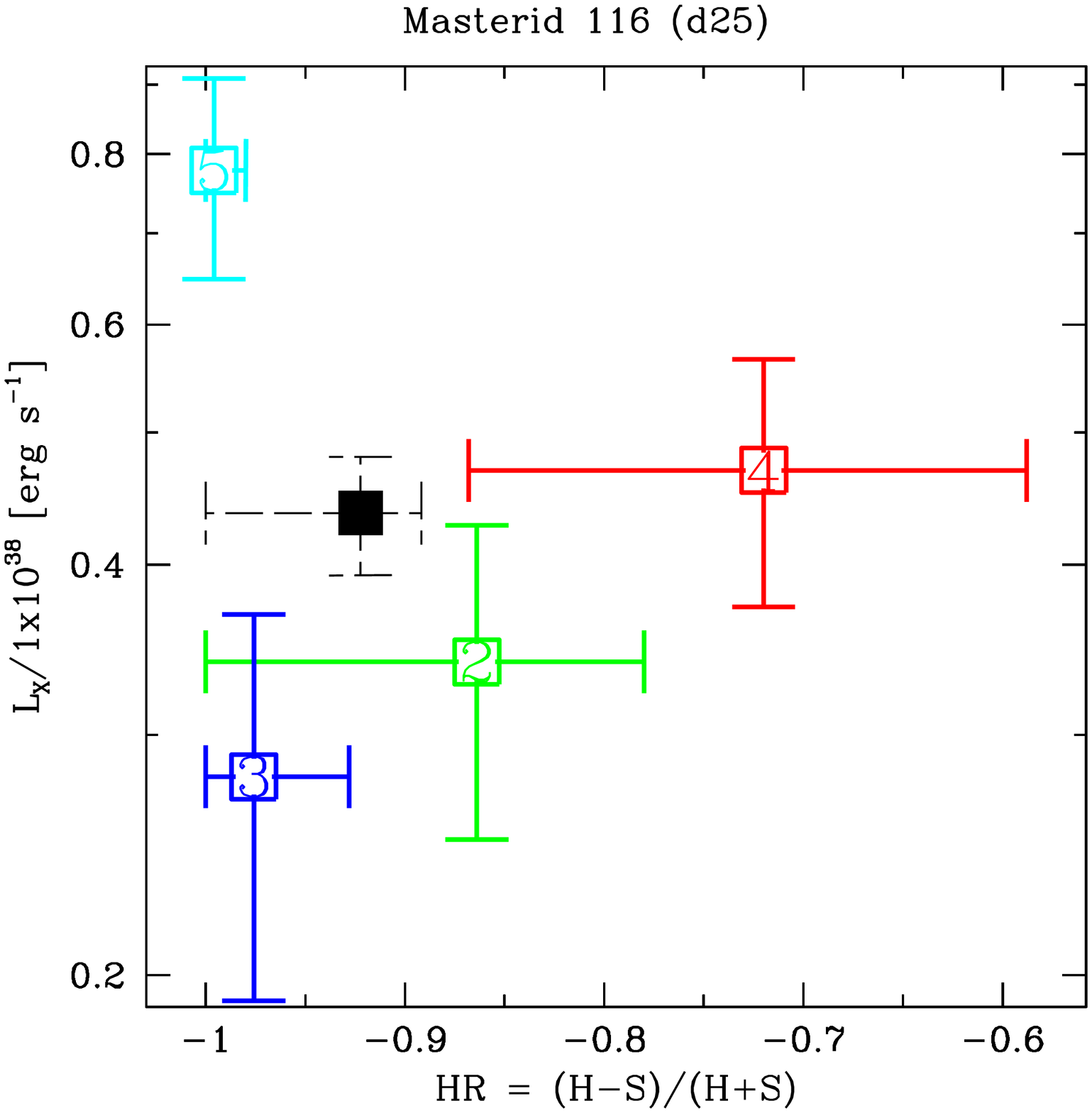}
  
  \end{minipage}
  \begin{minipage}{0.32\linewidth}
  \centering

    \includegraphics[width=\linewidth]{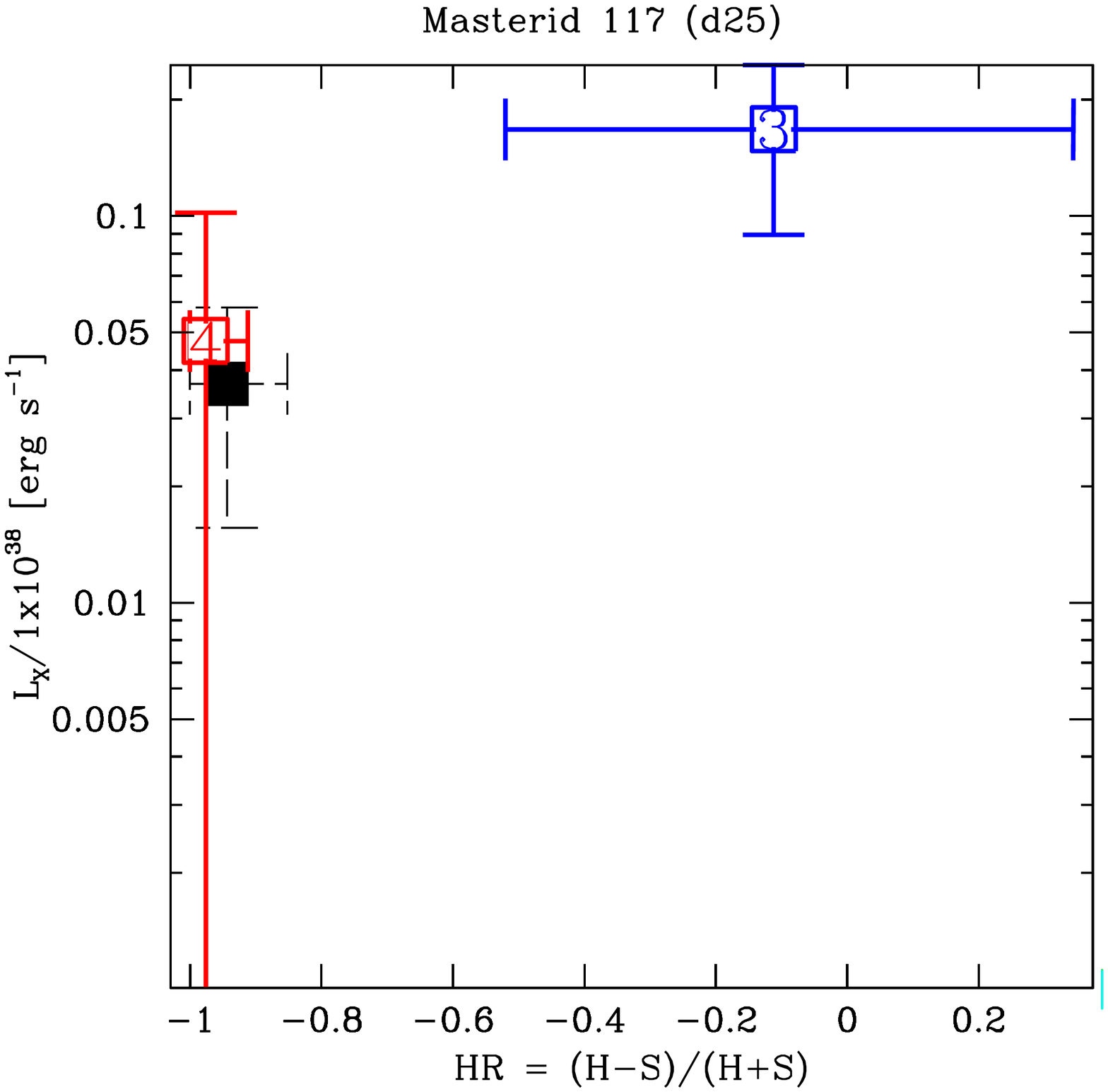}

\end{minipage}
\begin{minipage}{0.32\linewidth}
  \centering

    \includegraphics[width=\linewidth]{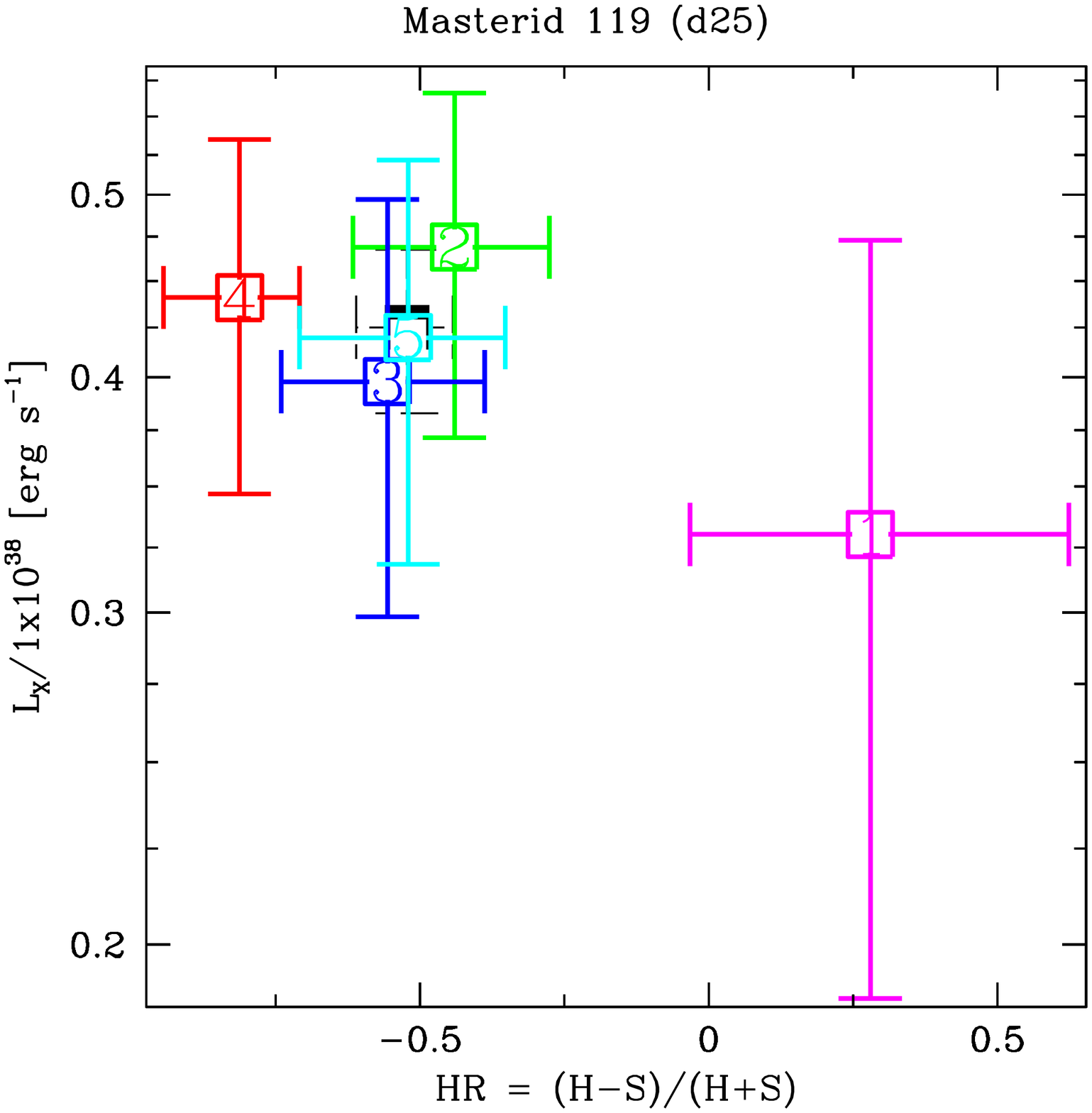}

 \end{minipage}
  
\end{figure}

\begin{figure}
  \begin{minipage}{0.32\linewidth}
  \centering
  
    \includegraphics[width=\linewidth]{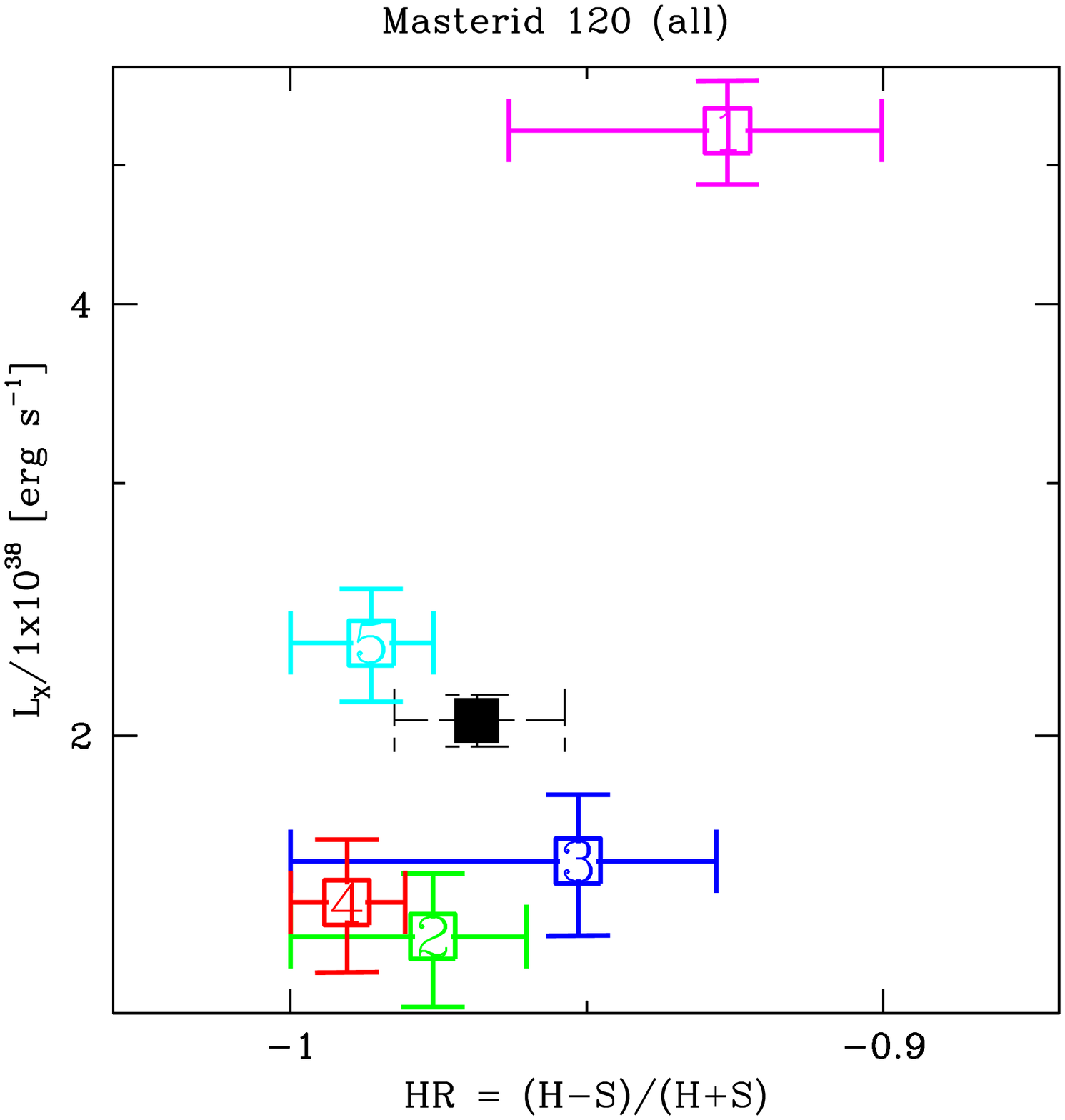}
  
  \end{minipage}
  \begin{minipage}{0.32\linewidth}
  \centering

    \includegraphics[width=\linewidth]{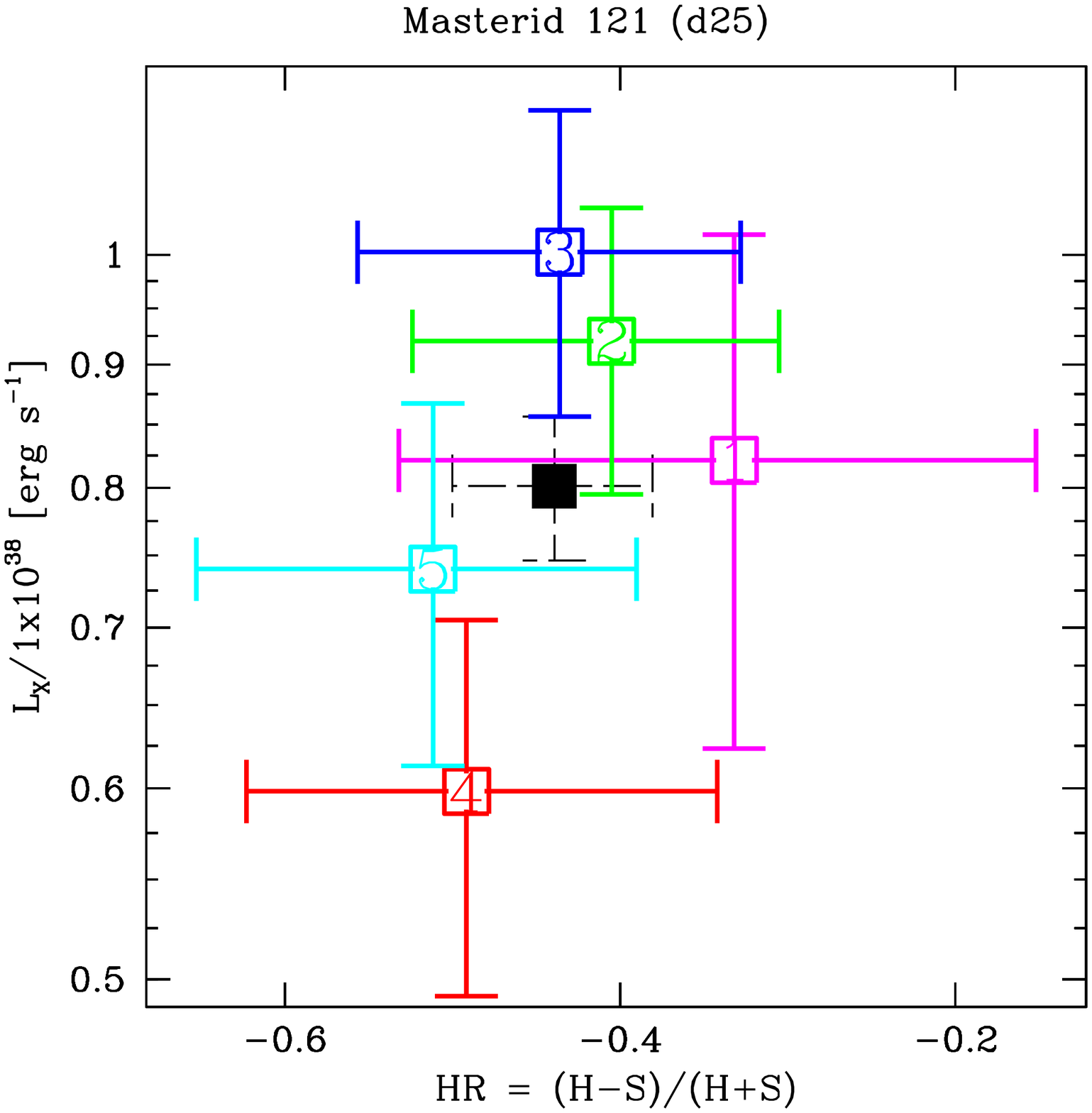}

\end{minipage}
\begin{minipage}{0.32\linewidth}
  \centering

    \includegraphics[width=\linewidth]{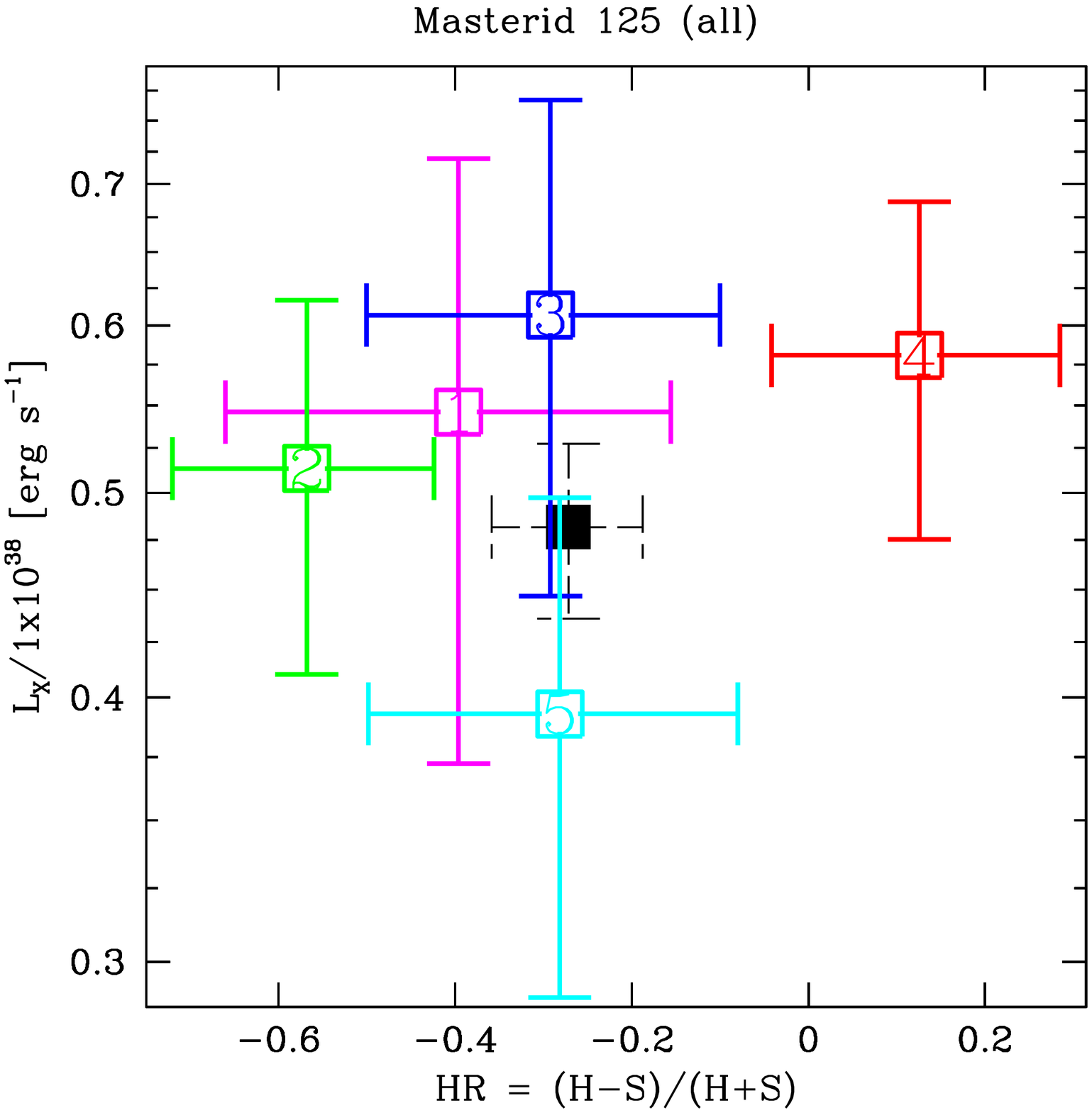}

 \end{minipage}

\begin{minipage}{0.32\linewidth}
  \centering
  
    \includegraphics[width=\linewidth]{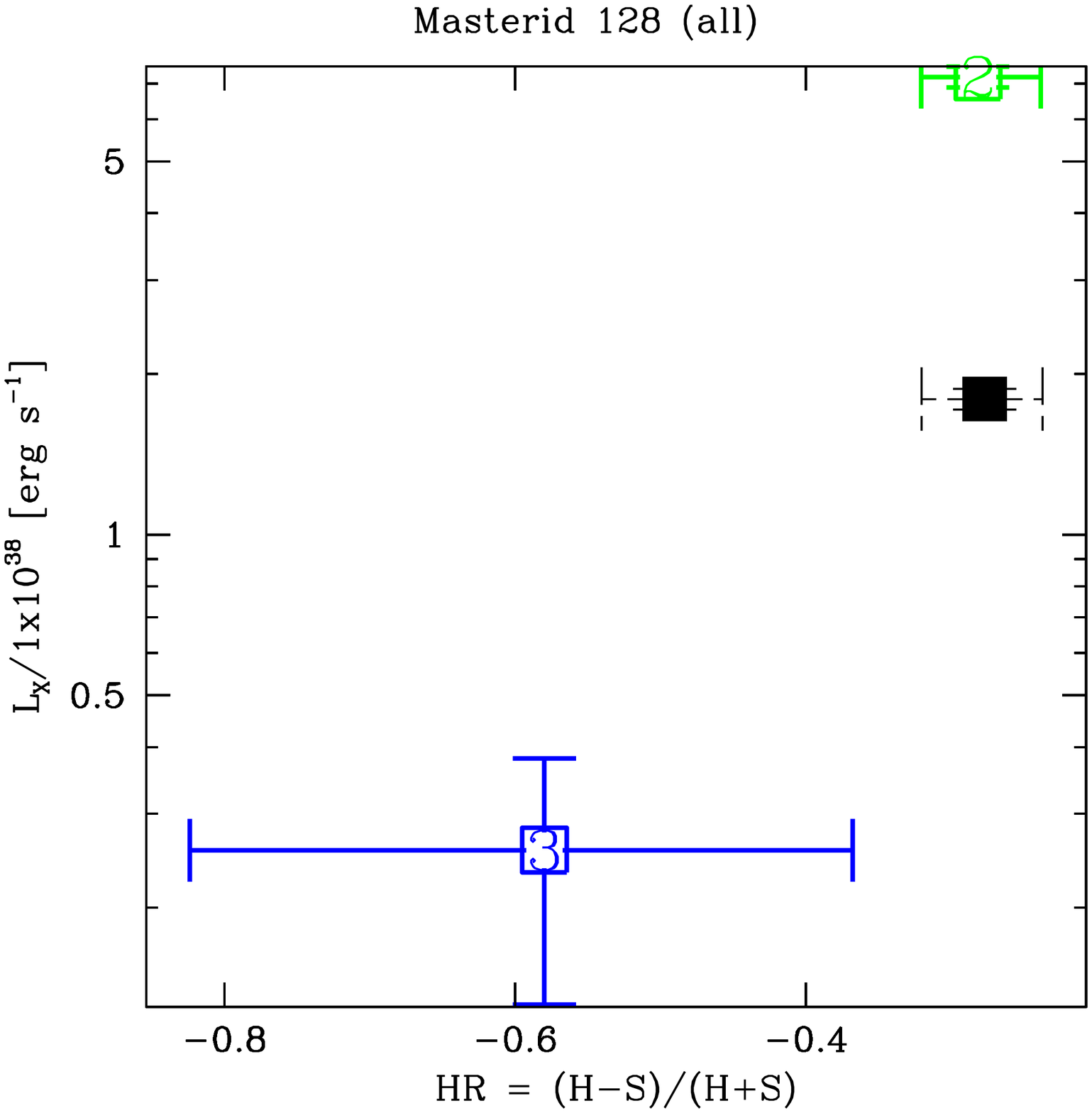}
  
  \end{minipage}
  \begin{minipage}{0.32\linewidth}
  \centering

    \includegraphics[width=\linewidth]{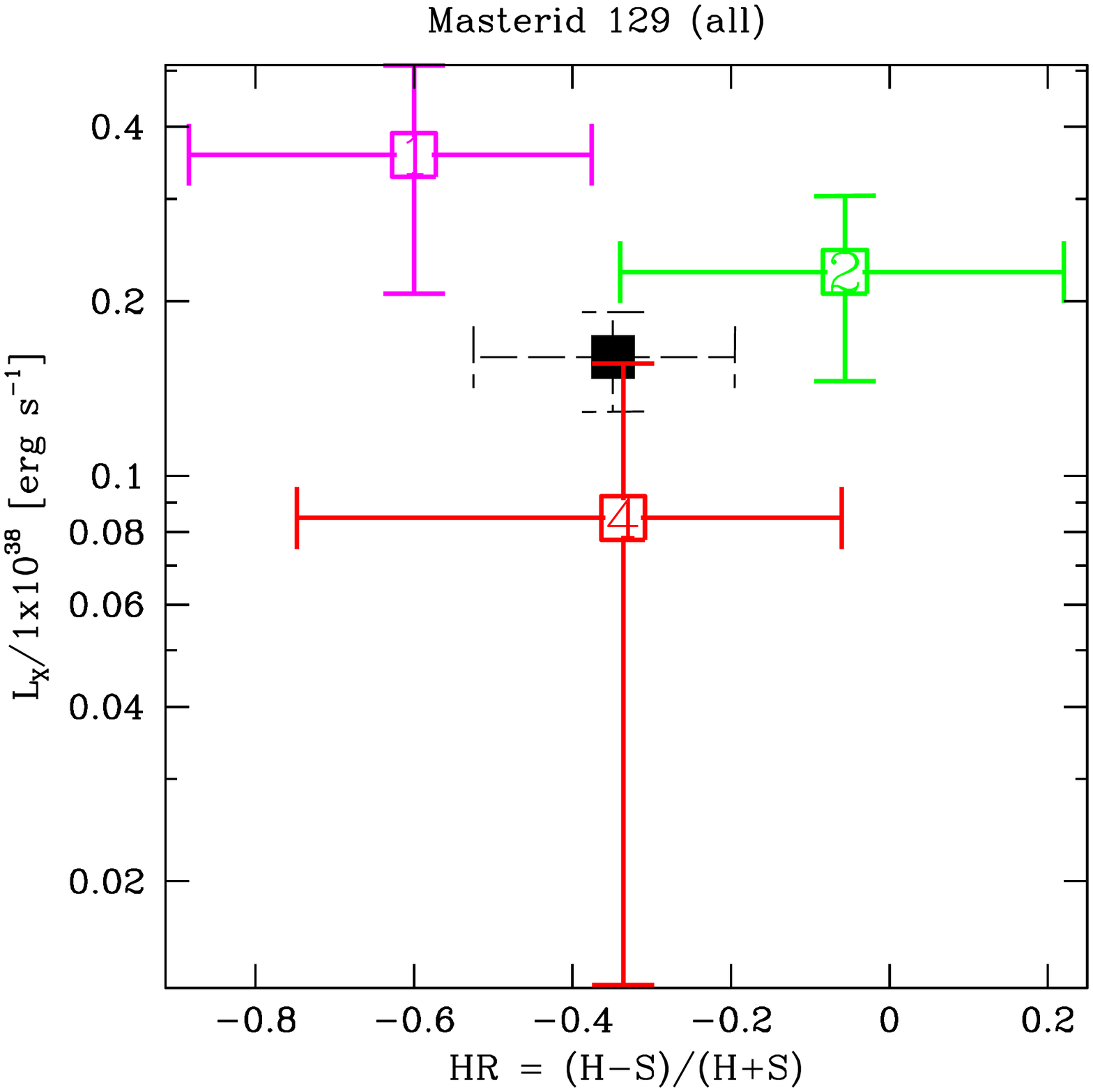}

\end{minipage}
\centering
\begin{minipage}{0.32\linewidth}
  \centering

    \includegraphics[width=\linewidth]{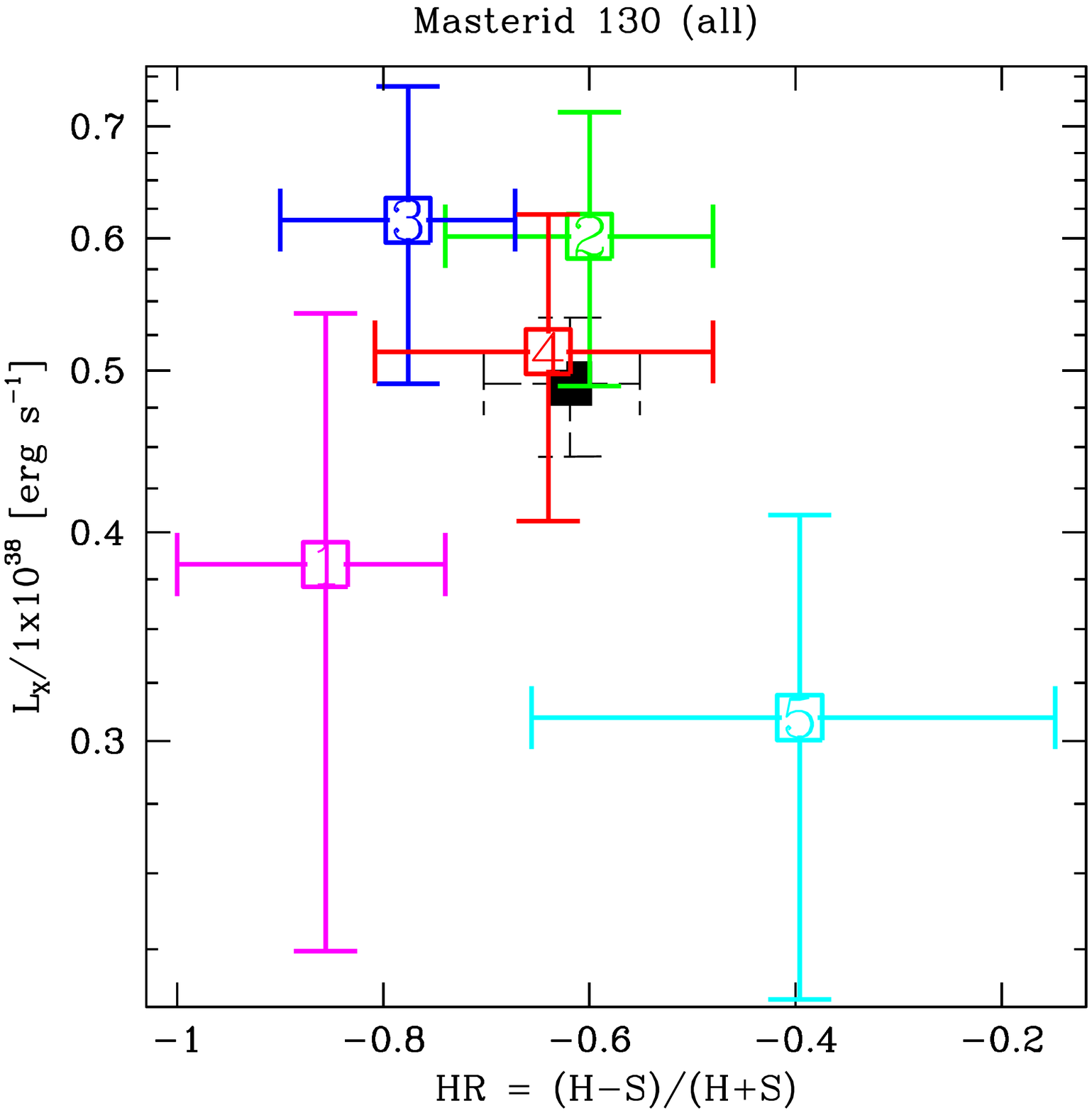}

 \end{minipage}

  \begin{minipage}{0.32\linewidth}
  \centering
  
    \includegraphics[width=\linewidth]{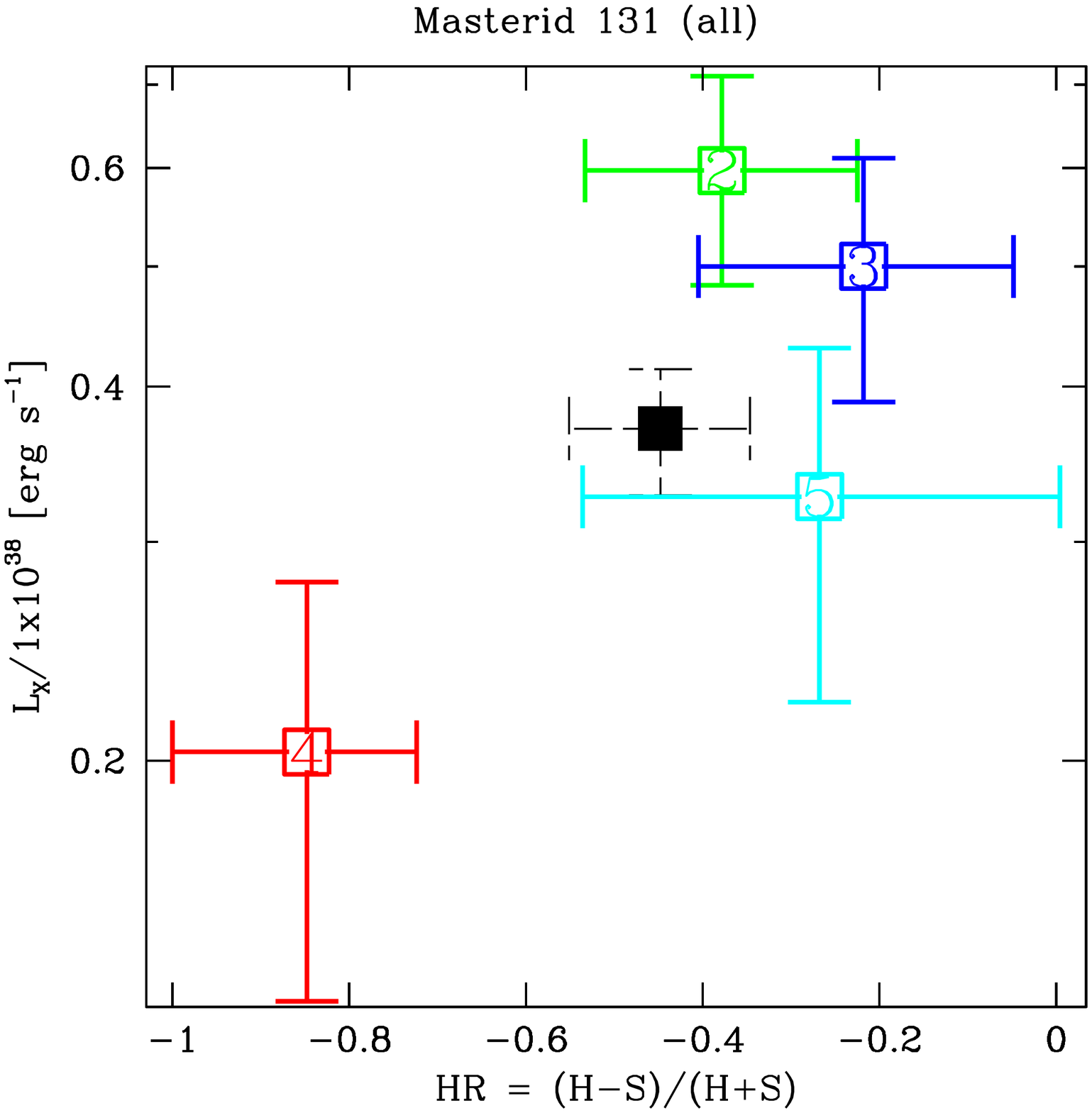}
  
  \end{minipage}
  \begin{minipage}{0.32\linewidth}
  \centering

    \includegraphics[width=\linewidth]{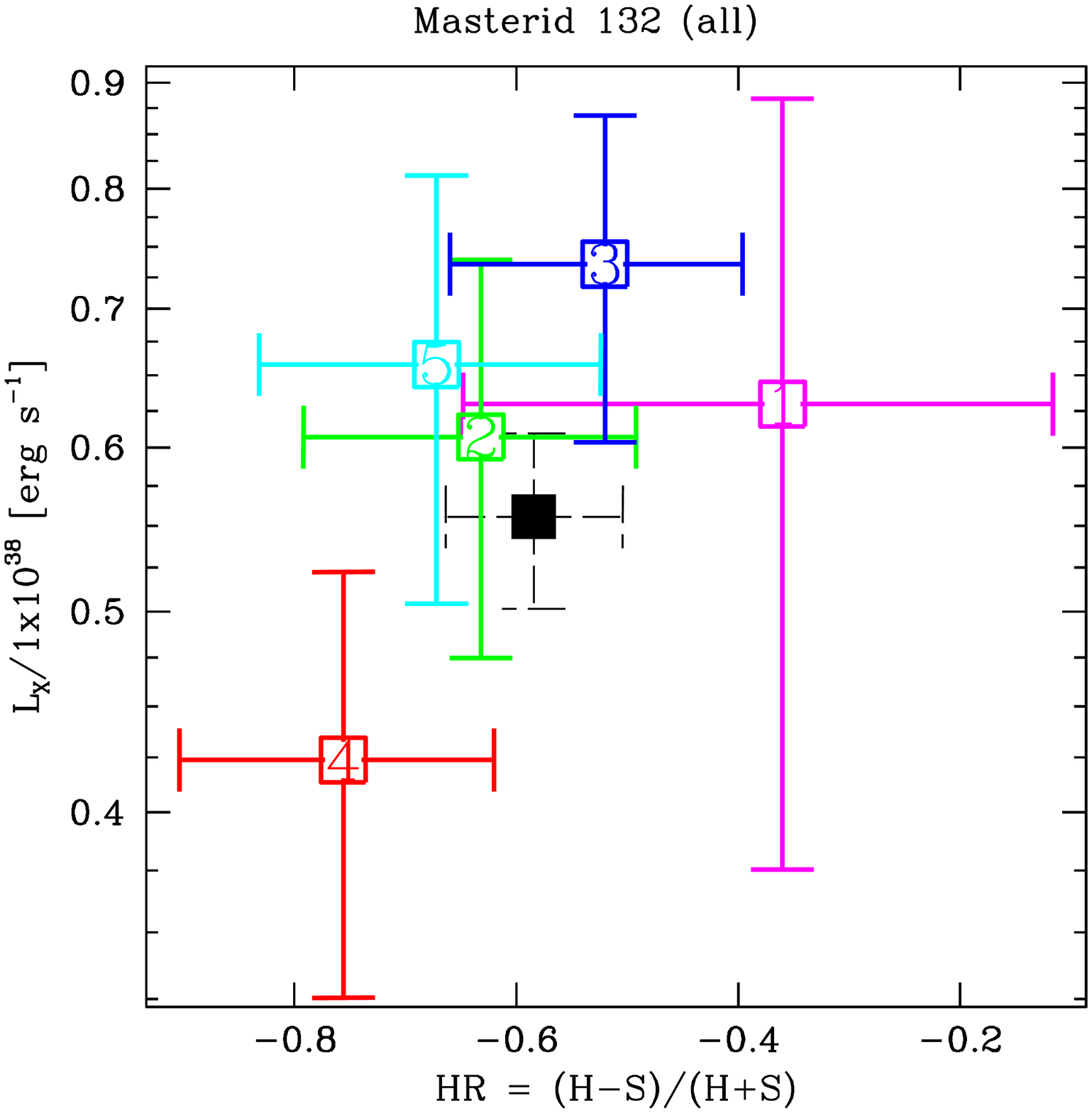}

\end{minipage}

\end{figure}

\begin{figure}
  \begin{minipage}{0.32\linewidth}
  \centering
  
    \includegraphics[width=\linewidth]{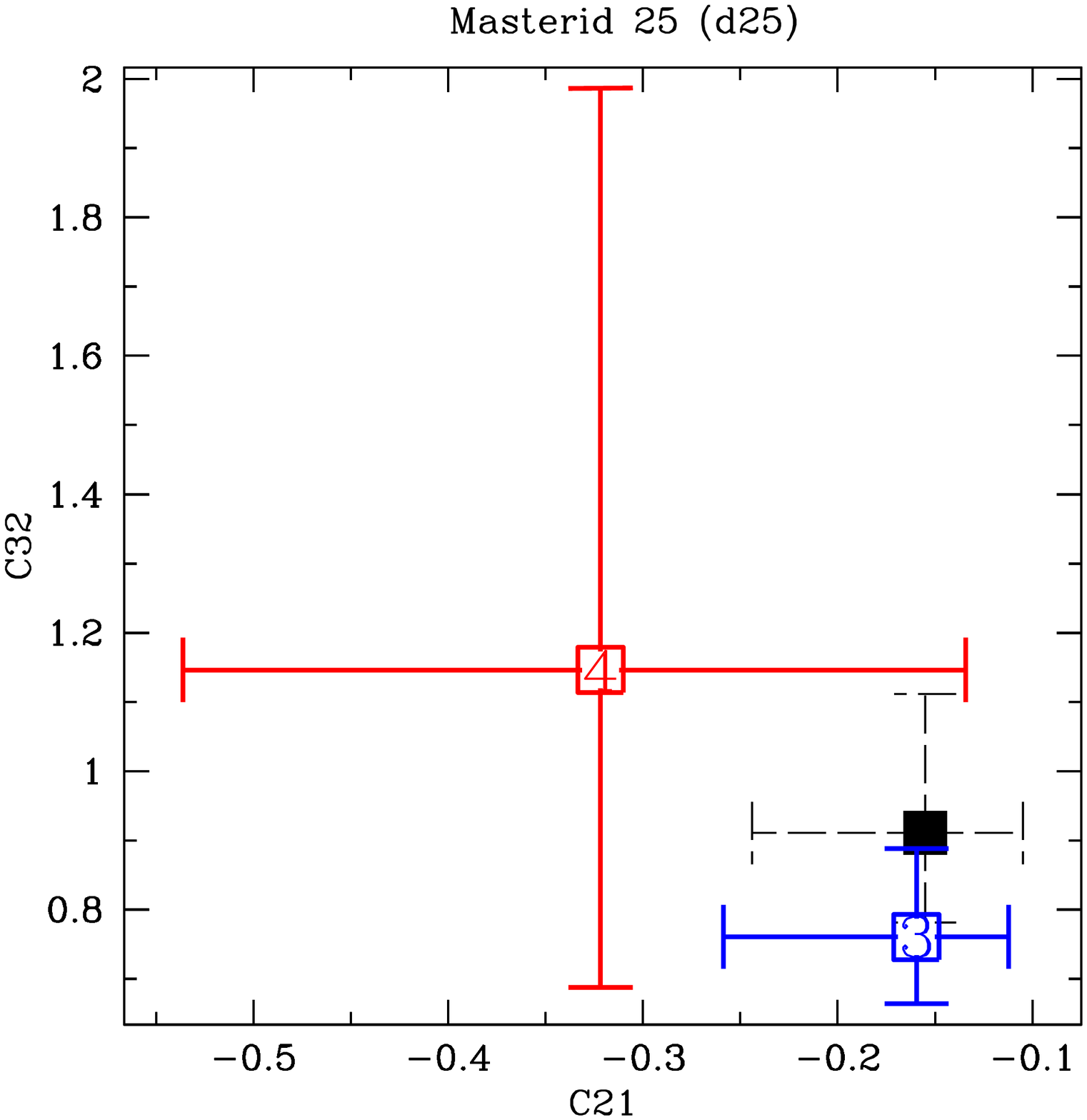}
  
  \end{minipage}
  \begin{minipage}{0.32\linewidth}
  \centering

    \includegraphics[width=\linewidth]{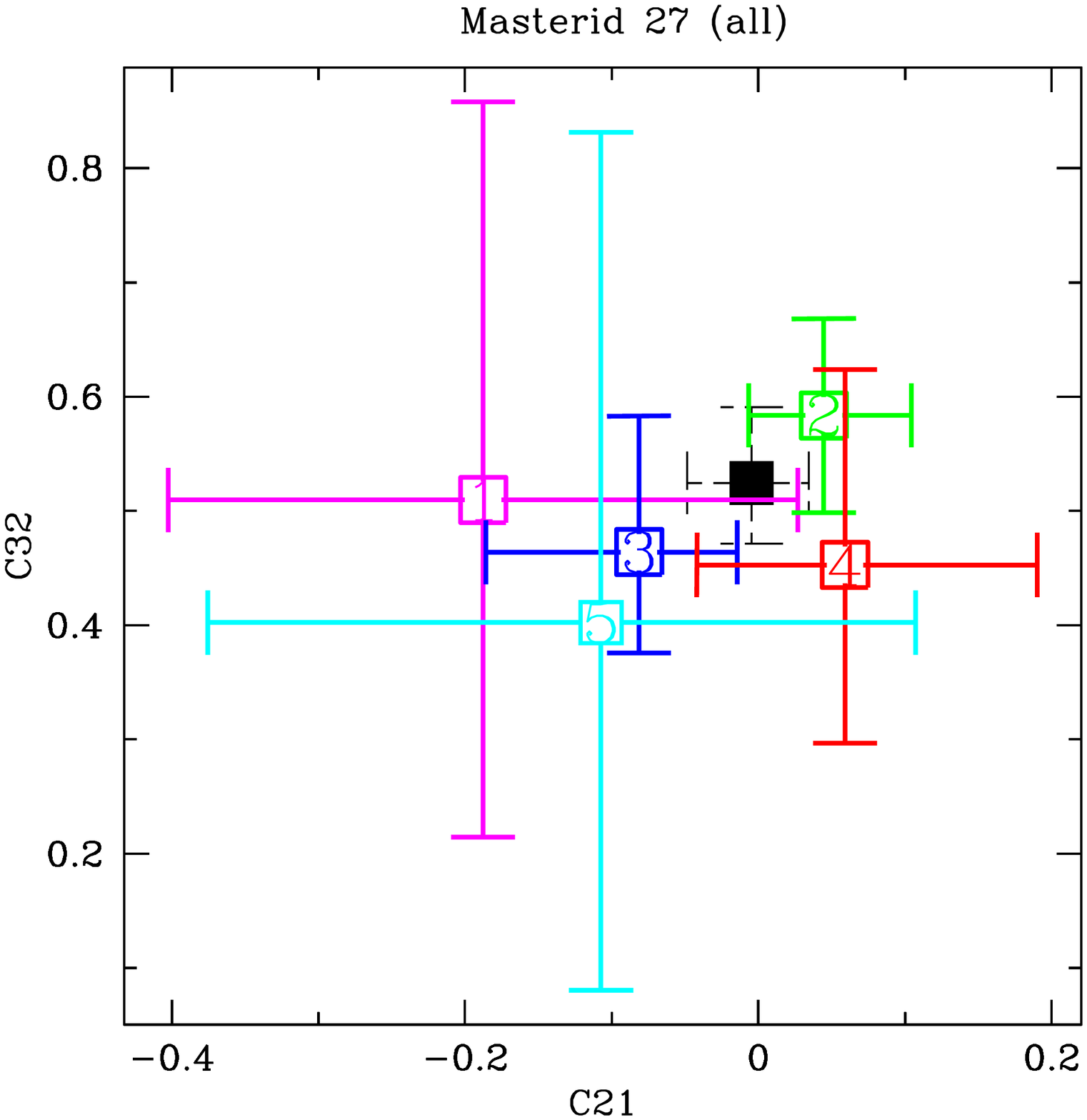}

\end{minipage}
\begin{minipage}{0.32\linewidth}
  \centering

    \includegraphics[width=\linewidth]{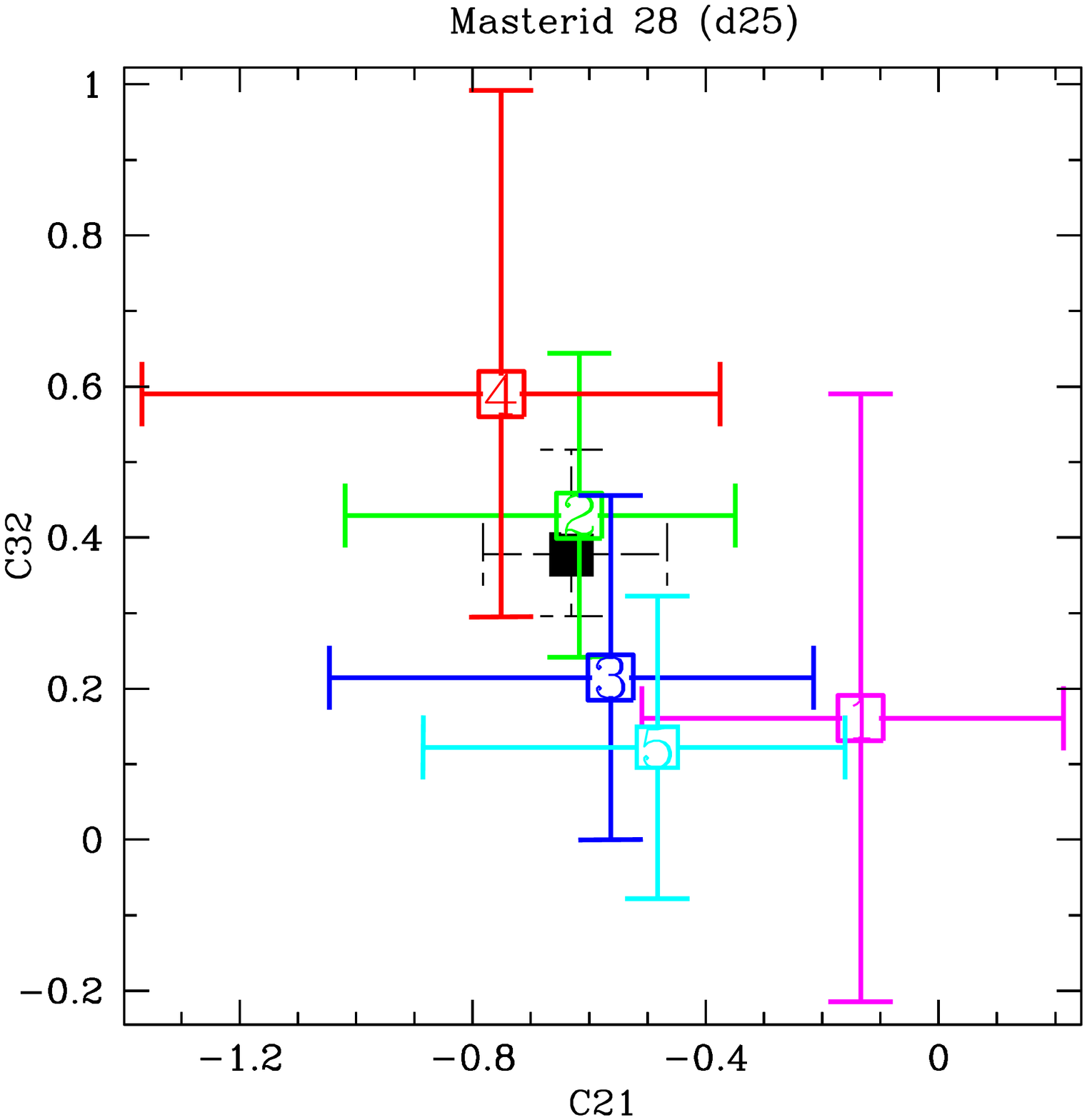}

 \end{minipage}

\begin{minipage}{0.32\linewidth}
  \centering
  
    \includegraphics[width=\linewidth]{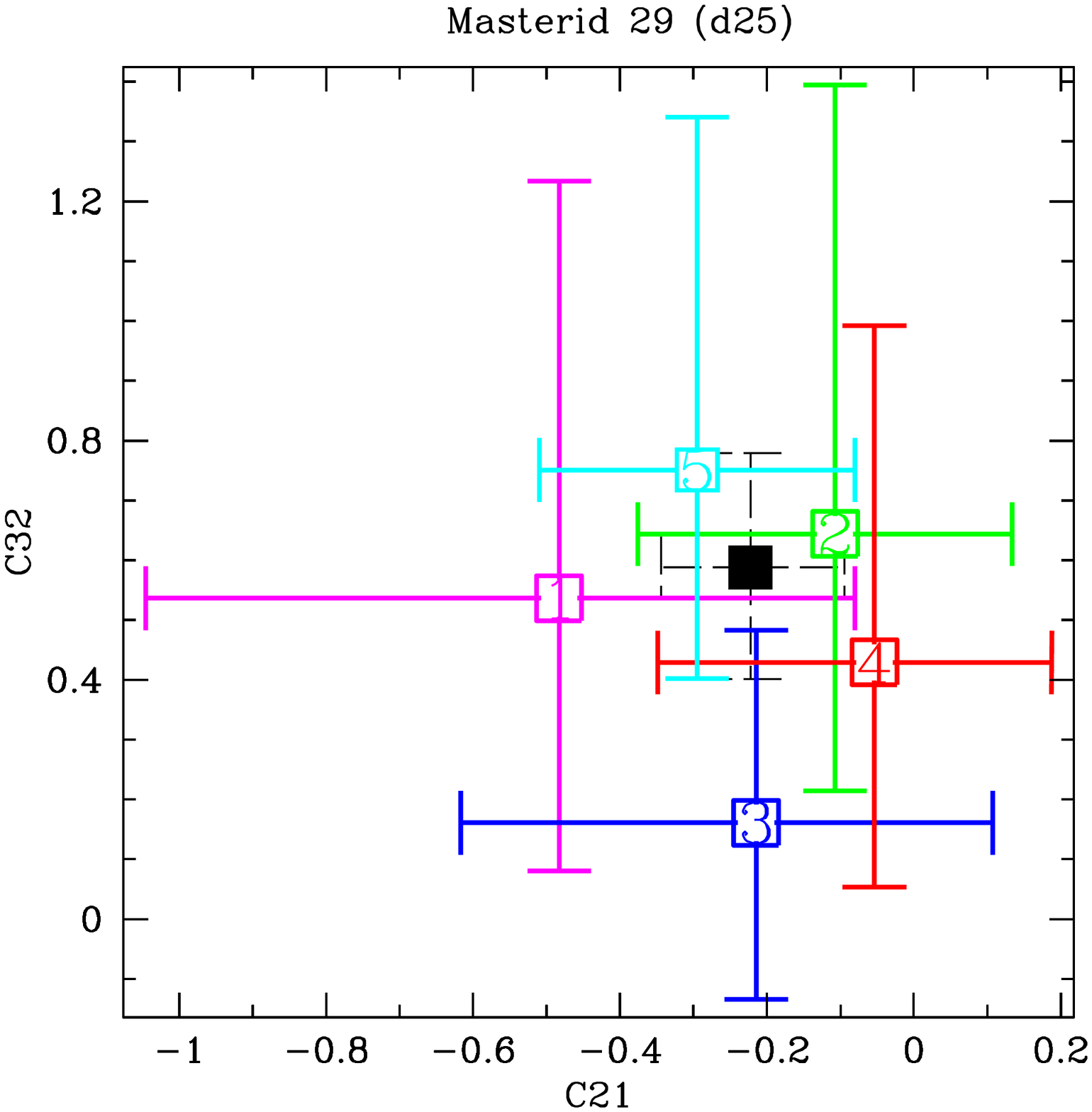}
  
  \end{minipage}
  \begin{minipage}{0.32\linewidth}
  \centering

    \includegraphics[width=\linewidth]{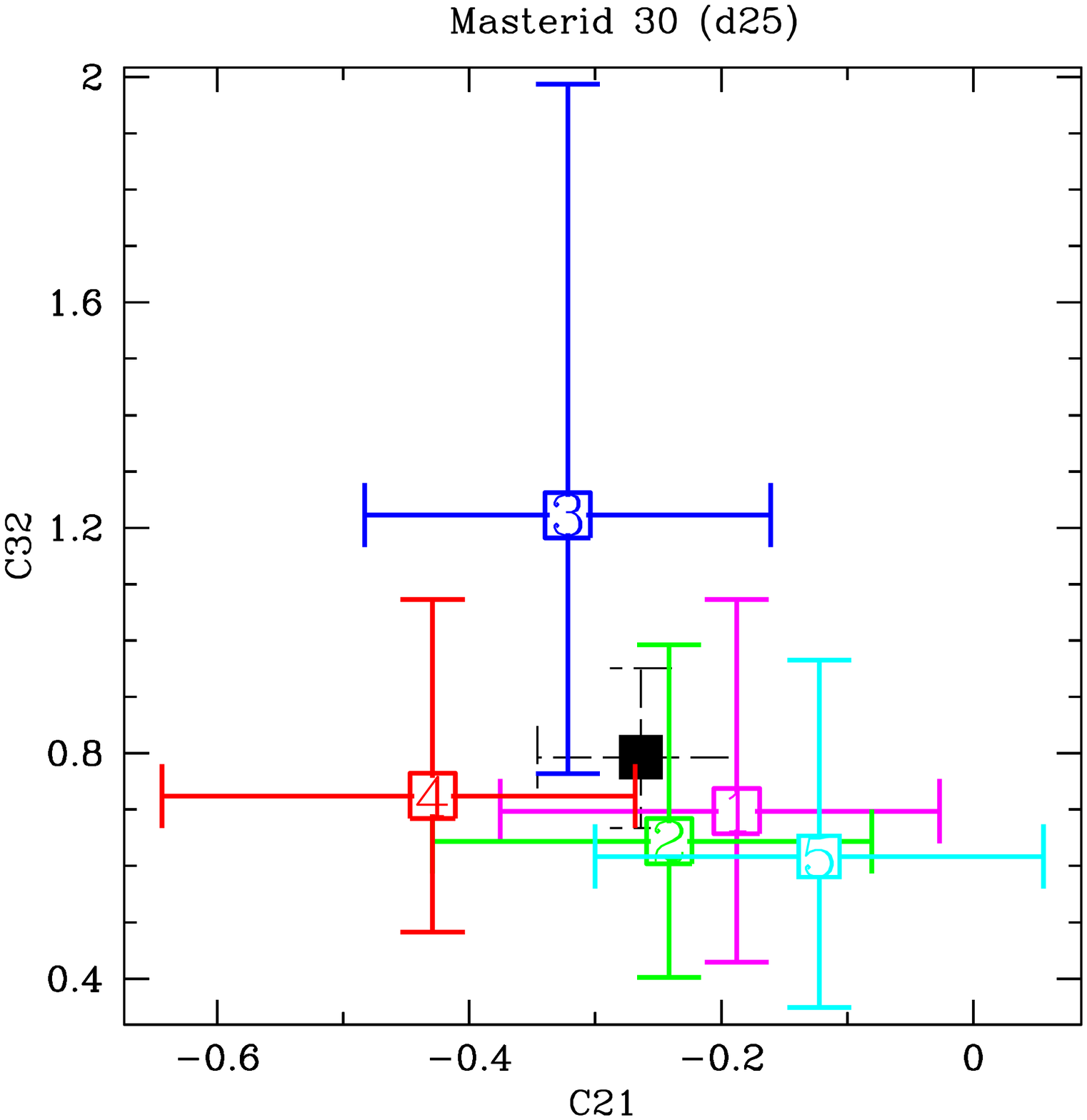}

\end{minipage}
\begin{minipage}{0.32\linewidth}
  \centering

    \includegraphics[width=\linewidth]{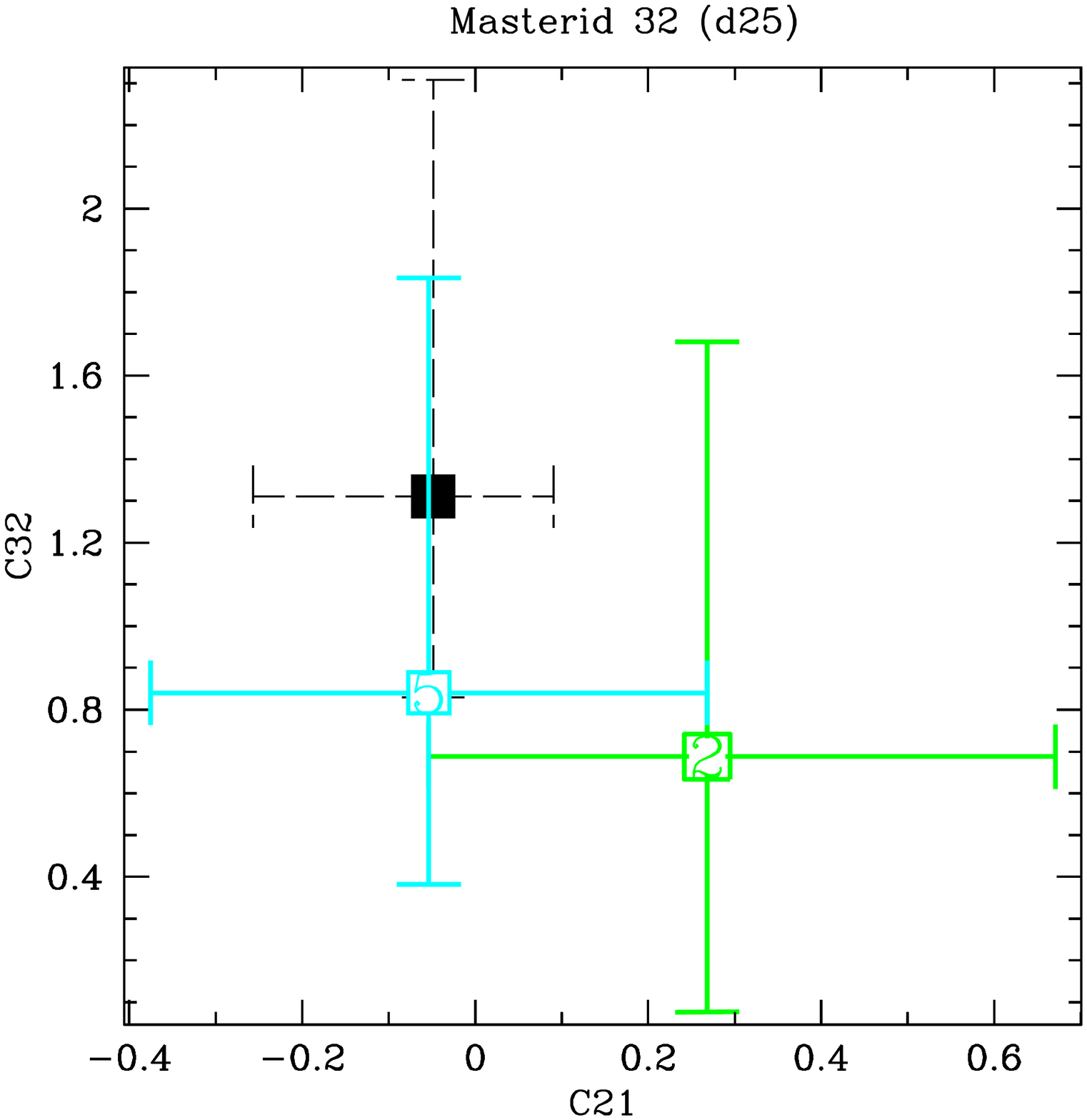}

 \end{minipage}

  \begin{minipage}{0.32\linewidth}
  \centering
  
    \includegraphics[width=\linewidth]{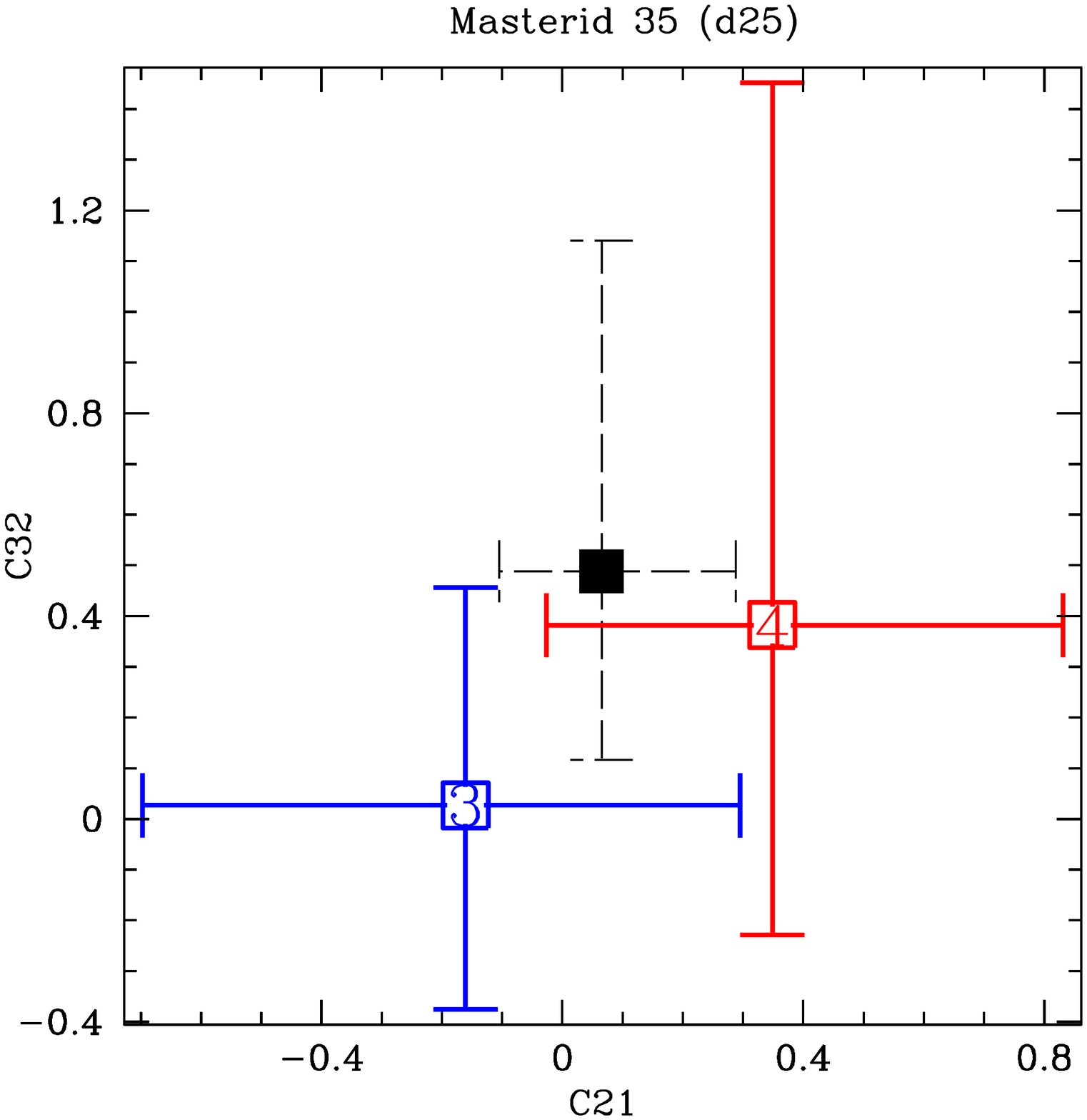}
  
  \end{minipage}
  \begin{minipage}{0.32\linewidth}
  \centering

    \includegraphics[width=\linewidth]{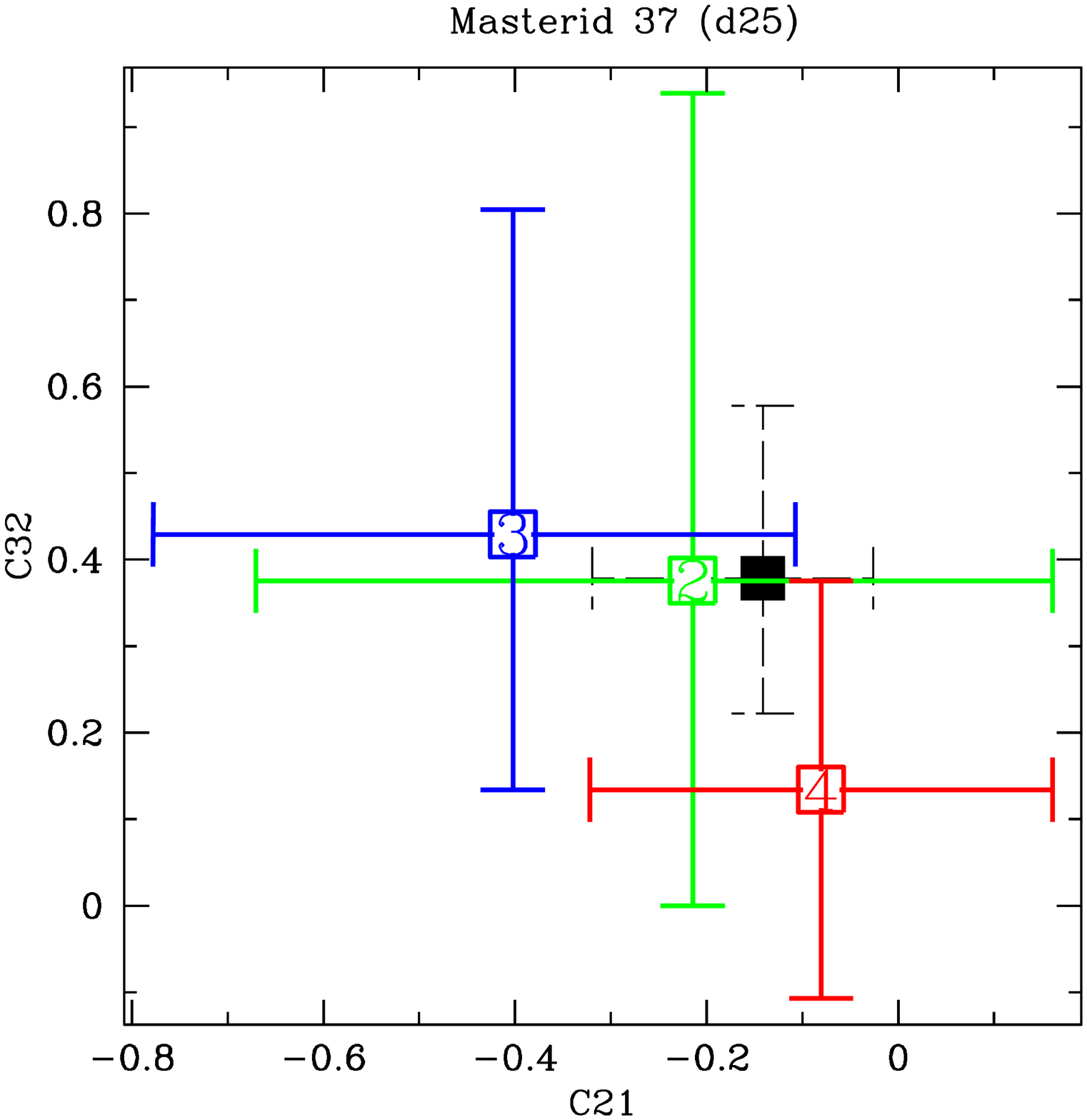}

\end{minipage}
\begin{minipage}{0.32\linewidth}
  \centering

    \includegraphics[width=\linewidth]{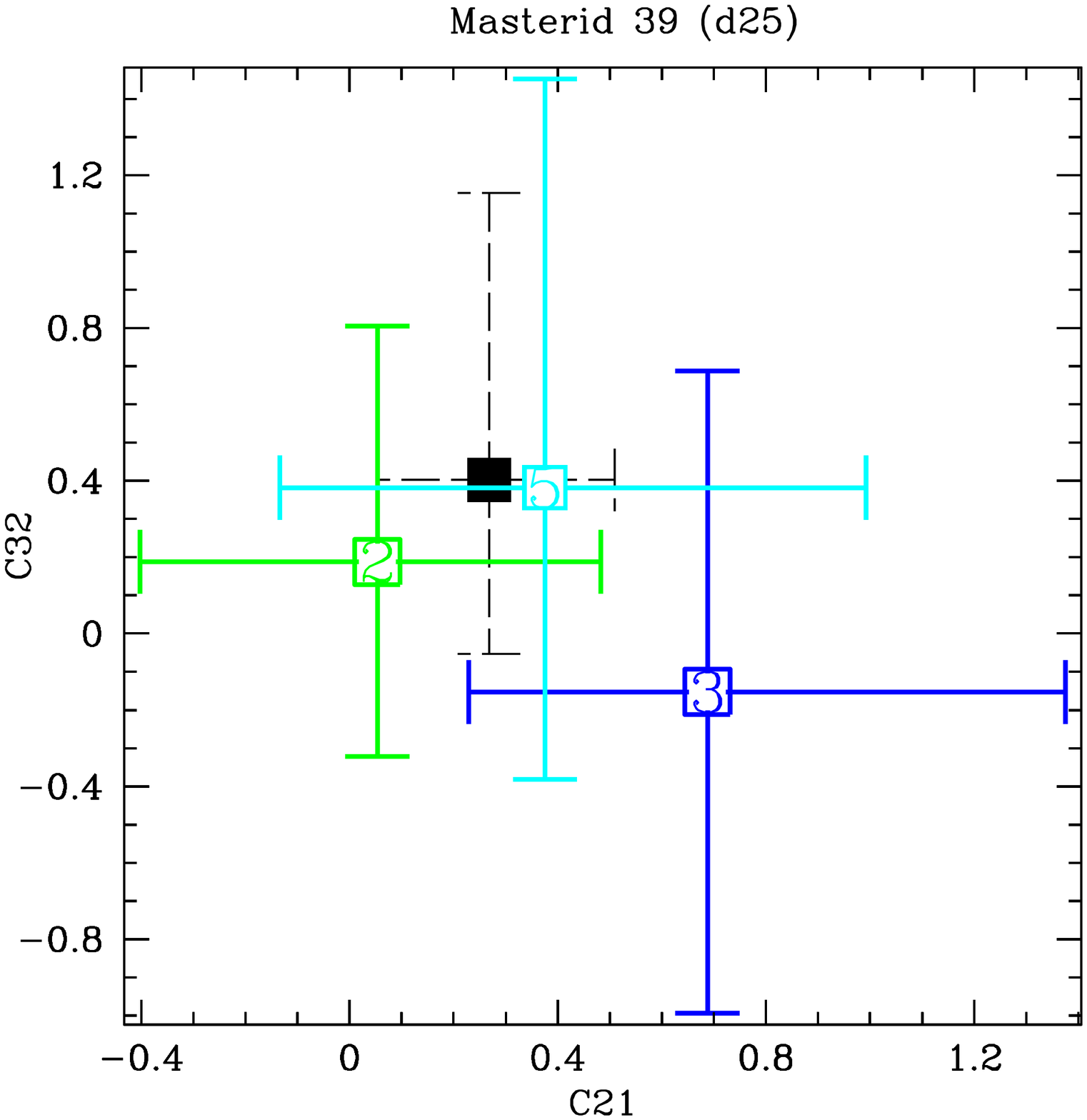}

 \end{minipage}

\begin{minipage}{0.32\linewidth}
  \centering
  
    \includegraphics[width=\linewidth]{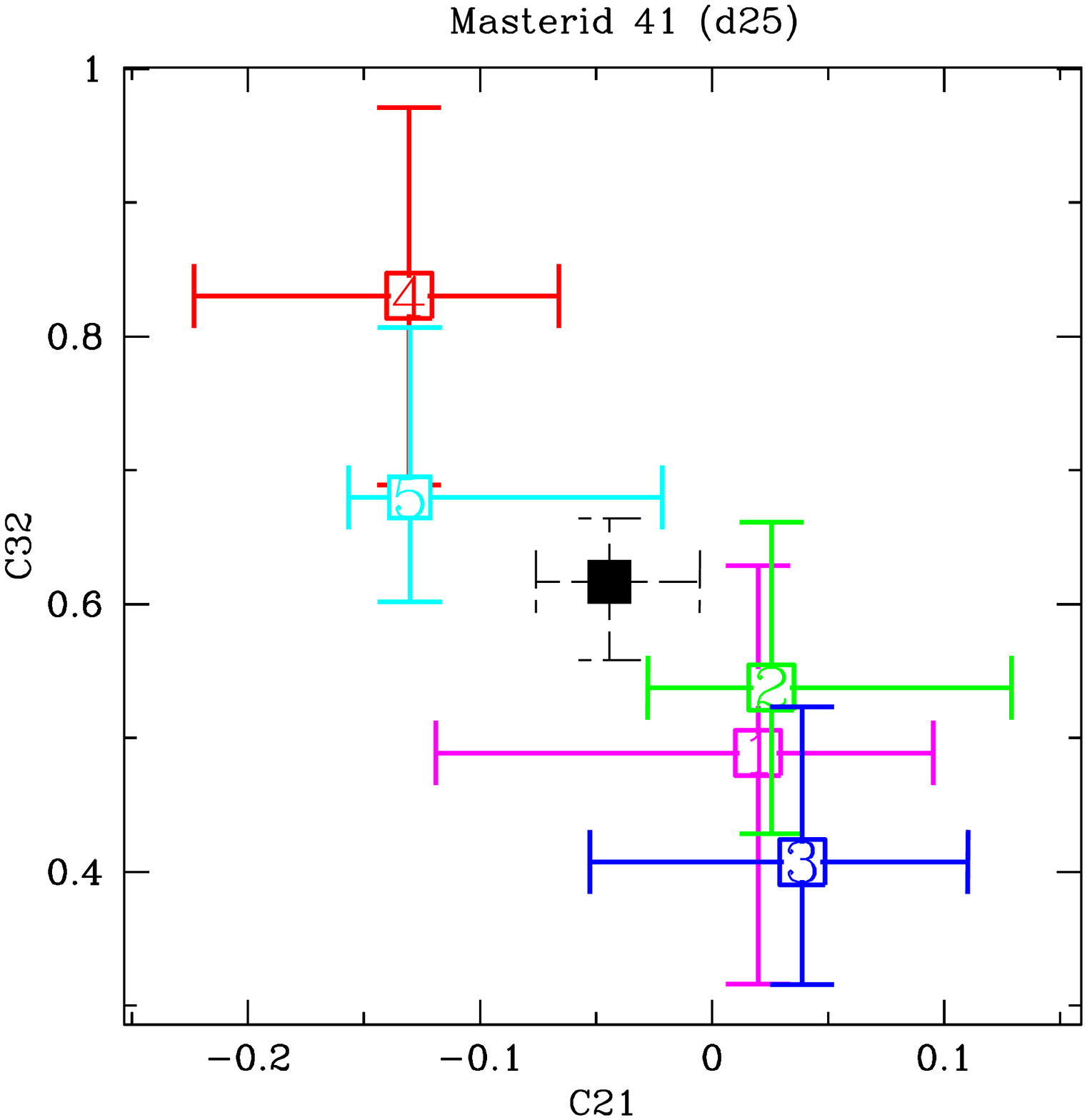}
  
  \end{minipage}
  \begin{minipage}{0.32\linewidth}
  \centering

    \includegraphics[width=\linewidth]{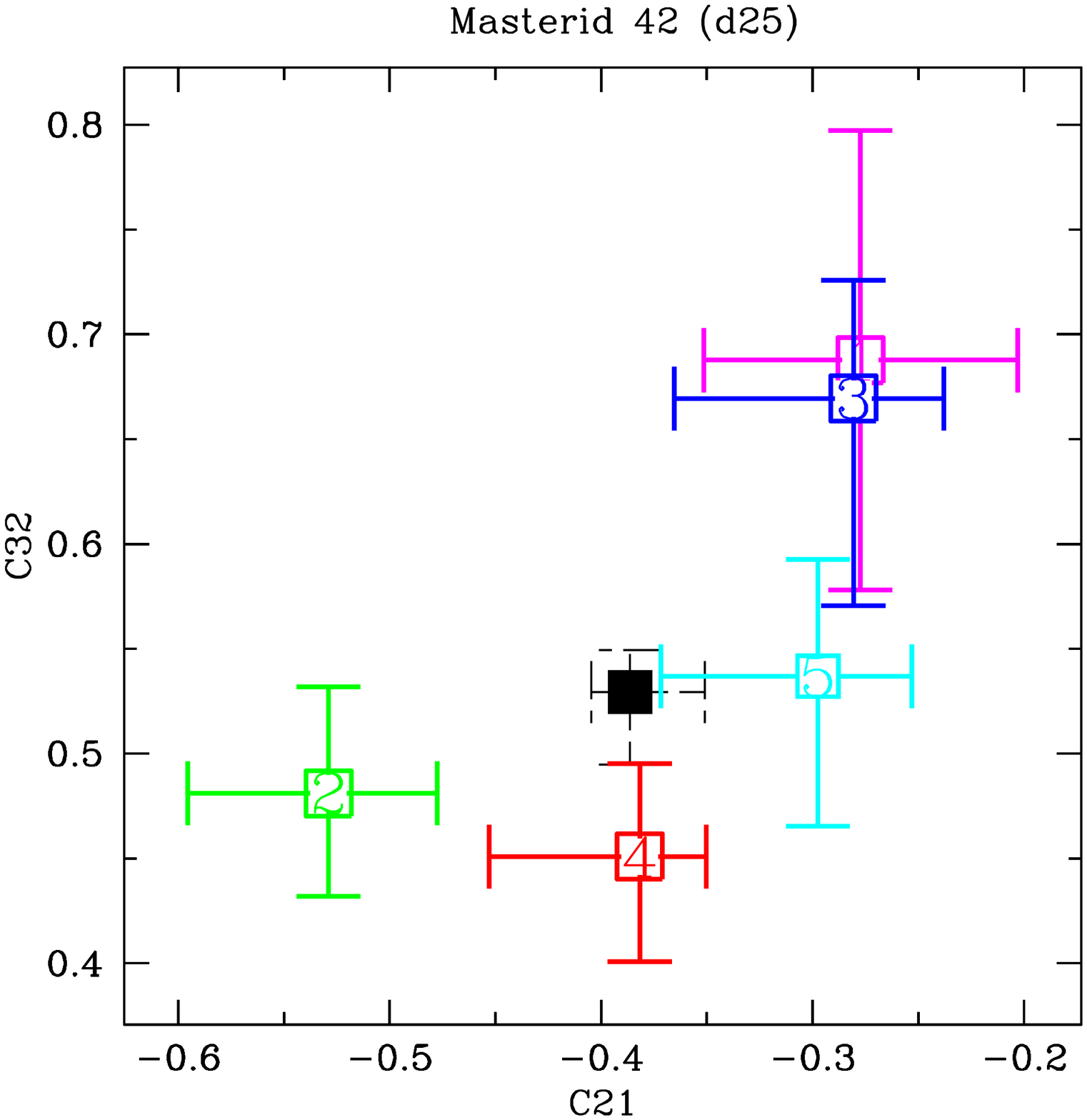}

\end{minipage}
\begin{minipage}{0.32\linewidth}
  \centering

    \includegraphics[width=\linewidth]{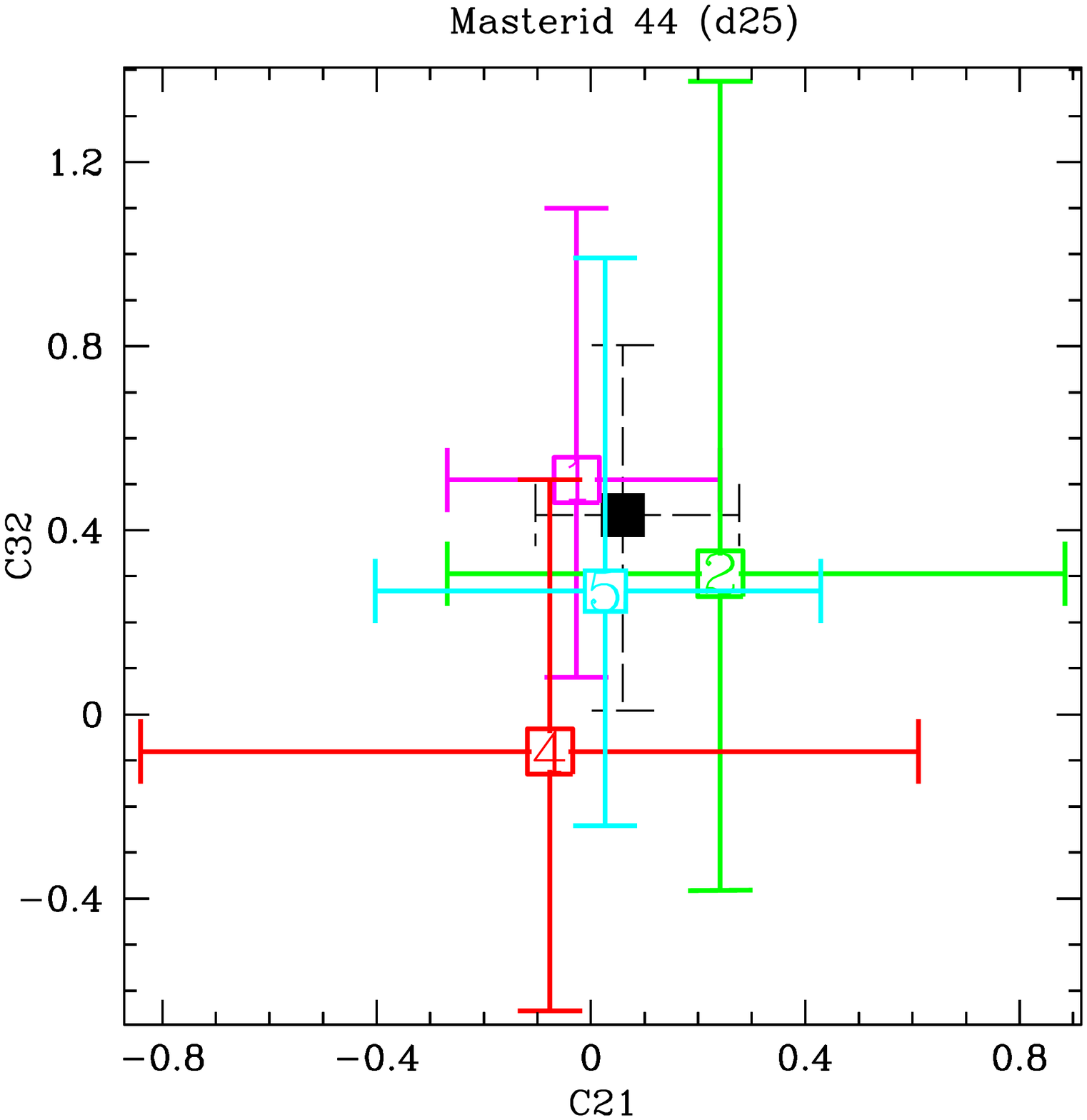}

 \end{minipage}
  
\end{figure}

\clearpage

\begin{figure}
  \begin{minipage}{0.32\linewidth}
  \centering
  
    \includegraphics[width=\linewidth]{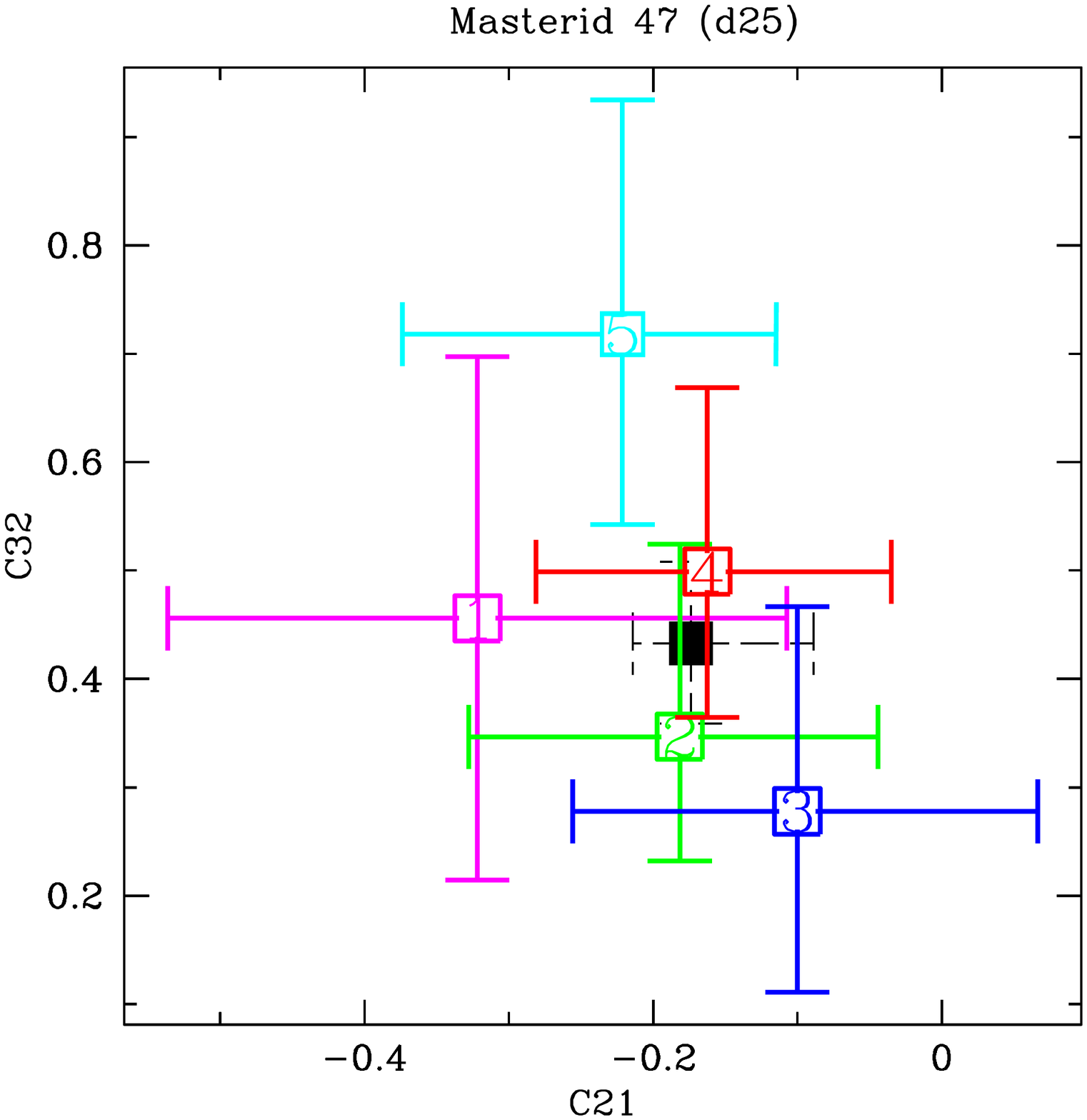}
  
  \end{minipage}
  \begin{minipage}{0.32\linewidth}
  \centering

    \includegraphics[width=\linewidth]{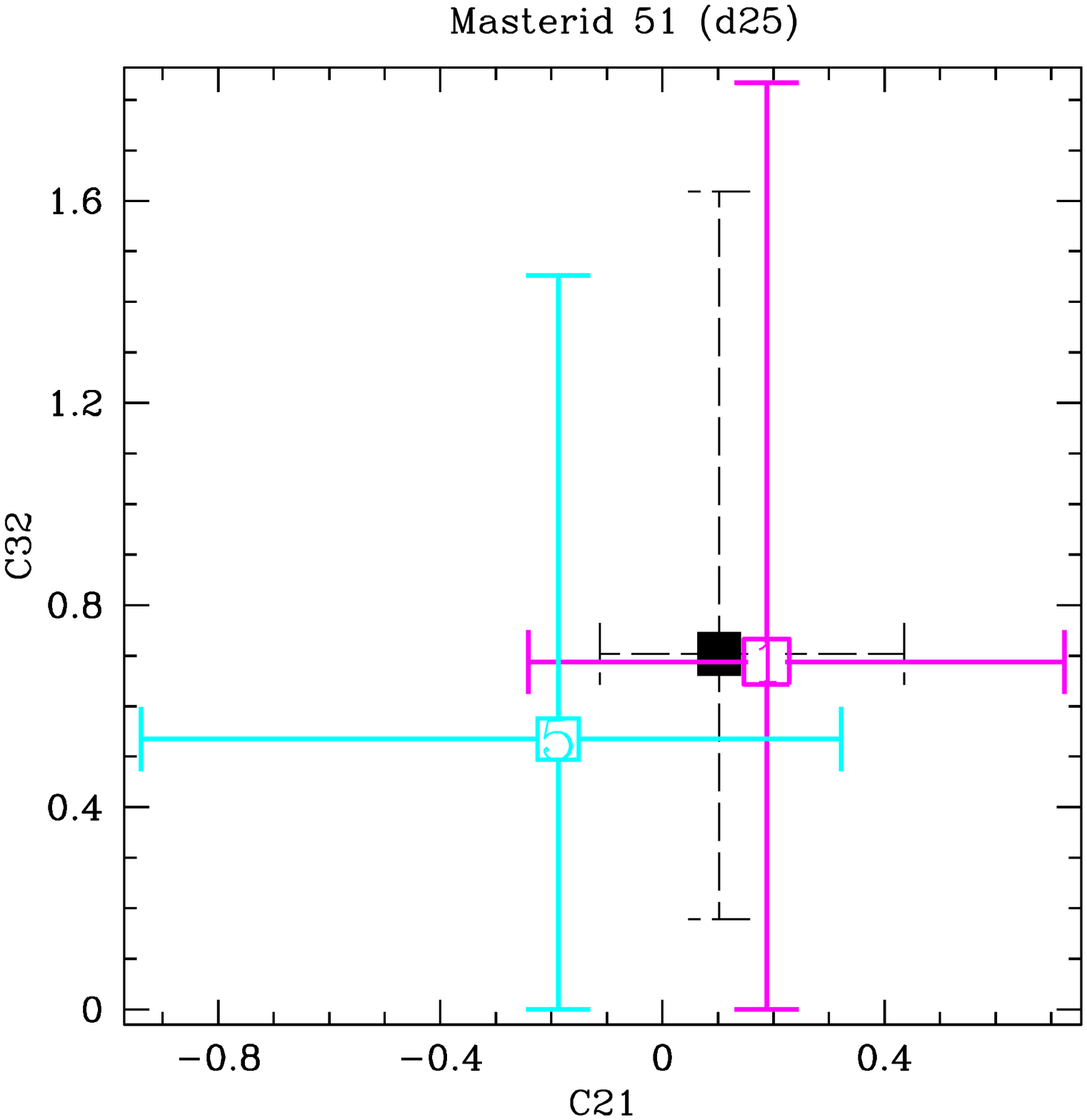}

\end{minipage}
\begin{minipage}{0.32\linewidth}
  \centering

    \includegraphics[width=\linewidth]{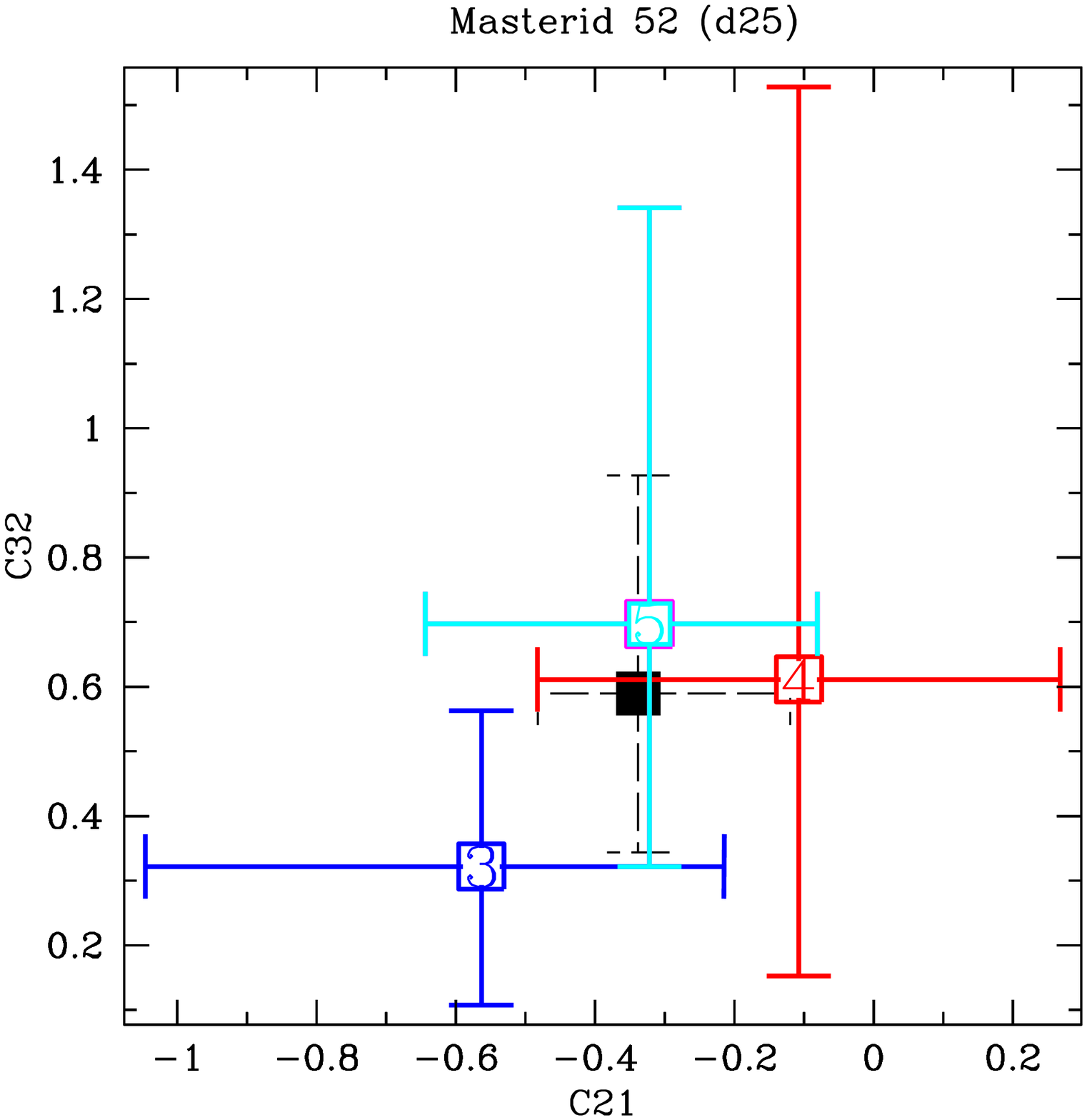}

 \end{minipage}

\begin{minipage}{0.32\linewidth}
  \centering
  
    \includegraphics[width=\linewidth]{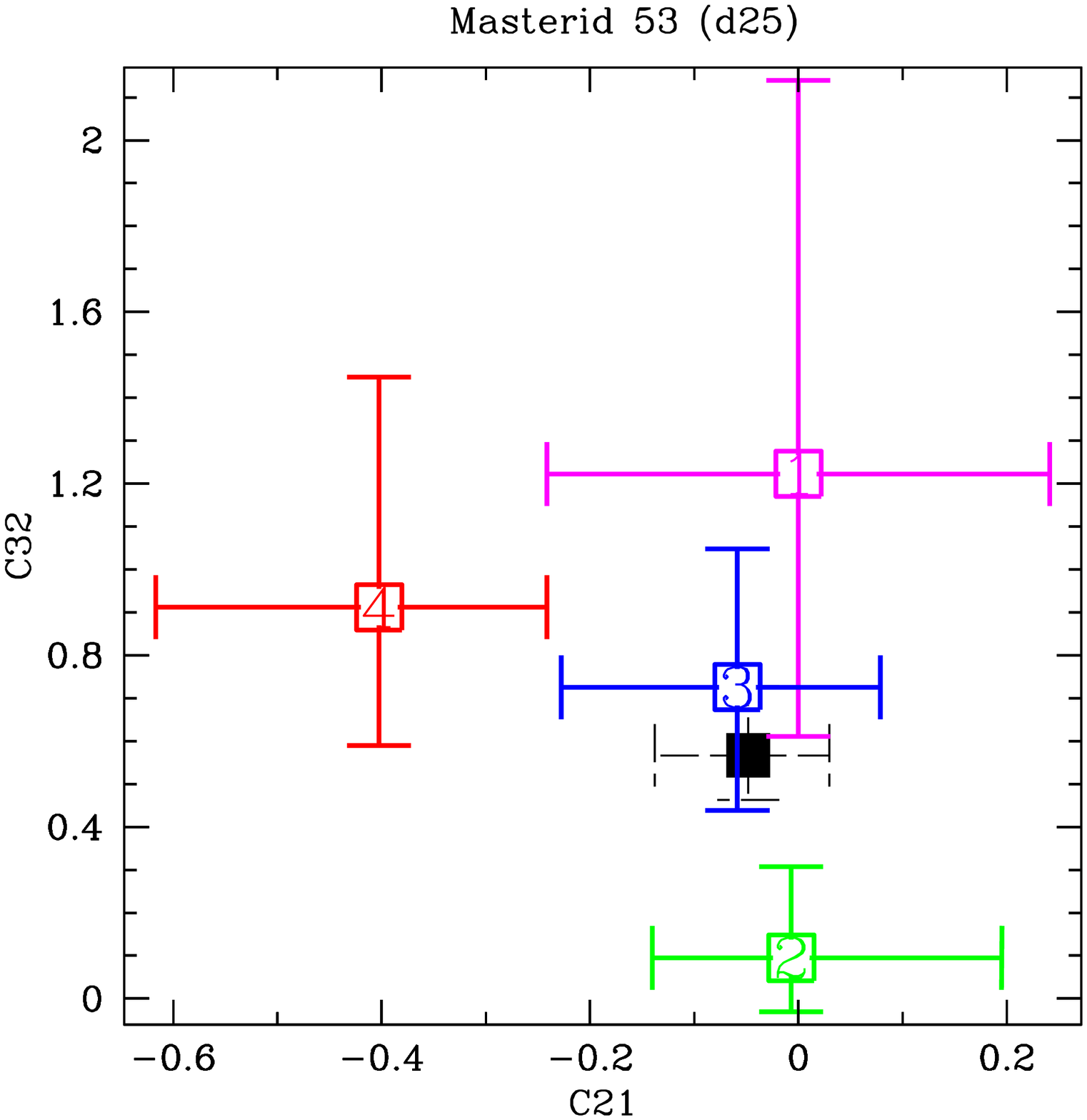}

  \end{minipage}
  \begin{minipage}{0.32\linewidth}
  \centering

    \includegraphics[width=\linewidth]{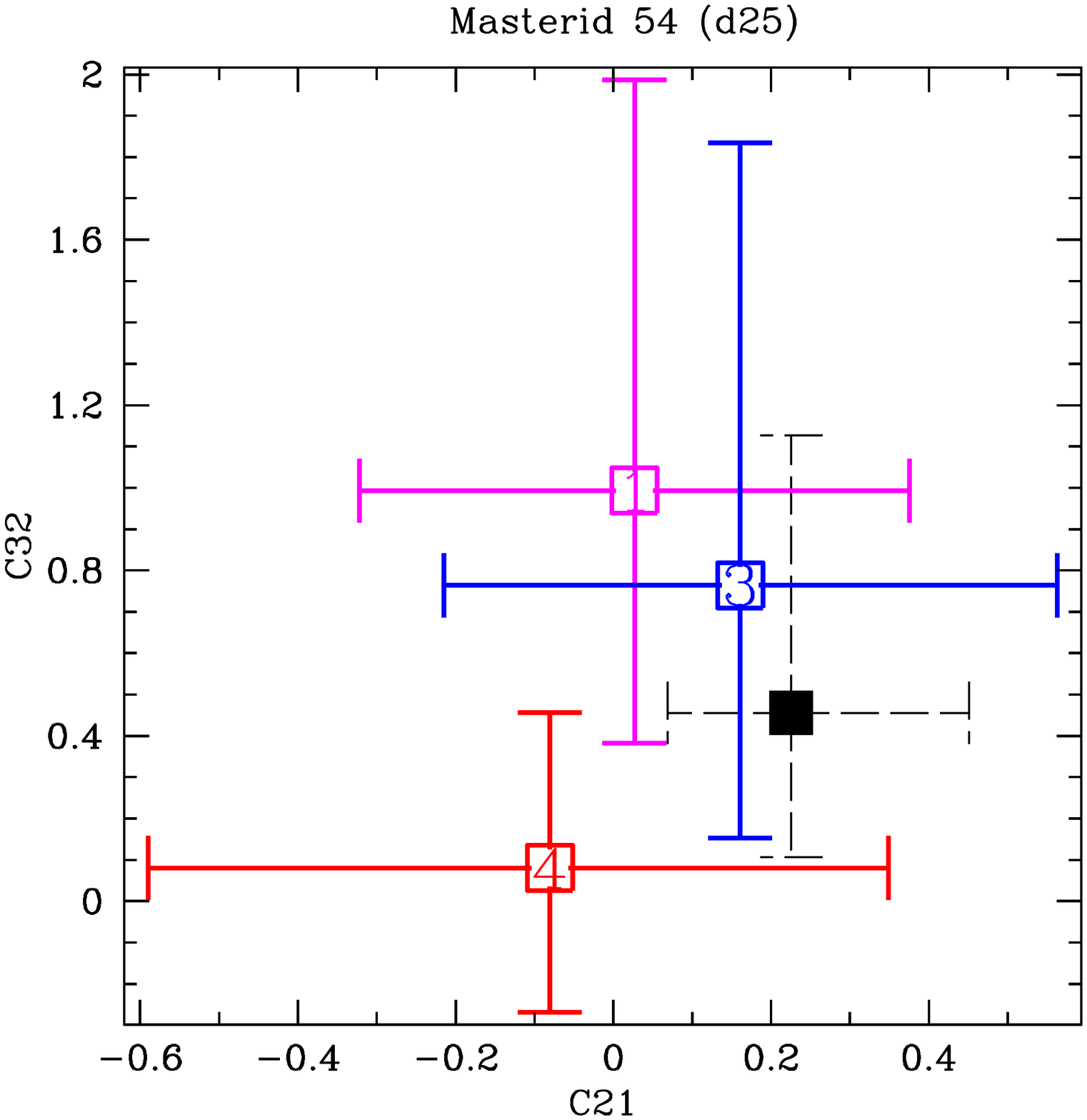}

\end{minipage}
\begin{minipage}{0.32\linewidth}
  \centering

    \includegraphics[width=\linewidth]{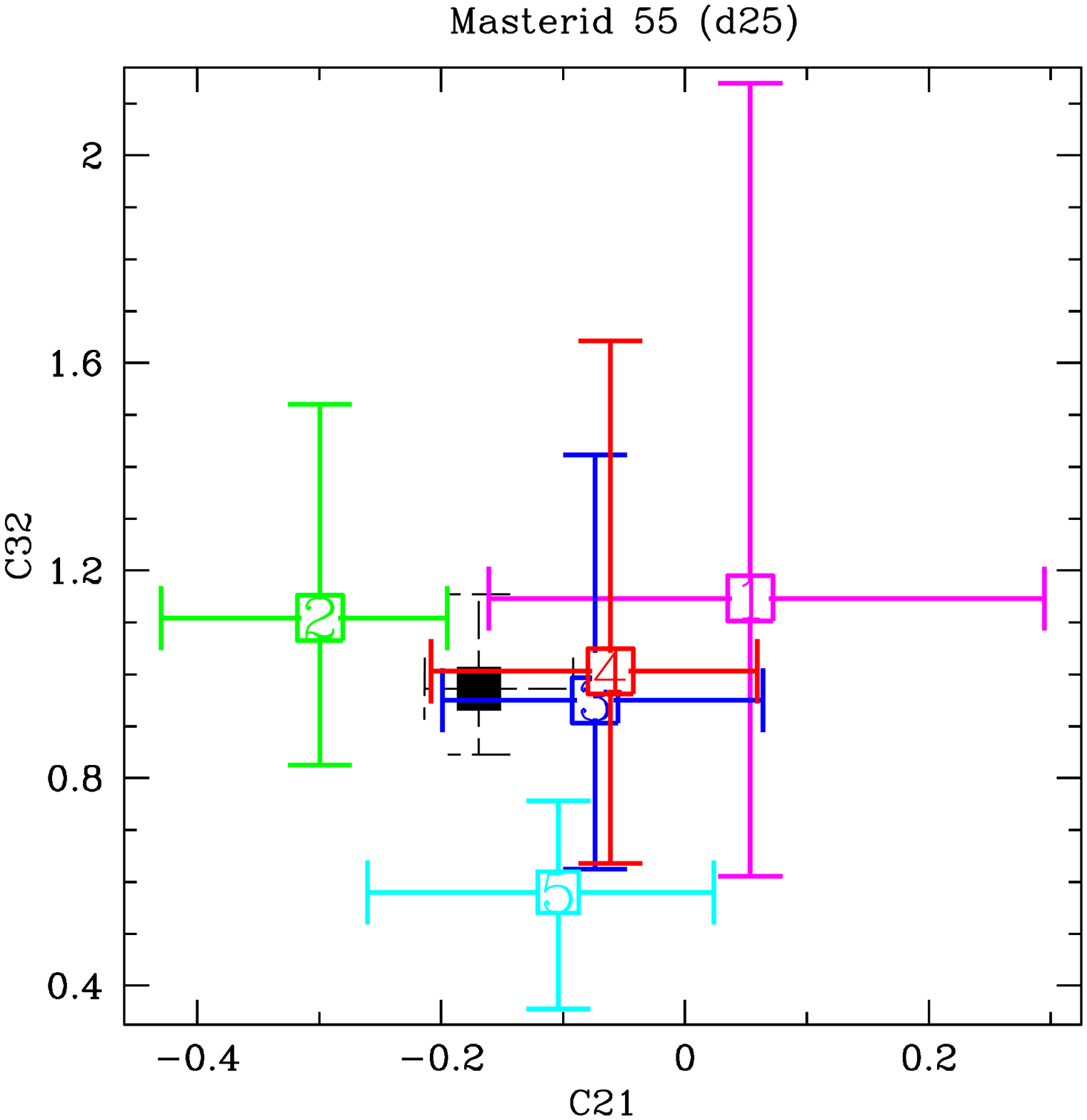}

 \end{minipage}

  \begin{minipage}{0.32\linewidth}
  \centering
  
    \includegraphics[width=\linewidth]{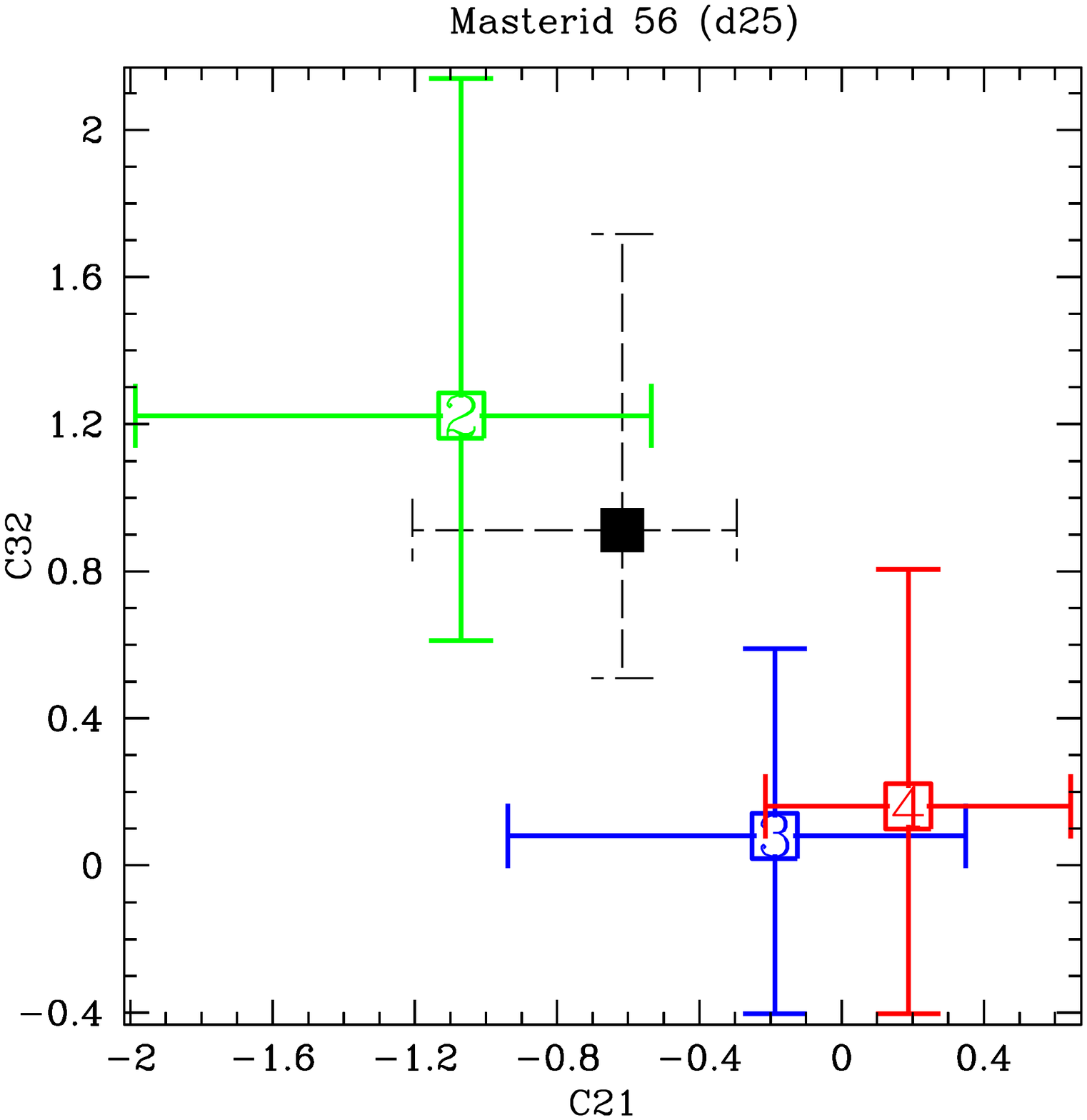}

  \end{minipage}
  \begin{minipage}{0.32\linewidth}
  \centering

    \includegraphics[width=\linewidth]{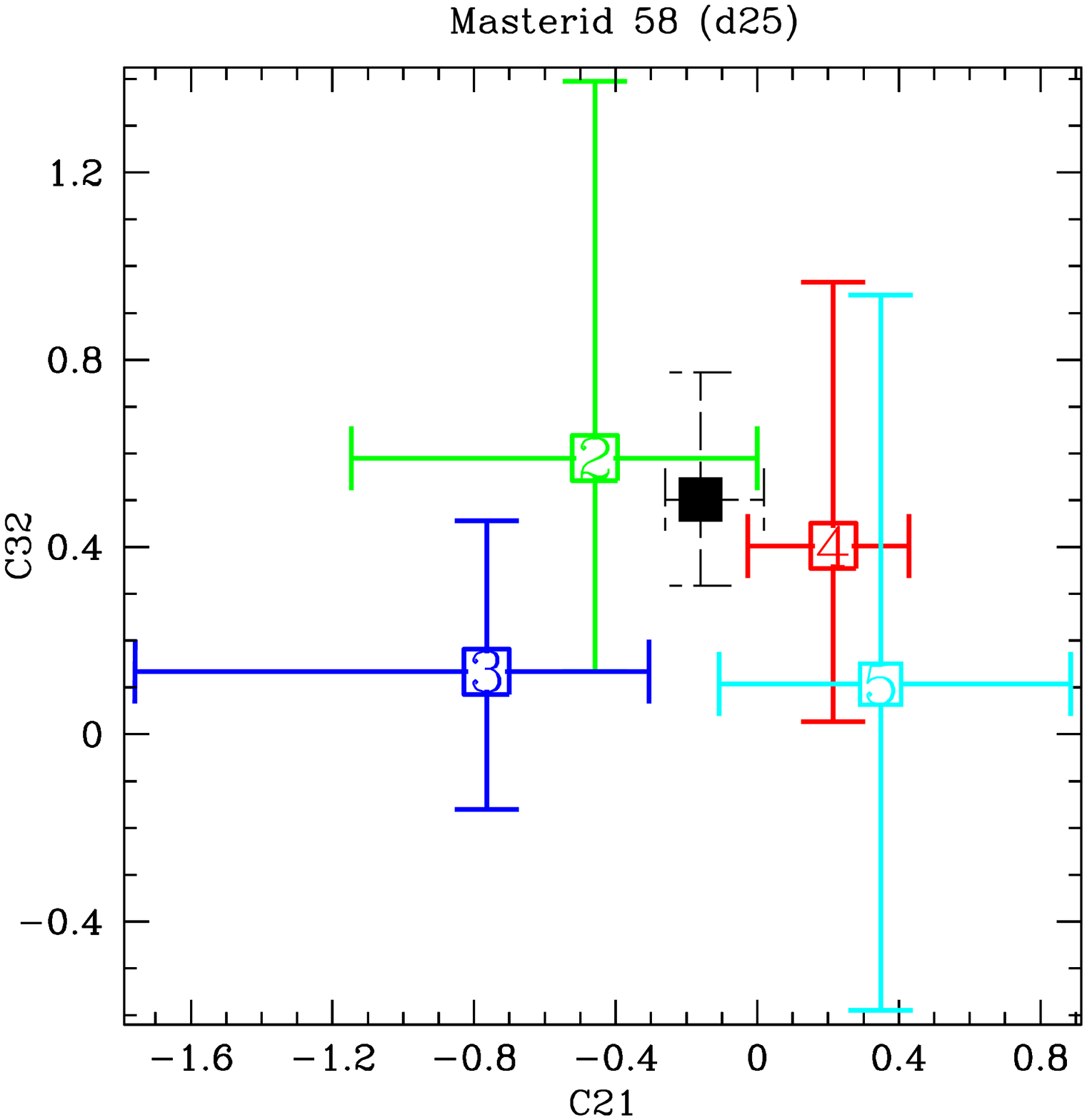}

\end{minipage}
\begin{minipage}{0.32\linewidth}
  \centering

    \includegraphics[width=\linewidth]{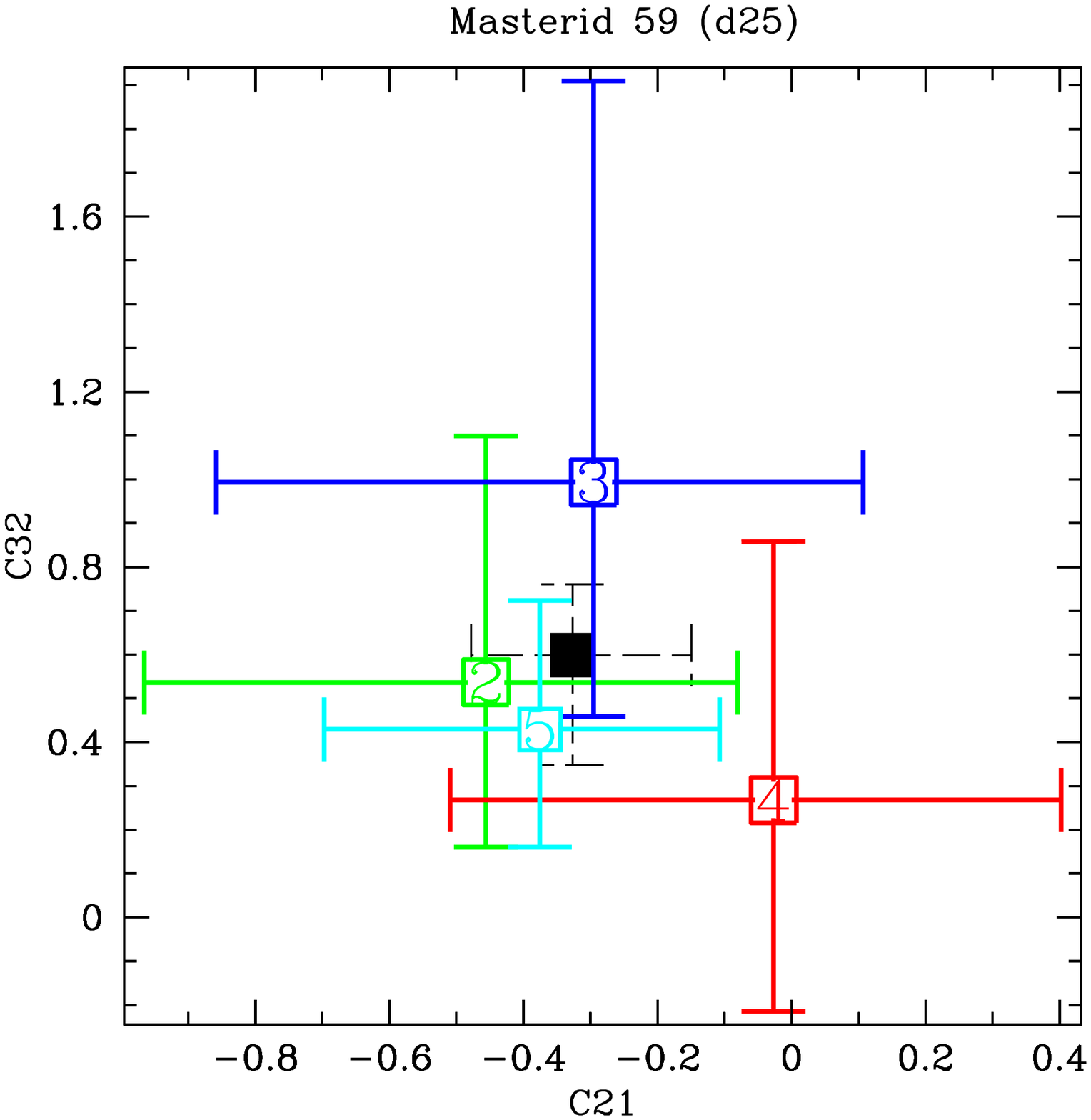}

 \end{minipage}

\begin{minipage}{0.32\linewidth}
  \centering
  
    \includegraphics[width=\linewidth]{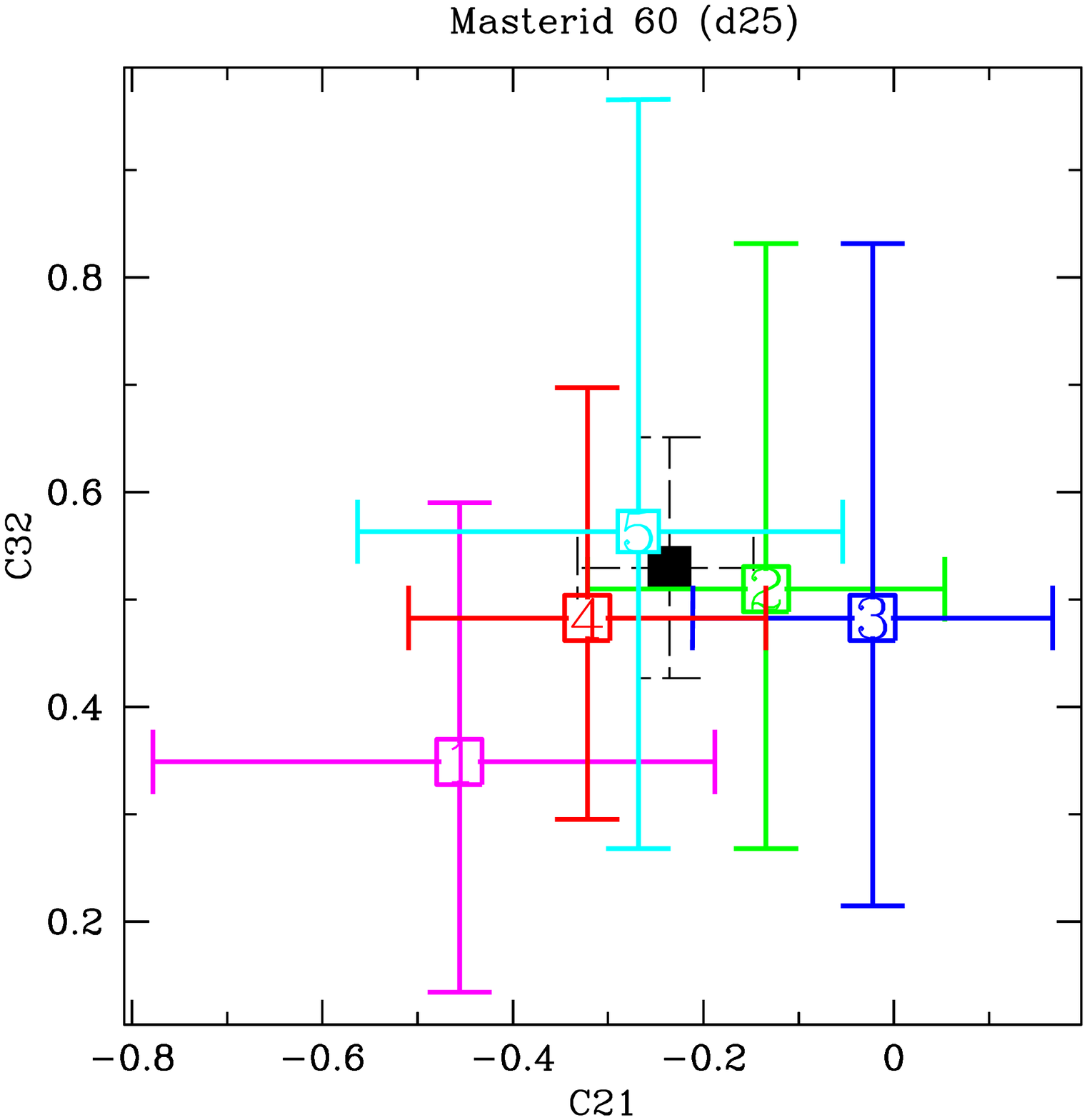}
  
  \end{minipage}
  \begin{minipage}{0.32\linewidth}
  \centering

    \includegraphics[width=\linewidth]{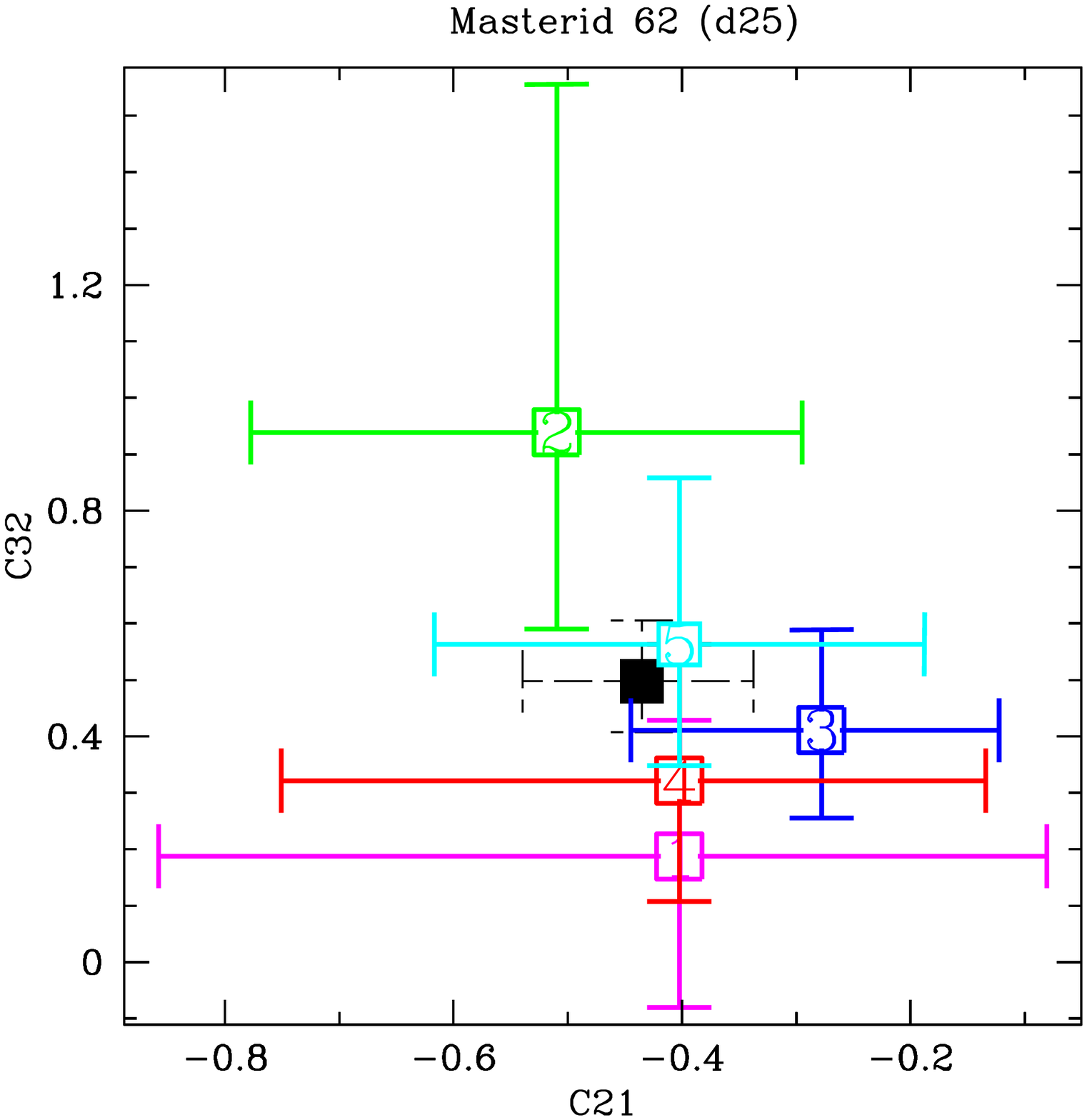}

\end{minipage}
\begin{minipage}{0.32\linewidth}
  \centering

    \includegraphics[width=\linewidth]{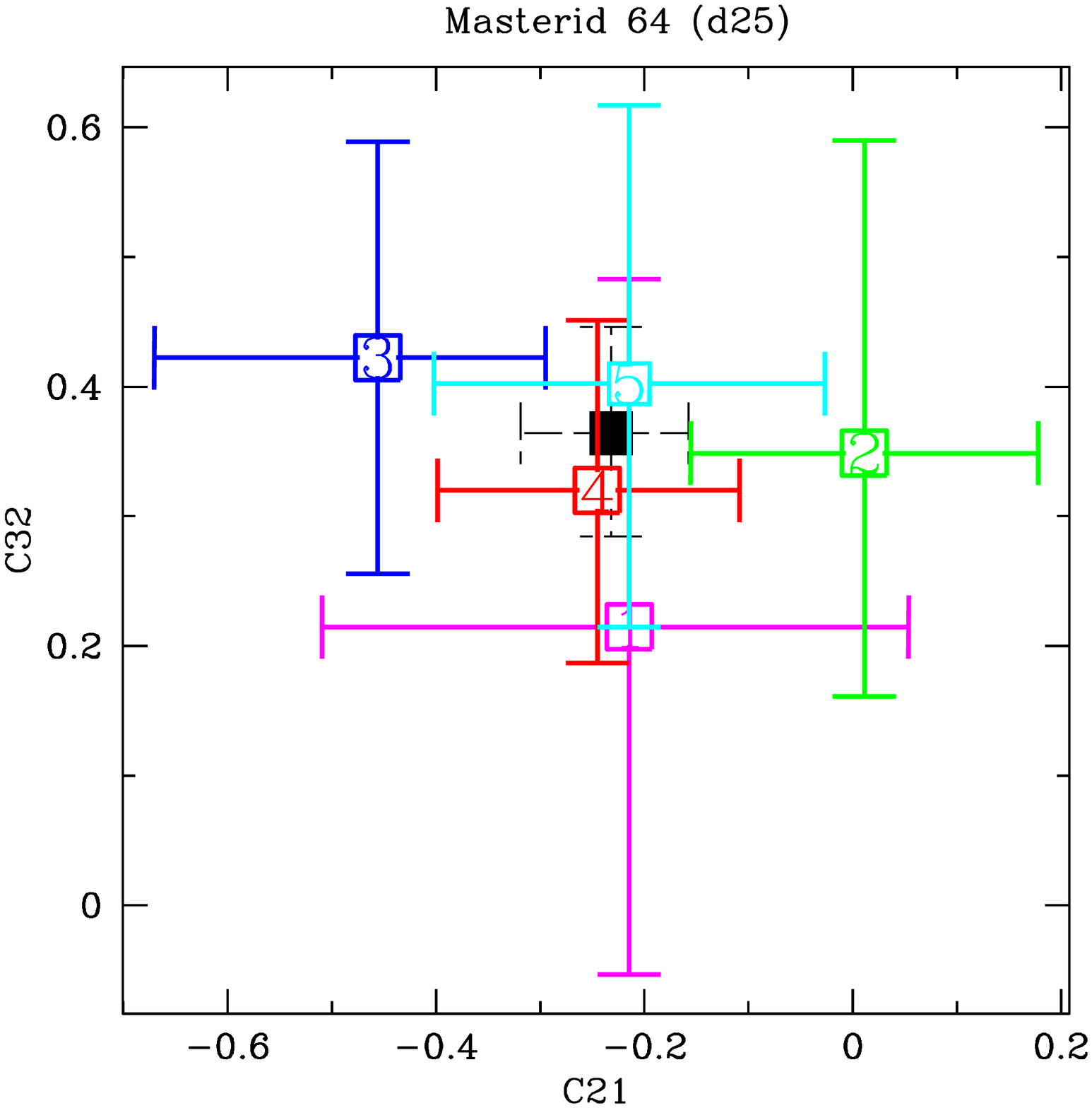}

 \end{minipage}
  
\end{figure}

\begin{figure}
  \begin{minipage}{0.32\linewidth}
  \centering
  
    \includegraphics[width=\linewidth]{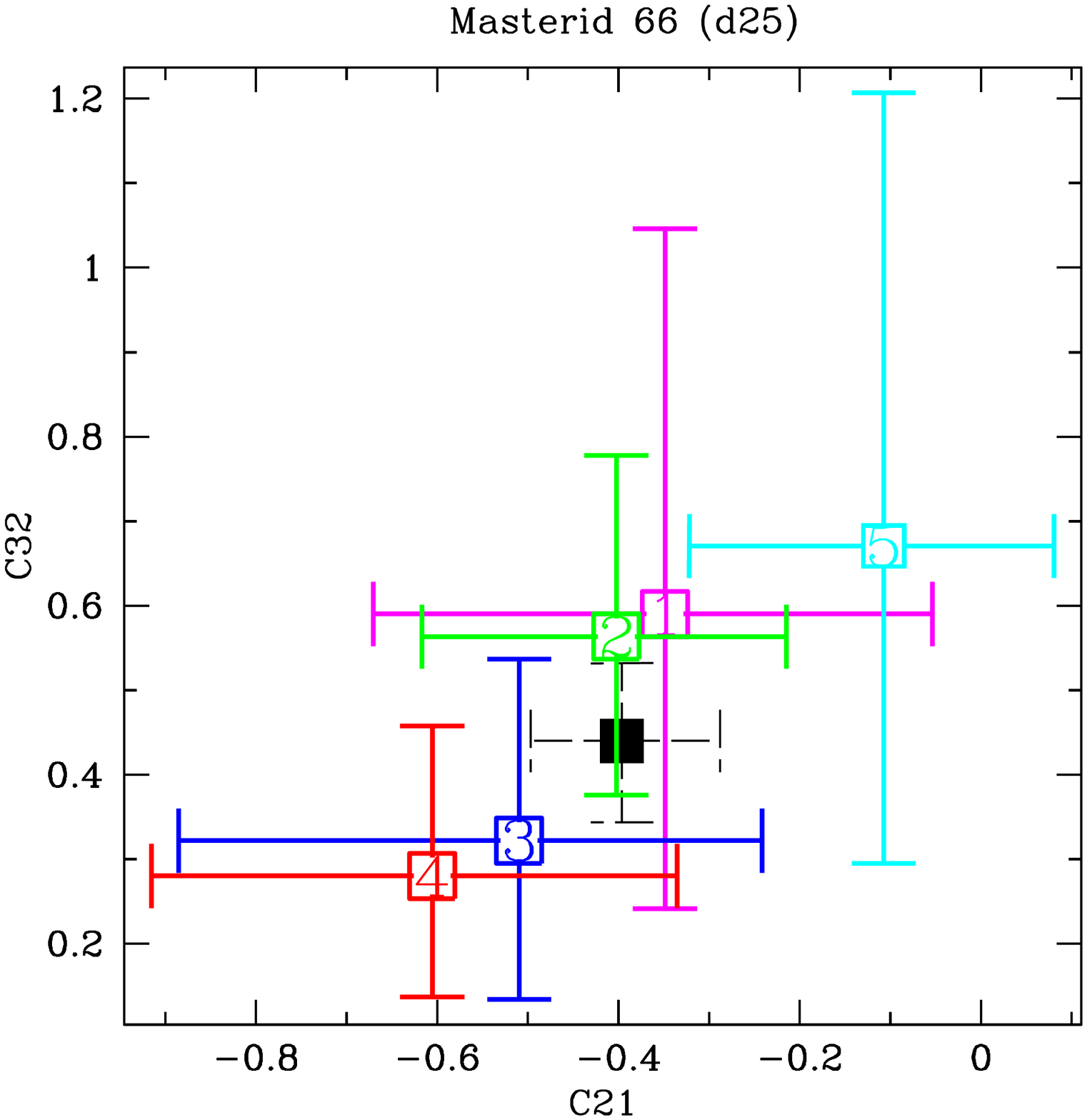}

  \end{minipage}
  \begin{minipage}{0.32\linewidth}
  \centering

    \includegraphics[width=\linewidth]{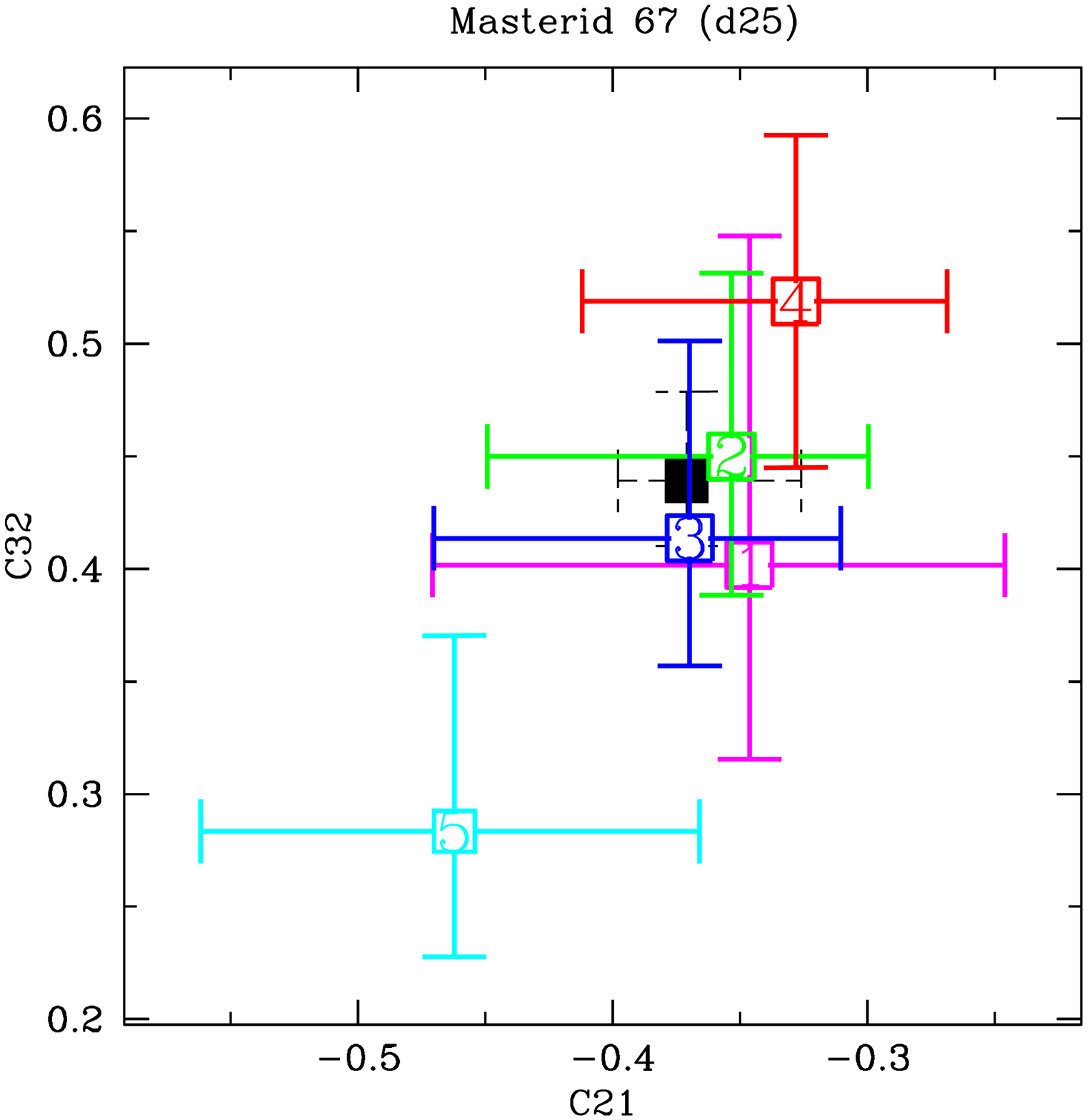}
 
\end{minipage}
\begin{minipage}{0.32\linewidth}
  \centering

    \includegraphics[width=\linewidth]{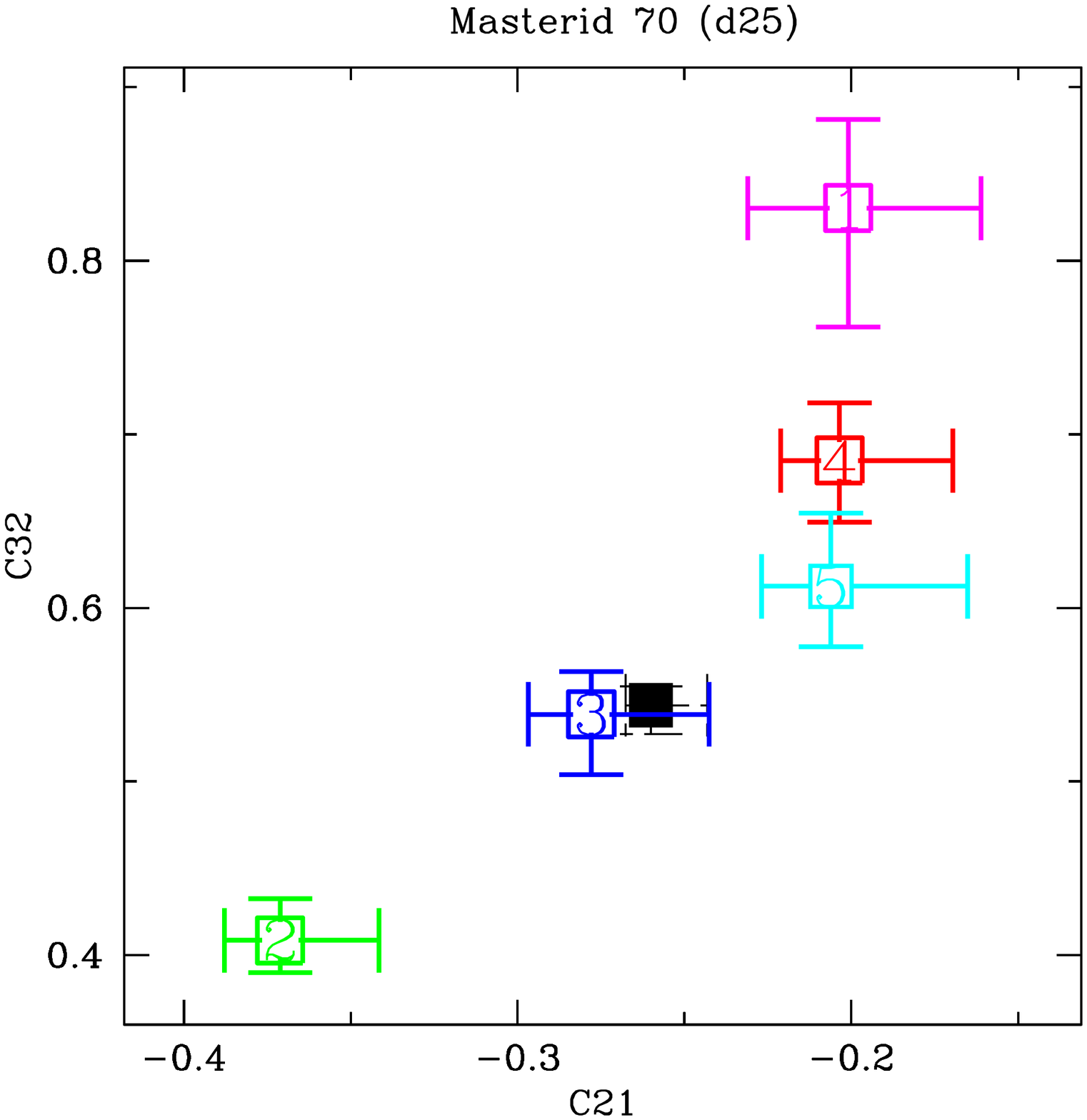}

 \end{minipage}

\begin{minipage}{0.32\linewidth}
  \centering
  
    \includegraphics[width=\linewidth]{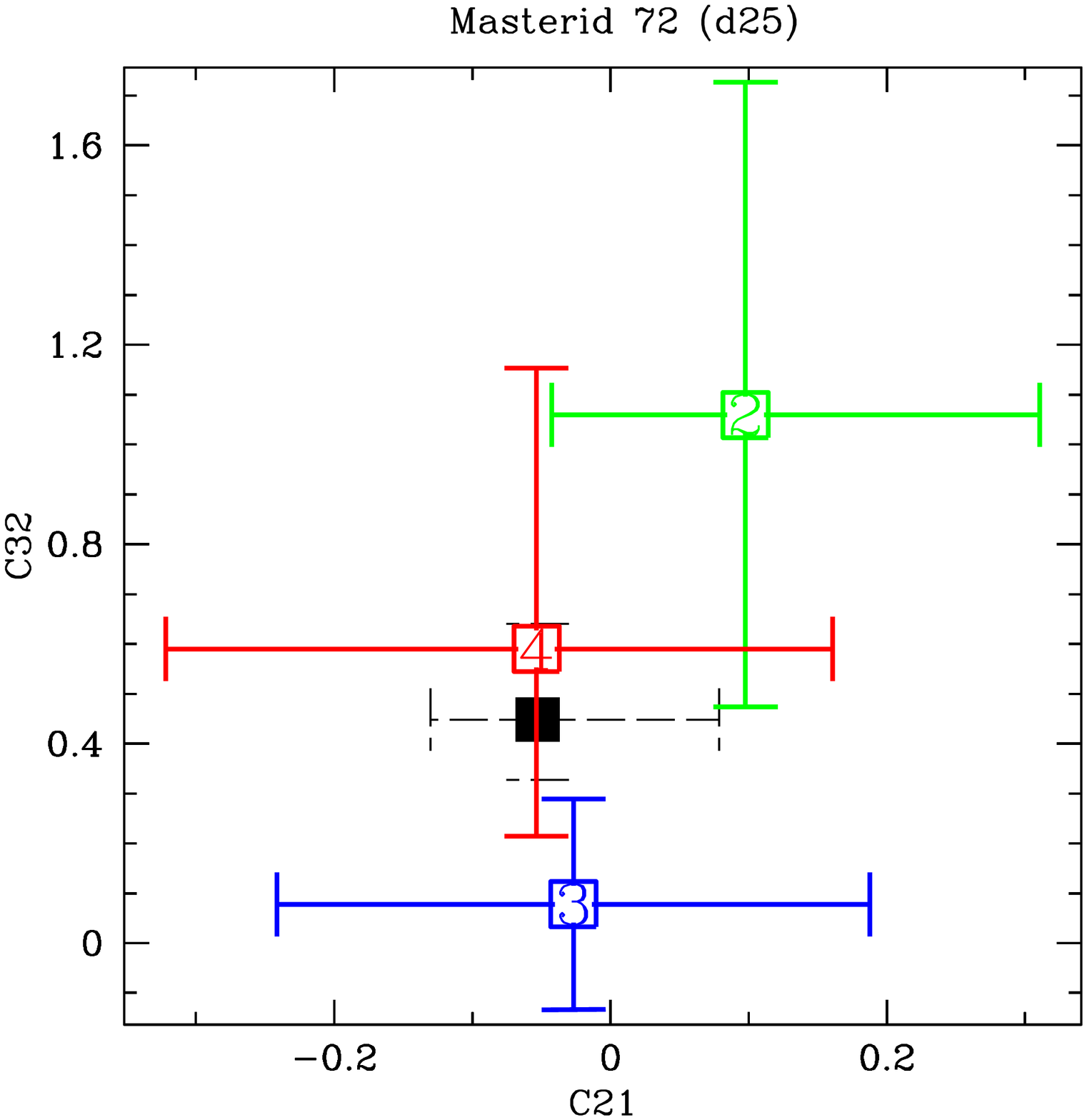}
  
  \end{minipage}
  \begin{minipage}{0.32\linewidth}
  \centering

    \includegraphics[width=\linewidth]{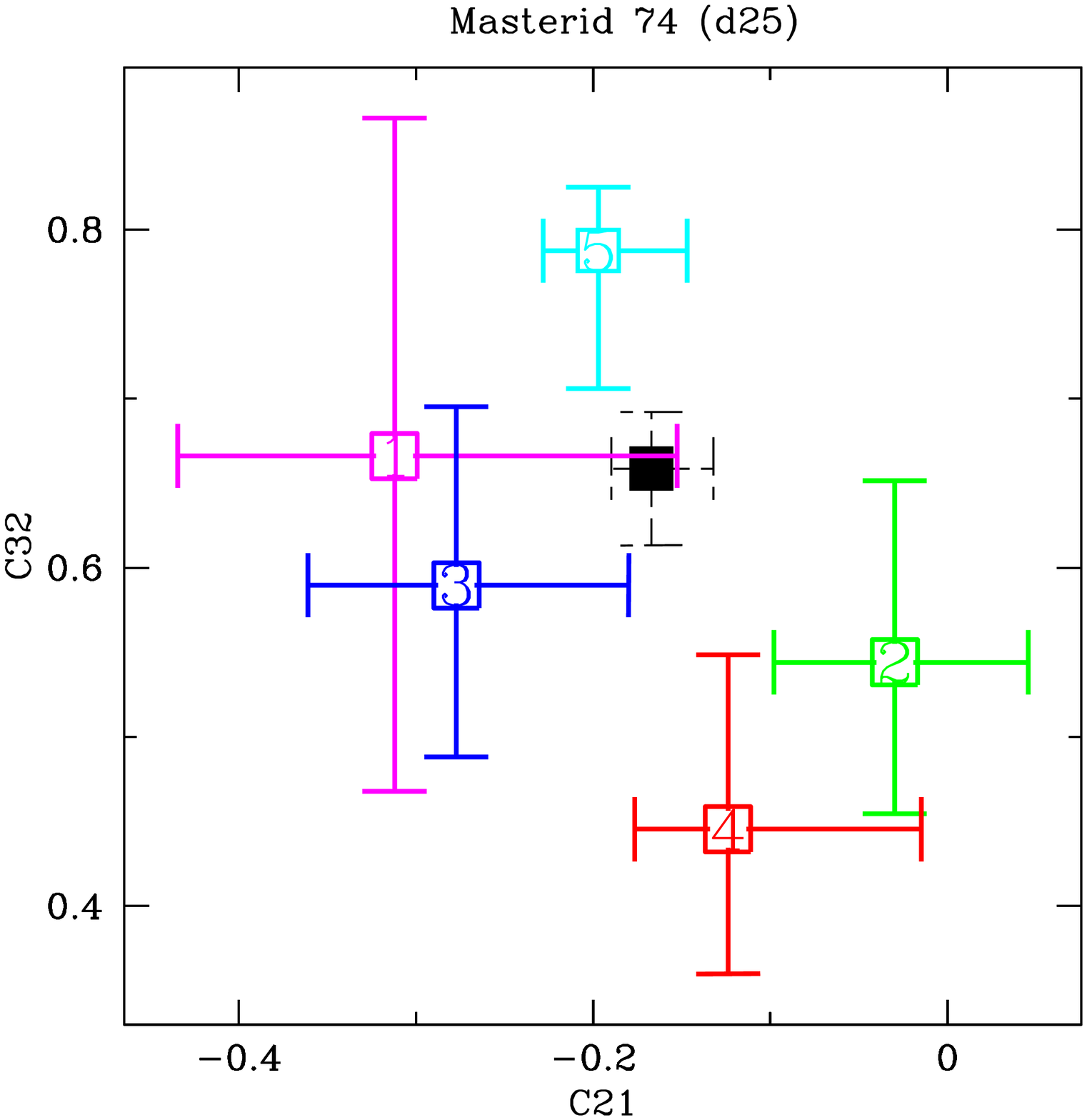}

\end{minipage}
\begin{minipage}{0.32\linewidth}
  \centering

    \includegraphics[width=\linewidth]{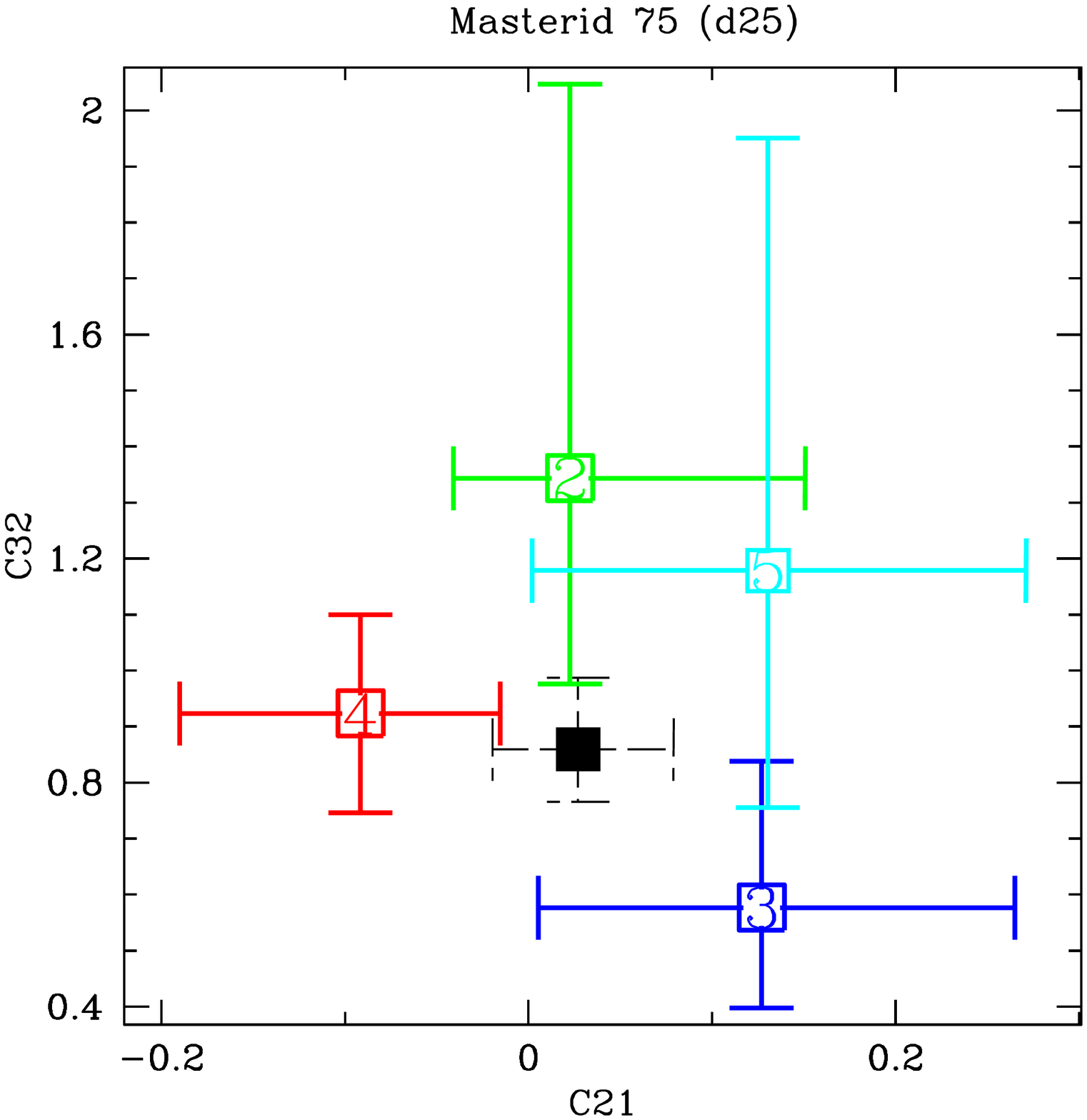}

 \end{minipage}

  \begin{minipage}{0.32\linewidth}
  \centering
  
    \includegraphics[width=\linewidth]{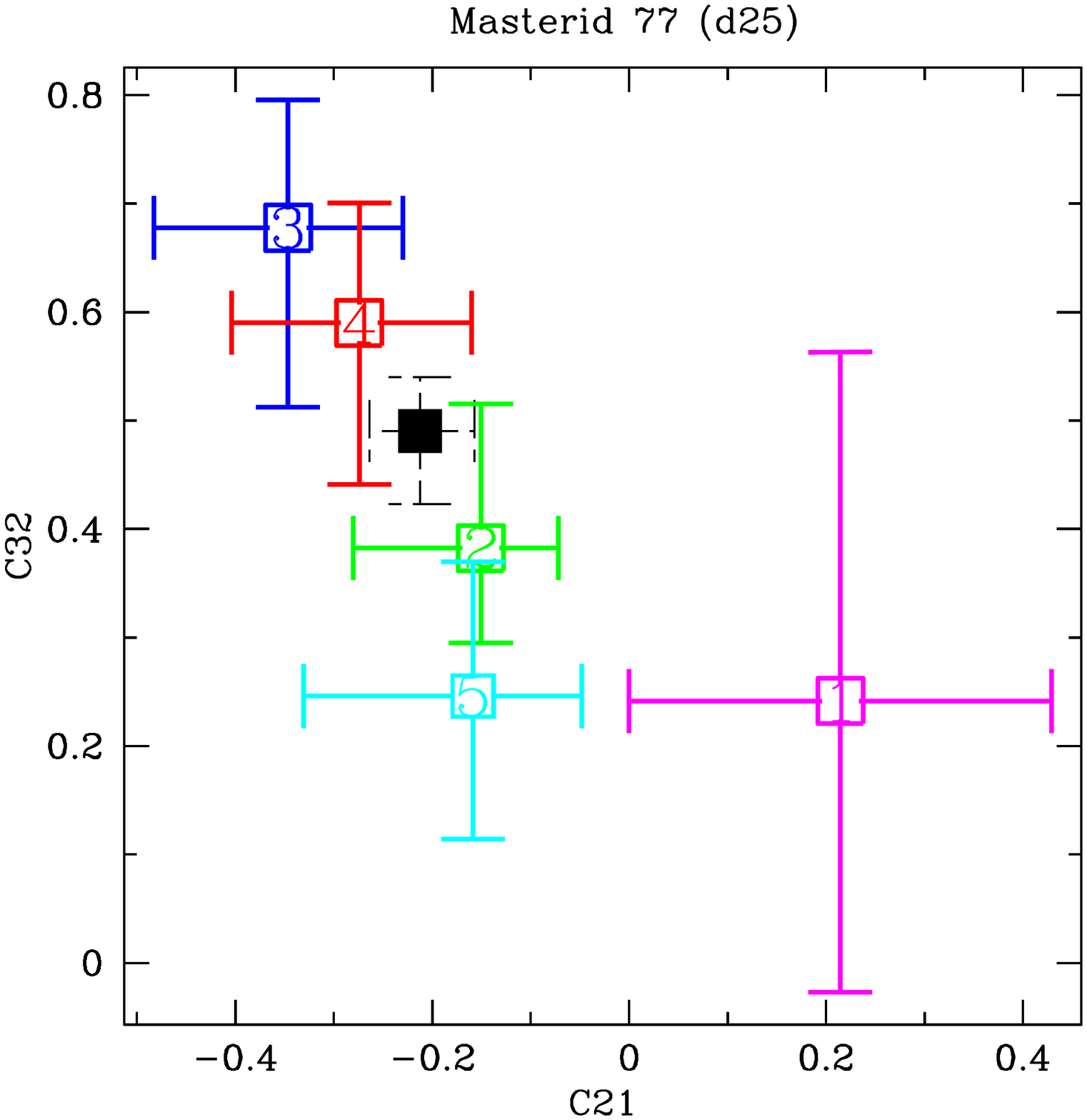}
  
  \end{minipage}
  \begin{minipage}{0.32\linewidth}
  \centering

    \includegraphics[width=\linewidth]{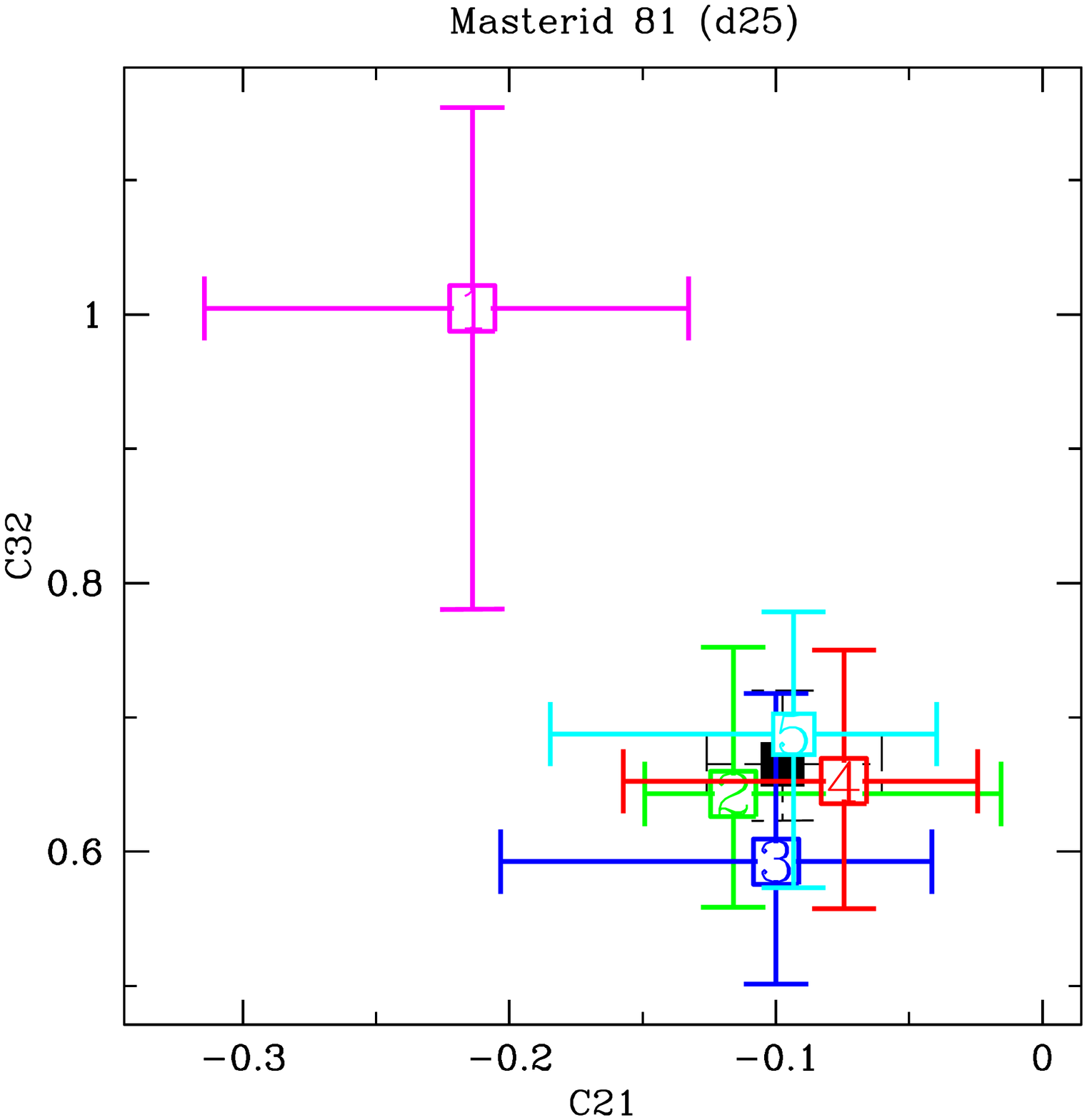}

\end{minipage}
\begin{minipage}{0.32\linewidth}
  \centering

    \includegraphics[width=\linewidth]{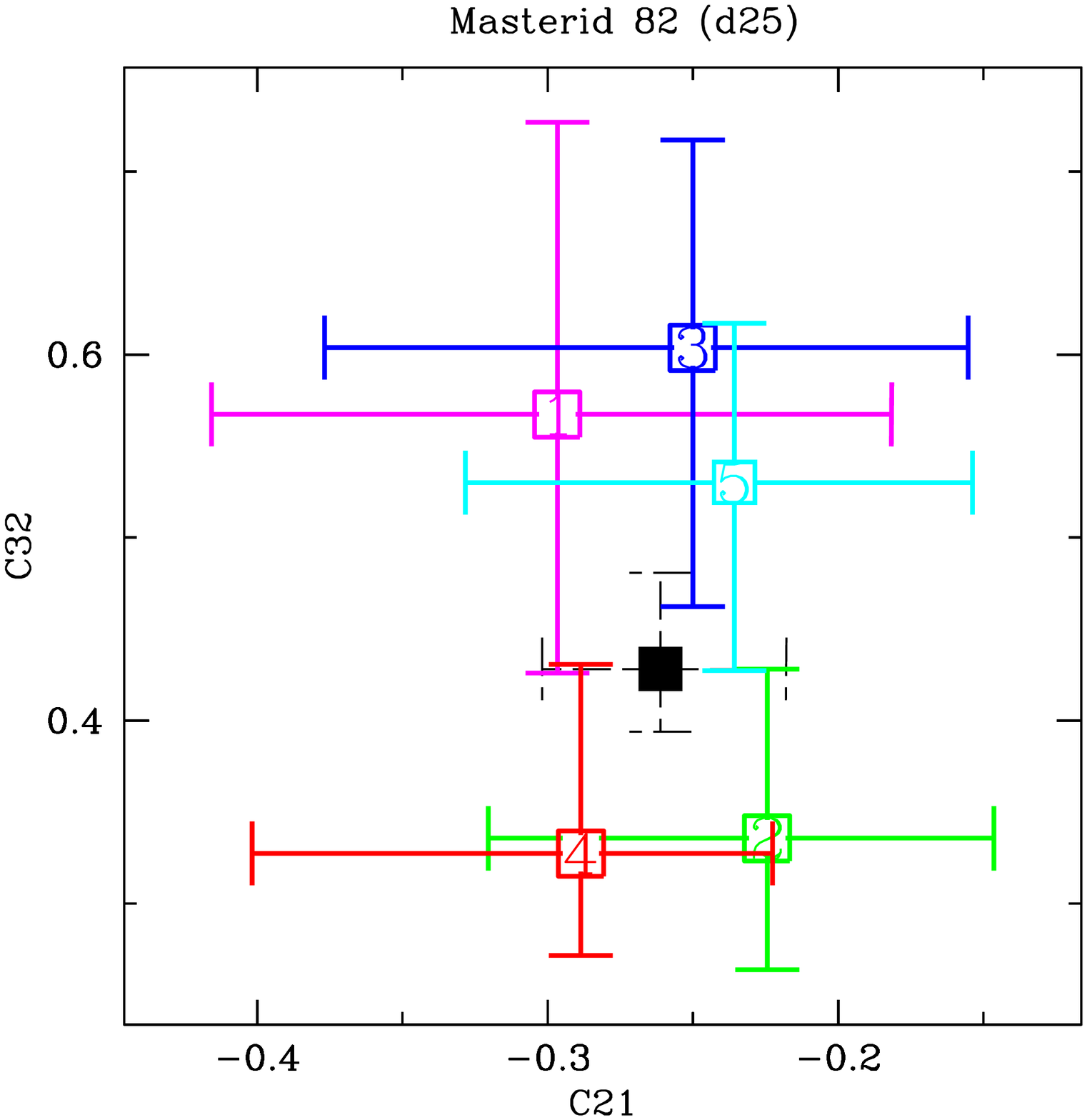}

 \end{minipage}

\begin{minipage}{0.32\linewidth}
  \centering
  
    \includegraphics[width=\linewidth]{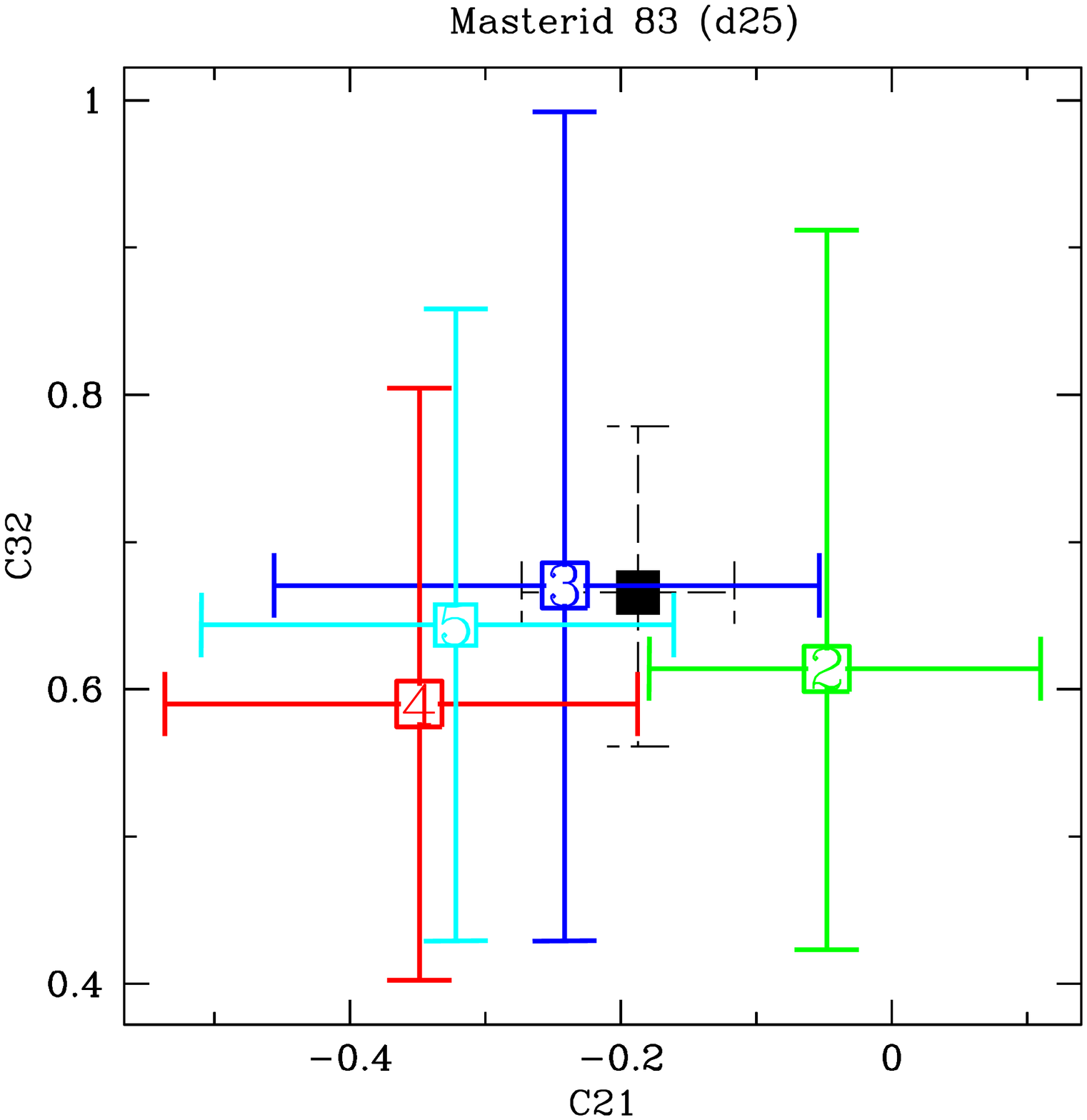}
  
  \end{minipage}
  \begin{minipage}{0.32\linewidth}
  \centering

    \includegraphics[width=\linewidth]{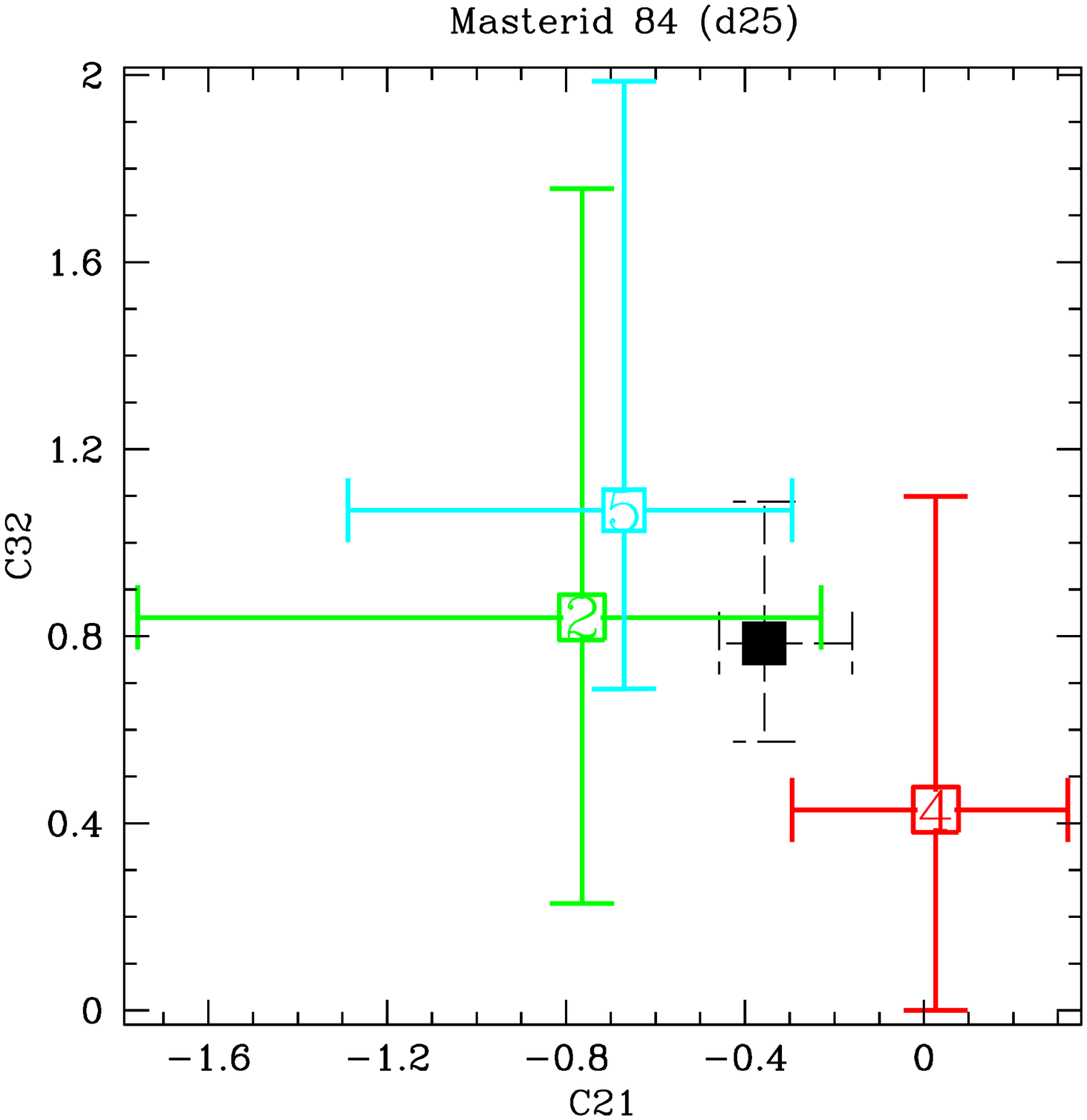}

\end{minipage}
\begin{minipage}{0.32\linewidth}
  \centering

    \includegraphics[width=\linewidth]{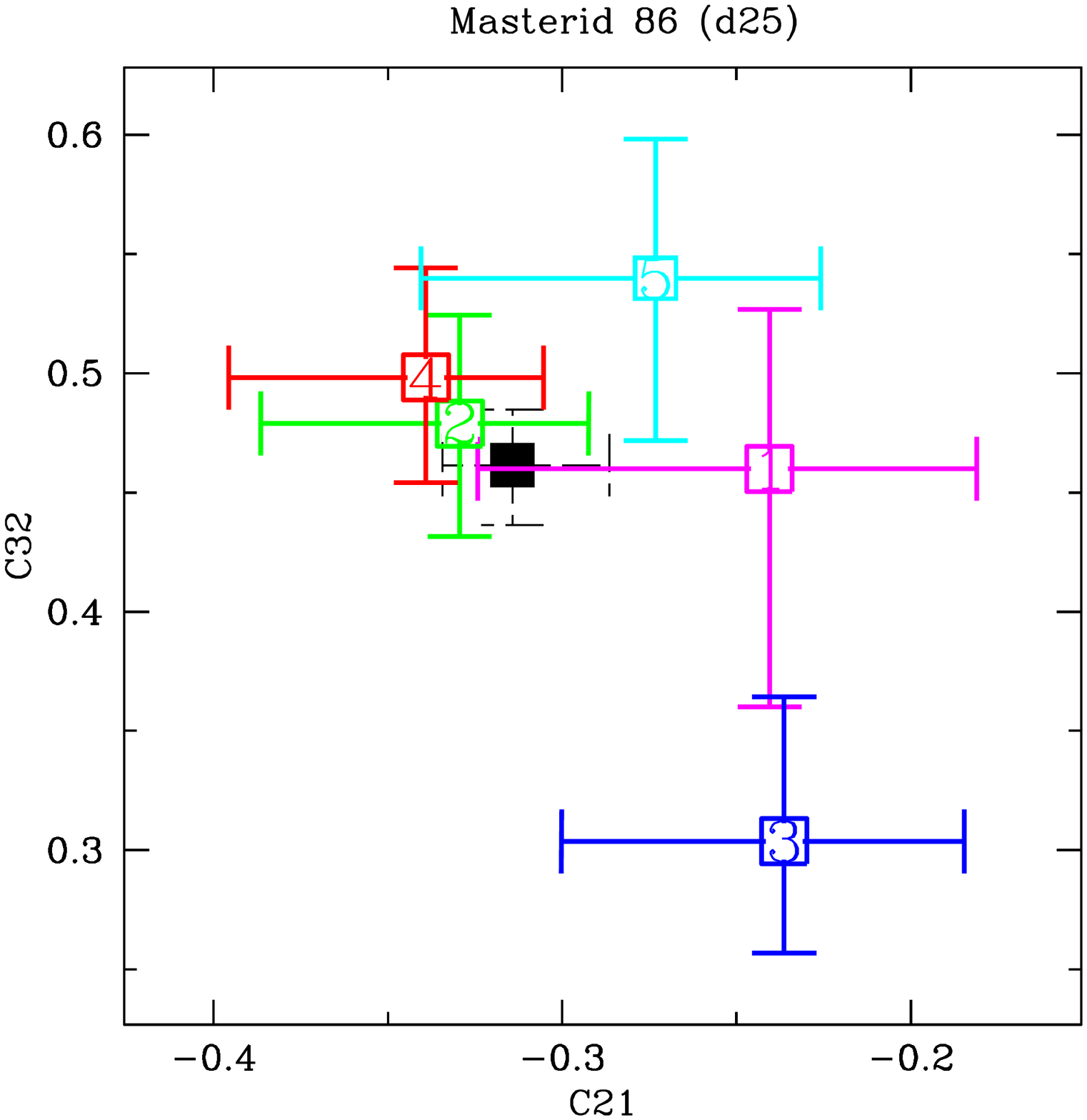}

 \end{minipage}
  
\end{figure}

\clearpage

\begin{figure}
  \begin{minipage}{0.32\linewidth}
  \centering
  
    \includegraphics[width=\linewidth]{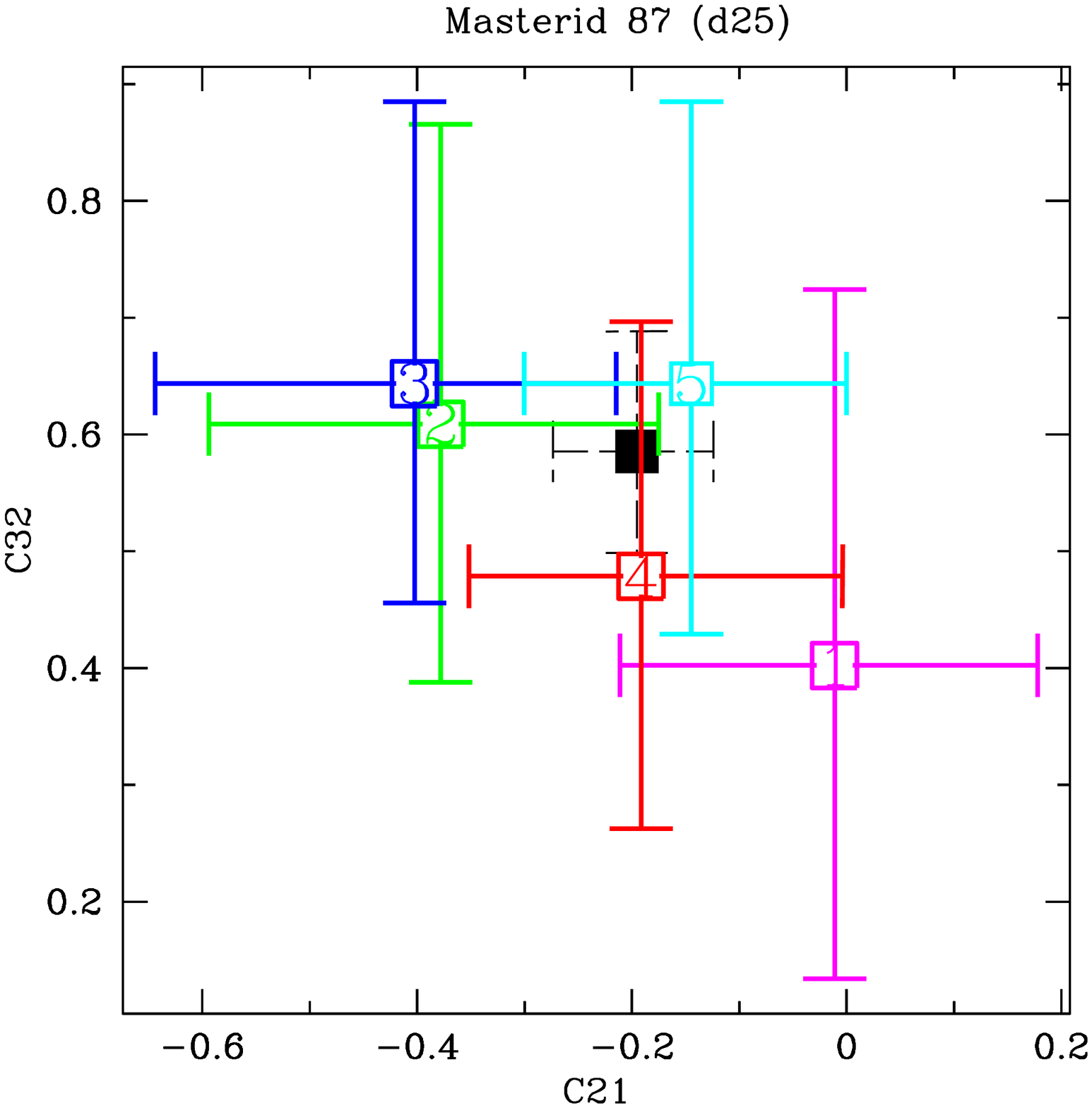}

  \end{minipage}
  \begin{minipage}{0.32\linewidth}
  \centering

    \includegraphics[width=\linewidth]{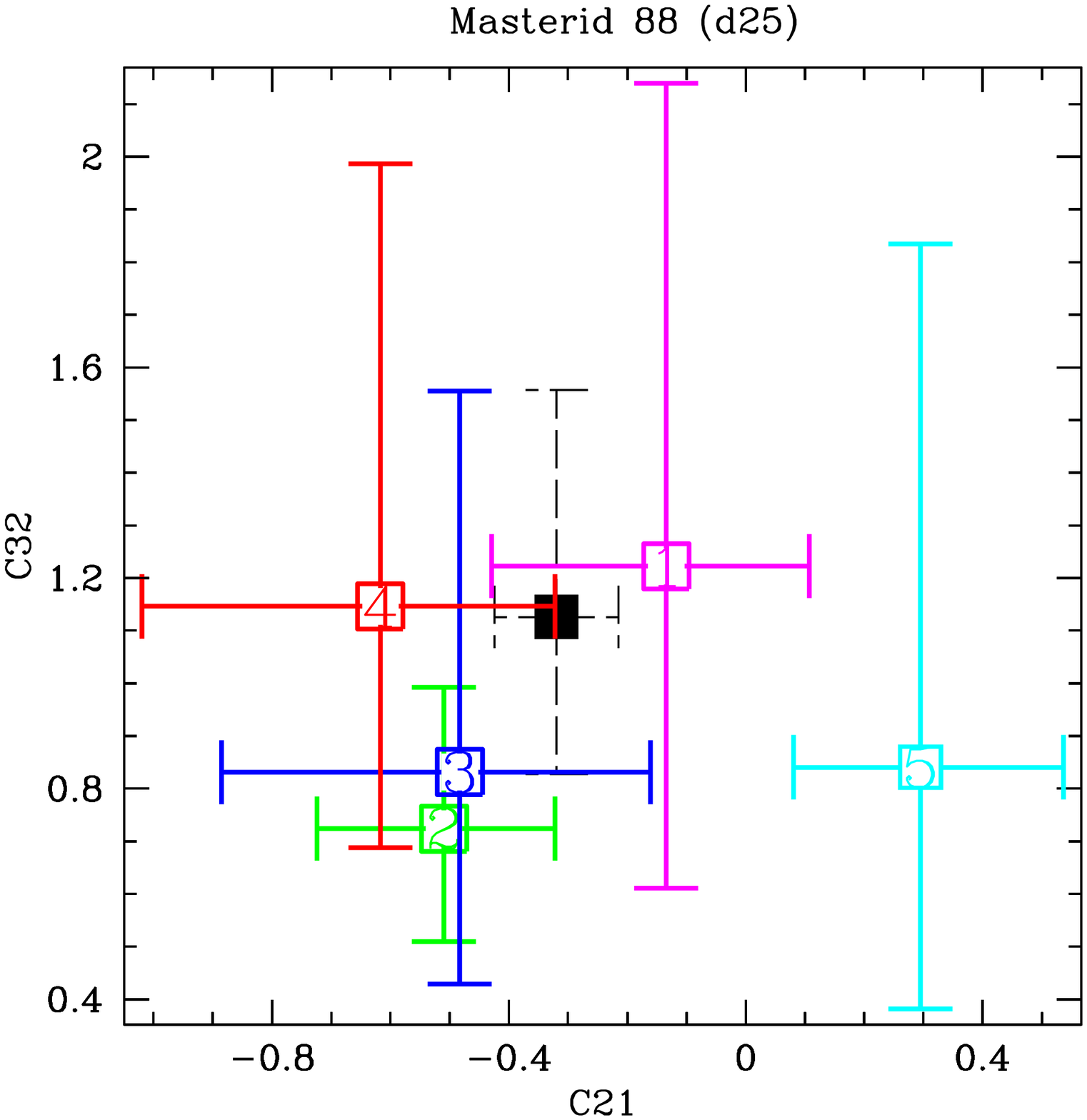}

\end{minipage}
\begin{minipage}{0.32\linewidth}
  \centering

    \includegraphics[width=\linewidth]{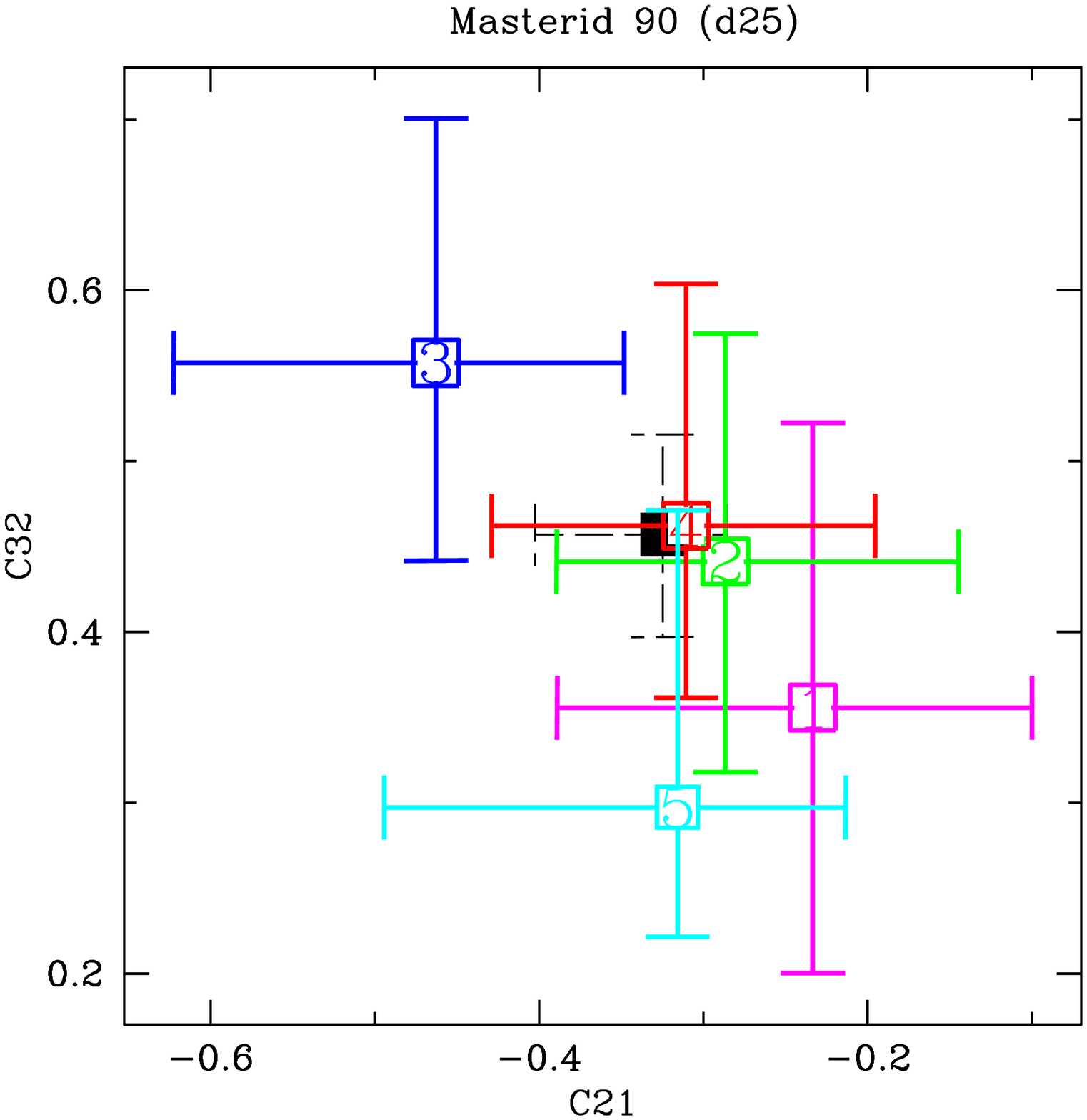}

 \end{minipage}

\begin{minipage}{0.32\linewidth}
  \centering
  
    \includegraphics[width=\linewidth]{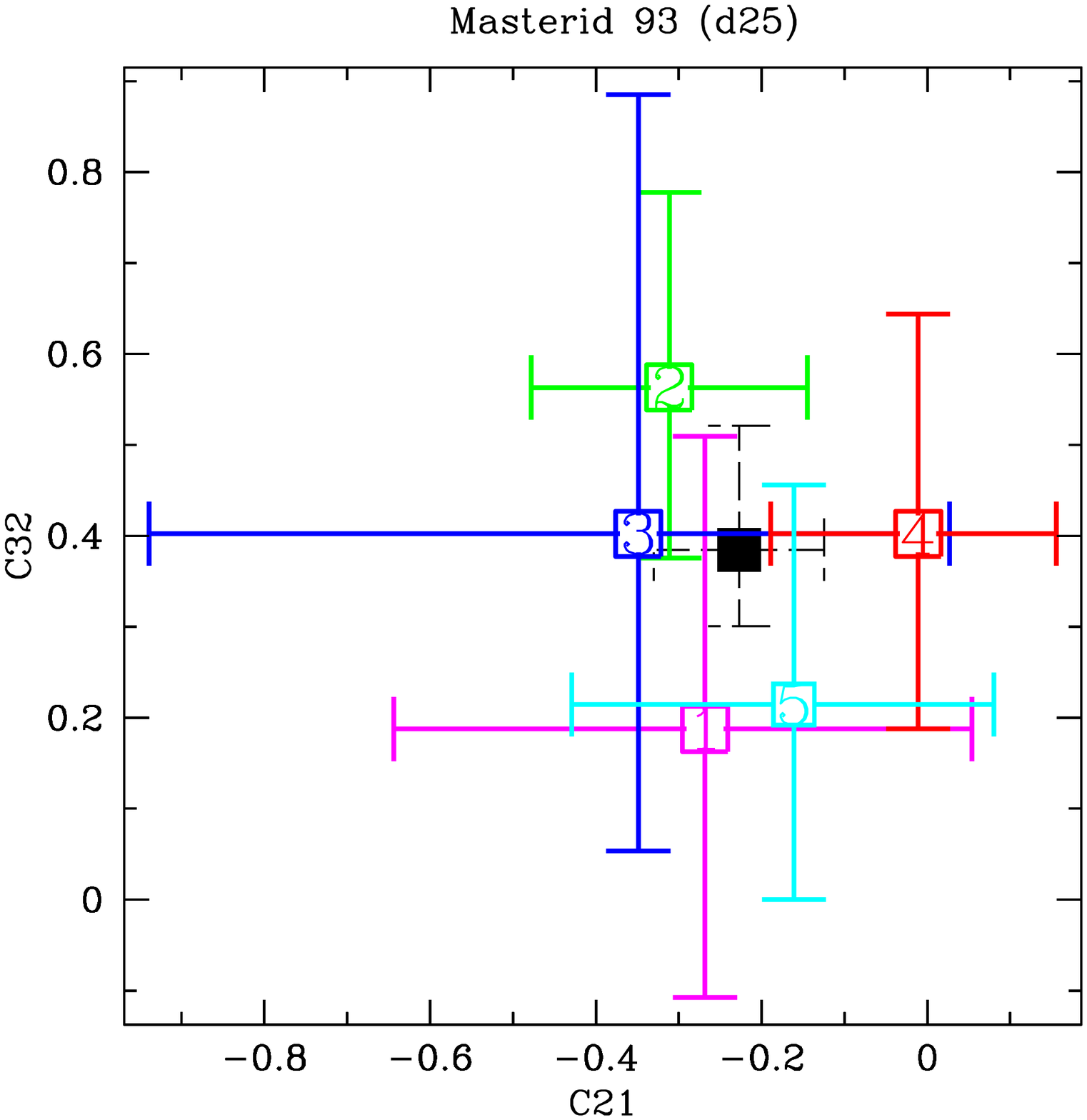}

  \end{minipage}
  \begin{minipage}{0.32\linewidth}
  \centering

    \includegraphics[width=\linewidth]{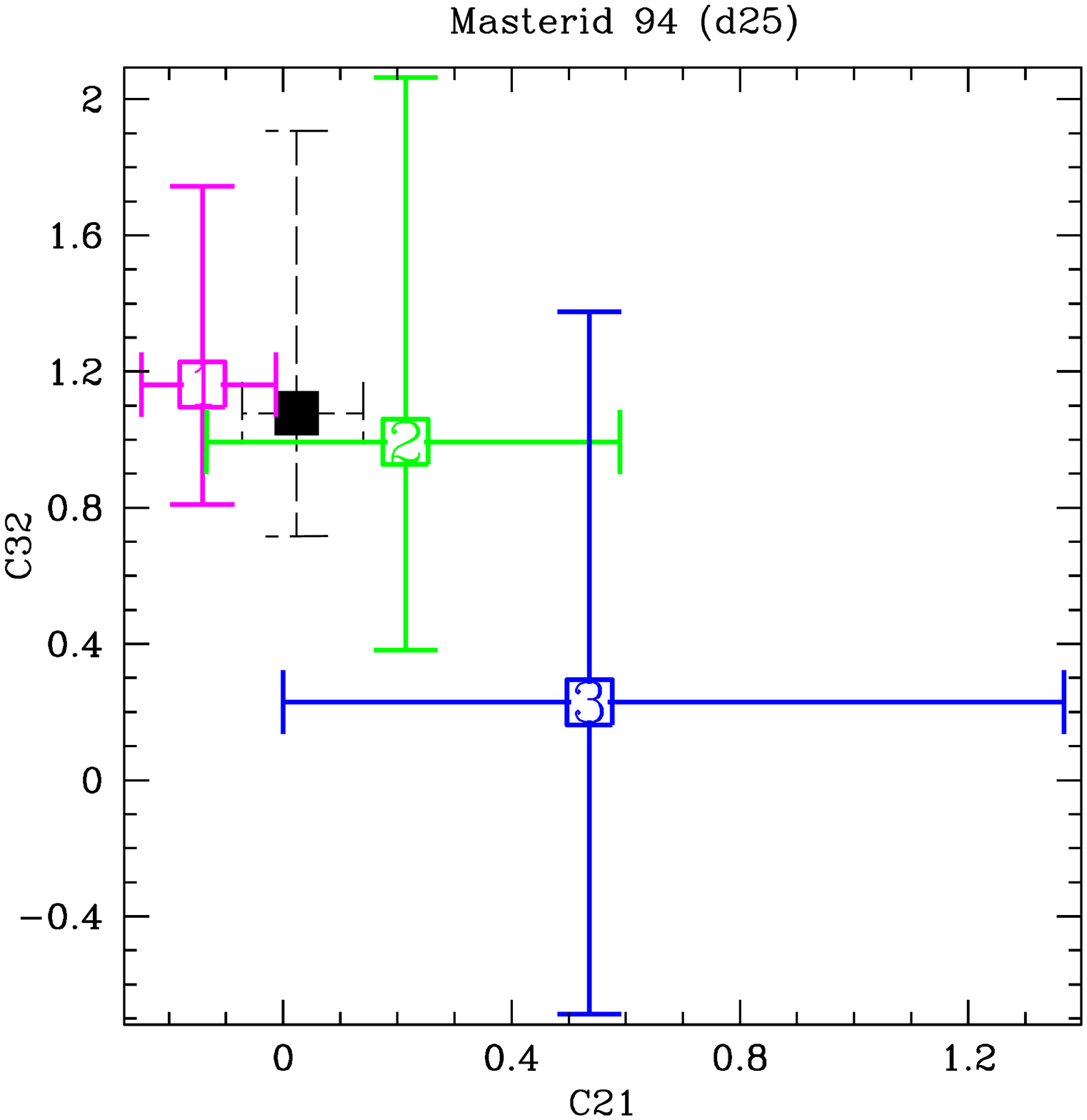}

\end{minipage}
\begin{minipage}{0.32\linewidth}
  \centering

    \includegraphics[width=\linewidth]{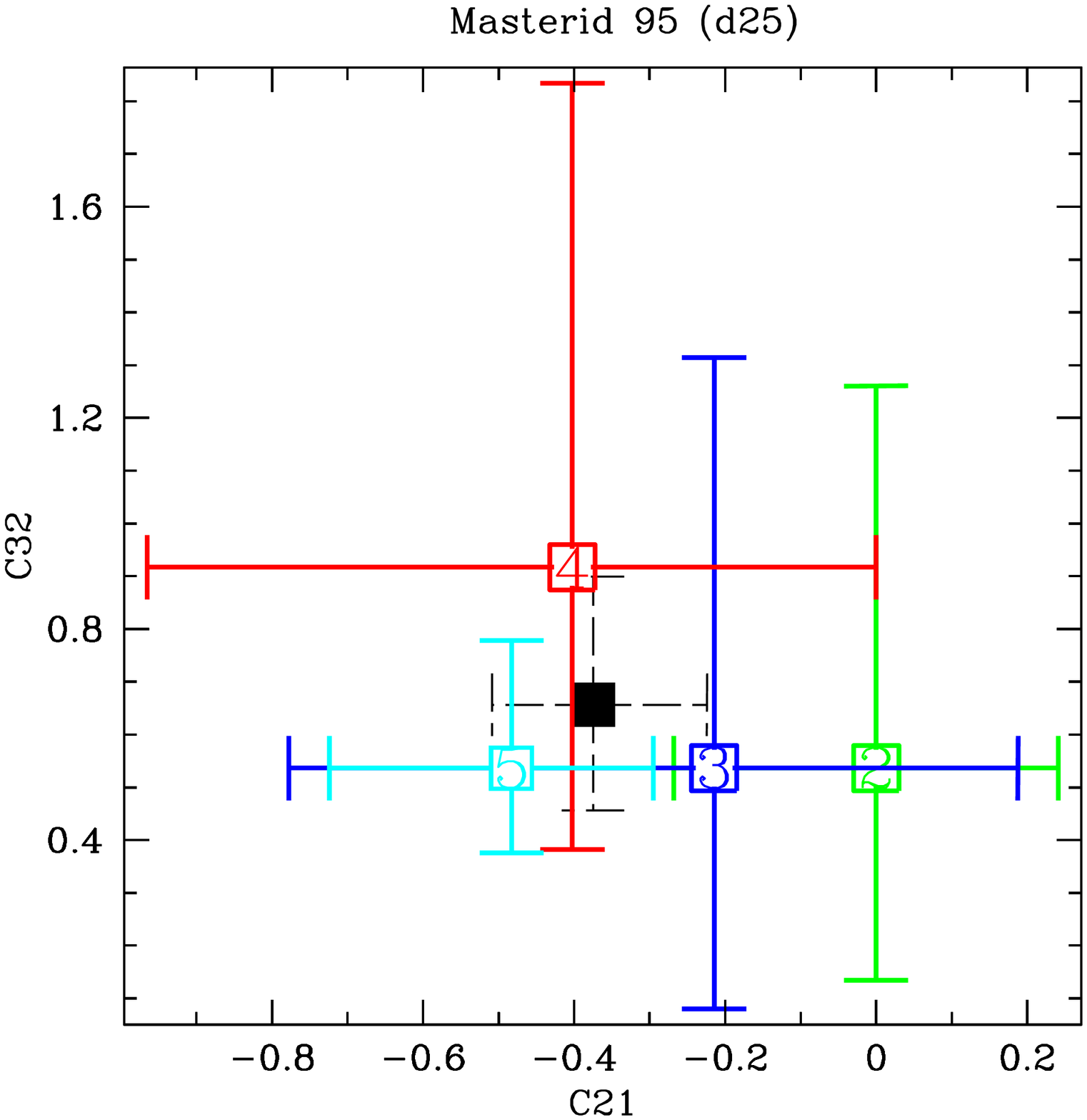}

 \end{minipage}

  \begin{minipage}{0.32\linewidth}
  \centering
  
    \includegraphics[width=\linewidth]{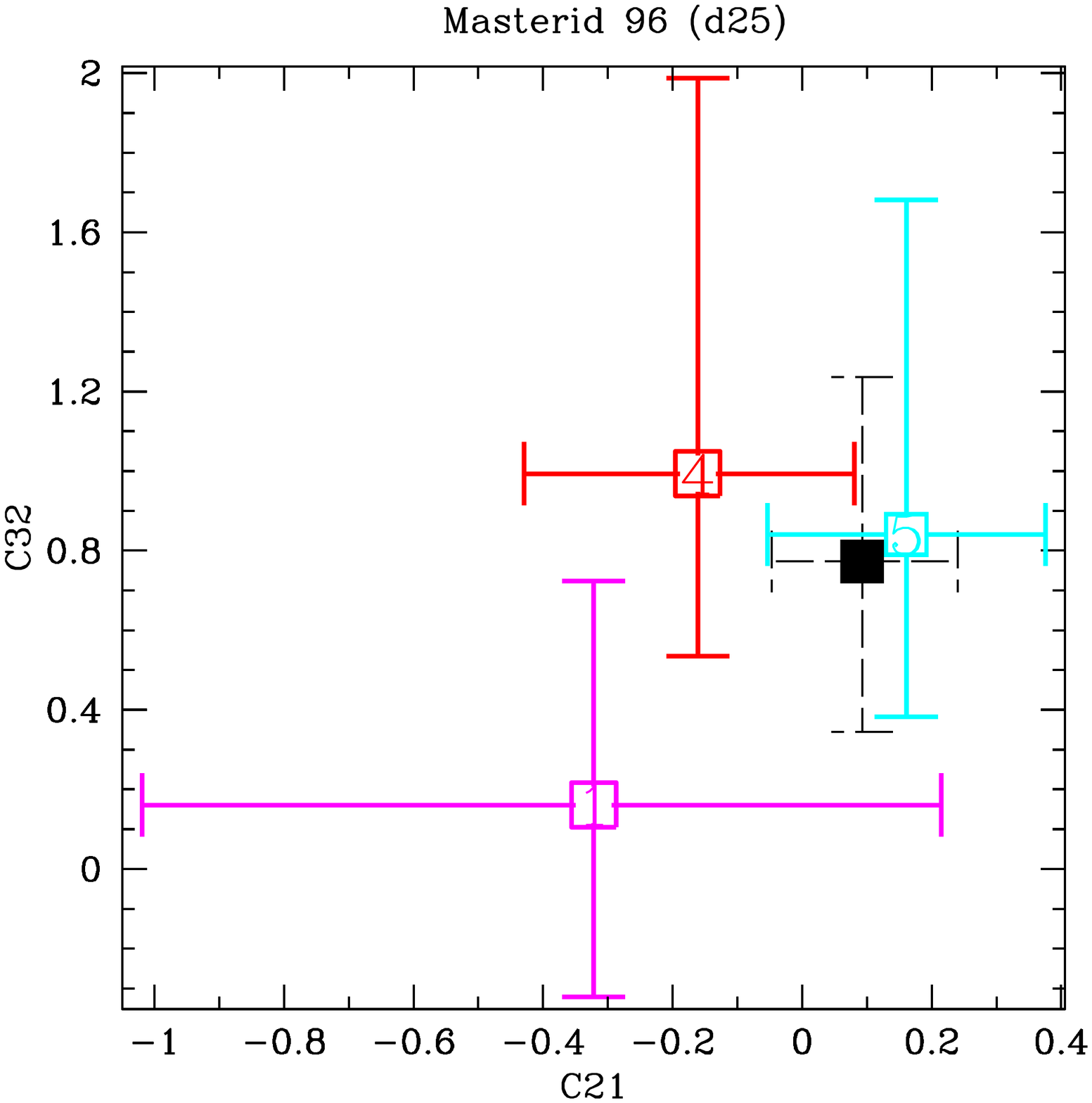}
  
  \end{minipage}
  \begin{minipage}{0.32\linewidth}
  \centering

    \includegraphics[width=\linewidth]{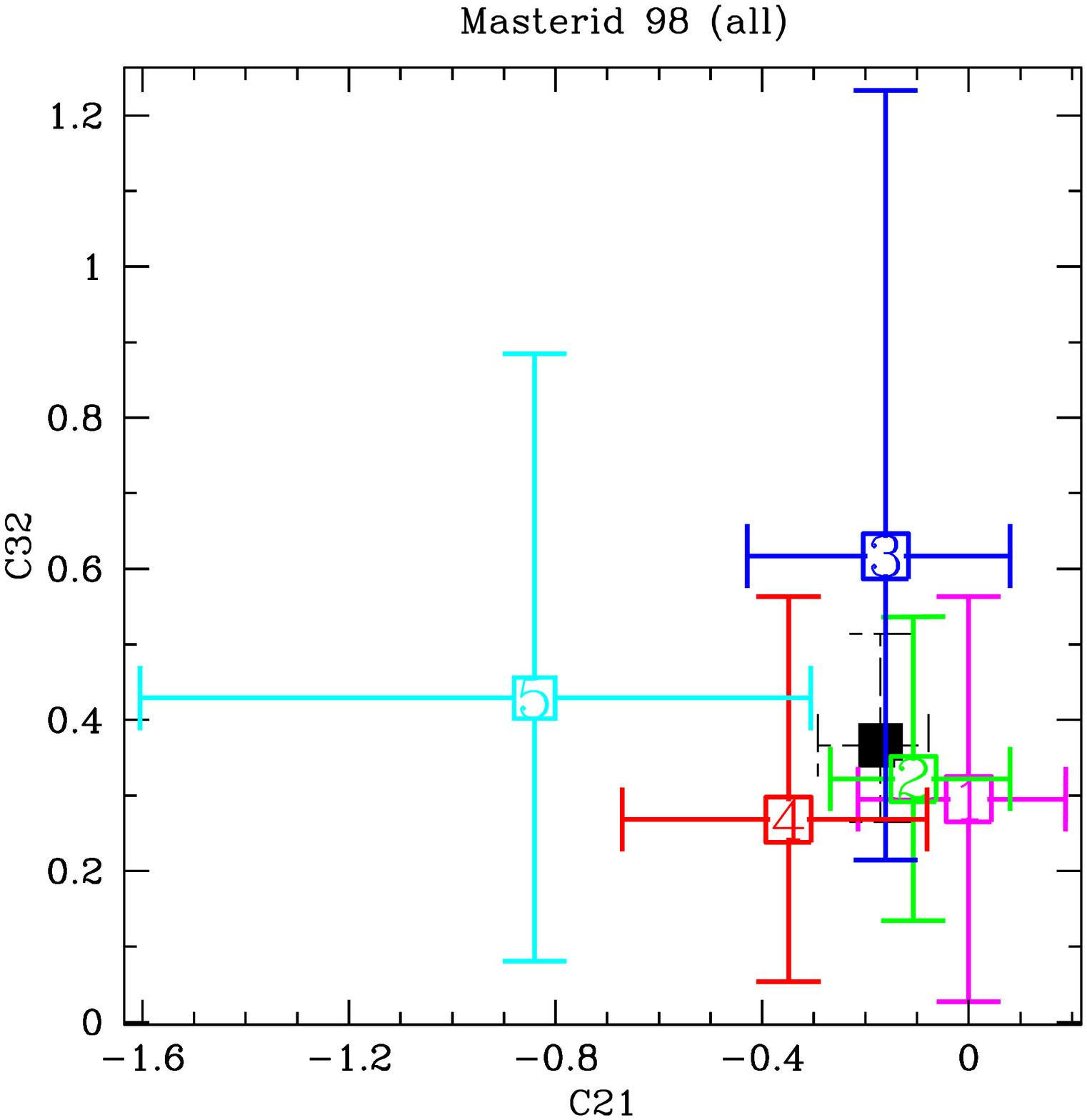}

\end{minipage}
\begin{minipage}{0.32\linewidth}
  \centering

    \includegraphics[width=\linewidth]{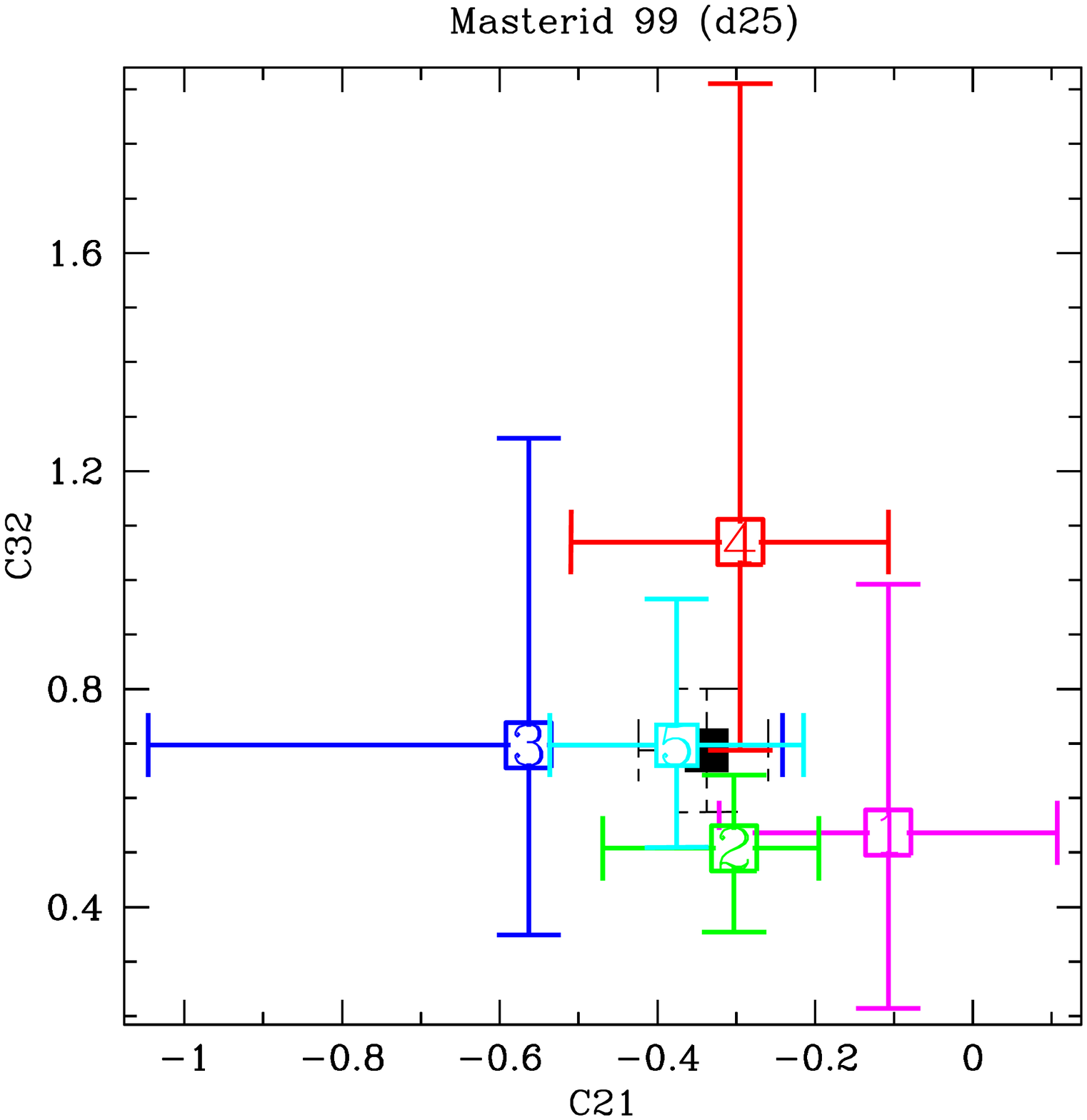}

 \end{minipage}

\begin{minipage}{0.32\linewidth}
  \centering
  
    \includegraphics[width=\linewidth]{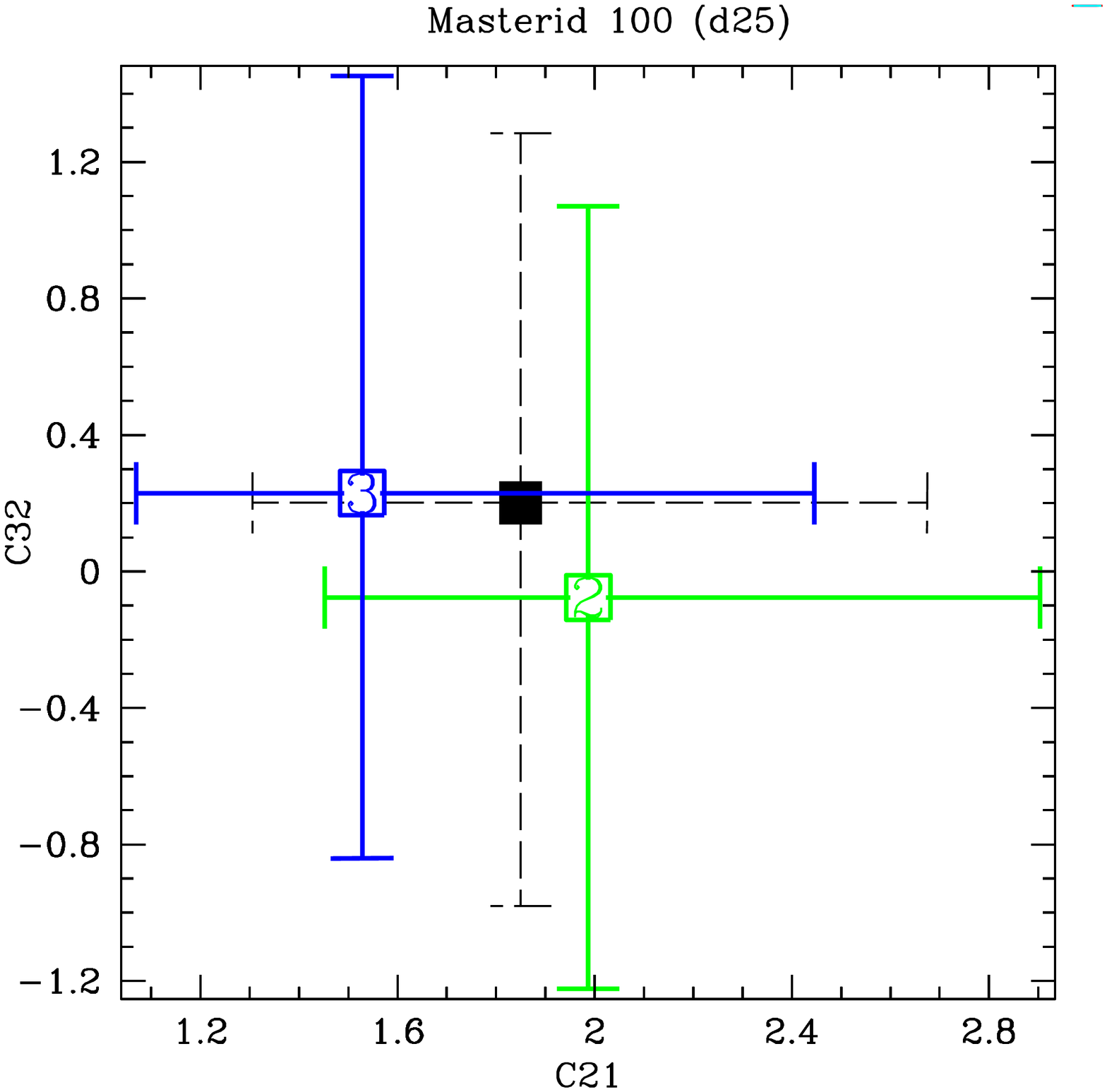}
  
  \end{minipage}
  \begin{minipage}{0.32\linewidth}
  \centering

    \includegraphics[width=\linewidth]{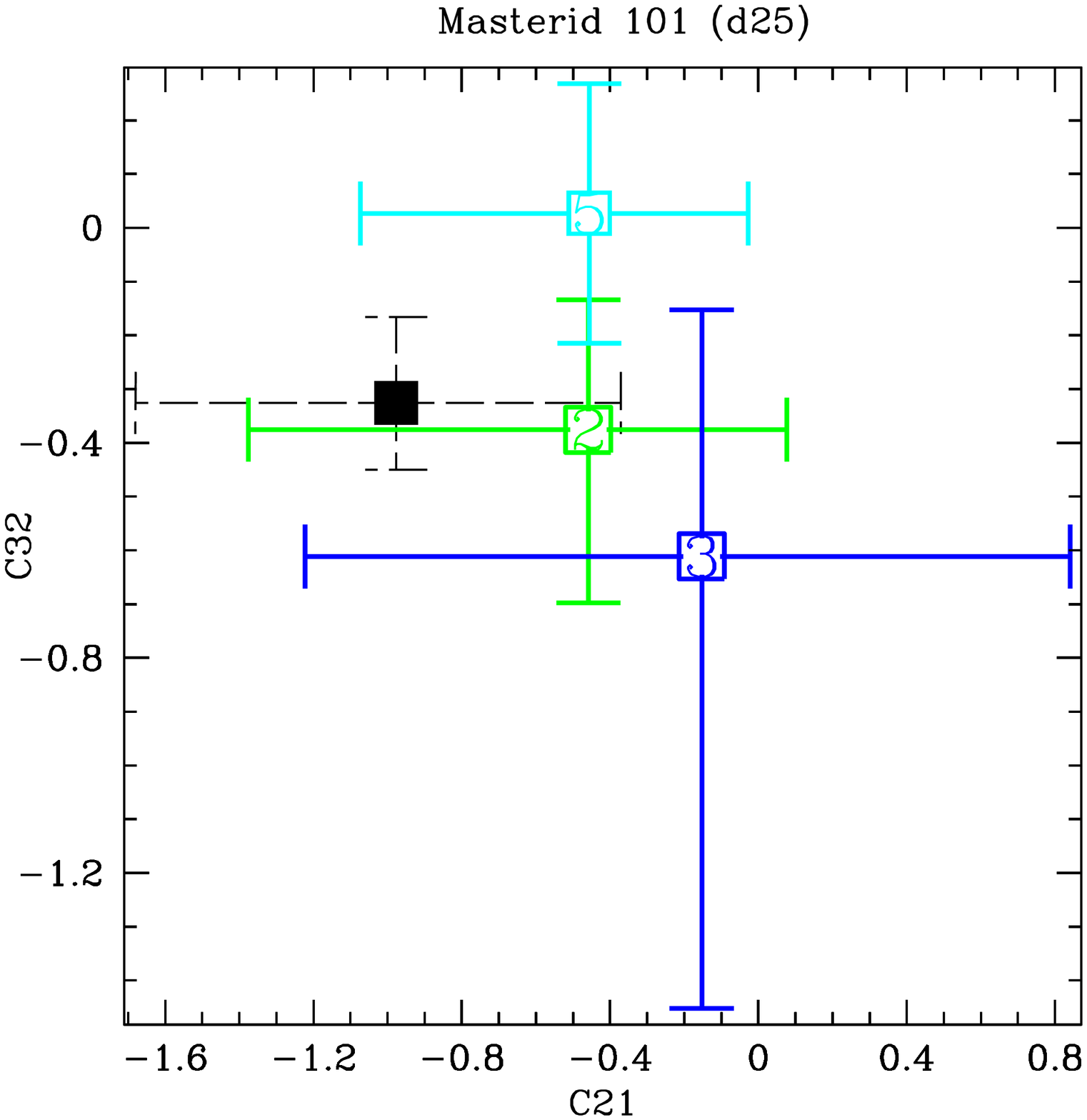}

\end{minipage}
\begin{minipage}{0.32\linewidth}
  \centering

    \includegraphics[width=\linewidth]{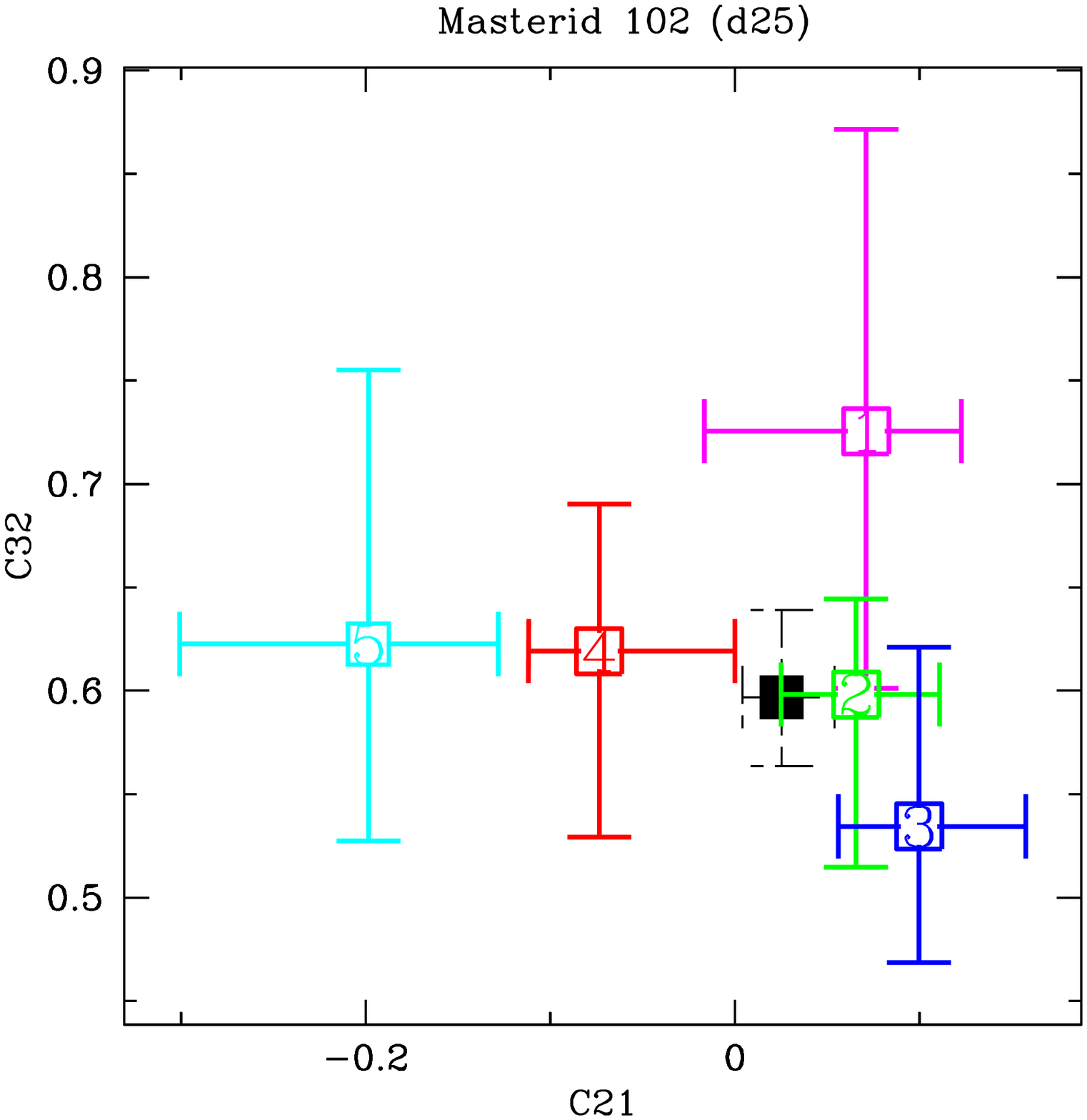}

 \end{minipage}
  
\end{figure}

\begin{figure}
  \begin{minipage}{0.32\linewidth}
  \centering
  
    \includegraphics[width=\linewidth]{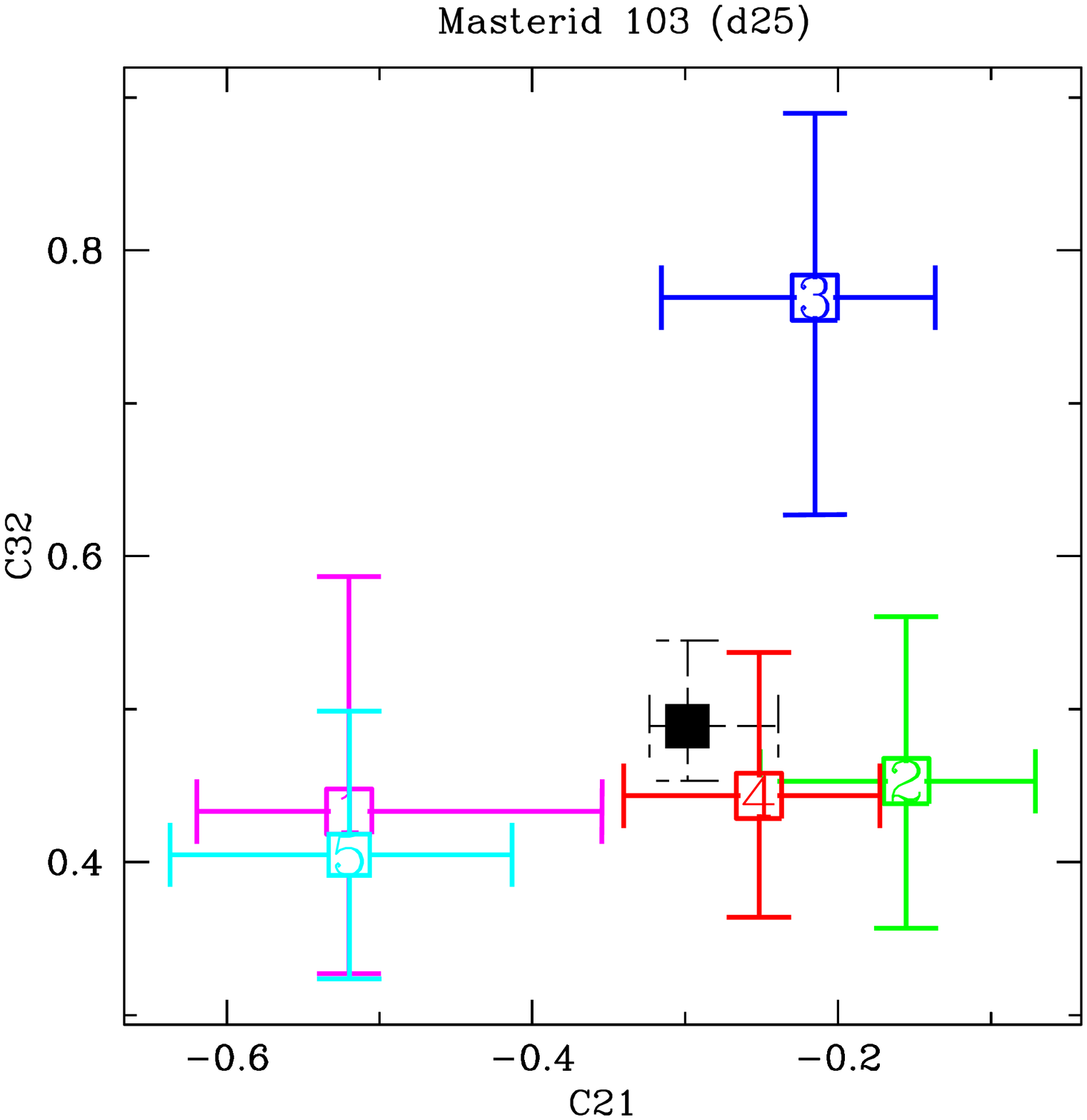}
  
  \end{minipage}
  \begin{minipage}{0.32\linewidth}
  \centering

    \includegraphics[width=\linewidth]{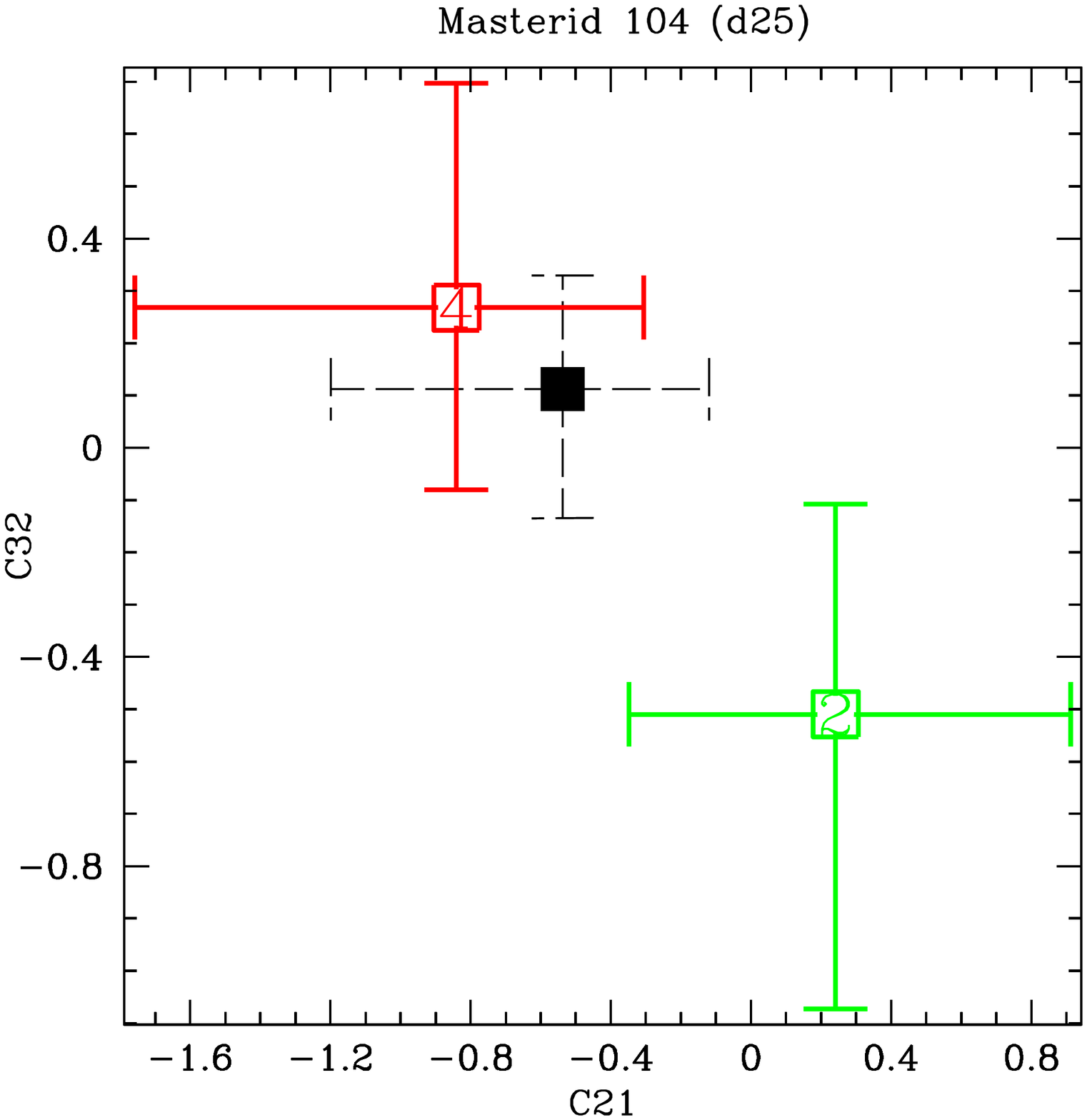}

\end{minipage}
\begin{minipage}{0.32\linewidth}
  \centering

    \includegraphics[width=\linewidth]{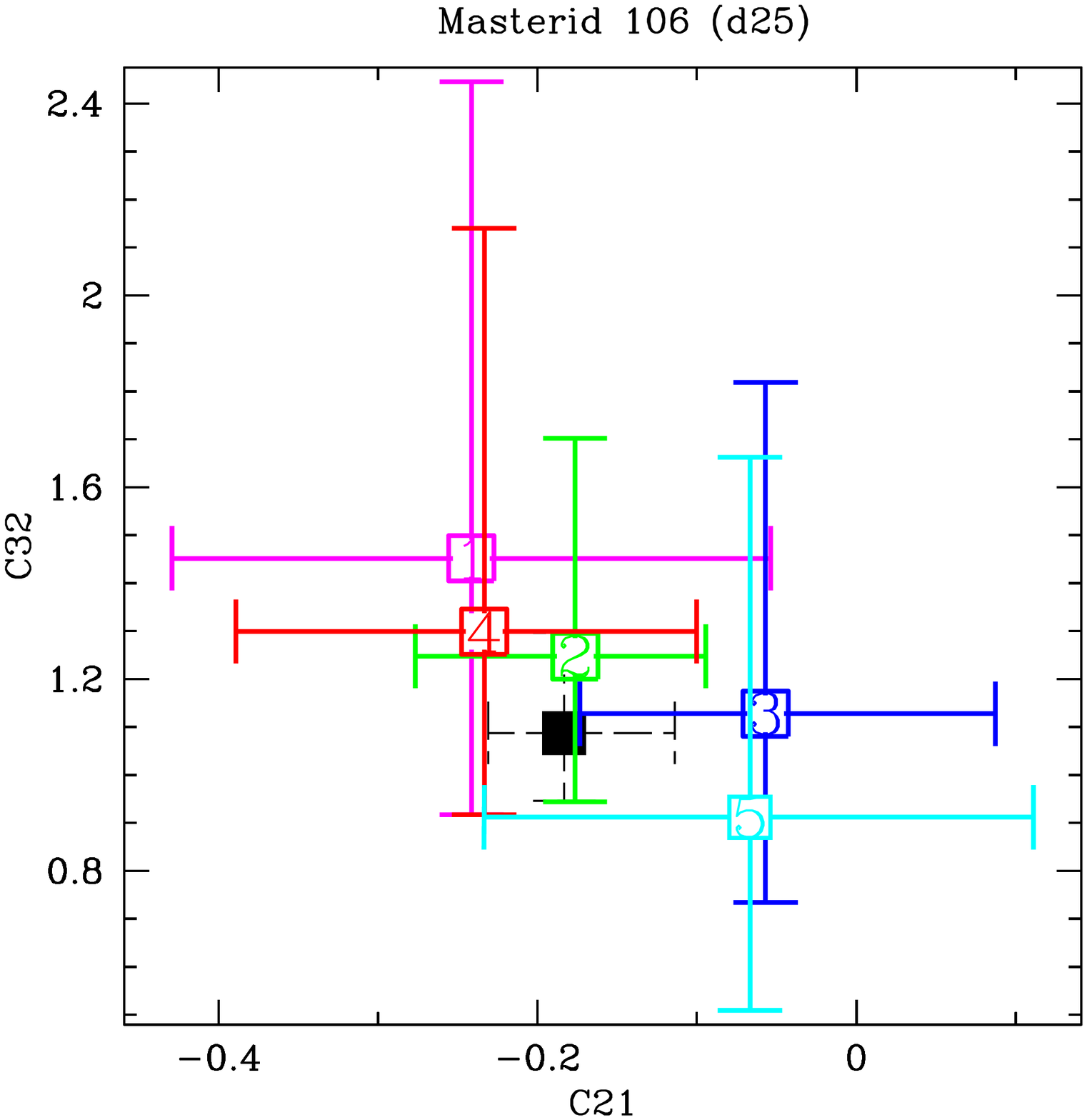}

 \end{minipage}

\begin{minipage}{0.32\linewidth}
  \centering
  
    \includegraphics[width=\linewidth]{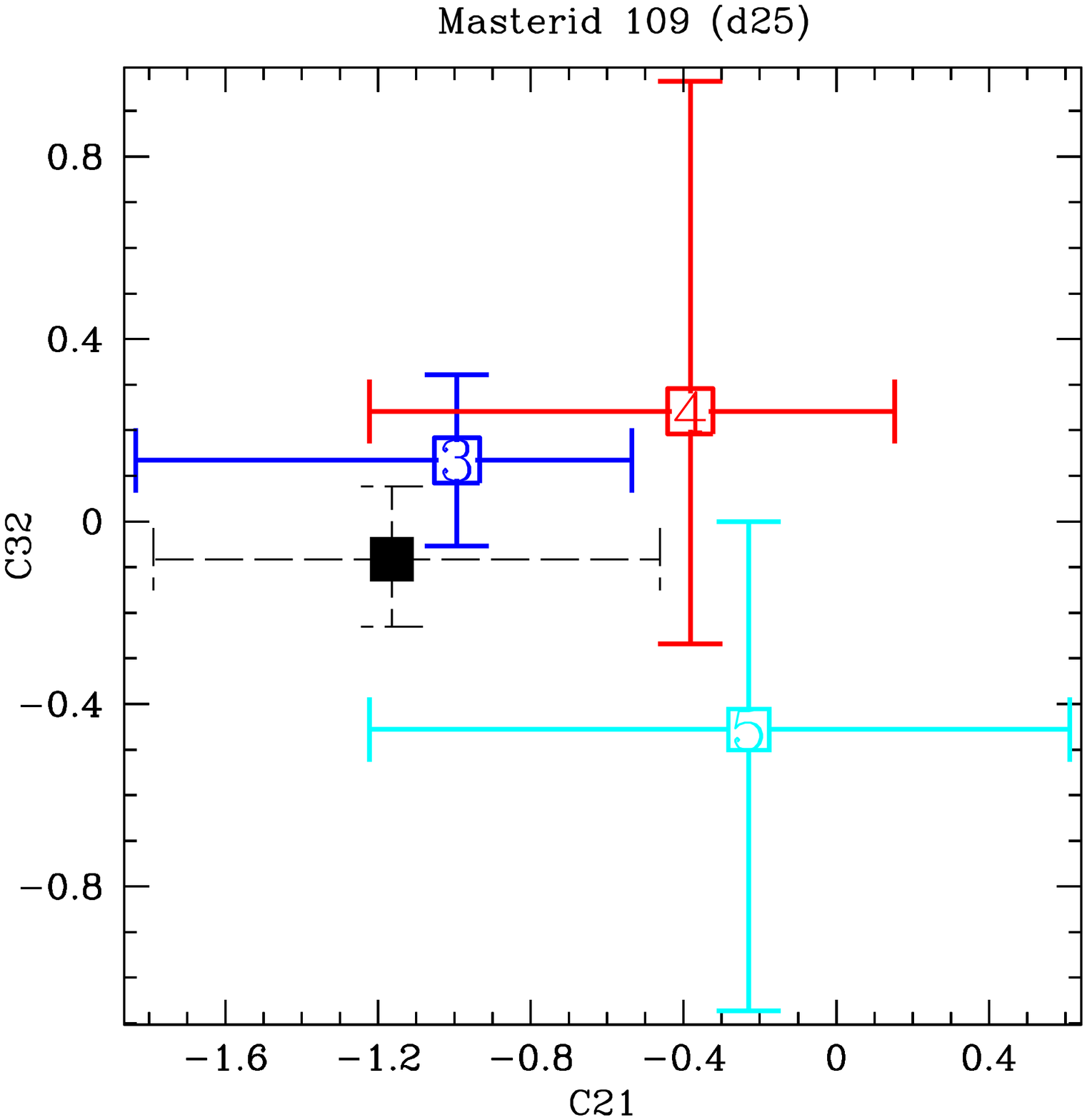}
  
  \end{minipage}
  \begin{minipage}{0.32\linewidth}
  \centering

    \includegraphics[width=\linewidth]{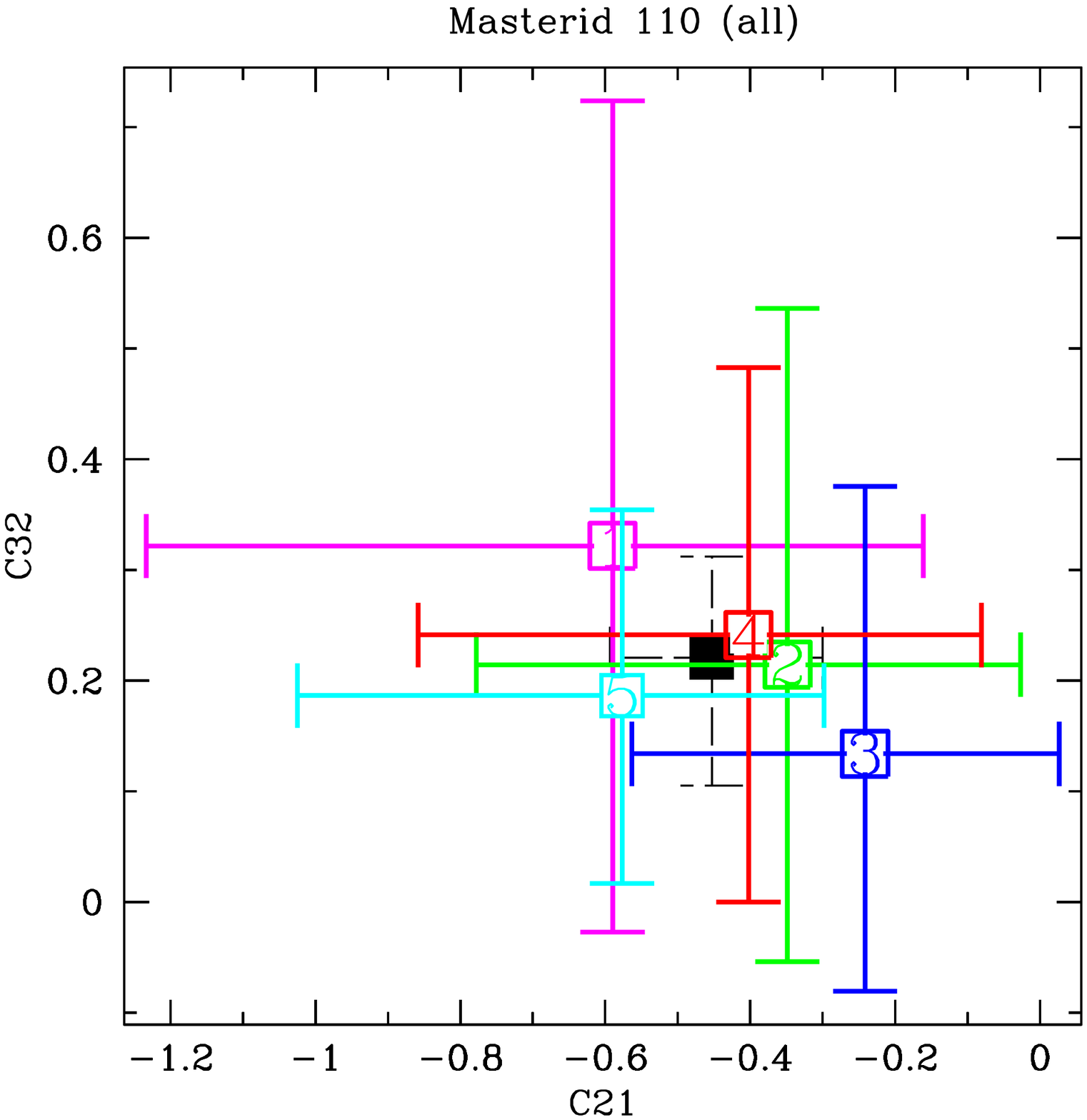}

\end{minipage}
\begin{minipage}{0.32\linewidth}
  \centering

    \includegraphics[width=\linewidth]{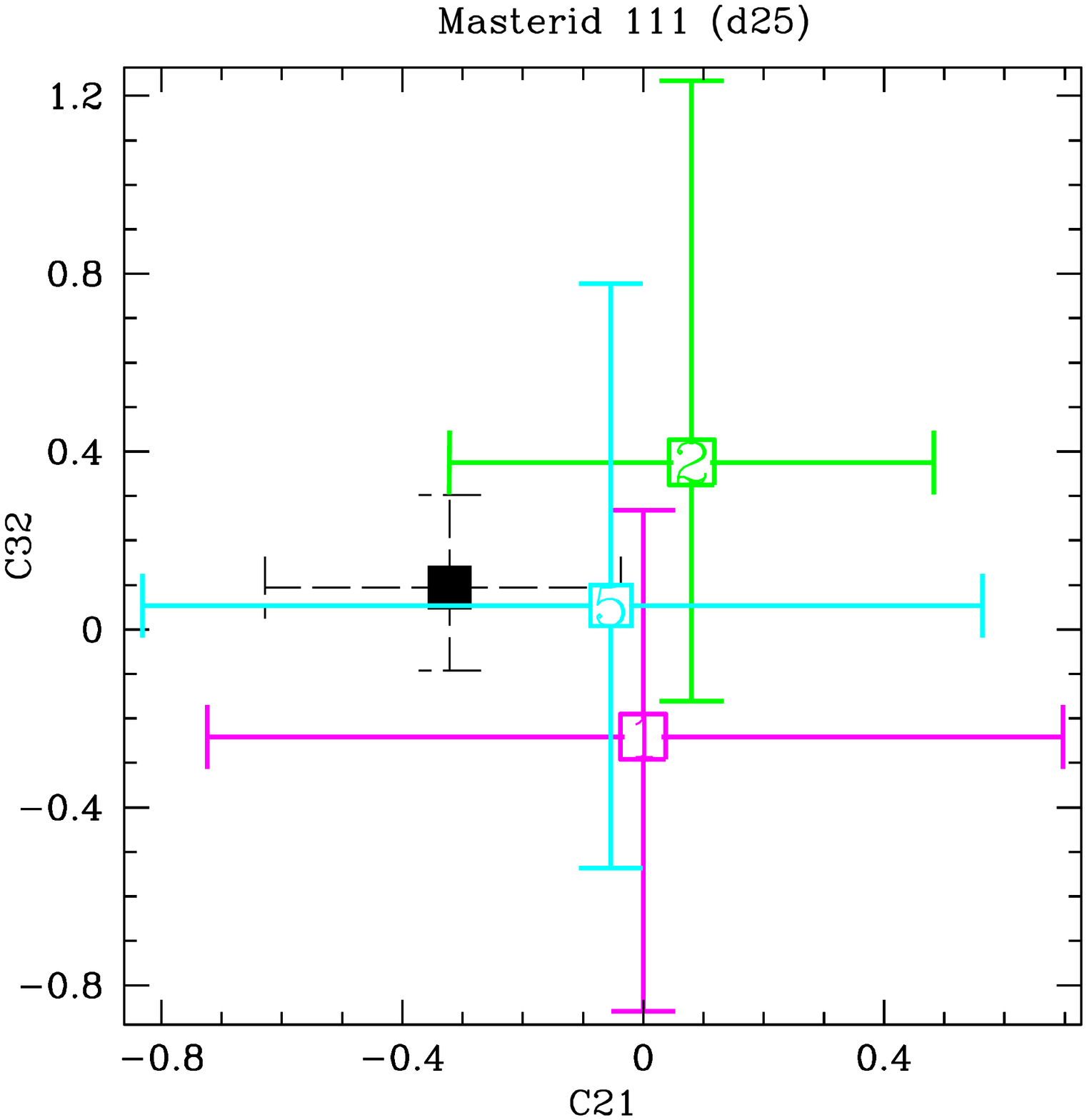}

 \end{minipage}

  \begin{minipage}{0.32\linewidth}
  \centering
  
    \includegraphics[width=\linewidth]{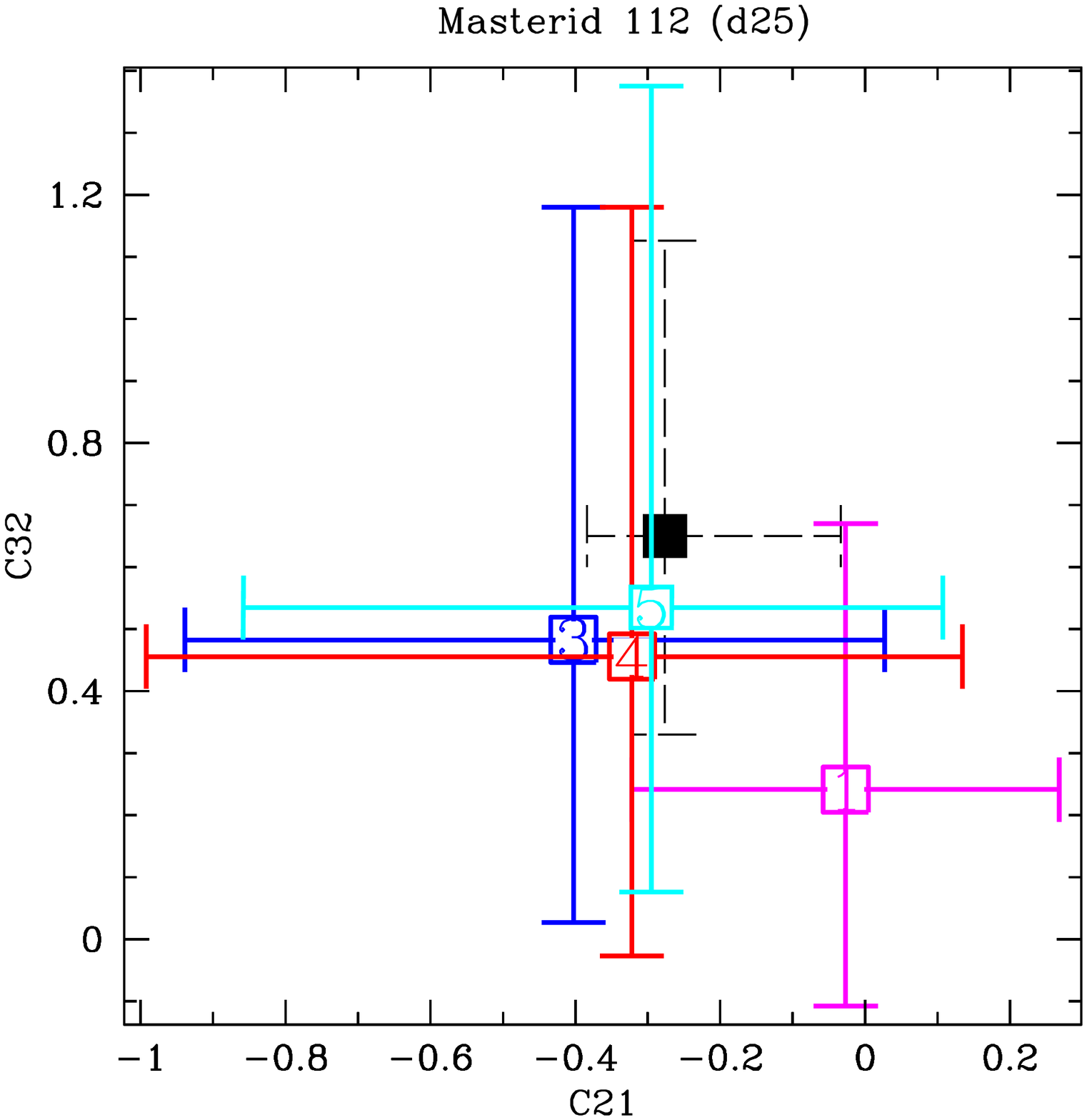}
  
  \end{minipage}
  \begin{minipage}{0.32\linewidth}
  \centering

    \includegraphics[width=\linewidth]{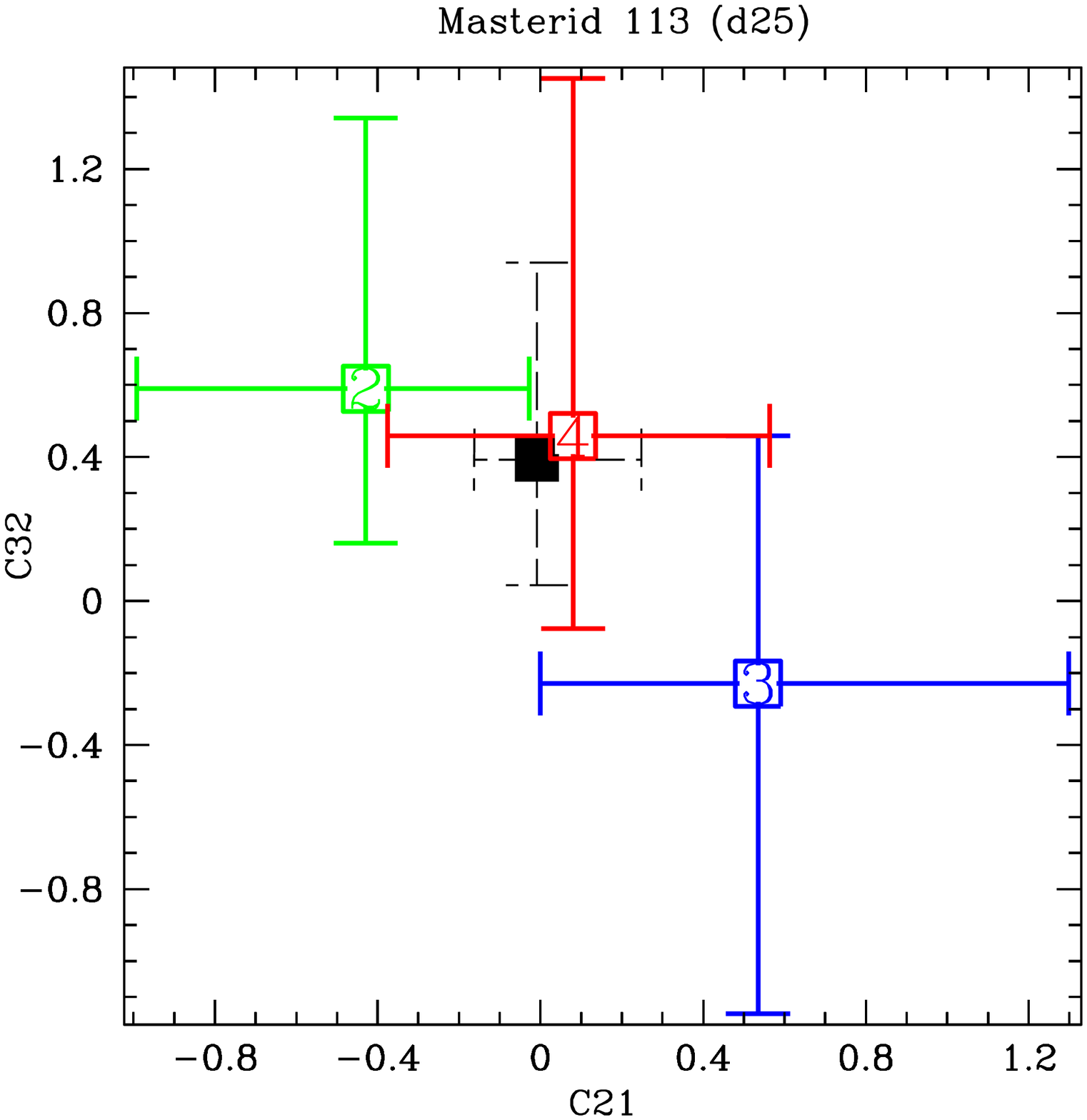}

\end{minipage}
\begin{minipage}{0.32\linewidth}
  \centering

    \includegraphics[width=\linewidth]{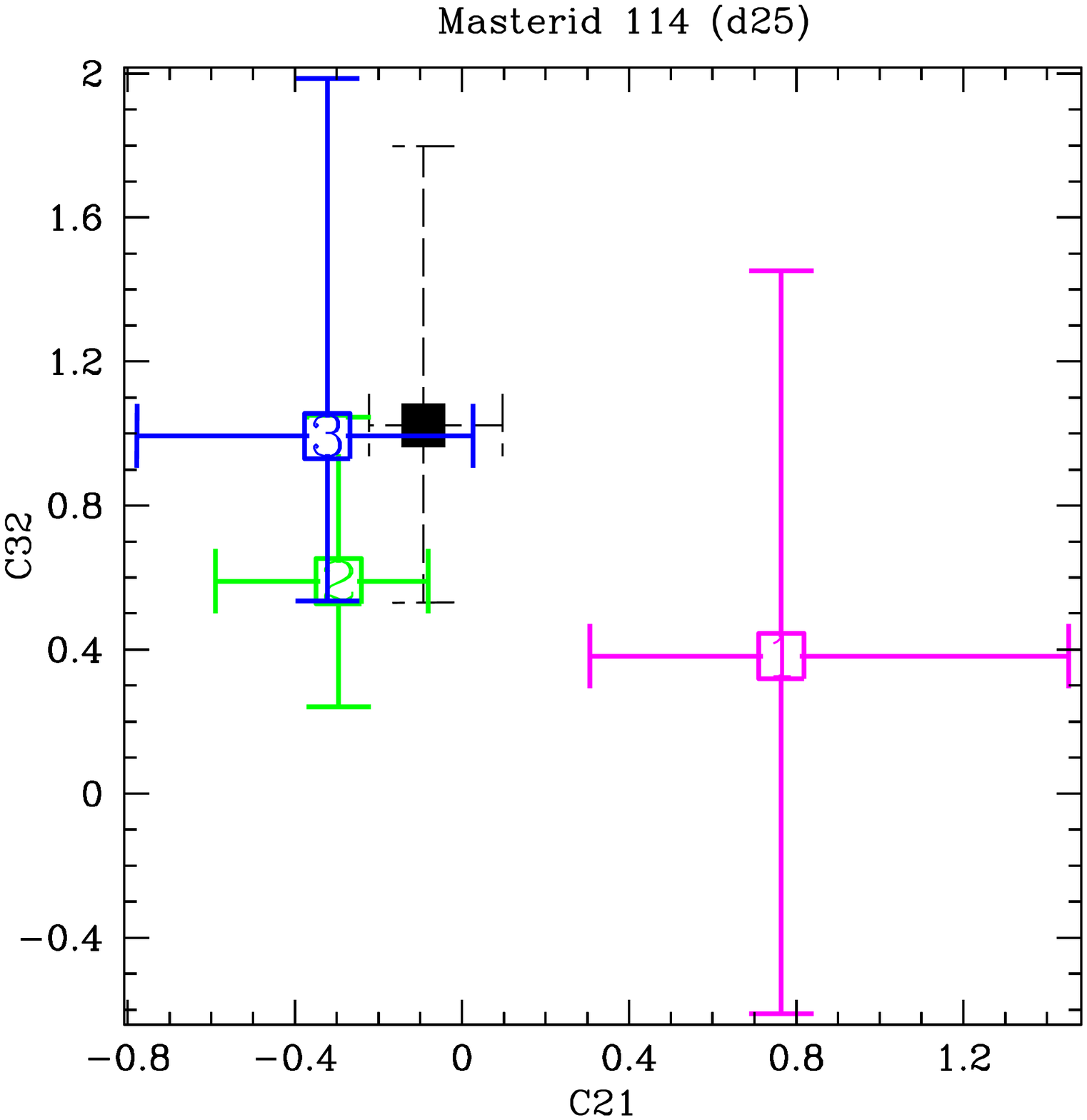}

 \end{minipage}

\begin{minipage}{0.32\linewidth}
  \centering
  
    \includegraphics[width=\linewidth]{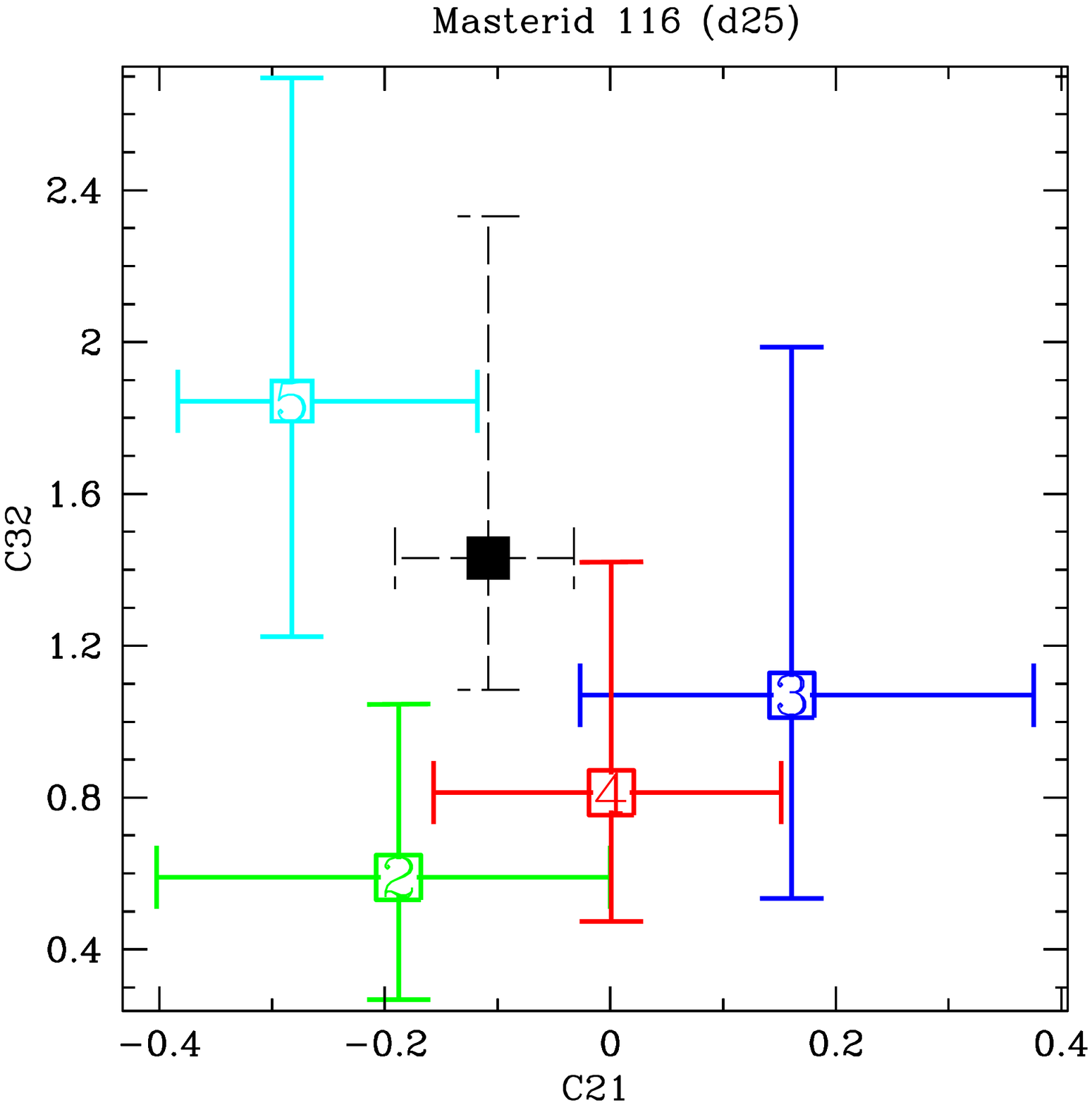}
  
  \end{minipage}
  \begin{minipage}{0.32\linewidth}
  \centering

    \includegraphics[width=\linewidth]{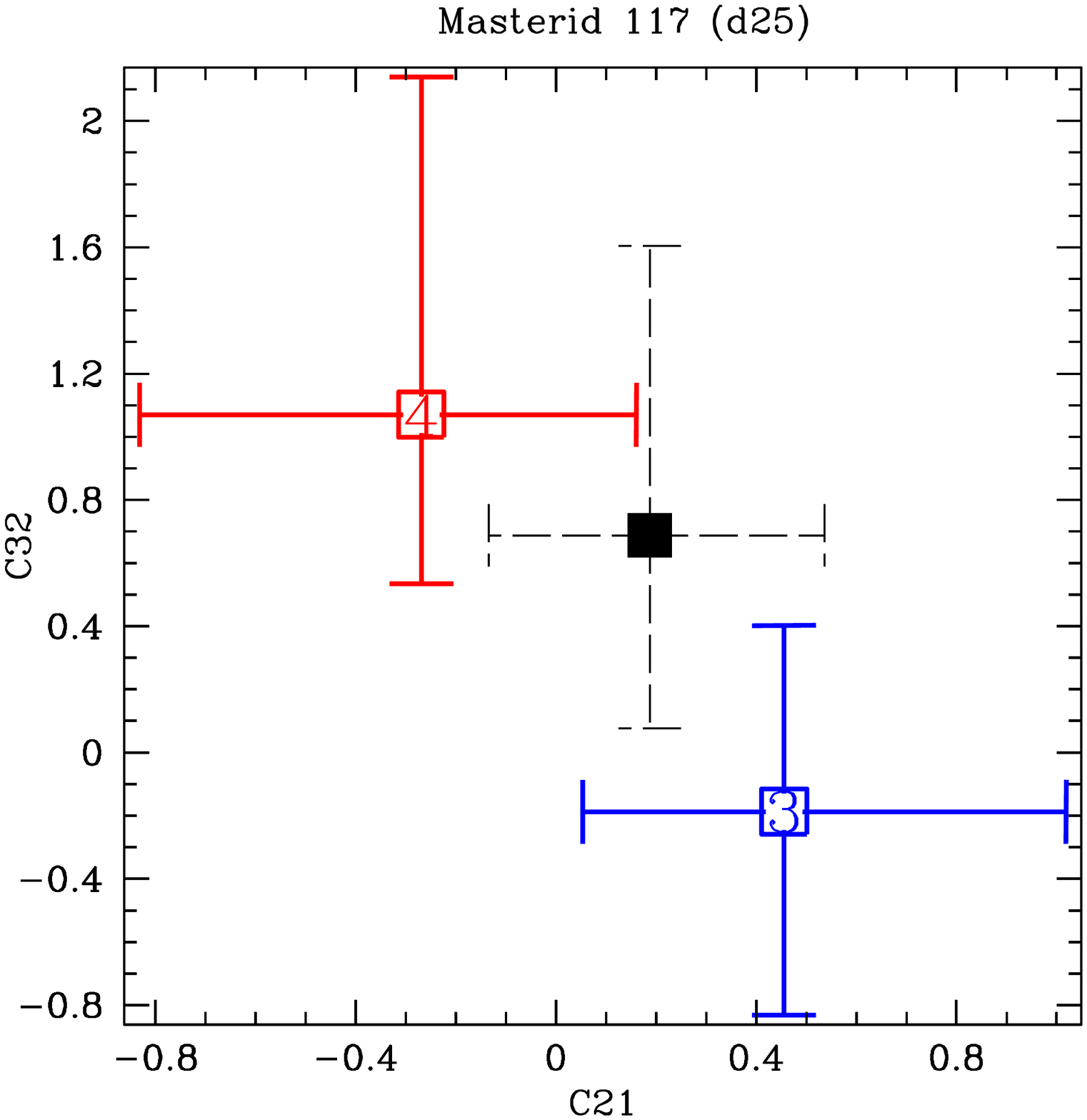}

\end{minipage}
\begin{minipage}{0.32\linewidth}
  \centering

    \includegraphics[width=\linewidth]{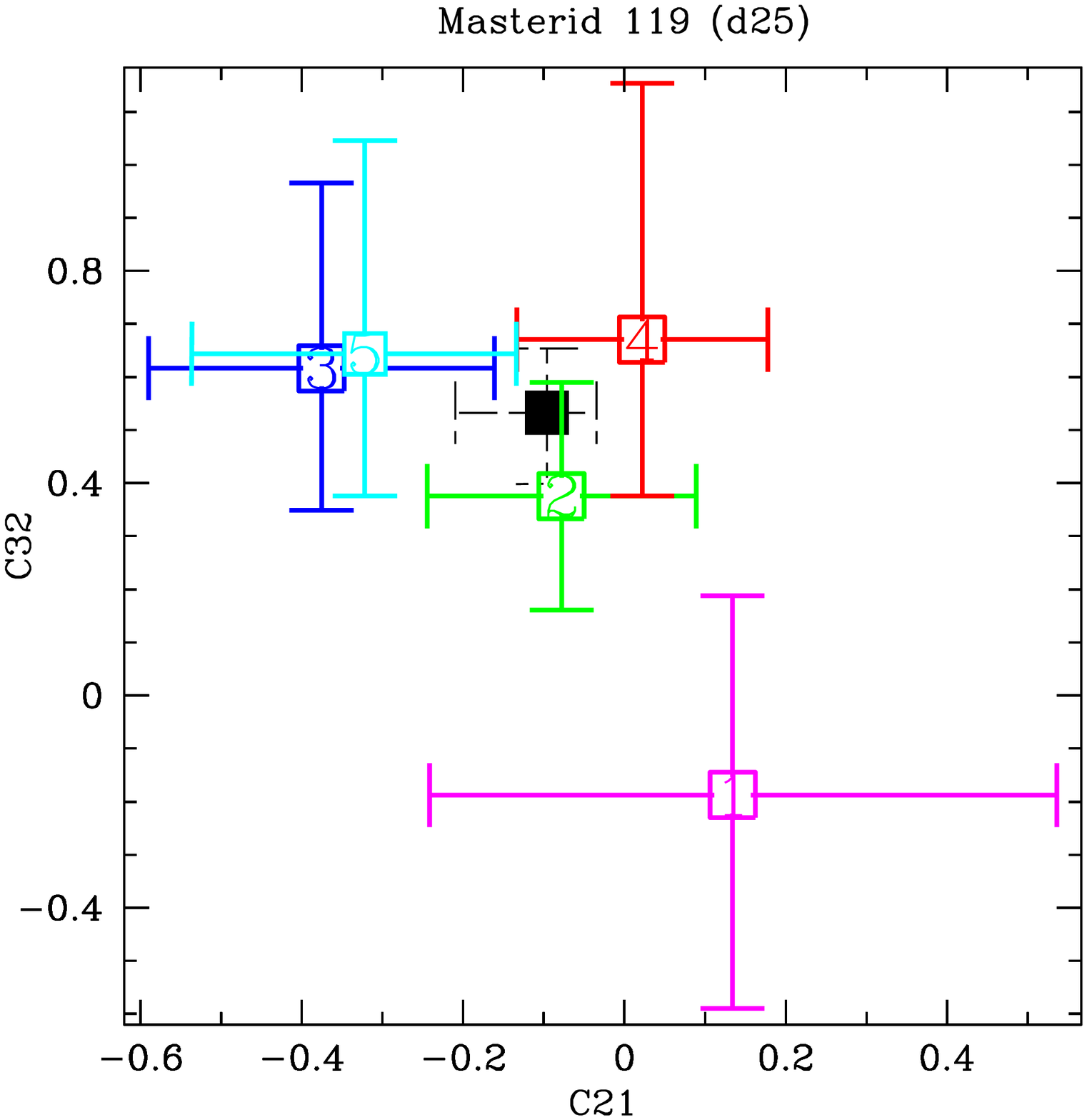}

 \end{minipage}
  
\end{figure}

\begin{figure}
  \begin{minipage}{0.32\linewidth}
  \centering
  
    \includegraphics[width=\linewidth]{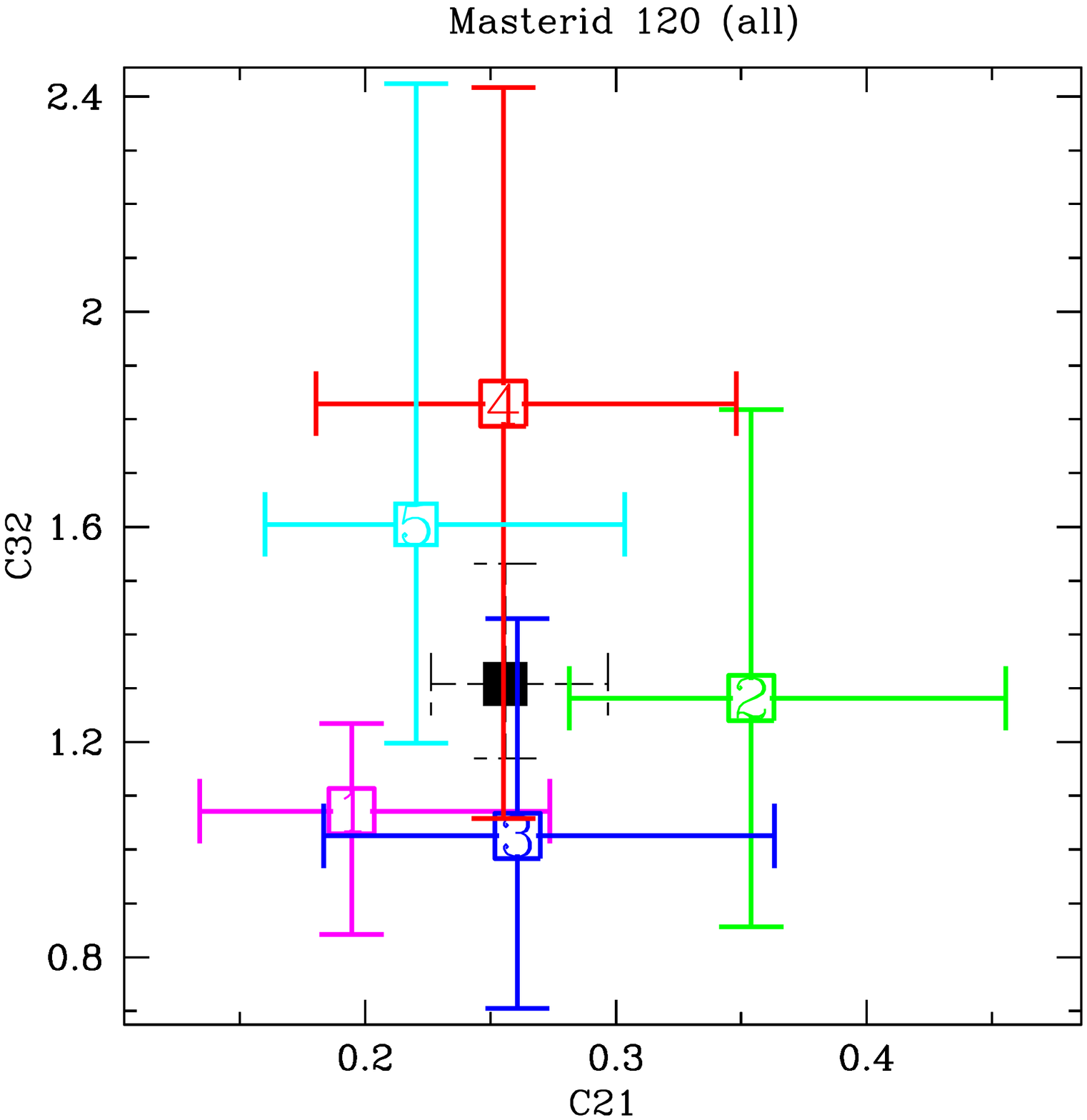}
  
  \end{minipage}
  \begin{minipage}{0.32\linewidth}
  \centering

    \includegraphics[width=\linewidth]{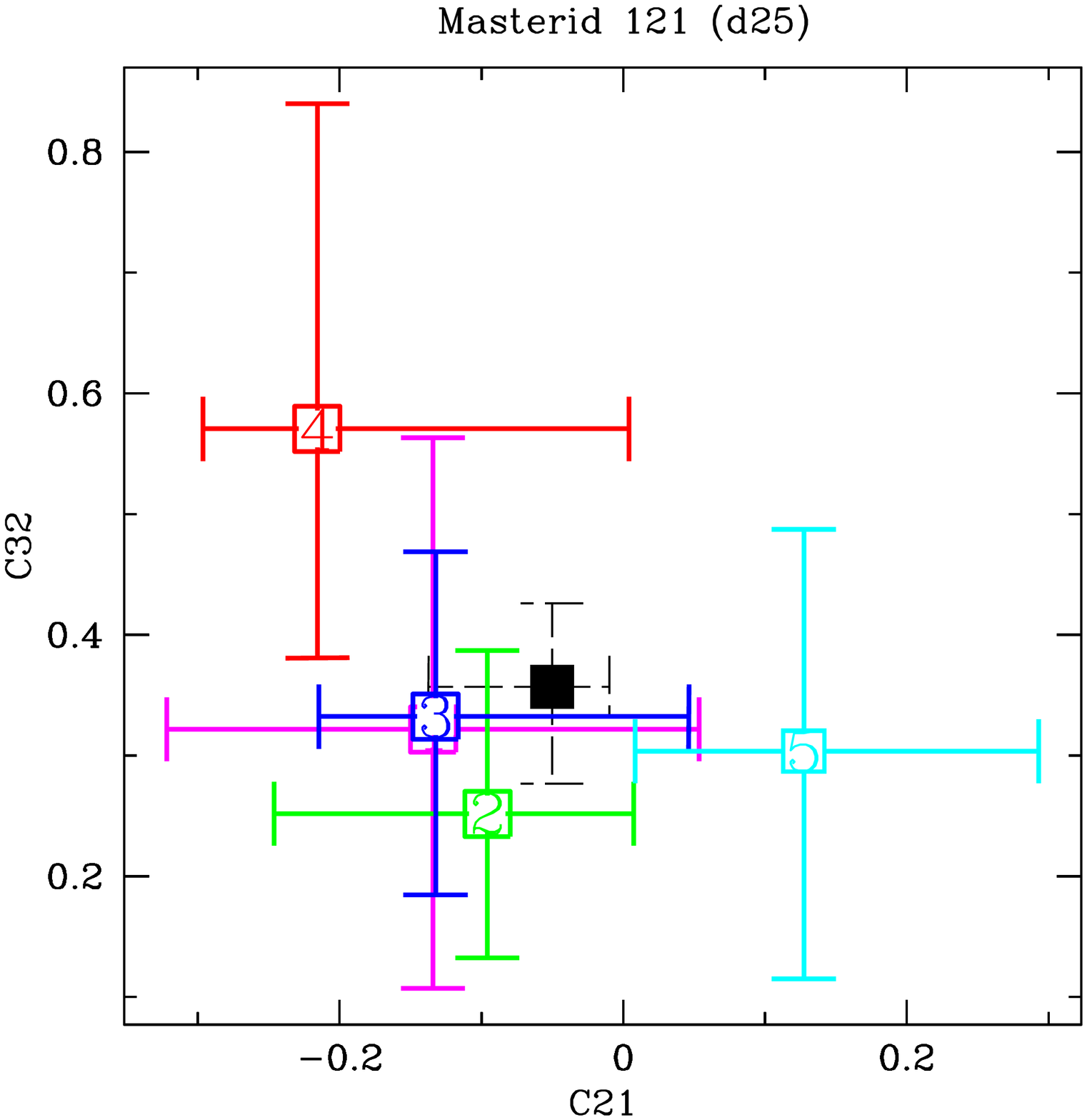}

\end{minipage}
\begin{minipage}{0.32\linewidth}
  \centering

    \includegraphics[width=\linewidth]{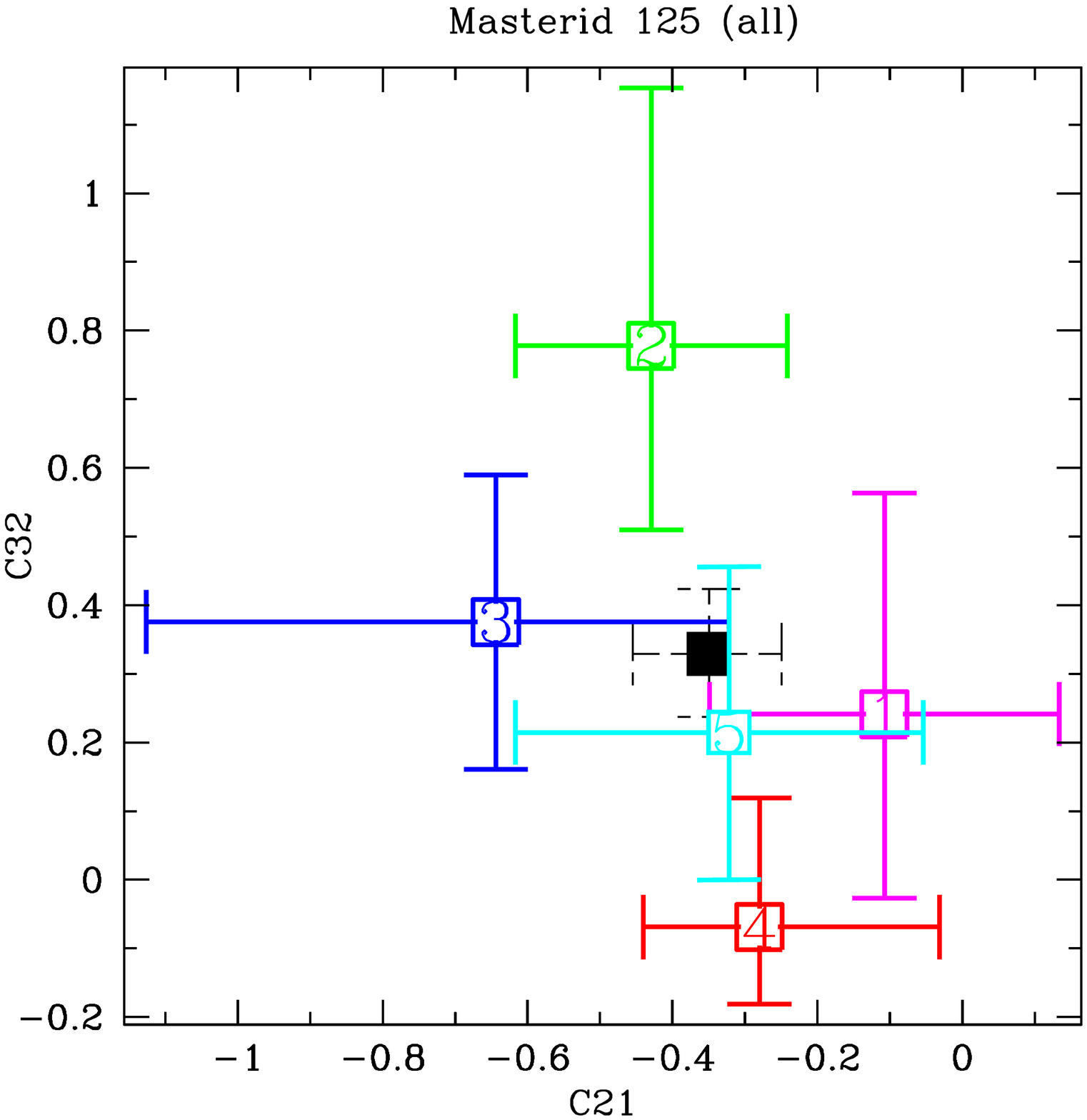}

 \end{minipage}

\begin{minipage}{0.32\linewidth}
  \centering
  
    \includegraphics[width=\linewidth]{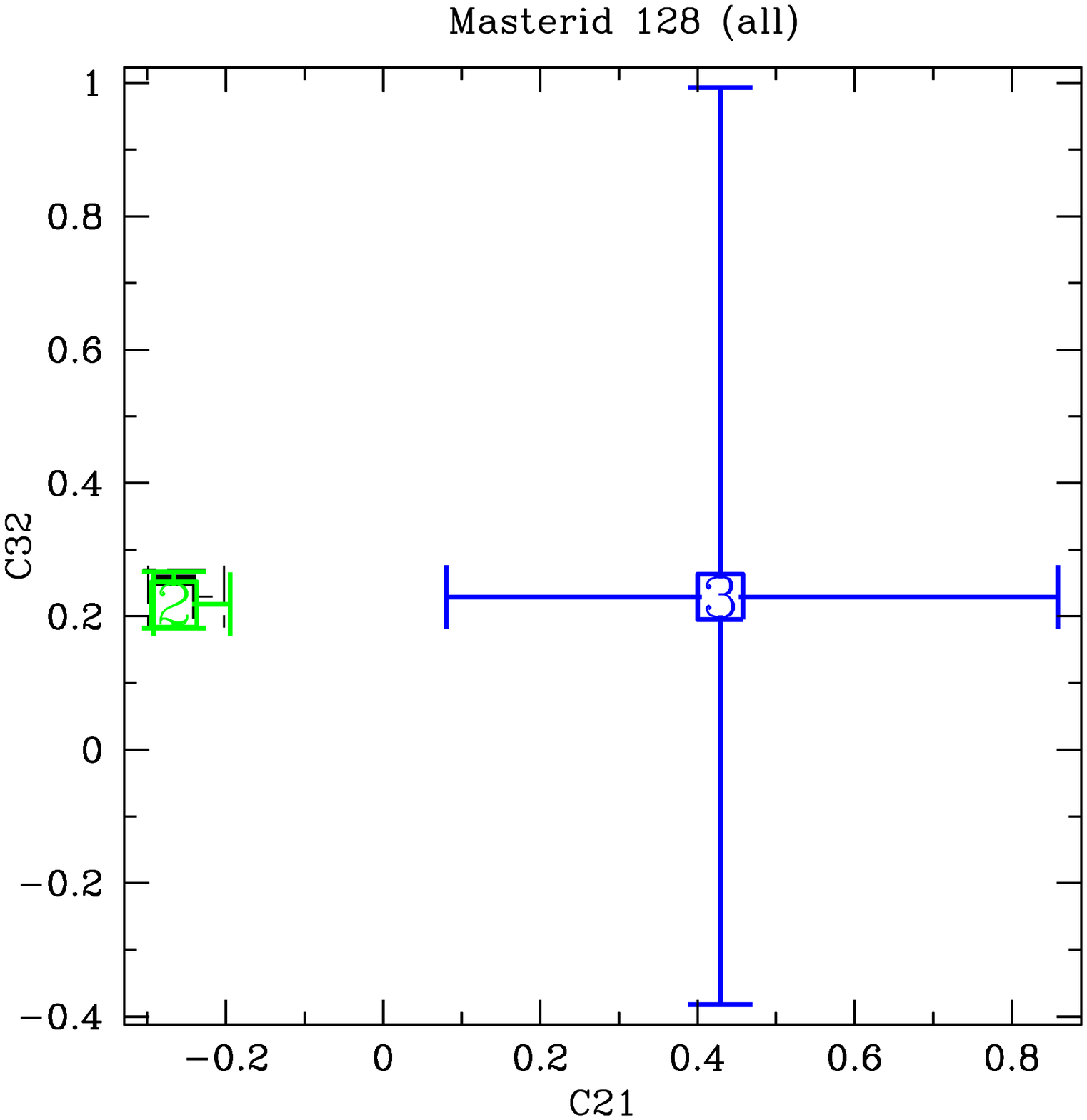}
  
  \end{minipage}
  \begin{minipage}{0.32\linewidth}
  \centering

    \includegraphics[width=\linewidth]{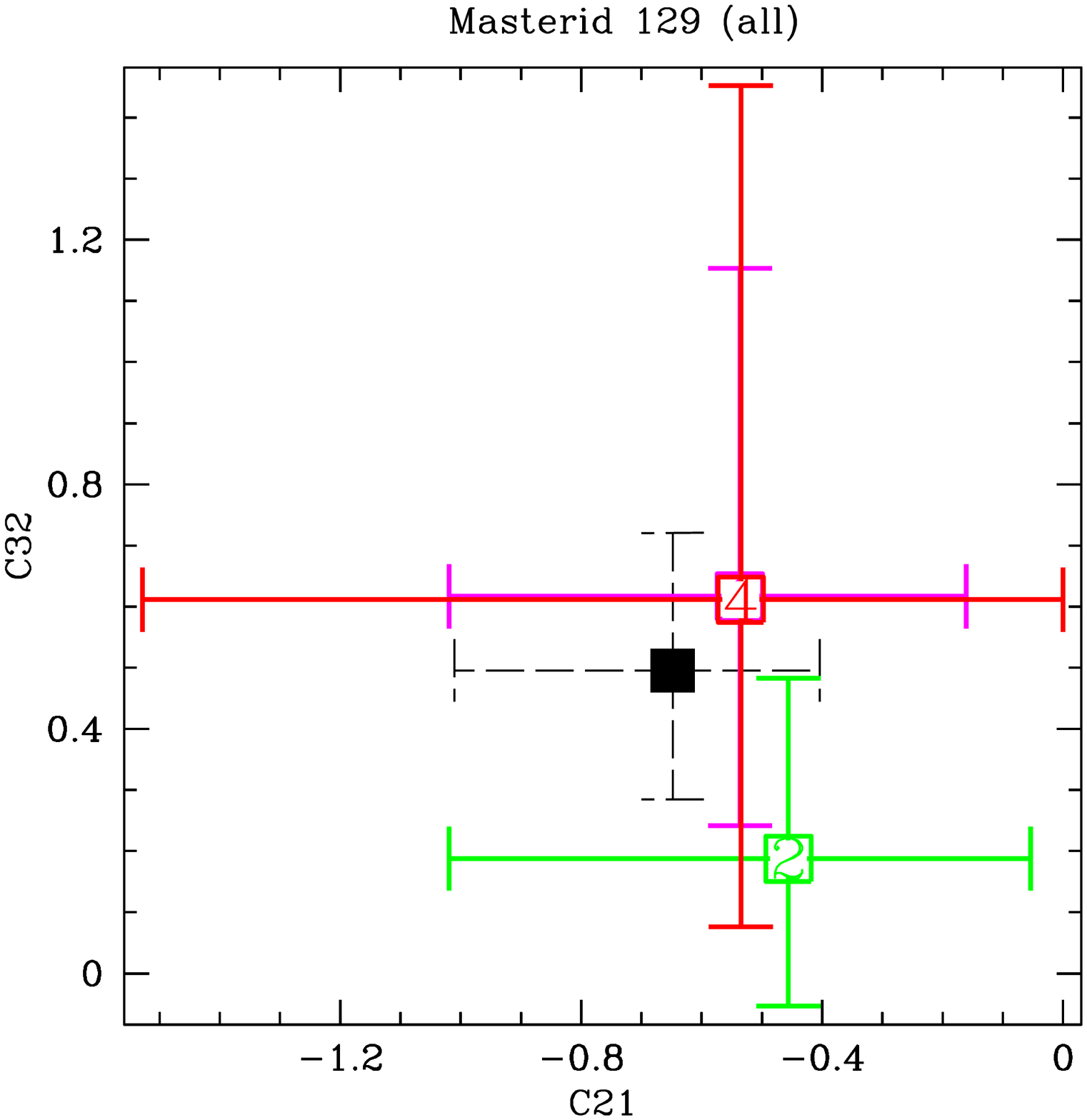}

\end{minipage}
\centering
\begin{minipage}{0.32\linewidth}
  \centering

    \includegraphics[width=\linewidth]{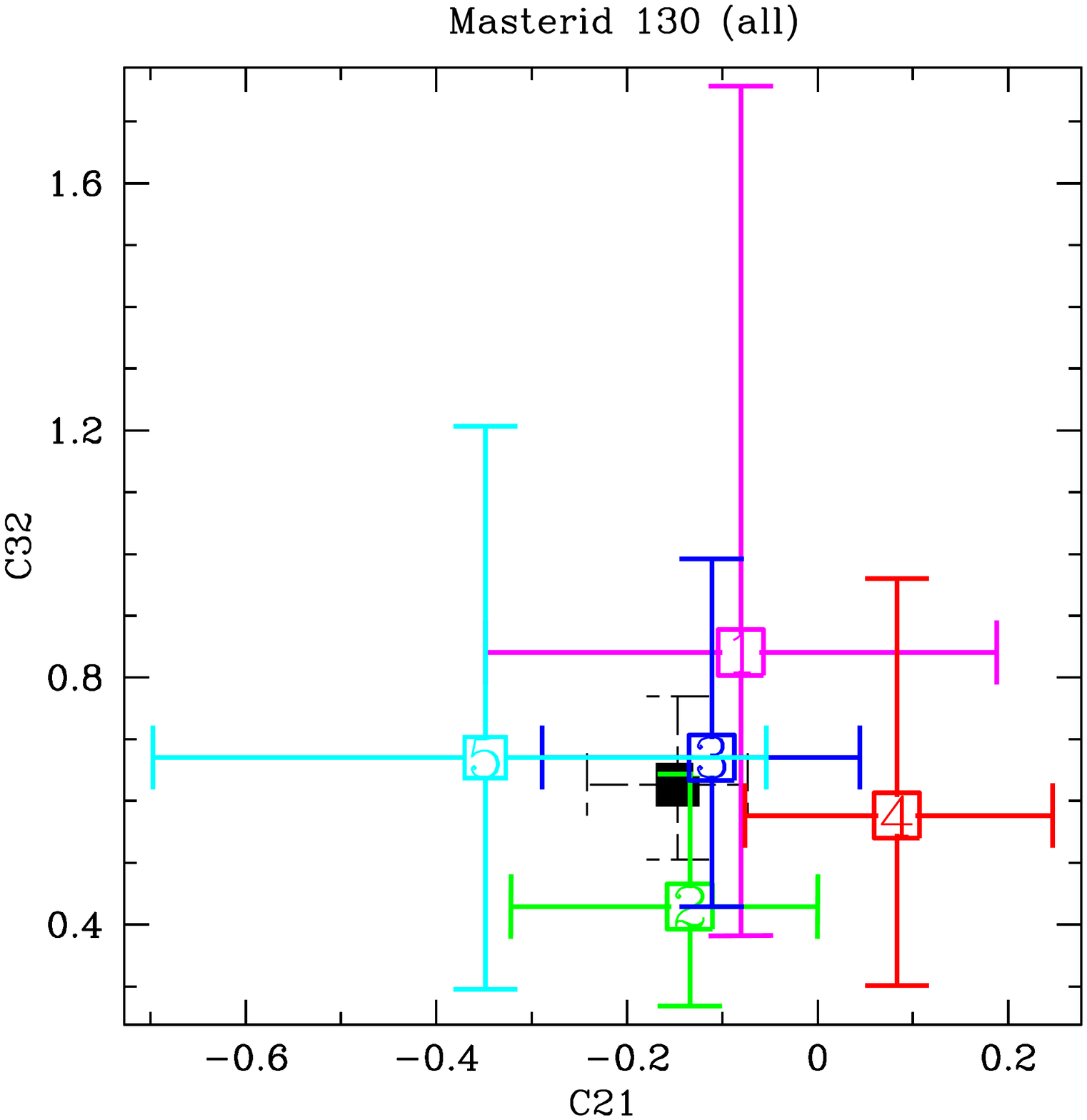}

 \end{minipage}

  \begin{minipage}{0.32\linewidth}
  \centering
  
    \includegraphics[width=\linewidth]{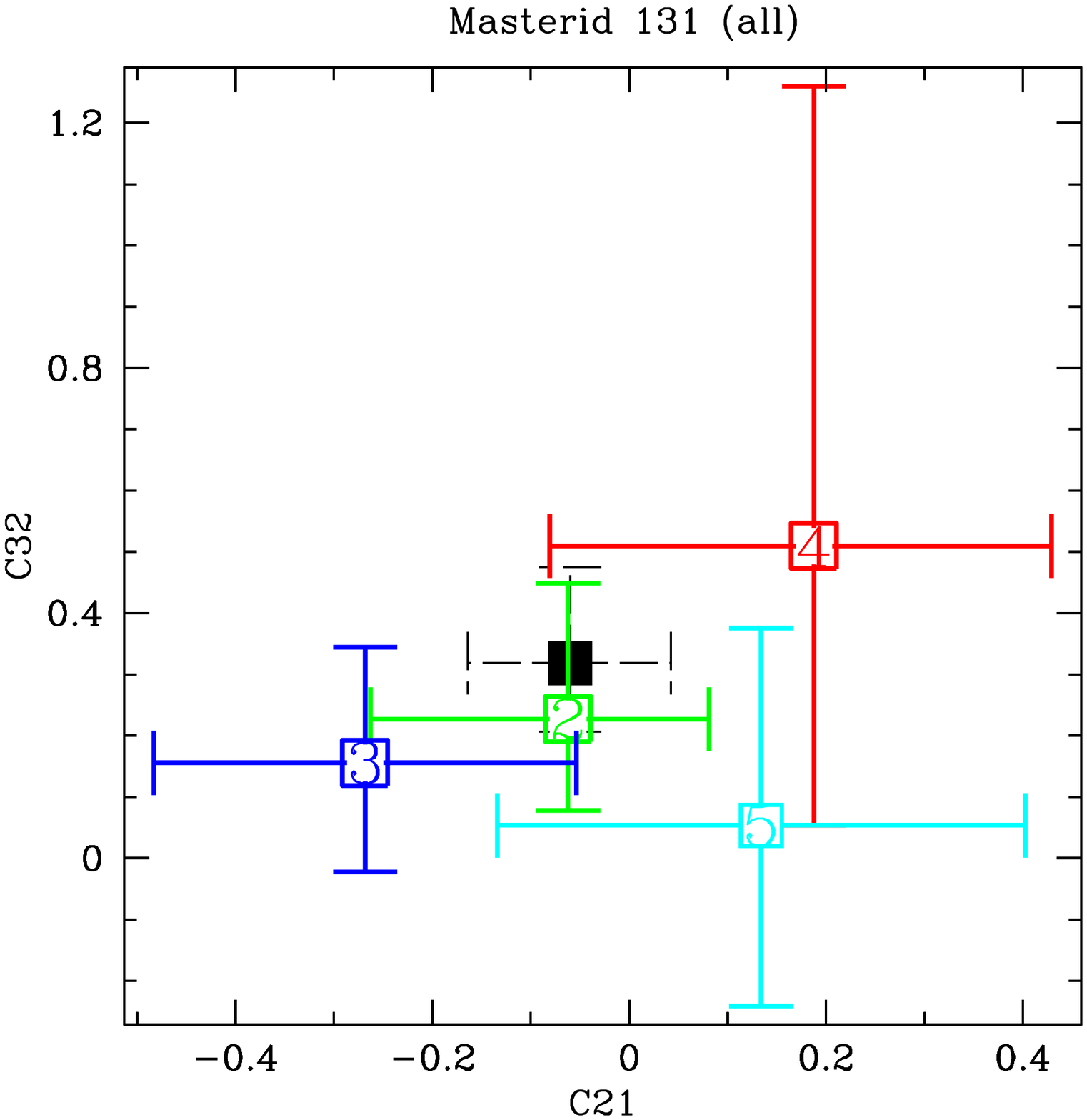}
  
  \end{minipage}
  \begin{minipage}{0.32\linewidth}
  \centering

    \includegraphics[width=\linewidth]{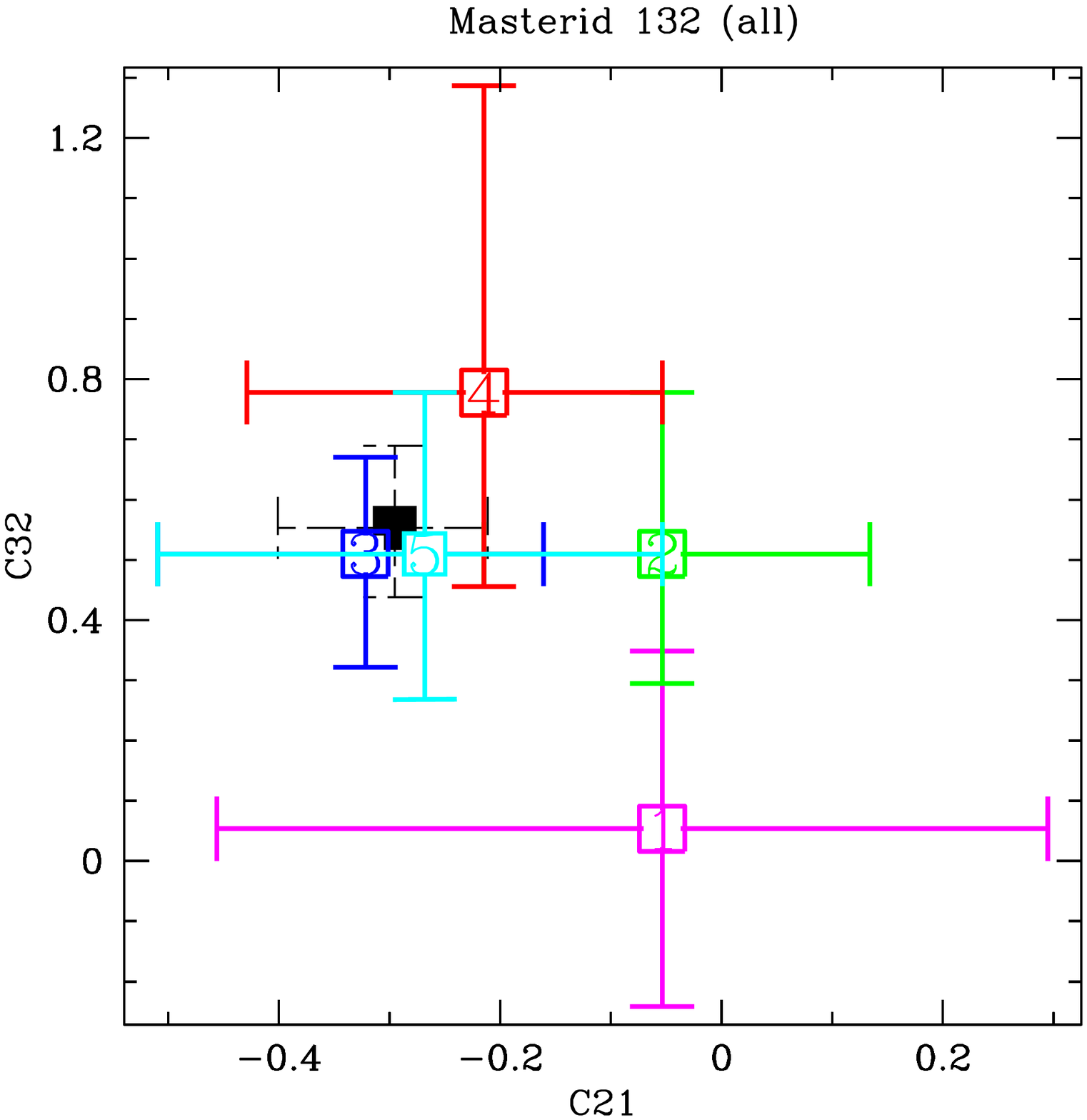}

\end{minipage}

\end{figure}

\end{document}